%% file: book.tex
\documentclass[spanningrule]{mycambridge7A} 

\input{ab}

\input{thmlist}

\usepackage{background}
\backgroundsetup{
contents={\parbox{1.6\textwidth}{%
{\small\centering
This material has been published by Cambridge University Press as
\emph{Entropy and Diversity: The Axiomatic Approach} by Tom Leinster.\\
This version is free to view and download for personal use only. 
Not for re-distribution, re-sale or use in derivative works.\\ 
Paperback:
\href{https://www.cambridge.org/9781108965576}{https://www.cambridge.org/9781108965576}. 
Hardback:
\href{https://www.cambridge.org/9781108832700}{https://www.cambridge.org/9781108832700}.\\
\copyright\ Tom Leinster 2020
\\ \ 
}}},
color=black,
opacity=1,
placement=bottom,
scale=1
}

\definecolor{myurlcolor}{rgb}{0.1,0.1,0.8}
\definecolor{mylinkcolor}{rgb}{0.05,0.05,0.4}
\usepackage{hyperref}
\hypersetup{colorlinks,
linkcolor=mylinkcolor,
citecolor=mylinkcolor,
urlcolor=mylinkcolor}

\usepackage{makeidx}
\makeindex

\renewcommand{\theenumi}{\roman{enumi}}
\renewcommand{\theenumii}{\alph{enumii}}

\usepackage[all]{xy}
\usepackage{multicol}

\begin{document}
\sloppy

\include{titlepages}

\frontmatter

\tableofcontents
\include{dependency}
\include{note}

\include{acknowledgements}

\mainmatter

\include{intro}

\include{ffe}

\include{shannon}

\include{rel}

\include{def}

\include{mns}

\include{sim}

\include{value}
\include{mm}

\include{prob}

\include{loss}

\include{p}

\include{cat}

\backmatter

\appendix

\include{proofs}

\include{condns}

\endappendix

\include{biblio}

\include{notn}
\input{see}
\printindex

\end{document}

%% file: ab.tex
\usepackage{latexsym}
\usepackage{amssymb,amsmath}
\usepackage{mathrsfs}
\usepackage{color}
\usepackage{txfonts}
\usepackage{stmaryrd}
\newcommand{\blank}{(\hspace*{1ex})}
\newcommand{\bref}[1]{(\ref{#1})}
\newcommand{\cat}[1]{\mathscr{#1}}
\newcommand{\fcat}[1]{\mathbf{#1}}
\newcommand{\ovln}[1]{\overline{#1}}
\newcommand{\twid}[1]{\widetilde{#1}}
\newcommand{\such}{:}
\newcommand{\without}{\setminus}
\newcommand{\epsln}{\varepsilon}
\newcommand{\Hom}{\mathrm{Hom}}
\newcommand{\op}{\mathrm{op}}
\newcommand{\Alg}{\fcat{Alg}}
\newcommand{\Cat}{\fcat{Cat}}
\newcommand{\Set}{\fcat{Set}}
\newcommand{\lbl}[1]{\label{#1}}
\newcommand{\oppairu}{\rightleftarrows}
\newcommand{\lwr}[1]{\mathbf{#1}}

\newcommand{\swnt}{\rotatebox{45}{$\Leftarrow$}\!}

\newcommand{\neeq}{\rotatebox{45}{$=$}}
\newcommand{\Mon}{\fcat{Mon}}
\newcommand{\demph}[1]{\textbf{\textup{#1}}}
\DeclareMathOperator{\End}{End}
\newcommand{\scat}[1]{\mathbf{#1}}
\newcommand{\iso}{\cong}
\newcommand{\nat}{\mathbb{N}}	

\newcommand{\pr}{\mathrm{pr}}
\newcommand{\blb}{{\scriptscriptstyle{\bullet}}}
\newcommand{\of}{\circ}
\newcommand{\sub}{\subseteq}
\newcommand{\cell}[4]{\put(#1,#2){\makebox(0,0)[#3]{#4}}}
\newcommand{\zmark}{\scriptstyle{\bullet}}
\newcommand{\One}{\mathbf{1}}

\newcommand{\slogan}[1]%
{\begin{center}\textit{#1}\end{center}}
\newcommand{\toby}[1]{\stackrel{#1}{\longrightarrow}}

\newcommand{\FinSet}{\fcat{FinSet}}
\newcommand{\Tp}{\fcat{Top}}
\newcommand{\Vect}{\fcat{Vect}}
\newcommand{\mg}[1]{\lvert#1\rvert}
\DeclareMathOperator{\Vol}{Vol}
\newcommand{\transp}{\mathrm{T}}
\newcommand{\ip}[2]{\langle #1, #2 \rangle}
\newcommand{\dee}{\,d}
\newcommand{\dx}{\dee x}
\newcommand{\incl}{\hookrightarrow}
\newcommand{\parpairu}{\rightrightarrows}
\newcommand{\pset}{\mathscr{P}}
\DeclareMathOperator{\im}{im}
\newcommand{\from}{\colon}
\newcommand{\C}{\mathbb{C}}
\newcommand{\R}{\mathbb{R}}
\newcommand{\Z}{\mathbb{Z}}
\newcommand{\Q}{\mathbb{Q}}
\newcommand{\vc}[1]{\mathbf{#1}} 
\newcommand{\p}{\vc{p}}
\newcommand{\w}{\vc{w}}
\newcommand{\relent}[2]{H(#1 \mathbin{\|} #2)}
\newcommand{\relEnt}[2]{H\bigl(#1 \mathbin{\|} #2\bigr)}

\newcommand{\melent}[2]{I(#1 \mathbin{\|} #2)}
\newcommand{\melEnt}[2]{I\bigl(#1 \mathbin{\|} #2\bigr)}
\newcommand{\Melent}[2]{I\Bigl(#1 \mathbin{\Big\|} #2\Bigr)}
\newcommand{\reldiv}[2]{D(#1 \mathbin{\|} #2)}
\newcommand{\relDiv}[2]{D\bigl(#1 \mathbin{\|} #2\bigr)}
\newcommand{\srelent}[3]{S_{#1}(#2 \mathbin{\|} #3)}

\newcommand{\srelEnt}[3]{S_{#1}\bigl(#2 \mathbin{\|} #3\bigr)}
\newcommand{\hrelent}[3]{H_{#1}(#2 \mathbin{\|} #3)}
\newcommand{\relenti}[2]{H^\binsym(#1 \mathbin{\|} #2)}
\newcommand{\relEnti}[2]{H^\binsym\bigl(#1 \mathbin{\|} #2\bigr)}
\newcommand{\crossent}[2]{H^\times(#1 \mathbin{\|} #2)}
\newcommand{\crossEnt}[2]{H^\times\bigl(#1 \mathbin{\|} #2\bigr)}
\newcommand{\crossenti}[2]{H^{\binsym\times}(#1 \mathbin{\|} #2)}
\newcommand{\crossdiv}[2]{D^\times(#1 \mathbin{\|} #2)}
\newcommand{\crossDiv}[2]{D^\times\bigl(#1 \mathbin{\|} #2\bigr)}
\newcommand{\condent}[2]{H(#1 \mathbin{|} #2)}
\newcommand{\condenti}[2]{H^\binsym(#1 \mathbin{|} #2)}

\newcommand{\given}{\mathbin{|}}
\DeclareMathOperator{\supp}{supp}
\newcommand{\rhow}{R}
\newcommand{\csuch}{\colon}
\newcommand{\Ex}{\mathbb{E}}
\newcommand{\mc}{\mathbin{\ast}}
\newcommand{\as}[1]{\textsf{#1}} 
\renewcommand{\emptyset}{\varnothing}
\newcommand{\binsym}{{(2)}}
\newcommand{\Hi}{H^\binsym}
\newcommand{\Si}{S^\binsym}
\newcommand{\hardref}[1]{#1}
\newcommand{\hlf}{\tfrac{1}{2}}
\newcommand{\measopd}{\Lambda}
\DeclareMathOperator{\Var}{Var}
\newcommand{\hsg}{H^{\textrm{HSG}}}
\newcommand{\FinProb}{\fcat{FinProb}}
\newcommand{\Fsym}{F_{\text{sym}}}
\newcommand{\cpd}{\sqcup}
\newcommand{\pmax}{\p_{\textup{max}}}
\newcommand{\Dmax}[1]{D_\textup{max}(#1)}
\newcommand{\Dmaxana}[1]{\Dmax{#1}}
\newcommand{\pext}[1]{\hat{#1}} 
\newcommand{\adjc}{\mathrel{\sim}}
\newcommand{\adjco}[2]{#1 \adjc #2}
\newcommand{\adjcoin}[3]{#1 \adjc #2 \text{ in } #3}
\newcommand{\gedge}{\!\!-\!\!}
\newcommand{\gcomp}[1]{\ovln{#1}}
\newcommand{\cliqueno}[1]{\omega(#1)}
\newcommand{\IP}{\fcat{IP}}
\DeclareMathOperator{\Ext}{Ext}
\newcommand{\Minkdim}{\dim_{\textup{M}}}
\newcommand{\ppi}{{\boldsymbol{\pi}}}
\newcommand{\ggamma}{{\boldsymbol{\gamma}}}
\newcommand{\ssigma}{{\boldsymbol{\sigma}}}
\newcommand{\fld}{K}

\newcommand{\finsetstyle}[1]{\mathcal{#1}}
\newcommand{\XX}{\finsetstyle{X}}
\newcommand{\YY}{\finsetstyle{Y}}
\newcommand{\Xx}{\finsetstyle{X}}
\newcommand{\Yy}{\finsetstyle{Y}}
\newcommand{\VV}{\finsetstyle{V}}
\newcommand{\WW}{\finsetstyle{W}}
\newcommand{\muc}[1]{{#1}^\star}
\newcommand{\fq}[1]{q_{#1}}
\newcommand{\Zp}{\Z/p\Z}
\newcommand{\Zps}{\Z/p^2\Z}
\newcommand{\Ztwo}{\Z/2\Z}
\newcommand{\Fmap}[2]{\langle #1 \rangle_{#2}}
\newcommand{\uqq}[1]{!_{#1}}
\newcommand{\uq}{!}
\newcommand{\splitdist}[1]{{#1}^{\#}}
\newcommand{\spltdist}[1]{#1^{\#}}
\newcommand{\Splitdist}[1]{{\bigl(#1\bigr)}^{\#}}
\newcommand{\fishsym}{{\textup{F}}}
\newlength{\tsk}
\newcommand{\lengths}{\setlength{\unitlength}{.9mm}%
\setlength{\fboxsep}{0pt}}
\newcommand{\valput}[3]{\cell{#1}{#2}{c}{#3}%
\put(#1,#2){\circle{6}}}

\newcommand{\dvd}{\mathrel{\mid}}
\newcommand{\ndvd}{\mathrel{\nmid}}
\newcommand{\introbreak}{%
\bigskip
\hfill $*$ \hfill $*$ \hfill $*$ \hfill
\bigskip}
\newcommand{\Pdot}{\scriptscriptstyle{\bullet}}
\newcommand{\descd}{\mathrel{\vartriangleleft}}
\newcommand{\CCJ}[1]{\text{CCJ}_{#1}}
\newcommand{\Hess}{\text{Hess}}
\newcommand{\ntn}[1]{\label{notn:#1}}   
\newcommand{\nref}[1]{\pageref{notn:#1}}
\newcommand{\nuse}[2]{#2, #1\\}         
\newcommand{\femph}[1]{\textbf{\textit{#1}}}
\newcommand{\dmph}[1]{\demph{#1}\index{#1}}
\newcommand{\swo}{\preceq}
\newcommand{\sigmaD}{\sigma}
\newcommand{\rhoD}{\rho}
\newcommand{\Rentp}{\mathbb{E}^{(p)}}
\newcommand{\Rat}[2]{\Delta_{#1}^{(#2)}}
\newcommand{\diagstack}[2]{*\txt{\ensuremath{#1}\\ \ensuremath{#2}}}
\renewcommand{\Pr}{\mathbb{P}}

%% file: thmlist.tex
\usepackage[amsmath,thmmarks]{ntheorem}

\newtheorem{thm}{Theorem}[section]
\newtheorem{propn}[thm]{Proposition}
\newtheorem{lemma}[thm]{Lemma}
\newtheorem{cor}[thm]{Corollary}

\newtheorem*{lemmaaff}{Lemma~\ref{lemma:aff}}
\newtheorem*{thmlf}{Theorem~\ref{thm:lf}}
\newtheorem*{lemmafff}{Lemma~\ref{lemma:fn-fin-fld}}

\theorembodyfont{\normalfont}

\newtheorem{defn}[thm]{Definition}
\newtheorem{example}[thm]{Example}
\newtheorem{examples}[thm]{Examples}
\newtheorem{remark}[thm]{Remark}
\newtheorem{remarks}[thm]{Remarks}

\theoremstyle{nonumberplain}
\theoremsymbol{\ensuremath{\square}}
\qedsymbol{\ensuremath{\square}}

\newtheorem{proof}{Proof}

\newcommand{\theoremtobeproved}{}
\newtheorem{pfoftheorem}{Proof of \theoremtobeproved}
\newenvironment{pfof}[1]
{
\renewcommand{\theoremtobeproved}{#1}
\begin{pfoftheorem}
}
{\end{pfoftheorem}}

%% file: titlepages.tex
\thispagestyle{empty}

{\centering
\vspace*{20mm}

\scalebox{1.2}{\textbf{\Huge Entropy and diversity}}\\[3mm]
{\huge The axiomatic approach}

\vspace*{14mm}

{\large T\,O\,M \, L\,E\,I\,N\,S\,T\,E\,R}\\
\textit{University \,of \,Edinburgh}

\vfill


\hspace*{-0.05\textwidth}%
\parbox{1.1\textwidth}{
\centering\large 
The printed version of this book is available in 
\href{https://www.cambridge.org/9781108965576}{paperback} or
\href{https://www.cambridge.org/9781108832700}{hardback}.\\
This arXiv version appears by agreement with Cambridge
University Press.\\
I thank CUP for offering this arrangement.%
}

}

\newpage

\thispagestyle{empty}

{\centering
\mbox{}
\vfill
\large 
\parbox{80mm}{\centering\itshape%
Much of this book was written in Catalonia
  during the turbulent years since 2017.  I dedicate it to all
  those in Catalonia who struggle for their democratic rights.}  
\vfill
\mbox{}}

%% file: dependency.tex
\chapter*{Interdependence of chapters}

A dotted line indicates that one chapter is helpful, but not essential, for
another. 

\vspace*{10mm}

\begin{center}
\includegraphics[width=1\textwidth]{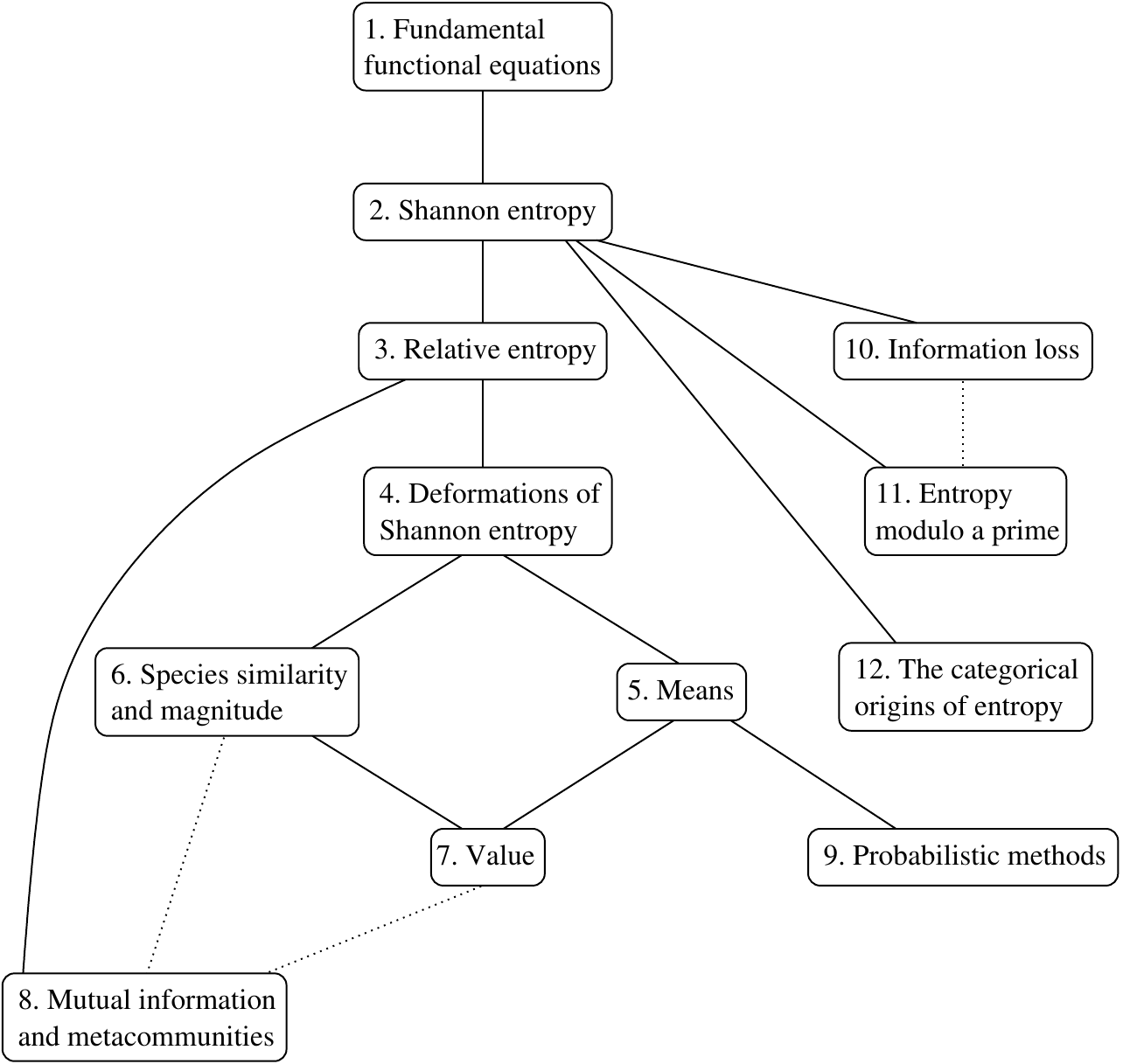}%
\end{center}

%% file: note.tex
\chapter*{Note to the reader}

This book began life as a seminar course on functional equations at the
University of Edinburgh in 2017, motivated by recent research on the
quantification of biological diversity.  The course attracted not only
mathematicians in subjects from stochastic analysis to algebraic topology,
but also participants from physics and biology.  In response, I did
everything I could to minimize the mathematical prerequisites.

I have tried here to retain the broad accessibility of the course.  At the
same time, I have not censored myself from including the many fruitful
connections with more advanced parts of mathematics.

These two opposing forces have been reconciled by confining the more
advanced material to separate chapters or sections that can easily be
omitted.  Chapter~\ref{ch:prob} requires some probability theory,
Chapter~\ref{ch:p} some abstract algebra, and Chapter~\ref{ch:cat} some
category theory, while Sections~\ref{sec:rel-misc}, \ref{sec:mag}
and~\ref{sec:mag-geom} also call on parts of geometry, analysis and
statistics.
However, the core narrative thread requires no more mathematics than a
first course in rigorous ($\epsln$-$\delta$) analysis.  Readers with this
background are promised that they are equipped to follow all the main
ideas and results.  The parts just listed, and any remarks that refer to
more specialized knowledge, can safely be omitted.

Moreover, those who regard themselves as wholly `pure' mathematicians will
find no barriers here.  Although much of this book is about the diversity
of ecological systems, no knowledge of ecology is needed.  Similarly, the
information theory that we use is introduced from the ground up.

In the middle parts of the book, many conditions on means and diversity
measures are defined: homogeneity, consistency, symmetry, etc.
Appendix~\ref{app:condns} contains a summary of this terminology for easy
reference.  There is also an index of notation.

%% file: acknowledgements.tex
\chapter*{Acknowledgements}

Many people have given me encouragement and the benefit of their insight,
wisdom, and expertise.  I am especially grateful to John Baez, Jim Borger,
Tony Carbery, Jos\'e Figueroa-O'Farrill, Tobias Fritz, Herbert Gangl, Heiko
Gimperlein, Dan Haydon, Richard Hepworth, Andr\'e Joyal, Joachim Kock,
Barbara Mable, Louise Matthews, Richard Reeve, Emily Roff, Mike Shulman,
Todd Trimble, Simon Willerton, and X\={\i}l\'{\i}ng Zh\={a}ng, as well as
Roger Astley at CUP.

My heartfelt thanks go to Christina%
\index{Cobbold, Christina} 
Cobbold and Mark%
\index{Meckes, Mark} 
Meckes, not only for many thought-provoking mathematical conversations over
the years, but also for taking the time to read drafts of this text, which
their perceptive and knowledgeable comments helped to improve.

I thank all of those involved in the original functional equations seminar
course in Edinburgh for their good humour and friendly participation.
Interactions during those seminars were formative for my understanding.  So
too were many conversations at \emph{The $n$-Category Caf\'e}, a research
blog that has been invaluable for my learning of new mathematics.

I owe a great deal to the Boyd Orr Centre for Population and Ecosystem
Health, an interdisciplinary research centre at the University of Glasgow,
of which I have been happy to have been a member since 2010.  It is a model
of the interdisciplinary spirit: welcoming, collaborative, informal, and
scientifically ambitious.  Warm thanks to all who foster that positive
atmosphere.

This work was supported at different times by an EPSRC Advanced Research
Fellowship, a BBSRC FLIP award (BB/P004210/1), and a Leverhulme Trust
Research Fellowship.  Critical research advances were made during a 2012
programme on the Mathematics of Biodiversity at the Centre
de Recerca Matem\`atica (CRM) in Barcelona, which was supported by the CRM,
a Spanish government grant, and a BBSRC International Workshop Grant; I was
also supported personally there by the Carnegie Trust for the Universities
of Scotland.  Finally, I thank the Department of Mathematics at the
Universitat Aut\`onoma de Barcelona for their hospitality during part of
the writing of this book.

Some parts of Section~\ref{sec:max} first appeared in Leinster and
Meckes~\cite{MDBB}, and are reproduced with the second author's permission.

Thanks to J\'er\^ome Euzenat, Lowenna Hooper, Ryosuke Iritani, Shashank
Ravichandran, and most especially Taichi Haruna for informing me of some
minor errors in earlier arXiv versions.

%% file: intro.tex
\chapter*{Introduction}

This book was born of research in category theory, brought to life by the
ongoing vigorous debate on how to quantify biological diversity, given
strength by information theory, and fed by the ancient field of functional
equations.  It applies the power of the axiomatic method to a biological
problem of pressing concern, but it also presents new advances in `pure'
mathematics that stand in their own right, independently of any
application.

The starting point is the connection between diversity and entropy.  We
will discover:
\begin{itemize}
\item
how Shannon entropy, originally defined for communications engineering, can
also be understood through biological diversity (Chapter~\ref{ch:shannon});

\item
how deformations of Shannon entropy express
a spectrum of viewpoints on the meaning of biodiversity
(Chapter~\ref{ch:def});

\item
how these deformations \emph{provably} provide the only reasonable
abundance-based measures of diversity (Chapter~\ref{ch:value});

\item
how to derive such results from characterization theorems for the
power means, of which we prove several, some new (Chapters~\ref{ch:mns}
and~\ref{ch:prob}). 
\end{itemize}
Complementing the classical techniques of these proofs is a large-scale
categorical programme, which has produced both new mathematics and new
measures of diversity now used in scientific applications.  For example, we
will find:
\begin{itemize}
\item
that many invariants of size from across the breadth of mathematics
(including cardinality, volume, surface area, fractional dimension, and
both topological and algebraic notions of Euler characteristic) arise
from one single invariant, defined in the wide generality of enriched
categories (Chapter~\ref{ch:sim});

\item
a way of measuring diversity that reflects not only the varying abundances
of species (as is traditional), but also the varying similarilities between
them, or, more generally, any notion of the values of the species
(Chapters~\ref{ch:sim} and~\ref{ch:value});

\item
that these diversity measures belong to the extended family of measures
of size (Chapter~\ref{ch:sim});

\item
a `best of all possible worlds': an abundance distribution on any given set
of species that maximizes diversity from an infinite number of viewpoints
simultaneously (Chapter~\ref{ch:sim});

\item
an extension of Shannon entropy from its classical context of finite sets
to distributions on a metric space or a graph (Chapter~\ref{ch:sim}),
obtained by translating the similarity-sensitive diversity measures into
the language of entropy.
\end{itemize}
Shannon entropy is a fundamental concept of information theory, but
information theory contains many riches besides.  We will mine them,
discovering:
\begin{itemize}
\item
how the concept of relative entropy not only touches subjects from
Bayesian inference to coding theory to Riemannian geometry, but also
provides a way of quantifying local diversity within a larger context
(Chapter~\ref{ch:rel});

\item
quantitative methods for identifying particularly unusual or atypical parts of
an ecological community (Chapter~\ref{ch:mm}, drawing on work of Reeve
et al.~\cite{HPD}).
\end{itemize}
The main narrative thread is modest in its mathematical prerequisites.  But
we also take advantage of some more specialized bodies of knowledge (large
deviation theory, the theory of operads, and the theory of finite fields),
establishing:
\begin{itemize}
\item
how probability theory can be used to solve functional equations
(Chapter~\ref{ch:prob}, following work of Aubrun and Nechita~\cite{AuNe});  

\item
a streamlined characterization of information loss, as a natural consequence
of categorical and operadic thinking (Chapters~\ref{ch:loss} and~\ref{ch:cat});

\item
that the concept of entropy is (provably) inescapable even in the
pure-mathematical heartlands of category theory, algebra and topology,
quite separately from its importance in scientific
applications (Chapter~\ref{ch:cat});

\item
the right definition of entropy for probability distributions whose
`probabilities' are elements of the ring $\Zp$ of integers modulo a
prime~$p$ (Chapter~\ref{ch:p}, drawing on work of
Kontsevich~\cite{KontOHL}). 
\end{itemize}
The question of how to quantify diversity is far more mathematically
profound than is generally appreciated.  This book makes the case that the
theory of diversity measurement is fertile soil for new
mathematics, just as much as the neighbouring but far more thoroughly
worked field of information theory.

\introbreak

What \emph{is} the problem of quantifying diversity?%
\index{diversity measure}%
\index{diversity}
Briefly, it is to take a biological community and extract from it a
numerical measure of its `diversity' (whatever that should mean).
This task is certainly beset with practical problems: for instance, field
ecologists recording woodland animals will probably observe the noisy, the
brightly-coloured and the gregarious more frequently than the quiet, the
camouflaged and the shy.  There are also statistical difficulties: if a
survey of one community finds $10$ different species in a sample of $50$
individuals, and a survey of another finds $18$ different species in a
sample of $100$, which is more diverse?

However, we will not be concerned with either the practical or the
statistical difficulties.  Instead, we will focus on a fundamental
conceptual problem: in an ideal world where we have complete, perfect data,
how can we quantify diversity in a meaningful and logical way?

In both the news media and the scientific literature, the most common
meaning given to the word `diversity' (or `biodiversity') is simply the
number of species present.  Certainly this is an important quantity.
However, it is not always very informative.  For instance, the number of
species of great ape\index{apes} on the planet is~$8$
(Example~\ref{eg:hill-apes}), but $99.99$\% of all great apes belong to
just one species: us.  In terms of global ecology, it is arguably more accurate
to say that there is effectively only one species of great ape.

An example illustrates the spectrum of possible interpretations of the
concept of diversity.  Consider two bird\index{birds} communities:
\[
\lbl{p:intro-birds}
\lengths
\begin{picture}(120,55)(0,2)
%
%
\cell{10}{8}{bl}{\includegraphics[height=8\unitlength]{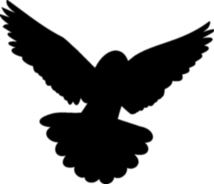}}
\cell{10}{16.2}{bl}{\includegraphics[height=8\unitlength]{birdF_std.png}}
\cell{10}{24.4}{bl}{\includegraphics[height=8\unitlength]{birdF_std.png}}
\cell{10}{32.6}{bl}{\includegraphics[height=8\unitlength]{birdF_std.png}}
\cell{10}{40.8}{bl}{\includegraphics[height=8\unitlength]{birdF_std.png}}
\cell{10}{49}{bl}{\includegraphics[height=8\unitlength]{birdF_std.png}}
\cell{20.5}{8}{bl}{\includegraphics[height=8\unitlength]{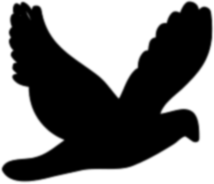}}
\cell{31}{8}{bl}{\includegraphics[height=8\unitlength]{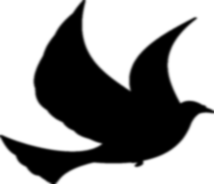}}
\cell{41.5}{8}{bl}{\includegraphics[height=8\unitlength]{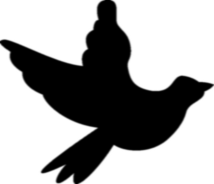}}
\cell{30}{2}{b}{A}
\cell{70}{8}{bl}{\includegraphics[height=8\unitlength]{birdF_std.png}}
\cell{70}{16.2}{bl}{\includegraphics[height=8\unitlength]{birdF_std.png}}
\cell{70}{24.4}{bl}{\includegraphics[height=8\unitlength]{birdF_std.png}}
\cell{80.5}{8}{bl}{\includegraphics[height=8\unitlength]{birdA_std.png}}
\cell{80.5}{16.2}{bl}{\includegraphics[height=8\unitlength]{birdA_std.png}}
\cell{80.5}{24.4}{bl}{\includegraphics[height=8\unitlength]{birdA_std.png}}
\cell{91}{8}{bl}{\includegraphics[height=8\unitlength]{birdG_std.png}}
\cell{91}{16.2}{bl}{\includegraphics[height=8\unitlength]{birdG_std.png}}
\cell{91}{24.4}{bl}{\includegraphics[height=8\unitlength]{birdG_std.png}}
\cell{85}{2}{b}{B}
\end{picture}
\]
In community~A, there are four species, but the majority of individuals
belong to a single dominant species.  Community~B contains the first three
species in equal abundance, but the fourth is absent.  Which community,
A~or~B, is more diverse?

One viewpoint%
\index{viewpoint!diversity@on diversity} 
is that the presence of \emph{species} is what matters.  Rare
species count for as much as common ones: every species is precious.  From
this viewpoint, community~A is more diverse, simply because more species
are present.  The abundances of species are irrelevant; presence or absence
is all that matters.

But there is an opposing viewpoint that prioritizes the balance of
\emph{communities}.  Common species are important; they are the ones that
exert the most influence on the community.  Community~A has a single very
common species, which has largely outcompeted the others, whereas
community~B has three common species, evenly balanced.  From this
viewpoint, community~B is more diverse.

These two viewpoints are the two ends of a continuum.  More precisely,
there is a continuous one-parameter family $(D_q)_{q \in [0, \infty]}$ of
diversity measures encoding this spectrum of viewpoints.  Low values of $q$
attach high importance to rare species; for example, $D_0$ measures
community~A as more diverse than community~B.  When $q$ is high, $D_q$ is
most strongly influenced by the balance of more common species; thus,
$D_\infty$ judges~B to be more diverse.  No single viewpoint is right or
wrong.  Different scientists adopt different viewpoints (that is, different
values of $q$) for different purposes, as the literature amply attests
(Examples~\ref{egs:hill}).

Long ago, it was realized that the concept of diversity is closely related
to the concept of entropy.  Entropy appears in dozens of guises across
dozens of branches of science, of which thermodynamics is probably the most
famous.  (The introduction to Chapter~\ref{ch:shannon} gives a long but
highly incomplete list.)  The most simple incarnation is Shannon entropy,
which is a real number associated with any probability distribution on a
finite set.  It is, in fact, the logarithm of the diversity measure $D_1$.
Most often, Shannon entropy is explained and understood through the theory
of coding; indeed, we provide such an explanation here.  But the diversity
interpretation provides a new perspective.

For example, the diversity measures $D_q$, known in ecology as the
Hill%
\index{Hill number}
numbers, are the exponentials of what information theorists know as the
R\'enyi%
\index{Renyi entropy@R\'enyi entropy} 
entropies.  From the very beginning of information theory, an
important role has been played by characterization theorems: results
stating that any measure (of information, say) satisfying a list of
desirable properties must be of a particular form (a scalar
multiple of Shannon entropy, say).  But what counts as a desirable
property depends on one's perspective.  We will prove that the Hill numbers
$D_q$ are, in a precise sense, the only measures of diversity with
certain natural properties (Theorem~\ref{thm:total-hill}).  This theorem
translates into a new characterization of the R\'enyi entropies, but it is
not one that necessarily would have been thought of from a purely
information-theoretic perspective.  

However, something is missing.  In the real world, diversity is understood
as involving not only the number and abundances of the species, but also
how \emph{different} they are.  (For example, this affects
conservation\index{conservation} policy; see the OECD quotation on
p.~\pageref{p:oecd-quote}.)  We describe the remedy in
Chapter~\ref{ch:sim}, defining a family of diversity measures that take
account of the varying similarity%
\index{similarity!species@of species} 
between species, while still incorporating the
spectrum of viewpoints discussed above.  This definition unifies into one
family a large number of the diversity measures proposed and used in the
ecological and genetics literature.

This family of diversity measures first appeared in a paper in
\emph{Ecology}~\cite{MDISS}, but it can also be understood and motivated
from a purely mathematical perspective.  The classical R\'enyi entropies
are a family of real numbers assigned to any probability distribution on a
finite \emph{set}.  By factoring in the differences or distances
between points (species), we extend this to a family of real numbers
assigned to any probability distribution on a finite
\emph{metric\index{metric!space} space}.
In the extreme case where $d(x, y) = \infty$ for all distinct points $x$
and $y$, we recover the R\'enyi entropies.  In this way, the
similarity-sensitive diversity measures extend the definition of R\'enyi
entropy from sets to metric spaces.

Different values of the viewpoint parameter $q \in [0, \infty]$ produce
different judgements on which of two distributions is the more diverse.
But it turns out that for any metric space (or in biological terms, any
set of species), there is a single distribution that maximizes%
\lbl{p:max-intro}
diversity from all viewpoints simultaneously.  For a generic finite metric
space, this maximizing distribution is unique.  Thus, almost every finite
metric space carries a canonical probability distribution (not usually
uniform).  The maximum%
\index{maximum diversity} 
diversity itself is also independent of $q$, and is therefore a numerical
invariant of metric spaces.  This invariant has geometric significance in
its own right (Section~\ref{sec:mag-geom}).

We go further.  One might wish to evaluate an ecological community in a way
that takes into account some notion of the values\index{value} of the
species (such as 
phylogenetic distinctiveness).  Again, there is a sensible family of
measures that does this job, extending not only the similarity-sensitive
diversity measures just described, but also further measures already
existing in the ecological literature.  The word `sensible' can be made
precise: as soon as we subject an abstract measure of the value of a
community to some basic logical requirements, it is forced to belong to a
certain one-parameter family $(\sigma_q)$ (Theorem~\ref{thm:val-char}),
which are essentially the R\'enyi \emph{relative} entropies.

Information theory also helps us to analyse the diversity of
metacommunities,\index{metacommunity} that is, ecological communities made
up of a number of smaller communities such as geographical regions.  The
established notions of relative entropy, conditional entropy and mutual
information provide meaningful measures of the structure of a metacommunity
(Chapter~\ref{ch:mm}).  But we will do more than simply translate
information theory into ecological language.  For example, the new
characterization of the R\'enyi entropies mentioned above is a byproduct of
the characterization theorem for measures of ecological value.  In this
way, the theory of diversity gives back to information theory.

\introbreak

The scientific importance of biological diversity goes far beyond the
obvious setting of conservation\index{conservation} of animals and plants.
Certainly such conservation efforts are important, and the need for
meaningful measures of diversity is well-appreciated in that context.  For
example, Vane-Wright%
\index{Vane-Wright, Richard} 
et al.~\cite{VWHW} wrote thirty years ago of the `agony of choice'
in conservation of flora and fauna, and emphasized how crucial it is to use
the right diversity measures.

But most life is microscopic.  Nee~\cite{NeeMTM}%
\index{Nee, Sean} 
argued in 2004 that
\begin{quote}
{}[w]e are still at the very beginning of a golden age of biodiversity
discovery, driven largely by the advances in molecular biology and a new
open-mindedness about where life might be found,%
\index{microbial systems}
\end{quote}
and that
\begin{quote}
all of the marvels in biodiversity's new bestiary are invisible.
\end{quote}
Even excluding exotic new discoveries of microscopic life, two recent lines
of research illustrate important uses of diversity measures at the
microbial level.

First, the extensive use of antimicrobial drugs on animals
unfortunate enough to be born into the modern meat industry is commonly
held to be a cause of antimicrobial resistance in pathogens affecting
humans.  However, a 2012 study of Mather%
\index{Mather, Alison} 
et al.~\cite{MMMR} suggests that the causality may be more complex.  By
analysing the diversity of antimicrobial%
\index{antimicrobial resistance}%
\index{microbial systems}
resistance in \emph{Salmonella} taken from animal populations on the one
hand, and from human populations on the other, the authors concluded that
the animal population is `unlikely to be the major source of resistance'
for humans, and that `current policy emphasis on restricting antimicrobial
use in domestic animals may be overly simplistic'.  The diversity measures
used in this analysis were the Hill numbers $D_q$ mentioned above and
central to this book.

Second, the increasing problem of obesity in humans has prompted research
into causes and treatments, and there is evidence of a negative
correlation between obesity and diversity of the gut%
\index{gut microbiome} 
microbiome (Turnbaugh et al.~\cite{THYC,TQFM}).  Almost all traditional
measures of diversity rely on a division of organisms into species or other
taxonomic groups, but in this case, only a fraction of the microbial
species concerned have been isolated and classified taxonomically.
Researchers in this field therefore use DNA sequence data, applying
sophisticated but somewhat arbitrary clustering algorithms to create
artificial species-like groups (`operational taxonomic units').  On the
other hand, the similarity-sensitive diversity measures mentioned above and
introduced in Chapter~\ref{ch:sim} can be applied directly to the sequence
data, bypassing the clustering step and producing a measure of genetic
diversity.  A test case was carried out in Leinster and
Cobbold~\cite{MDISS} (Example~4), with results that supported the
conclusions of Turnbaugh et al.

Despite the wide variety of uses of diversity measures in biology, none of
the mathematics presented in this text is intrinsically biological.%
\index{diversity measure!applications of}
Indeed, the mathematics of diversity was being developed as early as 1912
by the economist\index{economics} Corrado
Gini~\cite{Gini}%
\index{Gini, Corrado} 
(best known for the Gini
coefficient of disparity of wealth), and by the statistician Udny
Yule%
\index{Yule, G. Udny} 
in the 1940s for the analysis of lexical%
\index{lexical diversity}%
\index{diversity!lexical}
diversity in literature~\cite{Yule}.  Some of the diversity measures most
common in ecology have recently been used to analyse the ethnic and
sociological diversity of judges (Barton and Moran~\cite{BaMo}), and the
similarity-sensitive diversity measures that are the subject of
Chapter~\ref{ch:sim} have been used not only in multiple ecological
contexts (as listed after Example~\ref{eg:devries}), but also in
non-biological applications such as computer%
\index{computer network security} 
network security (Wang et al.~\cite{WZJS}).%
\index{diversity measure!applications of}

In mathematical terms, simple diversity measures such as the Hill numbers
are invariants of a probability distribution on a finite set.  The
similarity-sensitive diversity measures are defined for any probability
distribution on a finite set with an assigned degree of similarity between
each pair of points.  (This includes any finite metric space or graph.)
The value measures are defined for any finite set equipped with a
probability distribution and an assignment of a nonnegative value to each
element.  The metacommunity measures are defined for any probability
distribution on the cartesian product of a pair of finite sets.  Much of
this text is written using ecological terminology, but the mathematics is
entirely general.\lbl{p:general}

\introbreak

This work grew out of a general category-theoretic%
\index{category theory} 
study of size.\index{size} In many parts of mathematics, there is a
canonical notion of the size of the objects of study: sets have
cardinality, vector spaces have dimension, subsets of Euclidean space have
volume, topological spaces have Euler%
\index{Euler characteristic}
characteristic, and so on.  Typically, such measures of size satisfy
analogues of the elementary inclusion-exclusion and multiplicativity
formulas for counting finite sets:
\begin{align*}
\mg{X \cup Y}   &= \mg{X} + \mg{Y} - \mg{X \cap Y},     \\
\mg{X \times Y} &= \mg{X} \cdot \mg{Y}.
\end{align*}
(The interpretation of Euler characteristic as the topological analogue of
cardinality is not as well known as it should be; this is an insight of
Schanuel%
\index{Schanuel, Stephen} 
on which we elaborate in Section~\ref{sec:mag}.)  From a
categorical perspective, it is natural to seek a single invariant unifying
all of these measures of size.

Some unification is achieved by defining a notion of size for categories
themselves, called \emph{magnitude}\index{magnitude} or Euler
characteristic.  (Finiteness hypotheses are required, but will not be
mentioned in this overview.)  This definition already brings together
several established invariants of size~\cite{ECC}: cardinality of
sets, and the various notions of Euler characteristic for partially ordered
sets, topological spaces, and even orbifolds (whose Euler characteristics
are in general not integers).  The theory of magnitude of categories is
closely related to the theory of M\"obius--Rota inversion for partially
ordered sets~\cite{RotaFCT,NMI}.

But the decisive, unifying step is the generalization of the definition of
magnitude from categories to the wider class of \emph{enriched}%
\index{enriched category} 
categories~\cite{MMS}, which includes not only categories
themselves, but also metric spaces, graphs, and the additive categories
that are a staple of homological algebra.

The definition of the magnitude of an enriched category unifies still more
established invariants of size.  For example, in the representation theory
of associative algebras, one frequently considers the indecomposable
projective modules, which form an additive category.  The magnitude of that
additive category turns out to be the Euler form of a certain canonical
module, defined as an alternating sum of dimensions of $\Ext$ groups
(equation~\eqref{eq:ip}).  Magnitude for enriched categories can also be
realized as the Euler characteristic of a certain Hochschild-like homology%
\index{magnitude!homology}
theory of enriched categories, in the same sense that the Jones polynomial
for knots is the Euler characteristic of Khovanov homology~\cite{Khov}.
This was established in recent work led by Shulman~\cite{MHECMS}, building
on the case of magnitude homology for graphs previously developed by
Hepworth and Willerton~\cite{HeWi}. 

Since any metric\index{metric!space} space can be regarded as an enriched
category, the general definition of the magnitude of an enriched category
gives, in particular, a definition of the magnitude $\mg{X} \in \R$ of a
metric space $X$.  Unlike the other special cases just mentioned, this
invariant is essentially new.

Recent, increasingly sophisticated, work in analysis has connected
magnitude with classical invariants of geometric measure.  For example, for
a compact subset $X \sub \R^n$ satisfying certain regularity conditions, if
one is given the magnitude of all of the rescalings $tX$ of $X$ (for $t >
0$), then one can recover:
\begin{itemize}
\item 
the Minkowski\index{dimension} dimension of $X$ (one of the principal
notions of fractional dimension), a result proved by Meckes using results
in potential theory (Theorem~\ref{thm:mink});

\item
the volume\index{volume} of $X$, a result proved by Barcel\'o and Carbery
using PDE methods (Theorem~\ref{thm:bc});

\item
the surface%
\index{surface area} 
area of $X$, a result proved by Gimperlein and Goffeng using
global analysis (or more specifically, tools for computing heat trace
asymptotics; Theorem~\ref{thm:gg}).   
\end{itemize}
Gimperlein and Goffeng also proved an asymptotic
inclusion-exclusion%
\index{inclusion-exclusion principle}
principle:
\[
\mg{t(X \cup Y)} + \mg{t(X \cap Y)} - \mg{tX} - \mg{tY}
\to 0
\]
as $t \to \infty$, for sufficiently regular $X, Y \sub \R^n$
(Section~\ref{sec:mag-geom}).  This is another manifestation of the
cardinality-like nature of magnitude.

We have seen that every finite metric space $X$ has an unambiguous maximum
diversity $\Dmax{X} \in \R$, defined in terms of the similarity-sensitive
diversity measures (p.~\pageref{p:max-intro}).  We have also seen that $X$
has a magnitude $\mg{X} \in \R$.  These two real numbers are not in general
equal (ultimately because probabilities or species abundances are forbidden
to be negative),%
\index{negative!probability}
but they are closely related.  Indeed, $\Dmax{X}$ is always equal to the
magnitude of some \emph{subspace} of $X$, and in important families of
cases is equal to the magnitude of $X$ itself.  So, magnitude is closely
related to maximum diversity.  Indeed, this relationship was exploited by
Meckes%
\index{Meckes, Mark} 
to prove the result on Minkowski dimension.

There is a historical surprise.  Although this author arrived at the
definition of the magnitude of a metric space by the route of enriched
category theory, it had already arisen in earlier work on the
quantification of biodiversity.  In 1994, the environmental scientists
Andrew Solow%
\index{Solow, Andrew} 
and Stephen\label{p:sp-mag} Polasky%
\index{Polasky, Stephen} 
carried out a probabilistic analysis of the benefits of high biodiversity
(\cite{SoPo}, Section~4), and isolated a particular quantity that they
called the `effective%
\index{effective number!species@of species}
number of species'.  They did not
investigate it mathematically, merely remarking mildly that it `has
some appealing properties'.  It is exactly our magnitude.

\introbreak

Ecologists began to propose quantitative definitions of biological
diversity in the mid-twentieth century~\cite{SimpMD,WhitVSM}, setting in
motion more than sixty years of heated debate, with dozens of further proposed
diversity measures, hundreds of scholarly papers, at least one book devoted
to the subject~\cite{Magu}, and consequently, for some, despair (expressed
as early as 1971 in a famously-titled paper of Hurlbert~\cite{Hurl}).%
\index{Hurlbert, Stuart}  
Meanwhile, parallel debates were taking place in genetics and other
disciplines.

The connections between diversity measurement on the one hand, and
information theory and category theory on the other, are fruitful for both
mathematics and biology.  But any measure of biological diversity must be
justifiable in purely biological terms, rather than by borrowing authority
from information theory, category theory, or any other field.  The
ecologist E.~C.~Pielou%
\index{Pielou, Evelyn Chrystalla} 
warned against attaching ecological significance to diversity measures for
anything other than ecological reasons:
\begin{quote}
It should not be (but it is) necessary to emphasize that the object of
calculating indices of diversity is to solve, not to create, problems.  The
indices are merely numbers, useful in some circumstances but not in all.
[\ldots] Indices should be calculated for the light (not the shadow) they
cast on genuine ecological problems.
\end{quote}
(\cite{PielME}, p.~293).

In a series of incisive papers beginning in 2006, the
conservationist and botanist Lou Jost%
\index{Jost, Lou}
insisted that whatever diversity measures one uses, they must exhibit
\emph{logical behaviour}%
\index{diversity measure!logical behaviour of} 
\cite{JostED,JostPDI,JostGST,JostMBD}.  For
example, Shannon entropy is commonly used as a diversity measure by
practising ecologists, and it does behave logically if one is only using it
to ask whether one community is more or less diverse than another.  But as Jost
observed, any attempt to reason about percentage changes in diversity using
Shannon entropy runs into logical absurdities: Examples~\ref{eg:plague}
and~\ref{eg:oil} describe the plague that exterminates $90\%$ of species
but only causes a $17\%$ drop in `diversity', and the oil drilling that
simultaneously destroys \emph{and} preserves $83\%$ of the `diversity' of
an ecosystem.  It is, in fact, the \emph{exponential} of Shannon entropy
that should be used for this purpose.

In this sense, origin stories are irrelevant.  Inventing new diversity
measures is easy, and it is nearly as easy to tell a story of how a new
measure fits with some intuitive idea of diversity, or to justify it in
terms of its importance in some related discipline.  But if a measure does
not pass basic logical tests (as in Section~\ref{sec:prop-hill}), it is
useless or worse.

Jost noted that all of the Hill numbers $D_q$ do behave logically.  Again,
we go further: Theorem~\ref{thm:total-hill} states that the Hill numbers
are in fact the \emph{only} measures of diversity satisfying certain
logically fundamental properties.  (At least, this is so for the simple
model of a community in terms of species abundances only.)  This is the
ideal of the axiomatic approach: to prove results stating that if one
wishes to have a measure with such-and-such properties, then it can only be
one of \emph{these} measures.

Mathematically, such results belong to the field of functional equations.  We
review a small corner of this vast and classical theory, beginning with the
fact that the only measurable functions $f \from \R \to \R$ satisfying the
Cauchy functional equation $f(x + y) = f(x) + f(y)$ are the linear mappings
$x \mapsto cx$.  Building on classical results, we obtain new axiomatic
characterizations of a variety of measures of diversity, entropy and value.
We also explain a new method, pioneered by Aubrun%
\index{Aubrun, Guillaume}
and Nechita%
\index{Nechita, Ion} 
in
2011~\cite{AuNe}, for solving functional equations by harnessing the power
of probability theory.  This produces new characterizations of the $\ell^p$
norms and the power means.

Characterization theorems for the power%
\index{power mean} 
means are, in fact, the engine of
this book (Chapter~\ref{ch:mns}).  By definition, the power mean of order
$t$ of real numbers $x_1, \ldots, x_n$, weighted by a probability
distribution $(p_1, \ldots, p_n)$, is
\[
M_t(\vc{p}, \vc{x}) 
=
\Biggl( \sum_{i = 1}^n p_i x_i^t \Biggr)^{1/t}.
\]
The power means $(M_t)_{t \in \R}$ form a one-parameter family of
operations, and the central place that they occupy in this text is explained
by their relationship with several other important one-parameter families:
the Hill numbers, the R\'enyi entropies, the $q$-logarithms, the
$q$-logarithmic entropies (also known as Tsallis entropies), the value
measures of Chapter~\ref{ch:value}, and the $\ell^p$-norms.  We will prove
characterization theorems for all of these families, in each case finding a
short list of properties that determines them uniquely.

\introbreak

Much of this text can be described as `mathematical%
\index{mathematical anthropology} 
anthropology'.  The mathematical anthropologist begins by observing that
some group of scientists attaches great importance to a particular object
or concept: homotopy theorists talk a lot about simplicial sets, harmonic
analysts constantly use the Fourier transform, ecologists often count the
number of species present in a community, and so on.  The next step is to
ask: why do they attach such importance to that particular thing, not
something slightly different?  Is it the \emph{only} object that enjoys the
useful properties that it enjoys?  If not, why do they use the object they
use, and not some other object with those properties?  And if it \emph{is}
the only object with those properties, can we prove it?  For
example, 2008 work of Alesker, Artstein-Avidan and Milman~\cite{AAAM}
proved that the Fourier transform is, in fact, the only transform that enjoys
its familiar properties.

This is the animating spirit of the field of functional equations.
But there is another field that has been
enormously successful in mathematical anthropology: category%
\index{category theory} 
theory.  There, objects of mathematical interest are typically
characterized by universal%
\index{universal property} 
properties.  For instance, the tensor product $M \otimes N$ of modules $M$
and $N$ is the universal module equipped with a bilinear map $M \times N
\to M \otimes N$; the Hilbert space completion $\hat{X}$ of an inner
product space $X$ is the universal Hilbert space equipped with an isometry
$X \to \hat{X}$; the real interval $[0, 1]$ is the universal bipointed
topological space equipped with a map $[0, 1] \to [0, 1] \vee [0, 1]$
(Theorem~2.2 of Leinster~\cite{GTSS} and Theorem~2.5 of
Leinster~\cite{GSSO}, building on results of Freyd~\cite{FreARA}).  Any
universal property involves uniqueness at two levels: the literal
uniqueness of a connecting \emph{map}, and the fact that the universal
property characterizes the \emph{object} possessing it uniquely up to
isomorphism.  Thus, category theory is a potent tool for proving
characterization theorems.

We demonstrate this with a categorically-motivated characterization theorem
for entropy (Baez, Fritz and Leinster~\cite{CETIL}).  Briefly put, the
probability distributions on finite sets form an operad\index{operad}, we
construct a certain universal category acted on by that operad, and this
leads naturally to the concept of Shannon entropy.  The categorical
approach amounts to a shift of emphasis from the entropy of a probability
space (an object) to the amount of information lost by a deterministic
process (a map).

The moral of this result is that entropy is not just something for applied
scientists.  It emerges inevitably from a general categorical machine,
given as its inputs nothing more obscure than the real line and the
standard topological simplices.  In other words, even in algebra and
topology, entropy is inescapable.

To demonstrate the strength of the axiomatic approach, we finish by
applying it to an entity of purely mathematical interest: entropy modulo a
prime number.  The topic was first introduced as a curiosity by
Kontsevich,%
\index{Kontsevich, Maxim} 
as a byproduct of work on polylogarithms~\cite{KontOHL}.  Just as any real
probability distribution $\ppi = (\pi_1, \ldots, \pi_n)$ has a Shannon
entropy $H_\R(\ppi) \in \R$, one can define, for any
prime%
\index{entropy!modulo a prime} 
$p$ and `probabilities' $\pi_1, \ldots, \pi_n \in \Zp$, a kind of entropy
$H_p(\ppi) \in \Zp$.  The functional forms are quite different:
\[
\begin{array}{rcll}
H_\R(\pi_1, \ldots, \pi_n)      &
= &
\displaystyle
-\sum_{1 \leq i \leq n} \pi_i \log \pi_i
&\in \R, \\[1.5ex]
H_p(\pi_1, \ldots, \pi_n)       &
=&
\displaystyle
-\sum_{\substack{0 \leq r_1, \ldots, r_n < p\\ r_1 + \cdots + r_n = p}}
\frac{\pi_1^{r_1} \cdots \pi_n^{r_n}}{r_1! \cdots r_n!}
&\in \Zp.
\end{array}
\]
One would probably not guess that the second formula is the correct mod~$p$
analogue of the first.  However, the definition is fully justified by a
characterization theorem strictly analogous to the one that characterizes
real Shannon entropy.  And from the categorical perspective, there is a
strictly analogous characterization of information loss mod~$p$.  In short,
the apparatus developed for the real field can be successfully applied to
the field of integers modulo a prime.

\introbreak\pagebreak

Finally, this book aims to challenge outdated conceptions of what
applied%
\index{applied mathematics} 
mathematics can look like.  Too often, `applied mathematics' is
subconsciously understood to mean `methods of analysis applied to problems
of physics'.  (Or, worse, `applied' is taken to be a euphemism for
`unrigorous'.)  Those applications are certainly enormously important.
However, this excessively narrow interpretation ignores the glittering
array of applications of other parts of mathematics to other kinds of
problem.  It is mere historical accident that a researcher using PDEs in
the study of fluids is usually called an applied mathematician, but one
applying category theory to the design of programming languages is not.

Mathematicians are coming to appreciate that applications of their subject
to biology are enormously fruitful and, with the revolution in the
availability of genetic data, will only grow.  Mackey and Maini asked and
answered the question `What has mathematics done for
biology?'~\cite{MaMa},%
\index{Mackey, Michael}%
\index{Maini, Philip}
quoting the evolutionary biologist and slime mould specialist
John%
\index{Bonner, John} 
Bonner on the `rocking back and forth between the reality of experimental
facts and the dream world of hypotheses'.  They reviewed some major
contributions, including striking success stories in ecology, epidemiology,
developmental biology, physiology, and neuro-oncology.  But still, most of
the work cited there (and most of mathematical biology as a whole) uses
parts of mathematics traditionally thought of as `applied', such as
differential equations, dynamical systems, and stochastic analysis.

The reality is that many parts of mathematics conventionally called `pure'
are now being successfully applied in diverse contexts, both biological and
otherwise.  Knot theory has solved longstanding problems in genetic
recombination (Buck and Flapan~\cite{BuFlPKC,BuFlTCK}).  Group theory has
illuminated virus structure (Twarock, Valiunas and Zappa~\cite{TVZ}).
Topological data analysis, founded on the theory of persistent%
\index{persistent homology} 
homology and
calling on the power of algebraic topology, succeeded in identifying a
hitherto unknown subtype of breast cancer with a 100\% survival rate
(Nicolau, Levine and Carlsson~\cite{NLC}; see Lesnick~\cite{Lesn} for an
expository account).  Order theory, topos theory and classical logic have
all been employed in the quest for improved ways of specifying, modelling
and designing concurrent systems (Nygaard and Winskel~\cite{NyWi}; Joyal,
Nielsen and Winskel~\cite{JNW}; Hennessy and Milner~\cite{HeMi}).  And,
famously, number theory is used to both provide and undermine security of
communications on the internet (Hales~\cite{HaleNBD}).  All of these are
real applications of mathematics.  None is `applied mathematics' as
traditionally construed.

But applications are not the only product of applied mathematics.  It also
\emph{nourishes} the core of mathematics, providing new questions, answers,
and perspectives.  Mathematics applied to physics has done this from
Archimedes to Newton to Witten.  Reed~\cite{Reed} lists dozens of ways in
which mathematics applied to biology is doing it now.  The developments
surveyed in this book provide further evidence that a body of mathematics
can simultaneously be entirely rigorous, be applied effectively to another
branch of science, use parts of mathematics that do not fit the narrow
stereotype of `applied mathematics', and produce new results that are
significant and satisfying from a purely mathematical aesthetic.

%% file: ffe.tex
\chapter{Fundamental functional equations}
\lbl{ch:ffe}

Throughout this book, we will make contact with the venerable
subject of functional equations.  A functional%
\index{functional equation}
equation is an equation in an unknown function satisfied at all values of
its arguments; or more generally, it is an equation relating several
functions to each other in this way.

To set the scene, we give some brief indicative examples.  Viewing
sequences as functions on the set of positive integers, the
Fibonacci%
\index{Fibonacci sequence} 
sequence $(F_n)_{n \geq 1}$ satisfies the functional equation
\[
F_{n + 2} = F_n + F_{n + 1} 
\]
($n \geq 1$).  Together with the boundary conditions $F_1 = F_2 = 1$, this
functional equation uniquely characterizes the sequence.  But more
typically, one is concerned with functions of \emph{continuous} variables.
For instance, one might notice that the function
\[
\begin{array}{cccc}
f\from  &\R \cup \{\infty\}     &\to            &\R\cup\{\infty\}\\[1ex]
        &x                      &\mapsto        &
\displaystyle\frac{1}{1 - x}
\end{array}
\]
satisfies the functional equation
\begin{equation}
\lbl{eq:fe-triple}
f(f(f(x))) = x
\end{equation}
($x \in \R \cup \{\infty\}$).  The natural question, then, is whether $f$
is the \emph{only} function satisfying equation~\eqref{eq:fe-triple} for
all $x$.  In this case, it is not.  (This can be shown by constructing an
explicit counterexample or via the theory of M\"obius
transformations.)  So, it is then natural to seek the whole set of
solutions $f$, perhaps restricting the search to just those functions that
are continuous, differentiable, etc.

A more sophisticated example is the functional equation
%
\[
\zeta(1 - s) 
=
\frac{2^{1 - s}}{\pi^s} 
\cos\biggl(\frac{\pi s}{2}\biggr)\,\Gamma(s)\,\zeta(s)
\]
%
($s \in \C$) satisfied by the Riemann%
\index{Riemann, Bernhard!zeta function} 
zeta function $\zeta$ (Theorem~12.7 of Apostol~\cite{AposIAN}, for
instance).  Here $\Gamma$ is Euler's gamma function.  This functional
equation, proved by Riemann himself, is a fundamental property of the zeta
function.

In this chapter, we solve three classical, fundamental, functional
equations.  The first is Cauchy's equation on a function $f \from \R \to
\R$:
\[
f(x + y) = f(x) + f(y)
\]
($x, y \in \R$) (Section~\ref{sec:cauchy}).  Once we have solved this, we
will easily be able to deduce the solutions of related equations such as
\begin{equation}
\lbl{eq:fe-cauchy-log-intro}
f(xy) = f(x) + f(y)
\end{equation}
($x, y \in (0, \infty)$).  

The second is the functional equation
\[
f(mn) = f(m) + f(n)
\]
($m, n \geq 1$) on a \emph{sequence} $(f(n))_{n \geq 1}$.  Despite the
resemblance to equation~\eqref{eq:fe-cauchy-log-intro}, the shift from
continuous to discrete makes it necessary to develop quite different
techniques (Section~\ref{sec:log-seqs}).

Third and finally, we solve the functional equation
\[
f(xy) = f(x) + g(x)f(y)
\]
in two unknown functions $f, g\from (0, \infty) \to \R$.  The nontrivial,
measurable solutions $f$ turn out to be the constant multiples of the
so-called $q$-logarithms (Section~\ref{sec:q-log}), a one-parameter family
of functions of which the ordinary logarithm is just the best-known member.

\section{Cauchy's equation}
\lbl{sec:cauchy}

A function $f \from \R \to \R$ is \demph{additive}%
\index{additive function} 
if
\begin{equation}
\lbl{eq:additivity}
f(x + y) = f(x) + f(y)
\end{equation}
for all $x, y \in \R$.  This is 
\demph{Cauchy's%
\index{Cauchy, Augustin!functional equation} 
functional equation}, some of whose long history is recounted in
Section~2.1 of Acz\'el~\cite{AczeLFE}.  Let us say that $f$ is
\demph{linear}%
\index{linear function} 
if there exists $c \in \R$ such that
\[
f(x) = cx
\]
for all $x \in \R$.  Putting $x = 1$ shows that if such a constant $c$
exists then it must be equal to $f(1)$.

Evidently any linear function is additive.  The question is to what extent
the converse holds.  If we are willing to assume that $f$ is differentiable
then the converse is very easy:

\begin{propn}
\lbl{propn:add-diff}
Every differentiable additive function $\R \to \R$ is linear.
\end{propn}

\begin{proof}
Let $f \from \R \to \R$ be a differentiable additive function.
Differentiating equation~\eqref{eq:additivity} with respect to $y$ gives
\[
f'(x + y) = f'(y)
\]
for all $x, y \in \R$.  Taking $y = 0$ then shows that $f'$ is constant.
Hence there are constants $c, d \in \R$ such that $f(x) = cx + d$ for all
$x \in \R$.  Substituting this expression back into
equation~\eqref{eq:additivity} gives $d = 0$.
\end{proof}

However, differentiability is a stronger condition than we will want to
assume for our later purposes.  It is, in fact, unnecessarily strong.  In
the rest of this section, we prove that additivity implies linearity under
a succession of ever-weaker regularity conditions, starting with continuity
and finishing with mere measurability.

We begin with a lemma that needs no regularity conditions at all.

\begin{lemma}
\lbl{lemma:add-rat}
Let $f \from \R \to \R$ be an additive function.  Then $f(qx) = qf(x)$ for
all $q \in \Q$ and $x \in \R$.  
\end{lemma}

\begin{proof}
First, $f(0 + 0) = f(0) + f(0)$, so $f(0) = 0$.  Then, for all $x \in \R$,
\[
0 = f(0) = f(-x + x) = f(-x) + f(x),
\]
so $f(-x) = -f(x)$.

Let $x \in \R$.  By induction,
\begin{equation}
\lbl{eq:int-scalar}
f(nx) = nf(x)
\end{equation}
for all integers $n > 0$, and we have just shown that
equation~\eqref{eq:int-scalar} also holds when $n = 0$.  Moreover, when $n
< 0$,
\[
f(nx) 
=
f\bigl(-(-n)x\bigr)
=
-f\bigl((-n)x\bigr)
=
-(-n)f(x)
=
nf(x),
\]
using equation~\eqref{eq:int-scalar} for positive integers.
Hence~\eqref{eq:int-scalar} holds for all integers $n$.

Now let $x \in \R$ and $q \in \Q$.  Write $q = m/n$, where $m, n \in \Z$
with $n \neq 0$.  Then by two applications of
equation~\eqref{eq:int-scalar}, 
\[
f(qx)
=
\tfrac{1}{n} f(nqx)
=
\tfrac{1}{n} f(mx)
=
\tfrac{m}{n} f(x)
=
qf(x),
\]
as required.
\end{proof}

\begin{remark}
The same argument proves that any additive function between vector spaces
over $\Q$ is linear over $\Q$.  In the case of functions $\R \to \R$, our
question is whether (or under what conditions) $\Q$-linearity implies
$\R$-linearity, which here we are just calling `linearity'.
\end{remark}

Lemma~\ref{lemma:add-rat} enables us to improve
Proposition~\ref{propn:add-diff}, relaxing differentiability to continuity.
The following result was known to Cauchy himself (cited in Hardy,
Littlewood and P\'olya~\cite{HLP}, proof of Theorem~84).

\begin{propn}
\lbl{propn:add-cts}
Every continuous additive function $\R \to \R$ is linear.
\end{propn}

\begin{proof}
Let $f \from \R \to \R$ be a continuous additive function, and write $c =
f(1)$.  By Lemma~\ref{lemma:add-rat}, $f(q) = cq$ for all $q \in \Q$. Thus,
the two functions $f$ and $x \mapsto cx$ are equal when restricted to $\Q$.
But both are continuous, so they are equal on all of $\R$.
\end{proof}

It is now straightforward to relax continuity of $f$ to an apparently much
weaker condition:

\begin{propn}
\lbl{propn:add-cts-pt}
Every additive function $\R \to \R$ that is continuous at one or more point
is linear.
\end{propn}

In other words, every additive function is linear unless, perhaps, it is
discontinuous everywhere.

\begin{proof}
Let $f \from \R \to \R$ be an additive function continuous at a point $x
\in \R$.  By Proposition~\ref{propn:add-cts}, it is enough to show that $f$
is continuous.  Let $y, t \in \R$: then by additivity,
\[
f(y + t) - f(y)   
=
f(t)
=
f(x + t) - f(x)   
\to 0
\]
as $t \to 0$, as required.  
\end{proof}

Next we show that mere measurability suffices: every measurable additive
function is linear.

\begin{remark}
Readers unfamiliar with measure theory may wish to read the rest of this
remark then resume at Corollary~\ref{cor:add-transf}.
Measurability\index{measurability} is an extremely weak condition.  In the
usual logical framework for mathematics, there do exist nonmeasurable
functions and nonlinear additive functions (Remark~\ref{rmk:choice}).
However, every function that anyone has ever written down an explicit
formula for, or ever will, is measurable (by Remark~\ref{rmk:zf}).  So it
is not too dangerous to assume that every function is measurable and,
therefore, that every additive function is linear.
\end{remark}

There are several proofs that every measurable additive function is linear.
The first was published by Maurice%
\index{Frechet, Maurice@Fr\'echet, Maurice} 
Fr\'echet in his 1913 paper
`Pri la funkcia ekvacio $f(x + y) = f(x) + f(y)$' \cite{Frec}.  (Fr\'echet
wrote many papers in Esperanto,%
\index{Esperanto}
and served three years as the president of the Internacia Scienca Asocio
Esperantista.)  Here we give the proof by Banach~\cite{BanaSEF}.  It is
based on a standard measure-theoretic result of Lusin~\cite{Lusi}, which
makes precise Littlewood's%
\index{Littlewood, John Edensor} 
maxim that every measurable function is `nearly continuous'~\cite{Litt}.

Write $\lambda$\ntn{lambdaLeb} for Lebesgue measure on $\R$.

\begin{thm}[Lusin]
\lbl{thm:lusin}
\index{Lusin's theorem}
Let $a \leq b$ be real numbers, and let $f \from [a, b] \to \R$ be a
measurable function.  Then for all $\epsln > 0$, there exists a closed
subset $V \sub [a, b]$ such that $f|_V$ is continuous and $\lambda\bigl(
[a, b] \without V \bigr) < \epsln$.
\end{thm}

\begin{proof}
See Theorem~7.5.2 of Dudley~\cite{Dudl}, for instance.
\end{proof}

Following Banach, we deduce:

\begin{thm}
\lbl{thm:add-meas}
Every measurable additive function $\R \to \R$ is linear.
\end{thm}

\begin{proof}
Let $f \from \R \to \R$ be a measurable additive function.  
By Lusin's theorem, we can choose a closed set $V \sub [0, 1]$ such that
$f|_V$ is continuous and $\lambda(V) > 2/3$.  Since $V$ is compact, $f|_V$
is uniformly continuous.

By Proposition~\ref{propn:add-cts-pt}, it is enough to prove that $f$ is
continuous at $0$.  Let $\epsln > 0$.  We have to show that $|f(x)| <
\epsln$ for all $x$ in some neighbourhood of $0$.

By uniform continuity, we can choose $\delta > 0$ such that for $v, v' \in
V$,
\[
|v - v'| < \delta \implies |f(v) - f(v')| < \epsln.
\]
I claim that $|f(x)| < \epsln$ for all $x \in \R$ such that $|x| <
\min\{\delta, 1/3\}$.  Indeed, take such an $x$.  Then, writing $V - x = \{
v - x \such v \in V\}$, the inclusion-exclusion property of Lebesgue
measure $\lambda$ gives
\[
\lambda\bigl(V \cap (V - x)\bigr)
=
\lambda(V) + \lambda(V - x) - \lambda\bigl(V \cup (V - x)\bigr).
\]
Consider the right-hand side.  For the first two terms, we have $\lambda(V)
> 2/3$ and so $\lambda(V - x) > 2/3$.  For the last, if $x \geq 0$ then $V
\cup (V - x) \sub [-1/3, 1]$, if $x \leq 0$ then $V \cup (V - x) \sub [0,
  4/3]$, and in either case, $\lambda(V \cup (V - x)) \leq 4/3$.  Hence
\[
\lambda\bigl(V \cap (V - x)\bigr) > 2/3 + 2/3 - 4/3 = 0.
\]
In particular, $V \cap (V - x)$ is nonempty, so we can choose an element
$y$.  Then $y, x + y \in V$ with $|y - (x + y)| = |x| < \delta$, so $|f(y)
- f(x + y)| < \epsln$ by definition of $\delta$.  But since $f$ is
additive, this means that $|f(x)| < \epsln$, as required.
\end{proof}

The regularity condition can be weakened still further; see
Reem~\cite{Reem} for a recent survey.  However, measurability is as weak a
condition as we will need.

\begin{remark}
\lbl{rmk:choice}
Assuming the axiom of choice, there do exist additive functions $\R \to \R$
that are not linear.  To see this, first note that the real line $\R$ is a
vector space over $\Q$ in the evident way.  Choose a basis $B$ for $\R$
over $\Q$.  Choose an element $b$ of $B$, and let $\phi \from B \to \R$ be
the function taking value $1$ at $b$ and $0$ elsewhere.  By the universal
property of bases, $\phi$ extends uniquely to a $\Q$-linear map
$f \from \R \to \R$.

Certainly $f$ is additive.  On the other hand, we can show that $f$ is not
$\R$-linear (that is, not `linear' in the terminology of this section).
Indeed, any $\R$-linear function $\R \to \R$ either is identically zero or
vanishes nowhere except at $0$.  Now $f$ is not identically zero,
since $f(b) = \phi(b) = 1$.  But also, for any $b' \neq b$ in $B$,
we have $f(b') = \phi(b') = 0$ with $b' \neq 0$, so $f$ vanishes at some
point other than $0$.  Hence $f$ is a nonlinear, additive function $\R \to
\R$.
\end{remark}

\begin{remark}
\lbl{rmk:zf} 
It is consistent with the Zermelo--Fraenkel%
\index{Zermelo--Fraenkel axioms}%
\index{set theory} 
axioms of set theory (that is, ZFC without the axiom of choice) that all
functions $\R \to \R$ are measurable.  This is a 1970 theorem of
Solovay~\cite{SoloMST}.%
\index{Solovay, Robert}
If all functions $\R \to \R$ are measurable then
by Theorem~\ref{thm:add-meas}, all additive functions are linear.

On the other hand, the axiom of choice is also consistent with ZF.  If the
axiom of choice%
\index{axiom of choice} 
holds then by Remark~\ref{rmk:choice}, not all additive functions are
linear.

Hence, starting from ZF, one may consistently assume \emph{either} that
every additive function is linear \emph{or} that not every additive
function is linear.
\end{remark}

Theorem~\ref{thm:add-meas} classifies the measurable functions that convert
addition into addition.  One can easily adapt it to classify the
functions that convert addition into multiplication, multiplication into
multiplication, and so on:

\begin{cor}
\lbl{cor:add-transf}
\begin{enumerate}
\item 
\lbl{part:add-transf-exp}
Let $f \from \R \to (0, \infty)$ be a measurable function.  The following
are equivalent:
\begin{enumerate}
\item 
$f(x + y) = f(x)f(y)$ for all $x, y \in \R$;

\item
there exists $c \in \R$ such that $f(x) = e^{cx}$ for all $x \in \R$.
\end{enumerate}

\item
\lbl{part:add-transf-log}
Let $f \from (0, \infty) \to \R$ be a measurable function.  The following
are equivalent:
\begin{enumerate}
\item 
$f(xy) = f(x) + f(y)$ for all $x, y \in (0, \infty)$;

\item
there exists $c \in \R$ such that $f(x) = c \log x$ for all $x \in (0,
\infty)$. 
\end{enumerate}

\item
\lbl{part:add-transf-power}
Let $f \from (0, \infty) \to (0, \infty)$ be a measurable function.  The
following are equivalent:
\begin{enumerate}
\item 
$f(xy) = f(x)f(y)$ for all $x, y \in (0, \infty)$

\item
there exists $c \in \R$ such that $f(x) = x^c$ for all $x \in (0, \infty)$.
\end{enumerate}
\end{enumerate}
\end{cor}

\begin{proof}
For~\bref{part:add-transf-exp}, evidently~\hardref{(b)}
implies~\hardref{(a)}.  Assuming~\hardref{(a)}, define $g \from \R \to
\R$ by $g(x) = \log f(x)$.  Then $g$ is measurable and additive, so by
Theorem~\ref{thm:add-meas}, there is some constant $c \in \R$ such that
$g(x) = cx$ for all $x \in \R$.  It follows that $f(x) = e^{cx}$ for all $x
\in \R$, as required.

Parts~\bref{part:add-transf-log} and~\bref{part:add-transf-power} are proved
similarly, putting $g(x) = f(e^x)$ and $g(x) = \log f(e^x)$.
\end{proof}

\begin{remark}
\lbl{rmk:defn-log}
In this book, the notation $\log$ means the natural logarithm $\ln
= \log_e$.  However, the choice of base for logarithms is usually
unimportant, as it is in
Corollary~\ref{cor:add-transf}\bref{part:add-transf-log}: changing the base
amounts to multiplying the logarithm by a positive constant, which is
in any case absorbed by the free choice of the constant $c$.
\end{remark}

Theorem~\ref{thm:add-meas} also allows us to classify the additive
functions that are defined on only half of the real line.

\begin{cor}
\lbl{cor:cauchy-halfline}
Let $f \from [0, \infty) \to \R$ be a measurable function satisfying $f(x +
  y) = f(x) + f(y)$ for all $x, y \in [0, \infty)$.  Then there exists $c
    \in \R$ such that $f(x) = cx$ for all $x \in [0, \infty)$.    
\end{cor}

\begin{proof}
First we extend $f \from [0, \infty) \to \R$ to a measurable additive
  function $g \from \R \to \R$.  By the hypothesis on $f$, for all
  $a^+, a^-, b^+, b^- \in [0, \infty)$,
\[
a^+ - a^- = b^+ - b^-
\implies
f(a^+) - f(a^-) = f(b^+) - f(b^-).
\]
We can, therefore, consistently define a function $g \from \R \to \R$ by
\[
g(a^+ - a^-) = f(a^+) - f(a^-)
\]
($a^+, a^- \in [0, \infty)$).  To prove that $g$ is additive, let $x, y \in
  \R$, and choose $a^\pm, b^\pm \in [0, \infty)$ such that
\[
x = a^+ - a^-,
\qquad
y = b^+ - b^-.
\]
Then 
\[
x + y = (a^+ + b^+) - (a^- + b^-)
\]
with $a^+ + b^+, a^- + b^- \in [0, \infty)$.  Hence
\begin{align*}
g(x + y)        &
=
f(a^+ + b^+) - f(a^- + b^-)     \\
&
=
f(a^+) + f(b^+) - f(a^-) - f(b^-)       \\
&
=
f(a^+) - f(a^-) + f(b^+) - f(b^-)       \\
&
=
g(x) + g(y),
\end{align*}
as required.  To prove that $g$ is measurable, note that
\[
g(x) 
=
\begin{cases}
f(x)    &\text{if } x \geq 0,   \\
-f(-x)  &\text{if } x \leq 0
\end{cases}
\]
($x \in \R$), as if $x \geq 0$ then we can take $a^+ = x$ and $a^- = 0$
in the definition of $g$, and similarly for $x \leq 0$.  Since $f$ is
measurable, so is $g$.

By Theorem~\ref{thm:add-meas}, there exists a constant $c$ such that
$g(x) = cx$ for all $x \in \R$.  It follows that $f(x) = cx$ for all
$x \in [0, \infty)$.
\end{proof}

The techniques and results of this section can be assembled in several ways
to derive variant theorems.  Rather than attempting to catalogue all the
possibilities, we illustrate the point with two particular variants needed
later.

\begin{cor}
\lbl{cor:cauchy-log-01}
Let $f \from (0, 1] \to \R$ be a measurable function.  The following are
equivalent:
\begin{enumerate}
\item 
\lbl{part:cauchy-log-01-condns}
$f(xy) = f(x) + f(y)$ for all $x, y \in (0, 1]$;

\item
\lbl{part:cauchy-log-01-form}
there exists a constant $c \in \R$ such that $f(x) = c\log x$ for all $x
\in (0, 1]$.  
\end{enumerate}
\end{cor}

\begin{proof}
Trivially, \bref{part:cauchy-log-01-form}
implies~\bref{part:cauchy-log-01-condns}.  Now
assuming~\bref{part:cauchy-log-01-condns}, define $g \from [0, \infty) \to
\R$ by $g(u) = f(e^{-u})$.  Then $g$ is measurable and $g(u + v) = g(u) +
g(v)$ for all $u, v \in [0, \infty)$, so by
Corollary~\ref{cor:cauchy-halfline}, $g(u) = bu$ for some real constant
$b$.  It follows that $f(x) = -b\log x$ for all $x \in (0, 1]$, as
required.
\end{proof}

The moral of Corollary~\ref{cor:cauchy-log-01} is that for the Cauchy-like
functional equation $f(xy) = f(x) + f(y)$, there is no substantial
difference between solving it on the domain $(0, \infty)$ and solving it on
the domain $(0, 1]$ (or $[1, \infty)$, similarly).  But matters become very
different when we seek solutions on the discrete domain $\{1, 2, 3,
\ldots\}$, as we will discover in the next section.

\begin{remark}
\lbl{rmk:defn-inc}
In this text, we always use the terms `increasing' and `decreasing' in
their non-strict senses.  Thus, a function $f \from S \to \R$ on a subset
$S \sub \R$ is \demph{increasing}%
\index{increasing!function or sequence} 
if
\[
x \leq y \implies f(x) \leq f(y)
\]
($x, y \in S$), and \demph{decreasing}\index{decreasing} if $-f$ is
increasing.  It is \demph{strictly}%
\index{strictly increasing!function}%
\index{increasing!strictly}
increasing or decreasing%
\index{strictly decreasing function} 
if $x < y$ implies $f(x) < f(y)$
or $f(x) > f(y)$, respectively.  The same terminology applies to
sequences.%
%
\end{remark}

\begin{cor}
\lbl{cor:cauchy-mult-01}
Let $f \from (0, 1) \to (0, \infty)$ be an increasing function.  The
following are equivalent:
\begin{enumerate}
\item
\lbl{part:cauchy-mult-01-condns}
$f(xy) = f(x)f(y)$ for all $x, y \in (0, 1)$;

\item
\lbl{part:cauchy-mult-01-form}
there exists a constant $c \in [0, \infty)$ such that $f(x) = x^c$ for all
  $x \in (0, 1)$.
\end{enumerate}
\end{cor}

\begin{proof}
Trivially, \bref{part:cauchy-mult-01-form}
implies~\bref{part:cauchy-mult-01-condns}.
Assuming~\bref{part:cauchy-mult-01-condns}, define $g \from (0, \infty) \to
\R$ by $g(u) = -\log f(e^{-u})$.  Then $g(u + v) = g(u) + g(v)$ for all $u,
v \in (0, \infty)$, and $g$ is also increasing.

By the same argument as in the proof of Proposition~\ref{lemma:add-rat},
$g(qu) = qg(u)$ for all $q, u \in (0, \infty)$ with $q$ rational.  Define
$\twid{g} \from (0, \infty) \to \R$ by $\twid{g}(u) = g(1)u$.  Then $g(q) =
\twid{g}(q)$ for all $q \in (0, \infty) \cap \Q$.  Since $g$ is increasing
and $\twid{g}$ is either increasing or decreasing (depending on the sign of
$g(1)$), it follows that $\twid{g}$ is increasing.  But now $g, \twid{g}
\from (0, \infty) \to \R$ are increasing functions that are equal on the
positive rationals, so $g = \twid{g}$.  Hence $f(x) = x^{g(1)}$ for all $x
\in (0, 1)$.
\end{proof}

\section{Logarithmic sequences}
\lbl{sec:log-seqs}

A sequence $f(1), f(2), \ldots$ of real numbers is
\demph{logarithmic}%
\index{logarithmic sequence} 
if
\begin{equation}
\lbl{eq:log-seq}
f(mn) = f(m) + f(n)
\end{equation}
for all $m, n \geq 1$.  Certainly the sequence $(c \log n)_{n \geq 1}$ is
logarithmic, for any real constant $c$.  But in contrast to the situation
for functions $f \from (0, \infty) \to \R$ satisfying $f(xy) = f(x) + f(y)$
(Corollary~\ref{cor:add-transf}\bref{part:add-transf-log}), it is easy to
write down logarithmic sequences that are not of this simple form.  Indeed,
we can choose $f(p)$ arbitrarily for each prime $p$, and these choices
uniquely determine a logarithmic sequence, generally not of the form $(c
\log n)$.

However, there are reasonable conditions on a logarithmic sequence
$(f(n))$ guaranteeing that it is of the form $(c \log n)$.  One such
condition is that $f$ is increasing:
\[
f(1) \leq f(2) \leq \cdots.
\]
An alternative condition is that
\[
\lim_{n \to \infty} \bigl(f(n + 1) - f(n)\bigr) = 0.
\]
We will prove a single theorem implying both of these results.  But a
direct proof of the result on increasing sequences is short enough to be
worth giving separately, even though it is not logically necessary.

\begin{thm}[Erd\H{o}s]
\lbl{thm:erdos-inc}%
\index{Erdos, Paul@Erd\H{o}s, Paul}
Let $(f(n))_{n \geq 1}$ be an increasing sequence of real numbers.  The
following are equivalent:
\begin{enumerate}
\item 
\lbl{part:erdos-inc-condns}
$f$ is logarithmic;

\item
\lbl{part:erdos-inc-form}
there exists a constant $c \geq 0$ such that $f(n) = c \log n$ for all $n
\geq 1$. 
\end{enumerate}
\end{thm}

This was first proved by Erd\H{o}s~\cite{ErdoDFA}.  In fact, he
showed more: as is customary in number theory, he only required
equation~\eqref{eq:log-seq} to hold when $m$ and $n$ are relatively prime.
But since we will not need the extra precision of that result, we will not
prove it.  

The argument presented here follows Khinchin (\cite{Khin}, p.~11).

\begin{proof}
Certainly~\bref{part:erdos-inc-form} implies~\bref{part:erdos-inc-condns}.
Now assume~\bref{part:erdos-inc-condns}.  By the logarithmic property,
\[
f(1) = f(1 \cdot 1) = f(1) + f(1),
\]
so $f(1) = 0$.  Since $f$ is increasing, $f(n) \geq 0$ for all $n$.  If
$f(n) = 0$ for all $n$ then~\bref{part:erdos-inc-form} holds with $c = 0$.
Assuming otherwise, we can choose some $N > 1$ such that $f(N) > 0$.

Let $n \geq 1$.  For each integer $r \geq 1$, there is an integer
$\ell_r \geq 0$ such that
\[
N^{\ell_r} \leq n^r \leq N^{\ell_r + 1}
\]
(since $N > 1$).  As $f$ is increasing and logarithmic, 
\[
\ell_r f(N) \leq r f(n) \leq (\ell_r + 1)f(N),
\]
which since $f(N) > 0$ implies that 
\begin{equation}
\lbl{eq:inc-f}
\frac{\ell_r}{r} \leq \frac{f(n)}{f(N)} \leq \frac{\ell_r + 1}{r}.
\end{equation}
As $\log$ is also increasing and logarithmic, the same argument gives
\begin{equation}
\lbl{eq:inc-log}
\frac{\ell_r}{r} \leq \frac{\log n}{\log N} \leq \frac{\ell_r + 1}{r}.
\end{equation}
Inequalities~\eqref{eq:inc-f} and~\eqref{eq:inc-log} together imply that
\[
\left| \frac{f(n)}{f(N)} - \frac{\log n}{\log N} \right| 
\leq
\frac{1}{r}.
\]
But this conclusion holds for all $r \geq 1$, so 
\[
\frac{f(n)}{f(N)} = \frac{\log n}{\log N}.
\]
Hence $f(n) = c \log n$, where $c = f(N)/\log N$.  And since this is true
for all $n \geq 1$, we have proved~\bref{part:erdos-inc-form}.
\end{proof}

We now prove the unified theorem promised above.  Before stating it, let us
recall the concept of \demph{limit%
\index{limit inferior} 
inferior}.  Given a real sequence $(g(n))_{n \geq 1}$, define
\[
h(n) = \inf\bigl\{ g(n), g(n + 1), \ldots \bigr\} \in [-\infty, \infty)
\]
($n \geq 1$).  The sequence $(h(n))_{n \geq 1}$ is increasing and therefore
has a limit (perhaps $\pm\infty$), written as
\[
\liminf_{n \to \infty} g(n) = \lim_{n \to \infty} h(n) 
\in [-\infty, \infty].
\ntn{liminf}
\]
If the ordinary limit $\lim_{n \to \infty} g(n)$ exists then $\liminf_{n
  \to \infty} g(n) = \lim_{n \to \infty} g(n)$.  However, the limit
inferior exists whether or not the limit does. For instance, the sequence
$1, -1, 1, -1, \ldots$ has a limit inferior of $-1$, but no limit.

If $(f(n))$ is a sequence that either is increasing or
satisfies $f(n + 1) - f(n) \to 0$ as $n \to \infty$, then
\[
\liminf_{n \to \infty} \bigl(f(n + 1) - f(n)\bigr) \geq 0.
\]
The following theorem therefore implies both of the results mentioned above.

\begin{thm}[Erd\H{o}s, K\'atai, M\'at\'e]
\lbl{thm:erdos-liminf}%
\index{Erdos, Paul@Erd\H{o}s, Paul}%
\index{Katai, Imre@K\'atai, Imre}%
\index{Mate, Attila@M\'at\'e, Attila}
Let $(f(n))_{n \geq 1}$ be a sequence of real numbers such that
\[
\liminf_{n \to \infty} \bigl( f(n + 1) - f(n) \bigr) \geq 0.
\]
The following are equivalent:
\begin{enumerate}
\item 
\lbl{part:erdos-liminf-condns}
$f$ is logarithmic;

\item
\lbl{part:erdos-liminf-form}
there exists a constant $c$ such that $f(n) = c \log n$ for all $n \geq 1$.
\end{enumerate}
\end{thm}

This result was stated without proof by Erd\H{o}s in 1957~\cite{ErdoDAA},
then proved independently by K\'atai~\cite{Kata} and by
M\'at\'e~\cite{Mate}, both in 1967.  Again, the logarithmic condition can
be relaxed by only requiring that~\eqref{eq:log-seq} holds when $m$ and $n$
are relatively prime, but again, we have no need for this extra precision.

The proof below follows Acz\'el and Dar\'oczy's adaptation of
K\'atai's argument (Theorem~0.4.3 of~\cite{AcDa}).  The strategy is to put
$c = \liminf_{n \to \infty} f(n)/\log n$ and show that $f(N)/\log N = c$
for all $N$.

\begin{proof}
It is trivial that~\bref{part:erdos-liminf-form}
implies~\bref{part:erdos-liminf-condns}.  Now
assume~\bref{part:erdos-liminf-condns}.  I claim that for all $N \geq 2$,
\begin{equation}
\lbl{eq:liminf-eq}
\liminf_{n \to \infty} \frac{f(n)}{\log n}
=
\frac{f(N)}{\log N}.
\end{equation}
Let $N \geq 2$.  First we show that the left-hand side
of~\eqref{eq:liminf-eq} is less than or equal to the right.  For each $r
\geq 1$, the logarithmic property of $f$ implies that
\[
\frac{f(N^r)}{\log(N^r)}
=
\frac{rf(N)}{r\log N}
=
\frac{f(N)}{\log N}.
\]
Since $N^r \to \infty$ as $r \to \infty$, it follows from the definition of
limit inferior that
\[
\liminf_{n \to \infty} \frac{f(n)}{\log n}
\leq
\frac{f(N)}{\log N}.
\]
Now we prove the opposite inequality, 
\begin{equation}
\lbl{eq:liminf-claim}
\liminf_{n \to \infty} \frac{f(n)}{\log n} 
\geq
\frac{f(N)}{\log N}.
\end{equation}
Let $\epsln > 0$.  By hypothesis, we can choose $k \geq 1$ such that for
all $n \geq N^k$,
\begin{equation}
\lbl{eq:liminf-ineq}
f(n + 1) - f(n) \geq -\epsln.
\end{equation}
Any integer $n \geq N^k$ has a base $N$ expansion
\[
n = c_\ell N^\ell + \cdots + c_1 N + c_0
\]
with $c_0, \ldots, c_\ell \in \{0, \ldots, N - 1\}$, $c_\ell \neq 0$,
and $\ell \geq k$.  Then
\begin{align}
f(n)    &
\geq
f(c_\ell N^\ell + \cdots + c_1 N) - c_0 \epsln  
\lbl{eq:liminf-1}     \\
&
\geq
f(c_\ell N^\ell + \cdots + c_1 N) - N \epsln  
\lbl{eq:liminf-2}     \\
&
=
f(c_\ell N^{\ell - 1} + \cdots + c_1) + f(N) - N \epsln,  
\lbl{eq:liminf-3}
\end{align}
where inequality~\eqref{eq:liminf-1} follows from~\eqref{eq:liminf-ineq}
using induction and the fact that $\ell \geq k$,
inequality~\eqref{eq:liminf-2} holds because $c_0 \leq N$, and
equation~\eqref{eq:liminf-3} follows from the logarithmic property of $f$.
As long as $\ell - 1 \geq k$, we can apply the same argument again with
$c_\ell N^{\ell - 1} + \cdots + c_1$ in place of $n = c_\ell N^\ell +
\cdots + c_0$, giving
\[
f(c_\ell N^{\ell - 1} + \cdots + c_1)
\geq
f(c_\ell N^{\ell - 2} + \cdots + c_2) + f(N) - N\epsln
\]
and so
\[
f(n)
\geq
f(c_\ell N^{\ell - 2} + \cdots + c_2) + 2(f(N) - N\epsln).
\]
Repeated application of this argument gives
\[
f(n)    
\geq 
f(c_\ell N^{k - 1} + \cdots + c_{\ell - k + 1}) 
+ (\ell - k + 1)(f(N) - N\epsln).
\]
Hence, writing $A = \min\bigl\{f(1), f(2), \ldots, f(N^k)\bigr\}$, 
\begin{equation}
\lbl{eq:liminf-4}
f(n) 
\geq
A + (\ell - k + 1)(f(N) - N\epsln).
\end{equation}
In~\eqref{eq:liminf-4}, the only term on the right-hand side that depends
on $n$ is $\ell$, which is equal to $\lfloor \log_N n \rfloor$, and
$\lfloor \log_N n \rfloor/\log_N n \to 1$ as $n \to \infty$.  Hence
\begin{align*}
\liminf_{n \to \infty} \frac{f(n)}{\log_N n}    &
\geq
\liminf_{n \to \infty} \Biggl\{
\frac{A}{\log_N n} + \Biggl( 
\frac{\lfloor \log_N n \rfloor}{\log_N n} + \frac{-k + 1}{\log_N n} 
\Biggr)\bigl(f(N) - N\epsln\bigr)
\Biggr\}        \\
&
=
f(N) - N\epsln.
\end{align*}
This holds for all $\epsln > 0$, so 
\[
\liminf_{n \to \infty} \frac{f(n)}{\log_N n} 
\geq 
f(N).
\]
Since $\log_N n = (\log n)/(\log N)$, this proves the claimed
inequality~\eqref{eq:liminf-claim} and, therefore,
equation~\eqref{eq:liminf-eq}.

Putting $c = \liminf_{n \to \infty} f(n)/\log n \in \R$, we have $f(N) = c
\log N$ for all $N \geq 2$.  Finally, the logarithmic property of $f$
implies that $f(1) = 0$, so $f(1) = c \log 1$ too.
\end{proof}

\begin{cor}
\lbl{cor:erdos-lim}
Let $(f(n))_{n \geq 1}$ be a sequence such that
\begin{equation}
\lbl{eq:erdos-lim}
\lim_{n \to \infty} \bigl( f(n + 1) - f(n) \bigr) = 0.
\end{equation}
The following are equivalent:
\begin{enumerate}
\item 
\lbl{part:erdos-lim-condns}
$f$ is logarithmic;

\item
\lbl{part:erdos-lim-form}
there exists a constant $c$ such that $f(n) = c \log n$ for all $n \geq 1$.
\end{enumerate}
\qed
\end{cor}

To apply this corollary, we will need to be able to verify the limit
condition~\eqref{eq:erdos-lim}.  The following improvement lemma will be
useful.

\begin{lemma}
\lbl{lemma:seq-improvement}
Let $(a_n)_{n \geq 1}$ be a real sequence such that $a_{n + 1} -
\tfrac{n}{n + 1} a_n \to 0$ as $n \to \infty$.  Then $a_{n + 1} - a_n \to
0$ as $n \to \infty$. 
\end{lemma}

Our proof of Lemma~\ref{lemma:seq-improvement} follows that of
Feinstein~\cite{Fein}%
\index{Feinstein, Amiel}
(p.~6--7), and uses a standard result:

\begin{propn}[Ces\`aro]
\lbl{propn:cesaro}
\index{Ces\`aro, Ernesto!limit}
Let $(x_n)_{n \geq 1}$ be a real sequence, and for $n \geq 1$, write
\[
\ovln{x}_n = \tfrac{1}{n} (x_1 + \cdots + x_n).
\]
Suppose that $\lim\limits_{n \to \infty} x_n$ exists.  Then $\lim\limits_{n
  \to \infty} \ovln{x}_n$ exists and is equal to $\lim\limits_{n \to
  \infty} x_n$.
\end{propn}

\begin{proof}
This can be found in introductory analysis texts such as
Apostol~\cite{AposMA} (Theorem~12-48).
\end{proof}

\begin{pfof}{Lemma~\ref{lemma:seq-improvement}}
It is enough to prove that $a_n/(n + 1) \to 0$ as $n \to \infty$.  Write
$b_1 = a_1$ and $b_n = a_n - \tfrac{n - 1}{n} a_{n - 1}$ for $n \geq 2$;
then by hypothesis, $b_n \to 0$ as $n \to \infty$.  We have $n a_n = nb_n +
(n - 1)a_{n - 1}$ for all $n \geq 2$, so
\[
na_n = nb_n + (n - 1)b_{n - 1} + \cdots + 1b_1
\]
for all $n \geq 1$.  Dividing through by $n(n + 1)$ gives
\begin{align}
\frac{a_n}{n + 1}       &
=
\frac{1}{2} \cdot \frac{1}{\hlf n(n + 1)} 
(b_1 + b_2 + b_2 + b_3 + b_3 + b_3 + \cdots 
+ \underbrace{b_n + \cdots + b_n}_n)    
\nonumber       \\
&
=
\frac{1}{2} \cdot M_1(b_1, b_2, b_2, b_3, b_3, b_3, \ldots, 
\underbrace{b_n, \ldots, b_n}_n),
\lbl{eq:seq-imp-1}
\end{align}
where $M_1$ denotes the arithmetic mean.  Since $b_n \to 0$ as $n \to
\infty$, the sequence
\[
b_1, b_2, b_2, b_3, b_3, b_3, \ldots, 
\underbrace{b_n, \ldots, b_n}_n, \ldots
\]
also converges to $0$.  Proposition~\ref{propn:cesaro} applied to this
sequence then implies that
\[
M_1(b_1, b_2, b_2, b_3, b_3, b_3, \ldots, \underbrace{b_n, \ldots, b_n}_n)
\to 0 
\text{ as } n \to \infty.
\]
But by equation~\eqref{eq:seq-imp-1}, this means that $a_n/(n + 1) \to 0$
as $n \to \infty$, completing the proof.
\end{pfof}

\begin{remark}
Lemma~\ref{lemma:seq-improvement} can also be deduced from the
Stolz--Ces\`aro%
\index{Stolz--Ces\`aro theorem}%
\index{Ces\`aro, Ernesto!Stolz theorem@--Stolz theorem}
theorem (Section~3.1.7 of Mure\c{s}an~\cite{Mure}, for
instance).  This is a discrete analogue of l'H\^opital's rule, and states
that given a real sequence $(x_n)$ and a strictly increasing sequence
$(y_n)$ diverging to $\infty$, if
\[
\frac{x_{n + 1} - x_n}{y_{n + 1} - y_n} \to \ell
\]
as $n \to \infty$ then $x_n/y_n \to \ell$ as $n \to \infty$.
Lemma~\ref{lemma:seq-improvement} follows by taking $x_n = na_n$ and $y_n =
\hlf n(n + 1)$.  (I thank X\={\i}l\'{\i}ng Zh\={a}ng for this observation.)
\end{remark}

\section{The $q$-logarithm}
\lbl{sec:q-log}

The $q$-logarithms ($q \in \R$) form a continuous one-parameter family of
functions that include the ordinary natural logarithm as the case $q = 1$.
They can be regarded as deformations of the natural logarithm.  We will
show that as a family, they are characterized by a single functional
equation.

For $q \in \R$, the \demph{$q$-logarithm}\index{q-logarithm@$q$-logarithm}
is the function 
\[
\ln_q \from (0, \infty) \to \R
\ntn{lnq}
\]
defined by
\[
\ln_q(x) 
=
\int_1^x t^{-q} \dee t
\]
($x \in (0, \infty)$).  Thus, 
\[
\ln_1(x) = \log(x)
\]
and for $q \neq 1$,
\begin{equation}
\lbl{eq:q-log-general}
\ln_q(x)
=
\frac{x^{1 - q} - 1}{1 - q}.
\end{equation}
Then $\ln_q(x) \to \ln_1(x)$ as $q \to 1$, by 
l'H\^opital's rule.

Let $q \in \R$.  The $q$-logarithm shares with the natural logarithm the
property that 
\[
\ln_q(1) = 0.
\]
However, in general
\[
\ln_q(xy) \neq \ln_q(x) + \ln_q(y).
\]
One can see this without calculation: for by
Corollary~\ref{cor:add-transf}\bref{part:add-transf-log}, the only
measurable functions that transform multiplication into addition are the
multiples of the natural logarithm.  There is nevertheless a simple formula
for $\ln_q(xy)$ in terms of $\ln_q(x)$ and $\ln_q(y)$:
\[
\ln_q(xy)
=
\ln_q(x) + \ln_q(y) + (1 - q)\ln_q(x)\ln_q(y).
\]
Later, we will use a second formula for $\ln_q(xy)$:
\begin{equation}
\lbl{eq:q-log-mult}
\ln_q(xy)
=
\ln_q(x) + x^{1 - q} \ln_q(y).
\end{equation}
Similarly, in general
\[
\ln_q(1/x) \neq -\ln_q(x),
\]
but instead we have the following three formulas for $\ln_q(1/x)$:
\begin{align}
\ln_q(1/x)      &
=
\frac{-\ln_q(x)}{1 + (1 - q)\ln_q(x)}
\nonumber       \\
&
=
-x^{q - 1} \ln_q(x)    
\nonumber       \\
&
=
-\ln_{2 - q}(x).
\lbl{eq:q-log-reciprocal}
\end{align}
By~\eqref{eq:q-log-reciprocal}, replacing $\ln_q$ by the function $x
\mapsto -\ln_q(1/x)$ defines an involution $\ln_q \leftrightarrow \ln_{2 -
  q}$ of the family of $q$-logarithms, with a fixed point at the classical
logarithm $\ln_1$.  Finally, there is a quotient formula
\begin{equation}
\lbl{eq:q-log-qt}
\ln_q(x/y) 
=
y^{q - 1} \bigl(\ln_q(x) - \ln_q(y)\bigr),
\end{equation}
obtained from equation~\eqref{eq:q-log-mult} by substituting $y$ for $x$
and $x/y$ for $y$.

\begin{remark}
\index{q-logarithm@$q$-logarithm!history of}
The history of the $q$-logarithms as an \emph{explicit} object of study
goes back at least as far as a 1964 paper of Box and Cox in statistics
(Section~3 of~\cite{BoCo}).  The name `$q$-logarithm' appears to have been
introduced by Umarov, Tsallis and Steinberg in 2008~\cite{UTS}, working in
statistical mechanics.  

But as Umarov et al.\ warned, there is more than one system of
$q$-analogues of the classical notions of calculus.  For instance, there
is the system developed by the early twentieth-century clergyman
F.~H. Jackson~\cite{JackQFC} (a modern account of which can be found in Kac
and Cheung~\cite{KaCh}).  In particular, this has given rise to a different
notion of $q$-logarithm, as developed in Chung, Chung, Nam and
Kang~\cite{CCNK}.  Ernst~\cite{Erns} gives a full historical treatment of
the various branches of $q$-calculus.  In any case, none of the
developments just mentioned use the $q$-logarithms considered here.
%
%
\end{remark}

We now prove that the $q$-logarithms are characterized by a simple
functional equation.  The proof is essentially the argument behind
Theorem~84 in the classic text of Hardy,%
\index{Hardy, Godfrey Harold} 
Littlewood%
\index{Littlewood, John Edensor}
and P\'olya~\cite{HLP}%
\index{Polya, George@P\'olya, George}.

\begin{thm}
\lbl{thm:q-log}%
\index{q-logarithm@$q$-logarithm!characterization of} 
Let $f \from (0, \infty) \to \R$ be a measurable function.  The following
are equivalent:
\begin{enumerate}
\item 
\lbl{part:q-log-condns}
there exists a function $g \from (0, \infty) \to \R$ such that for all $x,
y \in (0, \infty)$,
\begin{equation}
\lbl{eq:q-log-fe}
f(xy) = f(x) + g(x) f(y);
\end{equation}

\item
\lbl{part:q-log-form}
$f = c\ln_q$ for some $c, q \in \R$, or $f$ is constant.
\end{enumerate}
\end{thm}

\begin{proof}
First suppose that~\bref{part:q-log-form} holds.  If $f = c\ln_q$ for some
$c, q \in \R$ then equation~\eqref{eq:q-log-fe} holds with $g(x) = x^{1 -
  q}$, by equation~\eqref{eq:q-log-mult}.  Otherwise, $f$ is constant,
so~\eqref{eq:q-log-fe} holds with $g \equiv 0$.

Now assume~\bref{part:q-log-condns}.  Since $f(xy) = f(yx)$,
equation~\eqref{eq:q-log-fe} implies that
\[
f(x) + g(x) f(y) = f(y) + g(y) f(x),
\]
or equivalently
\begin{equation}
\lbl{eq:q-log-sym}
f(x) \bigl(1 - g(y)\bigr) = f(y) \bigl(1 - g(x)\bigr),
\end{equation}
for all $x, y \in (0, \infty)$.  If $f \equiv 0$ then $f$ is constant
and~\bref{part:q-log-form} holds.  Assuming otherwise, we can choose $y_0
\in (0, \infty)$ such that $f(y_0) \neq 0$.  Taking $y = y_0$
in~\eqref{eq:q-log-sym} and putting $a = (1 - g(y_0))/f(y_0)$ gives
\begin{equation}
\lbl{eq:q-log-gf}
g(x) = 1 - af(x)
\end{equation}
($x \in \R$).  Since $f$ is measurable, so is $g$.  There are now two
cases: $a = 0$ and $a \neq 0$.

\textbf{Case 1: $a = 0$.}  Then $g \equiv 1$, so the original functional
equation~\eqref{eq:q-log-fe} states that $f(xy) = f(x) + f(y)$.  Since $f$
is measurable, Corollary~\ref{cor:add-transf}\bref{part:add-transf-log}
implies that $f = c \log = c \ln_1$ for some $c \in \R$.

\textbf{Case 2: $a \neq 0$.}  Then equation~\eqref{eq:q-log-gf}
can be rewritten as
\begin{equation}
\lbl{eq:q-log-fg}
f(x) = \tfrac{1}{a} (1 - g(x))
\end{equation}
($x \in (0, \infty)$).  Substituting this into the original functional
equation~\eqref{eq:q-log-fe} gives 
\begin{equation}
\lbl{eq:q-log-g-mult}
g(xy) = g(x)g(y)
\end{equation}
($x, y \in (0, \infty)$).  In particular, $g(x) = g(\sqrt{x})^2 \geq 0$ for
all $x \in (0, \infty)$.  There are now two subcases: $g$ either sometimes
vanishes or never vanishes.

If $g(x_0) = 0$ for some $x_0 \in (0, \infty)$ then
\[
g(x) = g(x_0)g(x/x_0) = 0
\]
for all $x \in (0, \infty)$, so $g \equiv 0$.  Hence by
equation~\eqref{eq:q-log-fg}, $f$ is constant.

Otherwise, $g(x) > 0$ for all $x \in (0, \infty)$.  Since $g$ is measurable
and satisfies the multiplicativity condition~\eqref{eq:q-log-g-mult},
Corollary~\ref{cor:add-transf}\bref{part:add-transf-power} implies that
there is some constant $t \in \R$ such that $g(x) = x^t$ for all $x \in (0,
\infty)$.  We have assumed that $f \not\equiv 0$, so $g \not\equiv 1$ (by
equation~\eqref{eq:q-log-fg}), so $t \neq 0$.  Hence
\[
f(x)
=
\tfrac{1}{a} (1 - x^t)
=
\tfrac{-t}{a}\ln_{1 - t}(x)
\]
for all $x \in (0, \infty)$, completing the proof.
\end{proof}

%% file: shannon.tex
\chapter{Shannon entropy}
\lbl{ch:shannon}

\begin{quote}
My greatest concern was what to call it. I thought of calling it
`information', but the word was overly used, so I decided to call it
`uncertainty'.  When I discussed it with John von Neumann, he had a better
idea.  Von Neumann told me, `You should call it entropy, for two
reasons. In the first place your uncertainty function has been used in
statistical mechanics under that name, so it already has a name.  In the
second place, and more important, no one knows what entropy really is, so
in a debate you will always have the advantage.'  
\hfill 
-- Claude Shannon%
\index{Shannon, Claude}%
\index{Neumann, John@von Neumann, John}
(quoted in~\cite{TrMc}, p.~180).
\end{quote}

Entropy appears in almost every branch of science.  The most casual
literature search quickly brings up works on entropy in
thermodynamics~\cite{Ferm}, quantum physics~\cite{PST}, communications
engineering~\cite{ShanMTC,SCBE}, information theory~\cite{MacKITI},
statistical inference~\cite{JaynWDWS,JaynPTL}, machine learning and
artificial intelligence~\cite{BoKoPe,Deng,Ratn}, malware
detection~\cite{BJS}, macroecology~\cite{Hart}, the quantification of
biological diversity~\cite{Magu}, biochemistry~\cite{MDV}, water network
engineering~\cite{GDS}, the theory of algorithms and
complexity~\cite{Gacs}, ergodic theory and dynamical
systems~\cite{Parr,Down}, algebraic dynamics~\cite{EvWa}, combinatorial
dynamics~\cite{ALM}, topological dynamics~\cite{AKM}, and climate
science~\cite{HaTe}.  (The references given are a random sample.)  The word
`entropy' has many meanings, all related, and is applied in more ways
still.

This chapter is an introduction to the simplest kind of entropy: the
Shannon entropy of a probability distribution on a finite set.  There are
several ways of interpreting Shannon entropy, and we develop two in depth.
The first is through coding theory (Section~\ref{sec:ent-coding}), an
interpretation that is very standard in the mathematical literature and
goes back to Shannon's seminal paper of 1948~\cite{ShanMTC}.  The second is
through the theory of diversity (Section~\ref{sec:ent-div}).  This is
much less well-known, and is one of the main themes of this book.

The single most important property of Shannon entropy is the chain rule,
which is a formula for the entropy of a composite distribution.  The major
theoretical goal of this chapter is to prove that Shannon entropy is
essentially the \emph{only} quantity that satisfies the chain rule.  To
that end, we begin by reviewing probability distributions and composition
of them (Section~\ref{sec:prob-fin}).  The chain rule itself is derived in
Section~\ref{sec:ent-defn}, along with other basic properties of Shannon
entropy, and is explained in terms of coding and diversity in the next
two sections.  In the final section, we prove the unique characterization
of Shannon entropy by the chain rule.

\section{Probability distributions on finite sets}
\lbl{sec:prob-fin}

Let $n \geq 1$.  A \demph{probability distribution}%
\index{probability distribution} 
on the finite set $\{1, \ldots, n\}$ is an $n$-tuple $\p = (p_1, \ldots,
p_n)$ of real numbers $p_i \geq 0$ such that $\sum p_i = 1$.

Of the various interpretations of probability distributions, one will be
especially important for us.

\begin{example}
\lbl{eg:prob-eco}
Consider an ecological community\index{community} of living organisms
classified into $n$ 
species.  Let $p_i$ be the relative abundance of
the $i$th species, where 
`relative'\index{abundance!relative} 
means that the abundances have been normalized so that $\sum p_i
= 1$.  Then the probability distribution $\p = (p_1, \ldots, p_n)$ is a
model of the community, albeit a very crude one.

Some remarks are in order.  First, the distinction between
species\index{species} is 
inexact and sometimes arbitrary.  Mayden~\cite{Mayd} lists~24 inequivalent
ways of defining `species' (further discussed in Hey~\cite{Hey}).  The
difficulty is most acute for microbes, many of which have not been
classified into species at all.  In practice, for microbes, scientists
sequence the DNA of their sample and use software that applies a clustering
algorithm, thus automatically creating `species' according to a pre-chosen
(and somewhat arbitrary) level of genetic similarity.  We will find a way
through this difficulty in Chapter~\ref{ch:sim}.

Second, the meaning of `abundance'%
\index{abundance!meaning of} 
is completely flexible.  In some contexts, it may be appropriate to simply
count individuals.  But when the organisms are of very different sizes, it
may be better to interpret the abundance of a species as the total mass of
the members of that species.  Or, for plants, the area of land covered by a
species may be a more appropriate measure than the number of individuals.

Third, as emphasized in the Introduction (p.~\pageref{p:general}), nothing
that we will say about `communities' or `species' is actually specific to
ecology: mathematically speaking, it is entirely general.
\end{example}

For $n \geq 1$, write 
\[
\Delta_n 
=
\bigl\{ \text{probability distributions on } \{1, \ldots, n\} \bigr\}.
\ntn{Deltan}
\]
Occasionally we will want to
include the case $n = 0$, and we put $\Delta_0 = \emptyset$.  The
\demph{support}\index{support} of $\p \in \Delta_n$ is
\[
\supp(\p) 
=
\bigl\{ 
i \in \{1, \ldots, n\} 
\such
p_i > 0 \}.
\ntn{supp}
\]
We say that $\p \in \Delta_n$ has \demph{full%
\index{full support} 
support} if $\supp(\p) = \{1, \ldots, n\}$, and write
\[
\lbl{p:simp-int}
\Delta_n^\circ 
=
\{ \p \in \Delta_n
\such
p_i > 0 \text{ for all } i \}
\ntn{Deltano}
\]
for the set of probability distributions of full support.  Finally,
\[
\vc{u}_n = (1/n, \ldots, 1/n)
\ntn{un}
\]
denotes the \demph{uniform%
\index{uniform distribution} 
distribution} on $n$ elements.  Geometrically, $\Delta_n$ is the standard
$(n - 1)$-dimensional simplex\index{simplex}, $\Delta_n^\circ$ is its
interior, and $\vc{u}_n$ is its centre.

\begin{example}
Consider a community consisting of species numbered $1, \ldots, n$, with
relative abundance distribution $\p \in \Delta_n$.  Then $\supp(\p)$ is the
set of species that are actually present in the community, and $\p \in
\Delta_n^\circ$ if and only if every species is present.  (A typical
situation in which some species are absent is a longitudinal study: if the
same site is surveyed every year over several years, it may be that in some
years, not every species is present.)  The uniform distribution $\vc{u}_n$
represents the situation in which all species are equally common.
\end{example}

We now define a fundamental operation: composition of probability
distributions (Figure~\ref{fig:comp}).  

\begin{defn}
\lbl{defn:comp-dist}
Let $n, k_1, \ldots, k_n \geq 1$ and let 
\[
\w \in \Delta_n, 
\
\p^1 \in \Delta_{k_1}, \ 
\ldots, \ 
\p^n \in \Delta_{k_n}.
\]
Write $\p^i = (p^i_1, \ldots, p^i_{k_i})$.  The
\demph{composite%
\index{composition!probability distributions@of probability distributions}
distribution} is
\begin{align*}
\w \of (\p^1, \ldots, \p^n)     &
=
(w_1 p^1_1, \ldots, w_1 p^1_{k_1}, 
\ \ldots, \ 
w_n p^n_1, \ldots, w_n p^n_{k_n})       
\ntn{comp}      \\
&
\in
\Delta_{k_1 + \cdots + k_n}.
\end{align*}
\end{defn}

\begin{figure}
\centering
\lengths
\begin{picture}(120,44)(0,-1)
\cell{60}{23}{c}{\includegraphics[height=40\unitlength]{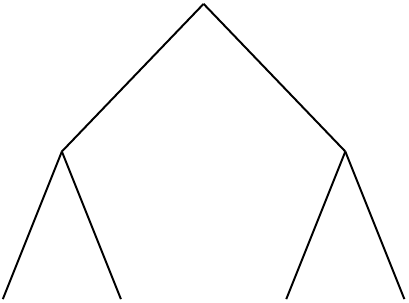}}
\cell{60}{35}{c}{$\vc{w}$}
\cell{60}{25}{c}{$\cdots\cdots$}
\cell{60}{10}{c}{$\cdots$}
\cell{41}{15}{c}{$\p^1$}
\cell{79}{15}{c}{$\p^n$}
\cell{41}{5}{c}{$\cdots$}
\cell{79}{5}{c}{$\cdots$}
\cell{42}{28}{c}{$w_1$}
\cell{78}{28}{c}{$w_n$}
\cell{33}{9}{c}{$p^1_1$}
\cell{51}{9}{c}{$p^1_{k_1}$}
\cell{71}{9}{c}{$p^n_1$}
\cell{88}{9}{c}{$p^n_{k_n}$}
\cell{32}{-1}{b}{$w_1 p^1_1$}
\cell{50}{-1}{b}{$w_1 p^1_{k_1}$}
\cell{70}{-1}{b}{$w_n p^n_1$}
\cell{89}{-1}{b}{$w_n p^n_{k_n}$}
\end{picture}
\caption{Composition of probability distributions.}
\lbl{fig:comp}
\end{figure}

\begin{example}
\lbl{eg:comp-coin}
\index{coin!die-card@-die-card process}
Flip a coin.  If it comes up heads, roll a die.\index{die}  If it comes up
tails, draw from a pack of cards.%
\index{cards, playing}  
Thus, the final outcome of the process is either a
number between~1 and~6 or a playing card.  There are, therefore, $6 + 52 =
58$ possible final outcomes.

Assuming that the coin toss, die roll, and card draw are all fair, the
probabilities of the $58$ possible outcomes are as shown in
Figure~\ref{fig:comp-coin}.
\begin{figure}
\centering
\lengths
\begin{picture}(120,45)(0,-2)
\cell{60}{23}{c}{\includegraphics[height=40\unitlength]{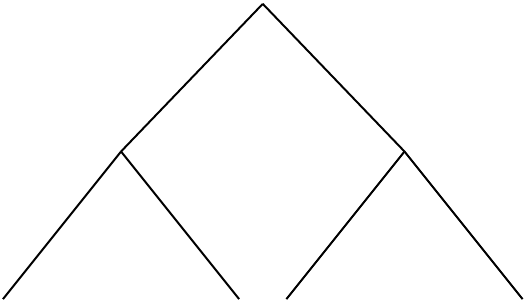}}
\cell{60}{30}{c}{Coin}
\cell{41}{12}{c}{Die}
\cell{79}{12}{c}{Cards}
\cell{41}{5}{c}{$\cdots$}
\cell{79}{5}{c}{$\cdots$}
\cell{43}{29}{c}{$\hlf$}
\cell{77}{29}{c}{$\hlf$}
\cell{26}{9}{c}{$\tfrac{1}{6}$}
\cell{56}{9}{c}{$\tfrac{1}{6}$}
\cell{64}{9}{c}{$\tfrac{1}{52}$}
\cell{95}{9}{c}{$\tfrac{1}{52}$}
\cell{25}{-2}{b}{$\tfrac{1}{12}$}
\cell{56}{-2}{b}{$\tfrac{1}{12}$}
\cell{64}{-2}{b}{$\tfrac{1}{104}$}
\cell{94}{-2}{b}{$\tfrac{1}{104}$}
\end{picture}
\caption{The composite distribution of
  Example~\ref{eg:comp-coin}.} 
\lbl{fig:comp-coin}
\end{figure}
That is, the final outcome has probability distribution
\[
\vc{u}_2 \of (\vc{u}_6, \vc{u}_{52}) 
=
\Bigl(
\underbrace{\tfrac{1}{12}, \ldots, \tfrac{1}{12}}_6, 
\underbrace{\tfrac{1}{104}, \ldots, \tfrac{1}{104}}_{52}
\Bigr).
\]
\end{example}

\begin{example}
\lbl{eg:comp-french}
\index{French language}
The French language is written with the same letters as English, but some
are sometimes decorated by an accent (diacritical mark).  For instance, the
letter \as{a} appears in the three forms \as{a} (no accent), \as{\`{a}} and
\as{\^{a}}, the letter \as{b} appears only as \as{b}, and the letter \as{c}
appears in the two forms \as{c} and \as{\c{c}}.  Let us make the
conventions that a \demph{letter}\index{letter} is one of \as{a}, \as{b},
\ldots, \as{z} and a \demph{symbol}\index{symbol} is a letter together
with, optionally, an accent.  Thus, the symbols are \as{a}, \as{\`a},
\as{\^a}, \as{b}, \as{c}, \as{\c{c}}, \ldots

Let $\vc{w} \in \Delta_{26}$ denote the frequency distribution of the
letters as used in written French.  For the sake of argument, let us
suppose that $w_1, w_2, w_3, \ldots, w_{26}$ have the values shown in
Figure~\ref{fig:comp-french}.  Suppose also that the letter \as{a} appears
without accent 50\% of the time, as \as{\`a} 25\% of the time, and as
\as{\^a} 25\% of the time, again as in the figure.  Write $\p^1 = (0.5,
0.25, 0.25)$, and similarly for $\p^2, \ldots, \p^{26}$.  Then the
frequency distribution of the symbols is the composite
\begin{multline*}
\vc{w} \of (\p^1, \ldots, \p^{26})\\
=
(0.05 \times 0.5, 0.05 \times 0.25, 0.05 \times 0.25, 0.02 \times 1, 
\ldots, 0.004 \times 1).
\end{multline*}

\begin{figure}
\centering
\lengths
\begin{picture}(120,38)(-5,4)
\cell{55}{23}{c}{\includegraphics[height=42\unitlength]{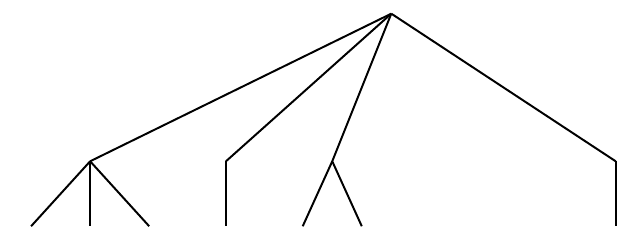}}
\cell{23}{22}{b}{\small 0.05}
\cell{41.5}{22}{b}{\small 0.02}
\cell{56}{22}{b}{\small 0.03}
\cell{102}{22}{b}{\small 0.004}
\cell{34}{27}{b}{\as{a}}
\cell{49}{27}{b}{\as{b}}
\cell{60}{27}{b}{\as{c}}
\cell{76}{27}{b}{$\cdots$}
\cell{92}{27}{b}{\as{z}}
\cell{7.5}{10}{b}{\as{a}}
\cell{14.5}{10}{b}{\colorbox{white}{\as{\`a}\vphantom{$l^l$}}}
\cell{21}{10}{b}{\as{\^a}}
\cell{40}{10}{b}{\as{b}}
\cell{53}{10}{b}{\as{c}}
\cell{61}{9.3}{b}{\as{\c{c}}}
\cell{108.5}{10}{b}{\as{z}}
\cell{2}{5}{b}{\small0.5}
\cell{14.5}{5}{b}{\colorbox{white}{\small0.25\vphantom{$l^l$}}}
\cell{28}{5}{b}{\small0.25}
\cell{40}{5}{b}{\small1}
\cell{49.5}{5}{b}{\small0.5}
\cell{64.5}{5}{b}{\small0.5}
\cell{108.5}{5}{b}{\small1}
\cell{76}{9}{b}{$\cdots$}
\end{picture}
\caption{The composite distribution for French symbols
  (Example~\ref{eg:comp-french}).} 
\lbl{fig:comp-french}
\end{figure}
\end{example}

\begin{example}
\lbl{eg:comp-islands}
\index{islands!composition of distributions@and composition of distributions}
Consider a group of $n$ islands.  Suppose that among all the species
living there, none is present on more than one island (as may in
principle be the case if the islands have been separate for a long enough
period of evolutionary time).  Write $k_i$ for the number of species on
the $i$th island, and $\p^i \in \Delta_{k_i}$ for their relative abundance
distribution.  Also write $\vc{w} \in \Delta_n$ for the relative sizes of
the $n$ islands, where 
`size'%
\index{size!community@of community} 
means the total abundance of organisms on each island.  Then the composite
\[
\vc{w} \of (\p^1, \ldots, \p^n) 
\in 
\Delta_{k_1 + \cdots + k_n}
\]
is the relative abundance distribution for the whole island group, with the
species on the first island listed first, then the species on the second
island, and so on.
\end{example}

\begin{example}
\lbl{eg:comp-genus}
Recall that in the standard taxonomic system, the next
level up from species is genus\index{genus} (plural: genera).  Take an
ecological 
community of $n$ genera, with relative abundances $\vc{w} = (w_1, \ldots,
w_n)$.  Let $\p^i$ be the relative abundance distribution of the species
within the $i$th genus.  Then the relative abundance distribution of the
species in the community is the composite $\vc{w} \of (\p^1, \ldots,
\p^n)$.
\end{example}

\begin{remark}
\lbl{rmk:comp-dist-opd}
Composition of probability distributions satisfies an associative%
\index{associativity}%
\index{composition!associativity of}
law: for
each $n, k_i, \ell_{i j} \geq 1$ and $\vc{w} \in \Delta_n$, $\p^i \in
\Delta_{k_i}$, $\vc{r}^{i j} \in \Delta_{\ell_{i j}}$,
\begin{multline*}
\Bigl( \vc{w} \of \bigl(\p^1, \ldots, \p^n\bigr) \Bigr) \of
\bigl(\vc{r}^{1 1}, \ldots, \vc{r}^{1 k_1}, 
\ \ldots, \ 
\vc{r}^{n 1}, \ldots, \vc{r}^{n k_n}\bigr)\\
=
\vc{w} \of \Bigl(
\p^1 \of \bigl(\vc{r}^{1 1}, \ldots, \vc{r}^{1 k_1}\bigr), \ \ldots, \ 
\p^n \of \bigl(\vc{r}^{n 1}, \ldots, \vc{r}^{n k_n}\bigr) 
\Bigr).
\end{multline*}
The unique distribution $\vc{u}_1$ on the one-element set acts as an
identity for composition:
\[
\p \of (\underbrace{\vc{u}_1, \ldots, \vc{u}_1}_n)
=
\p
=
\vc{u}_1 \of (\p)
\]
for all $n \geq 1$ and $\p \in \Delta_n$.  

These equations are straightforward to check.  In the language of abstract
algebra, they state that the sequence of sets $(\Delta_n)_{n \geq 0}$,
equipped with the operation of composition and the trivial distribution
$\vc{u}_1$, is an operad.%
\index{simplex!operad}%
\index{probability distribution!operad}%
\index{operad!simplex}
We explain and exploit this observation in Chapter~\ref{ch:cat}.
\end{remark}

Now consider the \emph{decomposition}\index{decomposition} problem: given
$\vc{r} \in \Delta_k$ and positive integers $n, k_1, \ldots, k_n$ such that
$\sum k_i = k$, do there exist distributions $\w \in \Delta_n$ and $\p^i
\in \Delta_{k_i}$ such that
\begin{equation}
\lbl{eq:decomp}
\w \of (\p^1, \ldots, \p^n) = \vc{r}?
\end{equation}
The answer is yes.  In fact, $\w$ and $\p^1, \ldots, \p^n$ are
very nearly uniquely determined, ambiguity only arising if
some of the probabilities $r_i$ are zero.  The exact
situation is as follows.

\begin{lemma}
\lbl{lemma:decomp}
Let $k \geq 1$ and $\vc{r} \in \Delta_k$.  Let $n, k_1, \ldots, k_n$ be
positive integers such that $k_1 + \cdots + k_n = k$.  Then there exist 
\[
\w \in \Delta_n, 
\ 
\p^1 \in \Delta_{k_1}, \ 
\ldots, \ 
\p^n \in \Delta_{k_n}
\]
such that equation~\eqref{eq:decomp} holds.  Moreover, $\vc{w}, \p^1,
\ldots, \p^n$ satisfy~\eqref{eq:decomp} if and only if
\begin{equation}
\lbl{eq:decomp-base}
w_i 
=
r_{k_1 + \cdots + k_{i - 1} + 1} 
+ \cdots + 
r_{k_1 + \cdots + k_{i - 1} + k_i}
\end{equation}
for each $i \in \{1, \ldots, n\}$ and 
\begin{equation}
\lbl{eq:decomp-fibre}
\vc{p}^i
=
\frac{1}{w_i} 
(r_{k_1 + \cdots + k_{i - 1} + 1}, \ldots, 
r_{k_1 + \cdots + k_{i - 1} + k_i})
\end{equation}
for each $i \in \supp(\w)$.  In particular, equation~\eqref{eq:decomp}
determines $\vc{w}$ uniquely.
\end{lemma}

\begin{proof}
Define $\vc{w}$ by equation~\eqref{eq:decomp-base}, define
$\vc{p}^i$ by equation~\eqref{eq:decomp-fibre} for each $i \in \supp(\w)$,
and for $i \not\in \supp(\w)$, let $\vc{p}^i$ be any element of
$\Delta_{k_i}$.  It is then trivial to verify equation~\eqref{eq:decomp}.

Conversely, suppose that $\vc{w}, \p^1, \ldots, \p^n$ are distributions
satisfying~\eqref{eq:decomp}.  Write 
$\p^i = (p^i_1, \ldots, p^i_{k_i})$.  We have 
\[
w_1 
= 
w_1 \bigl(p^1_1 + \cdots + p^1_{k_1}\bigr)
=
w_1 p^1_1 + \cdots + w_1 p^1_{k_1}
=
r_1 + \cdots r_{k_1},
\]
since $\p^1 \in \Delta_1$.  A similar argument holds for $w_2, \ldots,
w_n$, giving equation~\eqref{eq:decomp-base}, and
equation~\eqref{eq:decomp-fibre} then follows.
\end{proof}

Some further terminology illuminates this result, and will be
useful throughout.

\begin{defn}
\lbl{defn:pfwd}
Let $k, n \geq 1$, let 
\[
\pi \from \{1, \ldots, k\} \to \{1, \ldots, n\}
\]
be a map of sets, and let $\vc{r} \in \Delta_k$.  The
\demph{pushforward}\index{pushforward} of $\vc{r}$ along $\pi$ is the
distribution $\pi\vc{r} \in \Delta_n$\ntn{pfwd} with $i$th coordinate
\[
(\pi\vc{r})_i
=
\sum_{j \csuch \pi(j) = i} r_j
\]
($i \in \{1, \ldots, n\}$). 
\end{defn}

In the situation of Lemma~\ref{lemma:decomp}, consider the function
\[
\pi \from \{1, \ldots, k\} \to \{1, \ldots, n\}
\]
that maps the first $k_1$ elements of $\{1, \ldots, k\}$ to $1$, the next
$k_2$ elements to $2$, and so on.  Then part of the statement of the lemma
is that equation~\eqref{eq:decomp} determines $\vc{w}$ uniquely as $\vc{w}
= \pi\vc{r}$.

\begin{remark}
Definition~\ref{defn:pfwd} is a special case of the general
measure-theoretic notion of the pushforward $\pi_*\mu$ of a measure $\mu$
along a measurable map $\pi$.  (We omit the star.)  Our statements
about composition and decomposition on finite sets are trivial cases of a
general measure-theoretic theory of integration and
disintegration\index{disintegration}.   For a
summary of disintegration, see Section~3.2 of Dahlqvist, Danos, Garnier and
Kammar~\cite{DDGK}, or for a more comprehensive account, see around
Theorem~III.71 of Dellacherie and Meyer~\cite{DeMe}.
\end{remark}

An important special case of composition is the 
\demph{tensor\lbl{p:tensor}%
\index{tensor product} 
product}.  Given $\w \in \Delta_n$ and $\p \in \Delta_k$, define
\begin{align*}
\w \otimes \p   &
=
\w \of (\underbrace{\p, \ldots, \p}_n)  
\ntn{tensorprob}        \\
&
=
(w_1 p_1, \ldots, w_1 p_k, 
\ \ldots, \ 
w_n p_1, \ldots, w_n p_k)       \\
&
\in
\Delta_{nk}.
\end{align*}
Probabilistically, $\w \otimes \p$ is the joint distribution of two
independent random variables with distributions $\w$ and $\p$
respectively. 

\begin{example}
Consider a large ecological community~-- a
\demph{metacommunity}\index{metacommunity}~-- divided into $N$
subcommunities of relative sizes $w_1, \ldots, w_N$.  Write $S$ for the
number of species in the metacommunity, and $p_1, \ldots, p_S$ for their
relative abundances across the whole metacommunity.  There is an $S \times
N$ matrix representing how the organisms are distributed across the $S$
species and $N$ communities, with the $i$th row summing to $p_i$ and the
$j$th column summing to $w_j$.

If the metacommunity is homogeneous in the sense that the species
distributions in all the subcommunities are identical, then the $(i,
j)$-entry of this matrix is $w_j p_i$.  In that case, when the $SN$ entries
of the matrix are expressed as an $SN$-dimensional vector
(concatenating the columns in order), that
vector is exactly $\w \otimes \p$.
\end{example}

The tensor product of distributions has the usual algebraic properties of a
product: it satisfies the associativity%
\index{associativity!tensor product@of tensor product}%
\index{tensor product}
and identity laws
\[
(\w \otimes \p) \otimes \vc{r}
=
\w \otimes (\p \otimes \vc{r}),
\quad
\p \otimes \vc{u}_1 = \p = \vc{u}_1 \otimes \p.
\]
These follow from the equations in Remark~\ref{rmk:comp-dist-opd}.  For
$\vc{p} \in \Delta_n$ and $d \geq 1$, we write 
\[
\p^{\otimes d} 
= 
\underbrace{\p \otimes \cdots \otimes \p}_d
\in
\Delta_{n^d},
\ntn{tensorpwr}
\]
interpreted as $\vc{u}_1 \in \Delta_1$ if $d = 0$.

\section{Definition and properties of Shannon entropy}
\lbl{sec:ent-defn}

Let $\p = (p_1, \ldots, p_n)$ be a probability distribution on $n$
elements.  The \demph{Shannon%
\index{entropy!Shannon}%
\index{Shannon, Claude!entropy} 
entropy} of $\p$ is 
\[
H(\p)
=
- \sum_{i \in \supp(\p)} p_i \log p_i
=
\sum_{i \in \supp(\p)} p_i \log \frac{1}{p_i}.
\ntn{H}
\]
Equivalently, instead of restricting the sum to just those $i$ such that
$p_i > 0$, one can let $i$ run over all of $\{1, \ldots, n\}$, with the
conventions that
\[
0 \log 0 = 0 = 0 \log \frac{1}{0}.
\]
These conventions are justified by the facts that
\[
\lim_{p \to {0+}} p \log p
=
0
=
\lim_{p \to {0+}} p \log \frac{1}{p}.
\]

\begin{remark}
\lbl{rmk:ent-base}
Although we take $\log$ to denote the \emph{natural} logarithm
(Remark~\ref{rmk:defn-log}), changing the base of the logarithm simply
multiplies $H$ by a constant factor, and in this sense is unimportant.  In
information and coding theory, where one is typically concerned with
strings of binary digits, it is normal to take entropy to base~$2$.  We
write base~$2$%
\index{entropy!base-2} 
entropy as $\Hi$;\ntn{Hi} thus, $\Hi(\p) = H(\p)/\log 2$.
\end{remark}

Much of this chapter is devoted to explaining and interpreting Shannon
entropy, but we can immediately give several interpretations in brief:
\begin{description}
\item[Uniformity.]%
\index{uniform distribution}
For distributions $\p$ on a fixed number of elements, the entropy of $\p$
is greatest when $\p$ is uniform, and least when $\p$ 
is concentrated on a single element (Figure~\ref{fig:four-four-two} and
Lemma~\ref{lemma:ent-max-min} below).

\item[Information.]\index{information}
Regard $\log(1/p_i)$ as the amount of information gained by observing an
event of probability $p_i$.  For a near-inevitable event such as the sun
rising, $p_i \approx 1$ and so $\log(1/p_i) \approx 0$: knowing that the
sun rose this morning tells us nothing that we could not have predicted
with very high confidence beforehand.  The entropy $H(\p)$ is the average
amount of information gained per observation.  We develop this
interpretation in the pages that follow.

\item[Expected surprise.]\index{surprise}
Similarly, $\log(1/p_i)$ can be regarded as our surprise at observing an
event of probability $p_i$, and then $H(\p)$ is the expected surprise.  We
return to this viewpoint in Section~\ref{sec:q-log-ent}.

\item[Genericity.]%
\index{thermodynamics}%
\index{genericity}
In thermodynamics, a system in a state of high entropy is disordered, or
generic.  For instance, it is the usual state of a box of gas that every
cubic centimetre contains about the same number of molecules; this is a
high-entropy, generic, state.  If, by some unlikely chance, all the
molecules were concentrated into one cubic centimetre, this would be a
low-entropy and very non-generic state.

\item[The logarithm of diversity.]
Let $\p$ be a probability distribution modelling an ecological community,
as in Example~\ref{eg:prob-eco}.  In Section~\ref{sec:ent-div}, we will see
that $\exp(H(\p))$ is a sensible measure of the diversity of a community.
In later chapters, we will meet other types of entropy and show that their
exponentials are also meaningful measures of diversity.
\end{description}

\begin{figure}
\centering
\lengths
\begin{picture}(120,45)(-4,-2)
\cell{15}{5}{b}{\includegraphics[width=20\unitlength]{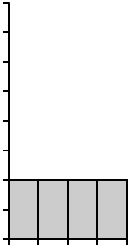}}
\cell{45}{5}{b}{\includegraphics[width=20\unitlength]{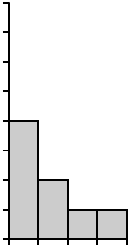}}
\cell{75}{5}{b}{\includegraphics[width=20\unitlength]{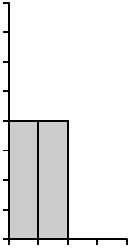}}
\cell{105}{5}{b}{\includegraphics[width=20\unitlength]{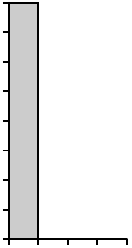}}
\cell{5}{6}{r}{$\scriptstyle 0$}
\cell{5}{15.2}{r}{$\tfrac{1}{4}$}
\cell{5}{24.2}{r}{$\tfrac{1}{2}$}
\cell{5}{33.5}{r}{$\tfrac{3}{4}$}
\cell{5}{42.2}{r}{$\scriptstyle 1$}
\cell{35}{6}{r}{$\scriptstyle 0$}
\cell{35}{15.2}{r}{$\tfrac{1}{4}$}
\cell{35}{24.2}{r}{$\tfrac{1}{2}$}
\cell{35}{33.5}{r}{$\tfrac{3}{4}$}
\cell{35}{42.2}{r}{$\scriptstyle 1$}
\cell{65}{6}{r}{$\scriptstyle 0$}
\cell{65}{15.2}{r}{$\tfrac{1}{4}$}
\cell{65}{24.2}{r}{$\tfrac{1}{2}$}
\cell{65}{33.5}{r}{$\tfrac{3}{4}$}
\cell{65}{42.2}{r}{$\scriptstyle 1$}
\cell{95}{6}{r}{$\scriptstyle 0$}
\cell{95}{15.2}{r}{$\tfrac{1}{4}$}
\cell{95}{24.2}{r}{$\tfrac{1}{2}$}
\cell{95}{33.5}{r}{$\tfrac{3}{4}$}
\cell{95}{42.2}{r}{$\scriptstyle 1$}
\cell{8.5}{4.5}{c}{$\scriptstyle 1$}
\cell{13}{4.5}{c}{$\scriptstyle 2$}
\cell{17.5}{4.5}{c}{$\scriptstyle 3$}
\cell{22}{4.5}{c}{$\scriptstyle 4$}
\cell{38.5}{4.5}{c}{$\scriptstyle 1$}
\cell{43}{4.5}{c}{$\scriptstyle 2$}
\cell{47.5}{4.5}{c}{$\scriptstyle 3$}
\cell{52}{4.5}{c}{$\scriptstyle 4$}
\cell{68.5}{4.5}{c}{$\scriptstyle 1$}
\cell{73}{4.5}{c}{$\scriptstyle 2$}
\cell{77.5}{4.5}{c}{$\scriptstyle 3$}
\cell{82}{4.5}{c}{$\scriptstyle 4$}
\cell{98.5}{4.5}{c}{$\scriptstyle 1$}
\cell{103}{4.5}{c}{$\scriptstyle 2$}
\cell{107.5}{4.5}{c}{$\scriptstyle 3$}
\cell{112}{4.5}{c}{$\scriptstyle 4$}
\cell{-1.5}{24}{c}{\rotatebox{90}{Probability}}
%
\cell{15}{-1.5}{b}{Entropy: $2$}
\cell{45}{-2}{b}{Entropy: $1\tfrac{3}{4}$}
\cell{75}{-1.5}{b}{Entropy: $1$}
\cell{105}{-1.5}{b}{Entropy: $0$}
\end{picture}
\caption{Four probability distributions on $\{1, 2, 3, 4\}$, and their
  entropies to base~$2$.}  \lbl{fig:four-four-two}
\end{figure}

\begin{examples}
\lbl{egs:ent-ufm}
Figure~\ref{fig:four-four-two} shows the base~$2$ entropies $\Hi(\p)$ of
four distributions $\p \in \Delta_4$.  For instance, the second is computed
as 
\begin{align*}
\Hi\Bigl(\tfrac{1}{2}, \tfrac{1}{4}, \tfrac{1}{8}, \tfrac{1}{8}\Bigr)     &
=
\tfrac{1}{2} \log_2 2 + \tfrac{1}{4} \log_2 4 
+ \tfrac{1}{8} \log_2 8 + \tfrac{1}{8} \log_2 8 
=
1 \tfrac{3}{4}.
\end{align*}
These examples illustrate the interpretation of entropy as uniformity.  The
highest entropy belongs to the first, uniform, distribution.  Each of the
four distributions on $\{1, 2, 3, 4\}$ is less uniform than its
predecessor, and, correspondingly, has lower entropy.
\end{examples}

We now set out the basic properties of entropy.  Here and later, we will
repeatedly use the following elementary fact about logarithms.

\begin{lemma}
\lbl{lemma:log-concave}
Let $\p \in \Delta_n$ and $x_1, \ldots, x_n \in (0, \infty)$.  Then
\[
\log \Biggl( \sum_{i = 1}^n p_i x_i \Biggr)
\geq
\sum_{i = 1}^n p_i \log x_i,
\]
with equality if and only if $x_i = x_j$ for all $i, j \in \supp(\p)$.
\end{lemma}

\begin{proof}
The function $\log \from (0, \infty) \to \R$ is strictly
concave\index{concavity}, since 
$\tfrac{d^2}{dx^2} \log x = -1/x^2 < 0$.  The result follows. 
\end{proof}

We now show that among all probability distributions on a finite set,
entropy is maximized by the uniform distribution and minimized by any
distribution of the form $(0, \ldots, 0, 1, 0, \ldots, 0)$.  

\begin{lemma}
\lbl{lemma:ent-max-min}
Let $n \geq 1$.
\begin{enumerate}
\item 
\lbl{part:ent-min}
$H(\p) \geq 0$ for all $\p \in \Delta_n$, with equality if and only if $p_i
  = 1$ for some $i \in \{1, \ldots, n\}$. 

\item
\lbl{part:ent-max}
$H(\p) \leq \log n$ for all $\p \in \Delta_n$, with equality if and only if
  $\p = \vc{u}_n$.
\end{enumerate}
\end{lemma}

\begin{proof}
Part~\bref{part:ent-min} follows from the fact that $\log(1/p_i) \geq 0$
for all $i \in \supp(\p)$, with equality if and only if $p_i = 1$.
For~\bref{part:ent-max}, Lemma~\ref{lemma:log-concave} gives
\[
H(\p)
=
\sum_{i \in \supp(\p)} p_i \log \frac{1}{p_i}
\leq
\log \Biggl( \sum_{i \in \supp(\p)} p_i \cdot \frac{1}{p_i} \Biggr)
=
\log\mg{\supp(\p)}
\leq
\log n.
\]
Again by Lemma~\ref{lemma:log-concave}, the first inequality is an equality
if and only if $\p$ is uniform on its support.  The second inequality is an
equality if and only if $\p$ has full support.  The result follows.
\end{proof}

It is often useful to express entropy in terms of the 
function
\[
\partial \from [0, 1] \to \R
\ntn{partial}
\]
defined by
\begin{equation}
\lbl{eq:defn-par}
\partial(x)
=
\begin{cases}
-x \log x       &\text{if } x > 0,      \\
0               &\text{if } x = 0.
\end{cases}
\end{equation}
Thus,
\begin{equation}
\lbl{eq:sh-par}
H(\p) = \sum_{i = 1}^n \partial(p_i)
\end{equation}
for all $n \geq 1$ and $\p \in \Delta_n$.  

\begin{lemma}
\lbl{lemma:ent-cts}
For each $n \geq 1$, the entropy function $H \from \Delta_n \to \R$ is
continuous. 
\end{lemma}

\begin{proof}
This follows from equation~\eqref{eq:sh-par} and the elementary fact
that $\partial$ is continuous.
\end{proof}

The operator $\partial$ is a nonlinear derivation\index{derivation}:
\begin{lemma}
\lbl{lemma:derivation}
$\partial(xy) = \partial(x)y + x\partial(y)$ for all $x, y \in [0, 1]$.
\qed
\end{lemma}

\begin{remark}
Up to a constant factor, $\partial$ is the only measurable function $d
\from [0, 1] \to \R$ satisfying $d(xy) = d(x)y + xd(y)$ for all $x, y$.
Indeed, taking $x = y = 0$ forces $d(0) = 0$, and the result follows
by applying Corollary~\ref{cor:cauchy-log-01} to the function $x \mapsto
d(x)/x$ on $(0, 1]$.
\end{remark}

We use Lemma~\ref{lemma:derivation} to prove the most important algebraic
property of Shannon entropy:

\begin{propn}[Chain rule]
\lbl{propn:ent-chain}
\index{chain rule!Shannon entropy@for Shannon entropy}
Let $\vc{w} \in \Delta_n$ and $\p^1 \in \Delta_{k_1}, \ldots, \p^n \in
\Delta_{k_n}$.  Then
\[
H\bigl(\vc{w} \of (\p^1, \ldots, \p^n)\bigr)
=
H(\vc{w}) + \sum_{i = 1}^n w_i H(\p^i).
\]
\end{propn}

\begin{proof}
Writing $\p^i = \bigl(p^i_1, \ldots, p^i_{k_i}\bigr)$ and using
Lemma~\ref{lemma:derivation}, we have
\begin{align*}
H\bigl(\vc{w} \of (\p^1, \ldots, \p^n)\bigr)    &
=
\sum_{i = 1}^n \sum_{j = 1}^{k_i} \partial \bigl(w_i p^i_j\bigr)   \\
&
=
\sum_{i} \sum_{j}
\bigl(\partial(w_i) p^i_j + w_i \partial\bigl(p^i_j\bigr)\bigr)   \\
&
=
\sum_{i} \partial(w_i) 
+
\sum_{i} w_i 
\sum_{j} \partial\bigl(p^i_j\bigr) \\
&
=
H(\vc{w}) + \sum_i w_i H(\p^i),
\end{align*}
as required.
\end{proof}

\begin{example}
\index{coin!die-card@-die-card process}
Consider again the coin-die-card process of Example~\ref{eg:comp-coin}.
How much information do we expect to gain from observing the final outcome
of the process?

Let us measure information by base~$2$ entropy, in bits.  The information
gained is as follows.
\begin{itemize}
\item
Whether the final outcome is a number between~1 and~6 or a card tells us
whether the coin came up heads or tails.  This gives us $\Hi(\vc{u}_2) = 1$
bit of information.

\item
With probability $1/2$, the outcome is the result of a die\index{die} roll,
which would give us $\Hi(\vc{u}_6) = \log_2 6$ bits of information.

\item
With probability $1/2$, the outcome is the result of a card%
\index{cards, playing} 
draw, which would give us $\Hi(\vc{u}_{52}) = \log_2 52$ bits of
information. 
\end{itemize}
Hence in total, the expected information gained from observing the outcome
of the composite process is
\[
\Hi(\vc{u}_2) + \hlf \Hi(\vc{u}_6) + \hlf \Hi(\vc{u}_{52})
=
1 + \hlf \log_2 6 + \hlf \log_2 52
\]
bits.  If we have reasoned correctly, this should be equal to the entropy
of the composite process, which is
\[
\Hi\bigl(\vc{u}_2 \of (\vc{u}_6, \vc{u}_{52})\bigr)
=
\Hi\Bigl(\underbrace{\tfrac{1}{12}, \ldots, \tfrac{1}{12}}_{6},
\underbrace{\tfrac{1}{104}, \ldots, \tfrac{1}{104}}_{52}\Bigr)
\]
bits.  The chain rule guarantees that these two numbers are, indeed, equal.
\end{example}

\begin{cor}
\lbl{cor:ent-log}
For all $\vc{w} \in \Delta_n$ and $\p \in \Delta_k$,
\begin{equation}
\lbl{eq:ent-log}
H(\vc{w} \otimes \p) = H(\vc{w}) + H(\vc{p}).
\end{equation}
\end{cor}

\begin{proof}
Take $\p^1 = \cdots = \p^n = \p$ in the chain rule.  
\end{proof}

In other words, $H$ has the logarithmic property of converting products
into sums.  Indeed, in the special case $\vc{w} = \vc{u}_n$ and $\p =
\vc{u}_k$, we have $\vc{w} \otimes \p = \vc{u}_{nk}$, so
equation~\eqref{eq:ent-log} is precisely the characteristic property of
the logarithm,
\[
\log(nk) = \log n + \log k.
\]
In the general case, equation~\eqref{eq:ent-log} states that the amount of
information gained by observing the outcome of a pair of
\emph{independent} events is equal to the information gained from the first
plus the information gained from the second.

\begin{remark}
\lbl{rmk:ent-chain-simp}
With the understanding that $H$ is symmetric in its arguments, the
chain%
\index{chain rule!forms of}
rule as stated in Proposition~\ref{propn:ent-chain} is equivalent to the
superficially less general statement that
\begin{equation}
\lbl{eq:ent-chain-simp1}
H\bigl(pw_1, (1 - p)w_1, w_2, \ldots, w_n\bigr)
=
H(\vc{w}) + w_1 H(p, 1 - p)
\end{equation}
for all $p \in [0, 1]$ and $\vc{w} \in \Delta_n$.  This is the special case
$k_1 = 2$, $k_2 = \cdots = k_n = 1$ of Proposition~\ref{propn:ent-chain},
and is sometimes known as the \demph{recursivity}\index{recursivity} of entropy
(Definition~1.2.8 of Acz\'el and Dar\'oczy~\cite{AcDa}) or the
\demph{grouping%
\index{grouping rule} 
rule} (Problem~4 of Chapter~2 of Cover and Thomas~\cite{CoTh1}).

The general chain rule of Proposition~\ref{propn:ent-chain} is also
equivalent to a different special case:
\begin{multline*}
H\bigl(wp_1, \ldots, wp_k, (1 - w)r_1, \ldots, (1 - w)r_\ell\bigr)
=\\
H(w, 1 - w) + w H(\p) + (1 - w) H(\vc{r})
\end{multline*}
for all $w \in [0, 1]$, $\p \in \Delta_k$, and $\vc{r} \in \Delta_\ell$.
This is the special case $n = 2$ of Proposition~\ref{propn:ent-chain}.

Both equivalences are routine inductions, carried out in
Appendix~\ref{sec:chain}.
\end{remark}

\section{Entropy in terms of coding}
\lbl{sec:ent-coding}
\index{entropy!coding@via coding}

The theory of coding provides a very concrete way of understanding the
concept of information.  The fundamental concepts and theorems of coding
theory were set out in Shannon's%
\index{Shannon, Claude} 
original 1948 paper~\cite{ShanMTC}, with rigour and detail added soon
afterwards by researchers such as Khinchin~\cite{Khin} and
Feinstein~\cite{Fein}.  This section presents parts of that early work, and
in particular, Shannon's source coding theorem.

The source%
\index{Shannon, Claude!source coding theorem}
coding theorem can be described informally as follows.  Take an alphabet of
symbols, say the English%
\index{English language} 
letters \as{a} to \as{z}, which occur with known frequencies $p_1, \ldots,
p_{26}$.  We want to design a scheme that encodes each letter as a finite
sequence of $0$s and $1$s.  Using this system, any message in English can
also be encoded as a sequence of $0$s and $1$s, by concatenating the codes
for the letters in the message.  Of course, we want our coding scheme to
have the property that the encoded message can be decoded unambiguously,
and it is also natural to want it to use as few bits as possible.  Roughly
speaking, the theorem is that in the most efficient coding scheme, the
number of bits needed per symbol is the base~$2$ entropy of the frequency
distribution $\p$.

We now give a more precise account.  In this section, entropy will always
be taken to base~$2$.  Details of everything that follows can be found in
introductions to information theory such as Cover and Thomas (\cite{CoTh1},
Chapter~5), MacKay (\cite{MacKITI}, Chapter~4), and Jones and
Jones~\cite{JoJo}.

Take an alphabet of $n$ symbols, with frequency distribution $\p \in
\Delta_n$; thus, in messages written using this alphabet, we expect the
symbols to be used in proportions $p_1, \ldots, p_n$.  A
\demph{code}\index{code} is an assignment to each $i \in \{1, \ldots, n\}$
of a finite sequence of bits (a \demph{code word}).  The $i$th code word
is, then, an element of the set $\{0, 1\}^{L_i}$ for some integer $L_i \geq
0$, and $L_i$ is called the \demph{word%
\index{word length}
length} of the $i$th symbol.  The expected word length of a symbol in our
alphabet is
\[
\sum_{i = 1}^n p_i L_i.
\]
We seek a code that minimizes the average word length, subject to the
natural constraint of unambiguous decodability (made precise shortly).

\begin{example}
\lbl{eg:code-naive}
Take an alphabet of four symbols \as{a}, \as{b}, \as{c}, \as{d}, with
frequency distribution $\p = (1/2, 1/4, 1/8, 1/8)$.  How should we encode
our symbols as strings of bits, in a way that uses as few bits as possible?

The basic principle is that common symbols should have short code words.
(The same principle guided the design of Morse%
\index{Morse code} 
code, where the most common letter, \as{e}, is encoded as a single dot, and
uncommon letters such as \as{z} use four dots or dashes.)  So let us encode
as follows:
\[
\as{a}: 0,
\quad
\as{b}: 10,
\quad
\as{c}: 110,
\quad
\as{d}: 111.
\]
For instance, 11110011010 represents \as{dbacb}.  The average word length
is
\[
\tfrac{1}{2} \cdot 1 +
\tfrac{1}{4} \cdot 2 + 
\tfrac{1}{8} \cdot 3 + 
\tfrac{1}{8} \cdot 3
=
1\tfrac{3}{4}.
\]
This is more efficient than the most naive coding system, which would
simply assign the four two-bit strings $00, 01, 10, 11$ to the four
symbols, for an average word length of $2$. 
\end{example}

A code is \demph{instantaneous}%
\index{instantaneous code} 
if none of the code words is a prefix (initial segment) of any other.
Thus, if $\delta_1 \cdots \delta_\ell$ and $\epsln_1 \cdots \epsln_m$ are
code words in an instantaneous code, with $\ell \leq m$, then $(\delta_1,
\ldots, \delta_\ell) \neq (\epsln_1, \ldots, \epsln_\ell)$.
This is the non-ambiguity condition, guaranteeing that any string of bits
produced by the system can only be decoded in one possible way.  

\begin{example}
The code of Example~\ref{eg:code-naive} is instantaneous.  But if we changed
the code word for \as{b} to $11$, the code would no longer be
instantaneous, since $11$ is a prefix of the code words for both \as{c} and
\as{d}.  Messages in this new code are not uniquely decodable; for
instance, the string $110$ could be decoded as either \as{c} or \as{ba}.
\end{example}

The average word length $1\tfrac{3}{4}$ of the code in
Example~\ref{eg:code-naive} happens to be equal to the entropy of the
frequency distribution of the symbols, calculated in
Example~\ref{egs:ent-ufm}.  In fact, it is not possible to find an
instantaneous code whose average word length is any shorter.  This is an
instance of part~\bref{part:sh1-nogo-bound} of the following result.

\begin{propn}
\lbl{propn:sh1-nogo}
Let $n, L_1, \ldots, L_n \geq 1$, and suppose that there exists an
instantaneous code on the alphabet $\{1, \ldots, n\}$ with word lengths
$L_1, \ldots, L_n$.  Then:
\begin{enumerate}
\item
\lbl{part:sh1-nogo-kraft}
$\displaystyle \sum_{i = 1}^n (1/2)^{L_i} \leq 1$;

\item
\lbl{part:sh1-nogo-bound}
$\displaystyle \sum_{i = 1}^n p_i L_i \geq \Hi(\p)$ for all $\p \in
\Delta_n$. 
\end{enumerate}
\end{propn}

Part~\bref{part:sh1-nogo-kraft}, together with
part~\bref{part:sh1-go-kraft} of Proposition~\ref{propn:sh1-go} below, is
known as \demph{Kraft's%
\index{Kraft's inequality} 
inequality} (Theorem~5.2.1 of Cover and Thomas~\cite{CoTh1}, for instance).

\begin{proof}
To prove~\bref{part:sh1-nogo-kraft}, we consider binary expansions $0.b_1
b_2 \ldots$ of elements of $[0, 1)$, where $b_i \in \{0, 1\}$.  We make the
convention that if $x \in [0, 1)$ has two binary expansions, one ending
with an infinite sequence of $0$s and the other with an infinite
sequence of $1$s, we choose the former.  In this way, each $x \in [0,
1)$ determines an infinite sequence of bits $b_1, b_2, \ldots$

Take an instantaneous code with word lengths $L_1, \ldots, L_n$.
For $i \in \{1, \ldots, n\}$, write 
\[
J_i 
=
\bigl\{ x \in [0, 1) \such \text{the binary expansion of } x
\text{ begins with the $i$th code word} \bigr\}.
\]
Then $J_i$ is a half-open interval of length $(1/2)^{L_i}$.  Since the code
is instantaneous, the intervals $J_1, \ldots, J_n$ are disjoint.  But since
they are all subsets of $[0, 1)$, their total length is at most $1$, giving
the desired inequality.

For~\bref{part:sh1-nogo-bound}, let $\p \in \Delta_n$.  By
Lemma~\ref{lemma:log-concave} and part~\bref{part:sh1-nogo-kraft}, 
\begin{align*}
\Hi(\p) - \sum_{i = 1}^n p_i L_i        &
=
\sum_{i \in \supp(\p)} p_i
\Bigl( \log_2 (1/p_i)
+ \log_2 \bigl((1/2)^{L_i} \bigr) \Bigr)   \\
&
=
\sum_{i \in \supp(\p)} 
p_i \log_2 \frac{(1/2)^{L_i}}{p_i}      \\
&
\leq
\log_2 \Biggl( 
\sum_{i \in \supp(\p)} p_i \cdot \frac{(1/2)^{L_i}}{p_i} 
\Biggr) \\
&
\leq
\log_2 \sum_{i = 1}^n (1/2)^{L_i}  \\
& 
\leq 
\log_2 1 = 0,
\end{align*}
as required.
\end{proof}

The frequency distribution of Example~\ref{eg:code-naive} has the
exceptional property that all the frequencies are powers of $1/2$.
In such cases, it is always possible to find an instantaneous code in which
the $i$th symbol is encoded in $\log_2(1/p_i)$ bits, so that the average
word length is exactly the entropy.  In the general case, this is not quite
possible; but it is nearly possible, as follows.

\begin{propn}
\lbl{propn:sh1-go}
Let $\p \in \Delta_n$.  Then:
\begin{enumerate}
\item 
\lbl{part:sh1-go-kraft}
there is an instantaneous code with word lengths $\lceil \log_2(1/p_1)
\rceil$, \ldots, $\lceil \log_2(1/p_n) \rceil$;

\item
\lbl{part:sh1-go-bound}
any such code has expected word length strictly less than
$\Hi(\p) + 1$.  
\end{enumerate}
\end{propn}

Here $\lceil x \rceil$ denotes the smallest integer greater than or equal
to $x$.  Codes with the property in~\bref{part:sh1-go-kraft} are called
\demph{Shannon%
\index{Shannon, Claude!code}
codes}.

\begin{proof}
For~\bref{part:sh1-go-kraft}, suppose without loss of generality that $p_1
\geq \cdots \geq p_n$.  For each $i \in \{1, \ldots, n\}$, put 
\[
L_i
=
\lceil \log_2(1/p_i) \rceil,
\quad
q_i 
=
(1/2)^{L_i}.
\]
In other words, $q_i$ is maximal among all powers of $1/2$ less than or
equal to $p_i$.  Now, $q_1, \ldots, q_i$ are all integer multiples of
$(1/2)^{L_i}$, so $q_1 + \cdots + q_{i - 1}$ and $q_1 + \cdots + q_i$ are
integer multiples of $(1/2)^{L_i}$ too.  It follows that the binary
expansions of the elements of the interval
\[
J_i = [q_1 + \cdots + q_{i - 1}, q_1 + \cdots + q_{i - 1} + q_i)
\]
all begin with the same $L_i$ bits, and, moreover, that no other element of
$[0, 1)$ begins with this bit-sequence.  (Here we use the same convention on
binary expansions as in the proof of Proposition~\ref{propn:sh1-nogo}.)
Take the $i$th code word to be this bit-sequence.  Since the intervals
$J_1, \ldots, J_n$ are disjoint, none of the code words is a prefix of any
other; that is, the code is instantaneous.

For~\bref{part:sh1-go-bound}, take a code as in~\bref{part:sh1-go-kraft},
again writing $L_i = \lceil \log_2(1/p_i)\rceil$.  We have
\[
L_i < \log_2(1/p_i) + 1
\]
for each $i \in \{1, \ldots, n\}$, so
\[
\sum_{i = 1}^n p_i L_i
=
\sum_{i \in \supp(\p)} p_i L_i
<
\sum_{i \in \supp(\p)} p_i \bigl( \log_2(1/p_i) + 1 \bigr)
=
\Hi(\p) + 1,
\]
as required.
\end{proof}

\begin{example}
Take the alphabet consisting of \as{a}, \as{b}, \as{c}, \as{d} with
frequencies $\p = (0.4, 0.3, 0.2, 0.1)$.  Following the construction in the
proof of Proposition~\ref{propn:sh1-go}, we round each frequency down to
the next power of $1/2$, giving
\[
(q_1, q_2, q_3, q_4)
=
\Bigl(\tfrac{1}{4}, \tfrac{1}{4}, \tfrac{1}{8}, \tfrac{1}{16}\Bigr)
=
\biggl( \Bigl(\tfrac{1}{2}\Bigr)^2, 
\Bigl(\tfrac{1}{2}\Bigr)^2, 
\Bigl(\tfrac{1}{2}\Bigr)^3, 
\Bigl(\tfrac{1}{2}\Bigr)^4 
\biggr).
\]
Thus, $(L_1, L_2, L_3, L_4) = (2, 2, 3, 4)$ and the intervals $J_i$ are as
follows, in binary notation:
\begin{align*}
J_1     &
=
\Bigl[0, \tfrac{1}{4}\Bigr)     
=
[0.00, 0.01),   \\
J_2     &
=
\Bigl[\tfrac{1}{4}, \tfrac{1}{2}\Bigr)     
=
[0.01, 0.10),   \\
J_3     &
=
\Bigl[\tfrac{1}{2}, \tfrac{5}{8}\Bigr)     
=
[0.100, 0.101),   \\
J_4     &
=
\Bigl[\tfrac{5}{8}, \tfrac{11}{16}\Bigr)     
=
[0.1010, 0.1011).
\end{align*}
We therefore encode as follows:
\[
\as{a}: 00,
\quad
\as{b}: 01, 
\quad
\as{c}: 100,
\quad
\as{d}: 1010.
\]
Short calculations show that
\[
\sum_{i = 1}^4 p_i L_i
=
2.4 
< 
2.846\ldots
=
\Hi(\p) + 1,
\]
as the proof of Proposition~\ref{propn:sh1-go} guarantees.

This is not the most efficient code. For instance, we could have encoded
\as{d} as 101 for a smaller average word length.  There are in fact
algorithms that construct for each $\p$ a code with the least possible
average word length, such as that of Huffman~\cite{Huff}%
\index{Huffman code}.  
But we will not need such precision here.
\end{example}

\begin{example}
Similarly, the code in Example~\ref{eg:code-naive} is the one constructed
by the algorithm in the proof of Proposition~\ref{propn:sh1-go}.
\end{example}

\begin{remark}
The bound $\Hi(\p) + 1$ in Proposition~\ref{propn:sh1-go} cannot be
improved to $\Hi(\p) + c$ for any constant $c < 1$.  For instance, if a
two-symbol alphabet has frequency distribution $\p = (0.99, 0.01)$ then
$\Hi(\p) \approx \Hi(1, 0) = 0$ (since $\Hi$ is continuous), but clearly
the average word length cannot be reduced to below~$1$.
\end{remark}

We now state a version of Shannon's source coding theorem.

\begin{thm}[Shannon]%
\index{Shannon, Claude!source coding theorem}
For an alphabet with frequency distribution $\p = (p_1, \ldots, p_n)$,
\[
\Hi(\p) \leq \inf \sum_{i = 1}^n p_i L_i < \Hi(\p) + 1,
\]
where the infimum is over all instantaneous codes on $n$ elements, with
$L_i$ denoting the $i$th word length.
\end{thm}

\begin{proof}
This is immediate from Propositions~\ref{propn:sh1-nogo}
and~\ref{propn:sh1-go}. 
\end{proof}

A crucial further insight of Shannon was that the upper bound $\Hi(\p) + 1$
can be reduced to $\Hi(\p) + \epsln$, for any $\epsln > 0$, as long as we
are willing to encode symbols in blocks\index{block} rather than one at a
time.  Informally, this works as follows.

For an alphabet with $n$ symbols, there are $n^{10}$ blocks of $10$
symbols.  Writing $\p \in \Delta_n$ for the frequency distribution of the
original alphabet and assuming that successive symbols in messages are
distributed independently, the frequency distribution of the $n^{10}$
blocks is $\p^{\otimes 10}$.  

Now treat each 10-symbol block as a unit, and consider ways of encoding
each block as a sequence of bits.  By Proposition~\ref{propn:sh1-go}, we
can find an instantaneous code for the blocks that uses an average of less
than $\Hi(\p^{\otimes 10}) + 1$ bits per block.  But $\Hi(\p^{\otimes 10})
= 10\Hi(\p)$ by Corollary~\ref{cor:ent-log}, so the average number of bits
per letter is less than
\[
\tfrac{1}{10} \Bigl( \Hi\bigl(\p^{\otimes 10}\bigr) + 1 \Bigr)
=
\Hi(\p) + \tfrac{1}{10}.
\]
In this way, by encoding symbols in large blocks rather than individually,
we can make the average number of bits per letter as close as we please to
the lower bound of $\Hi(\p)$.

(In applications, successive symbols are often not independent.  For
instance, in English, the letter pair \as{ch} is more frequent than
\as{hc}.  But it will follow from Remark~\ref{rmk:ub-coupling} that even if
they are not independent, the actual frequency distribution of the $n^{10}$
blocks has entropy at most $H(\p^{\otimes 10})$.  For that reason, the
argument above is valid even without the assumption of
independence.) 
\pagebreak

\begin{example}
Take a two-symbol alphabet \as{a}, \as{b} with frequency distribution $\p =
(0.6, 0.4)$.  Then $\Hi(\p) = 0.9709\ldots$\:\  We compute the average number
of bits per letter when encoding in larger and larger blocks, following the
code construction in the proof of Proposition~\ref{propn:sh1-go}.
\begin{itemize}
\item 
First encode one symbol at a time.  We round each $p_i$ down to the next
power of $1/2$, giving $\bigl((1/2)^1,
(1/2)^2\bigr)$.  Hence the average number of bits per symbol is
\[
0.6 \times 1 + 0.4 \times 2 = 1.4.
\]

\item
Now encode symbols in blocks of two.  The frequency distribution of
\as{aa}, \as{ab}, \as{ba}, \as{bb} is $(0.36, 0.24, 0.24, 0.16)$ (assuming
that successive symbols are distributed independently).  Following the same
algorithm, we round down to $\bigl((1/2)^2, (1/2)^3,
(1/2)^3, (1/2)^3\bigr)$ and obtain an average of
\[
0.36 \times 2 + 0.24 \times 3 + 0.24 \times 3 + 0.16 \times 3
=
2.64
\]
bits per two-symbol block, or equivalently an average of $1.32$ bits per
symbol.  This is an improvement on the original code.

\item
Similarly, encoding in three-symbol blocks gives an average of
$1.117\ldots$ bits per symbol, which is closer still to the
ideal of $\Hi(\p) \approx 0.971$ bits per symbol.
\end{itemize}
None of these three codes is as efficient as the naive code that assigns
the code words $0$ to \as{a} and $1$ to \as{b}, which has an average word
length of $1$.  But we can improve on that by encoding in large enough
blocks.  For instance, since
\[
0.971 + \tfrac{1}{35} < 1,
\]
we can attain an average word length of less than $1$ by coding blocks of
$35$ symbols at a time.
\end{example}

\begin{example}
\index{English language}
In written English, the base~$2$ entropy of the frequency distribution of
the $26$~letters of the alphabet is approximately 4.1 (Section~2 of
Shannon~\cite{ShanPEP}).  Thus, by using sufficiently large blocks, one can
encode English using about four bits per letter.  (It is as if English had
only $2^{4.1} \approx 17$ letters, used with equal frequency.)  This is
without taking advantage of the fact that \as{ch} occurs more often than
\as{hc}, for instance.  Using the non-independence of neighbouring letters
would enable us to reduce the number of bits still further, as detailed by
Shannon~\cite{ShanPEP} and later researchers.
\end{example}

A convenient fiction\lbl{p:fiction} when reasoning about entropy is that
for every probability distribution $\p$, there is an instantaneous code
with average word length $\Hi(\p)$.  This is not true unless all the
nonzero frequencies happen to be powers of $\hlf$, but it is approximately
true in the sense just described: we can come arbitrarily close by encoding
in sufficiently large blocks.  Let us call this (usually nonexistent) code an
\demph{ideal\lbl{p:ideal}%
\index{ideal code} 
code} for $\p$.

Ideal codes provide a way to understand the chain rule
(Proposition~\ref{propn:ent-chain}), as follows.

\begin{example}
\lbl{eg:ent-french}
\index{French language} 
Consider again the French language (Example~\ref{eg:comp-french}), which is
written with symbols such as \as{\`a} made up of a letter (in this case,
\as{a}) and an accent (in this case, \as{\`{\:}}).
Figure~\ref{fig:ent-french} shows a hypothetical frequency distribution
$\vc{w}$ of the letters, hypothetical frequency distributions
\[
\p^1 \in \Delta_3, \ 
\p^2 \in \Delta_1, \
\ldots, \
\p^{26} \in \Delta_1
\]
of the accents on each letter, and the base~$2$ entropy of each of the
distributions $\vc{w}, \p^1, \ldots, \p^{26}$.

\begin{figure}
\centering
\lengths
\begin{picture}(120,53)(-4,-3.5)
\cell{55}{23}{c}{\includegraphics[height=42\unitlength]{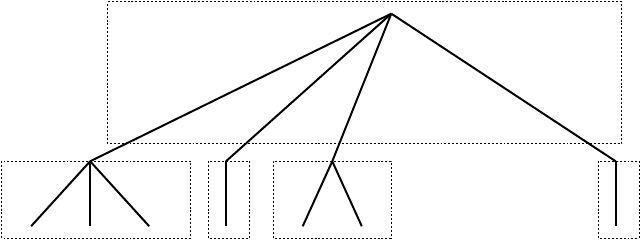}}
\cell{23}{22}{b}{\small 0.05}
\cell{41.5}{22}{b}{\small 0.02}
\cell{56}{22}{b}{\small 0.03}
\cell{102}{22}{b}{\small 0.004}
\cell{34}{27}{b}{\as{a}}
\cell{49}{27}{b}{\as{b}}
\cell{60}{27}{b}{\as{c}}
\cell{76}{27}{b}{$\cdots$}
\cell{92}{27}{b}{\as{z}}
\cell{7.5}{10}{b}{\as{a}}
\cell{14.5}{10}{b}{\colorbox{white}{\as{\`a}\vphantom{$l^l$}}}
\cell{21}{10}{b}{\as{\^a}}
\cell{40}{10}{b}{\as{b}}
\cell{53}{10}{b}{\as{c}}
\cell{61}{9.3}{b}{\as{\c{c}}}
\cell{108.5}{10}{b}{\as{z}}
\cell{2}{5}{b}{\small0.5}
\cell{14.5}{5}{b}{\colorbox{white}{\small0.25\vphantom{$l^l$}}}
\cell{28}{5}{b}{\small0.25}
\cell{40}{5}{b}{\small1}
\cell{49.5}{5}{b}{\small0.5}
\cell{64.5}{5}{b}{\small0.5}
\cell{108.5}{5}{b}{\small1}
\cell{76}{9}{b}{$\cdots$}
\cell{62.8}{45.5}{b}{Entropy $\Hi(0.05, 0.02, 0.03, \ldots, 0.004)$}
\cell{15.5}{0}{t}{Entropy $1.5$}
\cell{39}{0}{t}{Entropy $0$}
\cell{57}{0}{t}{Entropy $1$}
\cell{107}{0}{t}{Entropy $0$}
\end{picture}
\caption{The entropy of the French language
  (Example~\ref{eg:ent-french}).}
\lbl{fig:ent-french}
\end{figure}

To transmit a French symbol (such as \as{\`a}), we need to transmit both its
base letter (\as{a}) and its accent (\as{\`{\:}}).  Using ideal
codes, the average number of bits needed per symbol is as follows.  For
the base letter, we need $\Hi(\vc{w})$ bits.  The number of bits needed for
the accent depends on which letter it decorates:
\begin{itemize}
\item 
with probability $w_1$, the letter is \as{a}, and then the average number
of bits needed for the accent is $\Hi(\p^1)$;

\item 
with probability $w_2$, the letter is \as{b}, and then the average number
of bits needed for the accent is $\Hi(\p^2)$;
\end{itemize}
and so on.  Hence the average number of bits needed to encode the accent is
$\sum_{i = 1}^{26} w_i \Hi(\p^i)$.  The average number of bits needed per
symbol is the number for the base letter plus the number for the accent,
which is
\begin{equation}
\lbl{eq:ent-french-cR}
\Hi(\vc{w}) + \sum_{i = 1}^{26} w_i \Hi(\p^i).
\end{equation}
On the other hand, we saw in Example~\ref{eg:comp-french} that the overall
frequency distribution of the French symbols \as{a}, \as{\`a}, \as{\^a},
\as{b}, \ldots, \as{z} is $\vc{w} \of (\p^1, \ldots, \p^n)$, whose ideal
code uses
\begin{equation}
\lbl{eq:ent-french-cL}
\Hi\bigl(\vc{w} \of (\p^1, \ldots, \p^n)\bigr)
\end{equation}
bits per symbol.  If we have reasoned correctly then the
expressions~\eqref{eq:ent-french-cR} and~\eqref{eq:ent-french-cL} should be
equal.  The chain rule states that, indeed, they are.
\end{example}

\section{Entropy in terms of diversity}
\lbl{sec:ent-div}
\index{entropy!diversity@via diversity}

Entropies of various kinds have been used to measure biological diversity
for almost as long as diversity measures have been considered.  For
instance, among all the measures of diversity used by ecologists, one of the
most common is the Shannon entropy $H(\p)$.  Here $\p = (p_1, \ldots, p_n)$
is the relative abundance distribution of the community concerned, as in
Example~\ref{eg:prob-eco}.  For reasons that will be explained, when it
comes to measuring diversity, it is better to use the \emph{exponential} of
entropy than entropy itself.

Let us begin by considering intuitively what it means for a community of
$n$ species to be diverse, for a fixed value of $n$.  As described in the
Introduction, there is a spectrum of viewpoints on what the word
`diversity' should mean.  Loosely, though, diversity is low when most of
the population is concentrated into one or two very common species, and
high when the population is spread evenly across all species.  Another way
to say this is that diversity is low when an individual chosen at random
usually belongs to a common species, and high when an individual chosen at
random usually belongs to a rare species.  So, the diversity of
a community can be understood as the average rarity of an individual
belonging to it.

Since $p_i$ represents the relative abundance of the $i$th species, $1/p_i$
is a measure of its rarity\index{rarity} or specialness\index{specialness}.
We want to take the average rarity, and for now we will use the geometric
mean as our notion of average.  (Later, we will use different notions of
average.  The most important are the power means, which are introduced in
Section~\ref{sec:pwr-mns} and include the geometric mean.)  Thus, one
reasonable measure of the diversity of a community is the geometric mean of
the species rarities $1/p_1, \ldots, 1/p_n$, weighted by the species sizes
$p_1, \ldots, p_n$:
\[
\Biggl( \frac{1}{p_1} \Biggr)^{p_1}
\cdots
\Biggl( \frac{1}{p_n} \Biggr)^{p_n}.
\]
We therefore make the following definition.

\begin{defn}
\lbl{defn:div1}
\index{diversity!order 1@of order 1}
Let $n \geq 1$ and $\p \in \Delta_n$.  The \demph{diversity of order~$1$}
of $\p$ is 
\[
D(\p)
=
\frac{1}{p_1^{p_1} p_2^{p_2} \cdots p_n^{p_n}},
\ntn{D}
\]
with the convention that $0^0 = 1$.
\end{defn}

Equivalently,
\[
D(\p)
=
\prod_{i \in \supp(\p)} p_i^{-p_i}
=
e^{H(\p)}.
\]
In short: diversity is the exponential of entropy.

\begin{remarks}
\begin{enumerate}
\item
The meaning of `order~$1$' will be revealed in Section~\ref{sec:ren-hill}.
It is related to the different possible notions of average.  In this
section, `diversity' will always mean diversity of order~$1$.

\item
No choice of base is involved in the definition of $D$, in contrast to the
situation for $H$ (Remark~\ref{rmk:ent-base}).  For instance, $D(\p)$ is
equal to both $e^{H(\p)}$ and $2^{\Hi(\p)}$.
\end{enumerate}
\end{remarks}

Crucially, the word `diversity' refers only to the \emph{relative}, not
absolute, abundances.%
\index{abundance!relative vs.\ absolute}  
If half of a forest\index{forest!fire} burns down, or if a patient loses
$90\%$ of their gut%
\index{gut microbiome}%
\index{microbial systems}
bacteria, then it may be an
ecological or medical 
disaster; but assuming that the system is well-mixed, the diversity does
not change.  In the language of physics, diversity is an
intensive%
\index{intensive quantity} 
quantity
(like density or temperature) rather than an extensive%
\index{extensive quantity} 
quantity (like mass or heat), meaning that it is independent of the
system's size.

Lemma~\ref{lemma:ent-max-min} immediately implies:
\begin{lemma}
\lbl{lemma:div1-max-min}
Let $n \geq 1$.
\begin{enumerate}
\item 
\lbl{part:div1-min}
$D(\p) \geq 1$ for all $\p \in \Delta_n$, with equality if and only if $p_i
  = 1$ for some $i \in \{1, \ldots, n\}$. 

\item
\lbl{part:div1-max}
$D(\p) \leq n$ for all $\p \in \Delta_n$, with equality if and only if
  $\p = \vc{u}_n$.
\end{enumerate}
\qed
\end{lemma}

Similarly, the continuity of entropy (Lemma~\ref{lemma:ent-cts})
immediately implies:
\begin{lemma}
\lbl{lemma:div1-cts}
For each $n \geq 1$, the diversity function $D \from \Delta_n \to \R$ of
order~$1$ is continuous.  
\qed
\end{lemma}

Evidently
\[
D(\vc{u}_n) = n 
\]
for all $n \geq 1$.  This is a very important property for a diversity
measure, and we adopt the standard terminology for it:

\begin{defn}
\lbl{defn:div-eff}
Let $\bigl( E \from \Delta_n \to (0, \infty) \bigr)_{n \geq 1}$ be a
sequence of functions.  Then $E$ is an \demph{effective%
\index{effective number} 
number} if
$E(\vc{u}_n) = n$ for all $n \geq 1$. 
\end{defn}

Thus, $D$ is an effective number.  When the species are all present in
equal quantities, we think of the community as containing $n$ fully
present species and assign it a diversity value of $n$.  On the other hand,
if one species accounts for nearly $100\%$ of the community and all the
others are very rare, then the diversity value is barely more than~$1$ (by
Lemmas~\ref{lemma:div1-max-min}\bref{part:div1-min}
and~\ref{lemma:div1-cts}).  Effectively, there is barely more than one
species present.

For instance, if a community has a diversity of $18.2$, then the community
is slightly more diverse than a community of $18$ equally abundant species.
There are `effectively' slightly more than $18$ balanced species.

\begin{examples}
\lbl{egs:div1-ufm}
For the four distributions on $\{1, 2, 3, 4\}$ in
Examples~\ref{egs:ent-ufm}, the diversities are
\[
2^2 = 4, 
\quad
2^{7/4} \approx 3.364,
\quad
2^1 = 2,
\quad
2^0 = 1,
\]
respectively.  In particular, the community represented by the second
distribution is judged by $D$ to be somewhat more diverse than a community
of three species in equal proportions, but less diverse than a balanced
community of four species.
\end{examples}

Despite the popularity of Shannon entropy as a measure of biological
diversity, many ecologists have argued that it should be rejected in favour
of its exponential, including MacArthur~\cite{MacA} in 1965, Buzas and
Gibson~\cite{BuGi} in 1969, and Whittaker~\cite{WhitEMS} in 1972.  More
recently and more generally, Jost~\cite{JostED,JostPDI,JostMBD} has argued
convincingly that when measuring diversity, we should only use effective
numbers.  (That principle appears to be gaining acceptance, judging by the
editorial~\cite{ElliPD} of Ellison.)  The following example is adapted from
Jost~\cite{JostPDI}.%
\index{Jost, Lou}

\begin{example}
\lbl{eg:plague}%
\index{plague}%
\index{diversity measure!logical behaviour of}
Suppose that a plague strikes a continent of a million equally common
species, rendering $90\%$ of the species extinct and leaving the remaining
$10\%$ untouched.  How do $H$ and $D$ respond to this catastrophe?

The Shannon entropy $H$ drops by just
\[
1 - \frac{\log(10^5)}{\log(10^6)} = \frac{1}{6} \approx 17\%,
\]
suggesting a change of considerably smaller magnitude than the one that
actually occurred.  For comparison, if a community of four equally common
species loses only \emph{one} of its species, the rest remaining unchanged,
this causes a drop in Shannon entropy of
\[
1 - \frac{\log 3}{\log 4} \approx 21\%.
\]
So, if we judge by percentage change in Shannon entropy, losing $25\%$ of
four species destroys a greater proportion of the diversity than losing
$90\%$ of a million species.  Shannon entropy drops \emph{more} in the
situation where the species loss is \emph{less}.  So as an indicator of
change in diversity, percentage change in Shannon entropy is plainly
unsuitable. 

However, the effect of the plague on the diversity $D$ is to make it drop
by $90\%$ (from $10^6$ to $10^5$), because $D$ is an effective number.  And
for the same reason, in the four-species example, $D$ drops by $25\%$ (from
$4$ to $3$).  This is intuitively reasonable behaviour, faithfully
reflecting the scale of the change.
\end{example}

In information and coding theory, the logarithmic measure $H$ is the
more useful form, corresponding as it does to the number of bits per symbol
in an ideal code.  But for species diversity, it is the number of
species (not its logarithm) with which we reason most naturally.

We now consider the chain rule in terms of diversity.  Taking exponentials
in Proposition~\ref{propn:ent-chain} gives:

\begin{cor}
\lbl{cor:div1-chain}
Let $n, k_1, \ldots, k_n \geq 1$.  Then
\[
D\bigl(\vc{w} \of (\p^1, \ldots, \p^n)\bigr)
=
D(\vc{w}) \cdot \prod_{i = 1}^n D\bigl(\p^i\bigr)^{w_i}
\]
for all $\vc{w} \in \Delta_n$ and $\p^i \in \Delta_{k_i}$.
\qed
\end{cor}

The second factor on the right-hand side is the geometric mean of the
diversities $D(\p^1), \ldots, D(\p^n)$, weighted by $w_1, \ldots, w_n$.

The most important aspect of this result is not the specific formula, but
the fact that the diversity of the composite distribution depends only on
$\vc{w}$ and $D(\p^1), \ldots, D(\p^n)$, \emph{not} on $\p^1, \ldots, \p^n$
themselves.  This can be understood in either of the following ways.

\begin{example}
\lbl{eg:div1-chain-islands}
\index{islands!diversity of group of}
As in Example~\ref{eg:comp-islands}, consider a group of $n$ islands of
relative sizes $w_1, \ldots, w_n$, with no species shared between islands.
Let $d_i$ denote $D(\p^i)$, the diversity of the $i$th island.  Then the
diversity of the whole island group is
\begin{equation}
\lbl{eq:div1-chain-islands}
D(\vc{w}) \cdot d_1^{w_1} \cdots d_n^{w_n}.
\end{equation}
Thus, the diversity of the whole island group is determined by the
diversities and relative sizes of the islands.  It can be computed without
reference to the population distributions on each island.  
\end{example}

\begin{example}
\lbl{eg:div1-chain-genus}
\index{genus}
As in Example~\ref{eg:comp-genus}, consider a community of $n$ genera, with
the $i$th genus divided into $k_i$ species.  Let $\vc{w}$ denote the genus
distribution and $d_i$ the diversity of the species in the $i$th genus.
Then the species diversity of the whole community is again given
by~\eqref{eq:div1-chain-islands}.  For instance, if there are $2$~equally
abundant genera, with the first genus consisting of~$45$ species of equal
abundance and the second consisting of~$5$ species of equal abundance, then
the diversity of the whole community is
\[
D\bigl(\vc{u}_2 \of (\vc{u}_{45}, \vc{u}_5)\bigr)
=
D(\vc{u}_2) \cdot D(\vc{u}_{45})^{1/2} D(\vc{u}_5)^{1/2}
=
2 \sqrt{45}\sqrt{5} 
=
30.
\]
In other words, the whole community of $45 + 5 = 50$ species, which has
relative abundance distribution
\[
\Bigl(
\underbrace{\tfrac{1}{90}, \ldots, \tfrac{1}{90}}_{45},
\underbrace{\tfrac{1}{10}, \ldots, \tfrac{1}{10}}_{5}
\Bigr).
\]
has the same diversity as a community of~$30$ species of equal abundance.
\end{example}

Different chain rules will appear in Sections~\ref{sec:ren-hill}
and~\ref{sec:sim-props}, where we consider diversity of orders other
than~$1$.  But all share the crucial property that $D(\vc{w} \of (\p^1,
\ldots, \p^n))$ depends only on $\vc{w}$ and $D(\p^1), \ldots, D(\p^n)$.

We refer to this property of $D$ as
\demph{modularity}\lbl{p:D-mod}%
\index{modularity!diversity of order 1@of diversity of order 1}.  
The word is used here in the sense of modular software design, buildings or
furniture (as opposed to modular arithmetic or modules over a ring, say).
In this metaphor, the islands of Example~\ref{eg:div1-chain-islands} or the
genera of Example~\ref{eg:div1-chain-genus} are the `modules': when it
comes to computing the diversity of the whole assemblage, they are black
boxes whose internal features we do not need to know.

The logarithmic property of $H$ (Corollary~\ref{cor:ent-log}) translates
into a multiplicative property of $D$:
\begin{equation}
\lbl{eq:div1-mult}
D(\vc{w} \otimes \vc{p}) = D(\vc{w}) \cdot D(\vc{p})
\end{equation}
($n, k \geq 1$, $\vc{w} \in \Delta_n$, $\p \in \Delta_k$).  An important
special case is the \demph{replication\lbl{p:D-rep}%
\index{replication principle!diversity of order 1@for diversity of order 1} 
principle}:
\[
D(\vc{u}_n \otimes \p) = n D(\p)
\]
($n, k \geq 1$, $\p \in \Delta_k$).  In the language of
Example~\ref{eg:div1-chain-islands}, this principle states that given $n$
islands of equal size and the same species distributions, but with no
actual shared species, the diversity of the whole island group is $n$ times
the diversity of any individual island.

Another argument of Jost%
\index{Jost, Lou} 
(adapted from~\cite{JostMBD} and~\cite{JDWG}) makes a compelling case for
the importance of the replication principle:

\begin{example}
\index{oil company}%
\index{diversity measure!logical behaviour of}%
\lbl{eg:oil} An oil company is planning to carry out work on a group of
islands that will destroy all wildlife on half of the islands.%
\index{islands!oil drilling on}
Environmentalists are bringing a legal case to stop them.  What would be
the impact of the work on biodiversity?

Suppose that there are~$16$ equally-sized islands in the group, that there
are no species shared between islands, and that each island has diversity
$4$.  Then before the oil work, the diversity of the island group is
\[
16 \times 4 = 64.
\]
Afterwards, similarly, it will be $32$.  Thus, the diversity is reduced by
$50\%$.  This is intuitively reasonable, and is a consequence of the
replication principle for~$D$.

However, one of the most popular measures of diversity in ecology is
Shannon entropy (`many long-term investigations have chosen it as their
benchmark of biological diversity': Magurran~\cite{Magu},%
\index{Magurran, Anne}
p.~101).  The oil company's lawyers can therefore argue as follows.  Before
the works, the `diversity' (Shannon entropy) is $\log 64$, and afterwards,
it will be $\log 32$.  Thus, the proportion of diversity preserved is
\[
\frac{\log 32}{\log 64} 
=
\frac{5}{6} 
\approx
83\%.
\]
On the other hand, the environmentalists' lawyers can argue that the
islands whose wildlife is to be exterminated have a diversity of $\log
32$, out of a total of $\log 64$, so the proportion of diversity destroyed
will be 
\[
\frac{\log 32}{\log 64}
=
\frac{5}{6}
\approx 
83\%.
\]
So the oil company can truthfully claim that by the scientifically accepted
measure, $83\%$ of the diversity will be preserved, while the
environmentalists can just as legitimately claim that $83\%$ of the
diversity will be lost.  They cannot both be right, and, of course, both
are wrong: by any reasonable measure, $50\%$ of the diversity is preserved
and $50\%$ is lost.  The reason for the contradictory and illogical
conclusions is that Shannon entropy does not satisfy the replication
principle. 

Although this is an idealized hypothetical example, it is not hard to
see how a choice of diversity measure, far from being some obscure
theoretical issue, could have genuine environmental consequences.
\end{example}

Although the diversity measure $D$ does satisfy the replication principle,
and in that sense behaves logically, it has a glaring deficiency: it
takes no notice of the varying similarities%
\index{similarity!species@of species} 
between species.  A forest consisting of ten equally abundant
species of larch is intuitively less diverse than a forest of ten equally
abundant but highly varied tree species.  However, the measure $D$ gives
the same diversity to both.  The same criticism can be levelled at most of
the diversity measures used in ecology, and a remedy is presented in
Chapter~\ref{ch:sim}.

\section{The chain rule characterizes entropy}
\lbl{sec:ent-chain}
\index{chain rule!Shannon entropy@for Shannon entropy}

There are many characterizations of Shannon entropy, beginning with one in
the original paper by Shannon%
\index{Shannon, Claude}
himself (\cite{ShanMTC}, Theorem~2).  Here,
we prove a variant of one of the best-known such theorems, due
to Dmitry Faddeev~\cite{Fadd}.  

\begin{thm}[Faddeev]
\lbl{thm:faddeev}%
\index{Faddeev, Dmitry!entropy theorem}
Let $( I \from \Delta_n \to \R )_{n \geq 1}$ be a sequence of
functions.  The following are equivalent:
\begin{enumerate}
\item 
\lbl{part:faddeev-condns}
the functions $I$ are continuous and satisfy the chain rule
\[
I\bigl(\vc{w} \of (\p^1, \ldots, \p^n)\bigr)
=
I(\vc{w}) + \sum_{i = 1}^n w_i I(\p^i)
\]
($n, k_1, \ldots, k_n \geq 1$, $\vc{w} \in \Delta_n$, $\vc{p}^i \in
\Delta_{k_i}$);

\item
\lbl{part:faddeev-form}
$I = cH$ for some $c \in \R$.
\end{enumerate}
\end{thm}

In other words, up to a constant factor, entropy is uniquely characterized
by the chain rule and continuity.  We already know
that~\bref{part:faddeev-form} implies~\bref{part:faddeev-condns}; the
challenge is to show that~\bref{part:faddeev-condns}
implies~\bref{part:faddeev-form}. 

\begin{remarks}
\lbl{rmks:faddeev}
\begin{enumerate}
\item 
As noted in Remark~\ref{rmk:ent-base}, the appearance of the constant
factor should not be a surprise.  We could eliminate it by adding the axiom
that $I(\vc{u}_2) = \log 2$, for instance.

\item
\lbl{rmk:faddeev-sym} 
\index{symmetry in Faddeev-type theorems}
The theorem that Faddeev proved
in~\cite{Fadd} was slightly different.  He assumed that $I$ was symmetric,
that is, unchanged when the arguments $p_1, \ldots, p_n$ are permuted, but
he assumed only the superficially simpler form of the chain rule stated as
equation~\eqref{eq:ent-chain-simp1} (Remark~\ref{rmk:ent-chain-simp}).  As
noted in that remark, if we assume symmetry then the two forms
of the chain rule are equivalent via a straightforward induction
(Appendix~\ref{sec:chain}).  On the other hand, Theorem~\ref{thm:faddeev}
tells us that if we assume the chain rule in its general form then we do
not need symmetry.  This is not an obvious consequence of Faddeev's
original theorem.

\item
\lbl{rmk:faddeev-lee}
If we assume symmetry, the hypotheses of Faddeev's original theorem can be
weakened in a different direction, replacing
continuity by measurability.%
\index{measurability!entropy@of entropy}
This is a 1964 theorem of Lee~\cite{Lee}%
\index{Lee, Pan-Mook}.
We return to Lee's theorem at the end of
Chapter~\ref{ch:p}, but omit the proof.  

\item
\lbl{rmk:faddeev-reg}
It is not possible to prove a Faddeev-type theorem with no regularity
conditions at all (unless one drops the axiom of choice).  Indeed, let $f
\from \R \to \R$ be an additive nonlinear function, as in
Remark~\ref{rmk:choice}.  Then the assignment
\[
\p \mapsto - \sum_{i \in \supp(\p)} p_i f(\log p_i)
\]
satisfies the chain rule but is not a scalar multiple of Shannon
entropy. 
\end{enumerate}
\end{remarks}

The remainder of this section is devoted to the proof of
Theorem~\ref{thm:faddeev}.  \femph{For the rest of this section}, let $(I
\from \Delta_n \to \R)_{n \geq 1}$ be a sequence of continuous functions
satisfying the chain rule.

The strategy of the proof is to show that $I$ is proportional to $H$ on
successively larger classes of probability distributions.  First we prove
it for the uniform distributions $\vc{u}_n$, using the results on
logarithmic sequences in Section~\ref{sec:log-seqs}.  This forms the bulk
of the proof.  It is then relatively easy to extend the result to
distributions $\p$ for which each $p_i$ is a positive rational number, and
from there, by continuity, to all distributions.

We begin by studying the real sequence $(I(\vc{u}_n))_{n \geq 1}$.

\begin{lemma}
\lbl{lemma:fad-log}
\begin{enumerate}
\item 
\lbl{part:fad-two}
$I(\vc{u}_{mn}) = I(\vc{u}_m) + I(\vc{u}_n)$ for all $m, n \geq 1$.

\item
\lbl{part:fad-zero}
$I(\vc{u}_1) = 0$.
\end{enumerate}
\end{lemma}

\begin{proof}
By the chain rule, $I$ has the logarithmic property
\[
I(\vc{w} \otimes \vc{p}) =
I\bigl(\vc{w} \of (\p, \ldots, \p)\bigr) =
I(\vc{w}) + I(\vc{p})
\]
($\vc{w} \in \Delta_m$, $\vc{p} \in \Delta_n$).  In particular,
for all $m, n \geq 1$,
\[
I(\vc{u}_{mn}) 
= 
I(\vc{u}_m \otimes \vc{u}_n) 
=
I(\vc{u}_m) + I(\vc{u}_n),
\]
proving~\bref{part:fad-two}.  For~\bref{part:fad-zero}, take $m = n = 1$
in~\bref{part:fad-two}.
\end{proof}

As we saw in Section~\ref{sec:log-seqs}, the property $I(\vc{u}_{mn}) =
I(\vc{u}_m) + I(\vc{u}_n)$ alone does not tell us very much about the
sequence $(I(\vc{u}_n))$.  To take advantage of the results in that
section, we will need to prove some analytic condition on the sequence.
Specifically, we will show that $I(\vc{u}_{n + 1}) - I(\vc{u}_n) \to 0$ as
$n \to \infty$, then apply Corollary~\ref{cor:erdos-lim}.
\pagebreak

\begin{lemma}
\lbl{lemma:fad-10}
$I(1, 0) = 0$.
\end{lemma}

\begin{proof}
We compute $I(1, 0, 0)$ in two ways.  On the one hand, using the chain rule,
\[
I(1, 0, 0)
=
I\Bigl((1, 0) \of \bigl((1, 0), \vc{u}_1\bigr)\Bigr)
=
I(1, 0) + 1 \cdot I(1, 0) + 0 \cdot I(\vc{u}_1)
=
2I(1, 0).
\]
On the other, using the chain rule again and the fact that $I(\vc{u}_1) = 0$,
\[
I(1, 0, 0)
=
I\Bigl((1, 0) \of \bigl(\vc{u}_1, (1, 0)\bigr)\Bigr)
=
I(1, 0) + 1 \cdot I(\vc{u}_1) + 0 \cdot I(1, 0)
=
I(1, 0).
\]
Hence $I(1, 0) = 0$.
\end{proof}

\begin{lemma}
\lbl{lemma:diff-nearly-zero}
$I(\vc{u}_{n + 1}) - \tfrac{n}{n + 1} I(\vc{u}_n) \to 0$ as $n \to \infty$.
\end{lemma}

\begin{proof}
We have
\[
\vc{u}_{n + 1} 
=
\biggl( \frac{n}{n + 1}, \frac{1}{n + 1} \biggr) \of (\vc{u}_n, \vc{u}_1),
\]
so by the chain rule and the fact that $I(\vc{u}_1) = 0$,
\[
I(\vc{u}_{n + 1})
=
I\biggl( \frac{n}{n + 1}, \frac{1}{n + 1} \biggr)
+
\frac{n}{n + 1} I(\vc{u}_n).
\]
Hence
\[
I(\vc{u}_{n + 1}) - \frac{n}{n + 1} I(\vc{u}_n)
=
I\biggl( \frac{n}{n + 1}, \frac{1}{n + 1} \biggr)
\to 
I(1, 0)
=
0
\]
as $n \to \infty$, by continuity and Lemma~\ref{lemma:fad-10}.
\end{proof}

Now we can use the results of Section~\ref{sec:log-seqs}.

\begin{lemma}
\lbl{lemma:sh-ent-u}
There exists a constant $c \in \R$ such that $I(\vc{u}_n) = cH(\vc{u}_n)$
for all $n \geq 1$.
\end{lemma}

\begin{proof}
By Lemma~\ref{lemma:fad-log}\bref{part:fad-two}, the sequence
$(I(\vc{u}_n))$ is logarithmic.  By Lemmas~\ref{lemma:diff-nearly-zero}
and~\ref{lemma:seq-improvement}, $\lim_{n \to \infty} \bigl( I(\vc{u}_{n +
  1}) - I(\vc{u}_n)\bigr) = 0$.  Hence by Corollary~\ref{cor:erdos-lim},
there is some $c \in \R$ such that for all $n \geq 1$,
\[
I(\vc{u}_n) 
=
c \log n
=
c H(\vc{u}_n).
\]
\end{proof}

We now move to the second phase of the proof of Theorem~\ref{thm:faddeev}.
Let $c$ be the constant of Lemma~\ref{lemma:sh-ent-u} (which is uniquely
determined).  

\begin{lemma}
\lbl{lemma:fad-rational}
Let $\p \in \Delta_n$ with $p_1, \ldots, p_n$ rational and nonzero.  Then
$I(\p) = cH(\p)$. 
\end{lemma}

\begin{proof}
We can write 
\[
\p = \biggl( \frac{k_1}{k}, \ldots, \frac{k_n}{k} \biggr)
\]
for some positive integers $k_1, \ldots, k_n$, where $k = k_1 + \cdots +
k_n$.  Then
\[
\p \of (\vc{u}_{k_1}, \ldots, \vc{u}_{k_n}) = \vc{u}_k.
\]
Since $I$ satisfies the chain rule and $I(\vc{u}_r) = cH(\vc{u}_r)$ for
all $r \geq 1$, we have
\[
I(\p) + \sum_{i = 1}^n p_i \cdot cH(\vc{u}_{k_i}) = cH(\vc{u}_k).
\]
But since $cH$ satisfies the chain rule too, we also have
\[
cH(\p) + \sum_{i = 1}^n p_i \cdot cH(\vc{u}_{k_i}) = cH(\vc{u}_k).
\]
The result follows.
\end{proof}

The third and final phase of the proof is trivial: since the probability
distributions with positive rational probabilities are dense in the space
$\Delta_n$ of all probability distributions, and since $I$ and $cH$
are continuous functions agreeing on this dense set, they are equal
everywhere.  This proves Theorem~\ref{thm:faddeev}.

Like any result on entropy, Faddeev's theorem can be translated into
diversity terms.  In the following corollary, we eliminate the arbitrary
constant factor by requiring that $E$ be an effective number.  

\begin{cor}
\lbl{cor:fad-div}
Let $\bigl( E \from \Delta_n \to (0, \infty) \bigr)_{n \geq 1}$ be a
sequence of functions.  The following are equivalent:
\begin{enumerate}
\item 
\lbl{part:fad-div-condns}
the functions $E$ are continuous and satisfy the chain rule
\begin{equation}
\lbl{eq:E-div-chain}
E\bigl(\vc{w} \of (\p^1, \ldots, \p^n)\bigr)
=
E(\vc{w}) \cdot \prod_{i = 1}^n E(\p^i)^{w_i}
\end{equation}
($n, k_1, \ldots, k_n \geq 1$, $\vc{w} \in \Delta_n$, $\vc{p}^i \in
\Delta_{k_i}$), and $E$ is an effective number;

\item
\lbl{part:fad-div-form}
$E = D$.
\end{enumerate}
\end{cor}

\begin{proof}
By Faddeev's theorem applied to $\log E$, the sequences of continuous
functions $E$ satisfying the diversity chain rule~\eqref{eq:E-div-chain}
are exactly the real powers $D^c$ ($c \in \R$).  But the effective number
property (or indeed, the single equation $E(\vc{u}_2) = 2$) then forces $c
= 1$.
\end{proof}

%% file: rel.tex
\chapter{Relative entropy}
\lbl{ch:rel}
\index{relative entropy}

The notion of relative entropy allows us to compare two probability
distributions on the same space.  More specifically, for each pair of
probability distributions $\p, \vc{r}$ on the same finite set, there is
defined a real number $\relent{\p}{\vc{r}} \geq 0$, the entropy of $\p$
relative to $\vc{r}$.  It is zero just when $\p = \vc{r}$.  It extends the
definition of Shannon entropy, in the sense that the Shannon entropy of a
single distribution $\p$ on $\{1, \ldots, n\}$ is a function of
$\relent{\p}{\vc{u}_n}$, the entropy of $\p$ relative to the uniform
distribution.

Relative entropy goes by a remarkable number of names,%
\index{relative entropy!names for} 
attesting to its wide variety of interpretations and uses.  It is also
known as Kullback--Leibler information (as in \cite{ScheTS}, for instance),
Kullback--Leibler distance~\cite{CoTh1}, Kullback--Leibler
divergence~\cite{Joyc}, directed divergence~\cite{KullITS}, information
divergence~\cite{GrNi}, information deficiency~\cite{BJTV}, amount of
information~\cite{Reny}, discrimination information~\cite{KullKLD},
relative information~\cite{TaKu}, gain of information or information gain
(\cite{RenyPT}, Section~IX.4), discrimination distance~\cite{KantDDB}, and
error~\cite{Kerr}, among others.  This chapter provides multiple
explanations and applications of relative entropy, as well as a theorem
pinpointing what makes relative entropy uniquely useful.

Our first explanation of relative entropy is in terms of coding
(Section~\ref{sec:rel-coding}).  As we saw in Section~\ref{sec:ent-coding},
the Shannon entropy of $\p$ gives the average number of bits per
symbol needed to encode an alphabet with frequency distribution $\p$ in a
coding system optimized for that purpose.  In a similar sense,
$\relent{\p}{\vc{r}}$ measures the \emph{extra} number of bits per symbol
needed to encode an alphabet with frequencies $\p$ using a coding system
that was optimized for the frequency distribution $\vc{r}$.  In other
words, it is the penalty for using the wrong system.

The exponential of relative entropy is called relative diversity
(Section~\ref{sec:rel-div}).  Often we have a preconceived idea of what
an ordinary or default distribution of species is, and we judge how unusual
a community is relative to that expectation.  For instance, if we were
assessing the diversity of flowering plants in a particular region of
the island of Tasmania, we would naturally judge it by the standards of
Tasmania as a whole.  The relative diversity $\exp(\relent{\p}{\vc{r}})$
reflects the unusualness of a community with distribution $\p$ relative to
a reference distribution $\vc{r}$.

Section~\ref{sec:rel-misc} gives short accounts of roles played by relative
entropy in three other subjects.  In measure theory, we find that the
definition of relative entropy generalizes easily from finite sets to
arbitrary measurable spaces, while ordinary Shannon entropy does not.  The
slogan is: \emph{all%
\index{all entropy is relative} 
entropy is relative}.  In geometry, although $\relent{-}{-}$ does not
define a distance function on the set $\Delta_n$ of distributions, it turns
out that infinitesimally, it behaves like the square of a distance.  We can
extend this infinitesimal metric to a global metric in the manner of
Riemannian geometry.  In statistics, the second argument $\vc{r}$ of
$\relent{\p}{\vc{r}}$ should be thought of as a prior\index{prior}, and
maximizing likelihood can be reinterpreted as minimizing relative entropy.
The concept of relative entropy also gives rise to the notions of Fisher
information and the Jeffreys prior, an objective prior distribution in the
sense of Bayesian statistics.

We finish the chapter with a characterization theorem for relative entropy
(Section~\ref{sec:rel-char}), which first appeared in~\cite{SCRE}.  Just as
for Faddeev's characterization of ordinary entropy, the main characterizing
property is a chain rule.  And just as for ordinary entropy, many
characterization theorems for relative entropy have previously been proved;
but the one presented here appears to be the simplest yet.

\section{Definition and properties of relative entropy}
\lbl{sec:rel-defn}

This short section presents the definition and basic properties of
relative entropy, without motivation for now.  The later sections provide
multiple interpretations of, and justifications for, the definition.

\begin{defn}
\lbl{defn:rel-ent}
Let $n \geq 1$ and $\vc{p}, \vc{r} \in \Delta_n$.  The
\demph{entropy%
\index{relative entropy} 
of $\vc{p}$ relative to $\vc{r}$} is
\begin{equation}
\lbl{eq:defn-rel-ent}
\relent{\vc{p}}{\vc{r}}
=
\sum_{i \in \supp(\p)} p_i \log \frac{p_i}{r_i}.
\end{equation}
If there is some $i$ such that $p_i > 0 = r_i$ then
$\relent{\vc{p}}{\vc{r}}$ is defined to be $\infty$.
\end{defn}

In the literature, relative entropy is more often denoted by
$\reldiv{\vc{p}}{\vc{r}}$, but in this text, we reserve the letter $D$ for
measures of diversity.

\begin{example}
\lbl{eg:rel-ent-ufm}
Let $\p \in \Delta_n$.  Then
\begin{align*}
\relent{\p}{\vc{u}_n}   &
=
\sum_{i \in \supp(\p)} p_i \log(np_i)   \\
&
=
\log n - H(\p)  \\
&
=
H(\vc{u}_n) - H(\p).
\end{align*}
Thus, ordinary entropy is essentially a special case of relative entropy.
\end{example}

\begin{example}
\lbl{eg:rel-ent-large}
As well as sometimes taking the value $\infty$, relative entropy can take
arbitrarily large finite values (even for fixed $n$).  For instance, for $t
\in (0, 1)$, 
\[
\relEnt{\vc{u}_2}{(t, 1 - t)} 
=
\frac{1}{2} \log \frac{1}{2t} + \frac{1}{2} \log \frac{1}{2(1 - t)}
\to 
\infty
\]
as $t \to 0$.  
\end{example}

Unless $\vc{p} = \vc{r}$, there are some values of $i$ for which $p_i >
r_i$ and others for which $p_i < r_i$.  Hence, some of the summands
in~\eqref{eq:defn-rel-ent} are positive and others are negative.
Nevertheless:

\begin{lemma}
\lbl{lemma:rel-ent-pos-def}
$\relent{\vc{p}}{\vc{r}} \geq 0$, with equality if and only if $\vc{p} =
  \vc{r}$. 
\end{lemma}

\begin{proof}
If $p_i > 0 = r_i$ for some $i$ then $\relent{\vc{p}}{\vc{r}} = \infty$.
Suppose otherwise, so that $\supp(\p) \sub \supp(\vc{r})$.  Using
Lemma~\ref{lemma:log-concave},
\begin{align*}
\relent{\vc{p}}{\vc{r}} &
=
- \sum_{i \in \supp(\p)} p_i \log \frac{r_i}{p_i}       \\
&
\geq    
- \log \Biggl( \sum_{i \in \supp(\p)} p_i \frac{r_i}{p_i} \Biggr)       \\
&
\geq
- \log \Biggl( \sum_{i \in \supp(\vc{r})} r_i \Biggr)   \\
&
=
- \log 1 
=
0,
\end{align*}
with equality in the first inequality if and only if $r_i/p_i = r_j/p_j$
for all $i, j \in \supp(\p)$.  Equality holds in the second inequality if
and only if $\supp(\p) = \supp(\vc{r})$.  Hence for equality to hold
throughout, there must be some constant $\alpha$ such that $r_i = \alpha
p_i$ for all $i \in \supp(\p) = \supp(\vc{r})$.  But since $\sum_{i \in
  \supp(\p)} p_i = 1 = \sum_{i \in \supp(\vc{r})} r_i$, this forces $\alpha
= 1$ and so $\vc{p} = \vc{r}$.
\end{proof}

Lemma~\ref{lemma:rel-ent-pos-def} suggests that \emph{very} roughly
speaking, $\relent{\vc{p}}{\vc{r}}$ can be understood as a kind of
distance%
\index{relative entropy!metric@as metric} 
between $\vc{p}$ and $\vc{r}$.  However, relative entropy does not satisfy
the triangle inequality (Example~\ref{eg:rel-tri}).  Nor is it symmetric:%
\index{relative entropy!asymmetry of}
for as Examples~\ref{eg:rel-ent-ufm} and~\ref{eg:rel-ent-large} show,
$\relent{\p}{\vc{u}_2} \leq \log 2$ for all $\p \in \Delta_2$, whereas
$\relent{\vc{u}_2}{\p}$ can be arbitrarily large.  We will return to the
interpretation of relative entropy as a measure of distance in
Section~\ref{sec:rel-misc}.

We now list some of the basic properties of relative entropy.  Matters are
simplified if we restrict to just those pairs $(\vc{p}, \vc{r})$ such that
$\relent{\vc{p}}{\vc{r}} < \infty$.  For $n \geq 1$, write
\begin{align*}
\lbl{p:An}
A_n     &
=
\{ (\vc{p}, \vc{r}) \in \Delta_n \times \Delta_n \such
r_i = 0 \implies p_i = 0 \}     \\
&
=
\{ (\vc{p}, \vc{r}) \in \Delta_n \times \Delta_n \such
\supp(\vc{p}) \sub \supp(\vc{r}) \}.
\end{align*}
Then 
\[
\relent{\vc{p}}{\vc{r}} < \infty \iff (\vc{p}, \vc{r}) \in A_n.  
\]
So for each $n \geq 1$, we have the function
\begin{equation*}
\begin{array}[c]{cccc}
\relent{-}{-}\from      &A_n            &\to    &\R    \\
                        &(\vc{p}, \vc{r})&\mapsto&\relent{\vc{p}}{\vc{r}}.
\end{array}
\end{equation*}
This sequence of functions has the following properties, among others.
\begin{description}
\item[Measurability%
\index{measurability!relative entropy@of relative entropy}%
\index{relative entropy!measurability of}
in the second argument.]
For each fixed $\vc{p} \in \Delta_n$, the function
\[
\begin{array}{ccc}
\{ \vc{r} \in \Delta_n \such (\vc{p}, \vc{r}) \in A_n \}        &
\to     &
\R     \\
\vc{r}  &
\mapsto &
\relent{\vc{p}}{\vc{r}}
\end{array}
\]
is measurable.  Indeed, the function $\relent{-}{-} \from A_n \to \R$ is
continuous, but for the unique characterization of relative entropy proved
in Section~\ref{sec:rel-char}, measurability in the second argument is all
we will need.

\item[Permutation-invariance.]%
\index{permutation-invariance}%
\index{relative entropy!permutation-invariance of} 
The relative entropy $\relent{\p}{\vc{r}}$ is unchanged if the same
permutation is applied to the indices of both $\vc{p}$ and $\vc{r}$.  That
is,
\[
\relent{\p}{\vc{r}}
=
\relent{\p\sigma}{\vc{r}\sigma}
\]
for all $(\p, \vc{r}) \in A_n$ and permutations $\sigma$ of $\{1, \ldots,
n\}$, where
\begin{equation}
\lbl{eq:p-perm}
\p\sigma = (p_{\sigma(1)}, \ldots, p_{\sigma(n)})
\end{equation}
and similarly $\vc{r}\sigma$.  

\item[Vanishing.]%
\index{vanishing}%
\index{relative entropy!vanishing of}
$\relent{\p}{\p} = 0$ for all $\p \in \Delta_n$.

\item[Chain rule.]
\index{chain rule!relative entropy@for relative entropy}%
\index{relative entropy!chain rule for}
Let $n, k_1, \ldots, k_n \geq 1$ and
\[
(\w, \widetilde{\w}) \in A_n, \  
\bigl(\p^1, \widetilde{\p}^1\bigr) \in A_{k_1}, \ 
\ldots, \ 
\bigl(\p^n, \widetilde{\p}^n\bigr) \in A_{k_n}.
\]
Then
\begin{equation}
\lbl{eq:rel-ch}
\relEnt{\w \of (\p^1, \ldots, \p^n)}
{\widetilde{\w} \of (\widetilde{\p}^1, \ldots, \widetilde{\p}^n)}
=
\relent{\w}{\widetilde{\w}} + \sum_{i = 1}^n w_i
\relent{\p^i}{\widetilde{\p}^i}. 
\end{equation}
This is a straightforward check, similar to
Proposition~\ref{propn:ent-chain}.  Note that
\[
\bigl(\w \of (\p^1, \ldots, \p^n),
\widetilde{\w} \of (\widetilde{\p}^1, \ldots, \widetilde{\p}^n)\bigr)
\in 
A_{k_1 + \cdots + k_n},
\]
so the relative entropy of this pair is guaranteed to be finite.

As a special case, relative entropy has a logarithmic property:
\begin{equation}
\lbl{eq:rel-log}
\relent{\w \otimes \p}{\widetilde{\w} \otimes \widetilde{\p}}
=
\relent{\w}{\widetilde{\w}} + \relent{\p}{\widetilde{\p}}
\end{equation}
for all $(\w, \widetilde{\w}) \in A_n$ and $(\p, \widetilde{\p}) \in A_k$.
This follows from the chain rule by taking $k_i = k$, $\p^i = \p$ and
$\widetilde{\p}^i = \widetilde{\p}$ for all $i \in \{1, \ldots, n\}$.  
\end{description}

Just as for ordinary entropy, different choices of the base of the
logarithm in the definition of relative entropy only change it by a
constant factor.  We will see in Section~\ref{sec:rel-char} that up to a
constant factor, the four properties just listed characterize relative
entropy uniquely.

\section{Relative entropy in terms of coding}
\lbl{sec:rel-coding}
\index{relative entropy!coding@and coding}

We have already interpreted Shannon entropy in terms of coding
(Section~\ref{sec:ent-coding}).  Here we do the same for relative entropy.  

To help our understanding, let us regard a probability distribution $\p \in
\Delta_n$ as the frequency distribution of the $n$ symbols in some human
language, which we call \demph{language%
\index{language p@language $\p$}
$\p$}.  We make use of the convenient fiction introduced on
p.~\pageref{p:fiction}, imagining that there exists an ideal code for
language $\p$: a code whose average word length is exactly $\Hi(\p)$.  We
will suppose that the encoding is performed by a machine, called
\demph{machine%
\index{machine p@machine $\p$} 
$\p$}.  Although most distributions $\p$ have no ideal code, one can come
arbitrarily close (as in Section~\ref{sec:ent-coding}), and this justifies
the use of ideal codes as an explanatory device.

For $\p \in \Delta_n$, the ordinary base~$2$ entropy
\[
\Hi(\p) = \sum_{i \in \supp(\p)} p_i \log_2 \frac{1}{p_i}
\]
satisfies
\[
\Hi(\p)
=
\text{no.\ bits/symbol to encode language } \p \text{ using machine } \p.
\]
Now let $\p, \vc{r} \in \Delta_n$, with $\p$ and $\vc{r}$ viewed as the
frequency distributions of two languages on the same set of symbols.  Write
\[
\relenti{\p}{\vc{r}}
=
\sum_{i \in \supp(\p)} p_i \log_2 \frac{p_i}{r_i}
=
\frac{\relent{\p}{\vc{r}}}{\log 2}
\ntn{relenti}
\]
for the base~$2$ relative entropy.  We will interpret
$\relenti{\p}{\vc{r}}$ in terms of languages $\p$ and $\vc{r}$ and machines
$\p$ and $\vc{r}$.

To do this, first consider the quantity
%
\[
\crossenti{\p}{\vc{r}}
=
\sum_{i \in \supp(\p)} p_i \log_2 \frac{1}{r_i}.
\ntn{crossenti}
\]
%
Here $\log_2(1/r_i)$ is the number of bits that machine $\vc{r}$ uses to
encode the $i$th symbol.  (Of course, this is not usually an integer, but
recall the comments on ideal codes on p.~\pageref{p:ideal}.)  Hence
\[
\crossenti{\p}{\vc{r}}
=
\text{no.\ bits/symbol to encode language } \p \text{ using machine } \vc{r}.
\]
This quantity $\crossenti{\p}{\vc{r}}$, or its base~$e$ analogue
\begin{equation}
\lbl{eq:crossent}
\crossent{\p}{\vc{r}} 
= 
\sum_{i \in \supp(\p)} p_i \log \frac{1}{r_i}
=
\crossenti{\p}{\vc{r}}\cdot \log 2,
\end{equation}
is the \demph{cross%
\index{cross entropy} 
entropy} of $\p$ with respect to $\vc{r}$.

The relative, cross and ordinary entropies are related by the equation
\begin{equation}
\lbl{eq:rco}
\relent{\p}{\vc{r}} = \crossent{\p}{\vc{r}} - H(\p).
\end{equation}
Hence
\begin{align*}
\relenti{\p}{\vc{r}}    &
=
\crossenti{\p}{\vc{r}} - \Hi(\p)        
\\
&
=
\text{no.\ bits/symbol to encode language } \p \text{ using machine }
\vc{r}  
\\
&
\phantom{= \mbox{}}
{} - \text{no.\ bits/symbol to encode language } \p 
\text{ using machine } \p.
\end{align*}
So, for the task of encoding language $\p$, the relative entropy
$\relenti{\p}{\vc{r}}$ is the number of \emph{extra} bits needed if one
uses machine $\vc{r}$ instead of machine $\p$.  Machine $\p$ is ideal for
the job: it is optimized for exactly this purpose.  Relative entropy is,
then, the penalty for using the wrong machine.

This provides an intuitive explanation of why $\relent{\p}{\vc{r}}$ is
always nonnegative and why $\relent{\p}{\p} = 0$.  It also suggests why
relative entropy can be arbitrarily large, as in the following example.

\begin{examples}
\lbl{egs:relenti-large-sym}
\begin{enumerate}
\item
\lbl{eg:relenti-ls-large}
Consider an alphabet with $n = 2$ symbols.  Suppose that language $\p$ uses
the two symbols with equal frequency, and that in language $\vc{r}$ the
frequency distribution is $(2^{-1000}, 1 - 2^{-1000})$.  Then machine
$\vc{r}$ encodes the first symbol with a word of $1000$ bits.  Since
language $\p$ uses this symbol half the time, the average word length when
encoding language $\p$ using machine $\vc{r}$ is at least $500$ bits.  This
is drastically worse than when language $\p$ is encoded using the most
suitable machine, machine $\p$, which has an average word length of just
$1$ bit.  So the relative entropy $\relenti{\p}{\vc{r}}$ is at least $499$.

\item
The same example provides intuition for the fact that
\[
\relent{\p}{\vc{r}} \neq \relent{\vc{r}}{\p}.
\]
Machine $\p$ encodes the two symbols of the alphabet as the binary words
$0$ and $1$, of length $1$ each.  Hence the average number of bits used when
encoding language $\vc{r}$ (or indeed, any other language) in machine $\p$
is $1$.  So $\relenti{\vc{r}}{\p}$ is less than $1$, and is therefore much
smaller than the value of $\relenti{\p}{\vc{r}}$ derived
in~\bref{eg:relenti-ls-large}.
\end{enumerate}
\end{examples}

\begin{remark}
\index{cross entropy}
The name `cross entropy' has a tangled history.  It was introduced by Jack
Good%
\index{Good, Jack} 
in 1955 (\cite{GoodSTN}, Section~6), who defined it as in
equation~\eqref{eq:crossent} above and gave it its name.  But later, Good
used `cross entropy' as a synonym for relative entropy (\cite{GoodMEH},
p.~913), and others have done the same (Shore and Johnson~\cite{ShJo}, for
instance).  Nowadays the term is often used in the context of the cross
entropy method in operational research~\cite{DKMR}.  In the broadest terms,
this involves fixing a distribution $\vc{p}$ and minimizing
$\relent{\p}{\vc{r}}$, or equivalently $\crossent{\p}{\vc{r}}$, among all
$\vc{r}$ subject to certain constraints.  It makes no difference which one
minimizes, by equation~\eqref{eq:rco}.  From that point of view, the
concepts are essentially interchangeable, which has not helped to clarify
the terminological situation either.

This text uses the term with its original meaning, in part because relative
entropy already has an overabundance of synonyms.
\end{remark}

The chain rule for relative entropy (equation~\eqref{eq:rel-ch}) can also
be explained in terms of coding, as in the following example.

\begin{example}
\index{French language}%
\index{Swiss French}%
\index{Canadian French}
In Example~\ref{eg:ent-french}, we interpreted the chain rule for ordinary
Shannon entropy in terms of letters and their accents in French.  There are
many dialects of French, using the same letters and accents but slightly
different vocabulary, hence slightly different frequency distributions of
both letters and accents.  Here we consider Swiss and Canadian French,
which for brevity we call just `Swiss' and `Canadian'.

Define distributions $\vc{w}, \twid{\vc{w}}, \p^i, \twid{\p}^i$ as follows:
\begin{align*}
\vc{w} \in \Delta_{26}: &
\text{ frequency distribution of letters in Swiss}     \\
\twid{\vc{w}} \in \Delta_{26}:   &
\text{ frequency distribution of letters in Canadian}
\end{align*}
and then
\begin{align*}
\vc{p}^1 \in \Delta_3:  &
\text{ frequency distribution of accents on \as{a} in Swiss}      \\
\twid{\vc{p}}^1 \in \Delta_3:  &
\text{ frequency distribution of accents on \as{a} in Canadian}      \\
\vdots  \\
\vc{p}^{26} \in \Delta_1:  &
\text{ frequency distribution of accents on \as{z} in Swiss}      \\
\twid{\vc{p}}^{26} \in \Delta_1:  &
\text{ frequency distribution of accents on \as{z} in Canadian}.
\end{align*}
So, recalling the convention that a `symbol' is a letter plus a (possibly
nonexistent) accent,
\begin{align*}
\vc{w} \of \bigl(\vc{p}^1, \ldots, \vc{p}^{26}\bigr)      &
=
\text{frequency distribution of symbols in Swiss}  \\
\twid{\vc{w}} \of \bigl(\twid{\vc{p}}^1, \ldots, \twid{\vc{p}}^{26}\bigr)&
=
\text{frequency distribution of symbols in Canadian}.
\end{align*}
Now suppose that we encode Swiss using the Canadian machine.  How much
extra does this cost (in bits/symbol) compared to encoding Swiss using the
Swiss machine?

Since every symbol consists of a letter with an accent, we expect to have:
\begin{multline}
\lbl{eq:relent-french}
\text{mean extra cost per symbol}
=\\
\text{mean extra cost per letter}
+
\text{mean extra cost per accent}.
\end{multline}
The mean extra cost per symbol is
\[
\relEnti{\vc{w} \of \bigl(\vc{p}^1, \ldots, \vc{p}^{26}\bigr)}%
{\twid{\vc{w}} \of \bigl(\twid{\vc{p}}^1, \ldots, \twid{\vc{p}}^{26}\bigr)}.
\]
The mean extra cost per letter is
\[
\relenti{\vc{w}}{\twid{\vc{w}}}.
\]
The mean extra cost per accent is computed by conditioning on the letter
that it decorates.  Since it is Swiss rather than Canadian that we are
encoding, the probability of the $i$th letter occurring is $w_i$, so
the mean extra cost per accent is
\[
\sum_{i = 1}^{26} w_i \relEnti{\p^i}{\twid{\p}^i}.
\]
Hence the hoped-for equation~\eqref{eq:relent-french} predicts that
\[
\relEnti{\vc{w} \of \bigl(\vc{p}^1, \ldots, \vc{p}^{26}\bigr)}%
{\twid{\vc{w}} \of \bigl(\twid{\vc{p}}^1, \ldots, \twid{\vc{p}}^{26}\bigr)}
=
\relenti{\vc{w}}{\twid{\vc{w}}}
+
\sum_{i = 1}^{26} w_i \relEnti{\p^i}{\twid{\p}^i}.
\]
This is indeed true.  It is exactly the chain rule of
Section~\ref{sec:rel-defn}. 
\end{example}

\section{Relative entropy in terms of diversity}
\lbl{sec:rel-div}
\index{relative entropy!diversity@and diversity}

In Section~\ref{sec:ent-div}, we interpreted the exponential of Shannon
entropy as the diversity of a biological community.  Here we interpret the
exponential of relative entropy as a measure of how diverse or atypical one
community is when seen from the perspective of another.  This
interpretation elaborates on ideas of Reeve%
\index{Reeve, Richard} 
et al.~\cite{HPD}.

As in Section~\ref{sec:ent-div}, we consider communities of individuals
drawn from $n$ species, whose relative abundances define a probability
distribution on the set $\{1, \ldots, n\}$. 

\begin{defn}
Let $n \geq 1$ and $\p, \vc{r} \in \Delta_n$.  The \demph{diversity of $\p$
  relative to $\vc{r}$%
\index{relative diversity} 
(of order~$1$)} is 
\[
\reldiv{\p}{\vc{r}}
=
e^{\relent{\p}{\vc{r}}}
=
\prod_{i \in \supp(\p)} \biggl( \frac{p_i}{r_i} \biggr)^{p_i} 
\in 
[1, \infty].
\ntn{reldiv}
\]
\end{defn}

(We repeat the warning that although in the literature, the notation
$\reldiv{\p}{\vc{r}}$ is often used to mean relative \emph{entropy}, we
reserve the letter $D$ for diversity.)

By Lemma~\ref{lemma:rel-ent-pos-def}, $\reldiv{\p}{\vc{r}} \geq 1$, with
equality if and only if $\p = \vc{r}$.  

It is helpful to regard $\vc{r}$ as the distribution of a reference%
\index{community!reference}%
\index{reference community}
community (a community that one considers to be normal or the
default) and $\p$ as the distribution of the community in which we
are primarily interested.  As
we will see, $\reldiv{\p}{\vc{r}}$ measures how exotic or unusual this
other community is from the viewpoint of the reference community.

To explain this, it is helpful to begin with another quantity:
\begin{defn}
Let $n \geq 1$ and $\p, \vc{r} \in \Delta_n$.  The 
\demph{cross%
\index{cross diversity} 
diversity of $\p$ with respect to $\vc{r}$ (of order~$1$)} is
\[
\crossdiv{\p}{\vc{r}}
=
e^{\crossent{\p}{\vc{r}}}
=
\prod_{i \in \supp(\p)} 
\biggl( \frac{1}{r_i} \biggr)^{p_i}
\in
[1, \infty].
\ntn{crossdiv}
\]
\end{defn}

In Section~\ref{sec:ent-div}, the ordinary diversity of $\p$, 
\[
D(\p) = \prod_{i \in \supp(\p)} \biggl( \frac{1}{p_i} \biggr)^{p_i},
\]
was interpreted as follows: $1/p_i$ is the rarity\index{rarity} of the
$i$th species within the community, and $D(\p)$ is therefore the average
rarity of individuals in the community.  (In this case, `average' means
geometric mean.)
Cross diversity can be understood in a similar way.  If we use the second
community $\vc{r}$ as our reference point~-- the community by which others
are to be judged~-- then we naturally take the rarity or
specialness\index{specialness} of the $i$th species to be $1/r_i$ rather
than $1/p_i$.  Thus, $\crossdiv{\p}{\vc{r}}$ is the average rarity of
individuals in the first community, seen from the viewpoint of the second.

Since
\begin{equation}
\lbl{eq:rel-cross-div}
\reldiv{\p}{\vc{r}} 
=
\frac{\crossdiv{\p}{\vc{r}}}{D(\p)},
\end{equation}
the relative diversity measures how much \emph{more} diverse the first
community looks from the viewpoint of the second than from the viewpoint of
itself.  Some examples illuminate this interpretation.

\begin{example}
\lbl{eg:rel-div-same}
We have $\reldiv{\p}{\p} = 1$, which is the minimal
possible value of relative diversity: any community perceives itself as
completely normal.
\end{example}

\begin{example}
\lbl{eg:rel-div-liz} 
Let $\vc{p}$ and $\vc{r}$ be the relative abundance distributions of
reptiles in Portugal and Russia, respectively.  Geckos\index{geckos} are
commonplace in Portugal but rare in Russia.  Hence from the Russian
viewpoint, the ecology of Portugal seems exotic or atypical, in this
respect at least.

Mathematically, there are several values of $i$ (corresponding to species
of gecko) such that $r_i$ is small but $p_i$ is not.  This means
that the cross diversity contains some large factors, $(1/r_i)^{p_i}$, and
the relative diversity also contains some large factors, $(p_i/r_i)^{p_i}$.
Thus, both the cross diversity $\crossdiv{\p}{\vc{r}}$ and the relative
diversity $\reldiv{\p}{\vc{r}}$ are large, regardless of the diversity
$D(\p)$ of reptiles in Portugal.
\end{example}

\begin{example}
Taking the previous example to the extreme, if one or more species is
present in the test community $\vc{p}$ but absent in the reference
community $\vc{r}$ then $\reldiv{\p}{\vc{r}} = \infty$.
\end{example}

\begin{example}
\lbl{eg:rel-div-ufm} 
Suppose now that we judge communities from the
reference point of a community with a uniform distribution.  (This is in
some sense the canonical choice of reference, and is the one produced by
the maximum entropy method of statistics~\cite{JaynWDWS,BuMa}.)  The
cross diversity $\crossdiv{\p}{\vc{u}_n}$ is equal to $n$, regardless of
$\p$.  Hence equation~\eqref{eq:rel-cross-div} gives
\begin{equation}
\lbl{eq:rel-div-ufm}
\reldiv{\p}{\vc{u}_n}
=
\frac{n}{D(\p)}.
\end{equation}
This is also the exponential of the equation
\[
\relent{\p}{\vc{u}_n} = \log n - H(\p)
\]
derived in Example~\ref{eg:rel-ent-ufm}.

Equation~\eqref{eq:rel-div-ufm} implies that for a fixed number of species,
the diversity of a community relative to the uniform distribution is
inversely proportional to the intrinsic diversity of the community itself.
From the viewpoint of a community in which all $n$ species are balanced
equally, any variation from this balance looks unusual~-- and the more
unbalanced, the more unusual.  

As an illustration of the general point, house sparrows\index{sparrows} are
common throughout Britain, but a region of the country in which the
\emph{only} birds\index{birds} were house sparrows would be highly unusual.
Correspondingly, the relative diversity $\reldiv{\p}{\vc{r}}$ of that
region relative to the country would be high, even though the intrinsic
diversity $D(\p)$ of the region would take the minimum possible value, $1$.

By equation~\eqref{eq:rel-div-ufm} and Lemma~\ref{lemma:div1-max-min},
$\reldiv{\p}{\vc{u}_n}$ takes its minimal value, $1$, when $\vc{p} =
\vc{u}_n$.  It takes its maximal value, $n$, when
\[
\p = (0, \ldots, 0, 1, 0, \ldots, 0).
\]
That is, from the viewpoint of a completely balanced community, the most
unusual possible community is one consisting of a single species.
\end{example}

Often, we want to assess a community from the viewpoint of a larger
community that \emph{contains} it.  For instance, we are more likely to
study the diversity of plankton in the eastern Mediterranean Sea with reference
to the Mediterranean as a whole than with reference to the Arctic Ocean.

Consider, then, an ecological community with relative abundance
distribution $\vc{r} \in \Delta_n$, and a subcommunity consisting of some
of its organisms.  Write $\pi_i$ for the proportion of the community
consisting of individuals that belong to both the subcommunity and the
$i$th species.  Then $0 \leq \pi_i \leq r_i$.  The proportion of the whole
community made up by the subcommunity is $w = \sum \pi_i \leq 1$, and
the relative abundance distribution of the subcommunity is
\[
\p = (\pi_1/w, \ldots, \pi_n/w) \in \Delta_n.
\]
The inequality $\pi_i \leq r_i$ gives $w p_i \leq r_i$, or equivalently,
\begin{equation}
\lbl{eq:sc-bd}
\frac{p_i}{r_i} \leq \frac{1}{w},
\end{equation}
for all $i \in \supp(\vc{r})$.  Hence
\[
\reldiv{\p}{\vc{r}}
=
\prod_{i \in \supp(\p)} \biggl( \frac{p_i}{r_i} \biggr)^{p_i}
\leq
\prod_{i \in \supp(\p)} \biggl( \frac{1}{w} \biggr)^{p_i}
=
\frac{1}{w},
\]
giving
\begin{equation}
\lbl{eq:rel-div-sc-bounds}
1 \leq \reldiv{\p}{\vc{r}} \leq \frac{1}{w}.
\end{equation}
We now consider cases in which these bounds are attained.

\begin{examples}
\lbl{egs:rel-div-sc}
\begin{enumerate}
\item 
\lbl{eg:rel-div-sc-min}
By Lemma~\ref{lemma:rel-ent-pos-def}, $\reldiv{\p}{\vc{r}}$ attains its
lower bound of $1$ precisely when $\p = \vc{r}$.  This means that the
subcommunity is exactly typical or representative of the larger community;
it is minimally unusual.

\item
\lbl{eg:rel-div-sc-max}
The maximum $\reldiv{\p}{\vc{r}} = 1/w$ in~\eqref{eq:rel-div-sc-bounds} is
attained when $r_i = w p_i$ for all $i \in \supp(\p)$.  In the notation
above, this is equivalent to $\pi_i = r_i$ for all $i \in \supp(\p)$.  In
other words, the subcommunity is \demph{isolated}\index{isolated}: the
species occurring in the subcommunity occur nowhere else in the community.

If the isolated subcommunity is very small then its species distribution
appears highly unusual from the viewpoint of the whole community, and
correspondingly $\reldiv{\p}{\vc{r}} = 1/w$ is large.  But if, say, the
isolated subcommunity makes up $90\%$ of the whole community, then from the
viewpoint of the whole, the ecology of the subcommunity looks very typical.
So, it is intuitively reasonable that $\reldiv{\p}{\vc{r}} = 1/0.9$ is
close to the minimal possible value of $1$ .
\end{enumerate}
\end{examples}

The difference between relative diversity and cross diversity is
illustrated by the case of a uniform reference community
(Example~\ref{eg:rel-div-ufm}) and by the following example.

\begin{example}
\lbl{eg:three-scs} 
Consider a community with species distribution $\vc{r}$, containing a
subcommunity with species distribution $\vc{q}$, which in turn contains a
subcommunity with species distribution $\vc{p}$
(Figure~\ref{fig:three-scs}).
\begin{figure}
\centering
\lengths
\begin{picture}(120,30)
\cell{60}{15}{c}{\includegraphics[height=30\unitlength]{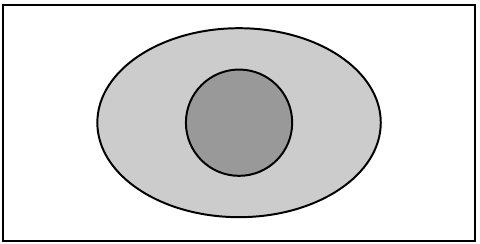}}
\cell{63}{18}{c}{$\vc{p}$}
\cell{71.5}{21}{c}{$\vc{q}$}
\cell{86}{26}{c}{$\vc{r}$}
\end{picture}
\caption{The three communities of Example~\ref{eg:three-scs}.}
\lbl{fig:three-scs}
\end{figure}
Suppose that the two subcommunities consist only of species that are rare
in the whole community, with $r_i = 1/100$ for all $i \in \supp(\vc{q})$.
The larger subcommunity consists of $50$ such species, and the smaller of
just one. 

For the smaller subcommunity,
\[
D(\p) = 1,
\quad
\crossdiv{\p}{\vc{r}} = 100,
\quad
\reldiv{\p}{\vc{r}} = 100.
\]
Indeed, $D(\p) = 1$ since the subcommunity contains just one species. For
the cross diversity, $1/r_i = 100$ for all $i \in \supp(\p)$, and
$\crossdiv{\p}{\vc{r}}$ is the geometric mean of $1/r_i$ over $i \in
\supp(\p)$, so $\crossdiv{\p}{\vc{r}} = 100$.  Then $\reldiv{\p}{\vc{r}} =
100/1 = 100$.

For the larger subcommunity,
\[
D(\vc{q}) = 50,
\quad
\crossdiv{\vc{q}}{\vc{r}} = 100,
\quad
\reldiv{\vc{q}}{\vc{r}} = 2,
\]
by a similar argument.  

This can be understood as follows.  From the viewpoint of the whole
community, the average rarity of the individuals in either subcommunity
is~$100$.  This is why both have a cross diversity of~$100$.  But the
larger subcommunity looks less unusual than the smaller one, because it
occupies more of the community and therefore resembles it more closely.
This is why the relative diversity of the larger subcommunity is lower.
\end{example}

\begin{remark}
\lbl{rmk:abg-rel} 
In ecology, there are concepts of alpha-\index{alpha-diversity},
beta-\index{beta-diversity} and gamma-diversity\index{gamma-diversity}.
\index{diversity!partitioning of}%
\index{partitioning of diversity}
The quantities $D(\p)$, $\reldiv{\p}{\vc{r}}$ and $\crossdiv{\p}{\vc{r}}$
are, respectively, kinds of alpha-, beta- and gamma-diversities, and
equation~\eqref{eq:rel-cross-div} is a version of the equation $\beta =
\gamma/\alpha$ that appears in the ecological literature (beginning with
Whittaker~\cite{WhitVSM},%
\index{Whittaker, Robert}
p.~321).

However, $D(\p)$, $\reldiv{\p}{\vc{r}}$ and $\crossdiv{\p}{\vc{r}}$ are
somewhat different from \mbox{alpha-,} beta- and gamma-diversity as usually
construed.  In the traditional ecological framework, a large community is
divided into a number of subcommunities, alpha-diversity is some kind of
average of the intrinsic diversities of the subcommunities, beta-diversity
is a measure of the variation between the subcommunities, and
gamma-diversity is, simply, the diversity of the whole.  Here,
non-traditionally, our beta-diversity (the relative diversity
$\reldiv{\p}{\vc{r}}$) and our gamma-diversity (the cross diversity
$\crossdiv{\p}{\vc{r}}$) express properties of an \emph{individual}
subcommunity with reference to the larger community.
This is one of the innovations introduced in recent work of Reeve%
\index{Reeve, Richard} 
et al.~\cite{HPD}, explored in depth in Chapter~\ref{ch:mm}.
\end{remark}

\section{Relative entropy in measure theory, geometry and statistics}
\lbl{sec:rel-misc}
\index{relative entropy!measure theory@and measure theory}%
\index{relative entropy!geometry@and geometry}%
\index{relative entropy!statistics@and statistics}

Here we give brief interpretations of relative entropy as seen from
specific standpoints in these three subjects.

\subsection*{Measure theory}

Let us attempt to generalize the notion of Shannon entropy from probability
distributions on a finite set to probability measures on an arbitrary
measurable space $\Omega$.  Starting from the definition
\[
H(\p) = - \sum_{i \in \supp(\p)} p_i \log p_i
\]
for finite sets, and reasoning purely formally, one might try to define the
entropy%
\index{entropy!measure space@on measure space} 
of a probability measure $\nu$ on $\Omega$ as
\[
H(\nu) = - \int_\Omega (\log \nu) \dee\nu.
\]
But this makes no sense, since there is no such function as `$\log \nu$'.

However, \emph{relative} entropy generalizes easily.  Indeed, given
probability measures $\nu$ and $\mu$ on $\Omega$, the
\demph{entropy of $\nu$ relative to $\mu$} is defined as
\begin{equation}
\lbl{eq:rel-ent-meas}
\relent{\nu}{\mu}
=
\int_\Omega \log \biggl( \frac{d\nu}{d\mu} \biggr) \dee\nu
\in
[0, \infty],
\end{equation}
where $d\nu/d\mu$ is
the Radon--Nikodym derivative.  (If $\nu$ is not absolutely
continuous with respect to $\mu$ then $d\nu/d\mu$ sometimes takes the value
$\infty$; but as in the finite case, we allow $\infty$ as a relative
entropy.)  

\begin{examples}
\lbl{egs:relent-meas}
\begin{enumerate}
\item
\lbl{eg:rm-dens}
Fix a measure $\lambda$ on $\Omega$, and take measures $\nu$ and $\mu$ on
$\Omega$ with densities $p$ and $r$ with respect to $\lambda$.  Thus, $d\nu
= p \dee\lambda$, $d\mu = r\dee\lambda$, and $d\nu/d\mu = p/r$.  It follows
that
\begin{equation*}
\relent{\nu}{\mu}
=
\int_{\supp(p)} p \log \biggl(\frac{p}{r}\biggr) \dee\lambda.
\end{equation*}
Provided that the choice of reference measure $\lambda$ is understood,
$\relent{\nu}{\mu}$ can be written as $\relent{p}{r}$.

\item
In particular, when $\Omega$ is a finite set with counting measure
$\lambda$, we recover Definition~\ref{defn:rel-ent}.
\end{enumerate}
\end{examples}

The measure-theoretic viewpoint also explains some earlier notation.  On
p.~\pageref{p:An}, we introduced the set $A_n$ of pairs $(\vc{p}, \vc{r})$
of probability distributions on $\{1, \ldots, n\}$ such that
$\relent{\p}{\vc{r}} < \infty$.  Regarding $\vc{p}$ and $\vc{r}$ as
measures on $\{1, \ldots, n\}$, the set $A_n$ consists of exactly the pairs
such that $\vc{p}$ is absolutely continuous with respect to $\vc{r}$.

The slogan
\slogan{all entropy is relative}%
\index{all entropy is relative}
is partly justified by the fact just established: relative entropy makes
sense in a wide measure-theoretic context in a way that ordinary entropy
does not.  A different justification is given in
Section~\ref{sec:all-ent-rel}.

There is, nevertheless, a useful concept of the entropy of a probability
distribution on Euclidean space.  Indeed, the
\demph{differential entropy}%
\index{entropy!differential}%
\index{differential entropy}
of a probability density function $f$ on $\R^n$ is defined as
%
\[
H(f)
=
-\int_{\supp(f)} f(x) \log f(x) \dx,
\]
%
and it is a fundamental fact that among all density functions with a given
mean and variance, the one with the maximal entropy is the
normal%
\index{normal distribution} 
distribution.  (This fact is closely related
to the central%
\index{central limit theorem} 
limit theorem, as explained in Johnson~\cite{JohnITC}, for instance.)
However, $H(f)$ is still a kind of relative entropy, since the integration
takes place with respect to Lebesgue measure $\lambda$.  Writing $\nu =
f\dee\lambda$ for the probability measure corresponding to the density $f$,
we have $f = d\nu/d\lambda$, hence
\[
H(f)
=
- \int \log \biggl( \frac{d\nu}{d\lambda} \biggr) \dee\nu.
\]
Formally, the right-hand side is the negative of the expression for
$\relent{\nu}{\lambda}$ given by equation~\eqref{eq:rel-ent-meas}, even
though $\lambda$ is not a \emph{probability} measure.

\subsection*{Geometry}
\index{relative entropy!metric@as metric}

We have tentatively evoked the idea that $\relent{\p}{\vc{r}}$ is some kind
of measure of distance, difference or divergence between the probability
distributions $\p$ and $\vc{r}$, and it is true that $\relent{\p}{\vc{r}}
\geq 0$ with equality if and only if $\p = \vc{r}$.  However, we have also
seen that relative entropy does not have one of the standard properties of
a distance function:
\[
\relent{\p}{\vc{r}} \neq \relent{\vc{r}}{\p}.
\index{relative entropy!asymmetry of}
\]
If that were the only problem, it would not be so bad: for as Lawvere,%
\index{Lawvere, F. William}
Gromov,%
\index{Gromov, Misha} 
and others have argued (\cite{LawvMSG}, p.~138--139
and~\cite{GromMSR}, p.~xv), and as anyone who has walked up and down a
hill\index{hill} 
already knows, there are useful notions of distance%
\index{metric!nonsymmetric} 
that are not symmetric.  A more serious problem is that relative entropy
fails the triangle inequality:%
\index{relative entropy!triangle inequality@and triangle inequality}

\begin{example}
\lbl{eg:rel-tri}
Define $\p, \vc{q}, \vc{r} \in \Delta_2$ by 
\[
\vc{p} = (0.9, 0.1),
\quad
\vc{q} = (0.2, 0.8),
\quad
\vc{r} = (0.1, 0.9).
\]
Then
\[
\relent{\vc{p}}{\vc{q}}
+
\relent{\vc{q}}{\vc{r}}
=
1.190\ldots
<
1.757\ldots
=
\relent{\vc{p}}{\vc{r}}.
\]
\end{example}

So, relative entropy only crudely resembles a distance function or metric
in the sense of metric spaces.

However, it is a highly significant fact that the \emph{square root} of
relative entropy is an \emph{infinitesimal} distance on the set of
probability distributions.  We explain this twice: first informally, then
in the language of Riemannian geometry.

Informally, let $\p \in \Delta_n^\circ$, and consider the relative entropy 
\[
\relent{\p + \vc{t}}{\vc{p}}
\]
for $\vc{t} \in \R^n$ close to $\vc{0}$ such that $\sum t_i = 0$.  (Then
$\p + \vc{t} \in \Delta_n^\circ$.)  We can expand $\relent{\p +
  \vc{t}}{\vc{p}}$ as a Taylor series in $t_1, \ldots, t_n$.  Since
$\relent{\p + \vc{t}}{\p}$ attains its minimum of $0$ at $\vc{t} = \vc{0}$,
the constant term in the Taylor expansion is $0$ and the terms in $t_1,
\ldots, t_n$ also vanish.  A straightforward calculation shows that, in
fact,
\[
\relent{\p + \vc{t}}{\p} 
=
\sum_{i = 1}^n \frac{1}{2p_i} t_i^2
+ \text{higher order terms}.
\]
Thus, up to a different scale factor $1/2p_i$ in each coordinate, relative
entropy locally resembles the square of Euclidean distance.  The same
is true with the arguments reversed:
\[
\relent{\p}{\p + \vc{t}}
=
\sum_{i = 1}^n \frac{1}{2p_i} t_i^2
+ \text{higher order terms}.
\]
So although $\relent{-}{-}$ is not symmetric%
\index{relative entropy!asymmetry of} 
in its two arguments, it is infinitesimally so, to second order.

These formulas suggest that we regard the square root of relative entropy,
rather than relative entropy itself, as a metric.  But again, it is not a
metric in the sense of metric spaces, because it fails the triangle%
\index{relative entropy!triangle inequality@and triangle inequality}
inequality.  The same $\p$, $\vc{q}$ and $\vc{r}$ as in
Example~\ref{eg:rel-tri} provide a counterexample:
\[
\sqrt{\relent{\vc{p}}{\vc{q}}}
+
\sqrt{\relent{\vc{q}}{\vc{r}}}
=
1.281\ldots
<
1.325\ldots
=
\sqrt{\relent{\vc{p}}{\vc{r}}}.
\]
%
%
%

Nevertheless, $\sqrt{\relent{-}{-}}$ can successfully be used as an
\emph{infinitesimal}%
\index{infinitesimal metric} 
metric.  Still speaking informally, the process is as follows.

Suppose that we are given a set $X \sub \R^n$ and a nonnegative real-valued
function $\delta$ defined on all pairs of points of $X$ that are
sufficiently close together. Then under suitable hypotheses on $\delta$, we
can define a metric $d$ on $X$.  First, define the length of any path
$\gamma$ in $X$ by finite approximations: plot a large number of
close-together points $\vc{x}_0, \ldots, \vc{x}_m$ along $\gamma$, use
$\sum_{r = 1}^m \delta(\vc{x}_{r - 1}, \vc{x}_r)$ as an approximation to
the length of $\gamma$, then pass to the limit.  The distance $d(\vc{x},
\vc{y}) \in [0, \infty]$ between two points $\vc{x}, \vc{y} \in X$ is
defined as the length of a shortest path between $\vc{x}$ and $\vc{y}$.
This $d$ is a metric in the sense of metric spaces.

Applied when $X = \Delta_n^\circ$ and $\delta = \sqrt{\relent{-}{-}}$, this
process gives a new metric $d$\ntn{dF1}%
\index{simplex!metric on} 
on the simplex.  `Have you ever seen anything like that?'\ asked Gromov%
\index{Gromov, Misha} 
(\cite{GromSS1}, Section~2).  As it turns out, $d$ is not so exotic.  Let
\[
S^{n - 1} = \Bigl\{ \vc{x} \in \R^n \such \sum x_i^2 = 1 \Bigr\}
\]
denote the unit $(n - 1)$-sphere.  It carries the geodesic metric $d_{S^{n
    - 1}}$, in which $d_{S^{n - 1}}(\vc{x}, \vc{y})$ is the length of a
shortest path between $\vc{x}$ and $\vc{y}$ on the sphere (an arc of a
great circle).  Any distribution $\p \in \Delta_n^\circ$ has a
corresponding point $\sqrt{\p} = (\sqrt{p_1}, \ldots, \sqrt{p_n})$ on $S^{n
  - 1}$.  And as we will see, the metric $d$ on $\Delta_n^\circ$ satisfies
\[
d(\vc{p}, \vc{r}) 
= 
\sqrt{2} d_{S^{n - 1}}\bigl(\sqrt{\vc{p}}, \sqrt{\vc{r}}\bigr)
\]
($\p, \vc{r} \in \Delta_n^\circ$).  So when the simplex is equipped with
this distance $d$, it is isometric to a subset of the sphere of radius
$\sqrt{2}$.  With different constant factors, $d(\p, \vc{r})$ is known
as the Fisher%
\index{Fisher, Ronald!distance} 
distance or Bhattacharyya%
\index{Bhattacharyya angle} 
angle between $\p$ and $\vc{r}$, as detailed below.

We now sketch the precise development.  The story told here is the
beginning of the subject of information%
\index{information geometry}
geometry, and we refer to the literature in that subject for details of
what follows.  The books by Ay, Jost, L{\^e} and
Schwachh{\"o}fer~\cite{AJLS} and Amari~\cite{AmarIGA} are comprehensive
modern introductions to information geometry.  Other important sources are
the earlier book of Amari and Nagaoka~\cite{AmNa}, the foundational 1983
paper of Amari~\cite{AmarFIG}, and the 1987 articles of
Lauritzen~\cite{Laur} and Rao~\cite{RaoDMP}.  The idea of converting an
infinitesimal distance-like function on a manifold into a genuine distance
function is developed systematically in Eguchi's theory of contrast
functions~\cite{EgucDGA,EgucGMC}, a summary of which can be found in
Section~3.2 of~\cite{AmNa}.

Let $M = (M, g)$ be a Riemannian%
\index{Riemannian manifold} 
manifold, and write $d$ for its geodesic distance function.  (We
temporarily adopt the Riemannian geometers' practice of using \dmph{metric}
to mean a Riemannian metric, and \dmph{distance} for a metric in the sense
of metric spaces.)  For each point $p \in M$, we have the function
\[
d(-, p)^2 \from M \to \R.
\]
It takes its minimum value, $0$, at $p$, and is smooth on a neighbourhood
of $p$.  We can therefore take its Hessian\index{Hessian} (with respect to
the Levi-Civita connection) at any point $x$ near $p$, giving a bilinear
form
\[
\Hess_x \bigl( d(-, p)^2 \bigr)
\]
on the tangent space $T_x M$.  In particular, we can take $x = p$, giving a
bilinear form on $T_p M$.  But of course, we already have another bilinear
form on $T_p M$, the Riemannian metric $g_p$ at $p$.  And up to a
constant factor, the two forms are equal:
\begin{equation}
\lbl{eq:loc-glob}
g_p
=
\frac{1}{2}
\Hess_p \bigl( d(-, p)^2 \bigr).
\end{equation}
This equation expresses the Riemannian
metric in terms of the geodesic distance (together with the connection).
That is, it expresses infinitesimal distance in terms of global distance.

(Equation~\eqref{eq:loc-glob} is proved by an elementary calculation,
although it is not often stated directly in the literature.  It can be derived
from more sophisticated results such as Theorem~6.6.1 of
Jost~\cite{JostRGG}, by taking the limit as $x \to p$ there, or
equation~(5) in Supplement~A of Pennec~\cite{PennBSA}.)

The idea now is that given any manifold $M$ with connection and any
function $\delta \from M \times M \to \R$ with primitive distance-like
properties, we can define a Riemannian metric $g$ on $M$ by
\begin{equation}
\lbl{eq:loc-glob-fake}
g_p
=
\frac{1}{2}
\Hess_p \bigl( \delta(-, p)^2 \bigr)
\end{equation}
($p \in M$).  Then, in turn, $g$ gives rise to a geodesic distance function
$d$ on $M$.  So, starting from a distance-like function $\delta$, we will
have derived a \emph{genuine} distance function $d$.  By
equations~\eqref{eq:loc-glob} and~\eqref{eq:loc-glob-fake}, $d$ and
$\delta$ are equal infinitesimally to second order, and $d$ is
entirely determined by the second-order infinitesimal behaviour of
$\delta$.

We apply this procedure to the open simplex $\Delta_n^\circ$, taking
$\delta$ to be the square root of relative entropy.  Each of the tangent
spaces of $\Delta_n^\circ$ is naturally identified with
\[
T_n = \Biggl\{ \vc{t} \in \R^n \such \sum_{i = 1}^n t_i = 0 \Biggr\},
\]
so $\Delta_n^\circ$ carries a canonical connection.  For each $\p \in
\Delta_n^\circ$, we define a bilinear form $g$ on $T_\p \Delta_n^\circ =
T_n$ by
\[
g(\vc{t}, \vc{u})
=
\frac{1}{2}
\Hess_\p \bigl( \relent{-}{\p} \bigr)
\]
($\vc{t}, \vc{u} \in T_n$).  By a straightforward calculation, this reduces
to 
\begin{equation}
\lbl{eq:FM-sum}
g(\vc{t}, \vc{u})
=
\sum_{i = 1}^n \frac{1}{2p_i} t_i u_i.
\end{equation}
This is a Riemannian%
\index{simplex!Riemannian structure on} 
metric on $\Delta_n^\circ$.  Without the factor of $1/2$, it is called the
\demph{Fisher%
\index{Fisher, Ronald!metric} 
metric}, $(\vc{t}, \vc{u}) \mapsto \sum t_i u_i/p_i$.

Now write
\[
S^{n - 1}_+ 
= 
S^{n - 1} \cap (0, \infty)^n
\]
for the positive orthant of the unit $(n - 1)$-sphere $S^{n - 1}$. 
There is a diffeomorphism of smooth manifolds
\[
\sqrt{\phantom{x}} \from \Delta_n^\circ \to S^{n - 1}_+
\]
defined by taking square roots in each coordinate.  Transferring the
standard Riemannian structure on $S^{n - 1}_+$ across this diffeomorphism
gives a Riemannian structure on $\Delta_n^\circ$.  Explicitly, since
$\tfrac{d}{dx}\sqrt{x} = 1/(2\sqrt{x})$, the induced
inner product $\ip{-}{-}$ on the tangent space $T_n$ at $\p \in
\Delta_n^\circ$ is given by
\begin{equation}
\lbl{eq:IP-sum}
\ip{\vc{t}}{\vc{u}} 
= 
\sum_{i = 1}^n \frac{t_i}{2\sqrt{p_i}} \frac{u_i}{2\sqrt{p_i}}
=
\sum_{i = 1}^n \frac{1}{4p_i} t_i u_i
\end{equation}
(as in Proposition~2.1 of Ay, Jost, L\^{e} and Schwachh\"ofer~\cite{AJLS}).
Equations~\eqref{eq:FM-sum} and~\eqref{eq:IP-sum} together give
$g(\vc{t}, \vc{u}) = 2\ip{\vc{t}}{\vc{u}}$.  Thus, the Riemannian manifold
$(\Delta_n^\circ, g)$ is isometric to $\sqrt{2}S^{n - 1}_+$, the positive
orthant of the $(n - 1)$-sphere of radius $\sqrt{2}$.

Like any Riemannian metric, $g$ induces a distance function.  The
isometry just established makes it easy to compute.  Indeed, we already
know the geodesic distance on $S^{n - 1}_+$ induced by its Riemannian
structure; it is given by 
\[
d_{S^{n - 1}}(\vc{x}, \vc{y})
= 
\cos^{-1}(\vc{x} \cdot \vc{y})
\in 
[0, \pi/2]
\]
($\vc{x}, \vc{y} \in S^{n - 1}_+$), where $\cdot$ denotes the standard
inner product on $\R^n$.  But by the previous paragraph, the geodesic
distance $d$ induced by the Riemannian metric $g$ on $\Delta_n^\circ$ is
given by
\[
d(\vc{p}, \vc{r})
=
\sqrt{2} d_{S^{n - 1}}\bigl(\sqrt{\vc{p}}, \sqrt{\vc{r}}\bigr)
\]
($\vc{p}, \vc{r} \in \Delta_n^\circ$).  Hence
\[
d(\vc{p}, \vc{r})
=
\sqrt{2} \cos^{-1} \Biggl( \sum_{i = 1}^n \sqrt{p_i r_i} \Biggr)
\in 
\bigl[0, \pi/\sqrt{2}\bigr].
\]
With different normalizations, this distance function has established
names: the \demph{Fisher%
\index{Fisher, Ronald!distance} 
distance} and the \demph{Bhattacharyya%
\index{Bhattacharyya angle} 
angle}~\cite{Bhat} between $\vc{p}$ and $\vc{r}$ are, respectively,
\[
2 \cos^{-1} \Biggl( \sum_{i = 1}^n \sqrt{p_i r_i} \Biggr),
\quad
\cos^{-1} \Biggl( \sum_{i = 1}^n \sqrt{p_i r_i} \Biggr).
\]
The Fisher distance is the geodesic distance induced by the
Fisher metric $(\vc{t}, \vc{u}) \mapsto \sum t_i u_i/p_i$, and makes
$\Delta_n^\circ$ isometric to the positive orthant of a sphere of radius
$2$.  The Bhattacharyya angle has the advantage that when it is used as a
distance function, $\Delta_n^\circ$ is isometric to a subset of the
\emph{unit} sphere.

In summary, relative entropy produces a notion of distance between two
probability distributions on a finite set, obeying the axioms of a metric
space.  If the square root of relative entropy is regarded as an
infinitesimal metric, then its global counterpart is (up to a constant) the
Fisher distance.

Further development of these ideas leads to the notion of a
statistical%
\index{statistical manifold} 
manifold.  Loosely, this is a Riemannian manifold whose points are to be
thought of as probability distributions (on some usually-infinite space).
We refer to the original paper of Lauritzen~\cite{Laur} and, again,
information geometry texts such as~\cite{AJLS} and~\cite{AmarIGA}.

\subsection*{Statistics}

Cross entropy and relative entropy arise naturally from elementary
statistical considerations, as follows.

Suppose that we make $k$ observations of elements drawn (by any method)
from $\{1, \ldots, n\}$, with outcomes
\[
x_1, \ldots, x_k \in \{1, \ldots, n\}.
\]
The \demph{empirical%
\index{empirical distribution} 
distribution} $\hat{\p}\ntn{emp} = (\hat{p}_1, \ldots, \hat{p}_n) \in
\Delta_n$ of the observations is given by
\[
\hat{p}_i 
=
\frac{\bigl| \bigl\{ j \in \{1, \ldots, k\} \such x_j = i \bigr\} \bigr|}%
{k},
\]
or equivalently, $\hat{\p} = \tfrac{1}{k} \sum_{j = 1}^k \delta_{x_j}$,
where $\delta_x$ denotes the point mass at $x$.  For example, if $n = 4$,
$k = 3$ and $(x_1, x_2, x_3) = (4, 1, 4)$, then $\hat{\p} = (1/3, 0, 0,
2/3)$.

Now let $\p \in \Delta_n$, and suppose that $k$ elements of $\{1, \ldots,
n\}$ are drawn independently at random according to $\p$.  The probability
$\Pr(x_1, \ldots, x_k)$ of observing $x_1, \ldots, x_k$ in that order is,
in fact, a function of the cross diversity or cross entropy of $\hat{\p}$
with respect to $\p$.  Indeed,
\begin{align}
\Pr(x_1, \ldots, x_k)   &
=
\prod_{j = 1}^k p_{x_j}
=
\prod_{i = 1}^n p_i^{|\{j \such x_j = i\}|}     
=
\prod_{i = 1}^n p_i^{k\hat{p}_i}       
\nonumber       \\
&
=
\crossdiv{\hat{\p}}{\p}^{-k}    
\nonumber       \\
&
=
\exp\bigl(-k\crossent{\hat{\p}}{\p}\bigr).
\nonumber
\end{align}

\begin{example}
Let $\p$ be a probability distribution on $\{1, \ldots, n\}$ with rational
probabilities: 
\[
\p = (k_1/k, \ldots, k_n/k)
\]
($k_i \geq 0$, $k = \sum k_i$).  Make $k$ observations using this
distribution.  What is the probability that the results observed are, in
order,
\[
\underbrace{1, \ldots, 1}_{k_1}, 
\ \ldots, \ 
\underbrace{n, \ldots, n}_{k_n}
\ ?
\]
The empirical distribution of those observations is just $\p$, so the
answer is 
\[
\crossdiv{\p}{\p}^{-k} = D(\p)^{-k} = e^{-kH(\p)}.
\]
So, when $k$ is fixed, the probability of obtaining these observations is a
decreasing function of the entropy of $\p$.  For instance, take $k = n$.
At one extreme, if $p_i = 1$ for some $i$, then the probability of the
observed results being $i, \ldots, i$ is maximal (with value $1$) and the
entropy is minimal (with value $0$).  At the other extreme, if $\p =
\vc{u}_n$, then the probability of the results being $1, \ldots, n$ is
small ($1/n^n$), corresponding to the fact that $\p$ has the maximal
possible entropy.
\end{example}

A standard situation in statistics is that we are in the presence of a
probability distribution that is unknown, but which we are willing to
assume is a member of a specific family $(\p_\theta)_{\theta \in \Theta}$.
We make some observations drawn from the distribution, then we attempt to
make inferences\index{inference} about the value of the unknown parameter
$\theta$. 

(In our current setting, $\Theta$ is any set and each $\p_\theta$ is a
distribution on $\{1, \ldots, n\}$.  But usually in statistics, $\Theta$ is
a subset of $\R^n$ and the set on which the distributions are defined is
infinite.  For instance, one may be interested in the family of all normal
distributions on $\R$, parametrized by pairs $(\mu, \sigma)$ where $\mu \in
\R$ is the mean and $\sigma \in \R^+$ is the standard deviation.)

How to make such inferences is one of the central questions of statistics.
The simplest way is the \demph{maximum%
\index{maximum likelihood}
likelihood} method, as follows.  Write
\[
\Pr(x_1, \ldots, x_k \given \theta) 
\]
for the probability of observing $x_1, \ldots, x_k$ when drawing from the
distribution $\p_\theta$.  The maximum likelihood method is this: given
observations $x_1, \ldots, x_k$, choose the value of $\theta$ that
maximizes $\Pr(x_1, \ldots, x_k \given \theta)$.

We have already shown that
\[
\Pr(x_1, \ldots, x_k \given \theta) 
=
\exp\bigl(-k \crossent{\hat{\p}}{\p_\theta} \bigr),
\]
so it follows from equation~\eqref{eq:rco} that
\[
\Pr(x_1, \ldots, x_k \given \theta) 
=
\exp\Bigl(-k\bigl( \relent{\hat{\p}}{\p_\theta} + H(\hat{\p})
\bigr)\Bigr).
\]
The term $H(\hat{\p})$ is fixed, in the sense of depending only on the
observed data and not on the unknown $\theta$.  The right-hand side is a
decreasing function of $\relent{\hat{\p}}{\p_\theta}$.  Thus, the
maximum likelihood method amounts to choosing $\theta$ to minimize the
relative entropy $\relent{\hat{\p}}{\p_\theta}$.  
Regarding $\relent{\hat{\p}}{\p}$ as a kind of difference or distance
between $\hat{\p}$ and $\p$ (with the caveats above), this means choosing
$\theta$ so that $\p_\theta$ is as close as possible to the observed
distribution $\hat{\p}$, as in Figure~\ref{fig:rel-ent-max-like}.

Further details and context can be found in
Csisz\'ar and Shields~\cite{CsSh}.  The method of minimizing relative
entropy has uniquely good properties, as was proved by Shore and
Johnson~\cite{ShJo} in a slightly different context to the one described
here. 

\begin{figure}
\centering
\lengths
\begin{picture}(120,40)
\cell{60}{20}{c}{\includegraphics[height=40\unitlength]{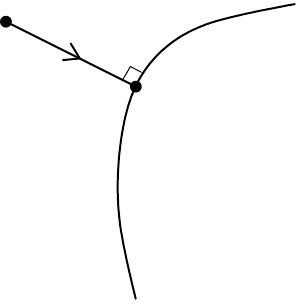}}
\cell{38}{38.5}{c}{$\hat{\p}$}
\cell{61.5}{27}{c}{$\p_\theta$}
\cell{64}{5}{c}{$(\p_\theta)_{\theta \in \Theta}$}
\end{picture}
\caption{Maximum likelihood and minimum relative entropy.}
\lbl{fig:rel-ent-max-like}
\end{figure}

\subsection*{Measure theory, geometry and statistics}

The connection between maximum likelihood and relative entropy involves
relative entropies $\relent{\hat{\p}}{\p_\theta}$ in which the arguments,
$\hat{\p}$ and $\p_\theta$, need not be close together in $\Delta_n$.
Nevertheless, we saw in the discussion of the Fisher metric that the
behaviour of $\relent{\p}{\vc{r}}$ when $\p$ and $\vc{r}$ are close is
especially significant.  More exactly, it is its infinitesimal behaviour to
second order that matters.  What follows is a brief further exploration of
this second-order behaviour, from a statistical perspective.

Let $\Omega$ be a measure space and let
$(f_\theta)_{\theta \in \Theta}$ be a smooth family of probability density
functions on $\Omega$, indexed over some real interval $\Theta$.  Fix
$\theta \in \Theta$.  The relative entropy
\[
\relent{f_\phi}{f_\theta} 
=
\int_\Omega f_\phi(x) \log \frac{f_\phi(x)}{f_\theta(x)} \dx
\]
($\phi \in \Theta$), defined as in
Example~\ref{egs:relent-meas}\bref{eg:rm-dens}, attains its minimum value
$0$ at $\phi = \theta$ (Figure~\ref{fig:fisher-curve}).
\begin{figure}
\centering
\lengths
\begin{picture}(120,33)(0,-3)
\cell{60}{15}{c}{\includegraphics[height=30\unitlength]{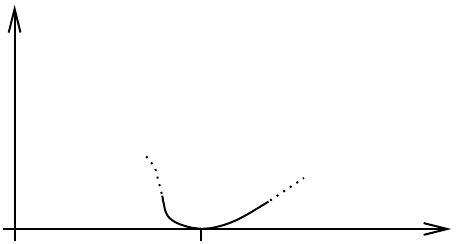}}
\cell{23}{25}{c}{$\relent{f_\phi}{f_\theta}$}
\cell{56.5}{-0.5}{t}{$\theta$}
\cell{83}{-0.5}{t}{$\phi$}
\end{picture}
\caption{Relative entropy for a parametrized family of probability
  distributions.  The
  Fisher information $I(\theta)$ is the second derivative of the graph at
  $\theta$, that is, the curvature there.}
\lbl{fig:fisher-curve}
\end{figure}
Thus, the function $\phi \mapsto \relent{f_\phi}{f_\theta}$ has both value
$0$ and first derivative $0$ at $\phi = \theta$.  The second derivative
measures how fast the distribution $f_\phi$ changes as $\phi$
varies near $\theta$.  It is called the
\demph{Fisher%
\index{Fisher, Ronald!information} 
information} $I(\theta)$ of our family at $\theta$:
\begin{equation}
\lbl{eq:fisher-info}
I(\theta)
=
\frac{\partial^2}{\partial\phi^2} \relent{f_\phi}{f_\theta} 
\biggr|_{\phi = \theta}.
\end{equation}
Substituting the definition of $\relent{f_\phi}{f_\theta}$
into~\eqref{eq:fisher-info} and performing some elementary calculations
leads to an explicit formula for the Fisher information:
\[
I(\theta) 
= 
\int_\Omega 
\frac{1}{f_\theta} 
\biggl( \frac{\partial f_\theta}{\partial \theta} \biggr)^2.
\]

Detailed discussions of Fisher information can be found in texts such as
Amari and Nagaoka~\cite{AmNa} (Section~2.2), where the definition is given
for families of distributions parametrized by \emph{several} real variables
$\theta_1, \ldots, \theta_n$, and Fisher information is put into the
context of the Fisher metric.  Here, we simply describe two uses of Fisher
information in statistics, remaining in the single-parameter
case.

The first is the Cram\'er--Rao%
\index{Cramer, Harald@Cram\'er, Harald!Rao bound@--Rao bound}%
\index{Rao, C. Radhakrishna!Cram\'er--Rao bound}
bound.  Suppose that we have an unbiased estimator $\hat{\theta}$ of the
parameter $\theta$.  The \demph{Cram\'er--Rao bound} for $\hat{\theta}$ is
a lower bound on its variance:
\begin{equation}
\lbl{eq:cr-bound}
\Var\bigl(\hat{\theta}\bigr) \geq \frac{1}{I(\theta)}
\end{equation}
(Cram\'er~\cite{Cram}, Rao~\cite{RaoIAA}).

This statement can be understood as follows.  Let $\theta$ denote the true
but unknown value of our parameter, which we are trying to infer from the
data.  If the Fisher information $I(\theta)$ at $\theta$ is small, then
$f_\phi$ changes only slowly when $\phi$ is near $\theta$.  Different
parameter values near $\theta$ produce similar distributions, so it is
difficult to infer the parameter value from observations with any degree of
accuracy.  The Cram\'er--Rao bound~\eqref{eq:cr-bound} formalizes this
intuition: since $1/I(\theta)$ is in this case large, any unbiased
estimator of $\theta$ must be imprecise, in the sense of having large
variance.  In contrast, if $f_\phi$ varies rapidly near $\phi = \theta$
then inferring\index{inference} $\theta$ from the data is easier, and it
may be possible to find a more precise unbiased estimator.

A second use of Fisher information is in the definition of the Jeffreys
prior\index{prior}.  A fundamental challenge in Bayesian statistics is how
to choose a prior distribution on the parameter space $\Theta$.  In
particular, one can ask for a universal method that takes as its input a
family $(f_\theta)_{\theta \in \Theta}$ of probability distributions and
produces as its output a canonical distribution on
$\Theta$, intended to be used as a prior.  In 1939, the
statistician Harold%
\index{Jeffreys, Harold} 
Jeffreys proposed using as a prior the density function
\[
\theta \mapsto \sqrt{I(\theta)},
\]
normalized (if possible) to integrate to $1$~\cite{JeffTP,JeffIFP}.  This is
the \demph{Jeffreys%
\index{Jeffreys, Harold!prior} 
prior}.

The Jeffreys prior has the crucial property of invariance%
\index{invariance under reparametrization}
under reparametrization.%
\index{reparametrization}
For example, suppose that one person works with the family $(f_\sigma)_{0
  \leq \sigma \leq 10}$ of normal distributions on $\R$ with mean $0$ and
standard deviation between $0$ and $10$, while another works with the
family $(g_V)_{0 \leq V \leq 100}$ of normal distributions with mean $0$
and variance between $0$ and $100$.  The difference between the two
families is obviously cosmetic, and if calculations based on the different
parametrizations resulted in different outcomes, something would be
seriously wrong.

But the Jeffreys prior behaves correctly.  The first person
can calculate the Jeffreys prior of their family to produce a probability
density function on $[0, 10]$, hence a probability measure $\nu_1$ on $[0,
  10]$.  The second person, similarly, obtains a probability measure
$\nu_2$ on $[0, 100]$.  The invariance property is that when $\nu_1$ is
pushed forward along the squaring map $[0, 10] \to [0, 100]$, the resulting
measure on $[0, 100]$ is equal to $\nu_2$.  In other words, the choice of
parametrization makes no difference to the Jeffreys prior.  

This is a very important logical property, and not all systems for
assigning a prior possess it.  For instance, suppose that we simply
assign the uniform prior to any family (Bernoulli and Laplace's principle%
\index{principle of insufficient reason}%
\index{insufficient reason, principle of}
of insufficient reason, discussed in Section~3 of Kass and
Wasserman~\cite{KaWa}).  Then invariance fails: in the example above, a
probability of $1/2$ is assigned to the standard deviation being less than
$5$, but a probability of $1/4$ to the variance being less than $25$.  This
is a fatal flaw.

A careful account of the Jeffreys prior, with historical and mathematical
context, can be found in Section~4.7 of Robert, Chopin and
Rousseau~\cite{RCR}.  This includes the full multi-parameter definition,
extending the single-parameter version to which we have confined ourselves
here.

\section{Characterization of relative entropy}
\lbl{sec:rel-char}

Here we show that relative entropy is uniquely characterized by the four
properties listed in Section~\ref{sec:rel-defn}, proving:
\begin{thm}
\lbl{thm:rel-char}
\index{relative entropy!characterization of}
Let $\bigl(\melent{-}{-} \from A_n \to \R\bigr)_{n \geq 1}$ be a sequence
of functions.  The following are equivalent:
\begin{enumerate}
\item 
\lbl{part:rel-char-props}
$\melent{-}{-}$ is measurable in the second argument,
permutation-invariant, and satisfies the vanishing and chain rules
(equation~\eqref{eq:rel-ch}, with $I$ in place of $H$);

\item
\lbl{part:rel-char-form}
$\melent{-}{-} = c\relent{-}{-}$ for some $c \in \R$.
\end{enumerate}
\end{thm}

Just as ordinary Shannon entropy has been the subject of many
characterization theorems, so too has relative entropy.
Theorem~\ref{thm:rel-char} and its proof first appeared in~\cite{SCRE}
(Theorem~II.1), and was strongly influenced by a categorical
characterization of relative entropy by Baez%
\index{Baez, John} 
and
Fritz~\cite{BaFr},%
\index{Fritz, Tobias}
which in turn built on work of Petz~\cite{Petz}.  It is also very close to
a result of Kannappan and Ng, although the proof is entirely different.
Historical commentary can be found in Remark~\ref{rmk:rel-char-hist}.

We now embark on the proof of Theorem~\ref{thm:rel-char}.

The four conditions in part~\bref{part:rel-char-props} are satisfied by
$\relent{-}{-}$ (as observed in Section~\ref{sec:rel-defn}), hence by
$c\relent{-}{-}$ for any scalar $c$.  Thus, \bref{part:rel-char-form}
implies~\bref{part:rel-char-props}.

\femph{For the rest of this section}, let $\melent{-}{-}$ be a sequence of
functions satisfying~\bref{part:rel-char-props}.  We have to prove that
$\melent{-}{-}$ is a scalar multiple of $\relent{-}{-}$.

Define a function $L \from (0, 1] \to \R$ by
\[
L(\alpha) = \melEnt{(1, 0)}{(\alpha, 1 - \alpha)}.
\]
(Since $\alpha > 0$, we have $\bigl((1, 0), (\alpha, 1 - \alpha)\bigr) \in
A_2$, so $L(\alpha) \in \R$ is well-defined.)  The idea is that if
$\melent{-}{-} = \relent{-}{-}$ then $L = -\log$. We will show that in
any case, $L$ is a scalar multiple of $\log$.

\begin{lemma}
\lbl{lemma:zeros}
Let $(\p, \vc{r}) \in A_n$ with $p_{k + 1} = \cdots = p_n = 0$, where $1
\leq k \leq n$.  Then $r_1 + \cdots + r_k > 0$ and
\[
\melent{\p}{\vc{r}} 
=
L(r_1 + \cdots + r_k) 
+
\melent{\p'}{\vc{r}'},
\]
where 
\[
\p' = (p_1, \ldots, p_k),
\qquad
\vc{r}' = \frac{(r_1, \ldots, r_k)}{r_1 + \cdots + r_k}.
\]
\end{lemma}

\begin{proof}
The case $k = n$ reduces to the statement that $L(1) = 0$, which follows
from the vanishing property.  Suppose, then, that $k < n$.  

Since $\p$ is a probability distribution with $p_i = 0$ for all $i > k$,
there is some $i \leq k$ such that $p_i > 0$, and then $r_i > 0$ since
$(\p, \vc{r}) \in A_n$.  Hence $r_1 + \cdots + r_k > 0$.  Let $\vc{r}'' \in
\Delta_{n - k}$ be the normalization of $(r_{k + 1}, \ldots, r_n)$ if $r_{k
  + 1} + \cdots + r_n > 0$, or choose $\vc{r}''$ arbitrarily in $\Delta_{n
  - k}$ otherwise.  (The set $\Delta_{n - k}$ is nonempty since $k < n$.)
Then by definition of composition,
\begin{align*}
\p      &
= 
(1, 0) \of (\p', \vc{r}''),  \\
\vc{r}  &
= 
(r_1 + \cdots + r_k, \, r_{k + 1} + \cdots + r_n) \of (\vc{r}', \vc{r}'').
\end{align*}
Hence by the chain rule, 
\[
\melent{\p}{\vc{r}}
=
L(r_1 + \cdots + r_k) +
1 \cdot \melent{\p'}{\vc{r}'} + 
0 \cdot \melent{\vc{r}''}{\vc{r}''},
\]
and the result follows. 
\end{proof}

\begin{lemma}
\lbl{lemma:two-add}
$L(\alpha\beta) = L(\alpha) + L(\beta)$ for all $\alpha, \beta \in (0,
1]$. 
\end{lemma}

\begin{proof}
By the chain rule, $\melent{-}{-}$ has the logarithmic property stated at
the end of Section~\ref{sec:rel-defn} (equation~\eqref{eq:rel-log}, with
$I$ in place of $H$).  Hence
\[
\melEnt{(1, 0) \otimes (1, 0)}{(\alpha, 1 - \alpha) \otimes (\beta, 1 -
  \beta)}
=
L(\alpha) + L(\beta).
\]
But also
\begin{align*}
&\melEnt{(1, 0) \otimes (1, 0)}%
{(\alpha, 1 - \alpha) \otimes (\beta, 1 - \beta)}       \\
&
=
\Melent{(1, 0, 0, 0)}%
{\bigl(\alpha\beta, \alpha(1 - \beta), 
(1 - \alpha)\beta, (1 - \alpha)(1 - \beta)\bigr)}       \\
&
=
L(\alpha\beta) + \melent{\vc{u}_1}{\vc{u}_1}    \\
&
=
L(\alpha\beta),
\end{align*}
by Lemma~\ref{lemma:zeros} (with $k = 1$) and the vanishing property.  
\end{proof}

We can now deduce:

\begin{lemma}
\lbl{lemma:rel-two-log}
There is some $c \in \R$ such that $L(\alpha) = -c\log\alpha$ for all
$\alpha \in (0, 1]$.
\end{lemma}

\begin{proof}
By hypothesis, $L$ is measurable, so this follows from
Lemma~\ref{lemma:two-add} and Corollary~\ref{cor:cauchy-log-01}.
\end{proof}

Our next lemma is an adaptation of the most ingenious part of Baez and
Fritz's argument (Lemma~4.2 of~\cite{BaFr}).

\begin{lemma}
\lbl{lemma:bf-full-supp}
Let $n \geq 1$ and $(\p, \vc{r}) \in A_n$.  Suppose that $\p$ has full
support.  Then $\melent{\p}{\vc{r}} = c\relent{\p}{\vc{r}}$.
\end{lemma}

\begin{proof}
Since $(\p, \vc{r}) \in A_n$, the distribution $\vc{r}$ also has full
support.  We can therefore choose some $\alpha \in (0, 1]$ such that $r_i -
  \alpha p_i \geq 0$ for all $i$.

We will compute the number
\[
x =
\melEnt{(p_1, \ldots, p_n, \underbrace{0, \ldots, 0}_n)}
{(\alpha p_1, \ldots, \alpha p_n, 
r_1 - \alpha p_1, \ldots, r_n - \alpha p_n)} 
\]
in two ways.  (The pair of distributions on the right-hand side belongs to
$A_{2n}$, so $x$ is well-defined.)  First, by Lemma~\ref{lemma:zeros}
and the vanishing property,
\[
x 
=
L(\alpha) + \melent{\p}{\p}
=
-c\log\alpha.
\]
Second, by permutation-invariance and then the chain rule,
\begin{align*}
x       &
=
\melEnt{(p_1, 0, \ldots, p_n, 0)}
{(\alpha p_1, r_1 - \alpha p_1, \ldots, \alpha p_n, r_n - \alpha p_n)} \\
&
=
\Melent{\p \of \bigl((1, 0), \ldots, (1, 0)\bigr)}
{\vc{r} \of 
\Bigl(
\Bigl(\alpha \tfrac{p_1}{r_1}, 1 - \alpha \tfrac{p_1}{r_1}\Bigr), \ldots, 
\Bigl(\alpha \tfrac{p_n}{r_n}, 1 - \alpha \tfrac{p_n}{r_n}\Bigr)
\Bigr)} \\
&
=
\melent{\p}{\vc{r}} + 
\sum_{i = 1}^n p_i L\Bigl(\alpha \tfrac{p_i}{r_i}\Bigr) \\
&
=
\melent{\p}{\vc{r}} - c\log \alpha - c\relent{\p}{\vc{r}}.
\end{align*}
Comparing the two expressions for $x$ gives the result.
\end{proof}

We have now proved that $\melent{\p}{\vc{r}} = c\relent{\p}{\vc{r}}$ when
$\p$ has full support.  It only remains to prove it for arbitrary $\p$. 

\begin{pfof}{Theorem~\ref{thm:rel-char}}
Let $(\p, \vc{r}) \in A_n$.  By permutation-invariance, we can assume that 
\[
p_1, \ldots, p_k > 0, 
\quad
p_{k + 1} = \cdots = p_n = 0,
\]
where $1 \leq k \leq n$.  Writing $\rhow = r_1 + \cdots + r_k$,
\[
\melent{\p}{\vc{r}}     
=
L(\rhow) 
+ \melEnt{(p_1, \ldots, p_k)}{\tfrac{1}{\rhow}(r_1, \ldots, r_k)}
\]
by Lemma~\ref{lemma:zeros}.  Hence by Lemmas~\ref{lemma:rel-two-log}
and~\ref{lemma:bf-full-supp},
\[
\melent{\p}{\vc{r}}     
=
-c\log R
+ c\relEnt{(p_1, \ldots, p_k)}{\tfrac{1}{\rhow}(r_1, \ldots, r_k)}.
\]
But by the same argument applied to $cH$ in place of $I$ (or by direct
calculation), we also have
\[
c\relent{\p}{\vc{r}}     
=
-c\log R
+ c\relEnt{(p_1, \ldots, p_k)}{\tfrac{1}{\rhow}(r_1, \ldots, r_k)}.
\]
The result follows.
\end{pfof}

\begin{remarks}
\lbl{rmks:rel-char-hyps}
\begin{enumerate}
\item 
Cross entropy satisfies all the properties listed in
Theorem~\ref{thm:rel-char}\bref{part:rel-char-props} except for vanishing,
which it does not satisfy.  Hence the vanishing axiom cannot be dropped
from the theorem.

\item
\lbl{rmk:rel-char-hyps-ch}
The chain%
\index{chain rule!forms of} 
rule can equivalently be replaced by a special case: 
\begin{multline*}
\Melent{\bigl(pw_1, (1 - p)w_1, w_2, \ldots, w_n\bigr)}%
{\bigl(\widetilde{p}\widetilde{w}_1, (1 - \widetilde{p})\widetilde{w}_1, 
\widetilde{w}_2, \ldots, \widetilde{w}_n\bigr)}      \\ 
=
\melent{\vc{w}}{\widetilde{\vc{w}}} 
+
w_1 \melEnt{(p, 1 - p)}{(\widetilde{p}, 1 - \widetilde{p})}
\end{multline*}
for all $(\vc{w}, \widetilde{\vc{w}}) \in A_n$ and $\bigl((p, 1 - p),
(\widetilde{p}, 1 - \widetilde{p})\bigr) \in A_2$.  Alternatively, it can
be replaced by a different special case:
\begin{multline*}
\melEnt{w\p \oplus (1 - w)\vc{r}}%
{\widetilde{w}\widetilde{\p} \oplus (1 - \widetilde{w})\widetilde{\vc{r}}}%
\\
=
\melEnt{(w, 1 - w)}{(\widetilde{w}, 1 - \widetilde{w})}
+
w\melent{\p}{\widetilde{\p}}
+
(1 - w)\melent{\vc{r}}{\widetilde{\vc{r}}}
\end{multline*}
for all $(\p, \widetilde{\p}) \in A_k$, $(\vc{r}, \widetilde{\vc{r}}) \in
A_\ell$, and $\bigl((w, 1 - w), (\widetilde{w}, 1 - \widetilde{w})\bigr)
\in A_2$.  Here we have used the notation
\[
w\p \oplus (1 - w)\vc{r}
=
\bigl(w p_1, \ldots, w p_k, (1 - w)r_1, \ldots, (1 - w)r_\ell\bigr).
\]
Both special cases are equivalent to the general case by elementary
inductions, as in Remark~\ref{rmk:ent-chain-simp} and
Appendix~\ref{sec:chain}.
\end{enumerate}
\end{remarks}

\begin{remark}
\lbl{rmk:rel-char-hist} 
The first characterization of relative entropy appears to have
been proved by R\'enyi%
\index{Renyi Alfred@R\'enyi, Alfred} 
in 1961 (\cite{Reny}, Theorem~4).  It relied on $\relent{\p}{\vc{r}}$
being defined not only for probability distributions $\p$ and $\vc{r}$, but
also for all `generalized'%
\index{generalized probability distribution}%
\index{probability distribution!generalized}
distributions (in which
the requirement that $\sum p_i = \sum r_i = 1$ is weakened to $\sum p_i,
\sum r_i \leq 1$).  The result does not translate easily into a
characterization of relative entropy for ordinary probability distributions
only.

Among the theorems characterizing relative entropy for ordinary probability
distributions, one of the first was that of Hobson~\cite{Hobs}%
\index{Hobson, Arthur}
in~1969.  His hypotheses were stronger than those of
Theorem~\ref{thm:rel-char}, for the same conclusion.  In common with
Theorem~\ref{thm:rel-char}, he assumed permutation-invariance, vanishing,
and the chain rule (in the second of the two equivalent forms given in
Remark~\ref{rmks:rel-char-hyps}\bref{rmk:rel-char-hyps-ch}).  But he also
assumed continuity in both variables (instead of just measurability in one)
and a monotonicity hypothesis unlike anything in
Theorem~\ref{thm:rel-char}.

In 1973, Kannappan%
\index{Kannappan, Palaniappan}
and Ng~\cite{KaNgMSF}%
\index{Ng, Che Tat}
proved a result very close to Theorem~\ref{thm:rel-char}.  They did not
explicitly \emph{state} that result in their paper, but the closing remarks
in another paper by the same authors~\cite{KaNgFEC} and the approach of a
contemporaneous paper by Kannappan and Rathie~\cite{KaRa} suggest
the intent.  The result resembling Theorem~\ref{thm:rel-char} was stated
explicitly in a 2008 article of Csisz\'ar (\cite{Csis}, Section~2.1), who
attributed it to Kannappan and Ng.

There are some small differences between the hypotheses of Kannappan and
Ng's theorem and those of Theorem~\ref{thm:rel-char}.  They assumed
measurability in both variables, whereas we only assumed measurability in
the second (and actually only used that $\melent{(1, 0)}{-}$ is
measurable).  On the other hand, they only needed the vanishing condition
for $\vc{u}_2$, whereas we needed it for all $\p$.  Like many authors on
functional equations in information theory, they used the chain rule in the
first of the equivalent forms in
Remark~\ref{rmks:rel-char-hyps}\bref{rmk:rel-char-hyps-ch}, under the name
of recursivity.

The proofs, however, are completely different.  Theirs was a tour de force
of functional equations, putting at its heart the so-called fundamental%
\index{fundamental equation of information theory}
equation of information theory (equation~\eqref{eq:feith}), and involving
the solution of such functional equations as
\[
f(x) + (1 - x) g\biggl( \frac{y}{1 - x} \biggr)
=
h(y) + (1 - y) j\biggl( \frac{x}{1 - y} \biggr)
\]
in four unknown functions.  The proof above bypasses these considerations
entirely. 
\end{remark}

%% file: def.tex
\chapter{Deformations of Shannon entropy}
\lbl{ch:def}
\index{deformation}
\index{entropy!deformations of}

Shannon entropy is fundamental, but it is not the only useful or natural
notion of entropy, even in the context of a single probability distribution
on a finite set.  In this chapter, we meet two one-parameter families of
entropies that both include Shannon entropy as a member
(Figure~\ref{fig:defs}).
\begin{figure}
\centering
\lengths
\begin{picture}(120,40)(7,-4)
\cell{60}{20}{c}{\includegraphics[height=30\unitlength]{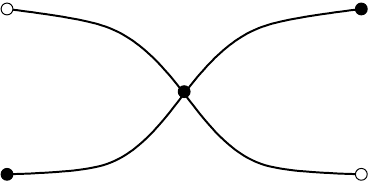}}
\cell{25}{34}{c}{$-\infty$}
\cell{25}{6}{c}{$-\infty$}
\cell{93}{34}{c}{$\infty$}
\cell{93}{6}{c}{$\infty$}
\cell{75}{28}{l}{R\'enyi entropies $H_q$}
\cell{75}{10}{l}{$q$-logarithmic entropies $S_q$}
\cell{60}{4}{b}{\vector(0,1){14}}
\cell{60}{0}{b}{Shannon entropy}
\cell{60}{-4}{b}{$H_1 = H = S_1$}
\end{picture}
\caption{Two families of deformations of Shannon entropy: the R\'enyi
  entropies $(H_q)_{q \in [-\infty, \infty]}$ and the $q$-logarithmic
  entropies $(S_q)_{q \in (-\infty, \infty)}$.}  
\lbl{fig:defs}
\end{figure}
Both are indexed by a real parameter $q$, and both have Shannon entropy as
the case $q = 1$.  Moving the value of $q$ away from $1$ can be thought of
as deforming Shannon entropy.  As in other mathematical contexts where
the word `deformation' is used, the undeformed object (Shannon
entropy) has uniquely good properties that are lost after deformation,
but the deformed objects nevertheless retain some of the original object's
features.  

We begin with the $q$-logarithmic entropies $(S_q)_{q \in \R}$, often
called `Tsallis entropies' (a misattribution detailed in
Remark~\ref{rmk:q-log-ent-name}).  The $q$-logarithmic entropies have been
used as measures of biological diversity, but should probably not be, as we
will see (Examples~\ref{egs:q-ent}).

Perhaps surprisingly, it is \emph{easier} to uniquely characterize the
entropy $S_q$ for $q \neq 1$ than it is in the Shannon case $S_1 = H$.
Moreover, the characterization theorem that we prove does not require any
regularity conditions at all, not even measurability.  The same goes for
the $q$-logarithmic relative entropy, which we also introduce and
characterize. 

After some necessary preliminaries on the classical topic of power means
(Section~\ref{sec:pwr-mns}), we introduce the other main family of
deformations of Shannon entropy: the R\'enyi entropies $(H_q)_{q \in
  [-\infty, \infty]}$ (Section~\ref{sec:ren-hill}).  The
$q$-logarithmic and R\'enyi entropies have exactly the same content: for
each finite value of $q$, there is a simple formula for $S_q(\p)$ in terms
of $H_q(\p)$, and vice versa.  But they have different and complementary
algebraic properties.  For instance, the $q$-logarithmic entropies satisfy
a simple chain rule similar to that for Shannon entropy, whereas the chain
rule for the R\'enyi entropies is more cumbersome.  On the other hand,
the R\'enyi entropies have the same log-like property as Shannon
entropy, 
\[
H_q(\p \otimes \vc{r}) = H_q(\p) + H_q(\vc{r}),
\]
but the $q$-logarithmic entropies do not.

The exponential of R\'enyi entropy, $D_q(\p) = \exp(H_q(\p))$, is known in
ecology as the Hill number of order $q$.  The Hill numbers are the most
important measures of biological diversity (at least, if we are using the
crude model of a community as a probability distribution on the set of
species).  Different values of $q$ reflect different aspects of a
community's composition, and graphing $D_q(\p)$ against $q$ enables one to
read off meaningful features of the community.  In
Sections~\ref{sec:ren-hill} and~\ref{sec:prop-hill}, we illustrate this
point and establish the properties that make the Hill numbers so suitable
as measures of diversity.

We finish by showing that the Hill number of a given order $q$ is uniquely
characterized by certain properties (Section~\ref{sec:hill-char-given}).
The same is therefore true of the R\'enyi entropies (since one is the
exponential of the other), although the properties appear more natural when
stated for the Hill numbers.  This is the first of two characterization
theorems for the Hill numbers that we will prove in this book.  The second
theorem characterizes the Hill numbers of \emph{unknown} orders, and we will
reach it in Section~\ref{sec:total-hill}.

\section{$q$-logarithmic entropies}
\lbl{sec:q-log-ent}

To obtain the definition of $q$-logarithmic entropy, we simply take the
definition of Shannon entropy and replace the logarithm by the
$q$-logarithm $\ln_q$ defined in Section~\ref{sec:q-log}.

\begin{defn}
Let $q \in \R$ and $n \geq 1$.  The \demph{$q$-logarithmic%
\index{q-logarithmic entropy@$q$-logarithmic entropy} 
entropy}
\[
S_q \from \Delta_n \to \R
\ntn{Sq}
\]
is defined by
\[
S_q(\p) 
= 
\sum_{i \in \supp(\p)} p_i \ln_q \biggl( \frac{1}{p_i} \biggr).
\]
\end{defn}

Thus, $S_1(\p)$ is the Shannon entropy $H(\p)$, and for $q \neq 1$,
\begin{equation}
\lbl{eq:q-log-explicit}
S_q(\p)
=
\frac{1}{1 - q} 
\Biggl(
\sum_{i \in \supp(\p)} p_i^q - 1
\Biggr).
\end{equation}

\begin{remark}
\lbl{rmk:q-ent-param}
We chose to generalize the expression $\sum p_i \log(1/p_i)$ for Shannon
entropy, but we could instead have used $-\sum p_i \log p_i$.  Since
$\ln_q(1/x) \neq -\ln_q(x)$, this would have given a different result.  But
by equation~\eqref{eq:q-log-reciprocal},
\[
-\sum_{i \in \supp(\p)} p_i \ln_q p_i
=
S_{2 - q}(\p),
\]
so this different choice only amounts to a different parametrization.
\end{remark}

The $q$-logarithmic entropy $S_q(\p)$ can be interpreted as expected
surprise.\index{surprise} Let $s \from [0, 1] \to \R \cup \{\infty\}$ be a
decreasing function such that $s(1) = 0$, thought of as assigning to each
probability $p$ the degree of surprise $s(p)$ that one would experience on
witnessing an event with that probability.  Then our expected surprise at
an event drawn from a probability distribution $\p = (p_1, \ldots, p_n)$ is
\[
\sum_{i \in \supp(\p)} p_i \cdot s(p_i).
\]
Expected surprise is a measure of uncertainty.  If $\p = (1, 0, \ldots, 0)$
then the expected surprise is $0$: the process of drawing from $\p$ is
completely predictable.  If $\p = \vc{u}_n$ then the expected surprise is
$s(1/n)$, which is an increasing function of $n$: the greater the number of
possibilities, the less predictable the outcome.

(Informally, the concept of expected surprise is familiar: someone who
lives in a stable environment will expect that most days, something may
mildly surprise them but nothing will astonish them.  The less stable the
environment, the greater the expected surprise.)

In these terms, $S_q(\p)$ is the expected surprise at an event drawn from
the distribution $\p$ when we use $p \mapsto \ln_q(1/p)$ as our surprise
function.  Figure~\ref{fig:surprise-fns} shows the surprise functions for
$q = 0, 1, 2, 3$.  For a general $q > 0$, we have
\[
0 \leq S_q(\p) \leq \ln_q(\vc{u}_n)
\]
for all $\p \in \Delta_n$, with $S_q(\p) = 0$ if and only if $\p = (0,
\ldots, 0, 1, 0, \ldots, 0)$ and $S_q(\p) = \ln_q(n)$ if and only if $\p =
\vc{u}_n$.  This will follow from the corresponding properties of the Hill
number $D_q$, once we have established
the relationship between the $q$-logarithmic entropies and the Hill
numbers (Remark~\ref{rmks:hill-dec}\bref{rmk:hill-dec-min}).

\begin{figure}
\lengths
\begin{picture}(120,53)(0,-3)
\cell{60}{24}{c}{\includegraphics[height=50\unitlength]{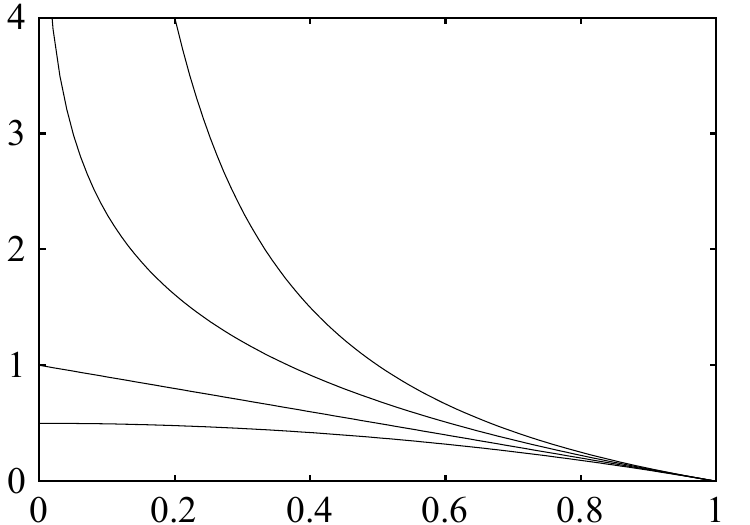}}
\cell{60}{-4}{b}{$p$}
\cell{17}{25}{c}{$\ln_q(1/p)$}
\cell{49}{40}{c}{$q = 0$}
\cell{39}{30}{c}{$q = 1$}
\cell{35}{16}{c}{$q = 2$}
\cell{33}{6}{c}{$q = 3$}
\end{picture}
\caption{The functions $p \mapsto \ln_q(1/p)$ for several values of $q$.}
\lbl{fig:surprise-fns}
\end{figure}

\begin{examples}
\lbl{egs:q-ent} 
In these examples, we regard $\p = (p_1, \ldots, p_n) \in \Delta_n$ as the
relative abundance distribution of $n$ species making up a biological
community.  Sometimes $S_q(\p)$ has been advocated as a measure of
diversity (as in Patil and Taillie~\cite{PaTaDCM}, Keylock~\cite{Keyl}, and
Ricotta and Szeidl~\cite{RiSzTUA}), but this is problematic, as now
explained.  
\begin{enumerate}
\item 
$S_0(\p) = |\supp(\p)| - 1$.  That is, the $0$-logarithmic entropy is one
  less than the number of species present.

\item
$S_1(\p) = H(\p)$.  The plague and oil company arguments of
  Examples~\ref{eg:plague} and~\ref{eg:oil} show why
$S_1$ should not be used as a diversity measure.  More generally, $S_q$
should not be used as a diversity measure either, for any value of $q$,
since it is not an effective number:
\[
S_q(\vc{u}_n) 
= 
\ln_q(n) 
\neq 
n.
\]
However, we will see in Section~\ref{sec:ren-hill} that $S_q$ can be
transformed into a well-behaved diversity measure, and that the result is
the Hill number of order $q$.

\item
\lbl{eg:q-ent-2}
The $2$-logarithmic entropy of $\p$ is
\[
S_2(\p)
=
1 - \sum_{i = 1}^n p_i^2
=
\sum_{i, j \csuch i \neq j} p_i p_j.
\]
This is the probability that two individuals chosen at random are of
different species.  In ecology, $S_2(\p)$ is associated with the names of
Edward H.~Simpson,%
\index{Simpson, Edward}
who introduced $S_2(\p)$ as an index of diversity in 1949~\cite{SimpMD},
and Corrado Gini,%
\index{Gini, Corrado}
who used $S_2(\p)$ in a wide-ranging 1912 monograph on economics,
statistics and demography~\cite{Gini}.  It is such a natural quantity that
it has been used in many different fields; it also has the advantage that
it admits an unbiased estimator.  These points are discussed in the 1982
note of Good~\cite{GoodDCM},%
\index{Good, Jack}
who wrote that `any statistician of this century who wanted a measure of
homogeneity would have taken about two seconds to suggest $\sum p_i^2$\,'.

Despite all this, $S_2(\p)$ has the defect of not being an effective
number.  Again, Section~\ref{sec:ren-hill} describes the remedy.
\end{enumerate}
\end{examples}

\begin{remark}
\index{q-logarithmic entropy@$q$-logarithmic entropy!naming of}
\lbl{rmk:q-log-ent-name} 
The $q$-logarithmic entropies have been discovered and rediscovered
repeatedly.  They seem to have first appeared in a 1967 paper on
information and classification by Havrda%
\index{Havrda, Jan}
and Charv\'at~\cite{HaCh},%
\index{Charv\'at, Franti{\v{s}}ek}
in a form adapted to base~$2$ logarithms:
\[
\Si_q(\p)
=
\frac{1}{2^{1 - q} - 1} \biggl( \sum p_i^q - 1 \biggr)
\ntn{Sqi}
\]
The constant factor is chosen so that $\Si_q(\p)$ converges to the base~$2$
Shannon entropy $\Hi(\p)$ as $q \to 1$, and so that $\Si_q(\vc{u}_2) = 1$
for all $q$.

Further work on the entropies $\Si_q$ was carried out in 1968 by
Vajda~\cite{Vajd} (with reference to Havrda and Charv\'at).  They were
rediscovered in 1970 by Dar\'oczy~\cite{DaroGIF} (without reference to
Havrda and Charv\'at), and were the subject of Section~6.3 of the 1975
book~\cite{AcDa} by Acz\'el and Dar\'oczy (with reference to all of the
above).

The base~$e$ entropies $S_q$ themselves seem to have appeared first in
Section~3.2 of a 1982 paper~\cite{PaTaDCM} by Patil%
\index{Patil, Ganapata P.}
and Taillie%
\index{Taillie, Charles} 
(with reference to Acz\'el and Dar\'oczy but none of the others), where
$S_q$ was proposed as an index of biological diversity.

In physics, meanwhile, the $q$-logarithmic entropies appeared in a 1971
article of Lindhard and Nielsen~\cite{LiNi} (according to Csiszar
\cite{Csis}, Section~2.4).  They also made a brief appearance in a review
article on entropy in physics by Wehrl (\cite{Wehr}, p.~247).  Finally,
they were rediscovered again in a 1988 paper on statistical physics by
Tsallis~\cite{TsalPGB}%
\index{Tsallis, Constantino} 
(with reference to none of the above).

Despite the twenty years of active life that the $q$-logarithmic entropies
had already enjoyed, it is after Tsallis that they are most commonly named.
The term `$q$-logarithmic entropy' is new, but has the benefits of being
descriptive and of not perpetuating a misattribution.
\end{remark}
\pagebreak

The chief advantage of the $q$-logarithmic entropies over the R\'enyi
entropies (introduced in Section~\ref{sec:ren-hill}) is that
they satisfy a simple chain rule:%
\index{chain rule!q-logarithm@for $q$-logarithmic entropy}
\begin{equation}
\lbl{eq:q-ent-chain}
S_q\bigl(\vc{w} \of (\p^1, \ldots, \p^n)\bigr)
=
S_q(\vc{w}) + \sum_{i \in \supp(\vc{w})} w_i^q S_q(\p^i)
\end{equation}
($q \in \R$, $\vc{w} \in \Delta_n$, $\vc{p}^i \in \Delta_{k_i}$).  This is
easily checked directly.  Alternatively, one can imitate the proof of the
chain rule for Shannon entropy (Proposition~\ref{propn:ent-chain}), replacing
$\partial$ by $\partial_q \from x \mapsto x \ln_q(1/x)$ and showing that
\[
\partial_q(xy) = \partial_q(x)y + x^q \partial_q(y).
\]
We will also prove a more general chain rule as
Proposition~\ref{propn:q-ent-sim-ch}.

The special case $\p^1 = \cdots = \p^n = \p$ gives
\begin{equation}
\lbl{eq:q-ent-mult}
S_q( \vc{w} \otimes \vc{p})
=
S_q(\vc{w}) + \Biggl( \sum_{i \in \supp(\vc{w})} w_i^q \Biggr) S_q(\p)  
\end{equation}
($q \in \R$, $\vc{w} \in \Delta_n$, $\p \in \Delta_k$).  In particular, the
symmetry present in the case $q = 1$,
\[
H(\vc{w} \otimes \p) = H(\vc{w}) + H(\p),
\]
disappears when we deform away from $q = 1$. This is the key to the
characterization theorem that follows.  

Before we state it, let us record one other property: $S_q$ is
\demph{symmetric}\lbl{p:q-ent-sym},%
\index{symmetric!entropy}%
\index{symmetric!diversity measure} 
meaning that
\begin{equation}
\lbl{eq:q-ent-sym}
S_q(\p)
=
S_q(\p\sigma)
\end{equation}
for all $q \in \R$, $\vc{p} \in \Delta_n$, and permutations $\sigma$ of
$\{1, \ldots, n\}$.

\begin{thm}
\lbl{thm:q-ent-char}
\index{q-logarithmic entropy@$q$-logarithmic entropy!characterization of}
Let $1 \neq q \in \R$.  Let $(I \from \Delta_n \to \R)_{n \geq 1}$ be a
sequence of functions.  The following are equivalent:
\begin{enumerate}
\item 
\lbl{part:q-ent-char-condns}
$I$ is symmetric and satisfies
\[
I( \vc{w} \otimes \vc{p})
=
I(\vc{w}) + \Biggl( \sum_{i \in \supp(\vc{w})} w_i^q \Biggr) I(\p)  
\]
for all $n, k \geq 1$, $\vc{w} \in \Delta_n$, and $\p \in \Delta_k$;

\item
\lbl{part:q-ent-char-form}
$I = cS_q$ for some $c \in \R$. 
\end{enumerate}
\end{thm}

This characterization of $q$-logarithmic entropy first appeared as
Theorem~III.1 of~\cite{SCRE}.  Notably, it needs no regularity conditions
whatsoever.  This is in contrast to the case $q = 1$ of Shannon entropy,
where some form of regularity is indispensable
(Remark~\ref{rmks:faddeev}\bref{rmk:faddeev-reg}).

\begin{proof}
By the observations just made, \bref{part:q-ent-char-form}
implies~\bref{part:q-ent-char-condns}.  Now
assume~\bref{part:q-ent-char-condns}.  By symmetry, $I(\vc{w} \otimes
\vc{p}) = I(\vc{p} \otimes \vc{w})$, so
\[
I(\vc{w}) + \Biggl( \sum_{i \in \supp(\vc{w})} w_i^q \Biggr) I(\vc{p})
=
I(\vc{p}) + \Biggl( \sum_{i \in \supp(\vc{p})} p_i^q \Biggr) I(\vc{w}),
\]
or equivalently,
\[
\Biggl( \sum_{i \in \supp(\vc{w})} w_i^q - 1 \Biggr) I(\vc{p})
=
\Biggl( \sum_{i \in \supp(\vc{p})} p_i^q - 1 \Biggr) I(\vc{w}),
\]
for all $\vc{w} \in \Delta_n$ and $\vc{p} \in \Delta_k$.  Take $\vc{w} =
\vc{u}_2$: then
\[
\bigl(2^{1 - q} - 1\bigr) I(\p)
=
\Biggl( \sum_{i \in \supp(\vc{p})} p_i^q - 1 \Biggr) I(\vc{u}_2)
\]
for all $\p \in \Delta_k$.  Since $q \neq 1$, we can define 
\[
c =
\frac{1 - q}{2^{1 - q} - 1}\,I(\vc{u}_2), 
\]
and then $I = cS_q$.
\end{proof}

\begin{remark}
\lbl{rmk:q-ent-char-hist} 
There have been several characterization theorems for the $q$-logarithmic
entropies.  One similar to Theorem~\ref{thm:q-ent-char} was published
by Dar\'oczy%
\index{Dar\'oczy, Zolt\'an} 
in 1970~\cite{DaroGIF}, and also appears as Theorem~6.3.9 of the book of
Acz\'el and Dar\'oczy~\cite{AcDa}.  In one sense it is stronger than
Theorem~\ref{thm:q-ent-char} (that is, has weaker hypotheses): where we
have assumed that $I \from \Delta_n \to \R$ is symmetric for all $n$,
Dar\'oczy assumed it only for $n = 3$.  On the other hand, Dar\'oczy's
theorem essentially assumed the full $q$-chain rule for $I(\vc{w} \of
(\p^1, \ldots, \p^n))$ (equation~\eqref{eq:q-ent-chain}), rather than just
the special case of $I(\vc{w} \otimes \p)$ that we used.

The word `essentially' here hides a historical wrinkle.  In
Remark~\ref{rmk:ent-chain-simp}, we noted that the chain rule for Shannon
entropy is equivalent to the special case
\[
H\bigl(pw_1, (1 - p)w_1, w_2, \ldots, w_n\bigr)
=
H(\vc{w}) + w_1 H(p, 1 - p),
\]
by a simple inductive argument.  Similarly, here, the $q$-chain rule of
equation~\eqref{eq:q-ent-chain} is equivalent to the special case
\begin{equation}
\lbl{eq:q-ent-chain-simp}
S_q\bigl(pw_1, (1 - p)w_1, w_2, \ldots, w_n\bigr)
=
S_q(\vc{w}) + w_1^q S_q(p, 1 - p),
\end{equation}
by the same simple inductive argument (given in Appendix~\ref{sec:chain}).
So, it is reasonable to regard~\eqref{eq:q-ent-chain}
and~\eqref{eq:q-ent-chain-simp} as equivalent.  But it was the special
case~\eqref{eq:q-ent-chain-simp}, not the general
case~\eqref{eq:q-ent-chain}, that was a hypothesis in Dar\'oczy's theorem.

The proof given by Dar\'oczy was entirely different, involving a
$q$-analogue of the `fundamental%
\index{fundamental equation of information theory} 
equation of information theory' (equation~\eqref{eq:feith}).

Other characterizations of $S_q$ have been proved, but using stronger
hypotheses than Theorem~\ref{thm:q-ent-char} to obtain the same conclusion
(such as the theorem in Section~2 of Suyari~\cite{Suya}, and Theorem~V.2
of Furuichi~\cite{Furu}).
\end{remark}

Just as ordinary entropy has a family of $q$-logarithmic deformations, so
too does relative entropy:

\begin{defn}
\lbl{defn:q-rel-ent}
Let $q \in \R$ and $\vc{p}, \vc{r} \in \Delta_n$.  The
\demph{$q$-logarithmic%
\index{q-logarithmic relative entropy@$q$-logarithmic relative entropy}
entropy of $\vc{p}$ relative to $\vc{r}$} is
\[
\srelent{q}{\vc{p}}{\vc{r}}
=
- \sum_{i \in \supp(\p)} p_i \ln_q \frac{r_i}{p_i}  
\in
[0, \infty].
\ntn{srelent}
\]
\end{defn}

Explicitly, $\srelent{1}{\vc{p}}{\vc{r}} = \relent{\vc{p}}{\vc{r}}$, and
for $q \neq 1$,
\[
\srelent{q}{\vc{p}}{\vc{r}}
=
\frac{1}{q - 1} \Biggl(
\sum_{i \in \supp(\p)} p_i^q r_i^{1 - q} - 1 
\Biggr).
\]
As for ordinary relative entropy $\relent{-}{-}$
(Section~\ref{sec:rel-defn}), we have
\[
\srelent{q}{\p}{\vc{r}} < \infty
\iff
(\vc{p}, \vc{r}) \in A_n.
\]

The definition of $q$-logarithmic relative entropy was given by
Rathie%
\index{Rathie, Pushpa N.} 
and Kannappan%
\index{Kannappan, Palaniappan}
in 1972~\cite{RaKa}.  (They used a version adapted to base~$2$ logarithms,
in the tradition of Havrda and Charv\'at described in
Remark~\ref{rmk:q-log-ent-name}.)  Their definition was taken up by Cressie
and Read in 1984 (\cite{CrRe}, Section~5), who used the base~$e$ version in
statistical work on goodness-of-fit tests.  It was rediscovered twice in
physics in 1998, by Shiino~\cite{Shii} and Tsallis~\cite{TsalGEB}
independently.

\begin{remark}
\lbl{rmk:q-rel-ent-param}
As in the definition of non-relative $q$-logarithmic entropy 
(Remark~\ref{rmk:q-ent-param}), there is a choice in how to generalize the
formula for ordinary relative entropy, given that $\ln_q(1/x) \neq
-\ln_q(x)$.  Again, making the other choice simply flips the
parametrization:
\[
\sum_{i \in \supp(\p)} p_i \ln_q \frac{p_i}{r_i}
=
\srelent{2 - q}{\vc{p}}{\vc{r}},
\]
by equation~\eqref{eq:q-log-reciprocal}.  The choice made in
Definition~\ref{defn:q-rel-ent} has the advantage that, as in the case $q =
1$, the relative entropy $\srelent{q}{\p}{\vc{u}_n}$ is a function of 
$S_q(\p)$ and $n$:
\begin{align*}
\srelent{q}{\p}{\vc{u}_n}       &
=
n^{q - 1} \bigl( \ln_q(n) - S_q(\p) \bigr)      \\
&
=
n^{q - 1} \bigl( S_q(\vc{u}_n) - S_q(\p) \bigr),
\end{align*}
as is easily checked.  
\end{remark}

Like its non-relative cousin, $q$-logarithmic relative entropy has an
extremely simple characterization.  It satisfies a chain rule
\begin{multline*}
\srelEnt{q}{\vc{w} \of (\p^1, \ldots, \p^n)}%
{\twid{\vc{w}} \of (\twid{\p}^1, \ldots, \twid{\p}^n)}\\
=
\srelent{q}{\vc{w}}{\twid{\vc{w}}} + 
\sum_{i \in \supp(\vc{w})} w_i^q \twid{w}_i^{1 - q}
\srelEnt{q}{\p^i}{\twid{\p}^i}
\end{multline*}
($\vc{w}, \twid{\vc{w}} \in \Delta_n$, $\p^i, \twid{\p}^i \in
\Delta_{k_i}$), which specializes to a multiplication rule
\begin{equation}
\lbl{eq:q-rel-ent-mult}
\srelent{q}{\vc{w} \otimes \p}{\twid{\vc{w}} \otimes \twid{\p}}
=
\srelent{q}{\vc{w}}{\twid{\vc{w}}} + 
\Biggl( 
\sum_{i \in \supp(\vc{w})} w_i^q \twid{w}_i^{1 - q}
\Biggr) 
\srelent{q}{\p}{\twid{\p}}
\end{equation}
($\vc{w}, \twid{\vc{w}} \in \Delta_n$, $\p, \twid{\p} \in \Delta_k$).
Moreover, $\srelent{q}{-}{-}$ is permutation-invariant in the same
sense as in the case $q = 1$ (Section~\ref{sec:rel-defn}).
Equation~\eqref{eq:q-rel-ent-mult} and permutation-invariance characterize
$\srelent{q}{-}{-}$ uniquely up to a constant factor:

\begin{thm}
\lbl{thm:q-rel-ent-char}
\index{q-logarithmic relative entropy@$q$-logarithmic relative entropy!characterization of}
Let $1 \neq q \in \R$.  Let $\bigl(\melent{-}{-} \from A_n \to \R\bigr)_{n
  \geq 1}$ be a sequence of functions.  The following are equivalent:
\begin{enumerate}
\item 
\lbl{part:q-rel-ent-char-condns}
$\melent{-}{-}$ is permutation-invariant and satisfies the multiplication
rule~\eqref{eq:q-rel-ent-mult} (with $I$ in place of $S_q$);

\item
\lbl{part:q-rel-ent-char-form}
$\melent{-}{-} = c\srelent{q}{-}{-}$ for some $c \in \R$.
\end{enumerate}
\end{thm}

This result first appeared as Theorem~IV.1 of~\cite{SCRE}.  Compared with
the characterization theorem for ordinary relative entropy
(Theorem~\ref{thm:rel-char}), it needs neither a regularity condition nor
the vanishing axiom.

\begin{proof}
The proof is very similar to that of Theorem~\ref{thm:q-ent-char}.
By the observations just made, \bref{part:q-rel-ent-char-form}
implies~\bref{part:q-rel-ent-char-condns}.  Now
assume~\bref{part:q-rel-ent-char-condns}.  By permutation-invariance,
\[
\melent{\vc{p} \otimes \vc{r}}{\twid{\vc{p}} \otimes \twid{\vc{r}}}
=
\melent{\vc{r} \otimes \vc{p}}{\twid{\vc{r}} \otimes \twid{\vc{p}}}
\]
for all $(\p, \twid{\p}) \in A_n$ and $(\vc{r}, \twid{\vc{r}})
\in A_k$.  So by the multiplication rule,
\[
\melent{\vc{p}}{\twid{\vc{p}}}
+ 
\biggl( \sum p_i^q \twid{p}_i^{1 - q} \biggr)
\melent{\vc{r}}{\twid{\vc{r}}}
=
\melent{\vc{r}}{\twid{\vc{r}}}
+ 
\biggl( \sum r_i^q \twid{r}_i^{1 - q} \biggr)
\melent{\p}{\twid{\p}},
\]
or equivalently,
\[
\biggl( \sum r_i^q \twid{r}_i^{1 - q} - 1 \biggr) 
\melent{\vc{p}}{\twid{\vc{p}}}
=
\biggl( \sum p_i^q \twid{p}_i^{1 - q} - 1 \biggr) 
\melent{\vc{r}}{\twid{\vc{r}}}.
\]
Take $\vc{r} = (1, 0)$ and $\twid{\vc{r}} = \vc{u}_2$: then
\[
(2^{q - 1} - 1) \melent{\vc{p}}{\twid{\vc{p}}}
=
\melent{(1, 0)}{\vc{u}_2} 
\biggl( \sum p_i^q \twid{p}_i^{1 - q} - 1 \biggr)
\]
for all $(\p, \twid{\p}) \in A_n$.  Since $q \neq 1$, we can put 
\[
c 
=
\frac{(q - 1)\melent{(1, 0)}{\vc{u}_2}}{2^{q - 1} - 1},
\]
and then $\melent{-}{-} = c \srelent{q}{-}{-}$.
\end{proof}

\begin{remark}
Other characterization theorems for $q$-logarithmic relative entropy have
been proved.  For example, Furuichi (\cite{Furu}, Section~IV) obtained the
same conclusion, but also assumed continuity and the full chain rule (or
more precisely, an equivalent special case, as in
Remark~\ref{rmk:q-ent-char-hist}) instead of just the multiplication
rule~\eqref{eq:q-rel-ent-mult}.
\end{remark}

\section{Power means}
\lbl{sec:pwr-mns}

We pause in our account of deformations of Shannon entropy to collect some
basic facts about power means (also called generalized means).  The reason
for doing this now is that the language and theory of power means make
possible a considerable streamlining of later material on R\'enyi entropies
and diversity measures.

The reader not interested in means for their own sake may wish to read 
Definition~\ref{defn:pwr-mn} and then jump ahead
to Section~\ref{sec:ren-hill}, referring back here only as necessary.  

This
section is essentially a list of properties satisfied by the power means,
together with the terminology for those properties.  A summary of the
terminology can also be found in Appendix~\ref{app:condns}.
Means are a classical topic of analysis, and almost everything in this
section can be found in Chapter~II of Hardy, Littlewood and P\'olya's
book~\cite{HLP}.

In what follows, $n$ denotes a positive integer.

\begin{defn}
\lbl{defn:pwr-mn}
Let $t \in [-\infty, \infty]$, $\p \in \Delta_n$, and $\vc{x} \in [0,
\infty)^n$.  The \demph{power%
\index{power mean} 
mean of order%
\index{order!power mean@of power mean}
$t$ of $\vc{x}$, weighted by $\p$}, is defined for $0 < t < \infty$ by
\begin{equation}
\lbl{eq:defn-pwr-mn}
M_t(\p, \vc{x})
=
\Biggl( \sum_{i \in \supp(\p)} p_i x_i^t \Biggr)^{1/t},
\end{equation}
for $-\infty < t < 0$ by
\begin{equation}
\lbl{eq:defn-pwr-mn-neg}
M_t(\p, \vc{x})
=
\begin{cases}
\displaystyle
\Biggl( \sum_{i \in \supp(\p)} p_i x_i^t \Biggr)^{1/t}  &
\text{if } x_i > 0 \text{ for all } i \in \supp(\p),    \\
\displaystyle
0       &
\text{otherwise},
\end{cases}
\end{equation}
and for the remaining values of $t$ by 
\begin{align*}
M_{-\infty}(\p, \vc{x}) &
=
\min_{i \in \supp(\p)} x_i,     \\
M_0(\p, \vc{x}) &
=
\prod_{i \in \supp(\p)} x_i^{p_i},      \\
M_\infty(\p, \vc{x})    &
=
\max_{i \in \supp(\p)} x_i.
\end{align*}
\end{defn}

The various exceptional cases in this definition are justified by
continuity, as detailed after the following examples.

\begin{examples}
\begin{enumerate}
\item 
The mean of order $1$ is the arithmetic\index{mean!arithmetic} mean $\sum
p_i x_i$ of $\vc{x}$ weighted by $\p$.

\item
The mean of order $0$ is the geometric\index{mean!geometric} mean of
$\vc{x}$ weighted by $\p$.

\item
The mean of order $-1$ is the harmonic\index{mean!harmonic} mean
\[
\frac{1}{\frac{p_1}{x_1} + \cdots + \frac{p_n}{x_n}}
\]
of $\vc{x}$ weighted by $\p$.
\end{enumerate}
\end{examples}

\begin{example}
Taking $\p = \vc{u}_n$ gives the \demph{unweighted}%
\index{power mean!unweighted}%
\index{mean!unweighted} 
(or uniformly weighted) power means $M_t(\vc{u}_n, \vc{x})$.
\end{example}

\begin{example}
For each $t \in [-\infty, \infty]$, the power mean $M_t$ has at least the
basic properties of an average:
\[
M_t\bigl(\p, (x, \ldots, x)\bigr) = x
\]
for all $\p \in \Delta_n$ and $x \in [0, \infty)$, and
\[
\min_{i \in \supp(\p)} x_i
\leq
M_t(\p, \vc{x})
\leq
\max_{i \in \supp(\p)} x_i
\]
for all $\p \in \Delta_n$ and $\vc{x} \in [0, \infty)^n$.  The rest of this
section is devoted to investigating the properties of power means in
greater depth.
\end{example}

We now prove three statements on the continuity of power means $M_t(\p,
\vc{x})$.  The first is on continuity in $\vc{x}$.

\begin{lemma}
\lbl{lemma:pwr-mns-cts-x}
\index{power mean!continuity of}
Let $t \in [-\infty, \infty]$ and $\p \in \Delta_n$.  Then the function
\[
M_t(\p, -) \from [0, \infty)^n \to [0, \infty)
\]
is continuous.
\end{lemma}

\begin{proof}
Let $\vc{x} \in [0, \infty)^n$.  From Definition~\ref{defn:pwr-mn}, it is
immediate that $M_t(\p, \vc{x})$ is continuous at $\vc{x}$ except perhaps
in the case where $t \in (-\infty, 0)$ and $x_i = 0$ for some $i \in
\supp(\p)$.  So, let $t \in (-\infty, 0)$ and suppose that, say, $x_1 =
0$ with $1 \in \supp(\p)$.  It suffices to show that $M_t(\p, \vc{y}) \to
0$ as $\vc{y} \to \vc{x}$ with $y_i > 0$ for all $i \in \supp(\p)$.  And
indeed, for such $\vc{y}$,
\begin{align*}
\bigl| M_t(\p, \vc{y}) \bigr|   &
=
\Biggl( \sum_{i \in \supp(\p)} p_i y_i^t \Biggr)^{1/t}  \\
&
\leq
\bigl( p_1 y_1^t \bigr)^{1/t}   \\
&
=
p_1^{1/t} y_1   \\
&
\to 
p_1^{1/t} x_1
=
0
\end{align*}
as $\vc{y} \to \vc{x}$, as required.
\end{proof}

The continuity properties of $M_t(\p, \vc{x})$ in $\p$ are more delicate.
Indeed, the power means of order $\leq 0$ are \emph{not} continuous in
$\p$: for when $t \leq 0$,
\[
M_t\bigl((\epsln, 1 - \epsln), (0, 1)\bigr) = 0
\]
for all $\epsln \in (0, 1]$, whereas
\[
M_t\bigl((0, 1), (0, 1)\bigr) = 1.
\]
Discontinuities do not only arise from zero values of $x_i$.  For
instance, $M_{-\infty}(\p, (1, 2))$ is not continuous in $\p$, since
\[
M_{-\infty}\bigl((\epsln, 1 - \epsln), (1, 2)\bigr) = 1,
\quad
M_{-\infty}\bigl((0, 1), (1, 2)\bigr) = 2
\]
for all $\epsln \in (0, 1]$.  There is a similar counterexample for
$M_\infty$.  But we do have the following.

\begin{lemma}
\lbl{lemma:pwr-mns-cts-px}
\index{power mean!continuity of}
\begin{enumerate}
\item 
\lbl{part:pwr-mns-cts-px-1}
For all $t \in [-\infty, \infty]$, the function
\[
M_t(-, -) \from \Delta_n^\circ \times [0, \infty)^n \to [0, \infty)
\]
is continuous.

\item 
\lbl{part:pwr-mns-cts-px-3}
For all $t \in (-\infty, \infty)$, the function
\[
M_t(-, -) \from \Delta_n \times (0, \infty)^n \to (0, \infty)
\]
is continuous.

\item 
\lbl{part:pwr-mns-cts-px-2}
For all $t \in (0, \infty)$, the function
\[
M_t(-, -) \from \Delta_n \times [0, \infty)^n \to [0, \infty)
\]
is continuous.
\end{enumerate}
\end{lemma}

\begin{proof}
Part~\bref{part:pwr-mns-cts-px-1} is immediate from the definition.  For
parts~\bref{part:pwr-mns-cts-px-3} and~\bref{part:pwr-mns-cts-px-2}, just
note that in the cases at hand, the formulas for $M_t$ are unchanged if $i$
is allowed to range over all of $\{1, \ldots, n\}$ instead of only
$\supp(\p)$.
\end{proof}

Our third and final continuity lemma states that power means are continuous
in their order.

\begin{lemma}
\lbl{lemma:pwr-mns-cts-t}
\index{power mean!continuity of}
Let $\p \in \Delta_n$ and $\vc{x} \in [0, \infty)^n$.  Then $M_t(\p,
  \vc{x})$ is continuous in $t \in [-\infty, \infty]$.
\end{lemma}

\begin{proof}
This is clear except perhaps at $t = 0$ and $t = \pm \infty$.  

For continuity at $t = 0$, first suppose that $x_i > 0$ for all $i \in
\supp(\p)$.  When $t$ is finite and nonzero, 
\begin{equation}
\lbl{eq:log-mn-hop}
\log M_t(\p, \vc{x})
=
\frac{\log \bigl(\sum_{i \in \supp(\p)} p_i x_i^t\bigr)}{t}.
\end{equation}
As $t \to 0$, 
\[
\log \Biggl(\sum_{i \in \supp(\p)} p_i x_i^t\Biggr)
\to
\log \Biggl(\sum_{i \in \supp(\p)} p_i\Biggr)
=
0,
\]
so we can apply l'H\^opital's rule to equation~\eqref{eq:log-mn-hop}, giving
\begin{align*}
\lim_{t \to 0} \log M_t(\p, \vc{x})     &
=
\lim_{t \to 0} 
\frac{\sum p_i x_i^t \log x_i}{\sum p_i x_i^t}  \\
&
=
\sum p_i \log x_i       \\
&
=
\log M_0(\vc{p}, \vc{x}),
\end{align*}
where all sums are over $i \in \supp(\p)$.  Hence the map $t \mapsto M_t(\p,
\vc{x})$ is continuous at $t = 0$.   

Now suppose that $x_i = 0$ for some $i \in \supp(\p)$.  By definition,
$M_t(\p, \vc{x}) = 0$ for all $t \leq 0$, so it suffices to show that
$M_t(\p, \vc{x}) \to 0$ as $t \to {0+}$.  For $t \in (0, \infty)$,
\[
0 
\leq 
M_t(\p, \vc{x}) 
=
\Biggl( \sum_{i \in \supp(\p)} p_i x_i^t \Biggr)^{1/t}
\leq 
M_\infty(\p, \vc{x}) \cdot
\Biggl(\sum_{i \in \supp(\p) \cap \supp(\vc{x})} p_i\Biggr)^{1/t}.
\]
But $\sum_{i \in \supp(\p) \cap \supp(\vc{x})} p_i < 1$, so our upper
  bound on $M_t(\p, \vc{x})$ converges to $0$ as $t \to {0+}$.  Hence also
  $M_t(\p, \vc{x}) \to 0$ as $t \to {0+}$, as required.

For continuity at $t = \infty$, suppose without loss of generality that
$\max_{i \in \supp(\p)} x_i$ is achieved at $i = 1$.  Then for $t \in (0,
\infty)$, 
\[
M_t(\p, \vc{x}) 
\leq
\Biggl( \sum_{i \in \supp(\p)} p_i x_1^t \Biggr)^{1/t}
=
x_1.
\]
On the other hand,
\[
M_t(\p, \vc{x}) 
\geq
\bigl( p_1 x_1^t \bigr)^{1/t}
=
p_1^{1/t} x_1
\to
x_1
\]
as $t \to \infty$.  Hence 
\[
M_t(\p, \vc{x}) \to x_1 = M_\infty(\p, \vc{x})
\]
as $t \to \infty$, as required.  The proof for $M_{-\infty}$ is similar.
\end{proof}

We now come to the celebrated inequality of the arithmetic and geometric
means:
\[
\frac{1}{n} \sum_{i = 1}^n x_i
\geq
\Biggl( \prod_{i = 1}^n x_i \Biggr)^{1/n}
\]
for all $x_1, \ldots, x_n \geq 0$.  This is a very special case of the
following classical and fundamental result (Theorem~9 of Hardy, Littlewood
and P\'olya~\cite{HLP}, for instance).  Recall from
Remark~\ref{rmk:defn-inc} that we use the word `increasing' in the
non-strict sense.

\begin{thm}
\lbl{thm:mns-inc-ord}
\index{power mean!increasing in order@is increasing in order}
Let $\p \in \Delta_n$ and $\vc{x} \in [0, \infty)^n$, with $x_i > 0$ for
  all $i \in \supp(\p)$.  Then the function
\[
\begin{array}{ccc}
[-\infty, \infty]       &\to            &[0, \infty)   \\
t                       &\mapsto        &M_t(\p, \vc{x})
\end{array}
\]
is increasing.  It is constant if $x_i = x_j$ for all $i, j \in \supp(\p)$,
and strictly increasing otherwise.
\end{thm}

\begin{proof}
If the coordinates $x_i$ of $\vc{x}$ have the same value $x$ for all $i \in
\supp(\p)$, then evidently $M_t(\p, \vc{x}) = x$ for all $t \in [-\infty,
  \infty]$.  Supposing otherwise, we have to prove that $M_t(\p, \vc{x})$
is strictly increasing in $t \in [-\infty, \infty]$.  We will prove that
$\tfrac{d}{dt} \log M_t(\p, \vc{x}) > 0$ for all $t \in (-\infty, 0) \cup
(0, \infty)$.  Since $M_t(\p, \vc{x})$ is continuous in $t \in [-\infty,
  \infty]$ (Lemma~\ref{lemma:pwr-mns-cts-t}), this suffices.  For real $t
\neq 0$, 
\begin{align}
\frac{d}{dt} \log M_t(\p, \vc{x})      &
=
\frac{d}{dt} \Biggl( \frac{\log \sum p_i x_i^t}{t} \Biggr)      
\nonumber       \\
&
=
\frac{t \bigl(\sum p_i x_i^t \log x_i\bigr)/
\bigl(\sum p_i x_i^t\bigr) - \log \sum p_i x_i^t}%
{t^2}   
\nonumber       \\
&
=
\frac{\sum p_i x_i^t \log x_i^t 
- \bigl(\sum p_i x_i^t\bigr) \log \sum p_i x_i^t}%
{t^2 \sum p_i x_i^t}    
\nonumber       \\
&
=
\frac{- \sum p_i \partial(x_i^t) + \partial \bigl(\sum p_i x_i^t\bigr)}%
{t^2 \sum p_i x_i^t},
\lbl{eq:log-M-conc}
\end{align}
where all sums are over $i \in \supp(\p)$ and $\partial(x) = -x\log x$ (as
in equation~\eqref{eq:defn-par}).  But $\partial''(x) = -1/x < 0$ for all
$x > 0$, so $\partial$ is strictly concave.  Hence by
equation~\eqref{eq:log-M-conc},
\[
\tfrac{d}{dt} \log M_t(\p, \vc{x}) \geq 0,
\]
with equality if and only if $x_i^t = x_j^t$ for all $i, j \in \supp(\p)$.
But $t \neq 0$, so equality only holds if $x_i = x_j$ for all $i, j \in
\supp(\p)$, contrary to our earlier assumption.  Hence the inequality is
strict, as required.
\end{proof}

There is a simple duality%
\index{power mean!duality for}%
\index{duality for power means}
law for power means: 
\begin{equation}
\lbl{eq:mn-duality}
M_{-t}(\p, \vc{x}) 
=
\frac{1}{M_t(\p, 1/\vc{x})}
\end{equation}
for all $t \in [-\infty, \infty]$, $\p \in \Delta_n$, and $\vc{x} \in (0,
\infty)^n$.  Here $1/\vc{x}$ denotes the vector $(1/x_1, \ldots,
1/x_n)$.  For instance, in the case $t = 1$, the harmonic mean is the
reciprocal of the arithmetic means of $1/x_1, \ldots, 1/x_n$.

\begin{remark}
\lbl{rmk:vec-op-conv}
Often in this text, we will want to perform coordinatewise algebraic
operations on vectors.\index{vector!operations}  For instance, given
$\vc{x}, \vc{y} \in \R^n$, we will use not only the (coordinatewise) sum
and difference $\vc{x} + \vc{y}$ and $\vc{x} - \vc{y}$, but also the
coordinatewise product and quotient
\[
\vc{x}\vc{y} = (x_1 y_1, \ldots, x_n y_n),
\qquad
\vc{x}/\vc{y} = (x_1/y_1, \ldots, x_n/y_n)
\ntn{vecops}
\]
(with the usual caveats regarding $y_i = 0$ in the latter case).  This is
just the standard notation for the product and quotient of real-valued
functions on a set $S$, applied to $S = \{1, \ldots, n\}$.
\end{remark}

We now run through some basic properties satisfied by the power means 
\[
\bigl( 
M_t \from \Delta_n \times [0, \infty)^n \to [0, \infty) 
\bigr)_{n \geq 1}
\]
of every order $t \in [-\infty, \infty]$.  For later purposes, it is useful
to set up the terminology in the generality of a sequence of functions
\[
\bigl( 
M \from \Delta_n \times I^n \to I
\bigr)_{n \geq 1},
\]
where $I$ is an arbitrary real interval.  The most important cases are $I =
[0, \infty)$ and $I = (0, \infty)$.

\begin{defn}
\lbl{defn:pwr-mn-elem}
Let $I$ be a real interval and let $(M \from \Delta_n \times I^n \to I)_{n
  \geq 1}$ be a sequence of functions.
\begin{enumerate}
\item 
\lbl{part:pme-sym}
$M$ is \demph{symmetric}%
\index{symmetric!weighted mean}
if $M(\p, \vc{x}) = M(\p\sigma, \vc{x}\sigma)$ for
  all $n \geq 1$, $\p \in \Delta_n$, $\vc{x} \in I^n$, and permutations
  $\sigma$ of $\{1, \ldots, n\}$, where $\p\sigma$ and $\vc{x}\sigma$ are
  defined as in equation~\eqref{eq:p-perm}.

\item
\lbl{part:pme-abs}
$M$ is \demph{absence-invariant}%
\index{absence-invariance!mean@of mean}
if whenever $\p \in \Delta_n$, $\vc{x} \in I^n$ and $1 \leq i \leq n$ with
$p_i = 0$, then
\[
M(\p, \vc{x})
=
M\bigl(
(p_1, \ldots, p_{i - 1}, p_{i + 1}, \ldots, p_n),
(x_1, \ldots, x_{i - 1}, x_{i + 1}, \ldots, x_n)
\bigr).
\]

\item
$M$ has the \demph{repetition}%
\index{repetition}
property if
whenever $\p \in \Delta_n$, $\vc{x} \in I^n$ and $1 \leq i < n$ with $x_i
= x_{i + 1}$, then
\begin{multline*}
M(\p, \vc{x}) 
\\
=
M\bigl(
(p_1, \ldots, p_{i - 1}, p_i + p_{i + 1}, p_{i + 2}, \ldots, p_n),
(x_1, \ldots, x_{i - 1}, x_i, x_{i + 2}, \ldots, x_n)
\bigr).
\end{multline*}
\end{enumerate}
\end{defn}

Absence-invariance states that $M$ behaves logically with respect to
elements $x_i$ that are absent (have zero weight): such elements might as
well be ignored.

\begin{lemma}
\lbl{lemma:pwr-mns-elem}
\index{power mean!symmetry of}
\index{power mean!absence-invariance of}
\index{power mean!repetition for} 
Let $t \in [-\infty, \infty]$.  Then $M_t$ has the symmetry,
absence-invariance and repetition properties.
\end{lemma}

A direct proof of this lemma is, of course, elementary, but it is
enlightening to derive all three properties from a single general law, as
follows.  Let
\[
f \from \{1, \ldots, m\} \to \{1, \ldots, n\}
\]
be a map of finite sets.  Any distribution $\p \in \Delta_m$ gives rise to
a pushforward distribution $f \p \in \Delta_n$
(Definition~\ref{defn:pfwd}).  On the other hand, any vector $\vc{x} \in
[0, \infty)^n$ can be pulled back along $f$ to give a vector $\vc{x}f \in
  [0, \infty)^m$, where
\[
(\vc{x}f)_i = x_{f(i)}
\ntn{xf}
\]
($i \in \{1, \ldots, m\}$).  

\begin{defn}
\lbl{defn:mn-nat}
Let $I$ be a real interval.  A sequence of functions $\bigl( M \from
\Delta_n \times I^n \to I \bigr)_{n \geq 1}$ is
\demph{natural}%
\index{naturality!weighted mean@of weighted mean} 
if  
\[
M(f\vc{p}, \vc{x}) = M(\vc{p}, \vc{x}f)
\]
for all $m, n \geq 1$, $\vc{p} \in \Delta_m$, $\vc{x} \in I^n$, and maps of
sets 
\[
f \from \{1, \ldots, m\} \to \{1, \ldots, n\}.
\]
\end{defn}

\begin{remark}
If we write $x_j$ as $\phi(j)$, so that $\phi$ is a function $\{1, \ldots,
n\} \to [0, \infty)$, then $\vc{x}f = \phi \of f$.  If we also write
$M(\p, -)$ as $\int - \dee\p$ then naturality states that
\[
\int \phi \dee(f \p) = \int (\phi \of f) \dee\p,
\]
the standard formula for integration under a change%
\index{change of variable}
of variable.  However, this notation is misleading: unlike an ordinary
integral, $M(\p, \vc{x})$ need not be linear in $\vc{x}$ (and is not when
$M = M_t$ for $t \neq 1$).
\end{remark}

\begin{lemma}[Naturality]
\lbl{lemma:pwr-mns-nat}
\index{power mean!naturality of}
For each $t \in [-\infty, \infty]$, the power mean $M_t$ on $[0, \infty)$
  is natural.
\end{lemma}

\begin{proof}
Take $\p$, $\vc{x}$, and $f$ as in Definition~\ref{defn:mn-nat}.  We have to
show that $M_t(f\p, \vc{x}) = M_t(\p, \vc{x}f)$.  First suppose that $t
\neq 0, \pm\infty$ and that $x_j > 0$ for all $j \in \{1, \ldots, n\}$.
Then
\begin{align*}
M_t(f\p, \vc{x})     &
=
\Biggl( \sum_{j \in \supp(f\p)} (f \p)_j x_j^t \Biggr)^{1/t}    \\
&
=
\Biggl( 
\sum_{\,j \in \supp(f\p)\ } \sum_{i \in f^{-1}(j)} p_i x_j^t 
\Biggr)^{1/t}   \\
&
=
\Biggl( \sum_{\,i \in \supp(\p)\,} p_i x_{f(i)}^t \Biggr)^{1/t}     \\
&
=
M_t(\p, \vc{x}f),
\end{align*}
as required.  The case where $x_j = 0$ for some values of $j$ follows by
continuity of $M_t(\p, \vc{x})$ in $\vc{x}$
(Lemma~\ref{lemma:pwr-mns-cts-x}), and the result for $t = 0$ and $t =
\pm\infty$ follows by continuity of $M_t$ in $t$
(Lemma~\ref{lemma:pwr-mns-cts-t}).  
\end{proof}

\begin{pfof}{Lemma~\ref{lemma:pwr-mns-elem}}\lbl{p:lpme-pf}
We use the naturality of the power means
for all three parts.  Write $\lwr{n} = \{1, \ldots, n\}$.  Symmetry follows
by taking $f$ to be a bijection $\lwr{n} \to \lwr{n}$.
Absence-invariance follows by taking $f$ to be the order-preserving
injection $\lwr{n - 1} \to \lwr{n}$ that omits $i$ from its image.  The
repetition property follows by taking $f$ to be the order-preserving
surjection $\lwr{n} \to \lwr{n - 1}$ that identifies $i$ with $i + 1$.
\end{pfof}

\begin{remark}
\lbl{rmk:defined-even-if-not}
The absence-invariance of the power means implies that $M_t(\p, \vc{x})$
is unaffected by the value of $x_i$ for coordinates $i$ such that $p_i =
0$.  Indeed, writing $\supp(\p) = \{i_1, \ldots, i_k\}$ with $i_1 < \cdots
< i_k$, we have
\[
M_t(\p, \vc{x})
=
M_t\bigl( (p_{i_1}, \ldots, p_{i_k}), (x_{i_1}, \ldots, x_{i_k})\bigr)
\]
for all $\vc{x}$, by absence-invariance and induction.  Hence
\[
M_t(\p, \vc{x}) = M_t(\p, \vc{y})
\]
whenever $\vc{x}, \vc{y} \in [0, \infty)^n$ with $x_i = y_i$ for all $i \in
\supp(\p)$.

Because of this, the expression $M_t(\p, \vc{x})$ has a clear meaning even
if $x_i$ is \emph{undefined} for some or all $i \not\in \supp(\p)$.  (We
can arbitrarily put $x_i = 0$ or $x_i = 17$ for all such $i$;
it makes no difference.)  For example, the expression
\[
M_t(\p, 1/\p)
\]
has a clear meaning for all $\p \in \Delta_n$, even if $p_i = 0$ for some
$i$; writing $\supp(\p) = \{i_1, \ldots, i_k\}$ as above, it is understood
to mean
\[
M_t\bigl( (p_{i_1}, \ldots, p_{i_k}), (1/p_{i_1}, \ldots, 1/p_{i_k})\bigr).
\]

\femph{We adopt the convention%
\index{power mean!undefined arguments@with undefined arguments}%
\index{undefined arguments}
throughout this text} that power means
$M_t(\p, \vc{x})$ are valid expressions even if $x_i$ is undefined for some
$i \not\in \supp(\p)$, and are to be interpreted as just described.  This
convention is strictly analogous to the standard interpretation of integral
notation $\int f \dee\mu$, for a function $f$ and a measure $\mu$: the
integral is unaffected by the value of $f$ off the support of $\mu$, and
has an unambiguous meaning even if $f$ is undefined there.
\end{remark}

A minimal requirement on anything called a mean is that the mean of
several copies of $x$ should be $x$:

\begin{defn}
\lbl{defn:w-cons}
Let $I$ be a real interval.  A sequence of functions $(M \from \Delta_n
\times I^n \to I)_{n \geq 1}$ is \demph{consistent}%
\index{consistent!weighted mean} 
if
\[
M\bigl(\p, (x, \ldots, x)\bigr) = x
\]
for all $n \geq 1$, $\vc{p} \in \Delta_n$, and $x \in I$.
\end{defn}

\begin{lemma}
\lbl{lemma:pwr-mns-con}
\index{power mean!consistency of}
For each $t \in [-\infty, \infty]$, the power mean $M_t$ 
is consistent.
\end{lemma}

\begin{proof}
Trivial.
\end{proof}

For $\vc{x}, \vc{y} \in \R^n$, write $\vc{x} \leq\ntn{leq} \vc{y}$ if $x_i
\leq y_i$ for all $i \in \{1, \ldots, n\}$.

\begin{defn}
\lbl{defn:w-isi}
Let $I$ be a real interval and let $( M \from \Delta_n \times I^n \to I)_{n
  \geq 1}$ be a sequence of functions.
\begin{enumerate}
\item
$M$ is \demph{increasing}%
\index{increasing!weighted mean}%
\index{mean!increasing}
if 
\[
\vc{x} \leq \vc{y}
\implies
M(\p, \vc{x}) \leq M(\p, \vc{y})
\]
for all $n \geq 1$, $\vc{p} \in \Delta_n$, and $\vc{x}, \vc{y} \in I^n$.

\item
$M$ is \demph{strictly increasing}%
\index{strictly increasing!weighted mean}%
\index{mean!strictly increasing}%
\index{increasing!strictly}
if 
\[
\bigl( \vc{x} \leq \vc{y} \text{ and } 
x_i < y_i \text{ for some } i \in \supp(\p) \bigr)
\implies
M(\p, \vc{x}) < M(\p, \vc{y})
\]
for all $n \geq 1$, $\vc{p} \in \Delta_n$, and $\vc{x}, \vc{y} \in I^n$. 
\end{enumerate}
\end{defn}

Whether the power mean $M_t$ is \emph{strictly} increasing depends on both
the order $t$ and whether the domain of definition is taken to be $[0,
  \infty)$ or $(0, \infty)$, as follows.

\begin{lemma}
\lbl{lemma:pwr-mns-inc}
\begin{enumerate}
\item
For all $t \in [-\infty, \infty]$, the power mean $M_t$ on $[0, \infty)$ is
  increasing. 

\item
For all $t \in (-\infty, \infty)$, the power mean $M_t$ on $(0, \infty)$ is
strictly increasing. 

\item
For all $t \in (0, \infty)$, the power mean $M_t$ on $[0, \infty)$ is strictly
  increasing.
\end{enumerate}
\end{lemma}

\begin{proof}
Elementary.
\end{proof}

\begin{remark}
\lbl{rmk:pwr-mns-not-si}
The careful statement of Lemma~\ref{lemma:pwr-mns-inc} is necessary because
of various limiting counterexamples.  The means $M_{\pm\infty}$ are not
strictly increasing on $(0, \infty)$, since, for instance,
\[
M_\infty\bigl(\vc{u}_2, (1, 3)\bigr) 
=
3 
= 
M_\infty\bigl(\vc{u}_2, (2, 3)\bigr).
\]
When $t \in [-\infty, 0]$, the mean $M_t$ is not strictly increasing on
$[0, \infty)$; for example,
\[
M_t\bigl( \vc{u}_2, (0, 1)\bigr)
=
0
=
M_t\bigl( \vc{u}_2, (0, 2)\bigr).
\]
\end{remark}

\begin{defn}
\lbl{defn:w-mn-hgs}
Let $I$ be a real interval closed under multiplication.  A sequence of
functions $(M \from \Delta_n \times I^n \to I)_{n \geq 1}$ is
\demph{homogeneous}%
\index{homogeneous!weighted mean} 
if 
\[
M(\p, c\vc{x}) = cM(\p, \vc{x})
\]
for all $n \geq 1$, $\p \in \Delta_n$, $c \in I$, and $\vc{x} \in I^n$.
\end{defn}

The hypothesis on $I$ guarantees that $M(\p, c\vc{x})$ is defined.

\begin{lemma}
\lbl{lemma:pwr-mns-hgs}
\index{power mean!homogeneity of}
For each $t \in [-\infty, \infty]$, the power mean $M_t$ on $[0, \infty)$
is homogeneous.  
\end{lemma}

\begin{proof}
Elementary.
\end{proof}

The most important algebraic property of the power means is a
chain rule.  Given vectors
\[
\vc{x}^1 = \bigl(x^1_1, \ldots, x^1_{k_1}\bigr) \in \R^{k_1},
\ \ldots, \ 
\vc{x}^n = \bigl(x^n_1, \ldots, x^n_{k_n}\bigr) \in \R^{k_n},
\]
write
\[
\vc{x}^1 \oplus \cdots \oplus \vc{x}^n
=
\bigl(x^1_1, \ldots, x^1_{k_1}, \ \ldots, \ x^n_1, \ldots, x^n_{k_n}\bigr)
\in 
\R^{k_1 + \cdots + k_n}.
\ntn{oplusvec}
\]

\begin{defn}
\lbl{defn:mns-chn}
Let $I$ be a real interval.  A sequence of functions $\bigl( M \from
\Delta_n \times I^n \to I \bigr)_{n \geq 1}$ satisfies the \demph{chain%
\index{chain rule!means@for means}
rule} if
\[
M\bigl(\vc{w} \of (\p^1, \ldots, \p^n), 
\vc{x}^1 \oplus\cdots\oplus \vc{x}^n\bigr)
=
M\Bigl(\vc{w},
\bigl( M(\p^1, \vc{x}^1), \ldots, M(\p^n, \vc{x}^n) \bigr)
\Bigr)
\]
for all $\vc{w} \in \Delta_n$, $\p^i \in \Delta_{k_i}$, and $\vc{x}^i \in
I^{k_i}$.
\end{defn}

\begin{propn}[Chain rule]
\lbl{propn:pwr-mns-chn}%
\index{chain rule!power means@for power means}%
\index{power mean!chain rule for}
For each $t \in [-\infty, \infty]$, the power mean $M_t$ on $[0, \infty)$
satisfies the chain rule.
\end{propn}

\begin{proof}
By the continuity of the power means in their second argument and in their
order (Lemmas~\ref{lemma:pwr-mns-cts-x} and~\ref{lemma:pwr-mns-cts-t}), it
is enough to prove the equation in Definition~\ref{defn:mns-chn} when
$x^i_j > 0$ for all $i, j$ and $0 \neq t \in \R$.  Then
\begin{align*}
M_t\bigl(\vc{w} \of (\p^1, \ldots, \p^n), 
\vc{x}^1 \oplus\cdots\oplus \vc{x}^n \bigr)   &
=
\Biggl\{ 
\sum_{i = 1}^n \sum_{j = 1}^{k_i} w_i p^i_j \bigl(x^i_j\bigr)^t     
\Biggr\}^{1/t}  \\
&
=
\Biggl\{
\sum_{i = 1}^n w_i M_t\bigl(\p^i, \vc{x}^i\bigr)^t        
\Biggr\}^{1/t}  \\
&
=
M_t \Bigl( \vc{w}, 
\bigl( M_t(\p^1, \vc{x}^1), \ldots, M_t(\p^n, \vc{x}^n) \bigr) 
\Bigr),
\end{align*}
as required.
\end{proof}

An important consequence of the chain rule is that in order to calculate
the mean of $\vc{x}^1 \oplus\cdots\oplus \vc{x}^n$ weighted by $\vc{w} \of
(\p^1, \ldots, \p^n)$, we only need to know $\vc{w}$ and the means
$M_t(\p^i, \vc{x}_i)$, not $\p^i$ and $\vc{x}^i$ themselves.  We
refer to this property as modularity, echoing the definition of
modularity for diversity measures (p.~\pageref{p:D-mod}).  (Modularity of this
kind has also been called \demph{quasilinearity},%
\index{quasilinear mean} 
as in Section~6.21 of Hardy, Littlewood and P\'olya~\cite{HLP}.)  Formally:

\begin{defn}
\lbl{defn:mns-mod}%
\index{modularity!weighted mean@of weighted mean}
Let $I$ be a real interval.  A sequence of functions $\bigl( M \from
\Delta_n \times I^n \to I \bigr)_{n \geq 1}$ is \demph{modular} if
\begin{align*}
&
M\bigl(\p^i, \vc{x}^i \bigr) 
= 
M\bigl(\twid{\p}^i, \twid{\vc{x}}^i\bigr) 
\text{ for all } i \in \{1, \ldots, n\}     \\
\implies
&
M\bigl( \vc{w} \of (\p^1, \ldots, \p^n), 
\vc{x}^1 \oplus\cdots\oplus \vc{x}^n  \bigr) =
M\bigl( \vc{w} \of (\twid{\p}^1, \ldots, \twid{\p}^n),
\twid{\vc{x}}^1 \oplus\cdots\oplus \twid{\vc{x}}^n  \bigr) 
\end{align*}
for all $n, k_1, \ldots, k_n, \twid{k}_1, \ldots, \twid{k}_n \geq 1$ and
$\vc{w} \in \Delta_n$, $\p^i \in \Delta_{k_i}$, $\twid{\p}^i \in
\Delta_{\twid{k}_i}$, $\vc{x}^i \in I^{k_i}$, $\twid{\vc{x}}^i \in
I^{\twid{k}_i}$.   
\end{defn}

\begin{cor}
\lbl{cor:pwr-mns-mod}
\index{power mean!modularity of}
For each $t \in [-\infty, \infty]$, the power mean $M_t$ on $[0, \infty)$
is modular. 
\qed
\end{cor}

As for diversity of order~$1$ (equation~\eqref{eq:div1-mult},
p.~\pageref{eq:div1-mult}), the chain rule also implies a
multiplicativity property.  For $\vc{x} \in \R^n$ and $\vc{y} \in \R^k$,
write
\begin{equation}
\lbl{eq:defn-real-tensor}
\vc{x} \otimes \vc{y}
=
(x_1 y_1, \ldots, x_1 y_k, 
\ \ldots, \ 
x_n y_1, \ldots, x_n y_k)
\in \R^{nk}.
\end{equation}
(To justify the notation: if the tensor product of vector spaces $\R^n
\otimes \R^k$ is identified with $\R^{nk}$ in the standard way, then the
vector usually written as $\vc{x} \otimes \vc{y} \in \R^n \otimes \R^k$
corresponds to what we are now writing as $\vc{x} \otimes \vc{y} \in
\R^{nk}$.)

\begin{defn}
\lbl{defn:w-mult}
Let $I$ be a real interval closed under multiplication.  A sequence of
functions $\bigl( M \from \Delta_n \times I^n \to I \bigr)_{n \geq 1}$ is
\demph{multiplicative}%
\index{multiplicative!weighted mean} 
if
\[
M(\vc{p} \otimes \vc{p}', \vc{x} \otimes \vc{x}')
=
M(\vc{p}, \vc{x}) M(\vc{p}', \vc{x}')
\]
for all $n, n' \geq 1$, $\vc{p} \in \Delta_n$, $\vc{p}' \in \Delta_{n'}$,
$\vc{x} \in I^n$, and $\vc{x}' \in I^{n'}$.  
\end{defn}

\begin{cor}
\lbl{cor:pwr-mns-mult}
\index{power mean!multiplicativity of}
For each $t \in [-\infty, \infty]$, the power mean $M_t$ on $[0, \infty)$
is multiplicative. 
\end{cor}

\begin{proof}
We apply the chain rule (Proposition~\ref{propn:pwr-mns-chn}) to the
composite distribution
\[
\p \of (\p', \ldots, \p') = \p \otimes \p'
\]
and the vector
\[
x_1 \vc{x}' \oplus \cdots \oplus x_n \vc{x}' = \vc{x} \otimes \vc{x}'.
\]
Doing this gives
\[
M_t(\p \otimes \p', \vc{x} \otimes \vc{x}')
=
M_t\Bigl( 
\p, 
\bigl(
M_t(\p, x_1\vc{x}'), \ldots, M_t(\p, x_n\vc{x}')
\bigr)
\Bigr).
\]
Hence by two uses of homogeneity,
\begin{align*}
M_t(\p \otimes \p', \vc{x} \otimes \vc{x}')     &
=
M_t\Bigl( 
\p, 
\bigl(
x_1 M_t(\p, \vc{x}'), \ldots, x_n M_t(\p, \vc{x}')
\bigr)
\Bigr)  \\
&
=
M_t(\p, \vc{x}) M_t(\p', \vc{x}').
\end{align*}
\end{proof}

The multiplicativity property is remarkably powerful, as we shall see in
Chapter~\ref{ch:prob}.

Finally, we record for later purposes a simple result connecting the power
means with the $q$-logarithms.

\begin{lemma}
\lbl{lemma:q-log-mean}
Let $q \in [0, \infty)$, $\p \in \Delta_n$, and $\vc{x} \in [0, \infty)^n$,
    with $x_i > 0$ for all $i \in \supp(\p)$.  Then
\[
\ln_q M_{1 - q} (\p, \vc{x}) = M_1(\p, \ln_q \vc{x}),
\]
where $\ln_q \vc{x} = (\ln_q x_1, \ldots, \ln_q x_n)$. 
\end{lemma}

\begin{proof}
Trivial algebraic manipulation.
\end{proof}

\section{R\'enyi entropies and Hill numbers}
\lbl{sec:ren-hill}

Historically, the first deformations of Shannon entropy were the R\'enyi
entropies~\cite{Reny}, defined as follows.

\begin{defn}
\index{Renyi entropy@R\'enyi entropy}
Let $q \in [-\infty, \infty]$, $n \geq 1$, and $\p \in \Delta_n$.  The
\demph{R\'enyi entropy of order%
\index{order!Renyi entropy@of R\'enyi entropy}
$q$} of $\p$ is
\begin{equation}
\lbl{eq:defn-renyi}
H_q(\p) = \log M_{1 - q}(\p, 1/\p),
\end{equation}
where $1/\p = (1/p_1, \ldots, 1/p_n)$.
\end{defn}

Here we use the convention introduced in
Remark~\ref{rmk:defined-even-if-not}, which covers the possibility that
$1/p_i$ is undefined for some values of $i$.

Explicitly,
\[
H_q(\p) 
=
\frac{1}{1 - q} \log \sum_{i \in \supp(\p)} p_i^q
\]
for $q \neq 1, \pm\infty$, and
\begin{align*}
H_{-\infty}(\p) &
=
-\log\min_{i \in \supp(\p)} p_i,        \\
H_1(\p) &
=
H(\p),  \\
H_\infty(\p)    &
=
-\log\max_{i \in \supp(\p)} p_i.
\end{align*}
By Lemma~\ref{lemma:pwr-mns-cts-t}, $H_q(\p)$ is continuous in $q$.

R\'enyi introduced these entropies in 1961~\cite{Reny}.  One of his
purposes in doing so was to point out that Shannon entropy is far from the
only useful quantity with the logarithmic property
\begin{equation}
\lbl{eq:ren-log}
H(\p \otimes \vc{r}) = H(\p) + H(\vc{r})
\end{equation}
($\vc{p} \in \Delta_n, \vc{r} \in \Delta_m$).  Indeed, $H_q$ has this same
property for all $q \in [-\infty, \infty]$.  This follows from the
multiplicativity of the power means (Corollary~\ref{cor:pwr-mns-mult}),
since
\begin{align*}
H_q(\vc{p} \otimes \vc{r})      &
=
\log M_{1 - q} \Bigl(\vc{p} \otimes \vc{r}, 
\tfrac{1}{\p} \otimes \tfrac{1}{\vc{r}} \Bigr)  \\
&
=
\log \Bigl(
M_{1 - q} \Bigl( \p, \tfrac{1}{\p} \Bigr)
M_{1 - q} \Bigl( \vc{r}, \tfrac{1}{\vc{r}} \Bigr)
\Bigr)  \\
&
=
H_q(\p) + H_q(\vc{r}).
\end{align*}
In this respect, the R\'enyi entropies resemble Shannon entropy more
closely than the $q$-logarithmic entropies do.  But there is a price to
pay.  Whereas the asymmetry of the multiplication formula for the
$q$-logarithmic entropies (equation~\eqref{eq:q-ent-mult}) could be exploited
to prove an extremely simple characterization theorem
(Theorem~\ref{thm:q-ent-char}), this avenue is not open to us for the
R\'enyi entropies.  We do prove a characterization theorem for the R\'enyi
entropy of any given order (Section~\ref{sec:hill-char-given}), but it is
more involved.

The $q$-logarithmic%
\index{q-logarithmic entropy@$q$-logarithmic entropy!Renyi entropy@and R\'enyi entropy}%
\index{Renyi entropy@R\'enyi entropy!q-logarithmic entropy@and $q$-logarithmic entropy}
and R\'enyi entropies each determine the other, since
both are invertible functions of $\sum p_i^q$.  Explicitly, 
\begin{align}
S_q(\p) &
=
\frac{1}{1 - q} \Bigl( \exp\bigl((1 - q) H_q(\p)\bigr) - 1 \Bigr),
\lbl{eq:q-log-ren}    \\
H_q(\p) &
=
\frac{1}{1 - q} \log \bigl( (1 - q) S_q(\p) + 1 \bigr)
\lbl{eq:ren-q-log}
\end{align}
for real $q \neq 1$, and 
\begin{equation}
\lbl{eq:shh}
S_1(\p) = H(\p) = H_1(\p).
\end{equation}
Equations~\eqref{eq:q-log-ren}--\eqref{eq:shh} can be written more
compactly as 
\begin{align}
S_q(\p) & 
= 
\ln_q (\exp H_q(\p)),   
\lbl{eq:q-log-ren-slick}      \\
H_q(\p) & 
= 
\log (\exp_q S_q(\p))
\lbl{eq:ren-q-log-slick}
\end{align}
($q \in \R$), where $\exp_q$ is the inverse%
\index{q-logarithm@$q$-logarithm!inverse of} 
function of $\ln_q$, given explicitly by
\[
\exp_q(y)
=
\begin{cases}
\bigl( 1 + (1 - q)y \bigr)^{1/(1 - q)}  &\text{if } q \neq 1,   \\
\exp(y)                                 &\text{if } q = 1.
\end{cases}
\ntn{expq}
\]
The transformations relating $S_q(\p)$ to $H_q(\p)$ are strictly
increasing, so maximizing or minimizing one is equivalent to maximizing or
minimizing the other.

\begin{remark}
When $q = \pm\infty$, the R\'enyi entropy $H_q(\p)$ is defined but the
$q$-logarithmic entropy $S_q(\p)$ is not.  It is straightforward to check
that
\[
\lim_{q \to \infty} S_q(\p) = 0
\]
for all $\p$, and 
\[
\lim_{q \to -\infty} S_q(\p)
=
\begin{cases}
0       &\text{if } p_i = 1 \text{ for some } i,        \\
\infty  &\text{otherwise}.
\end{cases}
\]
The only sensible way to define $S_\infty(\p)$ and $S_{-\infty}(\p)$ would
be as these limits; but then the definitions would be trivial, would take
infinite values in the latter case, and would break the result that
$H_q(\p)$ can be recovered from $S_q(\p)$.  We therefore leave
$S_{\pm\infty}(\p)$ undefined.
\end{remark}

\begin{remark}
\lbl{rmk:gen-def}
It is easy to manufacture other one-parameter families of entropies
extending the Shannon entropy: simply take the formula
\[
\frac{1}{1 - q} \log \sum_{i \in \supp(\p)} p_i^q
\]
defining R\'enyi entropy for $q \neq 1$, and replace $\log$ by some other
function $\lambda$.  In order that the limit as $q \to 1$ is $H(\p)$, the
requirements on $\lambda$ are that $\lambda(1) = 0$ and $\lambda'(1) = 1$.
The simplest function $\lambda$ with these properties is $\lambda(x) = x - 1$,
the linear approximation to $\log$ at $1$.  Indeed, taking this simplest
$\lambda$ gives exactly the $q$-logarithmic entropy.
\end{remark}

The \emph{exponentials} of the R\'enyi entropies turn out to have slightly
more convenient algebraic properties than the R\'enyi entropies themselves,
and are important measures of biological diversity.  We give the definition
and examples here, and describe their properties in the next section.

\begin{defn}
\lbl{defn:hill}
\index{Hill number}
Let $q \in [-\infty, \infty]$ and $\p \in \Delta_n$.  The \demph{Hill
  number of order%
\index{order!Hill number@of Hill number} 
$q$} of $\p$ is
\[
D_q(\p)
=
\exp H_q(\p)
=
M_{1 - q}(\p, 1/\p).
\ntn{Dq}
\]
We also call this the \demph{diversity%
\index{diversity!order q@of order $q$} 
of order%
\index{order!diversity measure@of diversity measure} 
$q$} of $\p$.
\end{defn}

Thus, the Hill number $D_q$ is related to the R\'enyi entropy $H_q$ and
$q$-logarithmic entropy $S_q$ by
\begin{equation}
\lbl{eq:dhs}
H_q = \log D_q,
\qquad
S_q = \ln_q D_q
\end{equation}
(by definition and equation~\eqref{eq:q-log-ren-slick}).  Explicitly, 
\begin{equation}
\lbl{eq:hill-gen}
D_q(\p) 
=
\Biggl( \sum_{i \in \supp(\p)} p_i^q \Biggr)^{1/(1 - q)}
\end{equation}
for $q \neq 1, \pm\infty$, and
\begin{align}
D_{-\infty}(\p) &
=
1\Big/\!\!\min_{i \in \supp(\p)} p_i,   
\lbl{eq:hill-minfty}    \\
D_1(\p) &
=
\prod_{i \in \supp(\p)} p_i^{-p_i}
=
D(\p),  
\lbl{eq:hill-order1}    \\
D_\infty(\p)    &
=
1\Big/\!\!\max_{i \in \supp(\p)} p_i.
\lbl{eq:hill-pinfty}
\end{align}
This definition of diversity of order $q$ extends the earlier definition of
diversity of order $1$ (Definition~\ref{defn:div1}), there written as $D$.

The quantities $D_q$ are named after the ecologist Mark%
\index{Hill Mark@Hill, Mark}
Hill~\cite{Hill}, who introduced them in 1973 as measures of diversity
(building on R\'enyi's work).  In Section~\ref{sec:total-hill}, we will
prove a theorem pinpointing what makes the Hill numbers uniquely suitable
as measures of diversity.  For now, the following explanation can be given.

Let $\p = (p_1, \ldots, p_n)$ be the relative abundance distribution of a
community.  As in Section~\ref{sec:ent-div}, $1/p_i$ measures the
rarity\index{rarity} or specialness\index{specialness} of the $i$th
species.  There, we took the geometric mean $\prod (1/p_i)^{p_i}$ of the
rarities as our measure of diversity.  But we could just as reasonably use
some other power mean $M_t(\p, 1/\p)$.  Reparametrizing as $q = 1 - t$,
this is exactly the Hill number $D_q(\p)$.

The Hill numbers are effective%
\index{effective number!Hill numbers are} 
numbers (Definition~\ref{defn:div-eff}):
\begin{equation}
\lbl{eq:hill-eff-num}
D_q(\vc{u}_n) = n
\end{equation}
for all $n \geq 1$ and $q \in [-\infty, \infty]$.  By
equation~\eqref{eq:dhs}, the quantities $D_q$, $H_q$ and $S_q$ are related
to one another by increasing, invertible transformations.  Thus, the Hill
numbers are the result of taking either the R\'enyi entropies $H_q$ or the
$q$-logarithmic entropies $S_q$ and converting them into effective numbers.
In the terminology originating in economics\index{economics}
(Bishop~\cite{Bish}, p.~789) and now also used in ecology
(Ellison~\cite{ElliPD}, for instance), $D_q$ is the \demph{numbers%
\index{numbers equivalent}
equivalent} of both $H_q$ and $S_q$.  

\begin{examples}
\lbl{egs:hill}
\begin{enumerate}
\item 
\lbl{eg:hill-sr}
The diversity or Hill number $D_0(\p)$ of order $0$ is simply
$\mg{\supp(\p)}$, the number of species present.  In ecology, this is
called the \demph{species%
\index{species!richness} 
richness}.  It is the most common measure of diversity in both the popular
media and the ecology literature, but makes no distinction between a rare
species and a common species, and says nothing about the balance between
the species present.

\item
We have already considered the diversity $D_1(\p)$ of order $1$
(Section~\ref{sec:ent-div}), which is the exponential of Shannon entropy.

\item
\lbl{eg:hill-2}
The diversity of $\p$ of order $2$ is 
\[
D_2(\p) = 1\bigg/\sum_{i = 1}^n p_i^2.
\]
Being the reciprocal of a quadratic form, it is especially convenient
mathematically.  It also has an intuitive probabilistic interpretation: if
we draw pairs of individuals at random from the community (with
replacement), $D_2(\p)$ is the expected number of trials needed in order to
obtain a pair of the same species.  Compare the probabilistic
interpretation of $S_2(\p)$ in Example~\ref{egs:q-ent}\bref{eg:q-ent-2}.

In ecology, $D_2(\p)$ is called the \demph{inverse%
\index{inverse Simpson concentration}%
\index{Simpson, Edward!inverse concentration} 
Simpson concentration}~\cite{SimpMD}.

\item
\lbl{eg:hill-bp}
The diversity 
\[
D_\infty(\p) 
=
1\Big/\!\max_{i \in \supp(\p)} p_i
\]
of order $\infty$ is known as the
\demph{Berger--Parker%
\index{Berger--Parker index} 
index}~\cite{BePa}.  It measures the extent to which the community is
dominated by a single species.  For instance, if one species has
outcompeted the others and makes up nearly $100\%$ of the community, then
$D_\infty(\p)$ is close to its minimum value of $1$.  At the opposite
extreme, if $\p = \vc{u}_n$ then no species is dominant and $D_\infty(\p)$
achieves its maximum value of $n$.  (General statements on maximization and
minimization of $D_q$ will be made in Lemma~\ref{lemma:div-max-min}.)
So while diversity of order $0$ gives rare species the same importance as
any other, diversity of order $\infty$ ignores them altogether.
\end{enumerate}
\end{examples}

\begin{example}
\lbl{eg:hsg}
Many of the diversity measures used in ecology are Hill numbers or
transformations of them.  Others can be expressed as combinations of
several Hill numbers.

For instance, Hurlbert~\cite{Hurl}%
\index{Hurlbert, Stuart!Smith--Grassle@--Smith--Grassle index}%
\index{expected number of species in sample}
and Smith and Grassle~\cite{SmGr} studied the expected number
$\hsg_m(\p)$\ntn{HSG} of different species represented in a random sample
(with replacement) of $m$ individuals.  Their measure turns out to be a
combination of Hill numbers of integer orders:
\[
\hsg_m(\p)
=
\sum_{q = 1}^m (-1)^{q - 1} \binom{m}{q} D_q(\p)^{1 - q}.
\]
This was first proved as Proposition~A8 in the appendix of Leinster and
Cobbold~\cite{MDISS}, and the proof is also given in
Appendix~\ref{sec:hsg} below. 
\end{example}

\begin{example}
The reciprocals of the Hill numbers have been used in
economics\index{economics} to measure concentration.\index{concentration}
One asks to what extent an industry or market is concentrated in the
hands of a small number of large players.  For example, if there are $n$
competing companies in an industry, with market shares $p_1, \ldots, p_n$,
then the concentration $1/D_q(\p)$ is maximized when one company has a
monopoly:
\[
\p = (0, \ldots, 0, 1, 0, \ldots, 0).
\]
See Hannah and Kay~\cite{HaKa} or Chakravarty and Eichhorn~\cite{ChEi}, for
instance. 
\end{example}

The parameter $q$ controls the sensitivity of the diversity measure $D_q$
to rare species, with higher values of $q$ corresponding to measures
\emph{less} sensitive to rare species.  Thus, $q$ is a
`viewpoint%
\index{viewpoint!parameter}
parameter', reflecting the importance that we wish to attach to rare
species.  For reasons to be explained, we usually restrict to
parameter values $q \geq 0$.

With the multiplicity of diversity measures that exist in the literature,
there is a risk of cherry-picking.\index{cherry-picking} Consciously or
not, a scientist might choose the measure that best supports the desired
conclusion.  There is also a risk of attaching too much importance to a
single number:
\begin{quote}
\index{Pielou, Evelyn Chrystalla}
The belief (or superstition) of some ecologists that a diversity index
provides a basis (or talisman) for reaching a full understanding of
community structure is totally unfounded
\end{quote}
(Pielou~\cite{PielED}, p.~19).  Both problems are mitigated by
systematically using \emph{all} the diversity measures $D_q$ ($0 \leq q
\leq \infty$).  The graph of $D_q(\p)$ against $q$ is called the
\demph{diversity%
\index{diversity profile}
profile} of $\p$, and plotting it displays all viewpoints%
\index{viewpoint!diversity@on diversity} 
simultaneously.

\begin{example}
\lbl{eg:hill-apes}
There are eight species of great ape\index{apes} in the world, but
$99.99\%$ of individual apes are humans.%
\index{human species} 
Figure~\ref{fig:apes} shows the absolute abundances of the eight species,
their relative abundances $p_i$, and their diversity profile.

That there are eight extant species is conveyed by the value $D_0(\p) = 8$
of the profile at $q = 0$. However, this single statistic hides the fact
that one of the species has all but totally outcompeted the others.  For
nearly any other value of the viewpoint parameter $q$, the diversity is
almost exactly $1$, reflecting the overwhelming dominance of a single
species.  For example, recall that $D_2(\p)$ is the reciprocal of the
probability that two individuals chosen at random belong to the same
species (Example~\ref{egs:hill}\bref{eg:hill-2}).  In this case, the
probability is very nearly $1$, so $D_2(\p)$ is only just greater
than $1$.

The very steep drop of the diversity profile at its left-hand end,
from $8$ to just above $1$, indicates that seven of the eight species are
exceptionally rare.
\end{example}

\begin{figure}
\centering
\begin{tabular}{|l|r|r|}
\hline
Species                 &Absolute abundance     &Relative abundance     \\
\hline
Human                   &7\,466\,964\,300       &0.99989926          \\
Bonobo                  &20\,000                &0.00000267          \\
Chimpanzee              &407\,500               &0.00005456          \\
Eastern gorilla         &4\,700                 &0.00000063          \\
Western gorilla         &200\,000               &0.00002678          \\
Bornean orangutan       &104\,700               &0.00001040          \\
Sumatran orangutan      &14\,600                &0.00000196          \\
Tapanuli orangutan      &800                    &0.00000011          \\
\hline
\end{tabular}\\

\bigskip

\lengths
\begin{picture}(120,64)(0,-4)
\cell{60}{30}{c}{\includegraphics[height=60\unitlength]{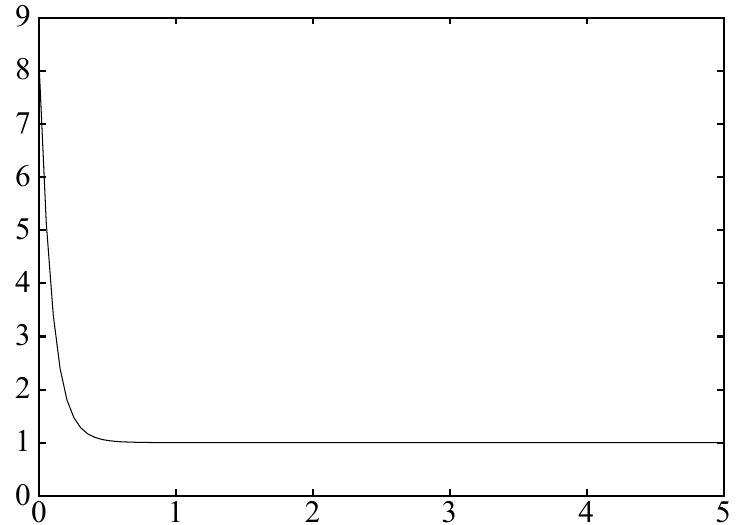}}
\cell{60}{-4}{b}{Viewpoint parameter, $q$}
\cell{15}{30}{c}{\rotatebox{90}{Diversity, $D_q(\p)$}}
\end{picture}

\caption{Abundances and species diversity profile of the estimated global
  distribution $\p$ of great apes (Hominidae).  Population estimates are
  all for 2016, with human data from United Nations~\cite{UNDE}, Tapanuli
  orangutan data from Nater et al.~\cite{Nate}, and all other data from the
  IUCN Red List of Threatened Species~\cite{AGMM,FHAF,HMOP,MBW,PRW,SWNU}.}  
\lbl{fig:apes}
\end{figure}

\begin{example}
\index{birds}
\lbl{eg:hill-birds}
Figure~\ref{fig:bird-profs-bars} shows the diversity profiles of the 
two bird communities of the Introduction (p.~\pageref{p:intro-birds}).
\begin{figure}
\lengths
\begin{picture}(61,50)(-5,-4)
\cell{28}{20}{c}{\includegraphics[height=40\unitlength]{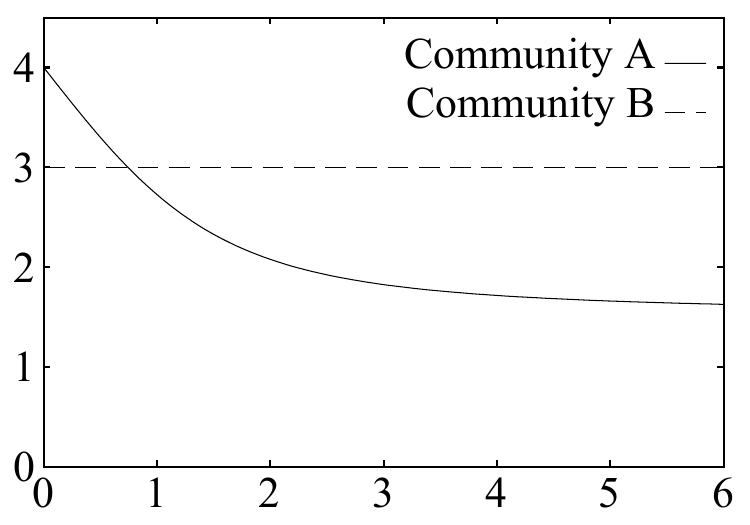}}
\put(30,30){\textcolor{white}{\rule{23.5\unitlength}{7.5\unitlength}}}
\cell{25}{15}{c}{Community A}
\cell{35}{30}{c}{Community B}
\cell{29}{-2}{c}{Viewpoint parameter, $q$}
\cell{-3}{21}{c}{\rotatebox{90}{Diversity, $D_q(\p)$}}
\end{picture}%
\hspace*{5mm}%
\lengths%
\begin{picture}(55,50)(-5,-4)
\cell{12}{24}{c}{\includegraphics[width=20\unitlength]{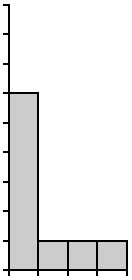}}
\cell{40}{24}{c}{\includegraphics[width=20\unitlength]{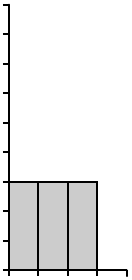}}
\cell{-3}{24}{c}{\rotatebox{90}{Probability}}
\cell{5.5}{2}{c}{$\scriptstyle 1$}
\cell{10}{2}{c}{$\scriptstyle 2$}
\cell{14.5}{2}{c}{$\scriptstyle 3$}
\cell{19}{2}{c}{$\scriptstyle 4$}
\cell{12}{-3}{b}{A}
\cell{1}{4}{c}{$\scriptstyle 0$}
\cell{1}{18}{c}{$\tfrac{1}{3}$}
\cell{1}{31}{c}{$\tfrac{2}{3}$}
\cell{1}{45}{c}{$\scriptstyle 1$}
\cell{33.5}{2}{c}{$\scriptstyle 1$}
\cell{38}{2}{c}{$\scriptstyle 2$}
\cell{42.5}{2}{c}{$\scriptstyle 3$}
\cell{47}{2}{c}{$\scriptstyle 4$}
\cell{40}{-3}{b}{B}
\cell{29}{4}{c}{$\scriptstyle 0$}
\cell{29}{18}{c}{$\tfrac{1}{3}$}
\cell{29}{31}{c}{$\tfrac{2}{3}$}
\cell{29}{45}{c}{$\scriptstyle 1$}
\end{picture}
\caption{The diversity profiles of the two hypothetical bird communities in
  the Introduction (p.~\pageref{p:intro-birds}).}  
\lbl{fig:bird-profs-bars}
\end{figure}
From the viewpoint%
\index{viewpoint!diversity@on diversity}
of low values of $q$, where rare species are given
nearly as much importance as common species, community~A is more diverse
than community~B.  For instance, at $q = 0$, community~A is more diverse
than community~B simply because it has more species.  But from the
viewpoint of high values, which give less importance to rare species,
community~B seems more diverse because it is better balanced.  In the
extreme, when $q = \infty$, we ignore all species except the most common,
and the dominance of the first species in community~A makes that community
much less diverse than the well-balanced community~B.

The flat profile of community~B indicates the uniformity of the species
present.  Generally, we have seen in the last two examples that the shape
of a diversity profile provides information on the community's structure.
For more on the interpretation of diversity profiles, see Example~1,
Example~2 and Figure~2 of Leinster and Cobbold~\cite{MDISS}.%
\index{Cobbold, Christina}
\end{example}

\begin{example}
Diversity profiles arising from experimental data often cross one another
(as in the last example), indicating that different viewpoints%
\index{viewpoint!diversity@on diversity}
on the importance of rare species lead to different judgements on which of
the communities is more diverse.  For example, Ellingsen tabulated
$D_0(\p)$, $D_1(\p)$ and $D_2(\p)$ for~16 distributions $\p$, corresponding
to the populations of certain marine organisms at~16 sites on the Norwegian
continental shelf (Table~1 of~\cite{Elli}).  There are $\binom{16}{2} =
120$ pairs of sites, and it can be deduced from the data that for at
least~$53$ of the~$120$ pairs, the profiles cross.

Typically, pairs of diversity profiles obtained from experimental data
cross at most once.  But it can be shown that in principle, there is no
upper bound on the number of times that a pair of diversity profiles can
cross. 
\end{example}

The ecological significance of the different judgements produced by
different diversity measures is discussed in the highly readable 1974 paper
of Peet~\cite{Peet}; see also Nagendra~\cite{Nage}.  More specifically,
diversity profiles of various types have long been discussed, beginning
with Hill%
\index{Hill Mark@Hill, Mark} 
himself in 1973~\cite{Hill}, and continuing with Patil%
\index{Patil, Ganapata P.} 
and Taillie~\cite{PaTaSDP,PaTaDCM}, Dennis and Patil~\cite{DePa},
T\'othm\'er\'esz~\cite{TothCDM}, Patil~\cite{PatiDP}, Mendes et
al.~\cite{META}, and others.
In political%
\index{politics}
science, $D_q(\p)$ has been used as a measure of the effective%
\index{effective number!political parties@of political parties}
number of parties in a parliamentary assembly, and diversity profiles have
been used to compare the political situations of different countries at 
different times (Laakso and Taagepera~\cite{LaTa}, especially
equation~[8] and Figure~1).

The next section establishes the mathematical
properties of the Hill numbers and, therefore, of diversity profiles.

\section{Properties of the Hill numbers}
\lbl{sec:prop-hill}

Here we establish the main properties of the Hill numbers, using what we
already know about properties of the power means.  Of course, any statement
about the Hill numbers can be translated into a statement about R\'enyi
entropies, since one is the logarithm of the other.  But here we work with
the Hill numbers, interpreting them in terms of diversity.

We have already noted that for each $q \in [-\infty, \infty]$, the Hill
number $D_q$ is an effective number: $D_q(\vc{u}_n) = n$.

Diversity profiles are always decreasing.  Intuitively, this is because
diversity decreases as less importance is attached to rare species.  The
precise statement is as follows.

\begin{propn}
\lbl{propn:div-dec}
\index{Hill number!monotonicity in order}
\index{diversity profile!decreasing@is decreasing}
Let $\p \in \Delta_n$.  Then $D_q(\p)$ is a decreasing function of $q \in
[-\infty, \infty]$.  It is constant if $\p$ is uniform on its support, and
strictly decreasing otherwise.
\end{propn}

\begin{proof}
Since $D_q(\p) = M_{1 - q}(\p, 1/\p)$, this follows from
Theorem~\ref{thm:mns-inc-ord}. 
\end{proof}

Figure~\ref{fig:bird-profs-bars} shows one strictly decreasing profile and one
that is constant (being uniform on its support).  Diversity profiles are
always continuous, by Lemma~\ref{lemma:pwr-mns-cts-t}.  

\begin{remark}
\lbl{rmk:prof-shapes}
\index{diversity profile!non-convex}
It is a curiosity that for most distributions $\p$ that arise
experimentally, the diversity profile of $\p$ appears to be convex.  (See
the works cited at the end of Section~\ref{sec:ren-hill}, for example.)
However, this is false for arbitrary $\p$.  
Figure~\ref{fig:prof-not-cvx} shows the diversity profile of the
distribution
\[
\p = 
\bigl(
\underbrace{10^{-6}, \ldots, 10^{-6}}_{999\,000}, 10^{-3}
\bigr)
\]
(adapted from an example of Willerton~\cite{WillITG}),%
\index{Willerton, Simon} 
which is evidently not convex.
\begin{figure}
\centering
\lengths
\begin{picture}(120,60)(0,-5)
\cell{60}{27}{c}{\includegraphics[width=80\unitlength]{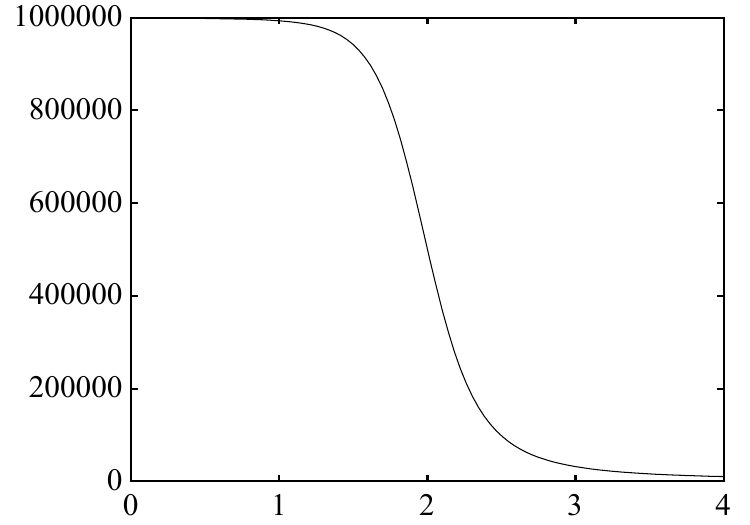}}
\cell{66}{-5}{b}{$q$}
\cell{15}{28}{c}{$D_q(\p)$}
\end{picture}
\caption{A non-convex diversity profile (Remark~\ref{rmk:prof-shapes}).}
\lbl{fig:prof-not-cvx}
\end{figure}
\end{remark}

For each parameter value $q > 0$, the maximum and minimum values of the
Hill number $D_q$, and the distributions at which they are attained, are
exactly the same as for the diversity $D_1 = D$ of order~$1$
(Lemma~\ref{lemma:div1-max-min}):

\begin{lemma}
\lbl{lemma:div-max-min}
\index{Hill number!bounds on}
Let $n \geq 1$ and $q \in [-\infty, \infty]$.
\begin{enumerate}
\item 
\lbl{part:div-min}
$D_q(\p) \geq 1$ for all $\p \in \Delta_n$, with equality if and only if
$p_i = 1$ for some $i \in \{1, \ldots, n\}$.

\item
\lbl{part:div-max}
If $q > 0$ then $D_q(\p) \leq n$ for all $\p \in \Delta_n$, with
equality if and only if $\p = \vc{u}_n$.
\end{enumerate}
\end{lemma}

\begin{proof}
For~\bref{part:div-min}, Proposition~\ref{propn:div-dec} implies that
\[
D_q(\p)
\geq
D_{\infty}(\p)
=
1\Big/\!\!\max_{i \in \supp(\p)} p_i 
\geq 
1.
\]
If the second inequality is an equality then $p_i = 1$ for some $i$.
Conversely, if $p_i = 1$ for some $i$ then $D_q(\p) = 1$.

For~\bref{part:div-max}, Proposition~\ref{propn:div-dec} implies that
\[
D_q(\p)
\leq 
D_0(\p)
=
|\supp(\p)|
\leq 
n,
\]
with equality in the first inequality if and only if $\p$ is uniform on its
support.  On the other hand, equality holds in the second inequality if and
only if $\p$ has full support.  Hence equality holds throughout if and only
if $\p = \vc{u}_n$.
\end{proof}

\begin{remarks}
\lbl{rmks:hill-dec}
\begin{enumerate}
\item 
\lbl{rmk:hill-dec-min}
It follows that for $q > 0$, the R\'enyi entropy $H_q(\p)$ is minimized
exactly when $\p$ is of the form $(0, \ldots, 0, 1, 0, \ldots, 0)$, with
value $0$, and maximized exactly when $\p = \vc{u}_n$, with value $\log n$.
Since the $q$-logarithmic entropy $S_q(\p)$ is an increasing invertible
transformation of $H_q(\p)$ (equations~\eqref{eq:q-log-ren-slick}
and~\eqref{eq:ren-q-log-slick}), it is minimized and maximized at these same
distributions, with minimum $0$ and maximum $ S_q(\vc{u}_n) = \ln_q(n)$.

\item
\lbl{rmk:hill-dec-neg}
\index{Hill number!negative order@of negative order}
\index{order!negative}
The Hill numbers of negative orders are \emph{not} maximized by the uniform
distribution.  Indeed, let $q < 0$, let $n \geq 2$, and take any non-uniform
distribution $\p \in \Delta_n$ of full support.
Then $D_0(\p) = |\supp(\p)| = n$, and the diversity profile of $\p$ is
strictly decreasing by Proposition~\ref{propn:div-dec}, so
\[
D_q(\p) > D_0(\p) = n = D_q(\vc{u}_n).
\]
Whatever the word `diverse' should mean, it is generally agreed that the
most diverse abundance distribution on a given set of species should be the
uniform distribution.  (At least, this should be the case for the crude
model of a community as simply a probability distribution, which we are using
here.  See also Section~\ref{sec:max}.)  For this reason, the Hill numbers
of negative orders are generally not used as measures of diversity.

On the other hand, the Hill numbers of negative orders measure
\emph{something}.  For instance, 
\[
D_{-\infty}(\p) = 1\Big/\!\min_{i \in \supp(\p)} p_i
= \max_{i \in \supp(\p)} (1/p_i)
\]
measures the rarity of the rarest species, giving a high value to any
community containing at least one species that is very rare.  This is a
meaningful quantity, even if it should not be called diversity.
\end{enumerate}
\end{remarks}

We now show that the Hill number $D_q(\p)$ of a given order $q$ is very
nearly continuous in $\p \in \Delta_n$, with the sole exception that $D_q$
is discontinuous at the boundary of the simplex when $q \leq 0$.  For
instance, species richness%
\index{species!richness} 
$D_0$ is discontinuous: in terms of the number of species present, a
relative abundance of $0.0001$ is qualitatively different from a relative
abundance of $0$.

\begin{defn}
\lbl{defn:hill-cts}
Let $\bigl(D\from \Delta_n \to (0, \infty)\bigr)_{n \geq 1}$ be a sequence
of functions.  Then $D$ is 
\demph{continuous}%
\index{continuous!diversity measure} 
if the function $D \from \Delta_n \to (0, \infty)$ is continuous for each
$n \geq 1$, and \demph{continuous%
\index{continuous!positive probabilities@in positive probabilities} 
in positive probabilities} if the restriction $D|_{\Delta_n^\circ}$ of $D$
to the open simplex is continuous for each $n \geq 1$.
\end{defn}

Continuity in positive probabilities means that small changes to the
abundances of the species \emph{present} cause only small changes in the
perceived diversity.  For example, $D_0$ is continuous in positive
probabilities, even though it is not continuous.

\begin{lemma}
\lbl{lemma:hill-cts}
\index{Hill number!continuity of}
\begin{enumerate}
\item 
\lbl{part:hill-cts-int}
For each $q \in [-\infty, \infty]$, the Hill number $D_q$ is continuous in
positive probabilities.

\item
\lbl{part:hill-cts-pos} 
For each $q \in (0, \infty]$, the Hill number $D_q$ is continuous.
\end{enumerate}
\end{lemma}

\begin{proof}
Part~\bref{part:hill-cts-int} is immediate from the explicit formulas for
$D_q$ (equations \eqref{eq:hill-gen}--\eqref{eq:hill-pinfty}), and
part~\bref{part:hill-cts-pos} follows from the observation that when $q >
0$, the formulas for $D_q$ are unchanged if we allow $i$ to range over all
of $\{1, \ldots, n\}$ instead of just $\supp(\p)$.
\end{proof}

Next we establish the algebraic properties of the Hill numbers, beginning
with the most elementary ones.  

\begin{defn}
\lbl{defn:hill-abs}
A sequence of functions $\bigl(D\from \Delta_n \to (0, \infty)\bigr)_{n
  \geq 1}$ is \demph{absence-invariant}%
\index{absence-invariance!diversity measure@of diversity measure} 
if whenever $\p \in \Delta_n$ and $1 \leq i \leq n$ with $p_i = 0$, then
\[
D(\p) = D(p_1, \ldots, p_{i - 1}, p_{i + 1}, \ldots, p_n).
\]
\end{defn}

Absence-invariance means that as far as $D$ is concerned, a species that
is absent might as well not have been mentioned.  

Recall from equation~\eqref{eq:q-ent-sym} that $D$ is said to be
symmetric if $D(\p\sigma) = D(\p)$ for all $\p \in \Delta_n$ and
permutations $\sigma$ of $\{1, \ldots, n\}$.  This means that the
diversity is unaffected by the order in which the species happen to be
listed.

\begin{lemma}
\lbl{lemma:hill-elem}
\index{Hill number!symmetry of}
\index{Hill number!absence-invariance of}
For each $q \in [-\infty, \infty]$, the Hill number $D_q$ of order $q$ is
symmetric and absence-invariant.
\end{lemma}

\begin{proof}
These statements follow from the symmetry and absence-invariance of the
power means (Lemma~\ref{lemma:pwr-mns-elem}).  Alternatively, they can be
deduced directly from the explicit formulas for $D_q$
(equations~\eqref{eq:hill-gen}--\eqref{eq:hill-pinfty}).
\end{proof}

\begin{remark}
\lbl{rmk:prof-perm}
\index{diversity profile!determines distribution}
By symmetry, $\p$ and $\p\sigma$ have the same diversity profile.  In fact,
the converse also holds: if $\p, \vc{r} \in \Delta_n$ have the same
diversity profile then $\p$ and $\vc{r}$ must be the same up to a
permutation.  This is proved in Appendix~\ref{sec:prof}.

Thus, the diversity profile of a relative abundance distribution contains
all the information about that distribution apart from which species is
which, packaged in a way that displays meaningful information about the
community's diversity.
\end{remark}

We finally come to the chain rule.  In Corollary~\ref{cor:div1-chain}, we
treated the case $q = 1$, showing that
\[
D_1\bigl(\vc{w} \of (\p^1, \ldots, \p^n)\bigr)
=
D_1(\vc{w}) \cdot \prod_{i = 1}^n D_1(\p^i)^{w_i}
\]
for all $\vc{w} \in \Delta_n$ and $\p^i \in \Delta_{k_i}$.  In
Example~\ref{eg:div1-chain-islands}, this formula was explained in terms of
a group of $n$ islands of relative sizes $w_i$ and diversities $d_i =
D_1(\p^i)$, with no shared species.  We now give the chain rule for general
$q$, in two different forms.

\begin{propn}[Chain rule, version~1]
\lbl{propn:hill-chn}%
\index{chain rule!Hill numbers@for Hill numbers}%
\index{Hill number!chain rule for}
Let $q \in [-\infty, \infty]$, $\vc{w} \in \Delta_n$, and $\p^1 \in
\Delta_{k_1}, \ldots, \p^n \in \Delta_{k_n}$.  Write $d_i = D_q(\p^i)$ and
$\vc{d} = (d_1, \ldots, d_n)$.  Then
\begin{align*}
D_q\bigl(\vc{w} \of (\p^1, \ldots, \p^n)\bigr)  &
=
M_{1 - q}(\vc{w}, \vc{d}/\vc{w})        \\
&
=
\begin{cases}
\bigl(\sum w_i^q d_i^{1 - q}\bigr)^{1/(1 - q)}    &
\text{if } q \neq 1, \pm\infty, \\
\max d_i/w_i    &
\text{if } q = -\infty, \\
\prod (d_i/w_i)^{w_i}   &
\text{if } q = 1,       \\
\min d_i/w_i    &
\text{if } q = \infty, 
\end{cases}
\end{align*}
where the sum, maximum, product, and minimum are over $i \in
\supp(\vc{w})$. 
\end{propn}

Here $\vc{d}/\vc{w} = (d_1/w_1, \ldots, d_n/w_n)$, as in
Remark~\ref{rmk:vec-op-conv}. 

\begin{proof}
We have
\begin{align*}
D_q\bigl(\vc{w} \of (\p^1, \ldots, \p^n)\bigr)  &
=
M_{1 - q}\Bigl(\vc{w} \of (\p^1, \ldots, \p^n), 
\tfrac{1}{w_1 \p^1} \oplus\cdots\oplus \tfrac{1}{w_n \p^n} \Bigr)       \\
&
=
M_{1 - q}\Bigl(\vc{w},
\Bigl( 
M_{1 - q}\Bigl(\p^1, \tfrac{1}{w_1 \p^1}\Bigr), \,\ldots,\,
M_{1 - q}\Bigl(\p^n, \tfrac{1}{w_n \p^n}\Bigl)    
\Bigr)
\Bigr) \\
&
=
M_{1 - q}\bigl( \vc{w}, (d_1/w_1, \ldots, d_n/w_n) \bigr),
\end{align*}
where the second equation follows from the chain rule for $M_{1 - q}$
(Proposition~\ref{propn:pwr-mns-chn}) and the last from the homogeneity of $M_{1
- q}$ (Lemma~\ref{lemma:pwr-mns-hgs}).  This proves the first equality
stated in the proposition, and the second follows from the explicit
formulas for the power means.
\end{proof}

There is an alternative form of the chain rule, for which we will need some
terminology.  Given a probability distribution $\vc{w} \in \Delta_n$ and a
real number $q$, the
\demph{escort%
\index{escort distribution}
distribution of order%
\index{order!escort distribution@of escort distribution} 
$q$} of $\vc{w}$ is the distribution \ntn{wq}$\vc{w}^{(q)} \in \Delta_n$
with $i$th coordinate
\[
w^{(q)}_i
=
\begin{cases}
w_i^q \bigg/\!\!\sum\limits_{j \in \supp(\vc{w})} w_j^q &
\text{if } i \in \supp(\vc{w}),     \\
0       &
\text{otherwise}.
\end{cases}
\]

\begin{lemma}
\lbl{lemma:mean-esc}
Let $q \in \R$, $\vc{w} \in \Delta_n$, and $\vc{d} \in [0, \infty)^n$.
Then
\[
M_{1 - q}(\vc{w}, \vc{d}/\vc{w})
=
D_q(\vc{w}) \cdot M_{1 - q}(\vc{w}^{(q)}, \vc{d}).
\]
\end{lemma}

\begin{proof}
For the case $q = 1$, note that
\[
M_0(\vc{w}, \vc{x}\vc{y}) =
M_0(\vc{w}, \vc{x}) M_0(\vc{w}, \vc{y})
\]
for all $\vc{x}, \vc{y} \in [0, \infty)^n$.  It follows that
\[
D_1(\vc{w}) \cdot M_0(\vc{w}^{(1)}, \vc{d})
=
M_0(\vc{w}, 1/\vc{w}) \cdot M_0(\vc{w}, \vc{d})
=
M_0(\vc{w}, \vc{d}/\vc{w}).
\]
On the other hand, for $1 \neq q \in \R$,
\begin{align*}  
M_{1 - q}(\vc{w}, \vc{d}/\vc{w})        &
=
\Biggl( \sum_{i \in \supp(\vc{w})} w_i^q d_i^{1 - q} \Biggr)^{1/(1 - q)}\\
&
=
D_q(\vc{w}) \cdot
\Biggl( 
\frac{\sum_{i \in \supp(\vc{w})} w_i^q d_i^{1 - q}}%
{\sum_{j \in \supp(\vc{w})} w_j^q}
\Biggr)^{1/(1 - q)}       \\
&
=
D_q(\vc{w}) \cdot M_{1 - q}(\vc{w}^{(q)}, \vc{d}),
\end{align*}
as required.
\end{proof}

The last two results immediately imply:

\begin{propn}[Chain rule, version~2]
\lbl{propn:hill-chn-esc}%
\index{chain rule!Hill numbers@for Hill numbers}%
\index{Hill number!chain rule for}
Let $q \in \R$, $\vc{w} \in \Delta_n$, and $\p^1 \in
\Delta_{k_1}, \ldots, \p^n \in \Delta_{k_n}$.  Write $d_i = D_q(\p^i)$ and 
$\vc{d} = (d_1, \ldots, d_n)$.  Then
\[
D_q\bigl(\vc{w} \of (\p^1, \ldots, \p^n)\bigr)
=
D_q(\vc{w}) \cdot M_{1 - q}(\vc{w}^{(q)}, \vc{d}).
\]
\qed
\end{propn}

\begin{remarks}
\lbl{rmks:hill-chn}
Here we provide context for the notion of escort distribution.
\begin{enumerate}
\item
The escort distributions of a distribution $\vc{w}$ form a 
one-parameter family
\[
\bigl( \vc{w}^{(q)} \bigr)_{q \in \R}
\]
of distributions, of which the original distribution $\vc{w}$ is the member
corresponding to $q = 1$.  The term `escort distribution' is taken from
thermodynamics\index{thermodynamics} (Chapter~9 of Beck and
Schl\"ogl~\cite{BeSc}).  There, one encounters expressions such as
\[
\frac{(e^{-\beta E_1}, \ldots, e^{-\beta E_n})}{Z(\beta)},
\]
where $Z(\beta) = e^{-\beta E_1} + \cdots + e^{-\beta E_n}$ is the
partition%
\index{partition function} 
function for energies $E_i$ at inverse temperature $\beta$.
Assuming without loss of generality that $\sum e^{-E_i} = 1$, the inverse
temperature $\beta$ plays the role of the parameter $q$.

\item
\lbl{rmk:hill-chn-vs}
The function $(q, \vc{w}) \mapsto \vc{w}^{(q)}$ is the scalar
multiplication of a real vector space structure on the interior
$\Delta_n^\circ$ of the simplex.%
\index{simplex!vector space structure on}
Addition is given by 
\[
(\vc{p}, \vc{r}) 
\mapsto
\frac{(p_1 r_1, \ldots, p_n r_n)}{p_1 r_1 + \cdots + p_n r_n},
\]
and the zero element is the uniform distribution $\vc{u}_n$.
Figure~\ref{fig:simp-vs} shows some one-dimensional linear subspaces of the
two-dimensional vector space $\Delta_3^\circ$.
\begin{figure}
\centering
\lengths
\includegraphics[width=70\unitlength]{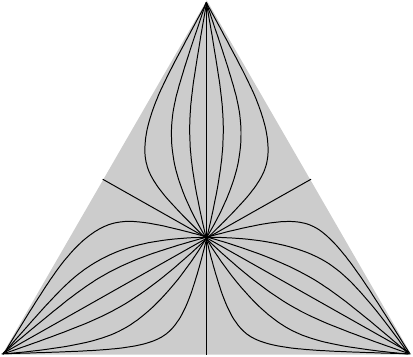}
\caption{Twelve one-dimensional linear subspaces of the open simplex
  $\Delta_3^\circ$ with the real vector space structure described in
  Remark~\ref{rmks:hill-chn}\bref{rmk:hill-chn-vs}.}
\lbl{fig:simp-vs}
\end{figure}
This vector space structure was used in the field of statistical inference
by Aitchison~\cite{Aitc},%
\index{Aitchison, John} 
and is sometimes named after him.  It can be understood algebraically as
follows.

Exponential and logarithm define a bijection between $\R$ and $(0,
\infty)$.  This induces a bijection between $\R^n$ and $(0, \infty)^n$, and
transporting the vector space structure on $\R^n$ across this bijection
gives a vector space structure on $(0, \infty)^n$.  Explicitly, addition in
the vector space $(0, \infty)^n$ is coordinatewise multiplication, the zero
element is $(1, \ldots, 1)$, and scalar multiplication by $q \in \R$ raises
each coordinate to the power of $q$.

Now take the linear subspace of $\R^n$ spanned by $(1, \ldots, 1)$.  The
corresponding subspace $W$ of $(0, \infty)^n$ is $\{ (\gamma, \ldots,
\gamma) \such \gamma \in (0, \infty) \}$, and we can form the quotient
vector space $(0, \infty)^n/W$.

An element of this quotient is an equivalence class of vectors $\vc{y} \in
(0, \infty)^n$, with $\vc{y}$ equivalent to $\vc{z}$ if and only if $\vc{y}
= \gamma\vc{z}$ for some $\gamma > 0$.  Geometrically, then, the
equivalence classes are the rays through the origin in the positive orthant
$(0, \infty)^n$.  Each ray contains exactly one element of the open simplex
\[
\Delta_n^\circ
=
\{ \vc{y} \in (0, \infty)^n \such y_1 + \cdots + y_n = 1 \}.
\]
This puts $(0, \infty)^n/W$ in bijection with $\Delta_n^\circ$, thus giving
$\Delta_n^\circ$ the structure of a vector space.  It is exactly the
vector space structure defined explicitly above.

\item
In statistical language, each linear subspace of the vector space
$\Delta_n^\circ$ is an exponential%
\index{exponential family} 
family of distributions on $\{1, \ldots, n\}$.  For example, the
one-dimensional subspace spanned by $\p \in \Delta_n^\circ$ is a
one-parameter exponential family with natural parameter $q \in \R$,
sufficient statistic $\log p_i$, and log-partition%
\index{partition function}
function $\log\bigl(\sum p_i^q\bigr)$.  More on this connection
can be found in Amari~\cite{AmarDGC}, Ay, Jost, L\^{e} and
Schwachh\"ofer (\cite{AJLS}, Section~2.8), and other information%
\index{information geometry}
geometry texts.
\end{enumerate}
\end{remarks}

As already discussed in the case $q = 1$
(Example~\ref{eg:div1-chain-islands}), the chain rule for the Hill numbers
has the important consequence that when computing the total diversity of a
group of islands%
\index{islands!diversity of group of} 
with no shared species, the only information one needs is the diversities
and relative sizes of the islands, not their internal make-up:

\begin{defn}
\lbl{defn:hill-mod}
A sequence of functions $\bigl(D \from \Delta_n \to (0, \infty) \bigr)_{n
  \geq 1}$ is \demph{modular}%
\index{modularity!diversity measure@of diversity measure}
if 
\begin{align*}
&
D\bigl(\p^i\bigr) = D\bigl(\twid{\p}^i\bigr) 
\text{ for all } i \in \{1, \ldots, n\}     \\
\implies
&
D\bigl( \vc{w} \of (\p^1, \ldots, \p^n) \bigr) =
D\bigl( \vc{w} \of (\twid{\p}^1, \ldots, \twid{\p}^n) \bigr) 
\end{align*}
for all $n, k_1, \ldots, k_n, \twid{k}_1, \ldots, \twid{k}_n \geq 1$ and
$\vc{w} \in \Delta_n$, $\p^i \in \Delta_{k_i}$, $\twid{\p}^i \in
\Delta_{\twid{k}_i}$.  
\end{defn}

In other words, $D$ is modular if $D\bigl( \vc{w} \of (\p^1, \ldots, \p^n)
\bigr)$ depends only on $\vc{w}$ and $D(\p^1), \ldots, D(\p^n)$. 

\begin{cor}[Modularity]
\lbl{cor:mod-hill}
\index{Hill number!modularity of}%
\index{modularity!Hill numbers@of Hill numbers}
For each $q \in [-\infty, \infty]$, the Hill number $D_q$ is modular.
\qed
\end{cor}

The chain rule has two further consequences.

\begin{defn}
\lbl{defn:hill-mult}
A sequence of functions $\bigl( D \from \Delta_n \to (0, \infty) \bigr)_{n
  \geq 1}$ is \demph{multiplicative}%
\index{multiplicative!diversity measure} 
if 
\[
D(\p \otimes \vc{r}) = D(\p) D(\vc{r})
\]
for all $m, n \geq 1$, $\p \in \Delta_m$, and $\vc{r} \in \Delta_n$. 
\end{defn}

\begin{cor}[Multiplicativity]
\index{Hill number!multiplicativity of}%
\index{multiplicative!Hill numbers are}
For each $q \in [-\infty, \infty]$, the Hill number $D_q$ is
multiplicative. 
\end{cor}

\begin{proof}
This follows from either 
the chain rule for the Hill numbers
or the logarithmic property of the
R\'enyi entropies (equation~\eqref{eq:ren-log}).
\end{proof}

\begin{defn}
\lbl{defn:hill-rep}
A sequence of functions $\bigl( D \from \Delta_n \to (0, \infty) \bigr)_{n
  \geq 1}$ satisfies the 
\demph{replication%
\index{replication principle} 
principle} if 
\[
D(\vc{u}_n \otimes \p) = nD(\p)
\]
for all $n, k \geq 1$ and $\p \in \Delta_k$.
\end{defn}

The oil company argument of Example~\ref{eg:oil} shows the fundamental
importance of the replication principle.  If $n$ islands have the same
relative abundance distribution $\p$, but on disjoint sets of species, the
diversity of the whole system should be $nD(\p)$.

\begin{cor}[Replication]
\index{Hill number!replication principle for}
\index{replication principle!Hill numbers@for Hill numbers}
For each $q \in [-\infty, \infty]$, the Hill number $D_q$ satisfies the
replication principle.
\end{cor}

\begin{proof}
This follows from multiplicativity and the fact that $D_q$ is an effective
number. 
\end{proof}

Accompanying the R\'enyi entropies, there is also a notion of R\'enyi
relative entropy (introduced in Section~3 of R\'enyi~\cite{Reny}).  We
defer discussion to Section~\ref{sec:value-rel}.

\section{Characterization of the Hill number of a given order}
\lbl{sec:hill-char-given}
\index{Hill number!characterization of}

In this book, we prove two characterization theorems for the Hill numbers.
The first states that for each given $q$, the unique function satisfying
certain conditions (which depend on $q$) is $D_q$.  The second states that
the only functions satisfying a different list of conditions (which make no
mention of $q$) are those belonging to the family $(D_q)_{q \in [-\infty,
    \infty]}$.  We prove the first characterization in this section, and
the second in Section~\ref{sec:total-hill}.

For the $q$-logarithmic entropies, we have already proved an analogue of
the first result (Theorem~\ref{thm:q-ent-char}).  We will not prove an
analogue of the second.  However, there is a theorem of this type due to
Forte and Ng, briefly discussed in Remark~\ref{rmk:total-hill-translate}.

Here we build on work of Routledge~\cite{Rout} to characterize the Hill
number $D_q$ for each given $q \in (0, \infty)$.  The restriction to positive
$q$ ensures that $D_q$ is continuous on all of $\Delta_n$ (by
Lemma~\ref{lemma:hill-cts}\bref{part:hill-cts-pos}).  

Recall that $D_q$ satisfies the chain rule
\begin{equation}
\lbl{eq:hill-chn-esc-char}
D_q \bigl( \vc{w} \of (\p^1, \ldots, \p^n) \bigr)
=
D_q(\vc{w}) \cdot
M_{1 - q} \Bigl(
\vc{w}^{(q)}, \bigl(D_q(\p^1), \ldots, D_q(\p^n)\bigr)
\Bigr),
\end{equation}
where $\vc{w} \in \Delta_n$ and $\p^i \in \Delta_{k_i}$
(Proposition~\ref{propn:hill-chn-esc}).
Let us reflect on equation~\eqref{eq:hill-chn-esc-char}, interpreting it in
terms of the island%
\index{islands!diversity of group of} 
scenario of Examples~\ref{eg:comp-islands} and~\ref{eg:div1-chain-islands}.
Equation~\eqref{eq:hill-chn-esc-char} can be interpreted as a decomposition
of the diversity of the island group into two factors: the variation
\emph{between} the islands (given by $D_q(\vc{w})$), and the average
variation or diversity \emph{within} the islands (given by the second
factor).  Recall that in the island scenario, there is no overlap of
species between islands, so the variation between the islands depends only
on the variation in sizes.

Now, suppose that we want to list some properties that a reasonable
diversity measure $D$ ought to satisfy.  One such property might be that
$D$ is decomposable in the sense just described: $D(\vc{w} \of (\p^1,
\ldots, \p^n))$ is equal to the variation $D(\vc{w})$ between islands
multiplied by the average of the diversities $D(\p^1), \ldots, D(\p^n)$
within each island.

But what could `average' reasonably mean?  We have already seen that the
power means have many good properties that we would expect of a notion of
avarage, and we will see in Chapter~\ref{ch:mns} that in a certain precise
sense, they are \emph{uniquely} good.  So, it is reasonable to take the
`average' to be some power mean $M_t$, and we can make the usual harmless
reparametrization $t = 1 - q$.

This reasoning suggests that our hypothetical diversity measure $D$ should
satisfy something like equation~\eqref{eq:hill-chn-esc-char}, with $D$ in
place of $D_q$.  Still, it does not explain why the average of the
within-island diversities should be calculated using the weighting
$\vc{w}^{(q)}$ on the islands, 
rather than some other weighting.  All that seems
clear is that the weighting should depend on the sizes of the islands only.
If we write the weighting as $\theta(\vc{w})$, then our conclusion is that
any reasonable diversity measure $D$ ought to satisfy the equation
\[
D \bigl( \vc{w} \of (\p^1, \ldots, \p^n) \bigr)
=
D(\vc{w}) \cdot
M_{1 - q} \Bigl(
\theta(\vc{w}), \bigl(D(\p^1), \ldots, D(\p^n)\bigr)
\Bigr)
\]
for some $q$ and some function $\theta \from \Delta_n \to \Delta_n$.  This
explains the most substantial of the hypotheses in our main result:

\begin{thm}
\lbl{thm:rout}
\index{Hill number!characterization of}
Let $q \in (0, \infty)$.  Let $\bigl( D \from \Delta_n \to (0, \infty)
\bigr)_{n \geq 1}$ be a sequence of functions.  The following are
equivalent: 
\begin{enumerate}
\item
\lbl{part:rout-condns}
the functions $D$ are continuous, symmetric and effective numbers, and for
each $n \geq 1$ there exists a function $\theta \from \Delta_n \to
\Delta_n$ with the following property: 
\[
D \bigl( \vc{w} \of (\p^1, \ldots, \p^n) \bigr)
=
D(\vc{w}) \cdot
M_{1 - q} \Bigl(
\theta(\vc{w}), \bigl(D(\p^1), \ldots, D(\p^n)\bigr)
\Bigr)
\]
for all $\vc{w} \in \Delta_n$, $k_1, \ldots, k_n \geq 1$, and $\p^i \in
\Delta_{k_i}$;

\item
\lbl{part:rout-form}
$D = D_q$.
\end{enumerate}
\end{thm}

Theorem~\ref{thm:rout} is a variation on a 1979 result of
Routledge%
\index{Routledge, Richard}
(Theorem~1 of the appendix to \cite{Rout}).

The rest of this section is devoted to the proof.  We already showed in
Section~\ref{sec:prop-hill} that~\bref{part:rout-form}
implies~\bref{part:rout-condns}.  Conversely, and \femph{for the rest of
  this section}, take $D$ and $\theta$ satisfying the conditions
of~\bref{part:rout-condns}.  By the standard abuse of notation, we use the
same letter $\theta$ for each of the functions $\theta \from \Delta_1 \to
\Delta_1$, $\theta \from \Delta_2 \to \Delta_2$, etc.  We have to prove
that $D = D_q$.

For $\p \in \Delta_n$, write
\[
\theta(\p) = \bigl( \theta_1(\p), \ldots, \theta_n(\p) \bigr).
\]
Our first lemma shows how $\theta_1$ can be expressed in terms of $D$.  We
temporarily adopt the notation
\[
\splitdist{\p}  
=
\p \of (\vc{u}_2, \vc{u}_1, \ldots, \vc{u}_1)   
=
\bigl( \hlf p_1, \hlf p_1, p_2, \ldots, p_n \bigr)
\ntn{hash}
\]
($\p \in \Delta_n$).

\begin{lemma}
\lbl{lemma:rout-th1}
For all $n \geq 1$ and $\p \in \Delta_n$,
\[
\theta_1(\p)
=
\frac{1}{\ln_q 2} \cdot \ln_q \frac{D(\splitdist{\p})}{D(\p)}.
\]
\end{lemma}

\begin{proof}
By the main hypothesis on $D$ and the effective number property,
\[
D(\splitdist{\p}) 
=
D(\p) \cdot M_{1 - q}\bigl( \theta(\p), (2, 1, \ldots, 1) \bigr).
\]
Hence by Lemma~\ref{lemma:q-log-mean},
\[
\ln_q \frac{D(\splitdist{\p})}{D(\p)} 
=
M_1 \bigl( \theta(\p), (\ln_q 2, \ln_q 1, \ldots, \ln_q 1) \bigr).
\]
But $\ln_q 1 = 0$, so the right-hand side is just $\theta_1(\p) \ln_q 2$.
\end{proof}

We use this lemma to compute the weighting $\theta(\vc{w}\of(\p^1, \ldots,
\p^n))$ of a composite distribution:

\begin{lemma}
\lbl{lemma:rout-theta-comp}
Let $\vc{w} \in \Delta_n$ and $\p^1 \in \Delta_{k_1}, \ldots, \p^n \in
\Delta_{k_n}$.  Then
\[
\theta_1\bigl( \vc{w} \of (\p^1, \ldots, \p^n) \bigr)
=
\frac{\theta_1(\vc{w}) D(\p^1)^{1 - q}}%
{\sum_{i = 1}^n \theta_i(\vc{w}) D(\p^i)^{1 - q}}
\,
\theta_1(\p^1).
\]
\end{lemma}

\begin{proof}
Write $d_i = D(\p^i)$ and $\spltdist{d_1} = D\Bigl(\splitdist{\p^1}\Bigr)$.
We have
\begin{align}
&
\theta_1\bigl(\vc{w} \of (\p^1, \ldots, \p^n)\bigr)
\nonumber
\\
&
=
\frac{1}{\ln_q 2} \cdot 
\ln_q 
\frac{D \Bigl(\Splitdist{\vc{w} \of (\p^1, \ldots, \p^n)}\Bigr)}%
{D\bigl( \vc{w} \of (\p^1, \ldots, \p^n)\bigr)}  
\lbl{eq:rtc-1}  \\
&
=
\frac{1}{\ln_q 2} \cdot 
\ln_q 
\frac{D \Bigl(\vc{w} \of 
\Bigl( \splitdist{\p^1}, \p^2, \ldots, \p^n\Bigr)
\Bigr)}%
{D\bigl( \vc{w} \of (\p^1, \p^2, \ldots, \p^n)\bigr)}  
\lbl{eq:rtc-2}  \\
&
=
\frac{1}{\ln_q 2} \cdot 
\ln_q 
\frac{M_{1 - q} \bigl(\theta(\vc{w}),
(\spltdist{d_1}, d_2, \ldots, d_n)
\bigr)}%
{M_{1 - q} \bigl( \theta(\vc{w}), (d_1, d_2, \ldots, d_n)\bigr)}  
\lbl{eq:rtc-3}  \\
&
=
\frac{1}{\ln_q 2} \cdot 
\frac{%
\ln_q M_{1 - q}\bigl(\theta(\vc{w}), (\spltdist{d_1}, d_2, \ldots, d_n)\bigr)
-
\ln_q M_{1 - q}\bigl(\theta(\vc{w}), (d_1, d_2, \ldots, d_n)\bigr)
}%
{M_{1 - q} \bigl( \theta(\vc{w}), (d_1, d_2, \ldots, d_n)\bigr)^{1 - q}}  
\lbl{eq:rtc-4}  \\
&
=
\frac{1}{\ln_q 2} \cdot 
\frac{%
M_1\bigl(
\theta(\vc{w}), (\ln_q \spltdist{d_1}, \ln_q d_2, \ldots)
\bigr)
-
M_1\bigl(
\theta(\vc{w}), (\ln_q d_1, \ln_q d_2, \ldots)
\bigr)
}%
{\sum_{i = 1}^n \theta_i(\vc{w}) d_i^{1 - q}}
\lbl{eq:rtc-5}  \\
&
=
\frac{1}{\ln_q 2} \cdot
\frac{
\theta_1(\vc{w}) (\ln_q \spltdist{d_1} - \ln_q d_1)
}%
{\sum_{i = 1}^n \theta_i(\vc{w}) d_i^{1 - q}}
\lbl{eq:rtc-6}  \\
&
=
\frac{1}{\ln_q 2} \cdot
\frac{
\theta_1(\vc{w}) d_1^{1 - q}
}%
{\sum_{i = 1}^n \theta_i(\vc{w}) d_i^{1 - q}}
\cdot
\ln_q \frac{\spltdist{d_1}}{d_1} 
\lbl{eq:rtc-7}  \\
&
=
\frac{\theta_1(\vc{w}) d_1^{1 - q}}%
{\sum_{i = 1}^n \theta_i(\vc{w}) d_i^{1 - q}}
\cdot
\theta_1(\p^1)
\lbl{eq:rtc-8},
\end{align}
where equations~\eqref{eq:rtc-1} and~\eqref{eq:rtc-8} follow from
Lemma~\ref{lemma:rout-th1}, equation~\eqref{eq:rtc-2} from the definition of
$\splitdist{\ }$, equation~\eqref{eq:rtc-3} from the main hypothesis on $D$,
equations~\eqref{eq:rtc-4} and~\eqref{eq:rtc-7} from the quotient formula
\[
\ln_q \frac{x}{y} 
=
\frac{\ln_q x - \ln_q y}{y^{1 - q}}
\]
for the $q$-logarithm (equation~\eqref{eq:q-log-qt}),
equation~\eqref{eq:rtc-5} from Lemma~\ref{lemma:q-log-mean} and the
definition of $M_{1 - q}$, and equation~\eqref{eq:rtc-6} from the
definition of the arithmetic mean $M_1$.
\end{proof}

We now deduce that the weightings must be the $q$-escort distributions:

\begin{lemma}
\lbl{lemma:rout-wts}
$\theta(\vc{w}) = \vc{w}^{(q)}$ for all $n \geq 1$ and $\vc{w} \in
  \Delta_n$. 
\end{lemma}

\begin{proof}
Following a familiar pattern, we prove this first when $\vc{w}$ is uniform,
then when the coordinates of $\vc{w}$ are positive and rational, and
finally for arbitrary $\vc{w}$.

For the case $\vc{w} = \vc{u}_n$, we have to prove that $\theta(\vc{u}_n) =
\vc{u}_n$.  By Lemma~\ref{lemma:rout-th1},
\[
\theta_1(\vc{u}_n)
=
\frac{1}{\ln_q 2}
\ln_q
\frac{D\bigl( 
\vc{u}_n \of (\vc{u}_2, \vc{u}_1, \vc{u}_1, \ldots, \vc{u}_1)
\bigr)}{D(\vc{u}_n)},
\]
and by the same argument,
\[
\theta_2(\vc{u}_n)
=
\frac{1}{\ln_q 2}
\ln_q
\frac{D\bigl( 
\vc{u}_n \of (\vc{u}_1, \vc{u}_2, \vc{u}_1, \ldots, \vc{u}_1)
\bigr)}{D(\vc{u}_n)}.
\]
By symmetry of $D$, the right-hand sides of these two equations are equal.
Hence $\theta_1(\vc{u}_n) = \theta_2(\vc{u}_n)$.  Similarly,
$\theta_i(\vc{u}_n) = \theta_j(\vc{u}_n)$ for all $i, j$, and so
$\theta(\vc{u}_n) = \vc{u}_n$. 

Now let $\vc{w} \in \Delta_n$ with 
\[
\vc{w} = (k_1/k, \ldots, k_n/k)
\]
for some positive integers $k_i$, where $k = \sum k_i$.  We have
\begin{equation}
\lbl{eq:rout-rat}
\vc{w} \of (\vc{u}_{k_1}, \ldots, \vc{u}_{k_n})
=
\vc{u}_k.
\end{equation}
Applying $\theta_1$ to both sides gives
\[
\frac{\theta_1(\vc{w}) k_1^{1 - q}}{\sum \theta_i(\vc{w}) k_i^{1 - q}}
\, \frac{1}{k_1}
=
\frac{1}{k},
\]
using Lemma~\ref{lemma:rout-theta-comp}, the effective number property of
$D$, and the previous paragraph.  This rearranges to
\[
\theta_1(\vc{w}) 
=
w_1^q \sum_{i = 1}^n \theta_i(\vc{w}) w_i^{1 - q}.
\]
By the same argument,
\[
\theta_j(\vc{w})
=
w_j^q \sum_{i = 1}^n \theta_i(\vc{w}) w_i^{1 - q}
\]
for all $j = 1, \ldots, n$.  The sum on the right-hand side is independent
of $j$, so $\theta(\vc{w})$ is a probability distribution proportional to
$\bigl(w_1^q, \ldots, w_n^q\bigr)$, forcing $\theta(\vc{w}) = \vc{w}^{(q)}$. 

Finally, we show that $\theta(\vc{w}) = \vc{w}^{(q)}$ for all $\vc{w} \in
\Delta_n$.  By Lemma~\ref{lemma:rout-th1} and the continuity hypothesis on
$D$, the map $\theta_1$ is continuous, and similarly for $\theta_2, \ldots,
\theta_n$.  Hence $\theta \from \Delta_n \to \Delta_n$ is continuous.  So
too is the map $\vc{w} \mapsto \vc{w}^{(q)}$.  But by the previous
paragraph, these last two maps are equal on the positive rational
distributions, so they are equal everywhere.
\end{proof}

\begin{pfof}{Theorem~\ref{thm:rout}}
First, consider distributions $\vc{w} = (k_1/k, \ldots, k_n/k)$ with
positive rational coordinates.  Apply $D$ to both sides of
equation~\eqref{eq:rout-rat}: then by Lemma~\ref{lemma:rout-wts} and the
effective number hypothesis on $D$,
\[
D(\vc{w}) \cdot M\bigl( \vc{w}^{(q)}, (k_1, \ldots, k_n ) \bigr)
=
k.
\]
But we can also apply $D_q$ to both sides of equation~\eqref{eq:rout-rat}:
then by the chain rule and the effective number property of $D_q$,
\[
D_q(\vc{w}) \cdot M\bigl( \vc{w}^{(q)}, (k_1, \ldots, k_n) \bigr)
=
k.
\]
Hence $D(\vc{w}) = D_q(\vc{w})$.  And by the continuity hypothesis on $D$
and the continuity property of $D_q$
(Lemma~\ref{lemma:hill-cts}\bref{part:hill-cts-pos}), it follows that
$D(\vc{w}) = D_q(\vc{w})$ for all $\vc{w} \in \Delta_n$.
\end{pfof}

%% file: mns.tex
\chapter{Means}
\lbl{ch:mns}
\index{mean}

The ideal of the axiomatic approach to diversity measurement is to be able
to say `any measure of diversity that satisfies properties X, Y and~Z must
be one of the following.'  Our later theorems of this type will
stand on the shoulders of characterization theorems for means.

The theory of means took shape in the first half of the twentieth century,
with the 1930 papers of Kolmogorov~\cite{KolmSNM,KolmONM}%
\index{Kolmogorov, Andrei} 
and Nagumo~\cite{Nagu}%
\index{Nagumo, Mitio} 
as well as Hardy,%
\index{Hardy, Godfrey Harold}
Littlewood%
\index{Littlewood, John Edensor}
and P\'olya's%
\index{Polya, George@P\'olya, George} 
seminal book \emph{Inequalities}~\cite{HLP}, first published in 1934.
(Acz\'el~\cite{AczeMV} describes the early history.)  But new results
continue to be proved.  The~2009 book by Grabisch, Marichal, Mesiar and Pap
lists some modern developments (\cite{GMMP}, Chapter~4), and most of the
characterization theorems in this chapter also appear to be new.

The arguments that we will use are entirely elementary and require no
specialist knowledge.  Nevertheless, the reader could omit almost all of
this chapter without affecting their ability to follow subsequent chapters.
The only parts needed later are the statements of Theorems~\ref{thm:w-inc}
and~\ref{thm:w-cts-inc}.

Compared to most of the literature on characterizations of means, the
results and proofs in this chapter have a particular flavour.  First, we
are mainly interested in the \emph{power} means, as opposed to the much
larger class of quasiarithmetic means (defined below).  That makes it
reasonable to assume a homogeneity axiom, which in turn means that we can
almost always do without continuity.  (The absence of continuity hypotheses
distinguishes our results from many others, such as those of Fodor and
Marichal~\cite{FoMa}.)  Second, we wish to include the end cases $M_\infty
= \max$ and $M_{-\infty} = \min$ of the power means, and the fact that
these means are not \emph{strictly} increasing alters considerably the
arguments that can be used.

A key role is played by what Tao%
\index{Tao, Terence} 
calls the `tensor%
\index{tensor power trick} 
power trick' (\cite{TaoSR}, Section~1.9), which can be described as
follows.  Take a set $X$ and two functions $F, G \from X \to \R^+$.
Suppose we want to prove that $F \leq G$, but have only been able to find a
constant $C$ (perhaps large) such that $F \leq C G$.  In general, there is
nothing more to be said.  However, suppose now that $X$ can be equipped
with a product that is preserved by both $F$ and $G$.  Let $x \in X$.  Then
for all $n \geq 1$,
\[
F(x) = F(x^n)^{1/n} \leq (C G(x^n))^{1/n} = C^{1/n} G(x),
\]
and letting $n \to \infty$ gives $F(x) \leq G(x)$, as desired. 

Trivial as it may seem, the tensor power trick can be wielded to powerful
effect.  Typically $X$ is taken to be a set of vectors or functions
equipped with the tensor product.  Tao~\cite{TaoSR}%
\index{Tao, Terence}
demonstrates the 
tensor power trick by using it to prove the Hausdorff--Young%
\index{Hausdorff--Young inequality} 
inequality, and notes that it also plays a part in Deligne's proof of the
Weil%
\index{Weil conjectures}
conjectures.  We will use it
in the proof of the pivotal Lemma~\ref{lemma:substance}.

This chapter begins with the classical theory of quasiarithmetic means,
which are just ordinary arithmetic means transported along a homeomorphism
(Section~\ref{sec:qa-mns}).  The bulk of the chapter
(Sections~\ref{sec:u-mns}--\ref{sec:ih-mns}) concerns general unweighted
means, and culminates in the four characterization theorems shown in
Table~\ref{table:mn-thms}.

\begin{table}
\centering
\normalsize
\begin{tabular}{|l|l|l|}
\hline
                &
strictly increasing             &increasing     \\
\hline          
$(0, \infty)$   &
$t \in (-\infty, \infty)$       &$t \in [-\infty, \infty]$      \\
                &
Theorem~\ref{thm:str-inc}       &Theorem~\ref{thm:inc}          \\
\hline
$[0, \infty)$   &
$t \in (0, \infty)$             &$t \in [-\infty, \infty]$      \\
                &
Theorem~\ref{thm:zero-str-inc}  &Theorem~\ref{thm:cts-inc}      \\
                &
                                &
\small(also assume continuous and nonzero)      \\
\hline
\end{tabular}
\caption{Summary of characterization theorems for symmetric, decomposable,
  homogeneous, unweighted means.  For instance, the top-left entry
  indicates that the strictly increasing such means on $(0, \infty)$ are
  exactly the unweighted power means $M_t$ of order $t \in (-\infty,
  \infty)$.  Table~\ref{table:w-mn-thms} (p.~\pageref{table:w-mn-thms})
  gives the corresponding results on weighted means.}%
\index{power mean!characterization of!unweighted on $(0, \infty)$}%
\index{power mean!characterization of!unweighted on $[0, \infty)$}%
\lbl{table:mn-thms}
\end{table}

Finally, in Section~\ref{sec:w-mns}, we develop a method for converting
characterization theorems for \emph{unweighted} means into characterization
theorems for \emph{weighted} means.  This method is applied to the four
theorems just mentioned.  One of the resulting four characterizations
of weighted means goes back to Hardy, Littlewood and P\'olya in
1934, while the others may be new.  

We will be defining a considerable amount of terminology for properties of
means.  Appendix~\ref{app:condns} contains a summary for convenient
reference.  The word `mean'\index{mean} in isolation will be used informally,
without precise definition.

\section{Quasiarithmetic means}
\lbl{sec:qa-mns}

Let $J$ be a real interval.  The arithmetic mean defines a sequence of
functions
\[
\bigl( M_1 \from \Delta_n \times J^n \to J \bigr)_{n \geq 1}.
\]
For any other set $I$ and bijection $\phi \from I \to J$, we can transport
the arithmetic mean on $J$ along $\phi$ to obtain a kind of mean on $I$.
We will focus on the case where $I$ is also an interval and $\phi$ is a
homeomorphism\index{homeomorphism} (that is, both $\phi$ and $\phi^{-1}$
are continuous), as follows.

\begin{defn}
\lbl{defn:qam}
Let $\phi \from I \to J$ be a homeomorphism between real intervals.  The
\demph{quasiarithmetic%
\index{quasiarithmetic mean}
mean} on $I$ generated by $\phi$ is the sequence of
functions 
\[
\bigl( M_\phi \from \Delta_n \times I^n \to I \bigr)_{n \geq 1}
\ntn{qam}
\]
defined by
\[
M_\phi(\p, \vc{x})  
=
\phi^{-1} \Biggl( \sum_{i = 1}^n p_i \phi(x_i) \Biggr)
\]
($\p \in \Delta_n$, $\vc{x} \in I^n$).
\end{defn}

The theory of quasiarithmetic means is classical, and most of the content
of this section can be found, more or less explicitly, in Chapter~III of
Hardy, Littlewood and P\'olya~\cite{HLP}.

\begin{remark}
In the literature, the terms `quasiarithmetic'%
\index{quasiarithmetic mean}
and `quasilinear'%
\index{quasilinear mean} 
are both
used, sometimes interchangeably, sometimes with the former reserved for the
unweighted case, and sometimes with the latter meaning what we call
modularity (Definition~\ref{defn:mns-mod}).  `Quasiarithmetic' has the
advantage of evoking the fact that a quasiarithmetic mean is just an
arithmetic mean disguised by a change of variable: 
the diagram
\[
\xymatrix{
\Delta_n \times I^n \ar[r]^-{M_\phi} \ar[d]_{1 \times \phi^n}    &
I \ar[d]^{\phi} \\
\Delta_n \times J^n \ar[r]_-{M_1}        &
J
}
\]
commutes. 
\end{remark}

\begin{example}
\lbl{eg:qa-pwr}
For real $t \neq 0$, the power mean $M_t$ on $(0, \infty)$ is the
quasiarithmetic mean $M_{\phi_t}$ generated by the homeomorphism
\[
\begin{array}{cccc}
\phi_t \from    &(0, \infty)    &\to            &(0, \infty)    \\
                &t              &\mapsto        &x^t.
\end{array}
\]
The geometric mean $M_0$ on $(0, \infty)$ is the quasiarithmetic mean
$M_{\phi_0}$ generated by the homeomorphism
\[
\phi_0 = \log \from (0, \infty) \to \R.
\]
Thus, all the power means of \emph{finite} order on $(0, \infty)$ are
quasiarithmetic. 
\end{example}

\begin{example}
The power means $M_{\pm\infty}$ on $(0, \infty)$ are not
quasiarithmetic, as we will prove in
Example~\ref{egs:u-mns-inc}\bref{eg:u-mns-inc-infty}. 
\end{example}

\begin{example}
The quasiarithmetic mean on $\R$ generated by the homeomorphism $\exp \from
\R \to (0, \infty)$ is given by
\[
M_{\exp}(\p, \vc{x})
=
\log \Biggl( \sum_{i = 1}^n p_i e^{x_i} \Biggr)
\]
($\p \in \Delta_n$, $\vc{x} \in \R^n$).  This is the
\demph{exponential%
\index{exponential mean}
mean}, whose special properties were established by Nagumo (\cite{Nagu},
p.~78; or for a modern account, see Theorem~4.15(i) of Grabisch, Marichal,
Mesiar and Pap~\cite{GMMP}).
\end{example}

The rest of this section is dedicated to three questions.  

First, when do two homeomorphisms out of an interval $I$ generate the same
quasiarithmetic mean on $I$?   

Second, among all quasiarithmetic means on $(0, \infty)$, how can we pick
out the power means $M_t$ ($t \in \R$)?  In other words, what special
properties do the power means possess?

Third (and imprecisely for now), given a mean on some large interval, if
its restrictions to smaller intervals are quasiarithmetic, is it
quasiarithmetic itself?

The answers to all three questions involve the notion of affine map.

\begin{defn}
Let $I$ be a real interval.  A function $\alpha \from I \to \R$ is
\demph{affine}%
\index{affine function} 
if 
\[
\alpha \bigl(px_1 + (1 - p)x_2\bigr) 
= 
p\alpha(x_1) + (1 - p)\alpha(x_2)
\]
for all $x_1, x_2 \in I$ and $p \in [0, 1]$.  
\end{defn}

\begin{lemma}
\lbl{lemma:aff}
Let $\alpha \from I \to J$ be a function between real intervals.  The
following are equivalent: 
\begin{enumerate}
\item
\lbl{part:aff-cvx}
$\alpha$ is affine;

\item
\lbl{part:aff-aff}
$\alpha\bigl( \sum \lambda_i x_i \bigr) = \sum \lambda_i \alpha(x_i)$ for
all $n \geq 1$, $x_1, \ldots, x_n \in I$ and $\lambda_1, \ldots,
\lambda_n \in \R$ such that $\sum \lambda_i = 1$ and $\sum \lambda_i x_i
\in I$;

\item 
\lbl{part:aff-explicit}
there exist constants $a, b \in \R$ such that $\alpha(x) = ax + b$ for all
$x \in I$; 

\item
\lbl{part:aff-cts}
$\alpha$ is continuous and $\alpha\bigl(\tfrac{1}{2}(x_1 + x_2)\bigr) =
\tfrac{1}{2}\bigl(\alpha(x_1) + \alpha(x_2)\bigr)$ for all $x_1, x_2 \in I$.
\end{enumerate}
\end{lemma}

Note that in~\bref{part:aff-aff}, some of the coefficients $\lambda_i$ may
be negative.

\begin{proof}
See Appendix~\ref{sec:aff}.
\end{proof}

By part~\bref{part:aff-explicit}, any affine map is either
injective or constant.  We will need the following elementary observation
on extension of affine maps to larger domains.  

\begin{defn}
A real interval is \demph{trivial}%
\index{trivial interval} 
if it has at most one element, and \demph{nontrivial}%
\index{nontrivial interval} 
otherwise.
\end{defn}

\begin{lemma}
\lbl{lemma:aff-ext}
Let $I \sub J$ be real intervals and let $\alpha \from I \to \R$ be an
affine map.  Then:
\begin{enumerate}
\item 
\lbl{part:aff-ext-ext}
there exists an affine map $\ovln{\alpha} \from J \to \R$ extending
$\alpha$;

\item
\lbl{part:aff-ext-inj}
if $\alpha$ is injective then we may choose $\ovln{\alpha}$ to be
injective;

\item
\lbl{part:aff-ext-unique}
if $I$ is nontrivial then $\ovln{\alpha}$ is uniquely determined by
$\alpha$.
\end{enumerate}
\end{lemma}

\begin{proof}
Choose $a, b \in \R$ such that $\alpha(x) = ax + b$ for all $x \in I$.
For~\bref{part:aff-ext-ext}, put $\ovln{\alpha}(y) = ay + b$ for $y \in J$.
For~\bref{part:aff-ext-inj}, if $\alpha$ is injective then we can choose
$a$ to be nonzero, so $\ovln{\alpha}$ is injective.
Part~\bref{part:aff-ext-unique} is trivial.
\end{proof}

We are now ready to answer the first question: when are two
quasiarithmetic%
\index{quasiarithmetic mean!equality of}
means equal?  

\begin{propn}
\lbl{propn:qa-eq}
Let 
\[
\xymatrix@R=1ex{
        &J      \\
I \ar[ru]^{\phi} \ar[rd]_{\phi'}        &       \\
        &J'
}
\]
be homeomorphisms between real intervals.  The following are equivalent:
\begin{enumerate}
\item 
\lbl{part:qa-eq-w}
$M_\phi = M_{\phi'} \from \Delta_n \times I^n \to I$ for all $n \geq 1$;

\item
\lbl{part:qa-eq-u}
$M_\phi(\vc{u}_n, -) = M_{\phi'}(\vc{u}_n, -) \from I^n \to I$ for all $n
\geq 1$;

\item
\lbl{part:qa-eq-fact}
the map $\phi' \of \phi^{-1} \from J \to J'$ is affine. 
\end{enumerate}
\end{propn}

This is Theorem~83 of the book~\cite{HLP} by Hardy,%
\index{Hardy, Godfrey Harold} 
Littlewood%
\index{Littlewood, John Edensor}
and P\'olya,%
\index{Polya, George@P\'olya, George}
who attribute it to Jessen and Knopp.

\begin{proof}
Trivially, \bref{part:qa-eq-w} implies~\bref{part:qa-eq-u}.  

Assuming~\bref{part:qa-eq-u}, we prove~\bref{part:qa-eq-fact}.  Write
$\alpha = \phi' \of \phi^{-1}$.  We will prove that $\alpha$ is affine
using Lemma~\ref{lemma:aff}\bref{part:aff-cts}.  Certainly $\alpha$ is
continuous.  Now let $y_1, y_2 \in J$.  We have
\[
M_\phi\Bigl(
\vc{u}_2, \bigl(\phi^{-1}(y_1), \phi^{-1}(y_2)\bigr)
\Bigr)
=
M_{\phi'}\Bigl(
\vc{u}_2, \bigl(\phi^{-1}(y_1), \phi^{-1}(y_2)\bigr)
\Bigr),
\]
or explicitly,
\[
\phi^{-1} \bigl(
\tfrac{1}{2} y_1 + \tfrac{1}{2} y_2
\bigr)
=
{\phi'}^{-1} \bigl(
\tfrac{1}{2} \phi'\phi^{-1}(y_1) + \tfrac{1}{2} \phi'\phi^{-1} (y_2)
\bigr).
\]
But this can be rewritten as
\[
\alpha\bigl(
\hlf(y_1 + y_2)
\bigr)
=
\hlf
\bigl( \alpha(y_1) + \alpha(y_2) \bigr),
\]
so condition~\bref{part:aff-cts} of Lemma~\ref{lemma:aff} holds and
$\alpha$ is affine.

Finally, assuming~\bref{part:qa-eq-fact}, we prove~\bref{part:qa-eq-w}.
Write $\alpha$ for the affine map $\phi' \of \phi^{-1} \from J \to J'$.
Then $\phi' = \alpha \of \phi$, so our task is to prove that
\[
M_{\alpha\of\phi}(\p, \vc{x})
=
M_\phi(\p, \vc{x})
\]
for all $n \geq 1$, $\p \in \Delta_n$, and $\vc{x} \in I^n$.  And indeed,
\begin{align*}
M_{\alpha\of\phi}(\p, \vc{x})   &
=
(\alpha \of \phi)^{-1} \Biggl(
\sum_{i = 1}^n p_i \alpha(\phi(x_i))
\Biggr) \\
&
=
\phi^{-1} \alpha^{-1} \alpha \Biggl(
\sum_{i = 1}^n p_i \phi(x_i)  
\Biggr) \\
&
=
M_\phi(\p, \vc{x}),
\end{align*}
using Lemma~\ref{lemma:aff}\bref{part:aff-aff} in the second equation.
\end{proof}

\begin{example}
\lbl{eg:aff-q-log} This example concerns the quasiarithmetic mean
$M_{\ln_q}$.  Strictly speaking, $M_{\ln_q}$ is undefined, as the
$q$-logarithm $\ln_q \from (0, \infty) \to \R$ is not surjective (hence not
a homeomorphism) unless $q = 1$.  However, we can change the codomain to
force it to be surjective; that is, we can consider the function
\[
\begin{array}{ccc}
(0, \infty)     &\to            &\ln_q(0, \infty)       \\
x               &\mapsto        &\ln_q(x),
\end{array}
\]
where $\ln_q(0, \infty)$ is the image of $\ln_q$.  This function, which by
abuse of notation we also write as $\ln_q$, \emph{is} a homeomorphism, and
its codomain is a real interval.  In this sense, we can speak of the
quasiarithmetic mean $M_{\ln_q}$.

For $q \neq 1$, the function $\ln_q \from (0, \infty) \to \ln_q(0, \infty)$
is the composite of homeomorphisms
\[
\xymatrix@R=1ex{
        &(0, \infty) \ar[dd]^\alpha     \\
(0, \infty) \ar[ru]^{\phi} \ar[rd]_{\ln_q}      &       \\
        &\ln_q(0, \infty),
}
\]
where 
\[
\phi(x) = x^{1 - q},
\quad
\alpha(y) = \frac{y - 1}{1 - q}.
\]
Here $\alpha$ is affine, so by
Proposition~\ref{propn:qa-eq} and Example~\ref{eg:qa-pwr},
\begin{equation}
\lbl{eq:M-ln-q}
M_{\ln_q} = M_{1 - q} \from 
\Delta_n \times (0, \infty)^n \to (0, \infty)
\end{equation}
($n \geq 1$).  This equation also holds for $q = 1$, by
Example~\ref{eg:qa-pwr}.  Hence it holds for all real $q$.

Equation~\eqref{eq:M-ln-q} can, of course, also be proved directly.  It is
equivalent to Lemma~\ref{lemma:q-log-mean}.
\end{example}

Next we answer the second question: among all quasiarithmetic means,
what distinguishes the power means?  The following result is Theorem~84 of
Hardy,%
\index{Hardy, Godfrey Harold} 
Littlewood%
\index{Littlewood, John Edensor}
and P\'olya~\cite{HLP}.%
\index{Polya, George@P\'olya, George}

\begin{thm}
\lbl{thm:qa-pwr}
\index{power mean!characterization of!weighted on $(0, \infty)$}
\index{quasiarithmetic mean!power means@and power means}
Let $J$ be a real interval and let $\phi \from (0, \infty) \to J$ be a
homeomorphism.  The following are equivalent:
\begin{enumerate}
\item 
\lbl{part:qa-pwr-u-condns}
$M_\phi(\vc{u}_n, c\vc{x}) = cM_\phi(\vc{u}_n, \vc{x})$ for all $n \geq 1$,
$\vc{x} \in (0, \infty)^n$, and $c \in (0, \infty)$;

\item
\lbl{part:qa-pwr-w-condns}
$M_\phi(\p, c\vc{x}) = cM_\phi(\p, \vc{x})$ for all $n \geq 1$, $\p \in
\Delta_n$, $\vc{x} \in (0, \infty)^n$, and $c \in (0, \infty)$;

\item
\lbl{part:qa-pwr-form}
$M_\phi = M_t$ for some $t \in \R$.
\end{enumerate}
\end{thm}

\begin{proof}
Trivially, \bref{part:qa-pwr-form} implies~\bref{part:qa-pwr-w-condns} and
\bref{part:qa-pwr-w-condns} implies~\bref{part:qa-pwr-u-condns}.

Now assume~\bref{part:qa-pwr-u-condns}; we prove~\bref{part:qa-pwr-form}.
By Proposition~\ref{propn:qa-eq}, we may assume that $\phi(1) = 0$: for if
not, replace $J$ by $J' = J - \phi(1)$ and $\phi$ by $\phi' = \phi -
\phi(1)$, which is a homeomorphism $(0, \infty) \to J'$ satisfying
$M_{\phi'} = M_\phi$ and $\phi'(1) = 0$.

For each $c > 0$, define $\phi_c \from (0, \infty) \to J$ by $\phi_c(x) =
\phi(cx)$.  Then $\phi_c$ is a homeomorphism, and for all $\vc{x} \in (0,
\infty)^n$,
\begin{align*}
M_{\phi_c}(\vc{u}_n, \vc{x})    &
=
\phi_c^{-1} \Biggl( 
\sum_{i = 1}^n \frac{1}{n} \phi_c(x_i) 
\Biggr) \\
&
=
\frac{1}{c} \phi^{-1} \Biggl( 
\sum_{i = 1}^n \frac{1}{n} \phi(cx_i)
\Biggr) \\
&
=
\frac{1}{c} M_\phi(\vc{u}_n, c\vc{x})   \\
&
=
M_\phi(\vc{u}_n, \vc{x}),
\end{align*}
where the last step used the homogeneity hypothesis
in~\bref{part:qa-pwr-u-condns}.  Hence by Proposition~\ref{propn:qa-eq},
there exist $\psi(c), \theta(c) \in \R$ such that $\phi_c = \psi(c)\phi +
\theta(c)$.

We have now constructed functions $\psi, \theta \from (0, \infty) \to \R$
such that 
\[
\phi(cx) = \psi(c) \phi(x) + \theta(c)
\]
for all $c, x \in (0, \infty)$.  Putting $x = 1$ and using $\phi(1) = 0$
gives $\theta = \phi$, so
\[
\phi(cx) = \phi(c) + \psi(c) \phi(x)
\]
for all $c, x \in (0, \infty)$.  Since $\phi$ is measurable and not
constant, the functional characterization of the $q$-logarithm
(Theorem~\ref{thm:q-log}) implies that $\phi = A\ln_q$ for some $A, q \in
\R$ with $A \neq 0$.  Hence $M_\phi = M_{\ln_q}$ by
Proposition~\ref{propn:qa-eq}.  But $M_{\ln_q} = M_{1 - q}$ by
Example~\ref{eg:aff-q-log}, so $M_\phi = M_{1 - q}$, as required.
\end{proof}

We now answer the third and final question: loosely, given a mean on a
large interval whose restriction to every small subinterval is
quasiarithmetic, is the original mean also quasiarithmetic?

The most important means for us are the power means, which are defined on
the unbounded interval $(0, \infty)$ or $[0, \infty)$.  However, some
results on means are most easily proved on closed bounded intervals.  The
following lemma allows us to leverage results on closed bounded intervals
to prove results on arbitrary intervals.  It states that whether a mean
on an arbitrary interval is quasiarithmetic is entirely determined by
its behaviour on closed bounded subintervals.

Our lemma concerns \emph{unweighted} means.  We will use
the abbreviated notation
\begin{equation}
\lbl{eq:w-u-abbr}
M_\phi(\vc{x}) = M_\phi(\vc{u}_n, \vc{x})
\end{equation}
for unweighted quasiarithmetic means, and we will say that a sequence of
functions $\bigl( M \from I^n \to I \bigr)_{n \geq 1}$ on a real interval
$I$ is a 
\demph{quasiarithmetic\lbl{p:is-qam}%
\index{quasiarithmetic mean} 
mean} if there exist an interval $J$ and a homeomorphism $\phi \from I \to
J$ such that $M$ is the unweighted quasiarithmetic mean $M_\phi$ generated
by $\phi$.

\begin{lemma}
\lbl{lemma:u-mn-extension}
Let $I$ be a real interval and let $(M \from I^n \to I)_{n \geq 1}$ be a
sequence of functions.  Suppose that $M$ restricts to a quasiarithmetic
mean on each nontrivial closed bounded subinterval of $I$.  Then $M$ is a
quasiarithmetic mean. 
\end{lemma}

\begin{proof}
If $I$ is trivial then so is the result.  Otherwise, we can write $I$ as
the union of an infinite nested sequence $I_1 \sub I_2 \sub \cdots$ of
nontrivial closed bounded subintervals.  By hypothesis, $M \from I^n \to I$
restricts to a function $M|_{I_r} \from I_r^n \to I_r$ for each $n, r \geq
1$, and the sequence of functions $(M|_{I_r} \from I_r^n \to I_r)_{n \geq
  1}$ is a quasiarithmetic mean for each $r \geq 1$.

We will construct, inductively, a nested sequence $J_1 \sub J_2 \sub
\cdots$ of real intervals and a sequence of homeomorphisms $\phi_r \from
I_r \to J_r$, each satisfying $M_{\phi_r} = M|_{I_r}$ and each extending
the last:
\[
\xymatrix@M+.5ex{
I_1 \ar@{^{(}->}[r] \ar[d]^{\phi_1}     &
I_2 \ar@{^{(}->}[r] \ar[d]^{\phi_2}     &
\cdots  \\
J_1 \ar@{^{(}->}[r] &
J_2 \ar@{^{(}->}[r] &
\cdots
}
\]

For the first step, since $M|_{I_1}$ is a quasiarithmetic mean, we can
choose an interval $J_1$ and a homeomorphism $\phi_1 \from I_1 \to J_1$
such that $M_{\phi_1} = M|_{I_1}$.

Now suppose inductively that $J_r$ and $\phi_r$ have been
defined for some $r \geq 1$, in such a way that $M_{\phi_r} = M|_{I_r}$.
Since $M|_{I_{r + 1}}$ is a quasiarithmetic mean, we can choose a real
interval $L_{r + 1}$ and a homeomorphism $\theta_{r + 1} \from I_{r + 1}
\to L_{r + 1}$ such that $M_{\theta_{r + 1}} = M|_{I_{r + 1}}$.  Then
$\theta_{r + 1}$ restricts to a homeomorphism of intervals $\theta'_{r + 1}
\from I_r \to \theta_{r + 1} I_r$, giving the top square of the commutative
diagram in Figure~\ref{fig:u-mn-extension}.
\begin{figure}
\[
\normalsize
\xymatrix@M+.5ex{\textstyle
I_r
\ar@/_4pc/[dd]_{\phi_r} \ar[d]_{\theta'_{r + 1}} \ar@{^{(}->}[r]        &
I_{r + 1}
\ar@/^4pc/[dd]^{\phi_{r + 1}} \ar[d]^{\theta_{r + 1}}   \\
\theta_{r + 1}I_r \ar@{^{(}->}[r] \ar[d]_{\alpha'_{r + 1}}      &
L_{r + 1} \ar[d]^{\alpha_{r + 1}}       \\
J_r \ar@{^{(}->}[r]     &
J_{r + 1}
}
\]
\caption{The inductive step in the proof of
  Lemma~\ref{lemma:u-mn-extension}.  The vertical and curved arrows are all
  homeomorphisms.} 
\lbl{fig:u-mn-extension}
\end{figure}

To construct the bottom square, we need to define $\alpha'_{r + 1}$, $J_{r
  + 1}$, and $\alpha_{r + 1}$.  Put $\alpha'_{r + 1} = \phi_r \of
{\theta'}_{r + 1}^{-1}$, which is a homeomorphism.  We have $M_{\theta_{r +
    1}} = M|_{I_{r + 1}}$ by definition of $\theta_{r + 1}$, so
\[
M_{\theta'_{r + 1}} = M|_{I_r} = M_{\phi_r}.
\]
Hence by Proposition~\ref{propn:qa-eq}, $\alpha'_{r + 1}$ is affine.  By
Lemma~\ref{lemma:aff-ext}, the affine injection $\alpha'_{r + 1}$ on
$\theta_{r + 1} I_r$ extends uniquely to an affine injection defined on the
larger interval $L_{r + 1}$.  Writing $J_{r + 1}$ for the image of this
extended function (which is an interval), this gives an affine
homeomorphism $\alpha_{r + 1} \from L_{r + 1} \to J_{r + 1}$ making the
bottom square of Figure~\ref{fig:u-mn-extension} commute.

Put $\phi_{r + 1} = \alpha_{r + 1} \of \theta_{r + 1}$.  Then $\phi_{r +
  1}$ is a homeomorphism since $\alpha_{r + 1}$ and $\theta_{r + 1}$ are.
Moreover, $M_{\phi_{r + 1}} = M_{\theta_{r + 1}}$ since $\alpha_{r + 1}$ is
affine.  But $M_{\theta_{r + 1}} = M|_{I_{r + 1}}$, so $M_{\phi_{r + 1}} =
M|_{I_{r + 1}}$, completing the inductive construction.

Finally, let $J$ be the interval $\bigcup_{r = 1}^\infty J_r$ 
and let $\phi \from I \to J$ be the unique function extending all of the
functions $\phi_r \from I_r \to J_r$.  Then $\phi$ is a homeomorphism
since every $\phi_r$ is.  Moreover, given $\vc{x} \in I^n$, we have
$\vc{x} \in I_r^n$ for some $r \geq 1$, so
\[
M_\phi(\vc{x})
=
M_{\phi_r}(\vc{x})
=
M|_{I_r}(\vc{x})
=
M(\vc{x}),
\]
where the middle equation is by construction of $\phi_r$ and the others are
immediate from the definitions.  Hence $M = M_\phi$ and $M$ is a
quasiarithmetic mean.
\end{proof}

\begin{remark}
Kolmogorov%
\index{Kolmogorov, Andrei} 
found an early characterization theorem for quasiarithmetic means on real
intervals~\cite{KolmSNM,KolmONM}.  He proved it for closed bounded
intervals, and asserted that his argument could be extended to closed
unbounded intervals with `only a minor modification' (\cite{KolmONM},
p.~144).  In fact, it can be extended to \emph{all} intervals.  Later
authors used results similar to Lemma~\ref{lemma:u-mn-extension} to prove
this and similar statements.  For example, the argument above is an
expansion of the argument on p.~291 of Acz\'el~\cite{AczeLFE}, and of part
of the proof of Theorem~4.10 of Grabisch, Marichal, Mesiar and
Pap~\cite{GMMP}.
\end{remark}

\section{Unweighted means}
\lbl{sec:u-mns}
\index{mean!unweighted}

In the next three sections, we focus exclusively on means that are
unweighted, that is, weighted by the uniform distribution $\vc{u}_n$.
Certainly this is a natural special case.  But the real reason for this
focus is that the results established will help us to prove theorems on
\emph{weighted} means (Section~\ref{sec:w-mns}), which in turn will be used
to prove unique characterizations of measures of value
(Section~\ref{sec:value-char}) and measures of diversity
(Section~\ref{sec:total-hill}).

The pattern of argument in this chapter is broadly similar to that in the
proof of Faddeev's theorem (Section~\ref{sec:ent-chain}).  There, given a
hypothetical entropy measure $I$ satisfying some axioms, most of the work
went into analysing the sequence $\bigl(I(\vc{u}_n)\bigr)_{n \geq 1}$,
which then made it relatively easy to find $I(\p)$ for distributions $\p$
with rational coordinates, and, in turn, for all $\p$.  Here, we
spend considerable time proving results on unweighted means $M(\vc{u}_n,
-)$.  This done, we will quickly be able to deduce results on weighted
means $M(\p, -)$, first for rational $\p$ and then for all $\p$.

For simplicity, we adopt the abbreviated notation
\[
M_t(\vc{x}) = M_t(\vc{u}_n, \vc{x})
\ntn{Mtx}
\]
($t \in [-\infty, \infty]$, $\vc{x} \in [0, \infty)^n$) for unweighted
  power means, as well as using the notation $M_\phi(\vc{x})$ as in
  equation~\eqref{eq:w-u-abbr}. 

Let $(M \from I^n \to I)_{n \geq 1}$ be a sequence of functions, where $I$
is either $(0, \infty)$ or $[0, \infty)$.  Over the next three sections, we
  answer the question:
\begin{quote}
\emph{What conditions on $M$ guarantee that it is one of the unweighted
  power means $M_t$?}  
\end{quote}
The question can be interpreted in several ways, depending on whether $I$
is $(0, \infty)$ or $[0, \infty)$, and also on whether we want to restrict
  the order $t$ of the power mean to be positive, finite, etc.  

We now list some of the conditions on $M$ that might reasonably be imposed.
For many of them, we have already considered similar conditions for
weighted means (Section~\ref{sec:pwr-mns}).  For Definitions
\ref{defn:u-sym}--\ref{defn:u-hgs}, let $I$ be a real interval and let $(M
\from I^n \to I)_{n \geq 1}$ be a sequence of functions.

\begin{defn}
\lbl{defn:u-sym}
$M$ is \demph{symmetric}%
\index{symmetric!unweighted mean} 
if $M(\vc{x}) = M(\vc{x}\sigma)$ for all $n \geq 1$, $\vc{x} \in I^n$, and
permutations $\sigma$ of $\{1, \ldots, n\}$.
\end{defn}

\begin{examples}
\lbl{egs:u-mns-sym}
All quasiarithmetic means are symmetric. So too are all the power means
$M_t$, including $M_\infty$ and $M_{-\infty}$ (which are not
quasiarithmetic).
\end{examples}

\begin{defn}
\lbl{defn:u-cons}
$M$ is \demph{consistent}%
\index{consistent!unweighted mean} 
(or \demph{idempotent}\index{idempotent}) if
\[
M(\underbrace{x, \ldots, x}_n) = x
\]
for all $n \geq 1$ and $x \in I$.
\end{defn}

\begin{example}
\lbl{eg:u-mns-cons}
All quasiarithmetic and power means are consistent.
\end{example}

\begin{defn}
\lbl{defn:u-isi}
$M$ is \demph{increasing}%
\index{increasing!unweighted mean}%
\index{mean!increasing}
if for all $n \geq 1$ and $\vc{x}, \vc{y} \in I^n$, 
\[
\vc{x} \leq \vc{y} \implies M(\vc{x}) \leq M(\vc{y}).
\]
It is \demph{strictly increasing}%
\index{strictly increasing!unweighted mean}%
\index{mean!strictly increasing}%
\index{increasing!strictly}
if for all $n \geq 1$ and $\vc{x}, \vc{y} \in I^n$,
\[
\vc{x} \leq \vc{y} \neq \vc{x} \implies M(\vc{x}) < M(\vc{y}).
\]
\end{defn}

\begin{example}
All quasiarithmetic means are strictly increasing.
\end{example}

\begin{examples}
\lbl{egs:u-mns-inc-pwr}
Given a sequence of functions $(M \from \Delta_n \times I^n \to I)_{n \geq
  1}$, if $M$ is increasing or strictly increasing in the sense of
Definition~\ref{defn:w-isi} then $\bigl( M(\vc{u}_n, -): I^n \to I\bigr)_{n
  \geq 1}$ is increasing or strictly increasing in the sense above.  In
particular, Lemma~\ref{lemma:pwr-mns-inc} implies that:
\begin{enumerate}
\item 
the unweighted power means $M_t$ of all orders $t \in [-\infty, \infty]$ on $[0,
  \infty)$ are increasing;

\item
\lbl{eg:u-mns-inc-pwr-fin}
the unweighted power means $M_t$ of \emph{finite} orders $t \in (-\infty,
\infty)$ on $(0, \infty)$ are strictly increasing;

\item
the unweighted power means $M_t$ of \emph{finite positive} orders $t \in
(0, \infty)$ on $[0, \infty)$ are strictly increasing.
\end{enumerate}
\end{examples}

\begin{examples}
\lbl{egs:u-mns-inc}
\begin{enumerate}
\item
\lbl{eg:u-mns-inc-infty}
The power means $M_\infty = \max$ and $M_{-\infty} = \min$ are increasing
but not strictly so, assuming that the interval $I$ is nontrivial.  (The
counterexample of Remark~\ref{rmk:pwr-mns-not-si} is easily adapted to
$I$.)  Hence $M_{\pm\infty}$ are not quasiarithmetic means.

\item
The power means $M_t$ of order $t \in [-\infty, 0]$ are not strictly
increasing on $[0, \infty)$ (again, as in Remark~\ref{rmk:pwr-mns-not-si}).
  So they are not quasiarithmetic on $[0, \infty)$, even though they are
    quasiarithmetic on $(0, \infty)$.
\end{enumerate}
\end{examples}

\begin{defn}
\lbl{defn:decomp}
$M$ is \demph{decomposable}\index{decomposable} if for all $n, k_1, \ldots,
k_n \geq 1$ and $x^i_j \in I$,
\[
M\bigl(x^1_1, \ldots, x^1_{k_1}, \ \ldots, \ x^n_1, \ldots, x^n_{k_n}\bigr)
=
M(\underbrace{a_1, \ldots, a_1}_{k_1}, 
\ldots, 
\underbrace{a_n, \ldots, a_n}_{k_n}),
\]
where $a_i = M\bigl(x^i_1, \ldots, x^i_{k_i}\bigr)$ for $i \in \{1, \ldots,
n\}$.
\end{defn}

We adopt the shorthand 
\begin{equation}
\lbl{eq:mc}
r \mc x 
=
\underbrace{x, \ldots, x}_r
\end{equation}
whenever $r \geq 1$ and $x \in \R$.  Thus, the decomposability equation
becomes
\[
M\bigl(x^1_1, \ldots, x^1_{k_1}, \ \ldots, \ x^n_1, \ldots, x^n_{k_n}\bigr)
=
M(k_1 \mc a_1, \ldots, k_n \mc a_n).
\]

Decomposability is an unweighted analogue\lbl{p:decomp-analogue} of the
chain rule for weighted means (Definition~\ref{defn:mns-chn}), as the
following examples show.

\begin{examples}
\lbl{egs:u-mns-decomp}
\begin{enumerate}
\item 
\lbl{eg:u-mns-decomp-pwr}
For each $t \in [-\infty, \infty]$, the power mean $M_t$ is decomposable.%
\index{power mean!decomposability of}
This can of course be shown by direct calculation, but instead we
prove it using earlier results on weighted power means.

Take $x^i_j$ and $a_i$ as in Definition~\ref{defn:decomp}.  Write
\[
k = k_1 + \cdots + k_n,
\quad
\p = (k_1/k, \ldots, k_n/k) \in \Delta_n.
\]
Then 
\[
\p \of (\vc{u}_{k_1}, \ldots, \vc{u}_{k_n}) = \vc{u}_k,
\]
so
\begin{align*}
&
M_t(x^1_1, \ldots, x^1_{k_1}, \ \ldots, \ x^n_1, \ldots, x^n_{k_n})     \\
&
=
M_t\bigl(
\p \of (\vc{u}_{k_1}, \ldots, \vc{u}_{k_n}),
\bigl(x^1_1, \ldots, x^1_{k_1}\bigr) 
\oplus\cdots\oplus 
\bigl(x^n_1, \ldots, x^n_{k_n}\bigr)
\bigr)  \\
&
=
M_t\bigl(
\p, (a_1, \ldots, a_n)
\bigr),
\end{align*}
by the chain rule for power means (Proposition~\ref{propn:pwr-mns-chn}).
On the other hand,
\begin{align*}
&
M_t(k_1 \mc a_1, \ldots, k_n \mc a_n)   \\
&
=
M_t\bigl(
\p \of (\vc{u}_{k_1}, \ldots, \vc{u}_{k_n}),
(k_1 \mc a_1) \oplus\cdots\oplus (k_n \mc a_n)
\bigr)  \\
&
=
M_t\Bigl(
\p, 
\bigl(
M_t(\vc{u}_{k_1}, k_1 \mc a_1), \ldots, M_t(\vc{u}_{k_n}, k_n \mc a_n)
\bigr)  
\Bigr)  \\
&
=
M_t\bigl(
\p, (a_1, \ldots, a_n)
\bigr),
\end{align*}
by the chain rule again and consistency of $M_t$.  Hence $M_t$ is
decomposable.

\item
In particular, $M_1$ is decomposable, from which it follows that all
quasiarithmetic means are decomposable.
\end{enumerate}
\end{examples}

\begin{remark}
In the literature, decomposability is often stated in the asymmetric form 
\[
M(x_1, \ldots, x_k, y_1, \ldots, y_\ell)
=
M(k \mc a, y_1, \ldots, y_\ell)
\]
($k, \ell \geq 1$, $x_i, y_j \in I$), where $a = M(x_1, \ldots, x_k)$.
(This was the form used by both Kolmogorov~\cite{KolmSNM,KolmONM} and
Nagumo~\cite{Nagu}, for instance.)  Under the mild assumptions that $M$ is
symmetric and consistent, this is equivalent to the definition above, by a
straightforward induction.
\end{remark}

\begin{defn}
\lbl{defn:u-mod}
$M$ is \demph{modular}
\index{modularity!unweighted mean@of unweighted mean} 
if for all 
\[
x^1_1, \ldots, x^1_{k_1}, \ \ldots,\ x^n_1, \ldots, x^n_{k_n} \in I,
\qquad
y^1_1, \ldots, y^1_{k_1}, \ \ldots,\ y^n_1, \ldots, y^n_{k_n} \in I
\]
such that
\[
M\bigl(x^i_1, \ldots, x^i_{k_i}\bigr) 
= 
M\bigl(y^i_1, \ldots, y^i_{k_i}\bigr)
\]
for each $i$, we have
\[
M\bigl(x^1_1, \ldots, x^1_{k_1}, \ \ldots,\ 
x^n_1, \ldots, x^n_{k_n}\bigr)
=
M\bigl(y^1_1, \ldots, y^1_{k_1}, \ \ldots,\ 
y^n_1, \ldots, y^n_{k_n}\bigr).
\]
\end{defn}

In other words, $M$ is modular if 
\[
M\bigl(x^1_1, \ldots, x^1_{k_1}, \ \ldots,\ x^n_1, \ldots, x^n_{k_n}\bigr)
\]
is determined by $k_1, \ldots, k_n$ and 
\[
M\bigl(x^1_1, \ldots, x^n_1\bigr), 
\ \ldots, \ 
M\bigl(x^n_1, \ldots, x^n_{k_n}\bigr).
\]
Evidently this is true if $M$ is decomposable.

\begin{defn}
\lbl{defn:u-hgs}
Suppose that $I$ is closed under multiplication.  Then $M$ is
\demph{homogeneous}%
\index{homogeneous!unweighted mean} 
if
\[
M(c\vc{x}) = c M(\vc{x})
\]
for all $n \geq 1$, $c \in I$, and $\vc{x} \in I^n$.
\end{defn}

\begin{examples}
\lbl{egs:u-mns-hgs}
All the power means are homogeneous.  But other quasiarithmetic means are
not, as Theorem~\ref{thm:qa-pwr} shows.
\end{examples}

It has already been mentioned that an important early result in the
theory of means was proved by Kolmogorov%
\index{Kolmogorov, Andrei} 
and Nagumo,%
\index{Nagumo, Mitio}  
independently in 1930~\cite{KolmSNM,Nagu}.  What they showed was that any
continuous, symmetric, consistent, strictly increasing, decomposable
sequence of functions $(M \from I^n \to I)_{n \geq 1}$ on a real interval
$I$ is a quasiarithmetic mean.

One of the purposes of this book is to prove characterization theorems
for diversity measures.  The measures that we characterize are closely
related to the power means $M_t$, where $t \in [-\infty, \infty]$.  In
particular, we want to include $M_{\pm\infty}$.  Since Kolmogorov and
Nagumo's theorem insists on a \emph{strictly} increasing mean, it is
inadequate for our purpose.  So, we follow a different path.

There is another difference between the results below and those of
Kolmogorov and Nagumo.  Our focus on \emph{power} means makes it natural to
impose a homogeneity condition (in the light of Theorem~\ref{thm:qa-pwr}).
It turns out that when homogeneity is assumed, the continuity condition in
the Kolmogorov--Nagumo theorem can be dropped.

In Section~\ref{sec:sih-mns}, we will characterize the power means of
finite orders $t \in (-\infty, \infty)$.  We will use this result in
Section~\ref{sec:ih-mns} to achieve our goal of characterizing the power
means of all orders $t \in [-\infty, \infty]$.  Our first steps are the same as the first steps of Kolmogorov's%
\index{Kolmogorov, Andrei} 
proof, and most of the lemmas in the remainder of this section
can be found in his paper~\cite{KolmSNM} (translated into English
as~\cite{KolmONM}).

Our first lemma concerns repetition of terms.

\begin{lemma}
\lbl{lemma:u-mn-rep}
Let $I$ be an interval and let $(M \from I^n \to I)_{n \geq 1}$ be a
symmetric, consistent, decomposable sequence of functions.  Then
\begin{equation}
\lbl{eq:u-mn-rep}
M(r \mc x_1, \ldots, r \mc x_n)
=
M(x_1, \ldots, x_n)
\end{equation}
for all $r, n \geq 1$ and $x_1, \ldots, x_n \in I$.
\end{lemma}

\begin{proof}
Write $a = M(x_1, \ldots, x_n)$.  By symmetry, the left-hand side of
equation~\eqref{eq:u-mn-rep} is equal to
\[
M(x_1, \ldots, x_n, \ \ldots, \ x_1, \ldots, x_n),
\]
with $rn$ terms in total.  By decomposability, this is equal to $M(rn \mc
a)$, which by consistency is equal to $a$.
\end{proof}

The next group of lemmas begins to answer the question: given a
quasiarithmetic mean $M$ on an interval $I$, how can we construct from $M$
a homeomorphism $\phi$ such that $M = M_\phi$?
Proposition~\ref{propn:qa-eq} tells us that there are many homeomorphisms
with this property.  But it also tells us that if $I$ is of the form $[a,
  b]$ for some real $a < b$, then there is a unique homeomorphism $\phi
\from [a, b] \to [0, 1]$ such that $\phi(a) = 0$, $\phi(b) = 1$, and $M =
M_\phi$.  The function $\psi$ constructed in the next lemma will turn out
to be the inverse of $\phi$, restricted to the rationals.

\begin{lemma}
\lbl{lemma:psi-rat}
Let $a, b \in \R$ with $a < b$.  Let $\bigl( M \from [a, b]^n \to [a, b]
\bigr)_{n \geq 1}$ be a symmetric, consistent, decomposable sequence of
functions. 
\begin{enumerate}
\item
\lbl{part:psi-rat-def}
There is a unique function
\[
\psi \from [0, 1] \cap \Q \to [a, b]
\]
satisfying
\[
\psi(r/s) = M \bigl( (s - r) \mc a, r \mc b \bigr)
\]
for all integers $0 \leq r \leq s$ with $s \geq 1$. 

\item
\lbl{part:psi-rat-ends}
$\psi(0) = a$ and $\psi(1) = b$.

\item
\lbl{part:psi-rat-eq}
For all $n \geq 1$ and $q_1, \ldots, q_n \in [0, 1] \cap \Q$,
\[
M \bigl( \psi(q_1), \ldots, \psi(q_n) \bigr)
=
\psi \Biggl( \frac{1}{n} \sum_{i = 1}^n q_i \Biggr).
\]

\item
\lbl{part:psi-rat-inc} 
If $M$ is increasing then so is $\psi$, and if $M$ is strictly increasing
then so is $\psi$.
\end{enumerate}
\end{lemma}

\begin{proof}
For~\bref{part:psi-rat-def}, uniqueness is immediate.  To prove existence,
we must show that different representations $r/s$ of the same rational
number give the same value of $M((s - r) \mc a, r \mc b)$.  Suppose that
$r/s = r'/s'$.  Then $s'r = sr'$, so using Lemma~\ref{lemma:u-mn-rep}
twice,
\begin{align*}
M\bigl((s - r) \mc a, r \mc b\bigr)       &
=
M\bigl(s'(s - r) \mc a, s'r \mc b\bigr)  \\
&
=
M\bigl(s(s' - r') \mc a, sr' \mc b\bigr) \\
&
=
M\bigl((s' - r') \mc a, r' \mc b\bigr).
\end{align*}
This proves~\bref{part:psi-rat-def}, and~\bref{part:psi-rat-ends} follows
from the formula for $\psi$ and consistency.

For~\bref{part:psi-rat-eq}, express $q_1, \ldots, q_n$ as fractions
over a common denominator, say $q_i = r_i/s$.  Then 
\begin{align}
M\bigl(
\psi(q_1), \ldots, \psi(q_n)
\bigr)  &
=
M\bigl(
s \mc \psi(q_1), \ldots, s \mc \psi(q_n)
\bigr)  &
\lbl{eq:pre-1}        \\
&
=
M\bigl(
(s - r_1) \mc a, r_1 \mc b, \ldots, (s - r_n) \mc a, r_n \mc b
\bigr)  
\lbl{eq:pre-2}        \\
&
=
M\bigl(
(ns - r_1 - \cdots - r_n) \mc a, (r_1 + \cdots + r_n) \mc b
\bigr)  
\lbl{eq:pre-3}
\\
&
=
\psi\biggl( \frac{r_1 + \cdots + r_n}{ns} \biggr)
=
\psi \Biggl( \frac{1}{n} \sum_{i = 1}^n q_i \Biggr),
\nonumber
\end{align}
where equation~\eqref{eq:pre-1} uses Lemma~\ref{lemma:u-mn-rep},
equation~\eqref{eq:pre-2} follows from decomposability and the definition
of $\psi$, and equation~\eqref{eq:pre-3} is by symmetry.

For~\bref{part:psi-rat-inc}, let $q, q' \in [0, 1] \cap \Q$ with $q < q'$.
We may write $q = r/s$ and $q' = r'/s$ for some integers $0 \leq r < r'
\leq s$ with $s \geq 1$.  Assuming that $M$ is increasing, 
\begin{align}
M(q)    &
=
M\bigl((s - r) \mc a, r \mc b\bigr)     
\nonumber       \\
&
=
M\bigl((s - r') \mc a, (r' - r) \mc a, r \mc b\bigr)       
\nonumber       \\
&
\leq
M\bigl((s - r') \mc a, (r' - r) \mc b, r \mc b\bigr)       
\lbl{eq:pri-2}  \\
&
=
M\bigl((s - r') \mc a, r' \mc b\bigr)
=
M(q'),
\nonumber
\end{align}
with strict inequality in~\eqref{eq:pri-2} if $M$ is strictly increasing. 
\end{proof}

We will chiefly be working with decomposable, homogeneous means on $(0,
\infty)$.  Such a mean is automatically consistent:

\begin{lemma}
\lbl{lemma:u-mn-dhc}
Let $\bigl( M \from (0, \infty)^n \to (0, \infty) \bigr)_{n \geq 1}$ be a
decomposable, homogeneous sequence of functions.  Then $M$ is consistent.
\end{lemma}

\begin{proof}
For $k \geq 1$, write $a_k = M(k \mc 1)$.  By decomposability, $a_k = M(k
\mc a_k)$.  (This follows from Definition~\ref{defn:decomp} by taking $n =
1$, $k_1 = k$, and $x^1_1 = \cdots = x^1_k = 1$.)  Hence by homogeneity,
$a_k = a_k M(k \mc 1) = a_k^2$, giving $a_k \in \{0, 1\}$.  But $M$ takes
values in $(0, \infty)$, so $a_k = 1$.  Thus, for all $\vc{x} \in (0,
\infty)^k$,
\[
M(k \mc x) = x a_k = x
\]
by homogeneity again.
\end{proof}

We will deduce that any symmetric such mean is multiplicative, in the
following sense.

\begin{defn}
\lbl{defn:u-mult}
Let $I$ be a real interval closed under multiplication.  A sequence of
functions $(M \from I^n \to I)_{n \geq 1}$ is \demph{multiplicative}%
\index{multiplicative!unweighted mean}
if 
\[
M(\vc{x} \otimes \vc{y}) = M(\vc{x}) M(\vc{y})
\]
for all $n, m \geq 1$, $\vc{x} \in I^n$, and $\vc{y} \in I^m$.
\end{defn}

For instance, if a \emph{weighted} mean is multiplicative in the sense of
Definition~\ref{defn:w-mult} then its unweighted part $\bigl(M(\vc{u}_n,
-)\bigr)_{n \geq 1}$ is multiplicative in the sense just defined.

\begin{lemma}
\lbl{lemma:u-mn-mult}
Let $\bigl( M \from (0, \infty)^n \to (0, \infty) \bigr)_{n \geq 1}$ be a
symmetric, decomposable, homogeneous sequence of functions.  Then $M$ is
multiplicative. 
\end{lemma}

\begin{proof}
By Lemma~\ref{lemma:u-mn-dhc}, $M$ is consistent.  Let $\vc{x} \in (0,
\infty)^n$ and $\vc{y} \in (0, \infty)^m$.  Writing
\[
b_i = M(x_i y_1, \ldots, x_i y_m),
\]
we have
\begin{equation}
\lbl{eq:umm}
M(\vc{x} \otimes \vc{y})
=
M(m \mc b_1, \ldots, m \mc b_n)
=
M(b_1, \ldots, b_n),
\end{equation}
by decomposability and Lemma~\ref{lemma:u-mn-rep} respectively.  But by
homogeneity, $b_i = x_i M(\vc{y})$.  Substituting this into~\eqref{eq:umm}
and using homogeneity again gives the result.
\end{proof}

Lemmas \ref{lemma:u-mn-rep}--\ref{lemma:u-mn-mult} are
largely taken from Kolmogorov~\cite{KolmSNM,KolmONM},%
\index{Kolmogorov, Andrei}
who, assuming that $M$ is continuous, went on to prove that the function
$\psi$ of Lemma~\ref{lemma:psi-rat} extends to a continuous function on
$[0, 1]$.  But this is where his path and ours diverge.

\section{Strictly increasing homogeneous means}
\lbl{sec:sih-mns}
\index{mean!strictly increasing}
\index{strictly increasing!unweighted mean}

Here we prove two theorems on strictly increasing, symmetric, decomposable,
homogeneous, unweighted means (Table~\ref{table:mn-thms}).  First we show
that on $(0, \infty)$, such means are exactly the power means of finite
order.  From this we deduce that on $[0, \infty)$, the means with these
properties are exactly the power means of finite \emph{positive} order.

To show that any sequence of functions $\bigl( M \from (0, \infty)^n \to
(0, \infty) \bigr)_{n \geq 1}$ with suitable properties is a power mean of
finite order, the main challenge is to show that $M$ is quasiarithmetic.
We do this by showing that the restriction of $M$ to each closed bounded
subinterval $K \subset (0, \infty)$ is quasiarithmetic, then invoking
Lemma~\ref{lemma:u-mn-extension}.
An important part of the proof that $M|_K$ is quasiarithmetic will be to
take the map
\[
\psi \from [0, 1] \cap \Q \to K
\]
provided by Lemma~\ref{lemma:psi-rat} and extend it to a map $[0, 1] \to
K$.  For this, we use a lemma of real analysis that has nothing
intrinsically to do with means.

\begin{lemma}
\lbl{lemma:cts-ext}
Let $\psi \from [0, 1] \cap \Q \to \R$ be a strictly increasing function.
Suppose that for all $z \in [0, 1)$,
\begin{equation}
\lbl{eq:cts-ext-1}
\sup \{ \psi(p) \such \text{rational } p \leq z \}
=
\inf \{ \psi(q) \such \text{rational } q > z\},
\end{equation}
and for all $z \in (0, 1]$, 
\begin{equation}
\lbl{eq:cts-ext-2}
\sup \{ \psi(p) \such \text{rational } p < z\}
=
\inf \{ \psi(q) \such \text{rational } q \geq z\}.
\end{equation}
Then $\psi$ extends uniquely to a continuous function $[0, 1] \to \R$, and
this extended function is strictly increasing.
\end{lemma}

Equations~\eqref{eq:cts-ext-1} and~\eqref{eq:cts-ext-2} can be understood
as follows.  Taken over rational $z$, they together state that the function
$\psi \from [0, 1] \cap \Q \to \R$ is continuous.  When $z$ is irrational,
both~\eqref{eq:cts-ext-1} and~\eqref{eq:cts-ext-2} reduce to the equation
\[
\sup \{ \psi(p) \such \text{rational } p < z \}
=
\inf \{ \psi(q) \such \text{rational } q > z \},
\]
which states that $\psi$ has no jump discontinuity at $z$.  Thus, the
result is that any continuous, strictly increasing
function on $[0, 1] \cap \Q$ extends to a function on $[0, 1]$ with
the same properties, as long as the original function has no jump
discontinuities. 

\begin{proof}
Uniqueness is immediate.  For existence, first note that
\begin{equation}
\lbl{eq:cts-ext-3}
\sup \{ \psi(p) \such \text{rational } p \leq z\}
=
\inf \{ \psi(q) \such \text{rational } q \geq z\}
\end{equation}
for all $z \in [0, 1]$.  Indeed, if $z \in \Q$ then both sides are equal to
$\psi(z)$ (since $\psi$ is increasing), and if $z \not\in \Q$
then~\eqref{eq:cts-ext-3} is equivalent to both~\eqref{eq:cts-ext-1}
and~\eqref{eq:cts-ext-2}.  Define a function $\ovln{\psi} \from [0, 1]
\to \R$ by taking $\ovln{\psi}(z)$ to be either side
of~\eqref{eq:cts-ext-3}.  Then $\ovln{\psi}|_{\Q} = \psi$.  

To see that $\ovln{\psi}$ is strictly increasing, let $z, z' \in [0, 1]$
with $z < z'$.  We can choose rational $q$ and $p$ such that $z \leq q < p
\leq z'$.  Then by definition of $\ovln{\psi}$ and the fact that $\psi$ is
strictly increasing, 
\[
\ovln{\psi}(z) \leq \psi(q) < \psi(p) \leq \ovln{\psi}(z'), 
\]
as required.

Finally, we show that $\ovln{\psi}$ is continuous.  Since $\ovln{\psi}$ is
increasing, it suffices to show that for all $z \in [0, 1)$,
\[
\ovln{\psi}(z) = \inf \bigl\{ \ovln{\psi}(w) \such w > z \bigr\},
\]
and for all $z \in (0, 1]$,
\[
\ovln{\psi}(z) = \sup \bigl\{ \ovln{\psi}(w) \such w < z \bigr\}.
\]
We prove just the first of these equations, the second being similar.  Let
$z \in [0, 1)$.  Then
\begin{align}
\inf \bigl\{ \ovln{\psi}(w) \such w > z \bigr\}     &
=
\inf \Bigl\{ 
\inf \{ \psi(q) \such \text{rational } q \geq w \} 
\such w > z
\Bigr\} 
\lbl{eq:cts-ext-4}    \\
&
=
\inf \bigcup_{w > z} 
\{ \psi(q) \such \text{rational } q \geq w \}    
\nonumber       \\
&
=
\inf \{ \psi(q) \such \text{rational } q > z \}
\nonumber       \\
&
=
\inf \{ \psi(q) \such \text{rational } q \geq z \}
\lbl{eq:cts-ext-5}    \\
&
=
\ovln{\psi}(z),
\lbl{eq:cts-ext-6}
\end{align}
where~\eqref{eq:cts-ext-4} and~\eqref{eq:cts-ext-6} are by definition of
$\ovln{\psi}$, and~\eqref{eq:cts-ext-5} follows from~\eqref{eq:cts-ext-1}
and~\eqref{eq:cts-ext-3}. 
\end{proof}

We now prove our characterization theorem for strictly increasing
unweighted means on $(0, \infty)$ (Figure~\ref{fig:mean-maps}). 

\begin{figure}
\normalsize
\[
\begin{array}{c}
\xymatrix@M+.5ex@C+1em{
(0, \infty)^n \ar[r]^M    &(0, \infty)      \\
K^n \ar@{^{(}->}[u] \ar[r]^{M|_K} \ar@/_1pc/[d]_{\phi^n}      &
K \ar@{^{(}->}[u] \ar@/_1pc/[d]_{\phi}        \\
[0, 1]^n \ar@/_1pc/[u]_{\psi^n} \ar[r]_{M_1}   &
[0, 1]  \ar@/_1pc/[u]_{\psi}    &
[0, 1] \cap \Q \ar@{_{(}->}[l] \ar@/_1pc/[lu]_\psi
}
\end{array}
\]
\caption{Maps involved in the proof of Theorem~\ref{thm:str-inc}.}
\lbl{fig:mean-maps}
\end{figure}

\begin{thm}
\lbl{thm:str-inc}
\index{power mean!characterization of!unweighted on $(0, \infty)$}
Let $\bigl( M \from (0, \infty)^n \to (0, \infty) \bigr)_{n \geq 1}$ be a
sequence of functions.  The following are equivalent:
\begin{enumerate}
\item 
\lbl{part:str-inc-condns}
$M$ is symmetric, strictly increasing, decomposable, and homogeneous;

\item
\lbl{part:str-inc-form}
$M = M_t$ for some $t \in (-\infty, \infty)$.
\end{enumerate}
\end{thm}

\begin{proof}
\bref{part:str-inc-form} implies~\bref{part:str-inc-condns} by Examples
\ref{egs:u-mns-sym}, \ref{egs:u-mns-inc-pwr}\bref{eg:u-mns-inc-pwr-fin},
\ref{egs:u-mns-decomp}\bref{eg:u-mns-decomp-pwr}, and~\ref{egs:u-mns-hgs}.

Now assume~\bref{part:str-inc-condns}.  The main part of the proof is to
show that $M$ restricts to a quasiarithmetic mean on each nontrivial closed
bounded subinterval $K \subset (0, \infty)$.  Let $K$ be such an interval.

First note that for each $n \geq 1$, the function $M \from (0, \infty)^n
\to (0, \infty)$ restricts to a function $M|_K \from K^n \to K$.  Indeed,
$M$ is consistent by Lemma~\ref{lemma:u-mn-dhc}, and increasing, so for all
$x_1, \ldots, x_n \in K$,
\[
\min \{ x_1, \ldots, x_n \}
\leq
M(x_1, \ldots, x_n)
\leq
\max \{ x_1, \ldots, x_n \},
\]
giving $M(x_1, \ldots, x_n) \in K$. 

Next we show that the sequence of functions $(M|_K \from K^n \to K)_{n \geq
  1}$ is a quasiarithmetic mean.  This sequence is symmetric, consistent,
strictly increasing and decomposable, since $M$ is.  Let $\psi \from [0, 1]
\cap \Q \to K$ be the function defined in Lemma~\ref{lemma:psi-rat}.  We
will extend $\psi$ to a continuous function $[0, 1] \to K$ using
Lemma~\ref{lemma:cts-ext}.  For this, we have to verify the hypotheses of
that lemma: that $\psi$ is strictly increasing (which is immediate from
Lemma~\ref{lemma:psi-rat}\bref{part:psi-rat-inc}) and that $\psi$ satisfies
equations~\eqref{eq:cts-ext-1} and~\eqref{eq:cts-ext-2}.  We prove
only~\eqref{eq:cts-ext-1}, the proof of~\eqref{eq:cts-ext-2} being 
similar.

Let $z \in [0, 1)$.  Put
\[
u
=
\sup \{ \psi(p) \such \text{rational } p \leq z \},
\qquad
v
=
\inf \{ \psi(q) \such \text{rational } q > z \}.
\]
We have to show that $u = v$.  Since $\psi$ is increasing, $u \leq v$.  It
remains to show that $u \geq v$.

Let $C > 1$.  We have $v \in K \subset (0, \infty)$, so $v > 0$, giving $Cv
> v$.  By definition of $v$, we can therefore choose a rational $q \in (z,
1]$ such that $\psi(q) \leq Cv$.  We can then choose a rational $p \in [0,
  z]$ such that $\tfrac{1}{2}(p + q) > z$.  By definition of $u$, we have
$\psi(p) \leq u \leq Cu$.  Now
\begin{align}
C M(u, v)       &
=
M(Cu, Cv)       
\lbl{eq:pfd-1}        \\
&
\geq
M(\psi(p), \psi(q))   
\lbl{eq:pfd-2}        \\
&
=
\psi\bigl(\tfrac{1}{2}(p + q)\bigr)       
\lbl{eq:pfd-3}        \\
&
\geq
v,
\lbl{eq:pfd-4}
\end{align}
where~\eqref{eq:pfd-1} is by homogeneity, \eqref{eq:pfd-2} is because $M$
is increasing, \eqref{eq:pfd-3} follows from
Lemma~\ref{lemma:psi-rat}\bref{part:psi-rat-eq}, and~\eqref{eq:pfd-4} is
because $\tfrac{1}{2}(p + q) \in (z, 1] \cap \Q$.  Hence $CM(u, v) \geq v$
for all $C > 1$, giving $M(u, v) \geq v$.  But then
\[
v = M(v, v) \geq M(u, v) \geq v
\]
(using consistency), so $M(v, v) = M(u, v)$.  Since $M$ is \emph{strictly}
increasing, this forces $u = v$, proving equation~\eqref{eq:cts-ext-1} in
Lemma~\ref{lemma:cts-ext}.

We have now shown that the function $\psi \from [0, 1] \cap \Q \to K$
satisfies the hypotheses of Lemma~\ref{lemma:cts-ext}.  By that lemma,
$\psi$ extends uniquely to a continuous, strictly increasing function $[0,
  1] \to K$, which we also denote by $\psi$.  The extended function $\psi$ is
endpoint-preserving by Lemma~\ref{lemma:psi-rat}\bref{part:psi-rat-ends},
and is therefore a homeomorphism.  Let $\phi \from K \to [0, 1]$ be its
inverse.

We will prove that $M|_K = M_{\phi}$, or equivalently that 
\begin{equation}
\lbl{eq:pfd3-0}
M\bigl(
\psi(z_1), \ldots, \psi(z_n)
\bigr)
=
\psi\bigl(
\tfrac{1}{n} (z_1 + \cdots + z_n)
\bigr)
\end{equation}
($z_i \in [0, 1]$).  Indeed, for all $z_1, \ldots, z_n \in [0, 1]$,
\begin{align}
M\bigl(
\psi(z_1), \ldots, \psi(z_n)
\bigr)  
&
\geq
\sup \Bigl\{
M\bigl(
\psi(q_1), \ldots, \psi(q_n)
\bigr)  \such
\text{rational } q_i \leq z_i
\Bigr\} 
\lbl{eq:pfd3-1}       \\
&
=
\sup \Bigl\{
\psi \bigl( \tfrac{1}{n} (q_1 + \cdots + q_n) \bigr)
\such
\text{rational } q_i \leq z_i
\Bigr\} 
\lbl{eq:pfd3-2}       \\
&
=
\psi \Bigl( 
\sup \Bigl\{
\tfrac{1}{n} (q_1 + \cdots + q_n)
\such
\text{rational } q_i \leq z_i
\Bigr\}
\Bigr)
\lbl{eq:pfd3-3}       \\
&
=
\psi\bigl(
\tfrac{1}{n} (z_1 + \cdots + z_n)
\bigr),
\lbl{eq:pfd3-4}
\end{align}
where~\eqref{eq:pfd3-1} holds because $M$ and $\psi$ are increasing,
\eqref{eq:pfd3-2} follows from
Lemma~\ref{lemma:psi-rat}\bref{part:psi-rat-eq}, equation~\eqref{eq:pfd3-3}
is a consequence of $\psi \from [0, 1] \to K$ being a strictly increasing
bijection, and~\eqref{eq:pfd3-4} is elementary.  So
\[
M \bigl( \psi(z_1), \ldots, \psi(z_n) \bigr)
\geq
\psi \bigl( \tfrac{1}{n} (z_1 + \cdots + z_n) \bigr).
\]
The same argument with the inequalities reversed and the suprema changed to
infima proves the opposite inequality, and equation~\eqref{eq:pfd3-0}
follows.  So $M|_K = M_\phi$, as claimed.

We have now shown that $M$ restricts to a quasiarithmetic mean on each
nontrivial closed bounded subinterval of $(0, \infty)$.  Hence by
Lemma~\ref{lemma:u-mn-extension}, $M$ itself is a quasiarithmetic mean.
But $M$ is homogeneous, so Theorem~\ref{thm:qa-pwr} now implies that $M =
M_t$ for some $t \in (-\infty, \infty)$.
\end{proof}

From this theorem about means on $(0, \infty)$, we deduce a 
theorem about means on $[0, \infty)$.

\begin{thm}
\lbl{thm:zero-str-inc}
\index{power mean!characterization of!unweighted on $[0, \infty)$}
Let $\bigl( M \from [0, \infty)^n \to [0, \infty) \bigr)_{n \geq 1}$ be a
sequence of functions.  The following are equivalent:
\begin{enumerate}
\item 
\lbl{part:zero-str-inc-condns}
$M$ is symmetric, strictly increasing, decomposable, and homogeneous;

\item
\lbl{part:zero-str-inc-form}
$M = M_t$ for some $t \in (0, \infty)$.
\end{enumerate}
\end{thm}

\begin{proof}
Certainly \bref{part:zero-str-inc-form}
implies~\bref{part:zero-str-inc-condns}, by Examples
\ref{egs:u-mns-sym}--\ref{egs:u-mns-hgs}.  Now
assume~\bref{part:zero-str-inc-condns}.  If $\vc{0} \neq \vc{x} \in [0,
\infty)^n$ then $M(\vc{x}) > M(\vc{0}) \geq 0$, so $M(\vc{x}) > 0$.
Hence $M$ restricts to a sequence of functions
\[
M|_{(0, \infty)} \from (0, \infty)^n \to (0, \infty).
\]
By Theorem~\ref{thm:str-inc}, $M|_{(0, \infty)} = M_t$ for some $t \in
(-\infty, \infty)$.  

To show that $t > 0$, note that 
\[
0 
< 
M(1, 0) 
\leq 
\inf_{\delta > 0} M(1, \delta)
= 
\inf_{\delta > 0} M_t(1, \delta) 
= 
M_t(1, 0),
\]
where in the last step we used the fact that $M_t$ is continuous
(Lemma~\ref{lemma:pwr-mns-cts-x}) and increasing.  Hence $M_t(1, 0) > 0$.
But $M_t(1, 0) = 0$ for all $t \in [-\infty, 0]$
(Definition~\ref{defn:pwr-mn}), so $t \in (0, \infty)$.

We now have to show that the equality $M(\vc{x}) = M_t(\vc{x})$, so far
proved to hold for all $\vc{x} \in (0, \infty)^n$, holds for all
$\vc{x} \in [0, \infty)^n$.

First I claim that 
\begin{equation}
\lbl{eq:zsi-0}
M(1, 0) = M_t(1, 0).  
\end{equation}
To prove this, we evaluate $M(2, 1, 0)$ in two ways.  Put $a = M(1, 0) >
0$.  Since $M$ is decomposable,
\begin{align*}
M(2, 1, 0)      &
=
M(2, a, a)      \\
&
=
M_t(2, a, a)    \\
&
=
\left( \tfrac{1}{3} (2^t + 2a^t) \right)^{1/t}.
\end{align*}
On the other hand, $M(2, 0) = 2a$ by homogeneity, so, using decomposability
again,
\begin{align*}
M(2, 1, 0)      &
=
M(1, 2, 0)      \\
&
=
M(1, 2a, 2a)    \\
&
=
M_t(1, 2a, 2a)  \\
&
=
\bigl( \tfrac{1}{3} (1 + 2^{t + 1} a^t) \bigr)^{1/t}.
\end{align*}
Equating these two expressions for $M(2, 1, 0)$ gives $a = (1/2)^{1/t}$, or
equivalently, $M(1, 0) = M_t(1, 0)$, as claimed.

Now we prove that $M(\vc{x}) = M_t(\vc{x})$ for all $n \geq 1$ and $\vc{x}
\in [0, \infty)^n$.  By symmetry, it suffices to prove this when
\[
\vc{x} = (x_1, \ldots, x_m, k \mc 0)
\]
for some $k, m \geq 0$ and $x_1, \ldots, x_m > 0$.  The proof is by
induction on $k$ for all $m$ simultaneously.  We already have the result
for $k = 0$.  Suppose now that $k \geq 1$ and the result holds for $k - 1$.
If $m = 0$ then the result is trivial (by homogeneity), so suppose
that $m \geq 1$.  We have
\begin{align}
&
M(x_1, \ldots, x_m, k \mc 0)    
\nonumber       \\
&
=
M\bigl(x_1, \ldots, x_{m - 1}, x_m, 0, (k - 1) \mc 0\bigr)&
\nonumber       \\
&
=
M\bigl(x_1, \ldots, x_{m - 1}, x_m M(1, 0), x_m M(1, 0), (k - 1) \mc 0\bigr)&
\lbl{eq:zsi-1}        \\
&
=
M\bigl(x_1, \ldots, x_{m - 1}, 
x_m M_t(1, 0), x_m M_t(1, 0), (k - 1) \mc 0\bigr)&
\lbl{eq:zsi-2}        \\
&
=
M_t\bigl(x_1, \ldots, x_{m - 1}, 
x_m M_t(1, 0), x_m M_t(1, 0), (k - 1) \mc 0\bigr)&
\lbl{eq:zsi-3}        \\
&
=
M_t\bigl(x_1, \ldots, x_{m - 1}, x_m, 0, (k - 1) \mc 0\bigr)  &
\lbl{eq:zsi-4}        \\
&
=
M_t(x_1, \ldots, x_m, k \mc 0),
\nonumber
\end{align}
where equations~\eqref{eq:zsi-1} and~\eqref{eq:zsi-4} are by
decomposability and homogeneity of $M$ and $M_t$, equation~\eqref{eq:zsi-2}
follows from equation~\eqref{eq:zsi-0}, and equation~\eqref{eq:zsi-3} is by
inductive hypothesis.  This completes the proof.
\end{proof}

\section{Increasing homogeneous means}
\lbl{sec:ih-mns}
\index{mean!increasing}
\index{increasing!unweighted mean}

The extremal cases $M_{\pm\infty}$ of the power means are neither strictly
increasing nor quasiarithmetic.  Both of these factors put $M_{\pm \infty}$
outside the ambit of many characterizations of means.  However, we will
prove characterization theorems that include $M_{\pm\infty}$, mostly so as
not to exclude the important Berger--Parker%
\index{Berger--Parker index}
index $D_\infty$ (Example~\ref{egs:hill}\bref{eg:hill-bp}) from
a later characterization theorem for diversity measures.  

We have already characterized the strictly increasing means, so our task
now is to characterize the means $M$ that are increasing but not strictly
so.  Assuming symmetry, we have
\[
M(x_1, \ldots, x_m, u) = M(x_1, \ldots, x_m, v)
\]
for some $x_i$, $u$, $v$ with $u \neq v$.  Our aim is to deduce from this
equation, and the usual other hypotheses on $M$, that $M$ is equal to
either $M_\infty = \max$ or $M_{-\infty} = \min$.  

\begin{lemma}
\lbl{lemma:u-mn-mc}
Let $I$ be a real interval.  Let $(M \from I^n \to I)_{n \geq 1}$ be a
symmetric, decomposable sequence of functions.  Let $m \geq 1$ and $x_1,
\ldots, x_m, u, v \in I$ with
\[
M(x_1, \ldots, x_m, u)
=
M(x_1, \ldots, x_m, v).
\]
Then
\[
M(x_1, \ldots, x_m, n \mc u)
=
M(x_1, \ldots, x_m, n \mc v)
\]
for all $n \geq 0$.
\end{lemma}

\begin{proof}
This is trivial for $n = 0$.  Suppose inductively that $n \geq 0$ and the
result holds for $n$.  Since $M$ is decomposable, it is modular
(Definition~\ref{defn:u-mod}).  Now
\begin{align}
M\bigl(x_1, \ldots, x_m, (n + 1) \mc u\bigr)    &
=
M(x_1, \ldots, x_m, n \mc u, u)
\nonumber       \\
&
=
M(x_1, \ldots, x_m, n \mc v, u)
\lbl{eq:u-mn-mc-2}    \\
&
=
M(x_1, \ldots, x_m, u, n \mc v)
\lbl{eq:u-mn-mc-3}    \\  
&
=
M(x_1, \ldots, x_m, v, n \mc v)
\lbl{eq:u-mn-mc-4}    \\  
&
=
M\bigl(x_1, \ldots, x_m, (n + 1) \mc v\bigr),
\nonumber
\end{align}
where~\eqref{eq:u-mn-mc-2} and~\eqref{eq:u-mn-mc-4} use modularity
and~\eqref{eq:u-mn-mc-3} is by symmetry.
\end{proof}

We deduce:

\begin{lemma}
\lbl{lemma:nonstrict-reduce}
Let $I$ be an interval and let $(M \from I^n \to I)_{n \geq 1}$ be a
sequence of functions that is symmetric, consistent, decomposable, and
increasing but not strictly so.  Then there exist $x, u, v \in I$ such that
$u \neq v$ but $M(x, u) = M(x, v)$.
\end{lemma}

\begin{proof}
By symmetry, there exist $n \geq 0$ and $x_1, \ldots, x_n, u, v \in I$ such
that $u \neq v$ and
\[
M(x_1, \ldots, x_n, u) = M(x_1, \ldots, x_n, v).
\]
By consistency, $n \geq 1$.  Writing $x = M(x_1, \ldots, x_n)$, we have
\[
M(n \mc x, u) = M(n \mc x, v)
\]
by decomposability.  Now
\[
M(x, u)
=
M(n \mc x, n \mc u)
=
M(n \mc x, n \mc v)
=
M(x, v),
\]
where the first and last equalities follow from Lemma~\ref{lemma:u-mn-rep}
and the second from Lemma~\ref{lemma:u-mn-mc}.
\end{proof}

Our next lemma contains the main substance of the argument that if $M$ is
increasing but not strictly so, then $M = M_{\pm\infty}$.

\begin{lemma}
\lbl{lemma:substance}
Let $\bigl(M \from (0, \infty)^n \to (0, \infty)\bigr)_{n \geq 1}$ be a
symmetric, increasing, decomposable, homogeneous sequence of functions.  If
there exist $x \in (0, \infty)$ and distinct $u, v \geq x$ such that $M(x,
u) = M(x, v)$, then there exist $a < b$ in $(0, \infty)$ such that $M(a, b)
= a$.
\end{lemma}

\begin{proof}
Take $x$, $u$ and $v$ as described.  We may assume without loss of
generality that $u < v$ and (by homogeneity) that $x = 1$.  We may
therefore choose a real number $C > 1$ and an integer $N \geq 1$ such that
$u \leq C^N < C^{N + 1} \leq v$.  We will prove that $M(1, C^N) = 1$.

Since $M$ is increasing, the hypothesis that $M(1, u) = M(1, v)$ implies
that
\[
M\bigl(1, C^N\bigr) = M\bigl(1, C^{N + 1}\bigr).  
\]
It follows from Lemma~\ref{lemma:u-mn-mc} that
\[
M\bigl(1, r \mc C^N\bigr) = M\bigl(1, r \mc C^{N + 1}\bigr)
\]
for all $r \geq 0$, then by homogeneity that
\begin{equation}
\lbl{eq:C-power-s}
M\bigl(C^s, r \mc C^{s + N}\bigr) 
=
M\bigl(C^s, r \mc C^{s + N + 1}\bigr) 
\end{equation}
for all $r, s \geq 0$.  

I claim that 
\begin{equation}
\lbl{eq:C-power-bound}
M\bigl(1, r \mc C^k\bigr) \leq C^N
\end{equation}
for all $k, r \geq 0$.  To prove this, first note that $M$ is consistent, by
Lemma~\ref{lemma:u-mn-dhc}.  When $k \leq N$, we have $C^k \leq C^N$, 
so~\eqref{eq:C-power-bound} holds because $M$ is consistent and increasing.
Now let $k \geq N$ and suppose inductively that~\eqref{eq:C-power-bound}
holds for $k$, for all $r$.  Then for all $r$,
\begin{align}
M\bigl(1, r \mc C^{k + 1}\bigr)   &
=
M\bigl(1, 1, 2r \mc C^{k + 1}\bigr)       
\lbl{eq:substance-1}  \\
&
\leq
M\bigl(1, C^{k - N}, 2r \mc C^{k + 1}\bigr)      
\lbl{eq:substance-2}  \\
&
=
M\bigl(1, C^{k - N}, 2r \mc C^k\bigr)     
\lbl{eq:substance-3}  \\
&
\leq
M\bigl(1, (2r + 1) \mc C^k\bigr)  
\lbl{eq:substance-4}  \\
&
\leq 
C^N,
\lbl{eq:substance-5}  
\end{align}
where equation~\eqref{eq:substance-1} follows from
Lemma~\ref{lemma:u-mn-rep}, inequality~\eqref{eq:substance-2} from the fact
that $C^{k - N} \geq 1$, equation~\eqref{eq:substance-3}
from~\eqref{eq:C-power-s} (with $s = k - N$) and decomposability,
inequality~\eqref{eq:substance-4} from the fact that $C^{k - N} \leq C^k$,
and inequality~\eqref{eq:substance-5} from the inductive hypothesis.  This
completes the induction and the proof of the claimed
inequality~\eqref{eq:C-power-bound}.

It follows from~\eqref{eq:C-power-bound} that
\begin{equation}
\lbl{eq:C-poly-bound}
M\bigl(1, C^{k_1}, \ldots, C^{k_r}\bigr) \leq C^N
\end{equation}
for all $r, k_1, \ldots, k_r \geq 0$.  Indeed, since $M$ is increasing, the
left-hand side of~\eqref{eq:C-poly-bound} is at most $M\bigl(1, r \mc
C^{\max k_i}\bigr)$, which by~\eqref{eq:C-power-bound} is at most $C^N$.

We finish by using the tensor%
\index{tensor power trick} 
power trick (Tao~\cite{TaoSR}, Section~1.9).  For all $r \geq 1$,
\[
M\bigl(1, C^N\bigr) 
=
M\Bigl( \bigl(1, C^N\bigr)^{\otimes r} \Bigr)^{1/r}
\]
by Lemma~\ref{lemma:u-mn-mult}.  Expanding the tensor power 
\[
\bigl(1, C^N\bigr)^{\otimes r} 
=
\bigl(1, C^N\bigr) \otimes \cdots \otimes \bigl(1, C^N\bigr)
\]
gives 
\begin{align*}
M\bigl(1, C^N\bigr)       &
=
M\Biggl( 
1, \binom{r}{1} \mc C^N, \ldots, \binom{r}{r - 1} \mc C^{(r - 1)N}, C^{rN}
\Biggr)^{1/r}   \\
&
\leq
C^{N/r}
\end{align*}
for all $r \geq 1$, by symmetry of $M$ and
inequality~\eqref{eq:C-poly-bound}.  Letting $r \to \infty$, this proves
that $M(1, C^N) \leq 1$.  But also
\[
M\bigl(1, C^N\bigr) \geq M(1, 1) = 1,
\]
since $M$ is increasing and consistent.  Hence $M(1, C^N) = 1$ with $C^N >
1$, completing the proof.
\end{proof}

The lemma just proved states that under certain hypotheses, $M(a, b) =
\min\{a, b\}$ for \emph{some} distinct numbers $a$ and $b$.  The next lemma
tells us that in that case, $M(a, b) = \min\{a, b\}$ for
\emph{all} $a$ and $b$. Better still, $M(x_1, \ldots, x_n) = \min x_i$ for
all $n \geq 1$ and all $x_i$.

\begin{lemma}
\lbl{lemma:min-two-all}
Let $\bigl(M \from (0, \infty)^n \to (0, \infty)\bigr)_{n \geq 1}$ be a
symmetric, increasing, decomposable, homogeneous sequence of functions.  If
$M(a, b) = a$ for some $0 < a < b$ then $M = M_{-\infty}$.
\end{lemma}

\begin{proof}
By homogeneity, we may assume that $a = 1$, so that $b > 1$ with $M(1, b) =
1$.

First I claim that $M(1, b^r) = 1$ for all $r \geq 0$.  By
Lemma~\ref{lemma:u-mn-dhc}, $M$ is consistent, which gives the case $r =
0$.  Now suppose inductively that $r \geq 1$ with $M(1, b^{r - 1}) = 1$.
By Lemma~\ref{lemma:u-mn-rep} and the fact that $M$ is increasing,
\begin{equation}
\lbl{eq:mta-A}
M(1, b^r)       
=
M(1, 1, b^r, b^r)   
\leq
M(1, b, b^r, b^r).
\end{equation}
By inductive hypothesis and homogeneity, $M(b, b^r) = b$.  Hence, using
decomposability twice,
\begin{equation}
\lbl{eq:mta-B}
M(1, b, b^r, b^r) = M(1, b, b, b^r) = M(1, b, b, b).
\end{equation}
But $M(1, b) = 1 = M(1, 1)$ by hypothesis and consistency, so by
Lemma~\ref{lemma:u-mn-mc}, 
\begin{equation}
\lbl{eq:mta-C}
M(1, b, b, b) = M(1, 1, 1, 1) = 1.
\end{equation}
Putting together \eqref{eq:mta-A}, \eqref{eq:mta-B} and~\eqref{eq:mta-C}
gives $M(1, b^r) \leq 1$.  But also $M(1, b^r) \geq M(1, 1) = 1$, so 
$M(1, b^r) = 1$, completing the induction and proving the claim.

Next I claim that
\begin{equation}
\lbl{eq:mta-1}
M(x, y) = \min\{x, y\}
\end{equation}
for all $x, y \in (0, \infty)$.  By homogeneity, it is enough to prove this
when $x = 1 \leq y$; then our task is to prove that $M(1, y) = 1$ for all
$y \geq 1$.  Certainly
\[
M(1, y) \geq M(1, 1) = 1.
\]
On the other hand, we can choose $r \geq 0$ such that $y \leq b^r$, and
then
\[
M(1, y) \leq M(1, b^r) = 1
\]
by the claim above.  Hence $M(1, y) = 1$, as claimed.

Finally, we prove that 
\[
M(x_1, \ldots, x_n) = \min\{x_1, \ldots, x_n\}
\]
for all $n \geq 1$ and $\vc{x} \in (0, \infty)^n$.  By symmetry, we may
assume that $x_1 = \min_i x_i$.  Then $M(x_1, x_i) = x_1$ for all $i$, by
equation~\eqref{eq:mta-1}.  Hence
\begin{align*}
M(x_1, x_2, x_3, x_4, \ldots, x_n)     &
=
M(x_1, x_1, x_3, x_4, \ldots, x_n)      \\
&
=
M(x_1, x_1, x_1, x_4, \ldots, x_n)      \\
&
=
\ldots  \\
&
=
M(x_1, x_1, x_1, x_1, \ldots, x_1)      \\
&
=
x_1,
\end{align*}
where the first equality follows from decomposability and the fact that
$M(x_1, x_2) = x_1$, the second from decomposability and the fact that
$M(x_1, x_3) = x_1$, and so on, while the last equality follows from
the consistency of $M$.  
\end{proof}

So far, we have focused on $M_{-\infty} = \min$ rather than $M_\infty
= \max$.  Of course, similar results hold for $M_\infty$ by reversing all
the inequalities, but the situation is handled most systematically by the following duality%
\index{duality for power means}%
\index{power mean!duality for} 
construction (Figure~\ref{fig:comm-dual}).
\begin{figure}
\[
\normalsize
\xymatrix{
(0, \infty)^n \ar[r]^M \ar[d]_{\rho^n}  &
(0, \infty) \ar[d]^\rho \\
(0, \infty)^n \ar[r]_{\ovln{M}} &
(0, \infty)
}
\]
\caption{Commutative diagram showing the relationship between a mean $M$
  and its dual $\ovln{M}$, where $\rho \from x \mapsto 1/x$ is the
  reciprocal map.}
\lbl{fig:comm-dual}
\end{figure}
Given a sequence of functions
$\bigl(M\from (0, \infty)^n \to (0, \infty)\bigr)_{n \geq 1}$, define
another such sequence $\ovln{M}$\ntn{dumn} by
\[
\ovln{M}(x_1, \ldots, x_n) 
=
\frac{1}{M\bigl(\frac{1}{x_1}, \ldots, \frac{1}{x_n}\bigr)}
\]
($x_1, \ldots, x_n \in (0, \infty)$).  For example,
equation~\eqref{eq:mn-duality} (p.~\pageref{eq:mn-duality}) implies that
\[
\ovln{M_t} = M_{-t}
\]
for all $t \in [-\infty, \infty]$.  Evidently $\ovln{\ovln{M}} = M$ for any
$M$.  

We will use without mention the following lemma, whose proof is trivial:
\begin{lemma}
Let $(M\from (0, \infty)^n \to (0, \infty))_{n \geq 1}$ be a sequence of
functions.  Then $M$ is symmetric, consistent, increasing, strictly
increasing, decomposable or homogeneous (respectively) if and only if
$\ovln{M}$ is.
\qed
\end{lemma}

The next result uses this duality.

\begin{propn}
\lbl{propn:min-max}
Let $\bigl(M \from (0, \infty)^n \to (0, \infty)\bigr)_{n \geq 1}$ be a
sequence of functions that is symmetric, decomposable, homogeneous, and
increasing but not strictly so.  Then $M = M_{\pm \infty}$.
\end{propn}

\begin{proof}
By Lemma~\ref{lemma:u-mn-dhc}, $M$ is consistent.  By
Lemma~\ref{lemma:nonstrict-reduce}, we can choose $x, u, v \in (0, \infty)$
with $u \neq v$ but $M(x, u) = M(x, v)$.  Without loss of generality, $u <
v$.  There are now three cases to consider, and we prove that in each,
$M(a, b) \in \{a, b\}$ for some $a < b$ in $(0, \infty)$.

\textbf{Case 1: $x \leq u < v$.}  By Lemma~\ref{lemma:substance}, $M(a,
b) = a$ for some $a < b$ in $(0, \infty)$.

\textbf{Case 2: $u < x < v$.}  We have 
\[
M(x, v) \geq M(x, x) = x, 
\]
but also 
\[
M(x, v) = M(x, u) \leq M(x, x) = x,
\]
so $M(x, v) = x$. Putting $a = x$ and $b = v$ gives $M(a, b) = a$ with $a <
b$. 

\textbf{Case 3: $u < v \leq x$.}  Then $1/x \leq 1/v < 1/u$ with
$\ovln{M}(1/x, 1/v) = \ovln{M}(1/x, 1/u)$.  Hence by
Lemma~\ref{lemma:substance} applied to $\ovln{M}$, there exist $B < A$ in
$(0, \infty)$ such that $\ovln{M}(B, A) = B$.  Putting $a = 1/A$ and $b =
1/B$ gives $M(b, a) = b$ with $a < b$.

So in all cases, we can choose $a < b$ in $(0, \infty)$ such that $M(a, b)
\in \{a, b\}$.  If $M(a, b) = a$ then $M = M_{-\infty}$ by
Lemma~\ref{lemma:min-two-all}.  Otherwise, $M(a, b) = b$, so $\ovln{M}(1/a,
1/b) = 1/b$ with $1/b < 1/a$.  Applying Lemma~\ref{lemma:min-two-all} to
$\ovln{M}$ gives $\ovln{M} = M_{-\infty}$, or equivalently, $M = M_\infty$.
\end{proof}

This brings us to our third characterization theorem for unweighted power
means, this time including the extremal cases $M_{\pm\infty}$.  

\begin{thm}
\lbl{thm:inc}
\index{power mean!characterization of!unweighted on $(0, \infty)$}
Let $\bigl(M \from (0, \infty)^n \to (0, \infty)\bigr)_{n \geq 1}$ be a
sequence of functions.  The following are equivalent:
\begin{enumerate}
\item
\lbl{part:inc-condns}
$M$ is symmetric, increasing, decomposable, and homogeneous;

\item
\lbl{part:inc-form}
$M = M_t$ for some $t \in [-\infty, \infty]$.
\end{enumerate}
\end{thm}

\begin{proof}
\bref{part:inc-form} implies~\bref{part:inc-condns} by Examples
\ref{egs:u-mns-sym}--\ref{egs:u-mns-hgs}.  Now
assume~\bref{part:inc-condns}.  If $M$ is strictly increasing then $M =
M_t$ for some $t \in (-\infty, \infty)$, by Theorem~\ref{thm:str-inc}.
Otherwise, $M = M_{\pm\infty}$ by Proposition~\ref{propn:min-max}.
\end{proof}

Our fourth and final characterization theorem for unweighted power means
captures all the power means $M_t$, including $M_{\pm\infty}$, on the
larger interval $[0, \infty)$.  It is an easy consequence of
  Theorem~\ref{thm:inc}, but comes at the cost of a significant extra
  hypothesis: that for each $n \geq 1$, the function
  $M \from [0, \infty)^n \to [0, \infty)$ is continuous.

We need one lemma in preparation:

\begin{lemma}
\lbl{lemma:u-mn-idhc}
Let $\bigl(M\from [0, \infty)^n \to [0, \infty)\bigr)_{n \geq 1}$ be a
sequence of functions, none of which is identically zero.  If $M$ is
increasing, decomposable, and homogeneous, then $M$ is consistent.
\end{lemma}

\begin{proof}
Let $n \geq 1$.  By the same argument as in Lemma~\ref{lemma:u-mn-dhc}, $M(n
\mc 1) \in \{0, 1\}$.  Suppose for a contradiction that $M(n \mc 1) = 0$.
Then by homogeneity, $M(n \mc x) = 0$ for all $x \in [0, \infty)$.  For all
  $\vc{x} \in [0, \infty)^n$, 
\[
M(\vc{x})
\leq
M\bigl(n \mc \max_i x_i\bigr)
=
0
\]
since $M$ is increasing.  Hence $M \from [0, \infty)^n \to [0, \infty)$ is
identically zero, contrary to our hypothesis.  Thus, $M(n \mc
1) = 1$.  It follows by homogeneity that $M$ is consistent.
\end{proof}

\begin{thm}
\lbl{thm:cts-inc}
\index{power mean!characterization of!unweighted on $[0, \infty)$}
Let $\bigl(M \from [0, \infty)^n \to [0, \infty)\bigr)_{n \geq 1}$ be a
sequence of functions.  The following are equivalent:
\begin{enumerate}
\item
\lbl{part:cts-inc-condns}
$M$ is symmetric, increasing, decomposable, homogeneous and continuous, and
none of the functions $M \from [0, \infty)^n \to [0, \infty)$ is
identically zero; 

\item
\lbl{part:cts-inc-form}
$M = M_t$ for some $t \in [-\infty, \infty]$.
\end{enumerate}
\end{thm}

\begin{proof}
It is straightforward that \bref{part:cts-inc-form}
implies~\bref{part:cts-inc-condns}, with the continuity coming from
Lemma~\ref{lemma:pwr-mns-cts-x}.  Now assume~\bref{part:cts-inc-condns}.

By Lemma~\ref{lemma:u-mn-idhc}, $M$ is consistent.  Since $M$ is also
increasing, $M(\vc{x}) \geq \min_i x_i > 0$ for all $\vc{x} \in (0,
\infty)^n$.  Hence $M$ restricts to a sequence of functions
\[
\bigl( 
M|_{(0, \infty)} \from (0, \infty)^n \to (0, \infty) 
\bigr)_{n \geq 1}.
\]
The functions $M|_{(0, \infty)}$ are symmetric, increasing, decomposable
and homogeneous, so by Theorem~\ref{thm:inc}, there exists $t \in [-\infty,
\infty]$ such that $M = M_t$ on $(0, \infty)$.  For each $n \geq 1$,
the functions
\[
M, M_t \from [0, \infty)^n \to [0, \infty)
\]
are continuous and are equal on the dense subset $(0, \infty)^n$, so they
are equal everywhere. 
\end{proof}

\begin{remarks}
\begin{enumerate}
\item
The continuity condition in Theorem~\ref{thm:cts-inc} cannot be dropped.
Indeed, take any $t \in (0, \infty]$ and define a function $M \from [0,
\infty)^n \to [0, \infty)$ for each $n \geq 1$ by
\[
M(\vc{x})
=
\begin{cases}
M_t(\vc{x})     &\text{if } \vc{x} \in (0, \infty)^n,   \\
0               &\text{otherwise.}
\end{cases}
\]
Then $M$ satisfies all the conditions of
Theorem~\ref{thm:cts-inc}\bref{part:cts-inc-condns} apart from continuity,
and is not a power mean.

\item
The hypothesis that none of the functions $M \from [0, \infty)^n \to [0,
    \infty)$ is identically zero cannot be dropped either.  Indeed, take
    any $t \in [-\infty, \infty]$ and any integer $k \geq 1$, and for
    $\vc{x} \in \R^n$, define
\[
M(\vc{x})
=
\begin{cases}
M_t(\vc{x})     &\text{if } n \leq k,   \\
0               &\text{if } n > k.
\end{cases}
\]
Then $M$ satisfies all the other conditions of
Theorem~\ref{thm:cts-inc}\bref{part:cts-inc-condns}, and is not a power
mean.
\end{enumerate}
\end{remarks}

\section{Weighted means}
\lbl{sec:w-mns}
\index{mean!weighted}

So far, this chapter has been directed towards characterization theorems
for unweighted means (Theorems~\ref{thm:str-inc}, \ref{thm:zero-str-inc},
\ref{thm:inc} and~\ref{thm:cts-inc}, summarized in
Table~\ref{table:mn-thms}).  But we can now deduce characterization
theorems for weighted means with comparatively little work.

We do this in three steps.  First, we record some elementary implications
between properties that a notion of weighted mean may or may not
satisfy, and between conditions on weighted and unweighted means.  Second,
we create a method for converting characterization theorems for unweighted
means into characterization theorems for weighted means.  Third, we apply
that method to the theorems just mentioned.  This produces four theorems
for weighted means, summarized in Table~\ref{table:w-mn-thms}.

\begin{table}
\centering
\normalsize
\begin{tabular}{|l|l|l|}
\hline
                &
strictly increasing             &increasing     \\
\hline          
$(0, \infty)$   &
$t \in (-\infty, \infty)$       &$t \in [-\infty, \infty]$      \\
                &
Theorem~\ref{thm:w-str-inc}     &Theorem~\ref{thm:w-inc}        \\
\hline
$[0, \infty)$   &
$t \in (0, \infty)$             &$t \in [-\infty, \infty]$      \\
                &
Theorem~\ref{thm:w-zero-str-inc}&Theorem~\ref{thm:w-cts-inc}    \\
                &
                                &
\small(also assume continuous in 2nd argument)  \\
\hline
\end{tabular}
\caption{Summary of characterization theorems for symmetric,
  absence-invariant, consistent, modular, homogeneous, weighted means.  For
  instance, the top-left entry indicates that the strictly increasing such
  means on $(0, \infty)$ are exactly the weighted power means $M_t$ of
  order $t \in (-\infty, \infty)$.  Table~\ref{table:mn-thms}
  (p.~\pageref{table:mn-thms}) gives the corresponding results on
  unweighted means.}  
\lbl{table:w-mn-thms} 
\index{power mean!characterization of!weighted on $(0, \infty)$}%
\index{power mean!characterization of!weighted on $[0, \infty)$}%
\end{table}

\begin{figure}
\centering
\lengths
\begin{picture}(120,32)(0,5)
\cell{61}{20}{c}{\includegraphics[width=122\unitlength]{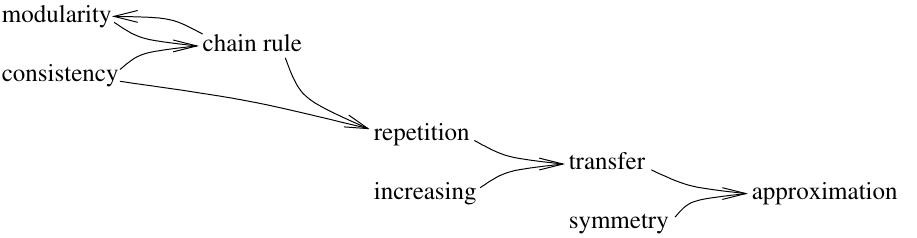}}
\cell{25}{35}{c}{\scriptsize trivial}
\cell{15}{29.5}{c}{\scriptsize\ref{lemma:cons-mod-chn}}
\cell{35.5}{24}{c}{\scriptsize\ref{lemma:cons-chn-rep}}
\cell{64}{14}{c}{\scriptsize\ref{lemma:transfer}}
\cell{88}{9.7}{c}{\scriptsize\ref{lemma:approx}}
\end{picture}
\caption{Implications between properties of weighted means (Lemmas
  \ref{lemma:cons-mod-chn}--\ref{lemma:approx}).  The labels on the arrows
  indicate lemma numbers.}  
\lbl{fig:mean-condns}
\end{figure}
The elementary implications (Lemmas
\ref{lemma:cons-mod-chn}--\ref{lemma:approx}) are shown in
Figure~\ref{fig:mean-condns}. In these lemmas, $I$ denotes a real interval
and $M$ denotes a sequence of functions $(M \from \Delta_n \times I^n \to
I)_{n \geq 1}$.  The properties of means mentioned there were all defined in
Section~\ref{sec:pwr-mns} (apart from transfer and approximation, defined
below) and are summarized in Appendix~\ref{app:condns}.

Plainly the chain rule implies modularity.  There is also a kind of
converse:

\begin{lemma}
\lbl{lemma:cons-mod-chn}
If $M$ is consistent and modular then $M$ satisfies the chain rule.
\end{lemma}

\begin{proof}
Let $\vc{w} \in \Delta_n$, let $\p^1 \in \Delta_{k_1}, \ldots, \p^n \in
\Delta_{k_n}$, and let $\vc{x}^1 \in I^{k_1}, \ldots, \vc{x}^n \in
I^{k_n}$, where $n, k_i \geq 1$ are integers.  Write $a_i = M(\p^i,
\vc{x}^i)$.  By consistency,
\[
M(\p^i, \vc{x}^i) = a_i = M\bigl(\vc{u}_1, (a_i)\bigr) 
\]
for each $i$.  Hence by modularity,
\[
M\bigl( 
\vc{w} \of (\p^1, \ldots, \p^n), 
\vc{x}^1 \oplus\cdots\oplus \vc{x}^n
\bigr)
=
M\bigl(
\vc{w} \of (\vc{u}_1, \ldots, \vc{u}_1),
(a^1) \oplus\cdots\oplus (a^n)
\bigr).
\]
But the right-hand side is $M\bigl(\vc{w}, (a_1, \ldots, a_n)\bigr)$, so
the result is proved.
\end{proof}

\begin{lemma}
\lbl{lemma:cons-chn-rep}
If $M$ is consistent and satisfies the chain rule then $M$ has the
repetition property.   
\end{lemma}

\begin{proof}
Let $\p \in \Delta_n$ and $\vc{x} \in I^n$; suppose that $x_i =
x_{i + 1}$ for some $i < n$.  We must prove that
\begin{align*}
M(\p, \vc{x}) &
=
M\bigl(
(p_1, \ldots, p_{i - 1}, p_i + p_{i + 1}, p_{i + 2}, \ldots, p_n),      \\
&
\hspace*{2.5em}
(x_1, \ldots, x_{i - 1}, x_i, x_{i + 2}, \ldots, x_n)
\bigr).
\end{align*}
For ease of notation, let us assume that $i = n - 1$.  (The general case is
similar.)  By Lemma~\ref{lemma:decomp},
\[
\p = 
(p_1, \ldots, p_{n - 2}, p_{n - 1} + p_n) \of
(\vc{u}_1, \ldots, \vc{u}_1, \vc{r})
\]
for some $\vc{r} \in \Delta_2$.  Then 
\begin{align*}
M(\p, \vc{x})
&
=
M \bigl(
(p_1, \ldots, p_{n - 2}, p_{n - 1} + p_n) \of
(\vc{u}_1, \ldots, \vc{u}_1, \vc{r}),   \\
&
\hspace*{2.5em}
(x_1) \oplus\cdots\oplus (x_{n - 2}) \oplus (x_{n - 1}, x_{n - 1})
\bigr),
\end{align*}
so by the chain rule and consistency,
\begin{align*}
M(\p, \vc{x})   &
=
M \Bigl(
(p_1, \ldots, p_{n - 2}, p_{n - 1} + p_n),      \\
&
\hspace*{2.5em}
\bigl( 
M(\vc{u}_1, (x_1)), \ldots, M(\vc{u}_1, (x_{n - 2})), 
M(\vc{r}, (x_{n - 1}, x_{n - 1}))
\bigr)
\Bigr)  \\
&
=
M \bigl(
(p_1, \ldots, p_{n - 2}, p_{n - 1} + p_n),
(x_1, \ldots, x_{n - 2}, x_{n - 1})
\bigr),
\end{align*}
as required.
\end{proof}

\begin{lemma}
\lbl{lemma:transfer}
If $M$ has the repetition property and is increasing, then $M$ also has the
\demph{transfer}%
\index{transfer property} 
property:
\[
M(\p, \vc{x})
\leq
M \bigl( 
(p_1, \ldots, p_{i - 1}, 
p_i - \delta, p_{i + 1} + \delta, 
p_{i + 2}, \ldots, p_n),
\vc{x}
\bigr)
\]
whenever $1 \leq i < n$, $\p \in \Delta_n$, $\vc{x} \in I^n$ with $x_i
  \leq x_{i + 1}$, and $0 \leq \delta \leq p_i$.
\end{lemma}

The transfer property states that a weighted mean increases when
weight is transferred from a smaller argument to a larger one.

\begin{proof}
As in the last proof, we may harmlessly assume that $i = n - 1$.  We have
\begin{align}
M(\p, \vc{x})   &
=
M\bigl(
(p_1, \ldots, p_{n - 2}, p_{n - 1} - \delta, \delta, p_n),
(x_1, \ldots, x_{n - 2}, x_{n - 1}, x_{n - 1}, x_n)
\bigr)  
\lbl{eq:trans-1}      \\
&
\leq
M\bigl(
(p_1, \ldots, p_{n - 2}, p_{n - 1} - \delta, \delta, p_n),
(x_1, \ldots, x_{n - 2}, x_{n - 1}, x_n, x_n)
\bigr)  
\lbl{eq:trans-2}      \\
&
=
M\bigl(
(p_1, \ldots, p_{n - 2}, p_{n - 1} - \delta, p_n + \delta),
\vc{x}
\bigr),
\lbl{eq:trans-3}
\end{align}
where~\eqref{eq:trans-1} and~\eqref{eq:trans-3} hold by repetition
and~\eqref{eq:trans-2} holds because $M$ is increasing.
\end{proof}

\begin{lemma}
\lbl{lemma:approx}
Suppose that $M$ is symmetric and has the transfer property.  Then $M$ also
has the following \demph{approximation}%
\index{approximation property}
property: for all $\p \in \Delta_n$, $\vc{x} \in I^n$ and $\delta > 0$,
there exist $\p^-, \p^+ \in \Delta_n$ such that all the coordinates of
$\p^-$ and $\p^+$ are rational,
\[
\max_i |p^-_i - p_i| < \delta,
\quad
\max_i |p^+_i - p_i| < \delta,
\]
and
\[
M(\p^-, \vc{x}) \leq M(\p, \vc{x}) \leq M(\p^+, \vc{x}).
\]
\end{lemma}

\begin{proof}
We just prove the existence of such a $\p^+$, the argument for $\p^-$ being
similar.  By symmetry, we may assume that $x_1 \leq \cdots \leq x_n$.

Choose $\delta_1 \in [0, \delta)$ with $0 \leq p_1 - \delta_1 \in \Q$.  By
the transfer property,
\[
M(\p, \vc{x})
\leq
M\bigl(
(p_1 - \delta_1, p_2 + \delta_1, p_3, \ldots, p_n), 
\vc{x}
\bigr).
\]
Next, choose $\delta_2 \in [0, \delta)$ such that $0 \leq p_2 + \delta_1 -
  \delta_2 \in \Q$.  By the transfer property,
\begin{multline*}
M\bigl(
(p_1 - \delta_1, p_2 + \delta_1, p_3, \ldots, p_n), 
\vc{x}
\bigr)
\\
\leq
M\bigl(
(p_1 - \delta_1, p_2 + \delta_1 - \delta_2, p_3 + \delta_2, p_4, 
\ldots, p_n), 
\vc{x}
\bigr).
\end{multline*}
Continuing in this way, we obtain $n - 1$ inequalities that together imply
that 
\[
M(\p, \vc{x})
\leq
M\bigl(
(p_1 - \delta_1, 
p_2 + \delta_1 - \delta_2, \ldots, 
p_{n - 1} + \delta_{n - 2} - \delta_{n - 1}, 
p_n + \delta_{n - 1}),
\vc{x}
\bigr).
\]
The result follows by taking $\p^+$ to be the distribution on the
right-hand side. 
\end{proof}

Many properties of weighted means $M(-, -)$ imply the corresponding
property of their unweighted counterparts $M(\vc{u}_n, -)$:

\begin{lemma}
Let $I$ be an interval and let $(M \from \Delta_n \times I^n \to I)_{n
  \geq 1}$ be a sequence of functions.  If $M$ is symmetric, consistent,
increasing, or strictly increasing (respectively), then so is the sequence of
functions $\bigl(M(\vc{u}_n, -) \from I^n \to I\bigr)_{n \geq 1}$.
Moreover, if $I$ is closed under multiplication and $M$ is homogeneous then
so is $(M(\vc{u}_n, -))_{n \geq 1}$.  
\end{lemma}

\begin{proof}
Trivial.
\end{proof}

It was stated on p.~\pageref{p:decomp-analogue} that decomposability is
an unweighted analogue of the chain rule.  The following lemma supports
that claim.

\begin{lemma}
\lbl{lemma:u-w-decomp}
Let $I$ be a real interval and let $(M \from \Delta_n \times I^n \to I)_{n
  \geq 1}$ be a sequence of functions that is consistent and satisfies the
chain rule.  Then $\bigl( M(\vc{u}_n, -)\from I^n \to I\bigr)_{n \geq 1}$
is decomposable.
\end{lemma}

\begin{proof}
Let $n, k_1, \ldots, k_n \geq 1$ and $\vc{x}^1 \in I^{k_1}, \ldots,
\vc{x}^n \in I^{k_n}$.  Write $a_i = M(\vc{u}_{k_i}, \vc{x}^i)$ and $k =
\sum k_i$.  We must show that
\[
M(\vc{u}_k, \vc{x}^1 \oplus\cdots\oplus \vc{x}^n)
=
M\bigl(\vc{u}_k, (k_1 \mc a_1, \ldots, k_n \mc a_n)\bigr).
\]
We have
\[
\vc{u}_k
=
(k_1/k, \ldots, k_n/k) \of (\vc{u}_{k_1}, \ldots, \vc{u}_{k_n}),
\]
so by the chain rule,
\begin{equation}
\lbl{eq:uwd-1}
M(\vc{u}_k, \vc{x}^1 \oplus\cdots\oplus \vc{x}^n)
=
M\bigl(
(k_1/k, \ldots, k_n/k), (a_1, \ldots, a_n)
\bigr).
\end{equation}
But by Lemma~\ref{lemma:cons-chn-rep}, $M$ has the repetition property,
which implies by induction that the right-hand side of~\eqref{eq:uwd-1} is
equal to 
\[
M\bigl(\vc{u}_k, (k_1 \mc a_1, \ldots, k_n \mc a_n)\bigr).
\]
This completes the proof.
\end{proof}

We now make a tool for converting theorems on unweighted means into
theorems on weighted means.

\begin{propn}
\lbl{propn:u-to-w}
\index{mean!weighted vs.\ unweighted}
Let $I$ be a real interval and let 
\[
\bigl( M, M' \from \Delta_n \times I^n \to I \bigr)_{n \geq 1}
\]
be two sequences of functions.  Suppose that:
\begin{enumerate}
\item 
both $M$ and $M'$ have the absence-invariance and repetition properties;

\item
$M$ is symmetric and increasing;

\item
for each $\vc{x} \in I^n$, the function $M'(-, \vc{x})$ is continuous on
the open simplex $\Delta_n^\circ$.
\end{enumerate}
Suppose also that
\[
M(\vc{u}_n, -) = M'(\vc{u}_n, -) \from I^n \to I
\]
for all $n \geq 1$.  Then $M = M'$.
\end{propn}

\begin{proof}
First we prove that $M(\p, -) = M'(\p, -)$ when the coordinates of $\p$ are
rational and nonzero.  Write
\[
\p = (k_1/k, \ldots, k_n/k),
\]
where $k_1, \ldots, k_n$ are positive integers and $k = \sum k_i$.  Let
$\vc{x} \in I^n$.  Then by the repetition property of $M$ and induction,
\begin{equation}
\lbl{eq:utw-1}
M(\p, \vc{x})
=
M\bigl( \vc{u}_k,
(k_1 \mc x_1, \ldots, k_n \mc x_n)
\bigr).
\end{equation}
The same argument applied to $M'$ gives
\begin{equation}
\lbl{eq:utw-2}
M'(\p, \vc{x})
=
M'\bigl( \vc{u}_k,
(k_1 \mc x_1, \ldots, k_n \mc x_n)
\bigr).
\end{equation}
But the right-hand sides of~\eqref{eq:utw-1} and~\eqref{eq:utw-2} are
equal by hypothesis, so $M(\p, \vc{x}) = M'(\p, \vc{x})$.

Now we show by induction on $n \geq 1$ that $M(\p, \vc{x}) = M'(\p,
\vc{x})$ for all $\p \in \Delta_n$ and $\vc{x} \in I^n$.  

For $n = 1$, we must have $\p = \vc{u}_1$, hence $M(\p, \vc{x}) = M'(\p,
\vc{x})$ by hypothesis.

Let $n \geq 2$ and assume the result for $n - 1$.  If $p_i =
0$ for some $i$ then $M(\p, \vc{x}) = M'(\p, \vc{x})$ by inductive
hypothesis and absence-invariance of $M$ and $M'$.  Suppose, then,
that $\p \in \Delta_n^\circ$.

Let $\epsln > 0$.  Since $M'(-, \vc{x})$ is continuous at $\p$, we can
choose $\delta \in (0, \min_i p_i)$ such that for $\vc{r} \in
\Delta_n^\circ$,
\[
\max_i |p_i - r_i| < \delta
\implies
|M'(\p, \vc{x}) - M'(\vc{r}, \vc{x})| < \epsln.
\]
By Lemma~\ref{lemma:transfer}, $M$ has the transfer property, so by
Lemma~\ref{lemma:approx}, $M$ also has the approximation property.  Choose
$\p^+$ as in Lemma~\ref{lemma:approx}; then
\begin{equation}
\lbl{eq:utw-3}
|M'(\p, \vc{x}) - M'(\p^+, \vc{x})| < \epsln.
\end{equation}
Also, since $\p \in \Delta_n^\circ$ and 
\[
\max_i |p_i - p^+_i| < \delta < \min_i p_i,
\]
we have $\p^+ \in \Delta_n^\circ$ too.  Now
\begin{align}
M(\p, \vc{x})   &
\leq
M(\p^+, \vc{x}) 
\lbl{eq:utw-4}        \\
&
=
M'(\p^+, \vc{x})
\lbl{eq:utw-5}        \\
&
<
M'(\p, \vc{x}) + \epsln,
\lbl{eq:utw-6}
\end{align}
where inequality~\eqref{eq:utw-4} is one of the defining properties of
$\p^+$, equation~\eqref{eq:utw-5} holds because the coordinates of $\p^+$
are rational and nonzero (using the first step of the proof), and
inequality~\eqref{eq:utw-6} follows from~\eqref{eq:utw-3}.  But
this holds for all $\epsln > 0$, so
\[
M(\p, \vc{x}) \leq M'(\p, \vc{x}).
\]
A very similar argument, using the distribution $\p^-$ of
Lemma~\ref{lemma:approx}, proves the opposite inequality.  Hence $M(\p,
\vc{x}) = M'(\p, \vc{x})$, completing the proof.
\end{proof}

We can now simply read off four characterization theorems for weighted
power means.  They are summarized in Table~\ref{table:w-mn-thms}, and are
derived from the four theorems on unweighted means shown in
Table~\ref{table:mn-thms}.

\begin{thm}
\lbl{thm:w-str-inc}
\index{power mean!characterization of!weighted on $(0, \infty)$}
Let $\bigl( M \from \Delta_n \times (0, \infty)^n \to (0, \infty) \bigr)_{n
  \geq 1}$ be a sequence of functions.  The following are equivalent:
\begin{enumerate}
\item 
\lbl{part:w-str-inc-condns}
$M$ is symmetric, absence-invariant, consistent, strictly increasing,
modular, and homogeneous;

\item
\lbl{part:w-str-inc-form}
$M = M_t$ for some $t \in (-\infty, \infty)$.
\end{enumerate}
\end{thm}

\begin{proof}
Part~\bref{part:w-str-inc-form} implies part~\bref{part:w-str-inc-condns}
by the results in Section~\ref{sec:pwr-mns}.
Now assume~\bref{part:w-str-inc-condns}.
The unweighted mean
\[
\bigl( 
M(\vc{u}_n, -) \from (0, \infty)^n \to (0, \infty)
\bigr)_{n \geq 1}
\]
is symmetric, strictly increasing, decomposable, and homogeneous (using
Lemmas~\ref{lemma:cons-mod-chn} and~\ref{lemma:u-w-decomp} for
decomposability).  Hence by Theorem~\ref{thm:str-inc}, there is some $t \in
(-\infty, \infty)$ such that
\[
M(\vc{u}_n, -) = M_t(\vc{u}_n, -)
\]
for all $n \geq 1$.  By Lemmas~\ref{lemma:cons-mod-chn}
and~\ref{lemma:cons-chn-rep}, $M$ has the repetition property.  Hence by the
previously-established properties of $M_t$, we can apply
Proposition~\ref{propn:u-to-w} with $M' = M_t$, giving $M = M_t$.
\end{proof}

Theorem~\ref{thm:w-str-inc} is essentially due to Hardy,%
\index{Hardy, Godfrey Harold} 
Littlewood%
\index{Littlewood, John Edensor}
and P\'olya~\cite{HLP}.%
\index{Polya, George@P\'olya, George}
Some minor details aside, it is the conjunction of
their Theorems~84 and~215, translated out of the language of Stieltjes
integrals and into elementary terms.  Section~6.21 of~\cite{HLP} gives
details.

\begin{thm}
\lbl{thm:w-zero-str-inc}
\index{power mean!characterization of!weighted on $[0, \infty)$}
Let $\bigl( M \from \Delta_n \times [0, \infty)^n \to [0, \infty) \bigr)_{n
  \geq 1}$ be a sequence of functions.  The following are equivalent:
\begin{enumerate}
\item 
$M$ is symmetric, absence-invariant, consistent, strictly increasing,
  modular, and homogeneous;

\item
$M = M_t$ for some $t \in (0, \infty)$.
\end{enumerate}
\end{thm}

\begin{proof}
This follows by exactly the same argument as for the last theorem, but
using Theorem~\ref{thm:zero-str-inc} instead of Theorem~\ref{thm:str-inc}.
\end{proof}

\begin{thm}
\lbl{thm:w-inc}
\index{power mean!characterization of!weighted on $(0, \infty)$}
Let $\bigl( M \from \Delta_n \times (0, \infty)^n \to (0, \infty) \bigr)_{n
  \geq 1}$ be a sequence of functions.  The following are equivalent:
\begin{enumerate}
\item 
$M$ is symmetric, absence-invariant, consistent, increasing, modular, and
  homogeneous;

\item
$M = M_t$ for some $t \in [-\infty, \infty]$.
\end{enumerate}
\end{thm}

\begin{proof}
This follows from Theorem~\ref{thm:inc} by the same argument.
\end{proof}

\begin{thm}
\lbl{thm:w-cts-inc}
\index{power mean!characterization of!weighted on $[0, \infty)$}
Let $\bigl( M \from \Delta_n \times [0, \infty)^n \to [0, \infty) \bigr)_{n
  \geq 1}$ be a sequence of functions.  The following are equivalent:
\begin{enumerate}
\item 
$M$ is symmetric, absence-invariant, consistent, increasing, modular,
  homogeneous, and continuous in its second argument;

\item
$M = M_t$ for some $t \in [-\infty, \infty]$.
\end{enumerate}
\end{thm}

\begin{proof}
This follows from Theorem~\ref{thm:cts-inc} by the same argument again,
this time also noting that by consistency, none of the functions
$M(\vc{u}_n, -)$ is identically zero.  
\end{proof}

We will use Theorem~\ref{thm:w-inc} to prove an axiomatic characterization
of measures of the value of a community (Section~\ref{sec:value-char}) and,
building on this, to characterize the Hill numbers
(Section~\ref{sec:total-hill}).

%% file: sim.tex
\chapter{Species similarity and magnitude}
\lbl{ch:sim}
\index{similarity!species@of species}

Alfred Russel Wallace, who in parallel with Charles Darwin discovered what
we now call the theory of evolution, spent much of the 1850s travelling in
tropical south-east Asia and South America.  On his return, he wrote
widely on what he had experienced, including the following description of 
the diversity of a tropical forest\index{forest!tropical} (\cite{Wall}, p.~65):
\begin{quote}
\index{Wallace, Alfred Russel}
If the traveller notices a particular species and wishes to find more like it,
he may often turn his eyes in vain in every direction.  Trees of varied forms,
dimensions, and colours are around him, but he rarely sees any one of them
repeated.  Time after time he goes towards a tree which looks like the one he
seeks, but a closer examination proves it to be distinct.  He may at length,
perhaps, meet with a second specimen half a mile off, or may fail altogether,
till on another occasion he stumbles on one by accident.
\end{quote}
One of Wallace's observations was that besides there being a large number
of species, mostly rare, there was also a great deal of similarity between
different species.  Clearly, any comprehensive account of the variety or
diversity of life has to incorporate the varying degrees of similarity
between species.  All else being equal, a community of species that are
closely related to one another should be judged as less diverse than if
they were highly dissimilar.

This is not an abstract concern.  The Organization for Economic
Co-operation and Development's guide to biodiversity%
\index{conservation}
for policy%
\index{politics}
makers recognizes this same point, stating that
\begin{quote}\index{OECD}
\lbl{p:oecd-quote} 
associated with the idea of diversity is the concept of \emph{distance},
i.e., some measure of the dissimilarity of the resources in question
\end{quote}
(\cite{OECD}, p.~25).  With global biodiversity now being lost at
historically unprecedented rates, it is crucial that politicians and
scientists speak the same language.  However, most conventional measures of
diversity, and all the ones discussed in this text so far, fail to take the
different dissimilarities between species into account.

Here we solve this problem, defining a system of measures that depend not
only on the relative abundances of the species, but also on the varying
similarity between them (Sections~\ref{sec:sim-basic}
and~\ref{sec:sim-props}).  It was first introduced in a 2012 article of
Leinster and Cobbold~\cite{MDISS}.%
\index{Cobbold, Christina}
We make no assumption about \emph{how} similarity is measured: it could be
genetic, phylogenetic, functional, etc., leading to measures of genetic
diversity, phylogenetic diversity, functional diversity, etc.  As such, the
system is adaptable to a wide variety of scientific needs.

More specifically, we will encode the similarities between species as a
real matrix $Z$, continuing to represent the relative abundances of the
species as a probability distribution $\p$.  With this model of a
community, we will define for each $q \in [0, \infty]$ a measure
$D_q^Z(\p)$ of the diversity of the community.  As for the Hill numbers,
the parameter $q$ controls%
\index{viewpoint!parameter} 
the extent to which the measure emphasizes the common species at the
expense of the rare ones.  Under the extreme hypothesis that different
species never have anything in common, $Z$ is the identity matrix $I$ and
the diversity $D_q^I(\p)$ reduces to the Hill number $D_q(\p)$.  In that
sense, these similarity-sensitive diversity measures generalize the Hill
numbers.

Let $\p$ be a probability distribution on a finite set. We saw in
Section~\ref{sec:ren-hill} that the Hill numbers $D_q(\p)$, the R\'enyi
entropies $H_q(\p)$ and the $q$-logarithmic entropies $S_q(\p)$ are all
simple increasing transformations of one another.  The same is true in the
more general context here. Thus, accompanying the similarity-sensitive
diversity measures $D_q^Z(\p)$ are similarity-sensitive R\'enyi entropies
$H_q^Z(\p)$ and $q$-logarithmic entropies $S_q^Z(\p)$.  Any metric on our
finite set gives rise naturally to a similarity matrix $Z$, as we shall
see.  So, we obtain definitions of the R\'enyi and $q$-logarithmic
entropies of a probability distribution on a finite metric space, extending
the classical definitions on a finite set.

How is diversity maximized?  For a fixed similarity matrix $Z$ (and in
particular, for a finite metric space), we can seek the probability
distribution $\p$ that maximizes the diversity or entropy of a given order
$q$.  As we saw in the special case of the Hill numbers, different values
of $q$ can lead to different judgements on which of two communities is the
more diverse.  So in principle, both the maximizing distribution and the
value of the maximum diversity depend on $q$.  However, it is a theorem
that neither does.  Every similarity matrix has an unambiguous maximum
diversity, independent of $q$, and a distribution that maximizes the
diversity of all orders $q$ simultaneously.  This is the subject of
Section~\ref{sec:max}.

The maximum diversity of a matrix $Z$ is closely related to another
quantity, the magnitude of a matrix.  The general concept of magnitude,
expressed in the formalism of enriched categories, brings together a wide
range of size-like invariants in mathematics, including cardinality, Euler
characteristic, volume, surface area, dimension, and other geometric
measures.  Sections~\ref{sec:mag} and~\ref{sec:mag-geom} are a broad-brush
survey of magnitude, and demonstrate that maximum diversity~-- far from
being tethered to ecology~-- has profound connections with 
fundamental invariants of geometry.

\section{The importance of species similarity}
\lbl{sec:sim-basic}

Here we introduce a family of measures of the diversity of an ecological
community that take into account the varying similarities between species, 
following work of Leinster and Cobbold~\cite{MDISS}.%
\index{Cobbold, Christina|(}  

These diversity measures will be almost completely neutral as to what
`similarity' means or how it is quantified, just as the diversity measures
discussed earlier were neutral as to the meaning of abundance
(Example~\ref{eg:prob-eco}).  The following examples illustrate some of the
ways in which similarity can be quantified.  In these
examples, the similarity%
\index{similarity!species@of species} 
$z$ between two species is measured on a scale of $0$ to $1$,
with $0$ representing complete dissimilarity and $1$ representing identical
species.

\begin{examples}
\lbl{egs:sim}
\begin{enumerate}
\item 
The similarity $z$ between two species can be interpreted as percentage
genetic%
\index{genetic diversity}
\index{diversity!genetic} 
similarity (in any of several senses; typically one would restrict
to a particular part of the genome).  With the rapid fall in the cost of
DNA sequencing, this way of quantifying similarity is increasingly common.
It can be used even when the taxonomic classification of the organisms
concerned is unclear or
incomplete, as is often the case for microbial%
\index{microbial systems}
communities (a problem discussed by Johnson~\cite{JohnUNA} and Watve and
Gangal~\cite{WaGa}, for instance).

\item
Functional
\index{functional diversity}%
\index{diversity!functional}
similarity can also be quantified.  For instance, suppose that we have a
list of $k$ functional traits satisfied by some species but not others. We
can then define the similarity $z$ between two species as $j/k$, where $j$
is the number of traits possessed by either both species or neither.  (For
an overview of functional diversity, see Petchey and
Gaston~\cite{PeGaFDB}.)

\item
\lbl{eg:sim-phylo}
Similarity can also be measured phylogenetically,%
\index{phylogenetic!diversity}
\index{diversity!phylogenetic} 
that is, in terms of an evolutionary tree.  For instance, $z$ can be
defined as the proportion of evolutionary time before the two species
diverged, relative to some fixed start time.

\item
\lbl{eg:sim-taxo}
In the absence of better data, we can measure similarity crudely
using taxonomy.%
\index{taxonomic diversity}%
\index{diversity!taxonomic}  
For instance, we could define the similarity $z$ between
two species by
\[
z
=
\begin{cases}
1       &\text{if the species are the same,}    \\
0.8     &\text{if the species are different but of the same genus,}     \\
0.5     &\text{if the species are of different genera but the same
  family,}      \\
0       &\text{otherwise},
\end{cases}
\]
or similarly for any other choice of constants and number of taxonomic
levels. 

\item
\lbl{eg:sim-naive}
More crudely still, we can define the similarity $z$ between two species by
\[
z
=
\begin{cases}
1       &\text{if the species are the same,}     \\
0       &\text{if the species are different}.
\end{cases}
\]
This definition embodies the assumption that different species never have
anything in common. Unrealistic as this is, we will see that it is implicit
in all of the measures of diversity defined in this book so far, and most
of the diversity measures common in the ecological literature.
\end{enumerate}
\end{examples}

Now consider a list of species, numbered as $1, \ldots, n$, and suppose
that we have fixed a way of quantifying the similarity between them.
We obtain an $n \times n$ matrix 
\[
Z = \bigl(Z_{ij}\bigr)_{1 \leq i, j \leq n},
\]
where $Z_{ij}$ is the similarity between species $i$ and $j$.

Formally, a real square matrix $Z$ is a
\demph{similarity%
\index{similarity!matrix} 
matrix} if $Z_{ij} \geq 0$ for all $i, j$ and $Z_{ii} > 0$ for all $i$.
The examples above suggest additional hypotheses: that $Z_{ij} \leq 1$ for
all $i, j$, that $Z_{ii} = 1$\lbl{p:Z1} for all $i$, and that $Z$ is
symmetric.  (Indeed, in the paper~\cite{MDISS} on which this section is
based, the term `similarity matrix' included the first two of these
additional hypotheses.)  But in most of what follows, we will not need
these extra assumptions, so we do not make them.

\begin{examples}
\lbl{egs:sim-mx}
\begin{enumerate}
\item 
The genetic, functional, phylogenetic and taxonomic similarity measures of
Examples~\ref{egs:sim} give genetic, functional, phylogenetic and taxonomic
similarity matrices $Z$, taking $Z_{ij}$ to be any of the quantities $z$
described there.

\item
\lbl{eg:sim-mx-naive}
The very crude similarities $z$ of
Example~\ref{egs:sim}\bref{eg:sim-naive}, where distinct species are taken
to be completely dissimilar, give the identity similarity matrix $Z = I$.
We will call this the \demph{naive%
\index{naive model} 
model} of a community.
\end{enumerate}
\end{examples}

\begin{example}
\lbl{eg:sim-matrix-met}
Given any finite metric space, with distance $d$ and points labelled as $1,
\ldots, n$, we obtain an $n \times n$ similarity matrix $Z$ by setting
\[
Z_{ij} = e^{-d(i, j)}.
\]
Thus, large distances correspond to small similarities.  In the extreme,
the metric defined by $d(i, j) = \infty$ for all $i \neq j$ corresponds to
the naive model.  (We allow $\infty$ as a distance in our metric spaces.)
Any taxonomic similarity matrix of the general type indicated in
Example~\ref{egs:sim}\bref{eg:sim-taxo} corresponds to an
\demph{ultrametric space},%
\index{ultrametric!space} 
that is, a metric space satisfying the stronger form
\[
d(i, k) \leq \max \{ d(i, j), d(j, k) \}
\]
of the triangle inequality.

From a purely mathematical viewpoint, this matrix $Z = \bigl(e^{-d(i,
  j)}\bigr)$ associated with a finite metric space is highly significant,
as we will discover when we come to the theory of magnitude
(Sections~\ref{sec:mag} and~\ref{sec:mag-geom}).  From a biological
viewpoint, we may find ourselves starting with a measure of inter-species
difference $\delta$ on a scale of $0$ to $\infty$ (as in Warwick and
Clarke~\cite{WaCl}, for instance), in which case the transformation $z =
e^{-\delta}$ converts it into a similarity $z$ on a scale of $0$ to $1$.
From both viewpoints, the choice of constant $e$ is arbitrary, and one
should consider replacing it by any other constant, or equivalently,
scaling the distance by a linear factor.  Again, this is a fundamental
point in the theory of magnitude, as demonstrated by the theorems in
Section~\ref{sec:mag-geom}.
\end{example}

\begin{example}
\lbl{eg:sim-matrix-gph} 
A symmetric similarity matrix whose entries are all $0$ or $1$ corresponds
to a finite reflexive graph%
\index{graph!adjacency matrix of} 
with no multiple edges.  Here, \demph{reflexive}%
\index{reflexive}%
\index{graph!reflexive}
means that there is an edge from each vertex to itself (a \demph{loop}).%
\index{loop in a graph}
The correspondence works as follows: labelling the vertices of the graph as
$1, \ldots, n$, we put $Z_{ij} = 1$ whenever there is an edge between $i$
and $j$, and $Z_{ij} = 0$ otherwise.  One says that $Z$ is the
\demph{adjacency%
\index{adjacency matrix} 
matrix} of the graph.  The reflexivity means that $Z_{ii} = 1$ for all
$i$.

No ecological relevance is claimed for this family of examples, but
mathematically it is a natural special case, and it sheds light on 
computational aspects of calculating maximum diversity
(Remark~\ref{rmk:no-quick-clique}).
\end{example}

Our earlier discussions of diversity modelled an ecological community
crudely as a finite probability distribution $\p = (p_1, \ldots, p_n)$.
Our new and less crude model of a community has two components: a relative
abundance distribution $\p \in \Delta_n$ and an $n \times n$ similarity
matrix $Z$.  We now build up to the definition of the diversity of a
community modelled in this way.

Treating $\p$ as a column vector, we can form the matrix product $Z\p$,
which has entries
\begin{equation}
\lbl{eq:Zpi}
(Z\p)_i
=
\sum_{j = 1}^n Z_{ij} p_j
\end{equation}
($1 \leq i \leq n$).  The quantity~\bref{eq:Zpi} is the expected similarity
between an individual of species $i$ and an individual chosen at random.
It can therefore be understood as the ordinariness\index{ordinariness} of
species $i$.  If the diagonal entries of $Z$ are all $1$ (as in every
example above) then
\begin{equation}
\lbl{eq:Zpi-lb1}
(Z\p)_i = \sum_j Z_{ij} p_j \geq Z_{ii} p_i = p_i.
\end{equation}
This inequality states that a species appears more ordinary when the
similarities between species are recognized than when they are
ignored. 

By inequality~\eqref{eq:Zpi-lb1}, any species that is highly abundant is
also highly ordinary: large $p_i$ implies large $(Z\p)_i$.  But even if
species $i$ is rare, its ordinariness $(Z\p)_i$ will be high if there is
some common species very similar to it.  The ordinariness of species $i$
will even be high if it is similar to several species that are each
individually rare, but whose total abundance is large. (For example, in
Wallace's%
\index{Wallace, Alfred Russel} 
tropical forest,%
\index{forest!tropical} 
many tree species have much higher ordinariness $(Z\p)_i$ than relative
abundance $p_i$.)  This makes intuitive sense: the more
thorny\lbl{p:thorny} bushes\index{bushes} a region contains, the more
ordinary any thorny bush will seem, even if its particular species is rare.

Judgements about what is `ordinary' depend on one's perception of
similarity.  If one wishes to make a strong distinction between different
species, one should use a similarity%
\index{similarity!matrix!choice of}
matrix $Z$ whose off-diagonal entries are small, and this will have the
effect of lowering the ordinariness of every species.

Since $(Z\p)_i$ measures how ordinary the $i$th species is, $1/(Z\p)_i$
measures how special it is.  In the case $Z = I$ (the naive model of
Example~\ref{egs:sim-mx}\bref{eg:sim-mx-naive}), this reduces to $1/p_i$,
which in Sections~\ref{sec:ent-div} and~\ref{sec:ren-hill} we called the
specialness\index{specialness} or rarity\index{rarity} of species $i$.  We
have now extended that concept to our more refined model.

When we modelled a community as a simple probability distribution, we
defined the diversity of a community to be the average specialness of an
individual within it. We do the same again now in our new model.

\begin{defn}
\lbl{defn:dqz}
Let $\p \in \Delta_n$, let $Z$ be an $n \times n$ similarity matrix, 
and let $q \in [0, \infty]$.  The \demph{diversity of
  $\p$ of order $q$},%
\index{similarity-sensitive!diversity}%
\index{order!diversity measure@of diversity measure}
with respect to $Z$, is 
\[
D_q^Z(\p)
=
M_{1 - q}(\p, 1/Z\p).
\ntn{DqZ}
\]
\end{defn}

Here, the vector $1/Z\p$ is defined as 
\[
\bigl( 1/(Z\p)_1, \ldots, 1/(Z\p)_n \bigr).
\]
Although there may be some values of $i$ for which $(Z\p)_i = 0$, this can
only occur when $p_i = 0$: for $Z_{ii} > 0$ by definition of similarity
matrix, so if $p_i > 0$ then
\[
(Z\p)_i = \sum_j Z_{ij} p_j \geq Z_{ii} p_i > 0.
\]
So by the convention in Remark~\ref{rmk:defined-even-if-not}, $M_{1 -
  q}(\p, 1/Z\p)$ is well-defined.  Explicitly,
\[
D_q^Z(\p)
=
\Biggl( 
\sum_{i \in \supp(\p)} p_i (Z\p)_i^{q - 1} 
\Biggr)^{1/(1 - q)}
\]
for $q \neq 1, \infty$, and 
\begin{align*}
D_1^Z(\p)       &
=
\prod_{i \in \supp(\p)} (Z\p)_i^{-p_i}
=
\frac{1}{(Z\p)_1^{p_1} \cdots (Z\p)_n^{p_n}},   \\
D_\infty^Z(\p)  &
=
\frac{1}{\max\limits_{i \in \supp(\p)} (Z\p)_i}.
\end{align*}
We could extend Definition~\ref{defn:dqz} to negative $q$, but it would be
misleading to call $D_q^Z(\p)$ `diversity'%
\index{diversity!negative order@of negative order}
when $q$ is negative, for the reasons given in
Remark~\ref{rmks:hill-dec}\bref{rmk:hill-dec-neg}.  We therefore restrict
to $q \in [0, \infty]$.

\begin{examples}
\lbl{egs:dqz}
Here we consider some special values of $Z$ and $q$, and in doing so
recover various earlier measures of diversity.

\begin{enumerate}
\item 
\lbl{eg:dqz-naive}
In the naive model $Z = I$, where distinct species are taken to be
completely dissimilar, $Z\p = \p$ and so $D_q^Z(\p)$ is just the Hill
number $D_q(\p)$.  In this sense, the Hill numbers implicitly use the
naive model of a community.

\item
For a general similarity matrix, the diversity of order $0$ is 
\[
D_0^Z(\p) = \sum_{i \in \supp(\p)} \frac{p_i}{(Z\p)_i}.
\]
This is a sum of contributions from all species present.  The contribution
made by the $i$th species, $p_i/(Z\p)_i$, is between $0$ and $1$, by
inequality~\eqref{eq:Zpi-lb1} (assuming that $Z_{ii} = 1$). It is large
when, relative to the size of the $i$th species, there are not many
individuals of other similar species~-- that is, when the $i$th species is
unusual.  We discuss the quantity $p_i/(Z\p)_i$ in greater depth in
Example~\ref{eg:value-dqz}.

\item
In the naive model, the diversity of order $\infty$ is the Berger--Parker%
\index{Berger--Parker index}
index
\[
D_\infty^I(\p) = D_\infty(\p) = 1\Big/\!\max_i p_i
\]
(Example~\ref{egs:hill}\bref{eg:hill-bp}).  It measures the dominance of
the most common species, the idea being that in a diverse community, no
species should be too dominant.  For a general similarity matrix, the
diversity
\[
D_\infty^Z(\p) = 1\Big/\!\!\max_{i \in \supp(\p)} (Z\p)_i
\]
of order $\infty$ can be interpreted in the same way, but now with
sensitivity to species similarity: $D_\infty^Z(\p)$ is low not only if
there is a single highly abundant species, but also if there is some highly
abundant \emph{cluster} of species.

\item
\lbl{eg:dqz-two}
The diversity of order $2$ is
\[
D_2^Z(\p) 
=
1\bigg/\!\sum_{i, j = 1}^n p_i Z_{ij} p_j
=
1\big/\p^\transp Z \p.
\]
(We continue to regard $\p$ as a column vector, so that its transpose
$\p^\transp$ is a row vector.)  The number $\p^\transp Z \p$ is the
expected similarity between a pair of individuals chosen at random.  This
is a measure of a community's \emph{lack} of diversity, and its reciprocal
$D_2^Z(\p)$ therefore measures diversity itself.

For instance, take a probability distribution $\p$ on the vertices of a
graph,%
\index{graph!diversity on} 
and let $Z$ be the adjacency matrix (as in
Example~\ref{eg:sim-matrix-gph}).  Then $D_2^Z(\p)$ is the
reciprocal of the probability that two vertices chosen at random are
\dmph{adjacent} (joined by an edge).  Equivalently, if pairs of vertices
are repeatedly chosen at random, $D_2^Z(\p)$ is the expected number of
trials needed in order to find an adjacent pair.
\end{enumerate}
\end{examples}

\begin{example}
\lbl{eg:dqz-integers}
By Example~\ref{egs:dqz}\bref{eg:dqz-two}, one can estimate the diversity
of order~$2$ of a community by sampling pairs of individuals at random,
recording the similarity between them, calculating the mean of these
similarities, then taking the reciprocal.  More generally, for any integer%
\index{diversity!integer order@of integer order}%
\index{order!integer}
$q \geq 2$, one can estimate $D_q^Z(\p)$ as follows.  Sample $q$
individuals at random from the community (with replacement).  Supposing
that they are of species $i_1, \ldots, i_q$, let us temporarily refer to
the product
\[
Z_{i_1 i_2} Z_{i_1 i_3} \cdots Z_{i_1 i_q}
\]
as their `group%
\index{group similarity} 
similarity'.  Let $\mu_q$ be the expected group similarity of $q$
individuals from the community.  Then
\[
D_q^Z(\p) = \mu_q^{1/(1 - q)}.
\]
This was first proved as Proposition~A3 of the appendix to~\cite{MDISS},
and the proof is also given here as Appendix~\ref{sec:q-int}.

For instance, in the naive model,%
\index{Hill number!integer order@of integer order}
$\mu_q$ is the probability that $q$ random individuals are all of the
same species, which is $\sum_i p_i^q$.  In this case, it is immediate that
$D_q(\p) = \mu_q^{1/(1 - q)}$.

This procedure for estimating the diversity of orders $2, 3, \ldots$ has
the advantage that it does not require the organisms to be classified into
species.  All we require is a measure of similarity between any pair of
individuals.  This is potentially very useful in studies of microbial%
\index{microbial systems}
systems, where there is often no complete taxonomic classification; all we
have is a way of measuring the similarity between two samples.  We can
estimate $\mu_q$, hence $D_q^Z(\p)$, by repeatedly drawing $q$ samples from
the community, recording their group similarity, and then taking the mean.
\end{example}

Both relative abundance and similarity can be quantified in whatever way is
appropriate to the scientific problem at hand.  
This
makes the diversity measures $D_q^Z(\p)$ highly versatile.  For example, if
the similarity coefficients $Z_{ij}$ are defined genetically then $D_q^Z$
measures genetic diversity,%
\index{genetic diversity}%
\index{diversity!genetic}
and in the same way, a phylogenetic,%
\index{phylogenetic!diversity}%
\index{diversity!phylogenetic}
functional%
\index{functional diversity}%
\index{diversity!functional}
or taxonomic%
\index{taxonomic diversity}%
\index{diversity!taxonomic}
similarity matrix will produce a measure of phylogenetic,
functional or taxonomic diversity.

The different diversity measures arising from different choices of
similarity matrix may produce opposing results.  This is a feature, not
a bug.  For instance, if over a period of time, a community undergoes an
increase in genetic diversity but a decrease in morphological%
\index{morphological diversity}%
\index{diversity!morphological}
diversity, the opposite trends are a point of scientific interest.

When selecting a similarity%
\index{similarity!matrix!choice of}
matrix, a useful observation is that if
\[
Z =
\begin{pmatrix}
1       &z      \\
z       &1
\end{pmatrix}
\]
then
\[
D_q^Z\bigl( \hlf, \hlf \bigr)
=
\frac{2}{1 + z},
\]
or equivalently%
\index{similarity!matrix!choice of}
\[
z
=
\frac{2}{D_q^Z\bigl(\hlf, \hlf\bigr)} - 1,
\]
for all $q \in [0, \infty]$.  So, deciding on the similarity $Z_{ij}$
between species~$i$ and~$j$ is equivalent to deciding on the diversity $d$
of a community consisting of species $i$ and $j$ in equal proportions:
\[
Z_{ij}
=
\frac{2}{d} - 1.
\]
Taking $d = 1$ embodies the viewpoint that this two-species community
consists of effectively only one species, giving a similarity coefficient
$Z_{ij} = 1$: the species are deemed to be identical.  At the opposite
extreme, if one decides that such a community should have diversity $2$
(`effectively $2$ species') for all $i$ and $j$, this produces the naive
matrix $Z = I$.

The flexibility afforded by the choice of similarity matrix may make it
tempting to reject the measures $D_q^Z$ in favour of the simpler Hill
numbers $D_q$, where no such choice is necessary.  However, it is a
mathematical fact that doing so amounts to choosing the naive%
\index{naive model} 
model $Z = I$ (Example~\ref{egs:dqz}\bref{eg:dqz-naive}), which
represents the extreme position that distinct species have nothing
whatsoever in common.  This always leads to an overestimate of diversity
(Lemma~\ref{lemma:dqz-range}).  The framework of similarity matrices forces
us to be transparent: using the naive similarity matrix $I$ is a
\emph{choice}, embodying ecological assumptions, just as much as for
any other similarity matrix.

The next example, adapted from~\cite{MDISS} (Example~3), demonstrates how
ecological judgements can be altered by taking species similarity into
account.  Extending the terminology of Chapter~\ref{ch:def}, we refer to
the graph of $D_q^Z(\p)$ against $q$ as a \demph{diversity%
\index{diversity profile} 
profile}.

\begin{example}
\lbl{eg:devries}
\index{DeVries, Philip}
DeVries et al.~(\cite{DML}, Table~5) counted butterflies\index{butterflies}
in the canopy and understorey at a certain site in the Ecuadorian rain
forest.\index{forest!Ecuadorian} In the subfamily Charaxinae, the
abundances were as shown in Table~\ref{table:devries}.
\begin{table}
\centering
\begin{tabular}{|l|r|r|}
\hline
Species &Abundance      &Abundance      \\
        &in canopy      &in understorey \\
\hline
\emph{Prepona laertes}          &15     &0      \\
\emph{Archaeoprepona demophon}  &14     &37     \\
\emph{Zaretis itys}             &25     &11     \\
\emph{Memphis arachne}          &89     &23     \\
\emph{Memphis offa}             &21     &3      \\
\emph{Memphis xenocles}         &32     &8      \\
\hline
\end{tabular}
\caption{Counts of butterflies of species in the subfamily Charaxinae, in
  the canopy and understorey of an Ecuadorian rain forest site
  (Example~\ref{eg:devries}; data from Table~5 of DeVries et
  al.~\cite{DML}).}  
\lbl{table:devries}
\end{table}
We will compare the diversity profiles of the canopy and the understorey in
two ways, once using the naive similarity matrix and once using a non-naive
matrix.

With the naive similarity matrix $I$, the diversity profiles are as shown in
Figure~\ref{fig:bfly-profiles}\hardref{(a)}.  The profile of the canopy
lies above that of the understorey until about $q = 5$, after which the two
profiles are near-identical.  So, whatever emphasis we may place on rare or
common species, the canopy is at least as diverse as the understorey.

Now let us compare the communities using a taxonomic similarity matrix.
Put
\[
Z_{ij}  =
\begin{cases}
1       &\text{if } i = j,      \\
0.5     &\text{if species $i$ and $j$ are different but of the same
  genus,}\\
0       &\text{otherwise}.
\end{cases}
\]
The resulting diversity profiles, shown in
Figure~\ref{fig:bfly-profiles}\hardref{(b)}, tell a different story.  For
most values of $q$, it is the understorey that is more diverse.  This can
be explained as follows.  Most of the canopy population belongs to the
three species in the \emph{Memphis} genus, so when we build into the model
the principle that species of the same genus tend to be somewhat similar,
the canopy looks less diverse than it did before.  On the other hand,
the understorey population does not contain large numbers of individuals of
different species but the same genus, so factoring in taxonomic similarity
does not cause its diversity to decrease so much.

\begin{figure}
\centering
\lengths
\begin{picture}(58,60)(0,-2)
\cell{33}{31}{c}{\includegraphics[width=53\unitlength]{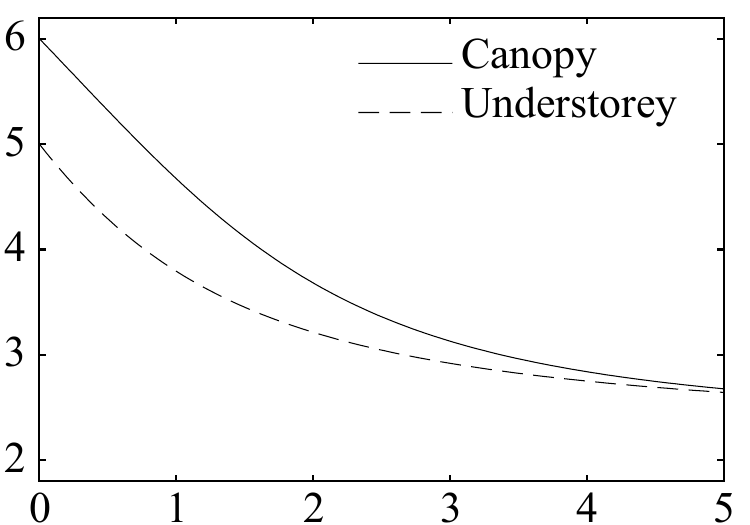}}
\cell{34}{8.5}{c}{Viewpoint parameter, $q$}
\cell{0}{32}{l}{\rotatebox{90}{Diversity, $D_q(\p)$}}
\cell{34}{-2}{b}{(a) Naive similarity}
\end{picture}%
\hspace*{4mm}%
\begin{picture}(58,60)(0,-2)
\cell{33}{31}{c}{\includegraphics[width=53\unitlength]{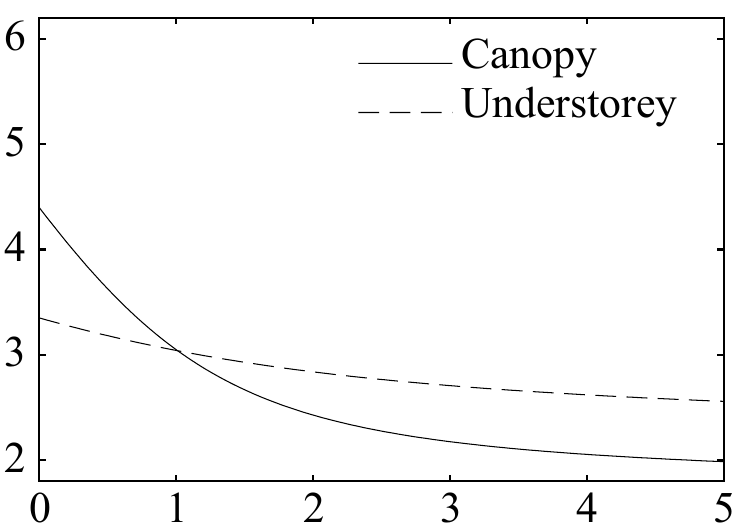}}
\cell{34}{8.5}{c}{Viewpoint parameter, $q$}
\cell{0}{32}{l}{\rotatebox{90}{Diversity, $D_q^Z(\p)$}}
\cell{34}{-2}{b}{(b) Taxonomic similarity}
\end{picture}%
\caption{Diversity profiles of butterflies in the canopy and understorey of
a rain forest site, using \hardref{(a)}~the naive similarity matrix $I$;
\hardref{(b)}~a taxonomic similarity matrix.  Graphs adapted from Figure~3
of Leinster and Cobbold~\cite{MDISS}.}
\lbl{fig:bfly-profiles}
\index{viewpoint!parameter}
\end{figure}
\end{example}

The measures $D_q^Z$, as well as unifying into one family many older
diversity measures, have also found application%
\index{diversity measure!applications of}
in a variety of ecological systems at many scales, from microbes%
\index{microbial systems} 
(Bakker et al.~\cite{BCMV}), fungi (Veresoglou et al.~\cite{VPDL}) and
crustacean zooplankton (Jeziorski et al.~\cite{JTYP}) to alpine plants
(Chalmandrier et al.~\cite{CMLT}) and large arctic predators (Bromaghin et
al.~\cite{BRBT}).  As one would expect, incorporating similarity has been
found to improve inferences\index{inference} about the diversity of natural
systems~\cite{VPDL}.  The measures have also been applied in non-biological
contexts such as computer%
\index{computer network security} 
network security (Wang et al.~\cite{WZJS}).

We now turn from diversity to entropy.  In the simpler context of
probability distributions $\p$ on a finite set, we defined three closely
related quantities for each parameter value $q$: the Hill number $D_q(\p)$,
the R\'enyi entropy $H_q(\p)$, and the $q$-logarithmic entropy $S_q(\p)$.
They are related to one another by increasing, invertible transformations:
\begin{align*}
H_q(\p) &= \log D_q(\p),        \\
S_q(\p) &= \ln_q D_q(\p)
\end{align*}
(equations~\eqref{eq:dhs}).
Given also a similarity matrix $Z$, we define the
\demph{similarity-sensitive R\'enyi entropy}%
\index{similarity-sensitive!Renyi entropy@R\'enyi entropy}%
\index{Renyi entropy@R\'enyi entropy!similarity-sensitive}
$H_q^Z(\p)$ and \demph{similarity-sensitive $q$-logarithmic entropy}%
\index{similarity-sensitive!q-logarithmic entropy@$q$-logarithmic entropy}%
\index{q-logarithmic entropy@$q$-logarithmic entropy!similarity-sensitive}
$S_q^Z(\p)$ by the same transformations:
\begin{align}
H_q^Z(\p)       &= \log D_q^Z(\p),      
\lbl{eq:defn-hqz}       \\
S_q^Z(\p)       &= \ln_q D_q^Z(\p).
\lbl{eq:defn-sqz}
\end{align}
In the first definition, $q \in [0, \infty]$, and in the second, $q \in
[0, \infty)$.

Let us be explicit.  For $q \neq 1, \infty$, the similarity-sensitive
R\'enyi entropy is
\[
H_q^Z(\p)
=
\frac{1}{1 - q} \log \sum_{i \in \supp(\p)} p_i (Z\p)_i^{q - 1},
\]
and in the exceptional cases,
\[
H_1^Z(\p)       
=
- \sum_{i \in \supp(\p)} p_i \log (Z\p)_i
\]
(generalizing the Shannon entropy) and 
\[
H_\infty^Z(\p)  
=
- \log \max_{i \in \supp(\p)} (Z\p)_i.  
\]
We now derive an explicit expression for $S_q^Z(\p)$.  By
Lemma~\ref{lemma:q-log-mean},
\[
S_q^Z(\p)
=
\ln_q M_{1 - q}(\p, 1/Z\p)
=
\sum_{i \in \supp(\p)} 
p_i \ln_q \frac{1}{(Z\p)_i}.
\]
Then applying the definition of $\ln_q$ gives 
\[
S_q^Z(\p) 
=
\frac{1}{1 - q}
\Biggl(
\sum_{i \in \supp(\p)} p_i (Z\p)_i^{q - 1} - 1
\Biggr)
\]
when $q \neq 1$, and
\[
S_1^Z(\p) 
=
H_1^Z(\p) 
= 
-\sum_{i \in \supp(\p)} p_i \log (Z\p)_i.
\]
Figure~\ref{fig:defs}, which schematically depicts the two families of
deformed entropies in the special case $Z = I$, applies equally to an
arbitrary similarity matrix~$Z$.

\begin{example}
The definitions above specialize to definitions of R\'enyi and
$q$-logarithmic entropies for a probability distribution on a finite metric
space.  Indeed, let $A = \{1, \ldots, n\}$ be a finite metric space%
\index{entropy!metric space@on metric space}%
\index{metric!entropy}  
and write $Z = \bigl(e^{-d(i, j)}\bigr)$, as in
Example~\ref{eg:sim-matrix-met}.  For any probability distribution $\p$ on
$A$, and for any parameter value $q$, we have an associated R\'enyi entropy
$H_q^Z(\p)$ and $q$-logarithmic entropy $S_q^Z(\p)$.  Naturally, these
quantities depend on the metric.  In the extreme case where $d(i, j) =
\infty$ for all $i \neq j$, we recover the standard definitions of the R\'enyi
and $q$-logarithmic entropies of a probability distribution on a finite
set.

(One can speculate about extending the results of classical information
theory to the metric context.  As usual, the elements of the set $A = \{1,
\ldots, n\}$ represent the source symbols and the distribution $\p$
specifies their frequencies, but now we also have a metric $d$ on the
source symbols.  It could be defined in such a way that $d(i, j)$ is small
when the $i$th and $j$th symbols are easily mistaken for one another, or
alternatively when one is an acceptable substitute for the other, for
applications such as the encoding of colour images.)
\end{example}

\begin{example}
For any similarity matrix $Z$, we can define a \demph{dissimilarity%
\index{dissimilarity matrix}
matrix}
$\Delta$ by $\Delta_{ij} = 1 - Z_{ij}$.  (Let us assume here that $Z_{ij}
\leq 1$ for all $i$ and $j$.)  In these terms, the $2$-logarithmic entropy
is
\[
S_2^Z(\p)
=
1 - \sum_{i, j} p_i Z_{ij} p_j
=
\sum_{i, j} p_i \Delta_{ij} p_j
=
\p^\transp \Delta \p.
\]
Thus, $S_2^Z(\p)$ is the dissimilarity between a pair of individuals
chosen at random.  This quantity, studied by the statistician
C.~R.~Rao~\cite{RaoDDC,RaoDIM}, is known as \demph{Rao's quadratic
  entropy}.%
\index{Rao, C. Radhakrishna!quadratic entropy}%
\index{quadratic entropy}

Of course, anything that can be expressed in terms of $Z$ can also be
expressed in terms of $\Delta$, and vice versa.  An important early step
towards similarity-sensitive diversity measures was taken by Ricotta%
\index{Ricotta, Carlo} 
and Szeidl~\cite{RiSzTUA},%
\index{Szeidl, Laszlo} 
who gave a version of the entropy $S_q^Z(\p)$ expressed in terms of a
dissimilarity matrix $\Delta$.
\end{example}

\begin{example}
\lbl{eg:dissim-gph}
Let $Z$ be the adjacency%
\index{adjacency matrix} 
matrix of a finite
reflexive graph%
\index{graph!diversity on} 
$G$ with vertex-set $\{1, \ldots, n\}$, as in
Example~\ref{eg:sim-matrix-gph}.  Write $i \adjc\ntn{adjc} j$ to mean that
vertices $i$ and $j$ are joined by an edge, and $i \not\adjc j$ otherwise.
Then the dissimilarity matrix $\Delta$ of the last example has entries $1$
for non-adjacent pairs and $0$ for adjacent pairs.  Hence the
$2$-logarithmic entropy of a probability distribution $\p$ on $G$ is given
by
\[
S_2^Z(\p) = \sum_{i, j \csuch i \not\adjc j} p_i p_j.
\]
This is the probability that two vertices chosen at random according to
$\p$ are \emph{not} joined by an edge.  Thus, entropy is high when vertices
of high probability tend not to be adjacent.  We will make more precise
statements of this type in Section~\ref{sec:max}, where we will solve the
problem of maximizing entropy on a set with similarities~-- and, in
particular, maximizing entropy on a graph.
\end{example}

\section{Properties of the similarity-sensitive diversity measures}
\lbl{sec:sim-props}
\index{similarity-sensitive!diversity}

Here we establish the algebraic and analytic properties of the
similarity-sensitive diversity measures $D_q^Z(\p)$, extending the results
already proved in Section~\ref{sec:prop-hill} for the Hill numbers (the
case $Z = I$).  Mathematically speaking, most of the properties of the
diversity measures are easy consequences of the properties of means.
However, they are given new significance by the ecological interpretation.

Each of the listed properties of $D_q^Z(\p)$ is a piece of evidence that
these measures behave logically,%
\index{diversity measure!logical behaviour of}
in the way that should be required of any diversity measure.  Contrast,
for instance, the behaviour of Shannon entropy in the oil company argument
of Example~\ref{eg:oil}.  Nearly all of the properties were first
established in the 2012 paper of Leinster and Cobbold~\cite{MDISS}.%
\index{Cobbold, Christina}

In the naive model, diversity profiles are either strictly decreasing or
constant (Proposition~\ref{propn:div-dec}), and we will show that this is
also true in general case.  But whereas in the naive model, the condition
for the profile to be constant is that all species present have equal
\emph{abundance} $p_i$, in the general case, the condition is that all
species present have equal \emph{ordinariness}\index{ordinariness}
$(Z\p)_i$ .

\begin{propn}
Let $Z$ be an $n \times n$ similarity matrix and let $\p \in \Delta_n$.
Then $D_q^Z(\p)$ is a decreasing%
\index{diversity profile!decreasing@is decreasing}
function of $q \in [0, \infty]$.  It is constant if $(Z\p)_i = (Z\p)_j$ for
all $i, j \in \supp(\p)$, and strictly decreasing otherwise.
\end{propn}

\begin{proof}
Since $D_q^Z(\p) = M_{1 - q}(\p, 1/Z\p)$, this follows from
Theorem~\ref{thm:mns-inc-ord}. 
\end{proof}

The more similar the species in a population are perceived to be, the less
the perceived diversity.  Our diversity measures conform to this intuition:

\begin{lemma}
\lbl{lemma:Z-dec}
Let $Z'$ and $Z$ be $n \times n$ similarity matrices with $Z'_{ij} \leq
Z_{ij}$ for all $i, j$.  Then $D_q^{Z'}(\p) \geq D_q^Z(\p)$ for all $\p \in
\Delta_n$ and $q \in [0, \infty]$.
\end{lemma}

\begin{proof}
Since $D_q^Z(\p) = M_{1 - q}(\p, 1/Z\p)$, this follows from the fact that
the power means are increasing (Lemma~\ref{lemma:pwr-mns-inc}).
\end{proof}

All of our examples of similarity matrices (Examples
\ref{egs:sim-mx}--\ref{eg:sim-matrix-gph}) have the additional properties
that all similarities are at most $1$ and the similarity of each species to
itself is $1$.  Assuming those properties, we can bound the range of
possible diversities:

\begin{lemma}[Range]
\lbl{lemma:dqz-range}
\index{diversity!bounds on}%
\index{diversity!range of}
Let $Z$ be an $n \times n$ similarity matrix such that $Z_{ij} \leq 1$ for
all $i, j$ and $Z_{ii} = 1$ for all $i$.  Then
\[
1 \leq D_q^Z(\p) \leq D_q(\p) \leq n
\]
for all $\p \in \Delta_n$ and $q \in [0, \infty]$.
\end{lemma}

\begin{proof}
Taking $Z' = I$ in Lemma~\ref{lemma:Z-dec} and using the hypotheses on $Z$
gives 
\[
D_q^Z(\p) \leq D_q^I(\p) = D_q(\p),
\]
and we already showed in Lemma~\ref{lemma:div-max-min}\bref{part:div-max}
that $D_q(\p) \leq n$.  It remains to prove that $D_q^Z(\p) \geq 1$.  For
each $i \in \{1, \ldots, n\}$, we have
\[
(Z\p)_i = \sum_{j = 1}^n Z_{ij} p_j.
\]
This is a mean, weighted by $\p$, of numbers $Z_{ij} \in [0, 1]$; hence
$(Z\p)_i \in [0, 1]$ and so $1/(Z\p)_i \geq 1$.  It follows that
\[
D_q^Z(\p) = M_{1 - q}(\p, 1/Z\p) \geq 1.
\]
\end{proof}

Fix a matrix $Z$ satisfying the hypotheses of Lemma~\ref{lemma:dqz-range}.
The minimum diversity $D_q^Z(\p) = 1$ is attained by any distribution in
which only one species is present:
\[
\p = (0, \ldots, 0, 1, 0, \ldots, 0).
\]
Much more difficult is to \emph{maximize} $D_q^Z(\p)$ for fixed $Z$ and
variable $\p$.  We do this in Section~\ref{sec:max}.

Since all of our examples of similarity matrices satisfy the hypotheses
of Lemma~\ref{lemma:dqz-range}, the corresponding diversities always lie in
the range $[1, n]$.  The maximum value of $n$ is attained just when $Z = I$
and $\p = \vc{u}_n$, by Lemmas~\ref{lemma:div-max-min}
and~\ref{lemma:dqz-range}. 

In the case $Z = I$, we interpreted $D_q(\p)$ as the effective number of
species in the community (Section~\ref{sec:ren-hill}).  The bounds in the
previous paragraph encourage us to interpret $D_q^Z(\p)$ as the effective%
\index{effective number!species@of species}
number of species for a general matrix $Z$ (at least if it satisfies the
hypotheses of Lemma~\ref{lemma:dqz-range}).  More precisely, $D_q^Z(\p)$ is
the effective number of \emph{completely \lbl{p:en-cd}dissimilar} species,
since a community of $n$ equally abundant, completely dissimilar species
has diversity~$n$.

It is nearly true that the diversity $D_q^Z(\p)$ is continuous in each of
$q$, $Z$ and $\p$.  The precise statement is as follows.

\begin{lemma}
\lbl{lemma:dqz-cts}
\index{diversity!continuity of}
\begin{enumerate}
\item 
\lbl{part:dqz-cts-q}
Let $Z$ be an $n \times n$ similarity matrix and $\p \in \Delta_n$.  Then
$D_q^Z(\p)$ is continuous in $q \in [0, \infty]$.

\item
\lbl{part:dqz-cts-Z}
Let $q \in [0, \infty]$ and $\p \in \Delta_n$.  Then $D_q^Z(\p)$ is
continuous in $n \times n$ similarity matrices $Z$.

\item
\lbl{part:dqz-cts-p}
Let $q \in (0, \infty)$ and let $Z$ be an $n \times n$ similarity matrix.
Then $D_q^Z(\p)$ is continuous in $\p \in \Delta_n$.
\end{enumerate}
\end{lemma}

Part~\bref{part:dqz-cts-q} states that diversity profiles are continuous,
just as in the naive model.

The first two parts follow immediately from results on power means, but the
third does not.  The subtlety is that the sum
\begin{equation}
\lbl{eq:div-explicit}
D_q^Z(\p)
=
\Biggl( 
\sum_{i \in \supp(\p)} p_i (Z\p)_i^{q - 1} 
\Biggr)^{1/(1 - q)}
\end{equation}
in the definition of diversity is taken only over $\supp(\p)$.  Thus, if
$p_i = 0$ then the contribution of the $i$th species to the sum is $0$.
However, if $p_i$ is nonzero but small then $(Z\p)_i$ may be small, which
if $q < 1$ means that $(Z\p)_i^{q - 1}$ is large; we need to show that,
nevertheless, $p_i(Z\p)_i^{q - 1}$ is close to $0$.

\begin{proof}
Part~\bref{part:dqz-cts-q} follows from Lemma~\ref{lemma:pwr-mns-cts-t}, and
part~\bref{part:dqz-cts-Z} from Lemma~\ref{lemma:pwr-mns-cts-x}.  

For part~\bref{part:dqz-cts-p}, we split into three cases: $q \in (1,
\infty)$, $q \in (0, 1)$, and $q = 1$.

If $q \in (1, \infty)$ then the sum in equation~\eqref{eq:div-explicit} can
equivalently be taken over $i \in \{1, \ldots, n\}$ (and the summands are
still well-defined), so the result is clear.

Now let $q \in (0, 1)$.  Define functions $\phi_1, \ldots, \phi_n \from
\Delta_n \to \R$ by
\[
\phi_i(\p) =
\begin{cases}
p_i (Z\p)_i^{q - 1}     &\text{if } p_i > 0,    \\
0                       &\text{otherwise}.
\end{cases}
\]
Then $D_q^Z(\p) = \bigl( \sum_{i = 1}^n \phi_i(\p) \bigr)^{1/(1 - q)}$, so
it suffices to show that each $\phi_i$ is continuous.

Fix $i \in \{1, \ldots, n\}$.  Write
\[
\Delta_n^{(i)} 
=
\{ \p \in \Delta_n \such p_i > 0 \}.
\]
Then $\phi_i$ is continuous on $\Delta_n^{(i)}$ and zero on its
complement, so all we have to prove is that if $\p \in \Delta_n$ with $p_i
= 0$ then $\phi_i(\vc{r}) \to 0$ as $\vc{r} \to \p$, and we may as well
constrain $\vc{r}$ to lie in $\Delta_n^{(i)}$.  We have $(Z\vc{r})_i \geq
Z_{ii} r_i$, so
\begin{equation}
\lbl{eq:sandwich-bigger}
0 
\leq 
\phi_i(\vc{r})
\leq 
r_i (Z_{ii} r_i)^{q - 1}
=
Z_{ii}^{q - 1} r_i^q
\end{equation}
(since $q < 1$).  Note that $Z_{ii} > 0$ by definition of similarity
matrix, so $Z_{ii}^{q - 1}$ is finite.  As $\vc{r} \to \p$, we have $r_i^q
\to p_i^q = 0$ (since $q > 0$).  Hence the
bounds~\eqref{eq:sandwich-bigger} give $\phi_i(\vc{r}) \to 0$, as required.

Finally, consider $q = 1$.  Define functions $\psi_1, \ldots, \psi_n \from
\Delta_n \to \R$ by
\[
\psi_i(\p) =
\begin{cases}
(Z\p)_i^{-p_i}          &\text{if } p_i > 0,    \\
1                       &\text{otherwise}.
\end{cases}
\]
Then $D_q^Z(\p) = \prod_{i = 1}^n \psi_i(\p)$, so it suffices to show that
each $\psi_i$ is continuous.

Fix $i \in \{1, \ldots, n\}$.  As in the
previous case, it suffices to show that if $\p \in \Delta_n$ with $p_i =
0$, then $\psi_i(\vc{r}) \to 1$ as $\vc{r} \to \p$ with $\vc{r} \in
\Delta_n^{(i)}$.  Writing $K = \max_j Z_{ij}$, we have
\[
Z_{ii} r_i
\leq 
(Z\vc{r})_i 
=
\sum_{j = 1}^n Z_{ij} r_j
\leq
\sum_{j = 1}^n K r_j
=
K,
\]
so
\begin{equation}
\lbl{eq:sandwich-one}
K^{-r_i}
\leq
\psi_i(\vc{r})
\leq 
Z_{ii}^{-r_i} r_i^{-r_i}.
\end{equation}
Now $K \geq Z_{ii} > 0$, so $K^{-r_i} \to 1$ and $Z_{ii}^{-r_i} \to 1$ as
$\vc{r} \to \p$.  Also, $\lim_{x \to 0+} x^x = 1$, so $r_i^{-r_i} \to 1$ as
$\vc{r} \to \vc{p}$.  Hence the bounds~\eqref{eq:sandwich-one} give
$\psi_i(\vc{r}) \to 1$ as $\vc{r} \to \vc{p}$.
\end{proof}

\begin{remark}
The cases $q = 0$ and $q = \infty$ were excluded from the
statement of Lemma~\ref{lemma:dqz-cts}\bref{part:dqz-cts-p} because
$D_q^Z(\p)$ is \emph{not} continuous in $\p$ when $q$ is $0$ or $\infty$.
We have already seen that $D_0^Z$ is discontinuous even in the
naive case $Z = I$, where $D_0^Z(\p) = D_0(\p)$ is the species richness
$\mg{\supp(\p)}$.  The diversity
\[
D_\infty^Z(\p) = 1\Big/\!\!\max_{i \in \supp(\p)} (Z\p)_i
\]
of order $\infty$ is continuous when $Z = I$, but not in general.
For example, let
\[
Z = 
\begin{pmatrix}
1       &1      &0      \\
1       &1      &1      \\
0       &1      &1
\end{pmatrix}
\]
(a similarity matrix that we will meet again in
Example~\ref{eg:graph-lin3}).  For $0 \leq t < 1/2$, put
\[
\p
=
\begin{pmatrix}
\hlf - t \\
2t      \\
\hlf - t
\end{pmatrix}.
\]
Then
\[
Z\p
=
\begin{pmatrix}
\hlf + t \\
1       \\
\hlf + t 
\end{pmatrix},
\]
so 
\[
D_\infty^Z(\p)
=
\begin{cases}
1       &\text{if } t > 0,      \\
2       &\text{if } t = 0.
\end{cases}
\]
Hence $D_\infty^Z$ is discontinuous. 

The idea behind this counterexample is that the second species is so
closely related to the other two that it appears more ordinary than them
($(Z\p)_2 = \max_i (Z\p)_i$) even if it
is very rare itself ($t$ is small).  However, if the second species
disappears entirely ($t = 0$) then its ordinariness $(Z\p)_2$ is excluded
from the maximum that defines $D_\infty^Z(\p)$, causing the discontinuity.
\end{remark}

Next we establish three properties of the measures that are logically%
\index{diversity measure!logical behaviour of}
fundamental.  We will deduce all of them from a naturality property (in the
categorical sense of natural transformations), following a strategy similar
to the one we used for power means (Section~\ref{sec:pwr-mns}).  Let
\begin{equation}
\lbl{eq:set-map}
\theta \from \{1, \ldots, m\} \to \{1, \ldots, n\}  
\end{equation}
be a map of sets ($m, n \geq 1$), let $\p \in \Delta_m$, and let $Z$ be an
$n \times n$ similarity matrix.  Then we obtain a pushforward distribution
$\theta \p \in \Delta_n$ (Definition~\ref{defn:pfwd}) and an $m \times m$
similarity matrix $Z\theta$\ntn{Ztheta} defined by
\[
(Z\theta)_{i i'} = Z_{\theta(i), \theta(i')}
\]
($i, i' \in \{1, \ldots, m\}$).  

\begin{lemma}[Naturality]
\lbl{lemma:div-nat}
\index{diversity!naturality of}%
\index{naturality!similarity-sensitive diversity@of similarity-sensitive diversity}
With $\theta$, $\p$ and $Z$ as above,
\[
D_q^{Z\theta}(\p) = D_q^Z(\theta\p)
\]
for all $q \in [0, \infty]$.
\end{lemma}

\begin{proof}
We use the naturality property of the
power means (Lemma~\ref{lemma:pwr-mns-nat}), which implies that
\begin{equation}
\lbl{eq:mns-div-nat}
M_{1 - q}(\theta\p, \vc{x})
=
M_{1 - q}(\p, \vc{x}\theta)
\end{equation}
for all $\vc{x} \in [0, \infty)^n$.  Let us adopt the convention that unless
  indicated otherwise, the indices $i$ and $i'$ range over $\{1, \ldots,
  m\}$ and the indices $j$ and $j'$ range over $\{1, \ldots, n\}$.  Then
\begin{align*}
\bigl( (Z\theta)\p \bigr)_i     &
=
\sum_{i'} (Z\theta)_{ii'} p_{i'} 
=
\sum_{i'} Z_{\theta(i), \theta(i')} p_{i'},     \\
\bigl( Z (\theta\p) \bigr)_j    &
=
\sum_{j'} Z_{jj'} (\theta\p)_{j'}
=
\sum_{j'} \sum_{i' \in \theta^{-1}(j')} Z_{jj'} p_{i'}
=
\sum_{i'} Z_{j, \theta(i')} p_{i'}.
\end{align*}
Hence 
\[
\bigl( (Z\theta) \p \bigr)_i 
=
Z(\theta\p)_{\theta(i)}
\]
for all $i$, or equivalently,
\begin{equation}
\lbl{eq:div-nat-mid}
(Z\theta) \p = \bigl( Z (\theta\p) \bigr) \theta.
\end{equation}
Now
\begin{align*}
D_q^Z(\theta\p) &
=
M_{1 - q}\Biggl(
\theta\p, \frac{1}{Z(\theta\p)}
\Biggr)  \\
&
=
M_{1 - q} \Biggl( 
\p, 
\frac{1}{Z(\theta\p)} \theta    
\Biggr) \\
&
=
M_{1 - q} \Biggl( 
\p, 
\frac{1}{\bigl(Z(\theta\p)\bigr)\theta} 
\Biggr),
\end{align*}
where the second equality follows from equation~\eqref{eq:mns-div-nat} and
the others are immediate.  Equation~\eqref{eq:div-nat-mid} now gives
\[
D_q^Z(\theta\p) 
=
M_{1 - q} \Biggl( \p, \frac{1}{(Z\theta)\p} \Biggr)       
=
D_q^{Z\theta}(\p),
\]
as required.
\end{proof}

From naturality, we deduce three elementary properties of the diversity
measures.  (In the special case of the Hill numbers, $Z = I$, the first two
already appeared as Lemma~\ref{lemma:hill-elem}.)  First, diversity is
independent of the order in which the species are listed:

\begin{lemma}[Symmetry]
\lbl{lemma:dqz-sym}
\index{symmetric!diversity measure}
\index{diversity!symmetry of}
Let $Z$ be an $n \times n$ similarity matrix, let $\p \in \Delta_n$, and
let $\sigma$ be a permutation of $\{1, \ldots, n\}$.  Define $Z'$ and $\p'$
by $Z'_{ij} = Z_{\sigma(i), \sigma(j)}$ and $p'_i = p_{\sigma(i)}$.  Then
$D_q^{Z'}(\p') = D_q^Z(\p)$ for all $q \in [0, \infty]$.  
\end{lemma}

\begin{proof}
By definition, $Z' = Z\sigma$ and $\p = \sigma\p'$, so the result follows
from Lemma~\ref{lemma:div-nat}.
\end{proof}

Diversity is also unchanged by ignoring any species with abundance $0$:

\begin{lemma}[Absence-invariance]
\lbl{lemma:dqz-abs}
\index{absence-invariance!diversity measure@of diversity measure}
\index{diversity!absence-invariance of}
Let $Z$ be an $n \times n$ similarity matrix, and let $\p \in \Delta_n$ with
$p_n = 0$.  Write $Z'$ for the restriction of $Z$ to the first $n - 1$
species, and write $\p' = (p_1, \ldots, p_{n - 1}) \in \Delta_{n - 1}$.
Then $D_q^{Z'}(\p') = D_q^Z(\p)$ for all $q \in [0, \infty]$. 
\end{lemma}

\begin{proof}
Let $\theta$ be the inclusion $\{1, \ldots, n - 1\} \incl \{1, \ldots,
n\}$.  Then $Z' = Z\theta$ and $\p = \theta\p'$, so the result follows from
Lemma~\ref{lemma:div-nat}. 
\end{proof}

Third and finally, if two species are identical, then merging them into one
leaves the diversity unchanged:

\begin{lemma}[Identical species]
\lbl{lemma:dqz-id}
\index{identical species}%
\index{diversity!identical species property}%
Let $Z$ be an $n \times n$ similarity matrix such that
\[
Z_{in} = Z_{i, n - 1}, 
\qquad
Z_{ni} = Z_{n - 1, i}
\]
for all $i \in \{1, \ldots, n\}$.  Let $\p \in \Delta_n$.  Write $Z'$ for
the restriction of $Z$ to the first $n - 1$ species, and define $\p' \in
\Delta_{n - 1}$ by
\[
p'_j
=
\begin{cases}
p_j             &\text{if } j < n - 1,  \\
p_{n - 1} + p_n &\text{if } j = n - 1.
\end{cases}
\]
Then $D_q^{Z'}(\p') = D_q^Z(\p)$ for all $q \in [0, \infty]$.
\end{lemma}

\begin{proof}
Define a function $\theta \from \{1, \ldots, n\} \to \{1, \ldots, n - 1\}$
by
\[
\theta(i)
=
\begin{cases}
i       &\text{if } i < n,       \\
n - 1   &\text{if } i = n.
\end{cases}
\]
Then $Z = Z'\theta$ and $\p' = \theta\p$, so the result follows from
Lemma~\ref{lemma:div-nat}. 
\end{proof}

The identical species property means that `a community of 100 species that
are identical in every way is no different from a community of only one
species' (Ives~\cite{Ives}%
\index{Ives, Anthony}, 
p.~102).

The boundaries between species%
\index{species!classification of} 
can be changeable and somewhat arbitrary, not only for microscopic life but
even for well-studied large mammals.  (For example, the classification of
the lemurs of Madagascar has changed frequently; see Mittermeier et
al~\cite{MGKG}.)  The challenge this poses for the quantification of
diversity has long been recognized.  Good wrote in 1982 of the need to
measure diversity in a way that resolves `the difficult ``species
problem''\,' and avoids `the Platonic all-or-none approach to the
definition of species' (\cite{GoodDCM},%
\index{Good, Jack}
p.~562).

Incorporating species similarity into diversity measurement, as we have
done, allows these challenges to be met.  In particular, our
measures behave reasonably when species are reclassified, as the following
example shows.

\begin{example}
\lbl{eg:reclassification}
This hypothetical example is from~\cite{MDISS}.
Consider a system of three totally dissimilar species with relative
abundances $\p = (0.1, 0.3, 0.6)$.  Suppose that on the basis of new
genetic evidence, the last species is reclassified%
\index{species!classification of} 
\index{reclassification of species}
into two separate species of equal abundance, so that the relative
abundances become $(0.1, 0.3, 0.3, 0.3)$.

If the two new species are assumed to be totally dissimilar to one another,
the diversity profile changes dramatically
(Figure~\ref{fig:reclassification}).  For example, the diversity of order
$\infty$ jumps by $100\%$, from $1.66\ldots$ to $3.33\ldots$\:\  Of course,
it is wholly unrealistic to assume that the new species are totally
dissimilar, given that until recently they were thought to be identical.  But
if, more realistically, the two new species are assigned a high
similarity, the diversity profile changes only slightly.
Figure~\ref{fig:reclassification} shows the profile based on similarities
$Z_{34} = Z_{43} = 0.9$ between the two new species (and $Z_{ij} = 0$ for
$i \neq j$ otherwise).

\begin{figure}
\centering
\lengths
\begin{picture}(120,54)(0,4)
\cell{63}{35}{c}{\includegraphics[width=118\unitlength]{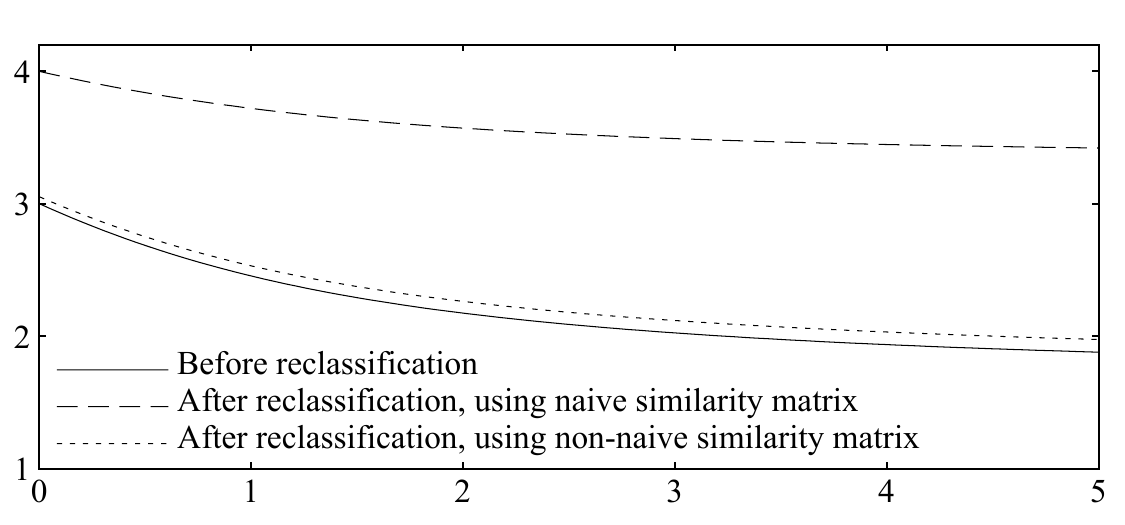}}
\cell{0}{35}{l}{\rotatebox{90}{Diversity of order $q$}}
\cell{65}{4}{b}{Viewpoint parameter, $q$}
\end{picture}
\caption{Diversity profiles of a hypothetical community before and after a
  species is reclassified (Example~\ref{eg:reclassification}).  (Figure
  adapted from~\cite{MDISS}, Figure~1.)}  
\lbl{fig:reclassification}
\end{figure}

This sensible behaviour is guaranteed by two features of the diversity
measures: the identical%
\index{identical species} 
species property and continuity in $Z$.
%
%
Indeed, if the two new species were deemed to be identical then the profile
would be unchanged.  So by continuity, if the new species are deemed to be
nearly identical then the profile is nearly unchanged.
\end{example}

For similar reasons, the diversity measures $D_q^Z$ behave reasonably under
changes of the level of resolution\index{resolution} in the data.  For
example, suppose that an initial, crude, survey of a community gathers
population abundance data at the genus level, a second survey records
abundance at the species level, and a third records abundance at the
subspecies level.  Provided that similarity is measured coherently, the
resulting three diversities will be comparable, in the sense of being
measured on the same scale.  The more fine-grained the data is, the more
variation becomes visible, so the diversity will be greater for the later
surveys.  But for the same reasons as in Example~\ref{eg:reclassification},
it will not jump disproportionately from one survey to the next.  There
will only be a large difference between the diversities calculated from the
first and second surveys if there is a large amount of variation within
genera.  Similarly, the difference between the diversities obtained from
the second and third surveys faithfully reflects the amount of
intraspecific variation.

In Propositions~\ref{propn:hill-chn} and~\ref{propn:hill-chn-esc}, we
proved two forms of the chain rule for the Hill numbers, interpreting them
as formulas for the diversity of a community spread across several
islands%
\index{islands!diversity of group of}
in terms of the diversities and relative sizes of those islands.  The
islands were assumed to have no species in common.  We now
derive two forms of the chain rule for the more general
similarity-sensitive diversity measures $D_q^Z(\p)$, under the stronger
assumption that the species on different islands are not only distinct, but
also completely dissimilar.

Thus, consider $n$ island communities with relative abundance
distributions $\p^1 \in \Delta_{k_1}, \ldots, \p^n \in \Delta_{k_n}$,
similarity matrices $Z^1, \ldots, Z^n$, and relative sizes $w_1, \ldots,
w_n$ (in the sense of Example~\ref{eg:comp-islands}).  The species
distribution of the whole group is, then,
\[
\vc{w} \of (\p^1, \ldots, \p^n) 
\in 
\Delta_k,
\]
where $k = k_1 + \cdots + k_n$.  Assuming that the species on different
islands are completely dissimilar, the $k \times k$ similarity matrix $Z$
for the whole group is the block sum
\[
Z
=
Z^1 \oplus \cdots \oplus Z^n
=
\begin{pmatrix}
Z^1     &0      &\cdots &0      \\
0       &Z^2    &\ddots &\vdots \\
\vdots  &\ddots &\ddots &0      \\
0       &\cdots &0      &Z^n
\end{pmatrix}.
\ntn{oplusmx}
\]
So, the diversity of the whole is
\[
D_q^Z\bigl( \vc{w} \of (\p^1, \ldots, \p^n) \bigr),
\]
and our task is to express this in terms of the islands' diversities
$D_q^{Z^i}(\p^i)$ and their relative sizes $w_i$.

\begin{propn}[Chain rule]
\lbl{propn:div-sim-ch}
\index{chain rule!diversity@for diversity}%
\index{diversity!chain rule for}
Let $q \in [0, \infty]$ and $n, k_1, \ldots, k_n \geq 1$.  For each $i \in
\{1, \ldots, n\}$, let $Z^i$ be a $k_i \times k_i$ similarity matrix and
let $\p^i \in \Delta_{k_i}$; also, let $\vc{w} \in \Delta_n$.  Write $Z =
Z^1 \oplus \cdots \oplus Z^n$ and $d_i = D_q^{Z^i}(\p^i)$.
\begin{enumerate}
\item 
\lbl{part:dsc-mn}
We have
\begin{align*}
D_q^Z\bigl(\vc{w} \of (\p^1, \ldots, \p^n)\bigr)        &
=
M_{1 - q}(\vc{w}, \vc{d}/\vc{w})        \\
&
=
\begin{cases}
\bigl(\sum w_i^q d_i^{1 - q}\bigr)^{1/(1 - q)}  &
\text{if } q \neq 1, \infty, \\
\prod (d_i/w_i)^{w_i}   &
\text{if } q = 1,       \\
\min d_i/w_i    &
\text{if } q = \infty,
\end{cases}
\end{align*}
where $\vc{d}/\vc{w} = (d_1/w_1, \ldots, d_n/w_n)$ and the sum,
product, and minimum are over all $i \in \supp(\vc{w})$. 

\item
\lbl{part:dsc-esc}
For $q < \infty$, 
\[
D_q^Z\bigl(\vc{w} \of (\p^1, \ldots, \p^n)\bigr) 
=
D_q(\vc{w}) \cdot M_{1 - q}(\vc{w}^{(q)}, \vc{d}),
\]
where $\vc{w}^{(q)}$ is the escort distribution defined after
Proposition~\ref{propn:hill-chn}. 
\end{enumerate}
\end{propn}

\begin{proof}
For~\bref{part:dsc-mn}, an elementary calculation shows that 
\[
Z\bigl(\vc{w} \of (\p^1, \ldots, \p^n)\bigr) 
=
w_1 (Z^1\p^1) \oplus\cdots\oplus w_n (Z^n\p^n).
\]
Hence, using the chain rule for the power means and then the homogeneity of
the power means,
\begin{align*}
&
D_q^Z\bigl(\vc{w} \of (\p^1, \ldots, \p^n)\bigr)        \\
&
=
M_{1 - q} \Biggl(
\vc{w} \of (\p^1, \ldots, \p^n),
\frac{1}{w_1(Z^1\p^1)} \oplus \cdots \oplus \frac{1}{w_n(Z^n\p^n)}
\Biggr)  \\
&
=
M_{1 - q} \Biggl(
\vc{w}, 
\Biggl(
M_{1 - q} \Biggl( \p^1, \frac{1}{w_1 (Z^1\p^1)} \Biggr),
\ldots,
M_{1 - q} \Biggl( \p^n, \frac{1}{w_n (Z^n\p^n)} \Biggr)
\Biggr)  
\Biggr)\\
&
=
M_{1 - q} \Biggl(
\vc{w}, \Biggl(\frac{d_1}{w_1}, \ldots, \frac{d_n}{w_n}\Biggr)
\Biggr).
\end{align*}
This proves the first equality in~\bref{part:dsc-mn}, and the second
follows from the explicit formulas for the power means.
Lemma~\ref{lemma:mean-esc} then gives~\bref{part:dsc-esc}.
\end{proof}

In particular, the diversity of the overall community depends only on the
sizes and diversities of the islands:

\begin{cor}[Modularity]
\index{modularity!similarity-sensitive diversity@of similarity-sensitive diversity}%
\index{diversity!modularity of}
In the situation of Proposition~\ref{propn:div-sim-ch}, the total
diversity $D_q^Z\bigl(\vc{w} \of (\p^1, \ldots, \p^n)\bigr)$ depends only
on $q$, $\vc{w}$, and $D_q^Z(\p^1)$, \ldots, $D_q^Z(\p^n)$.  
\qed
\end{cor}

A further consequence of the chain rule is also important.  Suppose that
the islands all have the same size and the same diversity, $d$.  (For
example, the islands will have the same diversity if they all have
the same species distributions, but on disjoint sets of
species; or formally, if $k_1 = \cdots = k_n$ and $\p^1 = \cdots = \p^n$.)
Then in the notation of Proposition~\ref{propn:div-sim-ch},
\[
\vc{d}/\vc{w}
=
\bigl(d/(1/n), \ldots, d/(1/n)\bigr)
=
(nd, \ldots, nd),
\]
so
\[
D_q^Z\bigl(\vc{w} \of (\p^1, \ldots, \p^n)\bigr) = nd.
\]
In other words, the diversity of a group of $n$ islands, each having
diversity $d$, is $nd$.  This is the
\demph{replication\lbl{p:dqz-rep}%
\index{replication principle}
principle} for the similarity-sensitive measures $D_q^Z$.  It generalizes
the replication principle for the Hill numbers, noted at the end of
Section~\ref{sec:prop-hill}.  The fact that our diversity measures satisfy
it means that they do not suffer from the problems described in the oil
company example (Example~\ref{eg:oil}).

Since the diversity $D_q^Z$, R\'enyi entropies $H_q^Z$, and $q$-logarithmic
entropies $S_q^Z$%
\index{similarity-sensitive!q-logarithmic entropy@$q$-logarithmic entropy}%
\index{q-logarithmic entropy@$q$-logarithmic entropy!similarity-sensitive}
are all related to one another by invertible transformations, the chain
rule for $D_q^Z$ can be translated into chain rules for $H_q^Z$ and
$S_q^Z$.  In the case of $S_q^Z$, it takes a simple form, generalizing the
chain rule for $q$-logarithmic entropy (equation~\eqref{eq:q-ent-chain}):

\begin{propn}[Chain rule]
\lbl{propn:q-ent-sim-ch}%
\index{chain rule!diversity@for diversity}%
\index{diversity!chain rule for}
Let $q \in [0, \infty)$.  For $\vc{w}$, $\vc{p}^i$, $Z^i$ and $Z$ as in 
Proposition~\ref{propn:div-sim-ch}, 
\[
S_q^Z\bigl( \vc{w} \of (\p^1, \ldots, \p^n) \bigr)
=
S_q(\vc{w}) + \sum_{i \in \supp(\vc{w})} w_i^q \cdot S_q^{Z^i}(\p^i).
\]
\end{propn}

\begin{proof}
Proposition~\ref{propn:div-sim-ch}\bref{part:dsc-mn} gives
\[
\ln_q\bigl( D_q^Z \bigl( \vc{w} \of (\p^1, \ldots, \p^n) \bigr) \bigr)
=
\ln_q \bigl( M_{1 - q} (\vc{w}, \vc{d}/\vc{w}) \bigr).
\]
By definition of $S_q^Z$ (equation~\eqref{eq:defn-sqz}) and
Lemma~\ref{lemma:q-log-mean}, an equivalent statement is that
\[
S_q^Z\bigl( \vc{w} \of (\p^1, \ldots, \p^n) \bigr)
=
\sum_{i \in \supp(\vc{w})} w_i \ln_q \frac{d_i}{w_i}.
\]
But writing $d_i/w_i = (1/w_i)d_i$
and applying the formula~\eqref{eq:q-log-mult} for the $q$-logarithm of a
product, the right-hand side is
\begin{align*}
&
\sum_{i \in \supp(\vc{w})} w_i \Biggl(
\ln_q \frac{1}{w_i} 
+ \Biggl( \frac{1}{w_i} \Biggr)^{1 - q} \ln_q d_i
\Biggr)        \\
&
=
\sum_{i \in \supp(\vc{w})} w_i \ln_q \frac{1}{w_i}
+
\sum_{i \in \supp(\vc{w})} w_i^q \ln_q d_i      \\
&
=
S_q(\vc{w}) + \sum_{i \in \supp(\vc{w})} w_i^q \cdot S_q^Z(\p^i).
\end{align*}
%
\end{proof}
\index{Cobbold, Christina|)}

\section{Maximizing diversity}
\lbl{sec:max}
\index{diversity!maximization of}
\index{Meckes, Mark|(}

Consider a community made up of organisms drawn from a fixed list of
species, whose similarities to one another are known.  Suppose that we can
control the abundances of the species within the community.  How should we
choose those abundances in order to maximize the diversity, and what is the
maximum diversity achievable?

In mathematical terms, fix an $n \times n$ similarity matrix $Z$.
The fundamental questions are these:
\begin{itemize}
\item 
Which distributions $\p$ maximize the diversity $D_q^Z(\p)$
of order $q$?

\item
What is the value of the maximum diversity, $\sup_{\p \in \Delta_n}
D_q^Z(\p)$? 
\end{itemize}
In principle, the answers to both questions depend on $q$.  After all, we
have seen that when comparing two abundance distributions, different values
of $q$ may produce different judgements on which of the distributions is
more diverse (as in Examples~\ref{eg:hill-birds} and~\ref{eg:devries}).
For instance, there seems no reason to suppose that a distribution
maximizing diversity of order $1$ will also maximize diversity of order $2$.
Similarly, we have seen nothing to suggest that the maximum diversity
$\sup_\p D_1^Z(\p)$ of order $1$ should be equal to the maximum diversity
of order~$2$.

However, it is a theorem that as long as $Z$ is symmetric, the answers to
both questions are indeed independent of $q$.  That is, every symmetric
similarity matrix has an unambiguous maximum diversity, and there is a
distribution $\p$ that maximizes $D_q^Z(\p)$ for all $q$ simultaneously.

This result was first stated and proved by Leinster~\cite{METAMB}.  An
improved proof and further results were given in a paper of Leinster and
Meckes~\cite{MDBB},%
\index{Meckes, Mark} 
from which much of this section is adapted.  We omit most proofs, referring
to~\cite{MDBB}.

Before stating the theorem, let us explore the maximum diversity problem
informally.

\begin{examples}
\lbl{egs:max-informal}
If there is only one species ($n = 1$) then the problem is trivial.  If
there are two then, assuming that $Z$ is symmetric, their roles are
interchangeable, so the distribution that maximizes diversity will clearly
be $(1/2, 1/2)$.

Now consider a three-species pond community consisting of two highly
similar species of frog and one species of newt.  If we ignore the
similarity between the species of frog and give the three species equal
status, then the maximizing distribution should be uniform: $(1/3, 1/3,
1/3)$.  But intuitively, this is not the distribution that maximizes
diversity, since it is $2/3$ frog and $1/3$ newt.  At the other extreme, if
we treat the two frog species as identical, then diversity is maximized
when there are equal quantities of frogs and newts (as in the two-species
example); so, the distribution $(1/4, 1/4, 1/2)$ should maximize diversity.
In reality, with a reasonable measure of similarities between species, the
distribution that maximizes diversity should be somewhere between these two
extremes.  We will see in Example~\ref{eg:max-specific-pond} that this is
indeed the case.
\end{examples}

\femph{For the rest of this section}, fix an integer $n \geq 1$ and an $n
\times n$ symmetric similarity matrix $Z$.  The symmetry hypothesis
matters, as we will see in Example~\ref{eg:nonsym}.

\begin{thm}[Maximum diversity]
\lbl{thm:max}%
\index{maximum diversity!theorem}
\begin{enumerate}
\item 
\lbl{part:max-dist}
There exists a probability distribution on $\{1, \ldots, n\}$ that
maximizes $D_q^Z$ for all $q \in [0, \infty]$ simultaneously.

\item
\lbl{part:max-value}
The maximum diversity $\sup_{\p \in \Delta_n} D_q^Z(\p)$ is independent of
$q \in [0, \infty]$.
\end{enumerate}
\end{thm}

\begin{proof}
This is Theorem~1 of~\cite{MDBB}.
\end{proof}

Let us say that a probability distribution $\p \in \Delta_n$ is
\demph{maximizing}%
\index{maximizing distribution}%
\index{probability distribution!maximizing} 
(with respect to $Z$) if $D_q^Z(\p)$ maximizes $D_q^Z$ for each $q \in [0,
  \infty]$.  Theorem~\ref{thm:max}\bref{part:max-value} immediately implies
that the diversity profile of a maximizing distribution is flat:

\begin{cor}
Let $\p$ be a maximizing distribution.  Then $D_q^Z(\p) = D_{q'}^Z(\p)$ for
all $q, q' \in [0, \infty]$.
\qed
\end{cor}

Theorem~\ref{thm:max} can be understood as follows
(Figure~\ref{fig:max-vis}\hardref{(a)}).  Each particular value of the
viewpoint%
\index{viewpoint!parameter}
parameter $q$ ranks the set of all distributions $\p$ in order of
diversity, with $\p$ placed above $\p'$ when $D_q^Z(\p) > D_q^Z(\p')$.
Different values of $q$ rank the set of distributions differently.
Nevertheless, there is a distribution $\pmax$ that is at the top of every
ranking.  This is the content of Theorem~\ref{thm:max}\bref{part:max-dist}.

Alternatively, we can visualize the theorem in terms of diversity profiles
(Figure~\ref{fig:max-vis}\hardref{(b)}).  Diversity profiles may cross,
reflecting the different priorities embodied by different values of $q$.
But there is at least one distribution $\pmax$ whose profile is above every
other profile; moreover, its profile is constant.  If diversity is seen as
a positive quality, then $\pmax$ is the best of all possible worlds.

\begin{figure}
\begin{center}
\lengths
\begin{picture}(51,39.5)(-0.5,0)
\cell{25}{-2}{t}{(a)}
\cell{10}{0}{b}{$q = 0$}
\cell{10}{4}{b}{\includegraphics[width=22\unitlength]{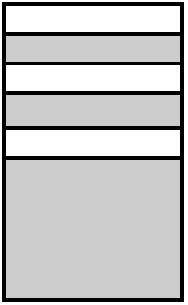}}
\cell{10}{37.3}{c}{$\pmax$}
\cell{10}{30.5}{c}{$\p$}
\cell{10}{22.7}{c}{$\p'$}
\cell{40}{0}{b}{$q = 2$}
\cell{40}{4}{b}{\includegraphics[width=22\unitlength]{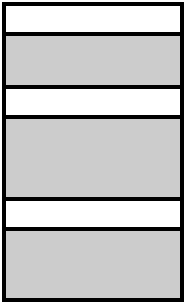}}
\cell{40}{37.3}{c}{$\pmax$}
\cell{40}{27.6}{c}{$\p'$}
\cell{40}{14.5}{c}{$\p$}
\end{picture}%
\hspace*{9\unitlength}%
\begin{picture}(60,37)
\cell{30}{-2}{t}{(b)}
\cell{30}{5}{b}{\includegraphics[width=60\unitlength]{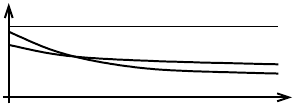}}
\cell{1}{15}{r}{$D_q^Z$}
\cell{30}{5}{t}{$q$}
\cell{10}{18.5}{c}{$\p$}
\cell{6}{14}{c}{$\p'$}
\cell{16}{23}{c}{$\pmax$}
\end{picture}
\end{center}
\caption{Visualizations of Theorem~\ref{thm:max}: (a)~in terms of how
  different values of $q$ rank the set of distributions, and (b)~in terms
  of diversity profiles.}
\lbl{fig:max-vis}
\end{figure}

Associated with the matrix $Z$ is a real number: the constant value
of any maximizing distribution.

\begin{defn}
The \demph{maximum%
\index{maximum diversity!matrix@of matrix} 
diversity} of the matrix $Z$ is \ntn{Dmax}$\Dmax{Z} =
\sup_{\p \in \Delta_n} D_q^Z(\p)$, for any $q \in [0, \infty]$.
\end{defn}

By Theorem~\ref{thm:max}\bref{part:max-value}, $\Dmax{Z}$ is independent of
$q$. 

Later, we will see how to compute the maximizing distributions and maximum
diversity of a matrix.  For now, we just note a trivial example:

\begin{example}
\lbl{eg:max-I} 
Let $Z$ be the $n \times n$ identity matrix $I$.  We have already seen that
$D_q^I(\p) = D_q(\p)$ is maximized when $\p$ is the uniform
distribution $\vc{u}_n$, and that the maximum value is $n$
(Lemma~\ref{lemma:div-max-min}\bref{part:div-max}).  It is a special case
of Theorem~\ref{thm:max}\bref{part:max-value} that this maximum value, $n$,
is independent of $q$.  In the notation just introduced, $\Dmax{I} = n$.
\end{example}

If a distribution $\p$ maximizes diversity of order $2$, must it also
maximize diversity of orders $1$ and $\infty$, for instance?  The answer
turns out to be yes:

\begin{cor}
\lbl{cor:irrelevance}
Let $\p \in \Delta_n$.  If $\p$ maximizes $D_q^Z$ for some $q \in (0,
\infty]$ then $\p$ maximizes $D_q^Z$ for all $q \in [0, \infty]$.
\end{cor}

\begin{proof}
This is Corollary~2 of~\cite{MDBB}.
\end{proof}

The significance of this result is that if we wish to find a distribution
that maximizes diversity of all orders $q$, all we need to do is to find
one that maximizes diversity of whichever nonzero order is most convenient.

The hypothesis that $q > 0$ cannot be dropped from
Corollary~\ref{cor:irrelevance}.  Indeed, take $Z = I$.  Then $D_0^I(\p)$
is species richness%
\index{species!richness} 
(the cardinality of $\supp(\p)$), which is maximized by any distribution
$\p$ of full support.  On the other hand, when $q > 0$, the diversity
$D_q^I(\p) = D_q(\p)$ is maximized only when $\p$ is uniform
(Lemma~\ref{lemma:div-max-min}\bref{part:div-max}).

\begin{remark}
Since the similarity-sensitive R\'enyi entropy $H_q^Z$ and
similarity-sensitive $q$-logarithmic entropy $S_q^Z$ are increasing
transformations of $D_q^Z$, the same distributions that maximize $D_q^Z$
for all $q$ also maximize $H_q^Z$ and $S_q^Z$ for all $q$.  And since
$H_q^Z = \log D_q^Z$, the maximum similarity-sensitive R\'enyi entropy,
$\sup_\p H_q^Z(\p)$, is also independent of $q$: it is simply $\log
\Dmax{Z}$.

In contrast, the maximum similarity-sensitive $q$-logarithmic entropy,
$\sup_\p S_q^Z(\p)$, is \emph{not} independent of $q$.  It is $\ln_q
\Dmax{Z}$, which varies with $q$.  This is one advantage%
\index{q-logarithmic entropy@$q$-logarithmic entropy!Renyi entropy@and R\'enyi entropy}%
\index{Renyi entropy@R\'enyi entropy!q-logarithmic entropy@and $q$-logarithmic entropy}
of the R\'enyi entropy (and its exponential) over the $q$-logarithmic
entropy.
\end{remark}

Theorem~\ref{thm:max} guarantees the existence of a maximizing distribution
$\pmax$, but does not tell us how to find one.  It also states that
$D_q^Z(\pmax)$ is independent of $q$, but does not tell us its value.  Our
next theorem repairs both omissions.  To state it, we need some definitions.

\begin{defn}
\lbl{defn:wtg}
A \dmph{weighting} on a matrix $M$ is a column vector $\vc{w}$ such
that $M\vc{w} = \begin{pmatrix}
  1\\ \vdots\\ 1 \end{pmatrix}$.
\end{defn}

\begin{lemma}
\lbl{lemma:wtg-cowtg}
Let $M$ be a matrix.  Suppose that $M$ and its transpose $M^\transp$
each have at least one weighting.  Then $\sum_i w_i$ is independent of the
choice of weighting $\vc{w}$ on $M$.
\end{lemma}

\begin{proof}
Let $\vc{w}$ and $\vc{w}'$ be weightings on $M$.  Choose a weighting $\vc{v}$
on $M^\transp$.  Then
\[
\sum_i w_i 
= 
(1 \ \cdots\ 1) \vc{w}
=
\bigl(M^\transp \vc{v}\bigr)^\transp \vc{w} 
=
\vc{v}^\transp M \vc{w}
=
\vc{v}^\transp \begin{pmatrix} 1\\ \vdots\\ 1 \end{pmatrix}
=
\sum_j v_j.
\]
Similarly, $\sum_i w'_i = \sum_j v_j$.  Hence $\sum_i w_i = \sum_i w'_i$.
\end{proof}

\begin{defn}
\lbl{defn:mag}
Let $M$ be a matrix such that both $M$ and $M^\transp$ have at least
one weighting.  Its \demph{magnitude}%
\index{magnitude!matrix@of matrix} 
$\mg{M}$\ntn{magmx} is $\sum_i w_i$, where $\vc{w}$ is any weighting on
$M$.
\end{defn}

By Lemma~\ref{lemma:wtg-cowtg}, the magnitude is independent of the choice
of weighting.  

\begin{remarks}
\lbl{rmks:defn-mag}
\begin{enumerate}
\item
\lbl{rmk:defn-mag-sym}
When $M$ is symmetric (the case of interest here), $\mg{M}$ is defined as
long as $M$ has at least one weighting.  

\item
\lbl{rmk:defn-mag-inv}
When $M$ is invertible, $M$ has exactly one weighting. Its entries are
the row-sums of $M^{-1}$.  Thus, $\mg{M}$ is the sum of all the entries of
$M^{-1}$.
\end{enumerate}
\end{remarks}

\begin{defn}
A vector $\vc{v} = (v_i)$ over $\R$ is
\demph{nonnegative}%
\index{vector!nonnegative}%
\index{nonnegative vector}
if $v_i \geq 0$ for all $i$, and \demph{positive}%
\index{vector!positive}%
\index{positive!vector} 
if $v_i > 0$ for all $i$.  
\end{defn}

For a nonempty subset $B \sub \{1, \ldots, n\}$, let $Z_B$\ntn{ZB} denote
the submatrix $(Z_{ij})_{i, j \in B}$ of $Z$.  This is also a symmetric
similarity matrix.  Suppose that we have a nonnegative weighting $\vc{w}$
on $Z_B$.  Then $\vc{w} \neq \vc{0}$, so $\sum_{j \in B} w_j \neq 0$.  We
can therefore define a probability distribution $\pext{\vc{w}}\ntn{pext} \in
\Delta_n$ by normalizing and extending by $0$:
\[
\pext{w}_i
=
\begin{cases}
w_i/\mg{Z_B}      &\text{if } i \in B,    \\
0               &\text{otherwise}
\end{cases}
\]
($i \in \{1, \ldots, n\}$).  

\begin{thm}[Computation of maximum diversity]
\lbl{thm:max-comp}%
\index{maximum diversity!computation of}%
\index{magnitude!maximum diversity@vs.\ maximum diversity}%
\index{maximum diversity!magnitude@vs.\ magnitude}
\begin{enumerate}
\item 
\lbl{part:max-comp-value}
The maximum diversity of $Z$ is given by
\begin{equation}
\lbl{eq:comp-val}
\Dmax{Z}
=
\max_B \mg{Z_B},
\end{equation}
where the maximum is over all nonempty $B \sub \{1, \ldots, n\}$ such that
$Z_B$ admits a nonnegative weighting.

\item
\lbl{part:max-comp-dist}
The set of maximizing distributions is
\[
\bigcup_B \{ 
\pext{\vc{w}} \such
\text{nonnegative weightings } \vc{w} \text{ on } B
\},
\]
where the union is over all $B$ attaining the maximum
in~\eqref{eq:comp-val}. 
\end{enumerate}
\end{thm}

\begin{proof}
This is Theorem~2 of~\cite{MDBB}.
\end{proof}

\begin{remark}
Let $B \sub \{1, \ldots, n\}$ be a subset attaining the maximum
in~\eqref{eq:comp-val}, and let $\vc{w}$ be a nonnegative weighting on
$Z_B$, so that $\pext{\vc{w}} \in \Delta_n$ is a maximizing distribution.
A short calculation shows that
\[
(Z\pext{\vc{w}})_i = \frac{1}{\mg{Z_B}}
\]
for all $i \in B$.  In particular, $(Z\pext{\vc{w}})_i$ is constant over $i
\in B$.  This can be understood as follows.  For the Hill numbers (the case
$Z = I$), the maximizing distribution takes the relative
abundances $p_i$ to be the same for all species $i$.  This is no longer
true when inter-species similarities are taken into account.  Instead, the
maximizing distributions have the property that the
\emph{ordinariness}\index{ordinariness} $(Z\p)_i$ is the same for all
species $i$ that are present.  

Determining which species are present%
\index{maximum diversity!elimination of species@and elimination of species}%
\index{elimination of species}%
\index{species!elimination of}
in a maximizing distribution is
not straightforward.  In particular, maximizing distributions do not always
have full support, a phenomenon discussed at the end of this section.
\end{remark}

Theorem~\ref{thm:max-comp} provides a finite-time algorithm for computing
the maximizing diversity of $Z$, as well as all its maximizing
distributions, as follows.

For each of the $2^n - 1$ nonempty subsets $B$ of $\{1, \ldots, n\}$,
perform some simple linear algebra to find the space of nonnegative
weightings on $Z_B$.  If this space is nonempty, call $B$ \demph{feasible}
and record the magnitude $\mg{Z_B}$.  Then $\Dmax{Z}$ is the maximum of all
the recorded magnitudes.  For each feasible $B$ such that $\mg{Z_B} =
\Dmax{Z}$, and each nonnegative weighting $\vc{w}$ on $Z_B$, the
distribution $\pext{\vc{w}}$ is maximizing.  This generates all of the
maximizing distributions.

This algorithm takes exponentially many steps in $n$, and
Remark~\ref{rmk:no-quick-clique} provides strong evidence that no algorithm
can compute maximum diversity in polynomial time.  But the situation is not
as hopeless as it might appear, for two reasons.

First, each step of the algorithm is fast, consisting as it does of solving
a system of linear equations.  For instance, using a standard laptop and a
standard computer algebra package, with no attempt at optimization, the
maximizing distributions of $25 \times 25$ matrices were computed in a few
seconds.  Second, for certain classes of matrices $Z$, the computing time
can be reduced dramatically, as we will see.

We first consider some examples, starting with the most simple cases.

\begin{example}
\lbl{eg:max-specific-two}
Take a $2 \times 2$ similarity matrix
\[
Z = 
\begin{pmatrix}
1       &z      \\
z       &1
\end{pmatrix},
\]
where $0 \leq z < 1$.  Let us run the algorithm just described.
\begin{itemize}
\item
First we determine for which nonempty $B \sub \{1, \ldots, n\}$ the submatrix
$Z_B$ has a nonnegative weighting, and record the magnitudes of those that
do.  

When $B = \{1\}$, the submatrix $Z_B$ is $(1)$; this has a unique
nonnegative weighting $\vc{w} = (1)$, so $\mg{Z_B} = 1$.  The same is true
for $B = \{2\}$.  When $B = \{1, 2\}$, we have $Z_B = Z$, which has a
unique nonnegative weighting 
\begin{equation}
\lbl{eq:2-wtg}
\vc{w} = \frac{1}{1 + z} \begin{pmatrix} 1\\ 1 \end{pmatrix}
\end{equation}
and magnitude $\mg{Z_B} = 2/(1 + z)$.  

\item
The maximum diversity of $Z$ is given by 
\[
\Dmax{Z} 
= 
\max\Biggl\{ 1, 1, \frac{2}{1 + z} \Biggr\}, 
\]
and $2/(1 + z) > 1$, so $\Dmax{Z} = 2/(1 + z)$.  The unique
maximizing distribution is the normalization of the
weighting~\eqref{eq:2-wtg}, which is the uniform distribution $\vc{u}_2$.
\end{itemize}
That the maximizing distribution is uniform conforms to the intuitive
expectation of Example~\ref{egs:max-informal}.  The computed value of
$\Dmax{Z}$ also conforms to the expectation that the maximum diversity
should be a decreasing function of the similarity between the species.
\end{example}

\begin{example}
\lbl{eg:max-specific-pond}
Now consider the three-species pond community of
Example~\ref{egs:max-informal}, with similarities as shown in
Figure~\ref{fig:pond}.  Implementing the algorithm or using
Proposition~\ref{propn:max-psd} below reveals that the unique maximizing
distribution is $(0.261, 0.261, 0.478)$
(to $3$ decimal places).  This confirms the intuitive guess of
Example~\ref{egs:max-informal}.

\begin{figure}
\begin{center}
\lengths
\begin{picture}(120,16)(0,-1)
\cell{0}{7}{l}{%
$Z 
= 
\begin{pmatrix} 
1       &0.9    &0.4    \\
0.9     &1      &0.4    \\
0.4     &0.4    &1
\end{pmatrix}$}
\thicklines
\put(41,-1.5){\includegraphics[width=80\unitlength]{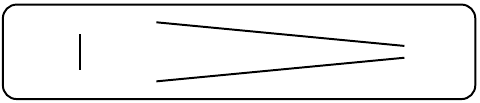}}
\cell{114}{7}{c}{newt}
\cell{66}{12}{r}{frog species A}
\cell{66}{2}{r}{frog species B}
\cell{49}{7}{l}{0.9}
\cell{89}{12}{c}{0.4}
\cell{89}{1.7}{c}{0.4}
\end{picture}%
\end{center}
\caption{Hypothetical three-species system.  Distances between species
  indicate degrees of dissimilarity between them (not to scale).}
\lbl{fig:pond}
\end{figure}
\end{example}

One of our standing hypotheses on $Z$ is symmetry.  Without it,
the main theorem fails in every respect:

\begin{example}
\lbl{eg:nonsym}
\index{symmetric!similarity matrix}
\index{similarity!matrix!nonsymmetric}
\index{maximum diversity!nonsymmetric similarity@with nonsymmetric similarity}
Let $Z = \Bigl( \begin{smallmatrix} 1 & 1/2 \\ 0 & 1 \end{smallmatrix}
\Bigr)$, which is a similarity matrix but not symmetric.  Consider a
distribution $\p = (p_1, p_2) \in \Delta_2$.  If $\p$ is $(1, 0)$ or $(0,
1)$ then $D_q^Z(\p) = 1$ for all $q$.  Otherwise,
\begin{align}
D_0^Z(\p)       &
=
3 - \frac{2}{1 + p_1},  
\lbl{eq:nonsym-0}
\\
D_2^Z(\p)       &
=
\frac{2}{3(p_1 - 1/2)^2 + 5/4}, 
\lbl{eq:nonsym-2}
\\
D_\infty^Z(\p)  &
=
\begin{cases}
\frac{1}{1 - p_1}       & 
\text{if } p_1 \leq 1/3,\\[.5ex]
\frac{2}{1 + p_1}       &
\text{if } p_1 \geq 1/3.
\end{cases}
\lbl{eq:nonsym-infty}
\end{align}
It follows 
that $\sup_{\p \in \Delta_2} D_0^Z(\p) = 2$.  However, no distribution
maximizes $D_0^Z$; we have $D_0^Z(\p) \to 2$ as $\p \to (1, 0)$, but
$D_0^Z(1, 0) = 1$.  Equations~\eqref{eq:nonsym-2}
and~\eqref{eq:nonsym-infty} imply that
\[
\sup_{\p \in \Delta_2} D_2^Z(\p) = 1.6,
\qquad
\sup_{\p \in \Delta_2} D_\infty^Z(\p) = 1.5,
\]
with unique maximizing distributions $(1/2, 1/2)$ and $(1/3, 2/3)$
respectively.

Thus, when $Z$ is not symmetric, the main theorem fails comprehensively:
the supremum $\sup_{\p \in \Delta_n} D_0^Z(\p)$ may not be attained; there
may be no distribution maximizing $\sup_{\p \in \Delta_n} D_q^Z(\p)$ for
all $q$ simultaneously; and that supremum may vary with $q$.
\end{example}

Perhaps surprisingly, nonsymmetric similarity matrices $Z$ do have
practical uses.  For example, it is shown in Proposition~A7 of the appendix
to Leinster and Cobbold~\cite{MDISS} that the mean phylogenetic diversity
measures of Chao,%
\index{Chao, Anne!Chiu--Jost@--Chiu--Jost phylogenetic diversity}
Chiu and Jost~\cite{CCJ} are a special case of the measures $D_q^Z(\p)$,
obtained by taking a particular $Z$ constructed from the phylogenetic tree
concerned.  This $Z$ is usually nonsymmetric, reflecting the asymmetry of
evolutionary time.  More generally, the case for dropping the symmetry
axiom for metric%
\index{metric!nonsymmetric} 
spaces was made by Lawvere%
\index{Lawvere, F. William} 
(p.~138--139 of~\cite{LawvMSG}), and Gromov%
\index{Gromov, Misha} 
has argued that symmetry `unpleasantly limits many applications' (p.~xv
of~\cite{GromMSR}).  So, the fact that the maximum diversity theorem fails
for nonsymmetric $Z$ is an important restriction.

Now consider finite, undirected graphs with no multiple edges (henceforth,
\demph{graphs}%
\index{graph!maximum diversity of} 
for short).  As in Example~\ref{eg:sim-matrix-gph}, any such graph
corresponds to a symmetric similarity matrix.  What, then, is the maximum
diversity of the adjacency matrix of a graph?

The answer requires some terminology.  Recall that vertices $x$ and $y$ of
a graph are said to be adjacent, written as $x \adjc y$, if there
is an edge between them.  (In particular, every vertex of a reflexive graph
is adjacent to itself.)  A set of vertices is \demph{independent}%
\index{independent!set of vertices in graph} 
if no two distinct vertices are adjacent.  The \demph{independence%
\index{independence!number}
number} $\alpha(G)$\ntn{indnum} of a graph $G$ is the number of vertices in
an independent set of greatest cardinality.

\begin{propn}
\lbl{propn:graph-main}
Let $G$ be a reflexive graph with adjacency matrix $Z$.  Then the maximum
diversity $\Dmax{Z}$ is equal to the independence number $\alpha(G)$.
\end{propn}

\begin{proof}
We will maximize the diversity of order $\infty$.  For any probability
distribution $\p$ on the vertex-set $\{1, \ldots, n\}$, 
\begin{equation}
\lbl{eq:graph-infty}
D_{\infty}^Z(\p)
=
1 \biggl/
\max_{i \in \supp(\p)} \sum_{j \csuch \adjco{i}{j}} p_j.
\end{equation}

First we show that $\Dmax{Z} \geq \alpha(G)$.  Choose an independent set
$B$ of cardinality $\alpha(G)$, and define $\p \in \Delta_n$ by
\[
p_i
=
\begin{cases}
1/\alpha(G)     &\text{if } i \in B,    \\
0               &\text{otherwise}.
\end{cases}
\]
For each $i \in \supp(\p) = B$, the sum on the right-hand side of
equation~\eqref{eq:graph-infty} is $1/\alpha(G)$.  Hence
$D_{\infty}^Z(\p) = \alpha(G)$, giving $\Dmax{Z} \geq \alpha(G)$.

Now we show that $\Dmax{Z} \leq \alpha(G)$.  By
equation~\eqref{eq:graph-infty}, an equivalent statement is that
for each $\p \in \Delta_n$, there is some $i \in \supp(\p)$ such that
\begin{equation}
\lbl{eq:gph-max-ind}
\sum_{j \csuch \adjco{i}{j}} p_j
\geq
\frac{1}{\alpha(G)}.
\end{equation}
Let $\p \in \Delta_n$.  Choose an independent
set $B \sub \supp(\p)$ with maximal cardinality among all independent
subsets of $\supp(\p)$.  Then every vertex in $\supp(\p)$ is adjacent to at
least one vertex in $B$, otherwise we could adjoin it to $B$ to make a
larger independent subset.  This gives the inequality
\[
\sum_{\,i \in B\,} \sum_{j \csuch \adjco{i}{j}} p_j
=
\sum_{\,i \in B\,} \sum_{j \in \supp(\p) \csuch \adjco{i}{j}} p_j
\geq
\sum_{j \in \supp(\p)} p_j
=
1.
\]
So we can choose some $i \in B$ such that $\sum_{j \csuch \adjco{i}{j}} p_j
\geq 1/\# B$, where $\#$ denotes cardinality.  But $\# B \leq
\alpha(G)$ since $B$ is independent, so the desired
inequality~\eqref{eq:gph-max-ind} follows.
\end{proof}

\begin{remark}
\lbl{rmk:graph-proof} 
\index{maximum diversity!elimination of species@and elimination of species}%
\index{elimination of species}%
\index{species!elimination of}
The first part of the proof (together with Corollary~\ref{cor:irrelevance})
shows that a maximizing distribution on a reflexive graph can be constructed by
taking the uniform distribution on some independent set of greatest
cardinality, then extending by zero to the whole vertex-set.  Except in the
trivial case of a graph with no edges between distinct vertices, this
maximizing distribution never has full support.
\end{remark}

\begin{example}
\lbl{eg:graph-lin3}
The reflexive graph $G =
\mbox{\ensuremath{\bullet\gedge\bullet\gedge\bullet}}$ (loops not shown) has
adjacency matrix $Z = \biggl( \begin{smallmatrix} 1 &1 &0 \\ 1& 1& 1 \\ 0&
  1& 1 \end{smallmatrix} \biggr)$.  The independence number of $G$ is $2$;
this, then, is the maximum diversity of $Z$.  There is a unique
independent set of cardinality $2$, and a unique maximizing distribution,
$(1/2, 0, 1/2)$.
\end{example}

\begin{example}
\lbl{eg:graph-lin4}
The reflexive graph
$\mbox{\ensuremath{\bullet\gedge\bullet\gedge\bullet\gedge\bullet}}$ also has
independence number $2$.  There are three independent sets of maximal
cardinality, so by Remark~\ref{rmk:graph-proof}, there are at least three
maximizing distributions,
\[
\bigl(\hlf, 0, \hlf, 0\bigr),
\qquad
\bigl(\hlf, 0, 0, \hlf\bigr),
\qquad
\bigl(0, \hlf, 0, \hlf\bigr),
\]
all with different supports.  (The possibility of multiple maximizing
distributions was also observed in the case $q = 2$ by Pavoine and
Bonsall~\cite{PaBo}.)
In fact, there are further maximizing distributions not constructed in the
proof of Proposition~\ref{propn:graph-main}, namely, $\bigl(\hlf, 0, t,
\hlf - t\bigr)$ and $\bigl(\hlf - t, t, 0, \hlf\bigr)$ for all $t \in (0,
\hlf)$.
\end{example}

\begin{example}
Kolmogorov's notion of the $\epsln$-entropy of a metric
space~\cite{KolmCAC} is approximately an instance of maximum diversity,
assuming that one is interested in its behaviour as $\epsln \to 0$ rather
than for individual values of $\epsln$.

Let $A$ be a finite metric space.  Given $\epsln > 0$, the
\demph{$\epsln$-covering%
\index{covering number} 
number} $N_\epsln(A)$ is the smallest number of closed $\epsln$-balls
needed to cover $A$.  But also associated with $\epsln$ is the graph%
\index{graph!maximum diversity of} 
$G_\epsln(A)$ whose vertices are the points of $A$ and with an edge between
$a$ and $b$ whenever $d(a, b) \leq \epsln$.  Write $Z_\epsln(A)$ for the
adjacency matrix of $G_\epsln(A)$.  From
Proposition~\ref{propn:graph-main}, it is not hard to deduce that
\[
N_\epsln(A) \leq \Dmax{Z_\epsln(A)} \leq N_{\epsln/2}(A)
\]
(Example~11 of~\cite{MDBB}).  

We have repeatedly seen that quantities called entropy tend to be the
logarithms of quantities called diversity.  Kolmogorov's
\demph{$\epsln$-entropy}%
\index{entropy!Kolmogorov}%
\index{Kolmogorov, Andrei!entropy}%
\index{metric!entropy} 
of $A$ is $\log N_\epsln(A)$, and, by the inequalities above, is closely
related to the logarithm of maximum diversity.
\end{example}

The moral of the proof of Proposition~\ref{propn:graph-main} is that by
performing the simple task of maximizing diversity of order $\infty$, we
automatically maximize diversity of all other orders.  Here is an example
of how this observation can be exploited.

Every graph $G$ has a \dmph{complement} $\gcomp{G}$, with the same
vertex-set as $G$; two vertices are adjacent in $\gcomp{G}$ if and only if
they are not adjacent in $G$.  Thus, the complement of a reflexive graph is
\demph{irreflexive}%
\index{irreflexive}%
\index{graph!irreflexive} 
(has no loops), and vice versa.  A set $B$ of vertices
in an irreflexive graph $X$ is a \dmph{clique} if all pairs of distinct
elements of $B$ are adjacent in $X$.  The \demph{clique
number} \ntn{clinum}$\cliqueno{X}$ of $X$ is the maximal cardinality of a
clique in $X$.  Thus, $\cliqueno{X} = \alpha\bigl(\gcomp{X}\bigr)$.

We now recover a result of Berarducci, Majer and Novaga (Proposition~5.10
of~\cite{BMN}).

\begin{cor}[Berarducci, Majer and Novaga]
\index{Berarducci, Alessandro}%
\index{Majer, Pietro}%
\index{Novaga, Matteo}%
%
Let $X$ be an irreflexive graph.  Then
\[
\sup_\p \sum_{(i, j) \csuch \adjco{i}{j}} p_i p_j
=
1 - \frac{1}{\cliqueno{X}},
\]
where the supremum is over probability distributions $\p$ on the vertex-set
of $X$ and the sum is over ordered pairs of adjacent vertices of $X$.
\end{cor}

\begin{proof}
Write $\{1, \ldots, n\}$ for the vertex-set of $X$, and $Z$ for the
adjacency matrix of the reflexive graph $\gcomp{X}$.  Then for all $\p \in
\Delta_n$, 
\begin{align*}
\sum_{(i, j) \csuch \adjcoin{i}{j}{X}} p_i p_j        &
=
\sum_{i, j = 1}^n p_i p_j - 
\sum_{(i, j) \csuch \adjcoin{i}{j}{\gcomp{X}}} p_i p_j  \\
&
=
1 - \sum_{i, j = 1}^n p_i Z_{ij} p_j    \\
&
=
1 - \frac{1}{D_2^Z(\p)}.
\end{align*}%
Hence by Proposition~\ref{propn:graph-main},
\[
\sup_{\p \in \Delta_n} \sum_{(i, j) \csuch \adjcoin{i}{j}{X}} p_i p_j    
=
1 - \frac{1}{\Dmax{\p}}
= 
1 - \frac{1}{\alpha\bigl(\gcomp{X}\bigr)}
=
1 - \frac{1}{\cliqueno{X}}.
\]
\end{proof}

It follows from this proof and Remark~\ref{rmk:graph-proof} that $\sum_{(i,
  j)\csuch \adjco{i}{j}} p_i p_j$ can be maximized as follows: take the
uniform distribution on some clique in $X$ of maximal cardinality, then
extend by zero to the whole vertex-set.  This distribution maximizes the
probability that two vertices chosen at random are adjacent, as in
Example~\ref{egs:dqz}\bref{eg:dqz-two}. 

\begin{remark}
\lbl{rmk:no-quick-clique}
Proposition~\ref{propn:graph-main} implies that computationally, finding
the maximum diversity of an arbitrary symmetric $n \times n$ similarity
matrix is at least as hard as finding the independence number of a
reflexive graph with $n$ vertices.  This is a very well-studied problem,
usually presented in its dual form (find the clique number of an
irreflexive graph) and called the 
\demph{maximum%
\index{maximum clique problem}%
\index{clique}
clique problem}~\cite{Karp}.  It is $\mathbf{NP}$-hard.  Hence, assuming
that $\mathbf{P} \neq \mathbf{NP}$, there is no polynomial-time algorithm
for computing maximum diversity, nor even for computing the support of a
maximizing distribution.
\end{remark}

We now return to general symmetric similarity matrices, addressing two
remaining questions: when are maximizing distributions unique, and when do
they have full support?

Recall that a real symmetric matrix $Z$ is \demph{positive%
\index{positive!definite} 
definite} if $\vc{x}^\transp
Z \vc{x} > 0$ for all $\vc{0} \neq \vc{x} \in \R^n$, and \demph{positive
  semidefinite}%
\index{positive!semidefinite} 
if $\vc{x}^\transp Z \vc{x} \geq 0$ for all $\vc{x} \in 
\R^n$.  Equivalently, $Z$ is positive definite if all its eigenvalues are
positive, and positive semidefinite if they are all nonnegative.  A
positive definite matrix is invertible and therefore has a unique weighting.

\begin{propn}
\lbl{propn:max-psd}
\begin{enumerate}
\item 
If $Z$ is positive semidefinite and has a nonnegative weighting $\vc{w}$,
then $\Dmax{Z} = \mg{Z}$ and $\vc{w}/\mg{Z}$ is a maximizing distribution.

\item
If $Z$ is positive definite and its unique weighting $\vc{w}$ is positive
then $\vc{w}/\mg{Z}$ is the unique maximizing distribution.
\end{enumerate}
\end{propn}

\begin{proof}
This is Proposition~3 of~\cite{MDBB}.
\end{proof}

In particular, if $Z$ is positive semidefinite and has a nonnegative
weighting, then computing its maximum diversity is trivial.

When $Z$ is positive definite and its unique weighting is positive,
its unique maximizing distribution eliminates no species.  Here are two
classes of such matrices $Z$.

\begin{example}
\lbl{eg:ultra-pos-def}
Call $Z$ \demph{ultrametric}\index{ultrametric!matrix} if $Z_{ik} \geq
\min\{Z_{ij}, Z_{jk}\}$ for all $i, j, k$ and $Z_{ii} > Z_{jk}$ for all $i,
j, k$ with $j \neq k$.  For instance, the matrix $Z = \bigl(e^{-d(i,
  j)}\bigr)$ of any ultrametric space is ultrametric; see
Example~\ref{eg:sim-matrix-met}.  If $Z$ is ultrametric then $Z$ is
positive definite with positive weighting, by Proposition~2.4.18 of
Leinster~\cite{MMS}.

(The positive definiteness of ultrametric matrices was also proved,
earlier, by Varga and Nabben~\cite{VaNa}, and a different proof still was
given in Theorem~3.6 of Meckes~\cite{MeckPDM}.  An earlier, indirect proof
of the positivity of the weighting can be found in Pavoine, Ollier and
Pontier~\cite{POP}.)

Such matrices arise in practice.  For instance, $Z$ is ultrametric if it is
defined from a phylogenetic or taxonomic tree as in
Examples~\ref{egs:sim}\bref{eg:sim-phylo} and~\bref{eg:sim-taxo}.  
\end{example}

\begin{example}
\lbl{eg:id-pos-def}
The identity matrix $Z = I$ is certainly positive definite with positive
weighting.  By topological arguments, there is a neighbourhood $U$ of $I$
in the space of symmetric matrices such that every matrix in $U$ also has
these properties.  (See the proofs of Propositions~2.2.6 and~2.4.6 of
Leinster~\cite{MMS}.)  

Quantitative versions of this result are also available.  For instance,
suppose that $Z_{ii} = 1$ for all $i, j$ and that $Z$ is \demph{strictly
  diagonally dominant},%
\index{diagonally dominant}%
\index{strictly diagonally dominant}
that is, $Z_{ii} > \sum_{j \neq i} Z_{ij}$ for all $i$.  Then $Z$ is
positive definite with positive weighting (Proposition~4 of Leinster and
Meckes~\cite{MDBB}).
\end{example}

In summary, if our similarity matrix $Z$ is ultrametric, or if it is close
to the matrix $I$ that encodes the naive model, then it enjoys many special
properties: the maximum diversity is equal to the magnitude, there is a
unique maximizing distribution, the maximizing distribution has full
support, and both the maximizing distribution and the maximum diversity can
be computed in polynomial time.

We saw in Examples~\ref{eg:graph-lin3} and~\ref{eg:graph-lin4} that for
some similarity matrices $Z$, no maximizing distribution has full
support.  Mathematically, this simply means that maximizing distributions
sometimes lie on the boundary of $\Delta_n$.  But ecologically, it may
sound shocking: is it reasonable that diversity can be increased by
\emph{eliminating} some species?%
\index{maximum diversity!elimination of species@and elimination of species}%
\index{elimination of species}%
\index{species!elimination of}

We argue that it is.  For example, consider a forest%
\index{forest!oak and pine} 
consisting of one species of oak and ten species of pine, with all species
equally abundant.  Suppose that an eleventh species of pine is added, with
the same abundance as all the existing species (Figure~\ref{fig:pine}).
This makes the forest even more heavily dominated by pine than it was
before, so it is intuitively reasonable that the diversity should decrease.
But now running time backwards, the conclusion is that if we start with a
forest containing the oak and all eleven pine species, then eliminating%
\index{elimination of species}%
\index{species!elimination of}
the eleventh should \emph{increase} the diversity.

\begin{figure}
\centering
\lengths
\begin{picture}(120,23)
\cell{60}{-.7}{b}{\includegraphics[width=77\unitlength]{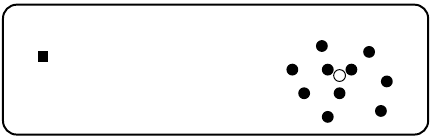}}
\cell{29.5}{17.2}{c}{oak}
\cell{82}{20}{c}{pine}
\end{picture}
\caption{Hypothetical community consisting of one species of oak
  ($\scriptscriptstyle\blacksquare$) and ten species of pine ($\bullet$), to
  which one further species of pine is then added ($\circ$).  Distances
  between species indicate degrees of dissimilarity (not to scale).}
\lbl{fig:pine}
\end{figure}

To clarify further, recall that diversity is defined in terms of the
\emph{relative}%
\index{abundance!relative vs.\ absolute} 
abundances only.  Thus, eliminating the $i$th species causes not only a
decrease in $p_i$, but also an increase in the other relative abundances
$p_j$.  If the $i$th species is particularly ordinary within the community
(like the eleventh species of pine), then eliminating it increases the
relative abundances of less ordinary species, resulting in a community that
is more diverse.

The instinct that maximizing diversity should not eliminate any species is
based on the assumption that the distinction between species is of high
value.  (After all, if two species were very nearly identical~-- or in the
extreme, actually identical~-- then losing one would be of little
importance.)  If one wishes to make that assumption, one must build it into
the model.  This is done by choosing a similarity%
\index{similarity!matrix!choice of} 
matrix $Z$ with a low similarity coefficient $Z_{ij}$ for each $i \neq j$.
Thus, $Z$ is close to the identity matrix $I$ (assuming that similarity is
measured on a scale of $0$ to $1$).  Example~\ref{eg:id-pos-def} guarantees
that in this case, there is a unique maximizing distribution and it does
not, in fact, eliminate any species.

The fact that maximizing distributions can eliminate some species has
previously been discussed in the ecological literature in the case of Rao's
quadratic entropy ($q = 2$): see Izs\'ak and Szeidl~\cite{IzSz},
Pavoine and Bonsall~\cite{PaBo}, and references therein.  The same
phenomenon was observed and explored by Shimatani in
genetics~\cite{ShimADD}, again in the case $q = 2$.

We finish by stating necessary and sufficient conditions for a symmetric
similarity matrix $Z$ to admit at least one maximizing distribution of full
support, so that diversity can be maximized without eliminating any
species.  We also state necessary and sufficient conditions for
\emph{every} maximizing distribution to have full support.  The latter
conditions are genuinely more restrictive.  For instance, if $Z =
\bigl( \begin{smallmatrix} 1& 1\\ 1& 1\\ \end{smallmatrix} \bigr)$ then
every distribution is maximizing, so some but not all maximizing
distributions have full support.

\begin{propn}
\lbl{propn:semi}
The following are equivalent:
\begin{enumerate}
\item
\lbl{part:semivec}
there exists a maximizing distribution for $Z$ of full support;

\item
\lbl{part:semidef}
$Z$ is positive semidefinite and has at least one positive weighting.
\end{enumerate}
\end{propn}

\begin{proof}
This is Proposition~5 of~\cite{MDBB}.
\end{proof}

\begin{propn}
The following are equivalent:
\begin{enumerate}
\item
\lbl{part:vec}
every maximizing distribution for $Z$ has full support;

\item
\lbl{part:unique}
$Z$ has exactly one maximizing distribution, which has full support;

\item
\lbl{part:def}
$Z$ is positive definite with positive weighting;

\item 
\lbl{part:crit}
$\Dmax{Z_B} < \Dmax{Z}$ for every nonempty proper subset $B$ of $\{1,
  \ldots, n\}$.
\end{enumerate}
\end{propn}

\begin{proof}
This is Proposition~6 of~\cite{MDBB}.
\end{proof}

Let us put our results on maximum diversity into context.  First, they
belong to the huge body of work on maximum entropy.  For example,
among all probability distributions on $\R$ with a given mean and variance,
the one with the maximum%
\index{maximum entropy}
entropy is the normal%
\index{normal distribution} 
distribution~\cite{LinnITP,BarrECL}.  Given the
fundamental nature of the normal distribution, this fact alone would
be motivation enough to seek maximum entropy distributions in other
settings (such as the one at hand), quite apart from the importance of
maximum entropy in thermodynamics, machine learning, and so on.

Second, the maximum diversity theorem (Theorem~\ref{thm:max}) is stated for
probability distributions on \emph{finite} sets equipped with a similarity
matrix, but it can be generalized to compact Hausdorff spaces $A$ equipped
with a suitable function $Z \from A \times A \to [0, \infty)$ measuring
similarity between points.  This more general theorem was recently proved
by Leinster and Roff~\cite{MEMS},%
\index{Roff, Emily}%
\index{entropy!metric space@on metric space}%
\index{maximum diversity!theorem}
along with a version of the computation theorem
(Theorem~\ref{thm:max-comp}).  It encompasses both the finite case and the
case of a compact metric space with similarities $Z(a, b) = e^{-d(a, b)}$.

Third, maximum diversity is closely related to the emerging invariant
known as magnitude, as now described.
\index{Meckes, Mark|)}

\section{Introduction to magnitude}
\lbl{sec:mag}
\index{magnitude}

In the solution to the maximum diversity problem, a supporting role was
played by the notion of the magnitude of a matrix
(Definition~\ref{defn:mag}).  Theorem~\ref{thm:max-comp} implies that the
maximum diversity of a symmetric similarity matrix $Z$ is always equal to
the magnitude of one of its principal submatrices $Z_B$, and
Examples~\ref{eg:ultra-pos-def} and~\ref{eg:id-pos-def} describe classes of
matrix for which the maximum diversity is actually equal to the magnitude.

The definition of magnitude was introduced without motivation, and may
appear to be nothing but a technicality.  But in fact, magnitude is
an answer to the following very broad conceptual challenge.

For many types of objects in mathematics, there is a canonical notion of
size.\index{size}  For example:
\begin{itemize}
\item 
Every set $A$ (finite, say) has a cardinality\index{cardinality} $\mg{A}$,
which satisfies the 
inclusion-exclusion%
\index{inclusion-exclusion principle} 
formula
\[
\mg{A \cup B} = \mg{A} + \mg{B} - \mg{A \cap B}
\]
(for subsets $A$ and $B$ of some larger set) and the
multiplicativity%
\index{multiplicative!cardinality is} 
formula 
\[
\mg{A \times B} = \mg{A} \cdot \mg{B}.
\]

\item 
Every measurable subset $A$ of Euclidean space has a volume\index{volume}
$\Vol(A)$, which satisfies similar formulas:
\begin{align*}
\Vol(A \cup B)  &
= 
\Vol(A) + \Vol(B) - \Vol(A \cap B),   \\
\Vol(A \times B)        & 
= 
\Vol(A) \cdot \Vol(B).
\end{align*}

\item 
Every sufficiently well-behaved topological space $A$ has an Euler
characteristic%
\index{Euler characteristic}
$\chi(A)$\ntn{ECsp}, which again satisfies
\begin{align*}
\chi(A \cup B)  & 
= 
\chi(A) + \chi(B) - \chi(A \cap B),   \\
\chi(A \times B)        &
=
\chi(A) \cdot \chi(B).
\end{align*}
(Here, inclusion-exclusion holds for subspaces $A$ and $B$ of some larger
space, under suitable hypotheses.  Technically, it is best to work
in the setting of either cohomology with compact supports, as in
Section~3.3 of Hatcher~\cite{Hatc}, or tame topology, as in Chapter~4 of
van den Dries~\cite{VDD} or Chapter~3 of Ghrist~\cite{GhriEAT}.)  The
insight that Euler%
\index{Euler characteristic} 
characteristic is the topological analogue of cardinality is
principally due to Schanuel, who compared Euler's investigation of spaces
of negative `cardinality' (Euler characteristic) with Cantor's
investigation of sets of infinite cardinality:
\begin{quote}
\index{Schanuel, Stephen}
Euler's analysis, which demonstrated that in counting suitably `finite'
spaces%
\index{set theory} 
one can get well-defined negative integers, was a revolutionary advance in
the idea of cardinal number~-- perhaps even more important than Cantor's
extension to infinite sets, if we judge by the number of areas in
mathematics where the impact is pervasive.
\end{quote}
(\cite{SchaNSE}, Section~3).
\end{itemize}

The close resemblance between these invariants suggests a challenge: find a
general notion of the size of a mathematical object, encompassing these
three invariants and others.  And this challenge has a solution: the
magnitude of an enriched category.%
\index{category theory}

Enriched categories are very general structures, and the theory of the
magnitude of an enriched category sweeps across many parts of mathematics,
most of them very distant from diversity measurement.  This section and the
next paint a broad-brush picture, omitting all details.  General references
for this material are Leinster and Meckes~\cite{MMSCG} and
Leinster~\cite{MMS}.

We begin with ordinary categories.  A finite category\index{category}
$\scat{A}$ consists of, first of all, a finite directed multigraph, that is,
a finite collection of objects $a_1, \ldots, a_n$ together with a finite
set $\Hom(a_i, a_j)$ for each $i$ and $j$, whose elements are to be thought
of as maps or arrows from $a_i$ to $a_j$ (Figure~\ref{fig:fin-cat}).  It is
also equipped with an associative operation of composition of maps and an
identity map on each object.  (See Mac~Lane~\cite{MacLCWM}, for instance.)

\begin{figure}
\centering
\lengths
\begin{picture}(120,23)
\cell{60}{10.5}{c}{\includegraphics[width=87\unitlength]{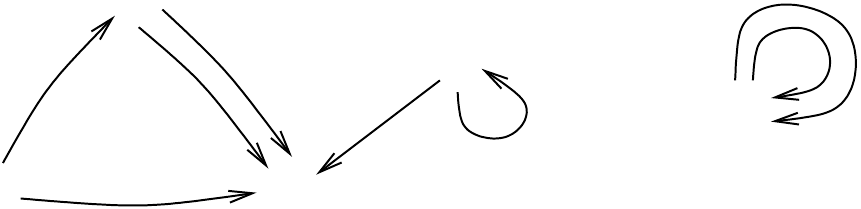}}
\cell{16}{2}{c}{\large$a_1$}
\cell{30}{21}{c}{\large$a_2$}
\cell{46}{2}{c}{\large$a_3$}
\cell{63}{14}{c}{\large$a_4$}
\cell{92}{10}{c}{\large$a_5$}
\end{picture}
\caption{The objects and maps of a finite category (identity maps not
  shown).}
\lbl{fig:fin-cat}
\end{figure}

Any finite category $\scat{A}$ gives rise to an $n \times n$ matrix
$Z_{\scat{A}}$ whose $(i, j)$-entry is $\mg{\Hom(a_i, a_j)}$, the number
of maps from $a_i$ to $a_j$.  The \demph{magnitude}%
\index{magnitude!category@of category}
$\mg{\scat{A}}\ntn{magcat} \in \Q$ of the category $\scat{A}$ is defined to
be the magnitude $\mg{Z_{\scat{A}}}$ of the matrix $Z_{\scat{A}}$, if it
exists.

Here we have used the notation $\mg{\,\cdot\,}$ for two purposes: first for
the cardinality of a finite set, then for the magnitude of a category.
This is deliberate.  In both cases, $\mg{\,\cdot\,}$ is a measure
of the size of the structure concerned.

For example, if $\scat{A}$ has no maps except for identities then
$Z_{\scat{A}}$ is the $n \times n$ identity matrix, so the magnitude
$\mg{\scat{A}}$ is the cardinality $n$ of its set of objects.  Less
trivially, any (small) category $\scat{A}$ has a classifying%
\index{classifying space} 
space $B\scat{A}$ (also called its nerve\index{nerve} 
or geometric%
\index{geometric realization} 
realization), which is a topological space constructed from $\scat{A}$ by
starting with one $0$-simplex for each object of $\scat{A}$, then pasting
in one $1$-simplex for each map in $\scat{A}$, one $2$-simplex for each
commutative triangle in $\scat{A}$, and so on (Segal~\cite{SegaCSS}).  It is a
theorem that
\begin{equation}
\lbl{eq:mag-classifying}
\mg{\scat{A}} = \chi(B\scat{A}),
\end{equation}
under finiteness conditions to ensure that the Euler%
\index{Euler characteristic} 
characteristic of $B\scat{A}$ is well-defined
(Proposition~2.11 of Leinster~\cite{ECC}).  So, the situation is similar to
group homology\index{homology}: the homology of a group $G$ can be
defined either through a direct algebraic formula (as for $\mg{\scat{A}}$)
or as the homology of its classifying space (as for $\chi(B\scat{A})$), and
it is a theorem that the two are equal.

\begin{example}
Let $\scat{A} = (\bullet \parpairu \bullet)$ (identity maps not shown).
Then  
\[
Z_{\scat{A}} 
= 
\begin{pmatrix}
1       &2      \\
0       &1  
\end{pmatrix},
\]
giving
\[
Z_{\scat{A}}^{-1}
=
\begin{pmatrix}
1       &-2     \\
0       &1  
\end{pmatrix}
\]
and so
\[
\mg{\scat{A}}
=
\mg{Z_{\scat{A}}}
=
1 + (-2) + 0 + 1
=
0.
\]
On the other hand, $B\scat{A} = S^1$, so $\chi(B\scat{A}) = 0$,
confirming equation~\eqref{eq:mag-classifying}.
\end{example}

Equation~\eqref{eq:mag-classifying} shows how, under hypotheses, magnitude
for categories can be derived from Euler%
\index{Euler characteristic} 
characteristic for topological spaces.  In the other direction, we can
derive topological Euler characteristic from categorical magnitude.  Let
$M$ be a finitely triangulated manifold.  Then associated with the
triangulation, there is a category $\scat{A}_M$ whose objects are the
simplices $s_1, \ldots, s_n$ of the triangulation, with one map $s_i \to
s_j$ whenever $s_i \sub s_j$, and with no maps $s_i \to s_j$ otherwise.
Then
\[
\chi(M) = \mg{\scat{A}_M}
\]
(Section~3.8 of Stanley~\cite{StanEC1} and Section~2 of
Leinster~\cite{ECC}).

The moral of the last two results is that topological Euler characteristic
and categorical magnitude each determine the other (under suitable
hypotheses).  Indeed, the magnitude of a category is often called its Euler
characteristic; see, for instance, Leinster~\cite{ECC}, Berger and
Leinster~\cite{ECCSDS}, Fiore, L\"uck and Sauer~\cite{FLSECC,FLSFOE},
Noguchi~\cite{NoguECA,NoguECC,NoguZFF}, and Tanaka~\cite{TanaECB}.

Further theorems connect the magnitude of a category to the Euler
characteristic of an orbifold (Proposition~2.12 of~\cite{ECC}) and to the
Baez--Dolan%
\index{Baez, John!Dolan cardinality@--Dolan cardinality}%
\index{Dolan, James}%
\index{groupoid cardinality}%
\index{cardinality!groupoid@of groupoid}
cardinality of a groupoid (Example~2.7 of~\cite{ECC} and
Section~3 of Baez and Dolan~\cite{BaDoFSF}), both of which are rational
numbers, not usually integers.  The notion of magnitude can also be seen as
an extension of the theory of M\"obius%
\index{Mobius--Rota inversion@M\"obius--Rota inversion} 
inversion for posets (most commonly
associated with the name of Rota~\cite{RotaFCT}),%
\index{Rota, Gian-Carlo}
which itself generalizes the classical M\"obius function of number theory;
see~\cite{ECC,NMI} for explanation.

The definition of the magnitude of a category involved $\mg{\Hom(a_i,
  a_j)}$, the cardinality of the set of maps from $a_i$ to $a_j$.  Thus, we
used the notion of the cardinality of a finite set to define the magnitude
of a finite category.  We can envisage that if $\Hom(a_i, a_j)$ were some
other kind of structure with a preexisting notion of size, a similar
definition could be made.  And indeed, this idea can be implemented in the
language of enriched categories, as follows.

A \demph{monoidal%
\index{monoidal category}
category} is, loosely speaking, a category $\cat{V}$ equipped with a
product operation satisfying reasonable conditions.  Section~VII.1 of
Mac~Lane~\cite{MacLCWM} gives the full definition, but the following
examples will be all that we need here.

\begin{examples}
Typical examples of monoidal categories are the category $\Set$ of sets
with the cartesian product $\times$ and the category $\Vect$ of vector
spaces with the tensor product $\otimes$.

A less obvious example is the category whose objects are the elements of
the interval $[0, \infty]$, with one map $x \to y$ whenever $x \geq y$, and
with no maps $x \to y$ otherwise.  Here we take $+$ as the `product'
operation.  (We could also take ordinary multiplication as the product, but
it is $+$ that will be of interest here.)
\end{examples}

Now fix a monoidal category $\cat{V}$, with product denoted by $\otimes$.
Loosely, a \demph{category enriched in $\cat{V}$},%
\index{enriched category}
or \demph{$\cat{V}$-category},%
\index{Vcategory@$\cat{V}$-category}
$\scat{A}$, consists of:
\begin{itemize}
\item 
a set $a, b, \ldots$ of \demph{objects} of $\scat{A}$;

\item
for each pair $(a, b)$ of objects of $\scat{A}$, an object $\Hom(a, b)$ of
$\cat{V}$;

\item
for each triple $(a, b, c)$ of objects of $\scat{A}$, a map 
\begin{equation}
\lbl{eq:enr-comp}
\Hom(a, b) \otimes \Hom(b, c) \to \Hom(a, c)
\end{equation}
in $\cat{V}$ (called \demph{composition} in $\scat{A}$),
\end{itemize}
subject to conditions.  (For the full definition, see Section~1.2 of
Kelly~\cite{KellBCE} or Section~6.2 of Borceux~\cite{BorcHCA2}.)

\begin{examples}
The following examples of enriched categories are depicted in
Figure~\ref{fig:mon-enr}.
\begin{enumerate}
\item 
A category enriched in $(\Set, \times)$ is just an ordinary category.  So,
an enriched category is not a category with special properties; it is
something \emph{more general} than a category.

\item
A category enriched in $(\Vect, \otimes)$ is a 
\demph{linear%
\index{linear category} 
category}: a category equipped with a vector space structure on each of the
sets $\Hom(a, b)$, in such a way that composition is bilinear.

\item
As first observed by Lawvere~\cite{LawvMSG},%
\index{Lawvere, F. William} 
any metric space $A$ can be viewed as a category $\scat{A}$ enriched in
$([0, \infty], +)$: the objects of $\scat{A}$ are the points of $A$, while
$\Hom(a, b) \in [0, \infty]$ is the distance $d(a, b)$, and the
composition~\bref{eq:enr-comp} is the triangle inequality
\[
d(a, b) + d(b, c) \geq d(a, c).
\]
\end{enumerate}
\end{examples}

Thus, categories, linear categories and metric spaces are all instances of
a single general concept: enriched category.  This enables constructions
and insights to be passed backwards and forwards between them, a strategy
that proves to have great power.

\begin{figure}
\centering
\lengths
\begin{picture}(120,72)(0,0)
\thicklines
\cell{30}{0}{bl}{\includegraphics[width=81\unitlength]{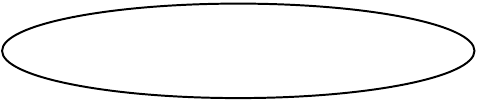}}
\cell{15}{11}{c}{\large monoidal}
\cell{15}{6}{c}{\large categories}
\cell{40}{9}{c}{$\zmark$}
\cell{42}{9}{l}{$\cat{V}$}
\cell{60}{12}{c}{$\zmark$}
\cell{62}{12}{l}{$\Set$}
\cell{75}{5}{c}{$\zmark$}
\cell{77}{5}{l}{$\Vect$}
\cell{95}{10}{c}{$\zmark$}
\cell{97}{10}{l}{$[0, \infty]$}
\cell{30}{72}{tl}{\includegraphics[width=81\unitlength]{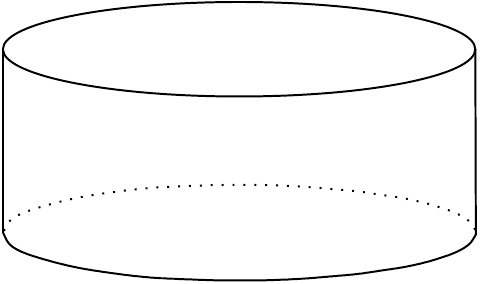}}
\cell{15}{49}{c}{\large enriched}
\cell{15}{44}{c}{\large categories}
\put(40,32){\line(0,1){31}}
\cell{41}{48}{l}{\rotatebox{-90}{$\cat{V}$-categories}}
\put(60,35){\line(0,1){31}}
\cell{61}{51}{l}{\rotatebox{-90}{categories}}
\put(75,28){\line(0,1){31}}
\cell{76}{43}{l}{\rotatebox{-90}{linear categories}}
\put(95,33){\line(0,1){31}}
\cell{96}{47}{l}{\rotatebox{-90}{metric spaces}}
\end{picture}
\caption{Schematic diagram of monoidal and enriched categories.}
\lbl{fig:mon-enr}
\end{figure}

In particular, it is straightforward to generalize the definition of the
magnitude of a finite category to finite enriched categories.  Let
$\cat{V}$ be a monoidal category equipped with a function
$\mg{\,\cdot\,}$\ntn{gauge} that assigns to each object $X$ of $\cat{V}$ an
element $\mg{X}$ of some ring.  This function $\mg{\,\cdot\,}$ is to play the
role of the cardinality of a finite set, and we therefore impose the
requirements that it is isomorphism-invariant and multiplicative:
\[
X \iso Y \implies \mg{X} = \mg{Y},
\qquad
\mg{X \otimes Y} = \mg{X} \cdot \mg{Y}.
\]
(Section~1.3 of Leinster~\cite{MMS} gives details.)  Then any
$\cat{V}$-category $\scat{A}$ with finitely many objects, $a_1, \ldots,
a_n$, gives rise to a matrix
\[
Z_{\scat{A}} = \bigl( \mg{\Hom(a_i, a_j)} \bigr)_{i, j}.
\]
The \demph{magnitude}%
\index{magnitude!enriched category@of enriched category} 
$\mg{\scat{A}}$ of $\scat{A}$ is defined to be the magnitude of
$Z_{\scat{A}}$, if it exists.

\begin{example}
If we begin with the monoidal category $\cat{V}$ of finite sets equipped
with the cartesian product, with the cardinality function
$\mg{\mbox{\ensuremath{\,\cdot\,}}}$ on finite sets, then we recover the
notion of the magnitude of a finite category.
\end{example}

\begin{example}
Let $\cat{V}$ be the category of finite-dimensional vector spaces over some
field, with the tensor product, and put $\mg{X} = \dim X$ for
finite-dimensional vector spaces $X$.  Then we obtain a notion of the
magnitude $\mg{\scat{A}}$ of a linear category $\scat{A}$ with finitely
many objects and finite-dimensional hom-spaces $\Hom(a, b)$.  By
definition, $\mg{\scat{A}}$ is the magnitude of the matrix
\[
\bigl( \dim \Hom(a, b) \bigr)_{a, b \in \scat{A}}.
\]
For instance, let $E$ be an associative algebra%
\index{associative algebra} 
over an algebraically closed field.  In the representation theory of
algebras, an important linear category associated with $E$ is $\IP(E)$, the
category of indecomposable%
\index{indecomposable module} 
projective%
\index{projective module} 
$E$-modules.  Under finiteness hypotheses on $E$, it is a theorem that
$\IP(E)$ has magnitude
\begin{equation}
\lbl{eq:ip}
\mg{\IP(E)} 
=
\sum_{n = 0}^\infty (-1)^n \dim \Ext_E^n(S, S),
\end{equation}
where $S$ is the direct sum of the simple $E$-modules (Theorem~1.1 of
Chuang,%
\index{Chuang, Joseph}
King%
\index{King, Alastair} 
and Leinster~\cite{MFDA}).  The
matrix $Z_{\IP(E)}$ is better known as the \demph{Cartan%
\index{Cartan matrix} 
matrix} of $E$.  The right-hand side of equation~\eqref{eq:ip} is called
the \demph{Euler%
\index{Euler form} 
form} $\chi(S, S)$ of the pair $(S, S)$, and is another manifestation of
the concept of Euler characteristic.
\end{example}

The examples so far of the magnitude of an enriched category have been
closely related to other, older invariants.  But when we apply the
definition to metric spaces, we obtain something new.

Let $\cat{V}$ be the monoidal category $[0, \infty]$, with product $+$.
For $x \in [0, \infty]$, define
\[
\mg{x} = e^{-x} \in \R.
\]
(Recall that $\mg{\,\cdot\,}$ is required to be `multiplicative', that is,
must convert the tensor product on $\cat{V}$ into multiplication.  In
this case, this means $\mg{x + y} = \mg{x} \cdot \mg{y}$, which by
Corollary~\ref{cor:add-transf}\bref{part:add-transf-exp} essentially forces
$\mg{x} = c^x$ for some constant $c$.)  Then we obtain a notion of the
magnitude of a finite $\cat{V}$-category, and in particular, of a finite
metric space.

In explicit terms, the definition is as follows.  Let $A = \{a_1, \ldots,
a_n\}$ be a finite metric\index{metric!space} space.  Form the $n \times n$
matrix
\[
Z_A = \bigl( e^{-d(a_i, a_j)} \bigr).
\]
Invert $Z_A$ (if possible); then the \demph{magnitude}%
\index{magnitude!metric space@of metric space}
$\mg{A}$ of $A$ is
the sum of all $n^2$ entries of $Z_A^{-1}$.

Here we have used Remark~\ref{rmks:defn-mag}\bref{rmk:defn-mag-inv} on the
magnitude of a matrix in terms of its inverse.  Since $Z_A$ is a square
matrix of real numbers, it is usually invertible, and in fact it is
\emph{always} invertible when $A$ is a subspace of Euclidean space
(Theorem~2.5.3 of Leinster~\cite{MMS} or Section~4 of
Meckes~\cite{MeckPDM}). 

\begin{examples}
\lbl{egs:mms}
\begin{enumerate}
\item 
The magnitude of the zero-point space is $0$, and the magnitude of the
one-point space is $1$.

\item
\lbl{eg:mms-two}
Consider the metric space $A$ consisting of two points distance $\ell$
apart:
\[
\xymatrix@C+1em{
\bullet \ar@{<->}[r]^{\displaystyle\ell} &\bullet
}
\]
Then 
\[
\mg{A}
=
\text{sum of entries of } 
\begin{pmatrix}
e^{-0}          &e^{-\ell}      \\
e^{-\ell}       &e^{-0}
\end{pmatrix}^{-1}
=
\frac{2}{1 + e^{-\ell}},
\]
as illustrated in Figure~\ref{fig:mms-two}.  

\begin{figure}
\centering
\lengths
\begin{picture}(120,38)(0,-2)
\cell{60}{20}{c}{\includegraphics[width=100\unitlength]{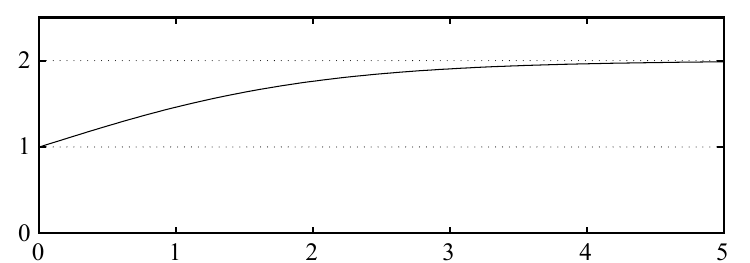}}
\cell{60}{0}{c}{$\ell$}
\cell{5}{18}{l}{$\mg{A}$}
\end{picture}
\caption{The magnitude of a two-point space $A$, with points distance
  $\ell$ apart (Example~\ref{egs:mms}\bref{eg:mms-two}).}
\lbl{fig:mms-two}
\end{figure}

This example can be understood as follows.  When $\ell$ is small, the two
points are barely distinguishable, and may appear to be only one point (at
poor resolution, for instance).  As $\ell$ increases, the two points
acquire increasingly separate identities, and correspondingly, the
magnitude increases towards $2$.  In the extreme, when $\ell = \infty$, the
two points are entirely separated and the magnitude is exactly $2$.  This
example and others suggest that we can usefully think of the magnitude of a
finite metric space as the `effective%
\index{effective number!points@of points} 
number of points', or, more fully, the effective number of
completely separate points.

\item
Let $A$ be a finite metric space in which all nonzero distances are
$\infty$.  Then $Z_A = I$ and $\mg{A}$ is just the cardinality of $A$.
This also fits with the interpretation of magnitude as the effective number
of points. 

\item
\lbl{eg:mms-long-thin}
This example is adapted from Willerton (\cite{WillSMS},%
\index{Willerton, Simon} 
Figure~1).  Let $A$ be a three-point space with the points arranged in a long
thin triangle, as in Figure~\ref{fig:long-thin}.  When $\ell$ is small, the
space appears to be just a single point, and the magnitude is close to $1$.
When $\ell$ is moderate, the space appears to have two points, and the
magnitude is about $2$.  When $\ell$ is large, the distinction between all
three points is clearly visible, and the magnitude is close to~$3$.

\begin{figure}
\centering
\lengths
\begin{picture}(50,45)(0,5)
\cell{26}{20}{c}{\includegraphics[width=48\unitlength]{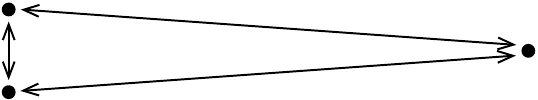}}
\cell{0}{20}{l}{$\ell$}
\cell{26}{25}{c}{$1000\ell$}
\cell{26}{15}{c}{$1000\ell$}
\cell{26}{7}{c}{$A$}
\end{picture}%
\hspace*{5mm}%
\begin{picture}(65,45)(-5,-5)
\cell{30}{20}{c}{\includegraphics[width=60\unitlength]{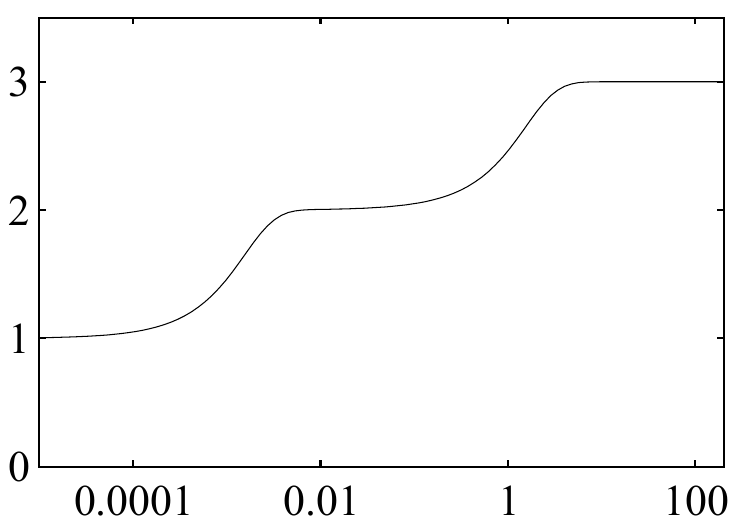}}
\cell{-5}{20}{l}{$\mg{A}$}
\cell{30}{-5}{b}{$\ell$}
\end{picture}
\caption{The magnitude of a certain three-point metric space $A$
  (Example~\ref{egs:mms}\bref{eg:mms-long-thin}). Note the logarithmic
  scale.}  
\lbl{fig:long-thin}
\end{figure}

Empirical data such as this suggests a connection between magnitude and
persistent%
\index{persistent homology} 
homology.  Indeed, results of Otter~\cite{Otte}%
\index{Otter, Nina} 
have begun to establish such a connection.  We return to this topic at the
end of the section.
\end{enumerate}
\end{examples}

Every metric space $A$ belongs to a one-parameter family $(tA)_{t >
  0}$\ntn{tA} of spaces, where $tA$ denotes $A$ scaled up by a factor of
$t$.  So, magnitude assigns to each finite metric space $A$ not just a
\emph{number} $\mg{A}$, but a (partially-defined) \emph{function}: its
\demph{magnitude%
\index{magnitude!function} 
function}
\[
\begin{array}{ccc}
(0, \infty)     &\to            &\R     \\
t               &\mapsto        &\mg{tA}.
\end{array}
\]
For instance, Figures~\ref{fig:mms-two} and~\ref{fig:long-thin} show the
magnitude functions of a certain two-point space and a certain three-point
space. 

\begin{example}
\lbl{eg:k23}
Magnitude functions can behave wildly.  Consider the complete bipartite%
\index{graph!bipartite}%
\index{bipartite graph}
graph $K_{2, 3}$ (Figure~\ref{fig:k23}), regarded as a metric space as
follows: the points of the space are the vertices of the graph, and the
distance between two vertices is the number of edges in a shortest path
between them.
\begin{figure}
\centering
\lengths
\begin{picture}(24,62)
\cell{12}{35}{c}{\includegraphics[width=24\unitlength]{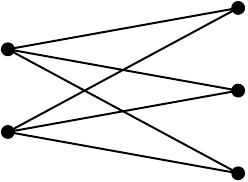}}
\cell{12}{21}{c}{$K_{2, 3}$}
\end{picture}%
\hspace*{7mm}%
\begin{picture}(89,62)(-9,-2)
\cell{42}{30}{c}{\includegraphics[width=80\unitlength]{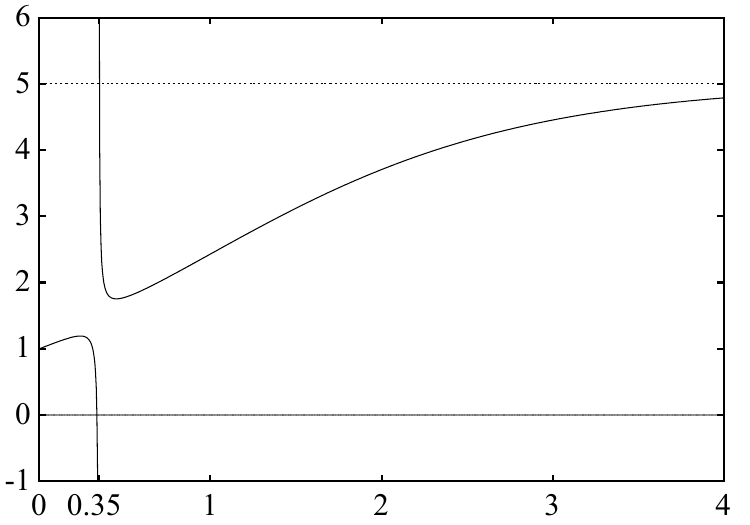}}
\cell{43}{-2}{b}{$t$}
\cell{-7}{32}{l}{$\mg{tK_{2,3}}$}
\end{picture}
\caption{The complete bipartite graph $K_{2, 3}$ and its magnitude function
(Example~\ref{eg:k23}).  The singularity is at $\log\sqrt{2} \approx 0.35$.}
\lbl{fig:k23}
\end{figure}

The magnitude function of $K_{2, 3}$ has several striking features: it
is sometimes negative, sometimes greater than the number of points,
sometimes undefined, and sometimes \emph{decreasing} in the scale factor
$t$.  Example~2.2.7 of Leinster~\cite{MMS} gives details.
\end{example}

However, the magnitude function of a finite metric space never behaves
\emph{too} badly.  It can be shown that the magnitude function has only
finitely many singularities (none for subspaces of Euclidean space), that
it is increasing for $t \gg 0$, and that $\mg{tA}$ converges to the
cardinality of $A$ as $t \to \infty$ (Proposition~2.2.6 of
Leinster~\cite{MMS}).  In particular, this last statement implies that the
magnitude function of a space knows its cardinality.

In Example~\ref{eg:k23}, we started from a graph, constructed the metric
space whose points are the vertices and whose distances are shortest
path-lengths, and considered the magnitude of that space.  This is a
construction of general interest, investigated in Leinster~\cite{MG}.  In
this context, we replace the real number $e^{-1}$ in the definition of
magnitude%
\index{magnitude!graph@of graph}%
\index{graph!magnitude of}
by a formal variable $x$.  The magnitude of a graph can then be
expressed as either a rational function or a power series in $x$ with
integer coefficients (Section~2 of~\cite{MG}).  For example, the graphs
\[
\includegraphics[height=2em]{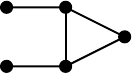}
\qquad
\includegraphics[height=2em]{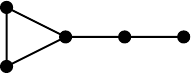}
\qquad
\includegraphics[height=2em]{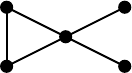}
\]
all have magnitude 
\[
\frac{5 + 5x - 4x^2}{(1 + x)(1 + 2x)}
=
5 - 10x + 16x^2 - 28x^3 + 52x^4 - 100x^5 + \cdots
\]
(Example~4.11 of~\cite{MG}).  The magnitude of a graph shares some
invariance properties with one of the most important graph invariants of
all, the Tutte%
\index{Tutte polynomial}%
\index{polynomial!Tutte}
polynomial.  For instance, it is invariant under Whitney%
\index{Whitney twist}
twists when the points of identification are adjacent.  But it is not a
specialization of the Tutte polynomial: it carries information that the
Tutte polynomial does not.

Graph magnitude satisfies multiplicativity and inclusion-exclusion
principles: 
\begin{align*}
\index{multiplicative!magnitude is}
\mg{G \times H} &
=
\mg{G} \cdot \mg{H},    \\
\index{inclusion-exclusion principle}
\mg{G \cup H}   &
=
\mg{G} + \mg{H} - \mg{G \cap H}
\end{align*}
(where the latter is under quite strict hypotheses), shown as Lemma~3.6
and Theorem~4.9 of Leinster~\cite{MG}.  As such, it has a reasonable claim
to being the graph-theoretic analogue of cardinality.

As additional evidence for this claim, Hepworth%
\index{Hepworth, Richard} 
and Willerton~\cite{HeWi}%
\index{Willerton, Simon}
constructed a graded homology%
\index{magnitude!homology}
theory of graphs%
\index{graph!magnitude homology of}
whose Euler characteristic is magnitude.  In more detail: since their
homology theory is graded, the Euler characteristic of a graph is not a
single number but a sequence of numbers, which when construed as a power
series is exactly the graph's magnitude.  Thus, their homology theory is a
categorification of magnitude in the same sense that the Khovanov%
\index{Khovanov homology}
homology of knots and 
links~\cite{Khov} is a categorification of the Jones%
\index{Jones polynomial} 
polynomial.  It is a finer invariant than magnitude, in that there are
graphs with the same magnitude but different homology groups (Gu~\cite{Gu},
Appendix~A; see also Summers~\cite{Summ}).

Not only the definition of magnitude for graphs, but also some theorems
about it, can be categorified.  For instance, Hepworth and Willerton proved
that the multiplicativity and inclusion-exclusion theorems for magnitude
lift to K\"unneth and Mayer--Vietoris theorems in homology.  In this sense,
known properties of the magnitude of graphs are shadows of functorial
results in homology.

Hepworth and Willerton's idea even works in the full generality of enriched
categories.  That is, the magnitude of an enriched category (a numerical
invariant) can be categorified to a graded homology theory for enriched
categories (an algebraic invariant).  As in the case of graphs,
`categorified' means that the Euler characteristic of the homology theory
is exactly magnitude.  This \demph{magnitude%
\index{magnitude!homology}
homology} for enriched categories was defined and developed in work led by
Shulman~\cite{MHECMS}.%
\index{Shulman, Michael}
It is a kind of Hochschild homology.

Since metric spaces are a special kind of enriched category, this
construction provides a new homology theory for metric spaces.  It is
genuinely metric rather than topological.  For example, the first magnitude
homology of a closed subset $X$ of $\R^n$ is trivial if and only if $X$ is
convex%
\index{convex!set}
(\cite{MHECMS}, Section~4).  Indeed, all of the magnitude homology groups
of a convex subset of $\R^n$ are trivial, a metric analogue of the
topological fact that the homology of a contractible space is
trivial.  This was proved independently by Kaneta and Yoshinaga
(Corollary~5.3 of~\cite{KaYo}) and by Jubin (Theorem~7.2 of~\cite{Jubi}).
Gomi~\cite{GomiMHG} states a slogan:
\index{Gomi, Kiyonori}
\begin{quote}
The more geodesics are unique, the more magnitude homology is trivial.
\end{quote}

Methods for computing the magnitude homology of metric spaces have recently
been developed and applied to calculate specific homology groups.  Gomi
developed spectral sequence techniques and used them to prove results on
the magnitude homology groups of circles (\cite{GomiSFM}, Section~4).
Kaneta and Yoshinaga~\cite{KaYo} showed that while ordinary topological
homology detects the existence of holes, magnitude homology detects the
\emph{diameter} of holes, in a sense made precise in their Theorem~5.7.
Asao proved that if a space contains a closed geodesic then its second
magnitude homology group is nontrivial (Theorem~5.3 of~\cite{Asao}), while
Gomi~\cite{GomiMHG} proved general results on the second and third
magnitude homology groups of metric spaces.

Magnitude homology is not the first homology theory of metric spaces: there
is also persistent%
\index{persistent homology}
homology, fundamental in the field of topological data analysis.  (For
expository accounts of persistent homology, see Ghrist~\cite{GhriBPT} or
Carlsson~\cite{Carl}.)  Otter~\cite{Otte}%
\index{Otter, Nina} 
has proved results relating the two homology theories, introducing for this
purpose a notion of `blurred magnitude homology'; see also Govc and
Hepworth~\cite{GoHe} and Cho~\cite{Cho}.

Finally, Hepworth~\cite{HepwMC} has introduced a theory of magnitude
\emph{cohomology}%
\index{magnitude!cohomology}%
\index{Hepworth, Richard}
for enriched categories.  It carries a product that formally resembles the
ordinary cup product, but is noncommutative.  For finite metric spaces,
magnitude cohomology is a complete invariant: the cohomology ring
of such a space determines it uniquely up to isometry.

\section{Magnitude in geometry and analysis}
\lbl{sec:mag-geom}

Most metric spaces of geometric interest are not finite.  The general
enriched-categorical concept of magnitude provides no definition of the
magnitude of an infinite metric space.  On the other hand, there are 
several plausible strategies for extending the definition of magnitude from
finite to compact metric spaces.  Meckes~\cite{MeckPDM,MeckMDC}%
\index{Meckes, Mark} 
showed that as long as the space satisfies a certain classical condition,
they all give the same outcome.

The condition is that the space must be of \demph{negative%
\index{negative!type} 
type}.  We do not need the original definition here, but Meckes refined old
results of Schoenberg~\cite{SchoMSP} to show that $A$ is of negative type
if and only if the matrix $Z_{tB}$ is positive%
\index{positive!definite}
definite for every finite $B \sub A$ and real
$t > 0$ (Theorem~3.3 of~\cite{MeckPDM}).  A great many spaces are of
negative type, including all subspaces of $\R^n$ with the Euclidean or
$\ell^1$ (taxicab) metric, all ultrametric spaces, real and complex
hyperbolic space, and spheres with the geodesic metric.  A list can be
found in Theorem~3.6 of~\cite{MeckPDM}.

The most direct way to state the extended definition of magnitude is as
follows.

\begin{defn}
Let $A$ be a compact metric space of negative type.  The \demph{magnitude}%
\index{magnitude!metric space@of metric space}
of $A$ is
\[
\mg{A} 
= 
\sup \{ \mg{B} \such \text{finite subsets } B \sub A \}
\in 
[0, \infty].
\ntn{magcpt}
\]
\end{defn}

Equivalently, one can choose a sequence $B_1 \sub B_2 \sub \cdots \sub A$
of finite subspaces $B_n$ of $A$ such that $\bigcup B_n$ is dense in $A$,
then put $\mg{A} = \lim_{n \to \infty} \mg{B_n}$.  Equivalently again, one
can define the magnitude of $A$ by the variational formula
\begin{equation}
\lbl{eq:mag-sup}
\mg{A}
=
\sup_\mu
\frac{\mu(A)^2}{\int_A \int_A e^{-d(a, b)} \dee\mu(a) \dee\mu(b)},
\end{equation}
where the supremum is over the finite signed Borel measures $\mu$ on $A$
for which the denominator is nonzero.  

This last characterization is related to yet another formulation.  A
\demph{weight%
\index{weight measure} 
measure} on $A$ is a finite signed Borel measure $\mu$ such
that
\[
\int_A e^{-d(a, b)} \dee\mu(b) = 1
\]
for all $a \in A$.  This definition was introduced by Willerton%
\index{Willerton, Simon} 
(\cite{WillMSS}, Section~1.1), and is the continuous analogue of the
notion of weighting (Definition~\ref{defn:wtg}).  If $\mu$ is a weight
measure on $A$ then $\mg{A} = \mu(A)$, by Theorem~2.3 of
Meckes~\cite{MeckPDM} or Proposition~5.3.6 of Leinster and
Meckes~\cite{MMSCG}.  However, not every compact metric space of negative
type admits a weight measure.  Most weightings are distributions of a more
general kind, defined in Meckes~\cite{MeckMDC}.%
\index{Meckes, Mark}

The equivalence of these and other definitions of magnitude was established
by Meckes~\cite{MeckPDM,MeckMDC},%
\index{Meckes, Mark}
using techniques of harmonic and functional analysis.

We now give some examples of compact spaces $A$ and their magnitude
functions $t \mapsto \mg{tA}$.

\begin{example}
\lbl{eg:boxy-line}
The magnitude function of a straight line $[0, \ell]$ of length $\ell$ is
given by 
\[
\mg{t \cdot [0, \ell]} = 1 + \hlf\ell\cdot t.
\]
Several proofs are known, as in Theorem~7 of Leinster and
Willerton~\cite{AMSES},%
\index{Willerton, Simon} 
Theorem~2 of Willerton~\cite{WillMSS}, and Proposition~3.2.1 of
Leinster~\cite{MMS}.  The easiest proof uses weight measures.  Let
$\delta_0$ and $\delta_\ell$ denote the point measures at $0$ and $\ell$,
let $\lambda_{[0, \ell]}$ denote Lebesgue measure on $[0, \ell]$, and put
\[
\mu = \hlf(\delta_0 + \lambda_{[0, \ell]} + \delta_\ell).
\]
It is easily verified that $\mu$ is a weight measure on $[0, \ell]$.  Hence
\[
\mg{[0, \ell]} = 1 + \hlf\ell,
\]
and so the magnitude function of $[0, \ell]$ is given by 
\[
\mg{t \cdot [0, \ell]} 
=
1 + \hlf\ell\cdot t.
\]
\end{example}

\begin{example}
\lbl{eg:boxy-boxes}
\index{multiplicative!magnitude is}
Magnitude is multiplicative with respect to the $\ell^1$%
\index{lone geometry@$\ell^1$ geometry} 
product of metric
spaces, that is, the product space with the metric given by adding the
distances in the two factors (Proposition~3.1.4 of Leinster~\cite{MMS}).
This has the following consequence.  Equip $\R^n$ with the $\ell^1$ metric:
\[
d(\vc{x}, \vc{y}) = \sum_{i = 1}^n \mg{x_i - y_i}
\]
($\vc{x}, \vc{y} \in \R^n$).  Then by the previous example, the magnitude
function of a rectangle 
\[
A = [0, \ell] \times [0, m] \sub \R^2
\]
is given by
\begin{align*}
\mg{tA} &
=
\bigl( 1 + \hlf\ell\cdot t \bigr) 
\bigl( 1 + \hlf m\cdot t \bigr)         \\
&
=
1 + \tfrac{1}{2} (\ell + m) \cdot t 
+ \tfrac{1}{4} \ell m \cdot t^2.
\end{align*}
Up to a constant factor, the coefficient of $t^2$ is the area of $A$, the
coefficient of $t$ is the perimeter of $A$, and the constant term is the
Euler characteristic of $A$.  Similar statements apply to
higher-dimensional boxes (Corollary~3.4.3 of Leinster~\cite{MMS}).

For rectangles, and for nonempty convex sets in general, the Euler
characteristic is always $1$.  As such, it may seem pretentious to call the
constant term the `Euler characteristic'.  This usage will be justified
shortly.
\end{example}

To begin to explain the geometric content of magnitude, we need to recall
the concept of intrinsic volumes (Klain and Rota~\cite{KlRo} or Section~4.1
of Schneider~\cite{SchnCBB2}), which with different normalizations are also
known as quermassintegrals or Minkowski functionals.

Consider all reasonable ways of measuring the size of compact convex
subsets of $\R^n$ (which in the present discussion will just be called
\demph{convex}%
\index{convex!set}
sets).  In the plane $\R^2$, there are at least three ways to measure a
set: take its area, its perimeter, or its Euler characteristic.  These are
$2$-, $1$-, and $0$-dimensional measures, respectively.  The general fact
is that there are $n + 1$ canonical ways of measuring convex subsets of
$\R^n$, which define functions
\[
V_0, \ldots, V_n \from \{\text{convex subsets of } \R^n\} \to \R.
\]
Here $V_i$ is $i$-dimensional, in the sense that $V_i(tA) = t^i V_i(A)$,
and $V_i(A)$\ntn{Vi} is called the $i$th \demph{intrinsic%
\index{intrinsic volume} 
volume} of $A$.

The $i$th intrinsic volume of a convex set $A \sub \R^n$ can be defined as
follows.  Choose at random an $i$-dimensional linear subspace $L$ of
$\R^n$, take the orthogonal projection $\pi_L(A)$ of $A$ onto $L$, then
take its $i$-dimensional volume $\Vol(\pi_L(A))$.  Up to a constant factor,
$V_i(A)$ is the expected value of $\Vol(\pi_L(A))$.

\begin{example}
Let $A$ be a convex subset of $\R^3$. Then $V_0(A)$ is $0$ if $A$ is empty,
and $1$ otherwise. (In both cases, $V_0(A)$ is the Euler%
\index{Euler characteristic}
characteristic of $A$.) The first intrinsic volume $V_1(A)$ is proportional
to the expected length of the projection of $A$ onto a random line, and is
called the \demph{mean%
\index{mean!width} 
width} of $A$.  The second intrinsic volume $V_2(A)$ is proportional to the
expected area of the projection of $A$ onto a random plane, and it is a theorem
of Cauchy%
\index{Cauchy, Augustin!surface area theorem}%
\index{surface area}
that this is proportional to the surface area of $A$ (Klain and
Rota~\cite{KlRo}, Theorem~5.5.2). Finally, $V_3(A)$ is just the volume of
$A$.
\end{example}

Each of the intrinsic volumes $V_i$ on convex sets is isometry-invariant,
continuous with respect to the Hausdorff metric, and a
\dmph{valuation}: $V_i(\emptyset) = 0$ and 
\[
V_i(A \cup B) = V_i(A) + V_i(B) - V_i(A \cap B)
\]
whenever $A$, $B$ and $A \cup B$ are convex.  The same is, therefore, true
of any linear combination of the intrinsic volumes.  A celebrated theorem
of Hadwiger~\cite{Hadw}%
\index{Hadwiger's theorem} 
states that such linear combinations are the \emph{only}
isometry-invariant continuous valuations on convex sets.

The intrinsic volumes can be adapted to more general classes of space and
to different geometries.  For instance, we can speak of the volume or
surface area of a sufficiently smooth subset of $\R^n$, and in that
context, the intrinsic volumes are closely related to curvature measures.
(See Section~2.1.1 of Alesker and Fu~\cite{AlFu} for a concise review of
the relationship, Morvan~\cite{Morv} or Gray~\cite{GrayT} for full accounts
of curvature measures, and Alesker~\cite{AlesTVM} for a survey of some more
recent developments.)  The intrinsic volumes can also be defined on any
finite union of convex sets (as in Klain and Rota~\cite{KlRo}).  At these
levels of generality, $V_0$ is no longer trivial; it is the Euler
characteristic. This justifies `Euler%
\index{Euler characteristic}
characteristic' as the right name for
$V_0$ even in the case of convex sets.

The next example uses a notion of
intrinsic volume adapted to $\R^n$ with the $\ell^1$ metric.

\begin{example}
\lbl{eg:ell1-cmc}
\index{lone geometry@$\ell^1$ geometry}
Generalizing Example~\ref{eg:boxy-boxes}, let $A \sub \R^n$ be a
\demph{convex\index{convex!body} body}, that is, a convex set with nonempty
interior.  Give $A$ the $\ell^1$ metric.  Then the magnitude function of
$A$ is the polynomial
\[
\mg{tA} 
=
\sum_{i = 0}^n \frac{1}{2^i} V'_i(A) \cdot t^i
\]
(Theorem~5.4.6(2) of Leinster and Meckes~\cite{MMSCG}).  Here
$V'_i(A)$\ntn{ivellone}%
\index{intrinsic volume!lone@$\ell^1$}
is the $\ell^1$ analogue of the $i$th intrinsic volume of $A$, discussed
in~\cite{MMSCG} and in Section~5 of Leinster~\cite{IGON}.  Explicitly, it is
the sum of the $i$-dimensional volumes of the projections of $A$ onto the
$i$-dimensional coordinate subspaces of $\R^n$.  The significance of $2^i$
is that the volume of the unit ball in the $1$-norm on $\R^i$ is $2^i/i!$.

So, for convex bodies in $\R^n$ equipped with the $\ell^1$ metric, the
magnitude function is a polynomial whose degree is the dimension and whose
$i$th coefficient is an $i$-dimensional geometric measure.
\end{example}

For the Euclidean rather than $\ell^1$ metric on $\R^n$, results on
magnitude are harder.  Until 2015, the only convex subset of $\R^n$ whose
magnitude was known was the line segment.  But a significant advance was made
by Barcel\'o and Carbery~\cite{BaCa}, who used PDE methods to prove:

\begin{thm}[Barcel\'o and Carbery]
\lbl{thm:bc}%
\index{Barcel\'o, Juan Antonio}%
\index{Carbery, Anthony}%
Let $n \geq 1$ be odd.  Then:
\begin{enumerate}
\item
\lbl{part:bc-rat}
the magnitude function $t \mapsto \mg{tB^n}$ of the $n$-dimensional unit
Euclidean ball $B^n$ is a rational function over $\Z$ of the radius $t$;%
\index{ball, magnitude of}

\item
\lbl{part:bc-low}
the magnitude functions of $B^1$, $B^3$ and $B^5$ are given by
\begin{align*}
\mg{tB^1}       &
=
1 + t,  \\
\mg{tB^3}       &
=
\frac{1}{3!}(6 + 12t + 6t^2 + t^3),
\\
\mg{tB^5}       &
=
\frac{1}{5!}
\frac{360 + 1080t + 525t^2 + 135t^4 + 18t^5 + t^6}{3 + t}.
\end{align*}
\end{enumerate}
\end{thm}

\begin{proof}
Part~\bref{part:bc-rat} is Theorem~4 of~\cite{BaCa}.  For
part~\bref{part:bc-low}, the formulas for $\mg{tB^3}$ and $\mg{tB^5}$ are
Theorems~2 and~3 of~\cite{BaCa}, and the formula for $\mg{tB^1}$ is
Example~\ref{eg:boxy-line} (not due to Barcel\'o and Carbery, but included
in the statement for completeness).
\end{proof}

In the $\ell^1$ metric on $\R^n$, the magnitude of a ball is a polynomial
in its radius, by Example~\ref{eg:ell1-cmc}.  In the Euclidean metric, it
is no longer a polynomial, but it is the next best thing: a rational
function.  Subsequent work of Willerton~\cite{WillMOBH,WillMOBP}%
\index{Willerton, Simon} 
identified exactly which rational function $\mg{tB^n}$ is, in terms of
Bessel%
\index{Bessel, Friedrich!polynomials}
polynomials and Hankel%
\index{Hankel determinants}
determinants.

Theorem~\ref{thm:bc} is stated under the hypothesis that $n$ is odd, a
condition imposed in order to put the proof into the realm of differential
rather than pseudodifferential equations.  The magnitude of
even-dimensional balls remains unknown.  Even the 2-dimensional disc $B^2$
has unknown magnitude, although numerical experiments suggest that it is a
certain quadratic polynomial in the radius (Willerton~\cite{WillHCC},
Section~3.2).

Barcel\'o and Carbery also proved a result on general compact sets
(Theorem~1 of~\cite{BaCa}):

\begin{thm}[Barcel\'o and Carbery]
\lbl{thm:bc-vol}%
\index{Barcel\'o, Juan Antonio}%
\index{Carbery, Anthony}%
\index{volume}%
\index{magnitude!determines volume}%
For all $n \geq 1$ and compact $A \sub \R^n$, 
\[
\Vol(A) 
= 
c_n \lim_{t \to \infty} \frac{\mg{tA}}{t^n},
\]
where the constant $c_n$ is $n! \Vol(B^n)$. 
\qed
\end{thm}

The volume of the Euclidean unit ball $B^n$ is given by a standard
classical formula, as in Propositions~6.2.1 and~6.2.2 of Klain and
Rota~\cite{KlRo}, for instance.

By Theorem~\ref{thm:bc-vol}, we can extract the volume of a set from its
magnitude function.  This substantiates the earlier claim that the general
notion of the magnitude of an enriched category encompasses the notion of
volume.

Better still, using methods of global analysis, Gimperlein and Goffeng
proved (Theorem~2(d) of~\cite{GiGo}):

\begin{thm}[Gimperlein and Goffeng]
\lbl{thm:gg}%
\index{Gimperlein, Heiko}%
\index{Goffeng, Magnus}%
\index{magnitude!determines surface area}%
\index{surface area}
Let $n \geq 1$ be odd, and let $A \sub \R^n$ be a bounded set with smooth
boundary such that $A$ is the closure of its interior.  Then the magnitude
function of $A$ has an asymptotic expansion
\[
\mg{tA}
\sim
\sum_{i = 0}^\infty m_i(A) t^{n - i}
\text{ as } t \to \infty,
\]
and up to a known constant factor (depending on $n$ and $i$ but not $A$),
the coefficient $m_i(A)$ is equal to the intrinsic volume $V_{n - i}(A)$
for $i = 0, 1, 2$.  
\qed
\end{thm}

Recent work of Gimperlein, Goffeng and Louca,%
\index{Louca, Nikoletta}
so far unpublished, removes the restriction that $n$ is odd.

For instance, $m_0(A) = \Vol(A)/n!\Vol(B^n)$ (as in
Theorem~\ref{thm:bc-vol}) and $m_1(A)$ is proportional to the $(n -
1)$-dimensional surface%
\index{surface area} 
area of $A$.  In the statement of Theorem~\ref{thm:gg}, the term `intrinsic
volume' has been extended beyond its usual context of convex sets.  A more
precise statement for $i = 2$ is that $m_2(A)$ is proportional to the
integral over $\partial A$ of the mean%
\index{mean!curvature}
curvature of $\partial A$ (which when $A$ is convex is itself proportional
to $V_{n - 2}(A)$).

The magnitude of a metric space does not satisfy the inclusion-exclusion
principle in the strongest conceivable sense, since otherwise, every
$n$-point space would have magnitude $n$.  But Gimperlein and Goffeng
showed that magnitude does satisfy inclusion-exclusion in an asymptotic
sense, using techniques related to the heat equation proof of the
Atiyah--Singer
%
%
index theorem and making essential use of \emph{complex} scale factors $t$.
Indeed, for subsets $A$, $B$ and $A \cap B$ of $\R^n$ satisfying the
regularity conditions of Theorem~\ref{thm:gg},
\[
\index{Gimperlein, Heiko}%
\index{Goffeng, Magnus}%
\index{inclusion-exclusion principle!asymptotic}%
\index{magnitude!inclusion-exclusion for}
\mg{t(A \cup B)} + \mg{t(A \cap B)} - \mg{tA} - \mg{tB}
\to 0
\text{ as } t \to \infty
\]
(Remark~3 of~\cite{GiGo}).  This is further evidence for the claim that
magnitude should be regarded as a measure of size.

Finally, we return to diversity.  Meckes%
\index{Meckes, Mark}
defined the \demph{maximum%
\index{maximum diversity}
diversity} of a compact space $A$ of negative type as
\[
\Dmaxana{A}
=
\sup_\mu
\frac{1}{\int_A \int_A e^{-d(a, b)} \dee\mu(a) \dee\mu(b)},
\]
which is similar to the formula~\eqref{eq:mag-sup} for magnitude, except
that now the supremum runs over only the Borel \emph{probability} measures
$\mu$, as opposed to all signed measures.  (In principle, the formula is
for the maximum diversity of order $2$, but Theorem~7.1 of Leinster and
Roff~\cite{MEMS} implies that the maximum diversity of every order is the same.)
Evidently $\Dmaxana{A} \leq \mg{A}$.

When $A$ is a subset of Euclidean space, $\Dmaxana{A}$ is equal to a
classical quantity, the Bessel%
\index{Bessel, Friedrich!capacity} 
capacity%
\index{capacity}
$C_{(n + 1)/2}(A)$.  As Meckes%
\index{Meckes, Mark}
showed, a deep result from the theory of capacities provides an upper bound
on $\mg{A}/\Dmaxana{A}$%
\index{magnitude!maximum diversity@vs.\ maximum diversity}%
\index{maximum diversity!magnitude@vs.\ magnitude}
in terms of $n$ alone (Corollary~6.2 of~\cite{MeckMDC}).
%
%
Thus, magnitude is never very different from this Bessel capacity.

Meckes~\cite{MeckMDC} exploited the connection between magnitude and
maximum diversity to extract information about the dimension of a compact
set $A \sub \R^n$ from its magnitude function.  We have already met some
families of spaces where the magnitude function is a polynomial whose
degree is the dimension (Example~\ref{eg:ell1-cmc}).  But here we allow
non-integer dimensions too.

One of the most important notions of fractional dimension is the
\demph{Minkowski}%
\index{dimension}%
\index{Minkowski, Hermann!dimension} 
or \demph{box-counting%
\index{box-counting dimension} 
dimension} (Section~3.1 of Falconer~\cite{Falc}).  The Minkowski dimension
of a subset of $\R^n$ is always greater than or equal to the Hausdorff
dimension, and equality often holds.  (See p.~43 of~\cite{Falc} for
a summary of how the two dimensions are related.)  For instance, both the
Minkowski and the Hausdorff dimensions of the middle-thirds Cantor set are
$\log 2/\log 3$.
Write $\Minkdim{A}$\ntn{dimM} for the
Minkowski dimension of a compact set $A \sub \R^n$, if defined.

Roughly speaking, the following result states that $\mg{tA}$ grows like
$t^{\Minkdim{A}}$ when $t$ is large.  Thus, we can can recover the
Minkowski dimension of a space from its magnitude function.  It is due to
Meckes (Corollary~7.4 of~\cite{MeckMDC}).

\begin{thm}[Meckes]
\lbl{thm:mink}%
\index{Meckes, Mark}
\index{magnitude!determines dimension}
Let $A$ be a compact subset of $\R^n$.  Then
\[
\Minkdim{A}
=
\lim_{t \to \infty} \frac{\log \mg{tA}}{\log t},
\]
with one side of the equation defined if and only if the other is.
\qed
\end{thm}

For instance, if $A$ is a subset of $\R^n$ with nonzero volume, then
$\mg{tA}$ grows like $t^n$ when $t$ is large, and by the volume theorem of
Barcel\'o and Carbery, the ratio $\mg{tA}/t^n$ converges to a known
constant times the volume of $A$.  When $A$ is the middle-thirds
Cantor%
\index{Cantor set} 
set, $\mg{tA}$ grows like $t^{\log 2/\log 3}$.  (In fact, the magnitude
function of the Cantor set also has a kind of hidden periodicity, as shown
in Section~3 of Leinster and Willerton~\cite{AMSES}.)  For convex subsets
of $\R^n$, more precise statements can be made; Meckes%
\index{Meckes, Mark}
bounds the magnitude function of a convex set by a polynomial whose
coefficients are proportional to its intrinsic volumes (\cite{MeckMIV},
Theorem~1). 

Theorem~\ref{thm:mink} demonstrates the usefulness of the concept of
maximum%
\index{maximum diversity!geometry and analysis@in geometry and analysis}
diversity for pure-mathematical purposes in geometry and analysis,
independently of any biological application.

%% file: value.tex
\chapter{Value}
\lbl{ch:value}
\index{value}

\begin{quote}
Quite apart from many theoretical and practical problems that continue to
affect the species concept and its application, is it appropriate for
conservation\index{conservation} purposes to regard all species as equal in
this manner?  To a conservationist, regardless of relative abundance, is
\emph{Welwitschia}\index{Welwitschia@\emph{Welwitschia}} equal to a species
of \emph{Taraxacum}?\index{Taraxacum@\emph{Taraxacum}} Is the
panda\index{panda} equivalent to one species of rat?\index{rat}\\ 
\ \quad%
\index{Vane-Wright, Richard}
\hfill -- Vane-Wright et al.~(\cite{VWHW}, p.~237)
\end{quote}

\noindent
Putting aside entropy and diversity, let us consider a very
general question:
\begin{quote}
\emph{What is the value of the whole in terms of its parts?}
\end{quote}
Although the question in this form is far too vague to admit a mathematical
treatment, we will see that once posed precisely, it has a complete answer.
From that answer, the concept of diversity arises automatically.  The
answer also leads to a unique characterization of the Hill numbers (or
equivalently, the R\'enyi entropies), more powerful than the
characterization theorem in Section~\ref{sec:hill-char-given}.

We will consider a `whole' divided into $n$ `parts' of relative sizes $p_1,
\ldots, p_n$, which are assigned values $v_1, \ldots, v_n$ respectively
(Figure~\ref{fig:value}).
\begin{figure}
\centering
\lengths
\begin{picture}(120,24)
\thicklines
\cell{60}{12}{c}{\includegraphics[height=24\unitlength]{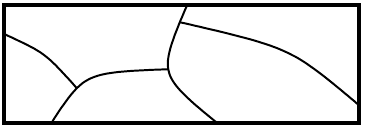}}
\cell{30}{10}{c}{$p_1$}
\valput{34}{6}{$v_1$}
\cell{40}{18}{c}{$p_2$}
\valput{46}{15}{$v_2$}
\cell{49}{7.5}{c}{$p_3$}
\valput{54}{4.5}{$v_3$}
\cell{73}{10}{c}{$p_4$}
\valput{79}{7}{$v_4$}
\cell{80}{19}{c}{$p_5$}
\valput{86}{16}{$v_5$}
\end{picture}
\caption{A whole divided into $5$ parts, with relative sizes $(p_1, \ldots,
  p_5) \in \Delta_5$ and values $(v_1, \ldots, v_5)$.}
\lbl{fig:value}
\end{figure}
The question is how to aggregate those values into a
single value $\sigma(\p, \vc{v})$ for the whole, measured in the same units
as the values $v_i$ of the parts.  This aggregation method should have
sensible properties.  For instance, if we put together two parts of equal
size and equal value, $v$, the result should have value $2v$.  

One simple method is to ignore the sizes of the parts and just sum their
values, so that
\[
\sigma(\p, \vc{v}) = v_1 + \cdots + v_n
\]
(or better, $\sigma(\p, \vc{v}) = \sum_{i \in \supp(\p)} v_i$).  But there
are many other possibilities.  In fact, we will define a one-parameter
family $(\sigma_q)$ of value measures.  They include as special cases the
Hill numbers $D_q$, the more general similarity-sensitive diversity
measures $D_q^Z$ of Chapter~\ref{ch:sim}, certain phylogenetic diversity
measures (due to Chao, Chiu and Jost~\cite{CCJ}), and, essentially, the
$\ell^p$\index{pnorm@$p$-norm} norms.  For example, when a community is
divided into $n$ species in proportions $p_1, \ldots, p_n$, and each
species is assigned the same value, $1$, the value of the whole according
to $\sigma_q$ is the Hill number of order $q$:
\[
\sigma_q\bigl(\p, (1, \ldots, 1)\bigr) = D_q(\p).
\]

In most of the cases just listed, the whole is taken to be an ecological
community and the parts are its species.  But there is an important
complementary situation, in which the whole is still a community but the
parts are taken to be subcommunities.  For instance, the community
might be divided geographically into regions, and we might attempt to
evaluate the community as a whole based on the sizes and values of those
regions.  In the case where value is interpreted as diversity, that is
exactly what we did when we derived the chain rule for diversity
(Propositions~\ref{propn:hill-chn} and~\ref{propn:div-sim-ch}).  Indeed,
the function $\sigma_q$ can be seen as an embodiment of the chain rules for
$D_q$ and $D_q^Z$, in a sense explained in Example~\ref{eg:value-ch}.

We begin by defining the value measures $\sigma_q$ and analysing some
special cases (Section~\ref{sec:value-defn}), with important examples from
both ecology and the analysis of social welfare.  We then introduce the
R\'enyi relative entropies, which are very closely related to the value
measures $\sigma_q$.  (The $q$-logarithmic relative entropies were already
covered in Section~\ref{sec:q-log-ent}.)  As a bonus, we use the R\'enyi
and $q$-logarithmic relative entropies to provide further evidence for the
canonical nature of the Fisher metric on probability distributions
(Remark~\ref{rmks:fisher-def}\bref{rmk:fisher-def-gods-joke}).

Using our earlier results on means, we then prove that the only value
measures with reasonable properties are those belonging to the family
$(\sigma_q)$ (Section~\ref{sec:value-char}).  From this we deduce that for
communities modelled as their relative abundance distributions,
the only reasonable measures of diversity are the Hill numbers
(Section~\ref{sec:total-hill}).

We have already proved a characterization theorem for the Hill%
\index{Hill number!difference between characterizations of}
numbers $D_q$ in Section~\ref{sec:hill-char-given}, showing that for a
\emph{fixed} $q$, if a diversity measure $D$ has certain properties
\emph{depending on $q$}, then it must be equal to $D_q$.  But in the
theorem proved in this chapter, there is no `$q$' mentioned in the
hypotheses, and the conclusion is that $D$ must be equal to $D_q$
\emph{\,for some $q$}.  In short, the earlier theorem characterized
the Hill numbers individually, but this theorem characterizes them as a
family.

\section{Introduction to value}
\lbl{sec:value-defn}

Here we consider sequences of functions
\[
\bigl( 
\sigma \from \Delta_n \times [0, \infty)^n \to [0, \infty) 
\bigr)_{n \geq 1},
\]
which will be referred to as \demph{value%
\index{value measure}
measures}.  We regard a pair $(\p, \vc{v}) \in \Delta_n \times [0,
\infty)^n$ as representing a whole made up of $n$ disjoint parts with
relative sizes $p_1, \ldots, p_n$ and values $v_1, \ldots, v_n$, and we
regard $\sigma(\p, \vc{v})$ as the value that $\sigma$ assigns to the
whole.

A special role is played by the family 
\[
\bigl( \sigma_q \bigr)_{q \in [-\infty, \infty]}
\ntn{sigmaq}
\]
of value\index{value measure} measures, defined by
\[
\sigma_q(\p, \vc{v}) = M_{1 - q} (\p, \vc{v}/\p)
\]
($n \geq 1$, $\p \in \Delta_n$, $\vc{v} \in [0, \infty)^n$).  The
convention adopted in Remark~\ref{rmk:defined-even-if-not} ensures that
$\sigma_q(\p, \vc{v})$ is always well-defined.
We call $\sigma_q$ the \demph{value measure of order~%
\index{value measure!order q@of order $q$}%
\index{order!value measure@of value measure}
$q$}.  Explicitly, when $q \neq 1, \pm \infty$,
\[
\sigma_q(\p, \vc{v})
=
\Biggl( 
\sum_{i \in \supp(\p)} p_i^q v_i^{1 - q} 
\Biggr)^{1/(1 - q)},
\]
unless $q > 1$ and $v_i = 0$ for some $i \in \supp(\p)$, in which case
$\sigma_q(\p, \vc{v}) = 0$.  For $q \in \{1, \pm\infty\}$, 
\begin{align*}
\sigma_{-\infty}(\p, \vc{v})    &
=
\max_{i \in \supp(\p)} \frac{v_i}{p_i}, \\
\sigma_1(\p, \vc{v})    &
=
\prod_{i \in \supp(\p)} \biggl( \frac{v_i}{p_i} \biggr)^{p_i},    \\
\sigma_{\infty}(\p, \vc{v})     &
=
\min_{i \in \supp(\p)} \frac{v_i}{p_i}.
\end{align*}

\begin{examples}
\lbl{egs:val-first}
\begin{enumerate}
\item 
\lbl{eg:val-first-indivs} 
Consider a set of $k$ individuals, divided into $n$ equivalence classes
(`parts'), with the $i$th part consisting of $k_i$ individuals.  Let $p_i =
k_i/k$ be the proportion of individuals in the $i$th part.  Let $v_1,
\ldots, v_n \in [0, \infty)$ be any values assigned to the parts.  Then
\[
\sigma_q(\p, \vc{v})
=
M_{1 - q}\Biggl( 
\p, \Biggl( \frac{kv_1}{k_1}, \ldots, \frac{kv_n}{k_n} \Biggr)
\Biggr),
\]
or equivalently,
\begin{equation}
\lbl{eq:val-indivs}
\sigma_q(\p, \vc{v})
=
k \cdot 
M_{1 - q}\Biggl( 
\p, \Biggl( \frac{v_1}{k_1}, \ldots, \frac{v_n}{k_n} \Biggr)
\Biggr).
\end{equation}
This can be understood as follows.  If the value $v_i$ of the $i$th part is
shared out evenly among its $k_i$ members, then the value%
\index{value!per individual}
per individual in the $i$th part is $v_i/k_i$.  Hence the mean value per
individual in the whole is
\[
M_{1 - q}\Biggl(
\p, \Biggl( \frac{v_1}{k_1}, \ldots, \frac{v_n}{k_n} \Biggr)
\Biggr).
\]
So, equation~\eqref{eq:val-indivs} states that
\[
\text{value of whole} = \text{number of individuals}
\times \text{mean value per individual}.
\]
This is the basic conceptual relationship between value measures and
means. 

\item
If in~\bref{eg:val-first-indivs} we interpret `mean' as arithmetic mean,
then we are in the case $q = 0$, and $\sigma_0$ is simply given by
\[
\sigma_0(\p, \vc{v}) = \sum_{i \in \supp(\p)} v_i
\]
(as in the introduction to this chapter).  But we have seen repeatedly in
this book that the \emph{arithmetic} mean is not the only useful kind.  The
other power means should always be considered alongside it, and in this
case, they give the whole family $(\sigma_q)$. 
\end{enumerate}
\end{examples}

\begin{remark}
\lbl{rmk:vals-mns}
The value measures $\sigma_q$ and the power means $M_t$ are sequences of
functions of the same type:
\[
\bigl( 
\sigma_q, M_t \from \Delta_n \times [0, \infty)^n \to [0, \infty)
\bigr)_{n \geq 1}.
\]
However, Example~\ref{egs:val-first}\bref{eg:val-first-indivs} makes clear
that there should be no overlap%
\index{value measure!mean@vs.\ mean}%
\index{mean!value measure@vs.\ value measure}
between the classes of value measures and means.  Indeed, a reasonable
value measure $\sigma$ should satisfy
\[
\sigma\bigl( \vc{u}_n, (v, \ldots, v)\bigr) = nv,
\]
whereas a minimal requirement of a mean $M$ is the consistency condition
\[
M\bigl( \vc{u}_n, (x, \ldots, x)\bigr) = x.
\]
So, no reasonable mean is a reasonable value measure.  We return to the
relationship between means and value measures in
Section~\ref{sec:value-char}.
\end{remark}

For positive parameters $q$, the value of the whole is never more than the
sum of the values of its parts:

\begin{lemma}
\lbl{lemma:sigma-sum-ineq}
For all $q \geq 0$, $\p \in \Delta_n$, and $\vc{v} \in [0, \infty)^n$,
\[
\sigma_q(\p, \vc{v}) \leq \sum_{i = 1}^n v_i.
\]
For $q > 0$, equality holds if and only if $\vc{v}$ is a scalar multiple of
$\p$.
\end{lemma}

So for fixed $\sum v_i$, the value of the whole is maximized when value
is spread evenly across the constituent parts, in proportion to their
sizes.

\begin{proof}
For all $q \geq 0$,
\[
\sigma_q(\p, \vc{v})    
=
M_{1 - q}(\p, \vc{v}/\p)        
\leq
M_1(\p, \vc{v}/\p) 
=
\sum_{i \in \supp(\p)} v_i      
\leq
\sum_{i = 1}^n v_i.
\]
Assuming now that $q > 0$, equality holds in the first inequality if and
only if $v_i/p_i$ is constant over $i \in \supp(\p)$ (by
Theorem~\ref{thm:mns-inc-ord}), and in the second if and only if $v_i = 0$
for all $i \not\in \supp(\p)$.  The result follows.
\end{proof}

The next two examples illuminate the meaning of the parameter $q$.
They concern the case where the parts are of equal size ($\p = \vc{u}_n$),
so that the value measures $\sigma_q$ are given by
\[
\sigma_q(\vc{u}_n, \vc{v}) = n \cdot M_{1 - q}(\vc{u}_n, \vc{v})
\]
($q \in [-\infty, \infty]$, $\vc{v} \in [0, \infty)^n$).  

\begin{example}
A classical question in welfare%
\index{welfare}
economics%
\index{economics} 
is how to take a group
of agents, each of which has an assigned utility,%
\index{utility}
and aggregate their individual utilities into a measure of the utility of
the group as a whole.  For instance, the agents might be the citizens of a
society, and the utility of a citizen might be their individual level of
welfare, wealth or well-being.  The challenge, then, is to combine them
into a single number representing the collective welfare of the society.
(As a general reference for all of this example, we refer to Section~1.2
and Chapter~3 of Moulin~\cite{Moul}.)

Specifically, fix $n$, and take a group of $n$ individuals with respective
utilities $v_1, \ldots, v_n \geq 0$.  A \demph{collective%
\index{collective utility function} 
utility function} assigns a real number $f(\vc{v})$ to each such tuple
$\vc{v} = (v_1, \ldots, v_n)$.  
For example,
\[
\sigma_q(\vc{u}_n, -) \from [0, \infty)^n \to \R
\]
is a collective utility function for each $q \in [-\infty, \infty]$.  

More important than the collective utility function $f$ itself is its
associated \demph{social%
\index{social welfare function}%
\index{ordering, social welfare}
welfare ordering}, which is the relation $\swo$ on $[0, \infty)^n$ defined
  by
\[
\vc{v} \swo \vc{v}' \iff f(\vc{v}) \leq f(\vc{v}').
\]
In the case of the welfare of the citizens of a society, $\vc{v} \swo
\vc{v}'$ is interpreted as the judgement that when the welfare levels of
the citizens are $v_1, \ldots, v_n$, society is in a poorer state than when
they are $v'_1, \ldots, v'_n$.

Of course, such judgements depend on a choice of collective utility
function $f$.  When $f = \sigma_q(\vc{u}_n, -)$, different values of $q$
correspond to different viewpoints, some of which are associated with
particular schools of political%
\index{politics} 
philosophy.%
\index{philosophy, political}
The case $q = 0$ is
\[
\sigma_0(\vc{u}_n, -) \from 
\vc{v} \mapsto \sum v_i,
\]
so that the collective welfare is simply the sum of the individual welfares.
This function is associated with classical utilitarianism,%
\index{utilitarianism} 
with its
roots in the philosophy of Jeremy Bentham%
\index{Bentham, Jeremy} 
and in John Stuart Mill's%
\index{Mill, John Stuart}
`sum total of happiness'.  When $q = \infty$, the collective utility
function is
\[
\sigma_\infty(\vc{u}_n, -) \from
\vc{v} \mapsto n \min v_i,
\]
so that
\[
\vc{v} \swo \vc{v}' \iff \min v_i \leq \min v'_i.
\]
This viewpoint on collective welfare is associated with the philosophy of
John Rawls:%
\index{Rawls, John}
a society should be judged by the welfare of its most
miserable citizen.  An intermediate position is $q = 1$, where
\[
\sigma_1(\vc{u}_n, -) \from 
\vc{v} \mapsto n \cdot \biggl( \prod v_i \biggr)^{1/n}
\]
and so
\[
\vc{v} \swo \vc{v}' \iff \prod v_i \leq \prod v'_i.
\]
In this context, the product operation $\vc{v} \mapsto \prod v_i$ is known
as the \demph{Nash%
\index{Nash collective utility function} 
collective utility function}, and has special properties not shared by any
other collective utility function (unsurprisingly, given the special role
played by the case $q = 1$ in the context of entropy).

An important property of collective utility functions is the
Pigou--Dalton%
\index{Pigou--Dalton principle}
principle.  In the language of wealth, this states that transferring a
small amount of wealth from a richer citizen to a poorer one is beneficial
to the overall welfare of society.  Formally, let $\vc{v} \in [0,
  \infty)^n$ and $i, j \in \{1, \ldots, n\}$ with $v_i < v_j$, and let $0
  \leq \delta \leq (v_j - v_i)/2$; define $\vc{v}' \in [0, \infty)^n$ by
\[
v'_k =
\begin{cases}
v_i + \delta &\text{if } k = i,      \\
v_j - \delta &\text{if } k = j,      \\
v_k     &\text{otherwise}.
\end{cases}
\]
The \demph{Pigou--Dalton principle} is that $\vc{v} \swo \vc{v'}$ for
all such $\vc{v}$, $i$, $j$, and $\delta$.

When $q \in [0, \infty]$, an elementary calculation shows that
$\sigma_q(\vc{u}_n, -)$ satisfies the Pigou--Dalton principle.  Thus,
redistribution%
\index{redistribution of wealth} 
is regarded positively.  On the other hand, the Pigou--Dalton principle
fails for all $q \in [-\infty, 0)$.  In fact, for $q \in (-\infty, 0)$,
  redistribution from richer to poorer always strictly \emph{decreases}
  overall welfare.  In the extreme case $q = -\infty$, the collective
  utility function is
\[
\sigma_{-\infty}(\vc{u}_n, -) \from \vc{v} \mapsto n \max_i v_i,
\]
so that the welfare of a society is proportional to the welfare of its most
privileged citizen.  (Recall that $n$ is fixed.)  Thus, from the viewpoint
of $q = -\infty$, collective welfare is optimized when all the wealth is
transferred to a single individual.  In the welfare economics literature,
negative values of $q$ are often excluded.

(The family $(\sigma_q(\vc{u}_n, -))$ of collective utility functions that
we have used is different from the family used in economics texts such as
Moulin~\cite{Moul}, but only superficially.  In the literature, it is
conventional to use the functions
\[
\begin{array}{rcll}
\vc{v}  &\mapsto        &\sum v_i^t     &(t \in (0, \infty)),   \\
\vc{v}  &\mapsto        &\sum \log v_i,         &               \\
\vc{v}  &\mapsto        &-\sum v_i^t  &(t \in (-\infty, 0)),
\end{array}
\]
whereas we have been using
\begin{equation}
\lbl{eq:val-ufm-exp}
\vc{v} \mapsto \sigma_q(\vc{u}_n, \vc{v})
=
\begin{cases}
n^{q/(q - 1)} \Bigl( \sum v_i^{1 - q} \Bigr)^{1/(1 - q)}      &
\text{if } 1 \neq q \in (-\infty, \infty),      \\
n \prod v_i^{1/n}       &
\text{if } q = 1.
\end{cases}
\end{equation}
But reparametrizing with $q = 1 - t$, the induced social welfare orderings
are identical.)
\end{example}

\begin{example}
In contexts such as collective welfare and diversity, it is natural to
restrict the parameter $q$ to be positive.  But for negative parameters
$q$, the value measures $\sigma_q$ also define something important, at
least when the parts are of equal size: the $\ell^p$\index{pnorm@$p$-norm}
norms.  Indeed, for $-\infty < q \leq 0$, equation~\eqref{eq:val-ufm-exp}
gives
\[
\sigma_q(\vc{u}_n, \vc{v})      
=
n^{q/(q - 1)} \|\vc{v}\|_{1 - q},
\]
where the norm $\|\cdot\|_{1 - q}$ is as defined in
Example~\ref{eg:norm-p}.
\end{example}

We now show that all of the diversity measures discussed in
previous chapters are encompassed by the value measures $\sigma_q$.

\begin{example}
\lbl{eg:value-hill}
Consider an ecological community made up of species with relative abundances
$p_1, \ldots, p_n$.  In the absence of other information, it is
natural to give all the species the same value, $1$.  We have
\[
\sigma_q\bigl( \p, (1, \ldots, 1)\bigr)
=
M_{1 - q}(\p, 1/\p)
=
D_q(\p),
\]
so the value assigned to the community by $\sigma_q$ is the Hill
number $D_q(\p)$.
\end{example}

\begin{example}
\lbl{eg:value-dqz} 
Now let us enrich our model of the community with an $n \times n$
similarity matrix $Z$.  Assume that the diagonal entries of $Z$ are all $1$
(as discussed on p.~\pageref{p:Z1}).  Based on this model, what value $v_i$
can we 
reasonably assign to each species?

In Section~\ref{sec:sim-basic}, we considered the quantity 
\[
(Z\p)_i
=
\sum_{j = 1}^n Z_{ij} p_j
\]
associated with the $i$th species.  This is the expected similarity between
an individual of species $i$ and an individual chosen from the community at
random.  We called $(Z\p)_i$ the ordinariness of species $i$, and
$1/(Z\p)_i$ its specialness.%
\index{specialness}  

This might seem to suggest using $1/(Z\p)_i$ as the value of the $i$th
species.  However, $1/(Z\p)_i$ is a measure of the specialness of an
\emph{individual} of the $i$th species, whereas $v_i$ is supposed to
measure the value of the $i$th part (species) \emph{as a whole}.  We
therefore define $v_i$%
\index{value!species@of a species}
to be the specialness per individual
in the species multiplied by the size of the species:
\[
v_i 
=
\frac{p_i}{(Z\p)_i}.
\]
When $Z$ is the naive similarity matrix $I$, this formula reduces to $v_i =
1$, as in Example~\ref{eg:value-hill}.  More generally, if species $i$ is
completely dissimilar to all other species ($Z_{ij} = 0$ for all $i \neq
j$) then $v_i = 1$.  In any case, $v_i \leq 1$, since $(Z\p)_i \geq p_i$ 
(inequality~\eqref{eq:Zpi-lb1}, p.~\pageref{eq:Zpi-lb1}).  Lower values
$v_i$ indicate that in comparison to the size of the $i$th species, there
are many individuals belonging to species similar to it.  This agrees with
the intuition that such a species contributes little to the diversity of
the whole.

With this definition of $\vc{v}$ as $\p/Z\p$, we recover the
similarity-sensitive diversity measures $D_q^Z$ of Chapter~\ref{ch:sim}:
\[
\sigma_q(\p, \p/Z\p)
=
M_{1 - q}(\p, 1/Z\p)
=
D_q^Z(\p),
\]
by definition of $D_q^Z$.
\end{example}

\begin{example}
\lbl{eg:value-ch} 
Now take a community of individuals that are not only classified into a
number of species (with similarities encoded in a matrix $Z$), but
also divided into $n$ disjoint subcommunities.  Thus, each individual
belongs to exactly one species and exactly one subcommunity.  We will
assume that the different subcommunities share no species, and that species
in different subcommunities are completely dissimilar (as in
Example~\ref{eg:div1-chain-islands} and Propositions~\ref{propn:hill-chn}
and~\ref{propn:div-sim-ch}, where the subcommunities were called
`islands').%
\index{islands!diversity of group of}

Write $w_i$ for the population size of the $i$th subcommunity relative to
the whole community, so that $\sum w_i = 1$.  Also write $d_i$ for the
diversity of order $q$ of the $i$th subcommunity.  Then by the chain rule
for the similarity-sensitive diversities
(Proposition~\ref{propn:div-sim-ch}), the diversity of order $q$ of the
whole community is
\[
\sigma_q(\vc{w}, \vc{d}).
\]
This is the fundamental relationship%
\index{value!diversity@and diversity}%
\index{diversity!value@and value}
between value and diversity.  If value is taken to mean diversity of order
$q$, then $\sigma_q$ correctly aggregates the values of the parts of a
community to give the value of the whole.
\end{example}

\begin{example}
In the ecological settings discussed, we have only ever considered the
\emph{relative} abundances of species.  But absolute%
\index{abundance!absolute}
abundances sometimes matter.  What happens if we measure the value of a
species within a community as its absolute abundance?

Consider a community of individuals divided into $n$ species, with absolute
abundances $A_1, \ldots, A_n$.  Writing $A = \sum A_i$, the relative
abundances are $p_i = A_i/A$.  For all $q \in [-\infty, \infty]$,
\begin{align*}
\sigma_q\bigl(\p, (A_1, \ldots, A_n)\bigr)      &
=
M_{1 - q}\Biggl( 
\p, \Biggl(\frac{A_1}{p_1}, \ldots, \frac{A_n}{p_n}\Biggr) 
\Biggr)  \\
&
=
M_{1 - q}\bigl( \p, (A, \ldots, A) \bigr)       \\
&
=
A.
\end{align*}
So, the value of the whole is simply the total abundance.

In this example, the value measures $\sigma_q$ give us no interesting new
quantity.  The answer to the question posed is trivial.  But it is also
reasonable: if the value of each part of a community is taken to be just
the number of individuals it contains, it is natural that the value of the
whole community is measured in that way too.
\end{example}

We conclude this introduction to value with a more substantial example.

\begin{example}
\lbl{eg:value-ccj}
\index{phylogenetic!diversity}%
\index{Chao, Anne!Chiu--Jost@--Chiu--Jost phylogenetic diversity}
Here we describe the phylogenetic diversity measures of Chao, Chiu and
Jost~\cite{CCJ} and show that they too are a special case of the value
measures $\sigma_q$.

A \demph{phylogenetic tree}\index{phylogenetic!tree} is a depiction of the
evolutionary history of a group of species, as in
Figure~\ref{fig:phylo-trees}.  (For a general guide to the subject, see
Lemey et al.~\cite{LSV}.)
\begin{figure}
\centering
\lengths
\begin{picture}(44,57)(0,-2)
\cell{22.9}{55}{t}{\includegraphics[width=41.8\unitlength]{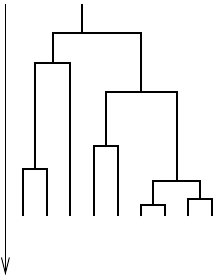}}
\cell{0}{28}{l}{\rotatebox{-90}{\normalsize time}}
\cell{6.5}{9}{b}{1}
\cell{11}{9}{b}{2}
\cell{15.6}{9}{b}{3}
\cell{20.1}{9}{b}{4}
\cell{24.7}{9}{b}{5}
\cell{29.2}{9}{b}{6}
\cell{33.8}{9}{b}{7}
\cell{38.3}{9}{b}{8}
\cell{43}{9}{b}{9}
\cell{22}{-2}{b}{(a)}
\end{picture}%
\hspace*{7\unitlength}%
\begin{picture}(38,57)(5.5,-2)
\cell{24.6}{55}{t}{\includegraphics[width=37.7\unitlength]{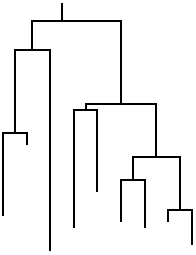}}
\cell{6.5}{9}{b}{1}
\cell{11}{23.5}{b}{2}
\cell{15.6}{3}{b}{3}
\cell{20.1}{7.5}{b}{4}
\cell{24.7}{14}{b}{5}
\cell{29.1}{8.5}{b}{6}
\cell{33.7}{7}{b}{7}
\cell{38.2}{8.5}{b}{8}
\cell{42.8}{3.5}{b}{9}
\cell{24.6}{-2}{b}{(b)}
\end{picture}%
\hspace*{7\unitlength}%
\begin{picture}(24,57)(0,3)
\cell{13.5}{31.5}{c}{\includegraphics[width=21\unitlength]{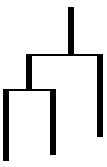}}
\cell{4}{12.5}{c}{1}
\cell{13.5}{14}{c}{2}
\cell{23}{17.5}{c}{3}
\cell{14.5}{43}{c}{\normalsize$b_1$}
\cell{6}{34}{c}{\normalsize$b_2$}
\cell{1.5}{22}{c}{\normalsize$b_3$}
\cell{16.5}{23}{c}{\normalsize$b_4$}
\cell{20.5}{28}{c}{\normalsize$b_5$}
\cell{13.5}{3}{b}{(c)}
\end{picture}%
\caption{Simple examples of phylogenetic trees, with present-day species
  labelled as $1, 2, \ldots$\:\  Tree~\hardref{(a)} is ultrametric, but
  trees~\hardref{(b)} and~\hardref{(c)} are not.  Tree~\hardref{(c)} has
  five branches, shown as thick lines and labelled as $b_1, \ldots, b_5$.}  
\lbl{fig:phylo-trees}
\end{figure}
The vertical axis indicates time, or some proxy for time; the horizontal
distances in the trees mean nothing.
Figure~\ref{fig:phylo-trees}\hardref{(a)} shows nine species descended from
a single species.  In that example, the tree is
\demph{ultrametric}\index{ultrametric!tree}, meaning that the tips of the
tree (the present-day species) are all at the same height.

Evolutionary history is often inferred from genetic data,
with the number of genetic mutations used as a means of estimating time.
Because the rate of genetic mutation is not constant (and for other
reasons), the trees produced in this way are generally not ultrametric.
Figure~\ref{fig:phylo-trees}\hardref{(b)} shows an example.  

From a phylogenetic tree, we can extract the following information:
\begin{itemize}
\item 
the set of present-day species, which we label as $1, \ldots, S$;

\item
the set $B$ of branches;

\item
the binary relation $\descd$\ntn{descd}, where for a present-day species $r$
and a branch $b$, we write $r \descd b$ if $r$ is descended from $b$;

\item
the length $L(b) \geq 0$ of each branch $b$.
\end{itemize}
These four pieces of information are the only aspects of a tree that we
will need for the present purposes.  For instance, the tree of
Figure~\ref{fig:phylo-trees}\hardref{(c)} has $S = 3$, $B = \{b_1, b_2,
b_3, b_4, b_5\}$, and
\begin{align*}
&1 \descd b_1, \ 1 \descd b_2, \ 1 \descd b_3, \\
&2 \descd b_1, \ 2 \descd b_2, \ 2 \descd b_4, \\
&3 \descd b_1, \ 3 \descd b_5.
\end{align*}
We do not require that the present-day species are all descended from a
common ancestor within the time span considered; that is, the `tree' may
actually consist of several disjoint trees (a
forest,\index{forest!mathematical} in mathematical terminology).

We will consider measures of community diversity based on two factors: a
phylogenetic tree for the species, and their present-day relative abundance
distribution $(\pi_1, \ldots, \pi_S)$.  To do this, we introduce some
notation.

For each branch $b$, write
\begin{equation}
\lbl{eq:ccj-abun}
\pi(b) = \sum_{r \csuch r \descd b} \pi_r,
\end{equation}
which is the total relative abundance of present-day species descended from
branch $b$.  So if the tree is ultrametric then whenever we draw a
horizontal line across the tree (representing a particular point $t$ in
evolutionary time), the sum of $\pi(b)$ over all branches $b$ intersecting
that line is $1$.  For any given point in evolutionary time, we therefore
have a probability distribution on the set of species present then,
although as Chao%
\index{Chao, Anne} 
et al.\ warn, `These abundances [$\pi(b)$] are not estimates of the actual
abundances of these ancestral species at time $t$, but rather measures of
their importance for the present-day assemblage' (\cite{CCJ},
Section~4(a)).

For each present-day species $r \in \{1, \ldots, S\}$, write
\[
L_r = \sum_{b \csuch r \descd b} L(b).
\]
This is the length of the lineage of species $r$ within the tree.  For the
tree to be ultrametric means that $L_1 = \cdots = L_S$.  Whether or not it
is ultrametric, we can define the average lineage length $\ovln{L}$ by any
of three equivalent formulas:
\[
\ovln{L}
=
\sum_r \pi_r L_r
=
\sum_{r, b \csuch r \descd b} \pi_r L(b)
=
\sum_b \pi(b) L(b).
\]
Hence $\ovln{L}$ is the expected lineage length of an individual chosen at
random from the present-day community.  

Chao, Chiu and Jost defined a phylogenetic diversity measure as follows.
For each time point $t$ in the period under consideration, they took the
abundance distribution described below equation~\eqref{eq:ccj-abun}.  They
then took its Hill number of order $q$, and formed the average of these
Hill numbers over all times $t$.  After some simplification, the result is
the diversity measure
\[
\CCJ{q}
=
\Biggl( \sum_b \frac{L(b)}{\ovln{L}} \pi(b)^q \Biggr)^{1/(1 - q)}
\]
for $1 \neq q \in [0, \infty)$, and 
\[
\CCJ{1}
=
\prod_b \pi(b)^{-{\textstyle\frac{L(b)}{\ovln{L}}} \pi(b)}.
\]
(Their derivation is on its surest footing when the tree is ultrametric.
Discussion of what can go wrong otherwise is in the supplement to
Chao et al.~\cite{CCJ} and in Example~A20 of the appendix to
Leinster and Cobbold~\cite{MDISS}.)  For example, the case $q = 0$ is
simply
\[
\CCJ{0} = \frac{1}{\ovln{L}} \sum_b L(b).
\]
Up to a factor of $\ovln{L}$, this is the total length of all the branches
in the tree, which is known as \demph{Faith's phylogenetic
  diversity}~\cite{Fait}.%
\index{Faith's phylogenetic diversity}

We now show that Chao, Chiu and Jost's measure $\CCJ{q}$ is a simple
instance of the value measure $\sigma_q$.  For this, we consider the
phylogenetic tree as our whole and the branches as its parts.
The value of a branch $b$ is defined as 
\[
v(b) = \frac{L(b)}{\ovln{L}},
\]
the proportion of evolutionary time over which the branch extends.  It is
purely a measure of the branch's historical duration, and is independent of
the abundances of present-day species.  We define the relative size $p(b)$
of the branch to be
\[
p(b) = \frac{\pi(b) L(b)}{\ovln{L}}.
\]
In other words, $p(b)$ is the product of $\pi(b)$, the proportion of
present-day individuals descended from branch $b$, and $L(b)/\ovln{L}$, the
relative length of the branch.  Then $\sum_b p(b) = 1$.

With these definitions, the value $\sigma_q(\p, \vc{v})$ of the community
is
\begin{align*}
\sigma_q(\p, \vc{v})    &
=
\Biggl( \sum_b p(b)^q v(b)^{1 - q} \Biggr)^{1/(1 - q)}  \\
&
=
\Biggl( 
\sum_b 
\frac{\pi(b)^q L(b)^q}{\ovln{L}^q}
\frac{L(b)^{1 - q}}{\ovln{L}^{1 - q}}
\Biggr)^{1/(1 - q)}     \\
&
=
\Biggl( 
\sum_b 
\frac{L(b)}{\ovln{L}} \pi(b)^q
\Biggr)^{1/(1 - q)}      \\
&
=
\CCJ{q}
\end{align*}
($q \neq 1, \infty$).  Similarly, $\sigma_1(\p, \vc{v}) = \CCJ{1}$.

The community value $\sigma_q(\p, \vc{v}) = \CCJ{q}$ is unitless, since the
individual branch values $v(b) = L(b)/\ovln{L}$ are unitless.  We could
alternatively put $v(b) = L(b)$, which might be measured in years or number
of mutations.  Then $\sigma_q(\p, \vc{v})$ would be measured in the same
units, and $\sigma_0(\p, \vc{v})$ would be exactly Faith's
phylogenetic diversity, without the factor of $1/\ovln{L}$.
\end{example}

In summary, the value measures $\sigma_q$ unify not only the Hill numbers
$D_q$, the similarity-sensitive diversity measures $D_q^Z$, and the
diversity of a community divided into completely dissimilar subcommunities
(Example~\ref{eg:value-ch}), but also some known measures of phylogenetic
diversity.

One could also assign a value to each species in a more literal,
utilitarian\index{utilitarianism} sense (perhaps
monetary).%
\index{economics}%
\index{value!monetary}%
\index{value!medicinal}
Solow%
\index{Solow, Andrew} 
and
Polasky%
\index{Polasky, Stephen} 
noted that `one justification for species
conservation\index{conservation} is that some species may provide a future
medical benefit' (\cite{SoPo}, p.~98), and analysed diversity from that
viewpoint.  This line of enquiry is worthwhile not only for the evident
scientific reasons, but also because it is how Solow and Polasky arrived at
the mathematically profound invariant now called magnitude (as related on
p.~\pageref{p:sp-mag}).  But we will not pursue it, instead making a
connection between value measures and established quantities in information
theory.

\section{Value and relative entropy}
\lbl{sec:value-rel}
\index{value!relative entropy@and relative entropy}
\index{relative entropy!value@and value}

The value measure $\sigma_q$ is a simple transformation of a classical
object of study, the R\'enyi relative entropy or R\'enyi divergence
(R\'enyi~\cite{Reny}, Section~3).  In this short section, we describe the
relationship between value, relative entropy, and some of the other
quantities that we have considered.  This provides useful context, although
nothing here is logically necessary for anything that follows.

For $q \in [-\infty, \infty]$ and probability
distributions $\p, \vc{r} \in \Delta_n$, the \demph{R\'enyi entropy of
  order $q$ of $\p$ relative to $\vc{r}$}%
\index{Renyi relative entropy@R\'enyi relative entropy}%
\index{order!Renyi relative entropy@of R\'enyi relative entropy}
is defined as 
\[
\hrelent{q}{\p}{\vc{r}}
=
\frac{1}{q - 1} \log \sum_{i \in \supp(\p)} p_i^q r_i^{1 - q}
\ntn{hrelent}
\]
when $q \neq 1, \pm \infty$, and in the exceptional cases by
\begin{align*}
\hrelent{-\infty}{\p}{\vc{r}}   &
=
\log\min_{i \in \supp(\p)} \frac{p_i}{r_i},    \\
\hrelent{1}{\p}{\vc{r}} &
=
\sum_{i \in \supp(\p)} p_i \log \frac{p_i}{r_i}
=
\relent{\p}{\vc{r}},    \\
\hrelent{\infty}{\p}{\vc{r}}    &
=
\log\max_{i \in \supp(\p)} \frac{p_i}{r_i}.
\end{align*}
In all cases, 
\[
\hrelent{q}{\p}{\vc{r}} 
= 
\log M_{q - 1}(\vc{p}, \vc{p}/\vc{r})
=
-\log M_{1 - q}(\vc{p}, \vc{r}/\vc{p})
\]
(by the duality equation~\eqref{eq:mn-duality},
p.~\pageref{eq:mn-duality}), giving
\begin{equation}
\lbl{eq:val-rel-ent}
\hrelent{q}{\p}{\vc{r}}
=
-\log \sigma_q(\p, \vc{r}).
\end{equation}
R\'enyi relative entropy can take the value $\infty$.  But as for classical
relative entropy ($q = 1$, p.~\pageref{p:An}), it is convenient to restrict
to pairs $(\vc{p}, \vc{r})$ such that $p_i = 0$ whenever $r_i = 0$; then
$\hrelent{q}{\vc{p}}{\vc{r}} < \infty$ for all $q$.

The R\'enyi relative entropies share with the classical version the basic
property that $\hrelent{q}{\p}{\p} = 0$ for all distributions $\p$.
When $q > 0$, they also share its positive definiteness property,
stated in the classical case as Lemma~\ref{lemma:rel-ent-pos-def}: 

\begin{lemma}
For all $q > 0$ and $\vc{p}, \vc{r} \in \Delta_n$,
\[
\hrelent{q}{\p}{\vc{r}} \geq 0,
\]
with equality if and only if $\vc{p} = \vc{r}$.
\end{lemma}

\begin{proof}
This follows from equation~\eqref{eq:val-rel-ent} and
Lemma~\ref{lemma:sigma-sum-ineq}, since $\sum_{i = 1}^n r_i = 1$.
\end{proof}

In the definition above of R\'enyi relative entropy, both arguments were
required to be probability distributions, whereas the second argument
$\vc{v}$ of the value measure $\sigma_q$ can be any vector of nonnegative
reals.  In fact, when R\'enyi introduced his relative entropies, he allowed
$\p$ and $\vc{r}$ to be `generalized%
\index{generalized probability distribution}%
\index{probability distribution!generalized}
probability distributions' (vectors of nonnegative reals summing to
\emph{at most} $1$), and he inserted a normalizing factor of $\sum p_i$
accordingly (Section~3 of~\cite{Reny}).  But we will consider relative
entropy only for pairs of genuine probability distributions.

Just as R\'enyi relative entropy of order $q$ is closely related to the
value measure $\sigma_q$, so too is $q$-logarithmic relative entropy
\[
\srelent{q}{\p}{\vc{r}}
=
- \sum_{i \in \supp(\p)} p_i \ln_q \frac{r_i}{p_i}
\]
(Definition~\ref{defn:q-rel-ent}).  The formula for $q$-logarithmic
relative entropy in terms of value is the same as the
formula~\eqref{eq:val-rel-ent} for R\'enyi relative entropy in terms of
value, but with the logarithm replaced by the $q$-logarithm:
\[
\srelent{q}{\p}{\vc{r}}
=
- \ln_q \sigma_q(\p, \vc{r})
\]
($-\infty < q < \infty$).  To prove this, we use
Lemma~\ref{lemma:q-log-mean}: 
\begin{align*}
\srelent{q}{\p}{\vc{r}}
&
=
- M_1 \bigl(\p, \ln_q(\vc{r}/\p) \bigr) \\
&
=
- \ln_q M_{1 - q} (\p, \vc{r}/\p) \\
&
=
- \ln_q \sigma_q(\p, \vc{r}).     
\end{align*}
Hence $\sigma_q(-,-)$, $\hrelent{q}{-}{-}$ and $\srelent{q}{-}{-}$ are
all simple transformations of one another.

R\'enyi relative entropy shares with ordinary relative entropy the
property that
\[
\hrelent{q}{\p}{\vc{u}_n}
=
H_q(\vc{u}_n) - H_q(\p)
\]
($q \in [-\infty, \infty]$, $\p \in \Delta_n$).  In this respect, R\'enyi
relative entropy has slightly more%
\index{Renyi relative entropy@R\'enyi relative entropy!q-logarithmic@and $q$-logarithmic relative entropy}%
\index{q-logarithmic relative entropy@$q$-logarithmic relative entropy!Renyi@and R\'enyi relative entropy}
convenient algebraic properties than
$q$-logarithmic relative entropy: compare the formula for
$\srelent{q}{\p}{\vc{u}_n}$ in Remark~\ref{rmk:q-rel-ent-param}.

\begin{remark}
\lbl{rmk:gen-def-rel}
In Remark~\ref{rmk:gen-def}, we observed that for any differentiable
function $\lambda\from (0, \infty) \to \R$ satisfying $\lambda(1) = 0$ and
$\lambda'(1) = 1$, the formula
\[
\frac{1}{1 - q} \, 
\lambda\,
\Biggl( \sum_{i \in \supp(\p)} p_i^q \Biggr)
\]
defines a one-parameter family of deformations of Shannon entropy, in the
sense that it converges to $H(\p)$ as $q \to 1$.  A similar statement holds
for relative entropy: for any such function $\lambda$, the generalized
relative entropy
\[
\frac{1}{q - 1} \,
\lambda\,
\Biggl( \sum_{i \in \supp(\p)} p_i^q r_i^{1 - q} \Biggr)
\]
converges to the ordinary relative entropy $\relent{\p}{\vc{r}}$ as $q \to
1$.  Taking $\lambda = \log$ gives R\'enyi relative entropy, and taking
$\lambda(x) = x - 1$ gives $q$-logarithmic relative entropy.
\end{remark}

\begin{remarks}
\lbl{rmks:fisher-def}
Here we relate the deformed relative entropies to the
Fisher%
\index{Fisher, Ronald!metric} 
metric on probability distributions.
\begin{enumerate}
\item
\lbl{rmk:fisher-def-gods-joke}
In Section~\ref{sec:rel-misc}, we showed that although the square root of
ordinary relative entropy is not a distance function on the open simplex
$\Delta_n^\circ$ (that is, not a metric in the sense of metric spaces), it
is an \emph{infinitesimal} metric in the Riemannian sense.  As we saw, it
is proportional to the Fisher metric, which itself is proportional to the
standard Riemannian metric on the positive orthant of the unit sphere,
transferred to $\Delta_n^\circ$ via the bijection $\p \leftrightarrow
\sqrt{\p}$.

It is natural to ask what happens if we apply the same procedure to the
R\'enyi relative entropy of order $q$, or the $q$-logarithmic relative
entropy, for some $q \neq 1$.  Do we obtain some new, deformed, Fisher-like
metric on $\Delta_n^\circ$?

The answer turns out to be no.  Using $\hrelent{q}{-}{-}$ or
$\srelent{q}{-}{-}$ instead of the ordinary relative entropy
$\relent{-}{-}$ simply multiplies the induced metric on $\Delta_n^\circ$ by
a constant factor of $q$.  More generally, the same is true of any family
of deformations of relative entropy of the type constructed in
Remark~\ref{rmk:gen-def-rel}. 
(We omit the proof, but it is similar to the argument for ordinary relative
entropy; compare also Section~2.7 of Ay, Jost, L{\^e} and
Schwachh{\"o}fer~\cite{AJLS} and Chapter~3 of Amari~\cite{AmarIGA}.)
It follows that the $q$-analogues of
Fisher%
\index{Fisher, Ronald!distance}  
distance and Fisher%
\index{Fisher, Ronald!information}
information (defined as in equation~\eqref{eq:fisher-info}) are
proportional to the classical Fisher distance and information, and that the
$q$-analogue of the Jeffreys%
\index{Jeffreys, Harold!prior}
prior is exactly equal to the classical notion.

The moral is that the Fisher metric on probability distributions is a very
stable, canonical concept.  However we may choose to deform relative
entropy, the induced metric is always essentially the same.

\item
The parameter value $q = 1/2$ plays a special role.  The R\'enyi%
\index{Renyi relative entropy@R\'enyi relative entropy!order 1@of order $1/2$}
and $q$-logarithmic%
\index{q-logarithmic relative entropy@$q$-logarithmic relative entropy!order 1@of order $1/2$}
relative entropies of order $1/2$ are
\[
\hrelent{1/2}{\p}{\vc{r}} = -2 \log \sum \sqrt{p_i r_i},
\qquad
\srelent{1/2}{\p}{\vc{r}} = 2 \biggl( 1 - \sum \sqrt{p_i r_i} \biggr).
\]
Both are symmetric in $\p$ and $\vc{r}$ (and $q = 1/2$ is the only
parameter value with this property).  In fact, both are
increasing, invertible transformations of the Fisher%
\index{Fisher, Ronald!distance}
distance
\[
d_\fishsym(\p, \vc{r}) 
= 
2 \cos^{-1} \biggl( \sum \sqrt{p_i r_i} \biggr).
\]
Thus, the R\'enyi relative entropy of order $1/2$ of a pair of
distributions determines the Fisher distance between them.  Similarly,
knowing either the $(1/2)$-logarithmic entropy of $(\p, \vc{r})$ or the
value of order $1/2$,
\[
\sigma_{1/2}(\p, \vc{r}) = \biggl( \sum \sqrt{p_i r_i} \biggr)^2,
\]
determines the Fisher distance between $\p$ and $\vc{r}$.
\end{enumerate}
\end{remarks}

\section{Characterization of value}
\lbl{sec:value-char}
\index{value measure!characterization of}

Here we show that the only value measures with reasonable properties are
those of the form $\sigma_q$ for some $q \in [-\infty, \infty]$.  

We defined the value measure $\sigma_q$ on the nonnegative half-line $[0,
  \infty)$, but it restricts to a sequence of functions
\[
\bigl( 
\sigma_q \from \Delta_n \times (0, \infty)^n \to (0, \infty) 
\bigr)_{n \geq 1}
\]
on the strictly positive reals. It is this family $(\sigma_q)_{q \in
  [-\infty, \infty]}$ that we will characterize.  A similar theorem on $[0,
  \infty)$ can be proved, at the cost of an extra hypothesis
  (Remark~\ref{rmk:val-char-nonneg}), but we will focus on strictly
  positive values.
Thus, we will identify a list of conditions on a sequence of functions
\begin{equation}
\lbl{eq:typ-val}
\bigl( 
\sigma \from \Delta_n \times (0, \infty)^n \to (0, \infty) 
\bigr)_{n \geq 1}
\end{equation}
that are satisfied by $\sigma_q$ for each $q \in [-\infty, \infty]$, but
not by any other $\sigma$.

We begin by describing those conditions.  

Recall that a weighted mean $M$ on $(0, \infty)$ is a sequence of functions
of the same type as a value measure on $(0, \infty)$:
\[
\bigl( 
M \from \Delta_n \times (0, \infty)^n \to (0, \infty) 
\bigr)_{n \geq 1}.
\]
Although the classes of reasonable means and reasonable value measures
are intended to be disjoint (Remark~\ref{rmk:vals-mns}), some of the
properties that one expects of a mean can also be expected of a value
measure.  We therefore reuse some of the terminology defined previously
for weighted means, and summarized in Appendix~\ref{app:condns}.  

In what follows, let $\sigma$ denote a sequence of
functions as in~\eqref{eq:typ-val}.
Then $\sigma$ may or may not have the following properties, all defined
previously in the context of weighted means.

\begin{description}
\item[Symmetry.]%
\index{symmetric!value measure}%
\index{value measure!symmetric}
For $\sigma$ to be symmetric
(Definition~\ref{defn:pwr-mn-elem}\bref{part:pme-sym}) means that the value
of the whole is independent of the order in which the parts are listed.

\item[Absence-invariance.]%
\index{absence-invariance!value measure@of value measure}%
\index{value measure!absence-invariant}
For $\sigma$ to be absence-invariant
  (Definition~\ref{defn:pwr-mn-elem}\bref{part:pme-abs}) means that a part
  that is absent ($p_i = 0$) makes no contribution to the value of the
  whole, and might as well be ignored.

\item[Increasing.]%
\index{increasing!value measure}%
\index{value measure!increasing}
For $\sigma$ to be increasing (Definition~\ref{defn:w-isi}) means that the
parts make a positive (or at least, nonnegative) contribution to the
whole: if the value of one part increases and the rest stay the same, this
does not cause the value of the whole to become smaller.

\item[Homogeneity.]%
\index{homogeneous!value measure}%
\index{value measure!homogeneous}
Homogeneity of $\sigma$ (Definition~\ref{defn:w-mn-hgs}) means that the
value of the whole and the values of the parts are measured in the same
units.  For instance, if the value of each part is measured in kilograms
then so is the value of the whole.  Converting to grams multiplies both by
$1000$.

\item[Chain rule.]%
\index{chain rule!value measures@for value measures}%
\index{value measure!chain rule for}
The chain rule for $\sigma$ (Definition~\ref{defn:mns-chn}) is the most
complicated of the properties that we will need, but it is logically
fundamental.  It states that
\begin{align}
\nonumber
&
\sigma\bigl( \vc{w} \of (\p^1, \ldots, \p^n),
\vc{v}^1 \oplus\cdots\oplus \vc{v}^n \bigr)
\\
\lbl{eq:val-ch}
=\ 
&
\sigma\Bigl(
\vc{w}, 
\bigl(
\sigma(\p^1, \vc{v}^1), \ldots, \sigma(\p^n, \vc{v}^n) 
\bigr)
\Bigr)  
\end{align}
for all $n, k_1, \ldots, k_n \geq 1$, $\vc{w} \in \Delta_n$, $\p^i \in
\Delta_{k_i}$, and $\vc{v}^i \in (0, \infty)^{k_i}$.  

\begin{figure}
\centering
\lengths
\begin{picture}(120,44)
\cell{60}{22}{c}{\includegraphics[height=44\unitlength]{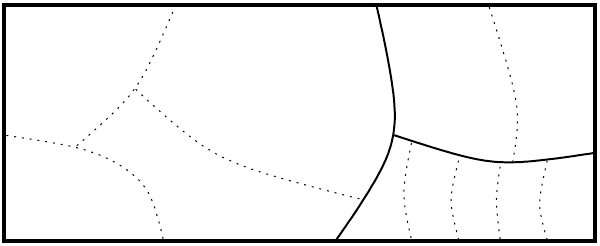}}
\cell{2}{22}{c}{\large$w_1$}
\cell{118}{30}{c}{\large$w_2$}
\cell{118}{9}{c}{\large$w_3$}
\cell{18}{35}{c}{$p^1_1$}
\valput{24}{32}{$v^1_1$}
\cell{55}{30}{c}{$p^1_2$}
\valput{61}{27}{$v^1_2$}
\cell{18}{10}{c}{$p^1_3$}
\valput{24}{7}{$v^1_3$}
\cell{43}{11}{c}{$p^1_4$}
\valput{49}{8}{$v^1_4$}
\cell{85}{31}{c}{$p^2_1$}
\valput{91}{28}{$v^2_1$}
\cell{102}{31}{c}{$p^2_2$}
\valput{108}{28}{$v^2_2$}
\cell{75.5}{10.5}{c}{$p^3_1$}
\valput{75.5}{4.5}{$v^3_1$}
\cell{83.5}{10.5}{c}{$p^3_2$}
\valput{83.5}{4.5}{$v^3_2$}
\cell{91.5}{10.5}{c}{$p^3_3$}
\valput{91.5}{4.5}{$v^3_3$}
\cell{99}{10.5}{c}{$p^3_4$}
\valput{99}{4.5}{$v^3_4$}
\cell{108}{10.5}{c}{$p^3_5$}
\valput{108}{4.5}{$v^3_5$}
\end{picture}
\caption{The chain rule for value measures, as in
  equation~\eqref{eq:val-ch}.  Here, the whole is divided into $n = 3$
  parts, the first part is divided into $k_1 = 4$ subparts, the second
  into $k_2 = 2$ subparts, and the third into $k_3 = 5$ subparts.}
\lbl{fig:val-ch}  
\end{figure}

This is a recursivity property (Figure~\ref{fig:val-ch}).  It means that
our method $\sigma$ of aggregating value behaves consistently when the
whole is divided into parts which are further divided into subparts.

Suppose, for example, that we are performing some evaluation of the
whole planetary landmass, and that we have already assigned a value to each
country.  We could first use $\sigma$ to compute the value of each
continent, then use $\sigma$ again on those continental values to compute
the global value.  This is the right-hand side of
equation~\eqref{eq:val-ch}, if $v^i_j$ denotes the value of the $j$th
country on the $i$th continent, $\p^i$ is the relative size distribution of
the countries on the $i$th continent, and $\vc{w}$ is the relative size
distribution of the continents.  Alternatively, we could
ignore the intermediate level of continents and use $\sigma$ to compute the
global value directly from the country values.  This gives the left-hand
side of equation~\eqref{eq:val-ch}.  The two methods for computing the
global value should give the same result, and the chain rule states that
they do.
\end{description}

We make two further definitions for value measures $\sigma$ on $(0,
\infty)$.

\begin{defn}
\lbl{defn:val-cipp}
$\sigma$ is \demph{continuous%
\index{continuous!positive probabilities@in positive probabilities} 
in positive probabilities} if for each $n \geq 1$ and $\vc{v} \in (0,
\infty)^n$, the function
\[
\begin{array}{cccc}
\sigma(-, \vc{v}):      &\Delta_n^\circ  &\to            &(0, \infty)    \\
                        &\p              &\mapsto        &\sigma(\p, \vc{v})
\end{array}
\]
on the open simplex is continuous.
\end{defn}

This condition only involves continuity in the \emph{sizes} of the parts
present, not their values.  The restriction to the interior of the simplex
means that we do not forbid value measures that make a sharp distinction
between presence and absence.

\begin{defn}
\lbl{defn:val-eff}
$\sigma$ is an \demph{effective%
\index{effective number}
number} if 
\[
\sigma\bigl( \vc{u}_n, (1, \ldots, 1) \bigr) = n
\]
for all $n \geq 1$.
\end{defn}

Assuming homogeneity, the effective number property is equivalent to 
\begin{equation}
\lbl{eq:val-eff-rep}
\sigma\bigl(\vc{u}_n, (v, \ldots, v)\bigr) = nv
\end{equation}
for all $n \geq 1$ and $v \in (0, \infty)$.  That is, if we put together
$n$ parts of equal size and equal value, $v$, the result has value $nv$.  

\begin{remark}
\lbl{rmk:val-undef}
Let $\sigma$ be an absence-invariant value measure.  Then $\sigma(\p,
\vc{v})$ is independent of $v_i$ for $i \not\in \supp(\p)$.  Indeed,
writing $\supp(\p) = \{i_1, \ldots, i_k\}$ with $i_1 < \cdots < i_k$,
absence-invariance implies that
\begin{equation}
\lbl{eq:val-supp}
\sigma(\p, \vc{v})
=
\sigma\bigl( (p_{i_1}, \ldots, p_{i_k}), (v_{i_1}, \ldots, v_{i_k}) \bigr).
\end{equation}
So we can consistently extend%
\index{value!undefined}%
\index{undefined arguments} 
the definition of $\sigma(\p, \vc{v})$ to pairs $(\p, \vc{v})$ where $v_i$
need not be within the permissible range $(0, \infty)$, or even defined at
all, when $i \not\in \supp(\p)$.  In that case, we define $\sigma(\p,
\vc{v})$ to be the right-hand side of equation~\eqref{eq:val-supp}.  This
convention is exactly analogous to the convention for means introduced in
Remark~\ref{rmk:defined-even-if-not}, and to the usual convention for
integrals of functions undefined on a set of measure zero.
\end{remark}

We now prove that the properties listed above uniquely characterize the
family of value measures $(\sigma_q)$.

\begin{thm}
\lbl{thm:val-char}
\index{value measure!characterization of}
Let $\bigl( \sigma \from \Delta_n \times (0, \infty)^n \to (0, \infty)
\bigr)_{n \geq 1}$ be a sequence of functions.  The following are
equivalent: 
\begin{enumerate}
\item 
\lbl{part:val-char-condns}
$\sigma$ is symmetric, absence-invariant, increasing, homogeneous,
continuous in positive probabilities and an effective number, and satisfies
the chain rule;

\item
\lbl{part:val-char-form}
$\sigma = \sigma_q$ for some $q \in [-\infty, \infty]$. 
\end{enumerate}
\end{thm}

\begin{proof}
To prove that~\bref{part:val-char-form}
implies~\bref{part:val-char-condns}, let $q \in [-\infty, \infty]$.  
That $\sigma_q$ is symmetric, absence-invariant, increasing, homogeneous, and
continuous in positive probabilities follows from
the definition
\[
\sigma_q(\p, \vc{v}) = M_{1 - q}(\p, \vc{v}/\p)
\]
of $\sigma_q$ and the corresponding properties of $M_{1 - q}$
(Lemmas~\ref{lemma:pwr-mns-elem}, \ref{lemma:pwr-mns-inc},
\ref{lemma:pwr-mns-hgs} and
\ref{lemma:pwr-mns-cts-px}\bref{part:pwr-mns-cts-px-1}).  That $\sigma_q$
is an effective number follows from the consistency of $M_{1 - q}$, and the
chain rule for $\sigma_q$ follows from the chain rule for $M_{1 - q}$
(Proposition~\ref{propn:pwr-mns-chn}).

Conversely, assume that $\sigma$ satisfies the conditions
in~\bref{part:val-char-condns}.  Define a sequence of functions
\[
\bigl( 
M \from \Delta_n \times (0, \infty)^n \to (0, \infty) 
\bigr)_{n \geq 1}
\]
by 
\[
M(\p, \vc{x}) = \sigma(\p, \p\vc{x})
\]
($\p \in \Delta_n$, $\vc{x} \in (0, \infty)^n$).  Although it may be that
$(\p\vc{x})_i = 0$ for some $i$, in which case $\sigma(\p, \p\vc{x})$ is
strictly speaking undefined, this can only happen when $p_i = 0$; hence
$\sigma(\p, \p\vc{x})$ can be interpreted according to the convention of
Remark~\ref{rmk:val-undef}.

We will prove that $M$ is a power mean.  We do this by showing that $M$
satisfies the hypotheses of Theorem~\ref{thm:w-inc}: $M$ is symmetric,
absence-invariant, increasing, homogeneous, modular, and consistent.  The
first four follow from the corresponding properties of $\sigma$.
It remains to prove that $M$ is modular and consistent.

For modularity, let $\vc{w} \in \Delta_n$, $\p^i \in \Delta_{k_i}$, and
$\vc{x}^i \in (0, \infty)^{k_i}$.  Using the chain rule and
homogeneity properties of $\sigma$, we find that
\begin{align*}
&
M\bigl( 
\vc{w} \of (\p^1, \ldots, \p^n), \vc{x}^1 \oplus\cdots\oplus \vc{x}^n
\bigr)  \\
&
=
\sigma\bigl(
\vc{w} \of (\p^1, \ldots, \p^n), 
w_1 \p^1 \vc{x}^1 \oplus\cdots\oplus w_n \p^n \vc{x}^n
\bigr)  \\
&
=
\sigma\Bigl(
\vc{w}, \bigl(
\sigma\bigl(\p^1, w_1 \p^1 \vc{x}^1\bigr), 
\ldots, 
\sigma\bigl(\p^n, w_n \p^n \vc{x}^n\bigr)
\bigr)
\Bigr)  \\
&
=
\sigma\Bigl(
\vc{w}, \bigl(
w_1 M(\p^1, \vc{x}^1), \ldots, w_n M(\p^n, \vc{x}^n)
\bigr)
\Bigr)  \\
&
=
M\Bigl(
\vc{w}, \bigl(
M(\p^1, \vc{x}^1), \ldots, M(\p^n, \vc{x}^n)
\bigr)
\Bigr).
\end{align*}
Hence $M$ satisfies the chain rule, and is therefore modular.

Proving that $M$ is consistent is equivalent, by homogeneity, to showing that
\[
\sigma(\p, \p) = 1
\]
for all $n \geq 1$ and $\p \in \Delta_n$.  We do this in three steps.

First suppose that the coordinates of $\p$ are positive and rational, so
that $\p = (k_1/k, \ldots, k_n/k)$ for some positive integers $k_i$ summing
to $k$.  Then
\[
\vc{u}_k = \p \of (\vc{u}_{k_1}, \ldots, \vc{u}_{k_n}),
\]
so by the chain rule for $\sigma$,
\[
\sigma(\vc{u}_k, k \mc 1)
=
\sigma\Bigl(
\p, \bigl(
\sigma(\vc{u}_{k_1}, k_1 \mc 1), \ldots, \sigma(\vc{u}_{k_n}, k_n \mc 1)
\bigr)
\Bigr).
\]
By the effective number property of $\sigma$, this means that
\[
k = \sigma\bigl( \p, (k_1, \ldots, k_n)\bigr).
\]
Dividing through by $k$ and using the homogeneity of $\sigma$ gives $1 =
\sigma(\p, \p)$, as required.

For the second step, let $\p$ be any point in $\Delta_n^\circ$.  Let
$\epsln > 0$.  Since $\sigma$ is continuous in positive probabilities,
there is some $\delta > 0$ such that for $\vc{r} \in \Delta_n^\circ$,
\begin{equation}
\lbl{eq:vc-imp}
\|\p - \vc{r}\| < \delta
\implies
\mg{\sigma(\p, \p) - \sigma(\vc{r}, \p)} < \epsln/2,
\end{equation}
where $\|\cdot\|$ denotes Euclidean length.  We can choose $\vc{r} \in
\Delta_n^\circ$ with rational coordinates such that
\[
\|\p - \vc{r}\| < \delta,
\quad
\max_i \frac{p_i}{r_i} \leq 1 + \frac{\epsln}{2},
\quad
\min_i \frac{p_i}{r_i} \geq 1 - \frac{\epsln}{2}.
\]
Since $\sigma$ is increasing and homogeneous,
\[
\sigma(\vc{r}, \p) 
\leq
\sigma\biggl(\vc{r}, \biggl(\max_i \frac{p_i}{r_i}\biggr) \vc{r}\biggr)
=
\biggl(\max_i \frac{p_i}{r_i}\biggr) \sigma(\vc{r}, \vc{r}),
\]
which by the first step gives
\[
\sigma(\vc{r}, \vc{p}) \leq \max_i \frac{p_i}{r_i} 
\leq 1 + \frac{\epsln}{2}.
\]
Similarly, $\sigma(\vc{r}, \p) \geq 1 - \epsln/2$, so
\[
\mg{\sigma(\vc{r}, \p) - 1} \leq \epsln/2.
\]
This, together with~\eqref{eq:vc-imp} and the triangle inequality, implies
that $\mg{\sigma(\p, \p) - 1} < \epsln$.  But $\epsln$ was arbitrary, so
$\sigma(\p, \p) = 1$.

Third and finally, take any $\p \in \Delta_n$.  Write $\supp(\p) = \{i_1,
\ldots, i_k\}$ with $i_1 < \cdots < i_k$, and write $\vc{r} =
(p_{i_1}, \ldots, p_{i_k}) \in \Delta_k^\circ$.  Then 
\[
\sigma(\p, \p) = \sigma(\vc{r}, \vc{r}) = 1,
\]
where the first equality holds for the reasons given in
Remark~\ref{rmk:val-undef}, and the second follows from the second step
above. 

This completes the proof that $M$ is consistent.  We have now shown
that $M$ satisfies the hypotheses of Theorem~\ref{thm:w-inc}.  By that
theorem, $M = M_{1 - q}$ for some $q \in [-\infty, \infty]$.  It follows
that
\[
\sigma(\p, \vc{v})
=
M_{1 - q}(\p, \vc{v}/\p)
=
\sigma_q(\p, \vc{v})
\]
for all $n \geq 1$, $\p \in \Delta_n$, and $\vc{v} \in (0, \infty)^n$.
\end{proof}

\begin{remark}
\lbl{rmk:val-char-nonneg} 
A similar characterization theorem can be proved for values in $[0,
  \infty)$ instead of $(0, \infty)$, using Theorem~\ref{thm:w-cts-inc} on
  means on $[0, \infty)$.  In this case, we have to strengthen the
    continuity requirement, also asking that $\sigma(\p, \vc{v})$ is
    continuous in $\vc{v}$ for each fixed $\p$. 
\end{remark}

Theorem~\ref{thm:val-char} can be translated into a characterization
theorem for either the R\'enyi relative entropies or the $q$-logarithmic
relative entropies, using the observations in Section~\ref{sec:value-rel}.
This translation exercise is left to the reader.

\section{Total characterization of the Hill numbers}
\lbl{sec:total-hill}

The axiomatic approach to diversity measurement is to specify mathematical
properties that we want the concept of diversity to possess, then to prove
a theorem classifying all the diversity measures with the specified
properties. 

Here we do this for the simple but very commonplace model of a community as
its relative abundance distribution $\p = (p_1, \ldots, p_n)$.  We prove
that any measure $\p \mapsto D(\p)$ satisfying a handful of intuitive
properties must be one of the Hill numbers $D_q$.  To do this, we use the
characterization theorem for value measures (Theorem~\ref{thm:val-char}).
The strategy is to construct from our hypothetical diversity
measure $D$ a value measure $\sigmaD$, apply Theorem~\ref{thm:val-char} to
show that $\sigmaD = \sigma_q$ for some $q$, and deduce from this that $D
= D_q$.

This is the second characterization theorem for the Hill numbers that we
have proved, and it is more powerful than the first
(Theorem~\ref{thm:rout}), in the sense%
\index{Hill number!difference between characterizations of}
that the hypotheses are simpler and have more direct ecological
explanations.  Another difference is that the previous theorem fixed a
parameter value $q$, whereas the one below characterizes $D_q$ for all $q$
simultaneously.  Further discussion of the differences can be found at the
end of the introduction to this chapter.

Consider, then, a sequence of functions
\[
\bigl( D \from \Delta_n \to (0, \infty) \bigr)_{n \geq 1},
\]
intended to measure the diversity $D(\p)$ of any community of $n$ species with
relative abundances $\p = (p_1, \ldots, p_n)$.  What properties would we
expect $D$ to possess?

We already discussed some desirable properties in
Section~\ref{sec:prop-hill}, arguing that any reasonable diversity measure
$D$ ought to be symmetric, absence-invariant, and continuous in positive
probabilities, and that it should obey the replication principle.  To fix
the scale on which we are working, we also ask that a community consisting
of only one species has diversity $1$.  Formally, $D$ is
\dmph{normalized}\lbl{p:hill-norm} if $D(\vc{u}_1) = 1$.

We impose one further condition on our hypothetical diversity measure.
Consider a pair of islands, perhaps with different population sizes, with
no species in common.  Replace the population of the first island by a
population of the same abundance but greater or equal diversity,
still sharing no species with the second island.  Then the diversity of the
two-island community should be greater than or equal to what it was
originally.

More generally, consider a group of several islands, perhaps with different
population sizes, with no species shared between islands.  Replace the
population of each island by a population of the same abundance but greater
or equal diversity, and still with no shared species between islands.  Then
the diversity of the whole island group should be greater than or equal to
what it was originally.  Although this condition is superficially stronger
than the special case described in the previous paragraph, it is equivalent
by induction.  We formalize it as follows.

\begin{defn}
\lbl{defn:mod-mono}
A sequence of functions $\bigl( D \from \Delta_n \to (0, \infty) \bigr)_{n
  \geq 1}$ is \demph{modular-monotone}%
\index{modularmonotone@modular-monotone}
if 
\begin{align*}
&
D(\p^i) \leq D(\twid{\p}^i) \text{ for all } i \in \{1, \ldots, n\}     \\
\implies\,
&
D\bigl( \vc{w} \of (\p^1, \ldots, \p^n) \bigr) \leq
D\bigl( \vc{w} \of (\twid{\p}^1, \ldots, \twid{\p}^n) \bigr) 
\end{align*}
for all $n, k_i, \twid{k}_i \geq 1$ and $\vc{w} \in \Delta_n$, $\p^i \in
\Delta_{k_i}$ and $\twid{\p}^i \in \Delta_{\twid{k}_i}$.  
\end{defn}

For comparison, recall that by definition,
$D$ is modular if and only if
\begin{align*}
&
D\bigl(\p^i\bigr) = D\bigl(\twid{\p}^i\bigr) 
\text{ for all } i \in \{1, \ldots, n\}     \\
\implies
&
D\bigl( \vc{w} \of (\p^1, \ldots, \p^n) \bigr) =
D\bigl( \vc{w} \of (\twid{\p}^1, \ldots, \twid{\p}^n) \bigr)
\end{align*}
(Definition~\ref{defn:hill-mod}).  Modular-monotonicity implies modularity
(Lemma~\ref{lemma:Dem}), and like modularity, it is a basic requirement for
a diversity measure.

\begin{example}
\lbl{eg:hill-mod-mono}
Let $q \in [-\infty, \infty]$.  The Hill number $D_q$ is modular-monotone,
since by the chain rule for $D_q$ (Proposition~\ref{propn:hill-chn}), 
\[
D_q\bigl(\vc{w} \of (\p^1, \ldots, \p^n)\bigr)
=
M_{1 - q}\Bigl( \vc{w},
\bigl( D_q(\p^1)/w_1, \ldots, D_q(\p^n)/w_n \bigr) 
\Bigr),
\]
and the power mean $M_{1 - q}$ is increasing.
\end{example}

We will prove:

\begin{thm}
\lbl{thm:total-hill}
\index{Hill number!characterization of}
Let $\bigl( D \from \Delta_n \to (0, \infty) \bigr)_{n \geq 1}$ be a
sequence of functions.  The following are equivalent:
\begin{enumerate}
\item 
\lbl{part:total-hill-condns}
$D$ is symmetric, absence-invariant, continuous in positive probabilities,
normalized and modular-monotone, and satisfies the replication principle;

\item
\lbl{part:total-hill-form}
$D = D_q$ for some $q \in [-\infty, \infty]$.
\end{enumerate}
\end{thm}

The rest of this section is devoted to the proof, and to a refinement of
the theorem that excludes negative values of $q$.  We have already shown
that~\bref{part:total-hill-form} implies~\bref{part:total-hill-condns}, so
it remains to prove the converse.

\femph{For the rest of this section}, let
\[
\bigl( D \from \Delta_n \to (0, \infty) \bigr)_{n \geq 1}
\]
be a sequence of functions satisfying the six conditions in
Theorem~\ref{thm:total-hill}\bref{part:total-hill-condns}.

We begin our proof by proving that the assumed properties of $D$ imply some
of the other desirable properties discussed in Section~\ref{sec:prop-hill}.

\begin{lemma}
\lbl{lemma:Dem}
$D$ is an effective number and modular.
\end{lemma}

\begin{proof}
For the effective number property, we have
\[
D(\vc{u}_n) = D(\vc{u}_n \otimes \vc{u}_1) = nD(\vc{u}_1) = n
\]
for each $n \geq 1$, by replication and normalization.

Modularity follows from modular-monotonicity, since in the notation of
Definition~\ref{defn:hill-mod}, if $D(\p^i) = D(\twid{\p}^i)$ then $D(\p^i)
\leq D(\twid{\p}^i) \leq D(\p^i)$.
\end{proof}

The next few results establish that $D$ is multiplicative.  This is
harder.  First we prove the weaker statement that $D(\p \otimes \vc{r})$
depends only on $D(\p)$ and $D(\vc{r})$.

\begin{lemma}
\lbl{lemma:mult-mod}
Let $\p \in \Delta_m$, $\p' \in \Delta_{m'}$, $\vc{r} \in \Delta_n$, and
$\vc{r}' \in \Delta_{n'}$.  Then
\[
D(\p) = D(\p'), \ D(\vc{r}) = D(\vc{r}')
\implies
D(\p \otimes \vc{r}) = D(\p' \otimes \vc{r}').
\]
\end{lemma}

\begin{proof}
Suppose that $D(\p) = D(\p')$ and $D(\vc{r}) = D(\vc{r}')$.  By definition
of $\otimes$ and modularity,
\begin{align*}
D(\p \otimes \vc{r})    &
=
D\bigl( \p \of (\vc{r}, \ldots, \vc{r})\bigr)   \\
&
=
D\bigl( \p \of (\vc{r'}, \ldots, \vc{r'})\bigr) \\
&
=
D(\p \otimes \vc{r}').
\end{align*}
By symmetry of $D$, the order of the factors in the tensor product is
irrelevant, so $D(\vc{p} \otimes \vc{r}') = D(\vc{p}' \otimes \vc{r}')$ by
the same argument.  The result follows.
\end{proof}

As the next step in showing that $D$ is multiplicative, we prove a
technical lemma (Figure~\ref{fig:seq-simp}).
\begin{figure}
\centering
\lengths
\begin{picture}(120,48)(0,-4)
\cell{60}{24}{c}{\includegraphics[height=40\unitlength]{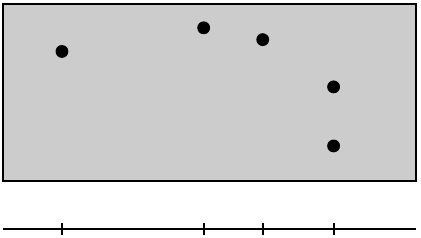}}
\cell{20}{28}{c}{\large$\Delta_n^\circ$}
\cell{35.8}{32}{c}{$\p^1$}
\cell{59.5}{35.5}{c}{$\p^2$}
\cell{69.5}{33.5}{c}{$\p^3$}
\cell{76}{34.5}{c}{$\ddots$}
\cell{83}{29.5}{l}{$\p'$}
\cell{83}{19}{l}{$\p$}
\cell{18}{5}{c}{\large$(0, \infty)$}
\cell{35.8}{0}{b}{\small$D(\p^1)$}
\cell{35.8}{-4}{b}{\small$\in \Q$}
\cell{59}{0}{b}{\small$D(\p^2)$}
\cell{59}{-4}{b}{\small$\in \Q$}
\cell{71}{0}{b}{\small$D(\p^3)$}
\cell{71}{-4}{b}{\small$\in \Q$}
\cell{76}{6}{c}{$\cdots$}
\cell{83.5}{0}{b}{\small$D(\p')$}
\cell{83.5}{-4}{b}{\small$=D(\p)$}
\end{picture}
\caption{Schematic illustration of Lemma~\ref{lemma:seq-simp}.}
\lbl{fig:seq-simp}
\end{figure}

\begin{lemma}
\lbl{lemma:seq-simp} 
Let $n \geq 1$ and $\p \in \Delta_n^\circ$.  Then there exists a sequence
$(\p^j)_{j = 1}^\infty$ in $\Delta_n^\circ$ converging to a point $\p' \in
\Delta_n^\circ$, such that $D(\p^j)$ is rational for all $j$ and $D(\p') =
D(\p)$.
\end{lemma}


\begin{proof}
We can choose a continuous map $\gamma \from [0, 1] \to \Delta_n^\circ$
such that $\gamma(0) = \vc{u}_n$ and $\gamma(1) = \p$.  (For example, take
$\gamma(t) = (1 - t)\vc{u}_n + t\p$.)  By continuity in positive
probabilities, $D\gamma[0, 1]$ is connected and is therefore a subinterval
of $(0, \infty)$.  It contains $D(\gamma(0))$, which by the effective
number property is $n$, and also contains $D(\gamma(1)) = D(\p)$.  Hence
$D\gamma[0, 1]$ contains all real numbers between $n$ and $D(\p)$.  Either
$D(\p) = n$ or $D(\p) \neq n$, and in either case, there is some sequence
$(d_j)_{j = 1}^\infty$ of rational numbers in $D\gamma[0, 1]$ that
converges to $D(\p)$ and is either increasing or decreasing.  (In the case
$D(\p) = n$, we can simply take $d_j = n$ for all $j$.)

Since $d_1 \in D\gamma[0, 1]$, we can choose $t_1 \in [0, 1]$ such that
$D(\gamma(t_1)) = d_1$.  Then by continuity in positive probabilities,
$D\gamma[t_1, 1]$ is an interval containing $d_1$ and $D(\gamma(1)) =
D(\p)$.  But $(d_j)$ is an increasing or decreasing sequence converging to
$D(\p)$, so the interval $D\gamma[t_1, 1]$ also contains $d_2$.  Hence we
can choose $t_2 \in [t_1, 1]$ such that $D(\gamma(t_2)) = d_2$.  Continuing
in this way, we obtain an increasing sequence $(t_j)_{j = 1}^\infty$ in
$[0, 1]$ with $D(\gamma(t_j)) = d_j$ for all $j \geq 1$.

Put $\p^j = \gamma(t_j) \in \Delta_n^\circ$ for each $j \geq 1$.  Then
$D(\p^j) = d_j \in \Q$ for all $j$.  Also put $t = \sup_j t_j \in [0, 1]$
and $\p' = \gamma(t) \in \Delta_n^\circ$.  Then $t_j \to t$ as $j \to
\infty$, so
\[
\p^j = \gamma(t_j) \to \gamma(t) = \p'
\]
as $j \to \infty$.  Since $D$ is continuous in positive probabilities, this
implies that $D(\p^j) \to D(\p')$ as $j \to \infty$.  But also $D(\p^j) =
d_j \to D(\p)$ as $j \to \infty$, by definition of the sequence $(d_j)$.
Hence $D(\p') = D(\p)$, as required.
\end{proof}

\begin{lemma}
\lbl{lemma:total-mult}
$D$ is multiplicative.
\end{lemma}

\begin{proof}
Let $\p \in \Delta_m$ and $\vc{r} \in \Delta_n$.  We have to show that
$D(\p \otimes \vc{r}) = D(\p)D(\vc{r})$.

First suppose that $D(\p)$ is rational, say $D(\p) = a/b$ for positive
integers $a$ and $b$.  Since $D$ is an effective number, $bD(\p) =
D(\vc{u}_a)$.  Hence by replication,
\begin{equation}
\lbl{eq:emm-5}
D(\vc{u}_b \otimes \p) = D(\vc{u}_a).
\end{equation}
Now
\begin{align}
bD(\p \otimes \vc{r})   &
=
D(\vc{u}_b \otimes \vc{p} \otimes \vc{r})
\lbl{eq:emm-1}        \\
&
=
D(\vc{u}_a \otimes \vc{r})     
\lbl{eq:emm-3}        \\
&
=
aD(\vc{r}),
\lbl{eq:emm-4}
\end{align}
where~\eqref{eq:emm-1} and~\eqref{eq:emm-4} follow from the replication
principle for $D$, 
and \eqref{eq:emm-3} follows from~\eqref{eq:emm-5} and
Lemma~\ref{lemma:mult-mod}.  Hence
\[
D(\p \otimes \vc{r})
=
(a/b) D(\vc{r})
=
D(\p) D(\vc{r}),
\]
as required.

Next we prove that $D(\p \otimes \vc{r}) = D(\p) D(\vc{r})$ in the case
that $\p \in \Delta_m^\circ$ and $\vc{r} \in \Delta_n^\circ$.  Choose a
sequence $(\p^j)$ in $\Delta_m^\circ$ converging to $\p' \in
\Delta_m^\circ$ as in Lemma~\ref{lemma:seq-simp}.  By the previous
paragraph,
\begin{equation}
\lbl{eq:emm-6}
D(\p^j \otimes \vc{r}) = D(\p^j) D(\vc{r})  
\end{equation}
for all $j \geq 1$.  Now $\p^j \otimes \vc{r} \in \Delta_{mn}^\circ$ for
all $j$, and $\p^j \otimes \vc{r} \to \p' \otimes \vc{r}$ as $j \to
\infty$.  Hence, taking the limit as $j \to \infty$ in
equation~\eqref{eq:emm-6} and using continuity in positive probabilities,
\[
D(\p' \otimes \vc{r}) = D(\p') D(\vc{r}).
\]
But $D(\p') = D(\p)$, so by Lemma~\ref{lemma:mult-mod}, 
\[
D(\p \otimes \vc{r}) = D(\p) D(\vc{r}),
\]
as required.  

Finally, we prove multiplicativity for an arbitrary $\p \in \Delta_m$ and
$\vc{r} \in \Delta_n$.  By symmetry, we may suppose that $\p = (p_1,
\ldots, p_{m'}, 0, \ldots, 0)$ with $p_1, \ldots, p_{m'} > 0$.  Write $\p'
= (p_1, \ldots, p_{m'}) \in \Delta_{m'}$, and similarly $\vc{r}' \in
\Delta_{n'}$.  By the previous paragraph, $D(\p' \otimes \vc{r}') = D(\p')
D(\vc{r}')$.  On the other hand, by absence-invariance, $D(\p') = D(\p)$
and $D(\vc{r}') = D(\vc{r})$.  Hence by Lemma~\ref{lemma:mult-mod}, $D(\p
\otimes \vc{r}) = D(\p) D(\vc{r})$, completing the proof.
\end{proof}

The plan for the rest of the proof of Theorem~\ref{thm:total-hill} is as
follows.  We wish to show that $D = D_q$ for some $q$.  We know that the
Hill number $D_q$ satisfies the chain rule
\[
D_q\bigl( \vc{w} \of (\p^1, \ldots, \p^n) \bigr)
=
\sigma_q \Bigl( \vc{w}, \bigl( D_q(\p^1), \ldots, D_q(\p^n) \bigr) \Bigr)
\]
(Example~\ref{eg:value-ch}).  Our diversity measure $D$ is modular, which
means that $D\bigl( \vc{w} \of (\p^1, \ldots, \p^n) \bigr)$ is some
function of $\vc{w}$ and $D(\p^1), \ldots, D(\p^n)$.  We will therefore be
able to define a function $\sigmaD$ by
\begin{equation}
\lbl{eq:hill-char-expl}
D\bigl(\vc{w} \of (\p^1, \ldots, \p^n)\bigr)
=
\sigmaD \Bigl( \vc{w}, \bigl( D(\p^1), \ldots, D(\p^n) \bigr) \Bigr).
\end{equation}
Roughly speaking, we then show that the assumed good properties of the
diversity measure $D$ imply good properties of $\sigmaD$, deduce from our
earlier characterization of value measures that $\sigmaD = \sigma_q$ for
some $q$, and conclude that $D = D_q$.

There is a subtlety.  In order to use the characterization of value
measures (Theorem~\ref{thm:val-char}), we need $\sigmaD$ to be defined on
all pairs $(\p, \vc{v})$ with $\p \in \Delta_n$ and $\vc{v} \in (0,
\infty)^n$, whereas equation~\eqref{eq:hill-char-expl} only defines
$\sigma(\p, \vc{v})$ on vectors $\vc{v}$ whose coordinates $v_i$ can be
expressed as values of the diversity measure $D$.  And it may happen that
some elements of $(0, \infty)$ do not arise as values of $D$.  Indeed, if
$D = D_q$ then $D_q(\vc{r}) \geq 1$ for all distributions $\vc{r}$.

For this reason, we now analyse the set of real numbers that arise as
diversities $D(\p)$.  Write
\[
\index{image of diversity measure}
\index{diversity measure!image of}
\im D 
=
\bigcup_{n = 1}^\infty D\Delta_n 
\sub 
(0, \infty).
\ntn{im}
\]
The case of the Hill numbers shows that the situation is not
entirely simple:

\begin{example}
\index{Hill number!image of}
\index{Hill number!range of}
For $q \in [-\infty, \infty]$, the Hill number $D_q$ has image
\begin{equation}
\lbl{eq:hill-im}
\im D_q
=
\begin{cases}
[1, \infty)             &\text{if } q > 0,      \\
\{1, 2, 3, \ldots \}    &\text{if } q = 0,      \\
\{1\} \cup [2, \infty)  &\text{if } q < 0.
\end{cases}
\end{equation}
The statement for $q > 0$ follows from the facts that $D_q(\p) \geq 1$ for
all $\p$ (Lemma~\ref{lemma:div-max-min}\bref{part:div-min}), $D_q$ is an
effective number (equation~\eqref{eq:hill-eff-num}), and $D_q$ is
continuous (Lemma~\ref{lemma:hill-cts}\bref{part:hill-cts-pos}).  For $q =
0$, the result is immediate, since $D_0(\p) = \mg{\supp(\p)}$.

Now let $q < 0$.  Since diversity profiles are decreasing
(Proposition~\ref{propn:div-dec}),
\[
D_q(\p) \geq \mg{\supp(\p)}
\]
for all $\p$.  If $\mg{\supp(\p)} = 1$ then $\p = (0, \ldots, 0, 1, 0,
\ldots, 0)$ and so $D_q(\p) = 1$.  Otherwise, $\mg{\supp(\p)} \geq 2$, so
$D_q(\p) \in [2, \infty)$.  Hence $\im D_q \sub \{1\} \cup [2, \infty)$.
    To prove the opposite inclusion, first note that both $1 =
    D_q(\vc{u}_1)$ and $2 = D_q(\vc{u}_2)$ belong to $\im D_q$.  An
    elementary calculation shows that
\[
D_q(t, 1 - t) \to \infty \text{ as } t \to {0+}.
\]
Since $D_q \from \Delta_2^\circ \to (0, \infty)$ is continuous 
(Lemma~\ref{lemma:hill-cts}\bref{part:hill-cts-int}), $D_q
\Delta_2^\circ$ is an interval that contains $2$ and is unbounded above.
Hence $D_q \Delta_2^\circ \supseteq [2, \infty)$, completing the proof of
  the last clause of equation~\eqref{eq:hill-im}.
\end{example}

\begin{lemma}
\lbl{lemma:imD-mult}
$\im D$ is closed under multiplication.
\end{lemma}

\begin{proof}
This follows from the multiplicativity of $D$
(Lemma~\ref{lemma:total-mult}). 
\end{proof}

\begin{lemma}
\lbl{lemma:big-divs-arise}
Suppose that $D \neq D_0$.  Then $\im D \supseteq [L, \infty)$ for some $L
  > 0$. 
\end{lemma}

\begin{proof}
If $D\Delta_n^\circ$ is a one-element set for each $n \geq 1$ then by the
effective number property, $D\Delta_n^\circ = \{n\}$ for each $n$.  Hence
by absence-invariance, $D = D_0$, a contradiction.

We can therefore choose $n \geq 1$ such that $D\Delta_n^\circ$ has more
than one element, which by continuity in positive probabilities implies
that $D\Delta_n^\circ$ is a nontrivial interval.  Since $D$ is an effective
number, this interval contains $n$.  Now $n \neq 1$ (since
$D\Delta_1^\circ$ is trivial), so $n \geq 2$, so $\im D \cap [1, \infty)$
  contains a nontrivial interval.  Since both $\im D$ and $[1, \infty)$ are
    closed under multiplication, so is $\im D \cap [1, \infty)$.

It is now enough to prove that any subset $B$ of $[1, \infty)$ that is
closed under multiplication and contains a nontrivial interval must contain
$[L, \infty)$ for some $L \geq 1$.  Indeed, since $B$ contains a nontrivial
interval, $B \supseteq [b, b^{1 + 1/r}]$ for some real $b > 1$ and
positive integer $r$.  Since $B$ is closed under multiplication, it
is closed under positive integer powers, so for every integer $m \geq r$,
\[
B
\supseteq
[b^m, b^{m + m/r}]
\supseteq
[b^m, b^{m + 1}].
\]
Hence
\[
B 
\supseteq 
\bigcup_{m \geq r} [b^m, b^{m + 1}]
=
[b^r, \infty),
\]
using $b > 1$ in the last step.
\end{proof}

We now construct from $D$ a value measure $\sigmaD$.  The construction
proceeds in two steps.  First, since $D$ is modular, we can consistently
define a sequence of functions
\[
\bigl( 
\rhoD \from \Delta_n \times (\im D)^n \to \im D
\bigr)_{n \geq 1}
\]
by
\[
\rhoD\Bigl( \vc{w}, \bigl(D(\p^1), \ldots, D(\p^n)\bigr) \Bigr)
=
D\bigl( \vc{w} \of (\p^1, \ldots, \p^n) \bigr)
\]
for all $n, k_1, \ldots, k_n \geq 1$, $\vc{w} \in \Delta_n$ and $\p^i \in
\Delta_{k_i}$.  Second, we extend $\rhoD$ to a sequence of functions
defined on not just $\Delta_n \times (\im D)^n$, but the whole of $\Delta_n
\times (0, \infty)^n$:

\begin{lemma}
\lbl{lemma:val-ext} 
Suppose that $D \neq D_0$.  Then there is a unique homogeneous sequence of
functions
\[
\bigl( 
\sigmaD \from \Delta_n \times (0, \infty)^n \to (0, \infty) 
\bigr)_{n \geq 1}
\]
such that
\[
\sigmaD\Bigl( \vc{w}, \bigl(D(\p^1), \ldots, D(\p^n)\bigr) \Bigr)
=
D\bigl( \vc{w} \of (\p^1, \ldots, \p^n) \bigr)
\]
for all $n, k_1, \ldots, k_n \geq 1$, $\vc{w} \in \Delta_n$, and $\p^i \in
\Delta_{k_i}$.
\end{lemma}

In brief, there is a unique homogeneous extension of $\rhoD$ from
$\im D$ to $(0, \infty)$. 

\begin{proof}
We begin by establishing a homogeneity property of $\rhoD$:
\begin{equation}
\lbl{eq:val-ext-0}
\rhoD(\vc{w}, c\vc{x})
=
c \rhoD(\vc{w}, \vc{x})
\end{equation}
for all $\vc{w} \in \Delta_n$, $\vc{x} \in (\im D)^n$, and $c \in \im D$.
(The left-hand side is well-defined since $\im D$ is closed under
multiplication, by Lemma~\ref{lemma:imD-mult}.)  To prove this, for
each $i \in \{1, \ldots, n\}$, choose $\p^i \in \Delta_{k_i}$ such that
$D(\p^i) = x_i$, and choose $\vc{r} \in \Delta_m$ such that $D(\vc{r}) =
c$.  Then
\begin{align}
\rhoD(\vc{w}, c\vc{x}) &
=
\rhoD\Bigl( \vc{w}, \bigl( 
D(\p^1)D(\vc{r}), \ldots, D(\p^n)D(\vc{r}) \bigr)\Bigr)
\lbl{eq:val-ext-1}    \\
&
=
\rhoD\Bigl( \vc{w}, \bigl( 
D(\p^1 \otimes \vc{r}), \ldots, D(\p^n \otimes \vc{r}) \bigr)\Bigr)
\lbl{eq:val-ext-2}    \\
&
=
D\bigl( \vc{w} \of 
(\p^1 \otimes \vc{r}, \ldots, \p^n \otimes \vc{r}) 
\bigr)
\lbl{eq:val-ext-3}    \\
&
=
D\Bigl( \bigl( \vc{w} \of (\p^1, \ldots, \p^n) \bigr) \otimes \vc{r} \Bigr)
\lbl{eq:val-ext-4}    \\
&
=
D\bigl( \vc{w} \of (\p^1, \ldots, \p^n) \bigr) D(\vc{r})        
\lbl{eq:val-ext-5}    \\
&
=
c\rhoD(\vc{w}, \vc{x}),
\lbl{eq:val-ext-6}
\end{align}
where equation~\eqref{eq:val-ext-1} is by definition of $\p^i$ and
$\vc{r}$, equations~\eqref{eq:val-ext-2} and~\eqref{eq:val-ext-5} are by
multiplicativity of $D$ (Lemma~\ref{lemma:total-mult}),
equations~\eqref{eq:val-ext-3} and~\eqref{eq:val-ext-6} are by definition
of $\rhoD$, and~\eqref{eq:val-ext-4} is by associativity of composition of
distributions (Remark~\ref{rmk:comp-dist-opd}).  This proves the claimed
homogeneity equation~\eqref{eq:val-ext-0}.

We now prove the uniqueness and existence stated in the lemma.

\paragraph*{Uniqueness}
Let $\p \in \Delta_n$ and $\vc{v} \in (0, \infty)^n$.  By
Lemma~\ref{lemma:big-divs-arise}, $\im D$ contains all sufficiently large
real numbers, so we can choose $c \in (0, \infty)$ such that $c\vc{v}
\in (\im D)^n$.  Then $\rho(\p, c\vc{v})$ is defined, and any sequence of
homogeneous functions $\sigmaD$ extending $\rhoD$ satisfies
\[
\sigmaD(\p, \vc{v})
=
\tfrac{1}{c} \rhoD(\p, c\vc{v}).
\]
This proves uniqueness.

\paragraph*{Existence}
First I claim that for all $\p \in \Delta_n$, $\vc{v} \in (0, \infty)^n$,
and $c, d \in (0, \infty)$ such that $c\vc{v}, d\vc{v} \in (\im D)^n$, 
\begin{equation}
\lbl{eq:val-ext-7}
\tfrac{1}{c} \rhoD(\p, c\vc{v})
=
\tfrac{1}{d} \rhoD(\p, d\vc{v}).
\end{equation}
Indeed, since $\im D$ contains all sufficiently large real numbers, we can
choose $a > 0$ such that $ac, ad \in \im D$.  Then
\[
ad \cdot \rhoD(\p, c\vc{v})
=
\rhoD(\p, acd\vc{v})
\]
by the homogeneity property~\eqref{eq:val-ext-0} of $\rhoD$.  Similarly, 
\[
ac \cdot \rhoD(\p, d\vc{v})
=
\rhoD(\p, acd\vc{v}).
\]
Combining the last two equations gives equation~\eqref{eq:val-ext-7}, as
claimed. 

It follows that there is a unique sequence of functions 
\[
\bigl( 
\sigmaD \from \Delta_n \times (0, \infty)^n \to (0, \infty) 
\bigr)_{n \geq 1}
\]
satisfying
\begin{equation}
\lbl{eq:val-ext-8}
\sigmaD(\p, \vc{v}) = \tfrac{1}{c} \rhoD(\p, c\vc{v})
\end{equation}
whenever $\p \in \Delta_n$, $\vc{v} \in (0, \infty)^n$ and $c \in (0,
\infty)$ with $c\vc{v} \in (\im D)^n$.  

It remains to prove that $\sigmaD$ is homogeneous.  Let $\p \in \Delta_n$,
$\vc{v} \in (0, \infty)^n$, and $a \in (0, \infty)$; we must show that
\begin{equation}
\lbl{eq:val-ext-9}
\sigmaD(\p, a\vc{v}) = a \sigmaD(\p, \vc{v}).
\end{equation}
Choose $d \in (0, \infty)$ such that $ad\vc{v}, d\vc{v} \in (\im D)^n$.  By
the claim just proved,
\[
\tfrac{1}{ad} \rhoD(\p, ad\vc{v})
=
\tfrac{1}{d} \rhoD(\p, d\vc{v}),
\]
or equivalently,
\[
\tfrac{1}{d} \rhoD(\p, d \cdot a\vc{v})
=
a \cdot \tfrac{1}{d} \rhoD(\p, d\vc{v}).
\]
But by the defining property~\eqref{eq:val-ext-8} of $\sigmaD$, this is
exactly the desired equation~\eqref{eq:val-ext-9}.
\end{proof}

\begin{example}
Consider the case $D = D_q$.  We have
\[
\sigma_q\Bigl( 
\vc{w}, \bigl( D_q(\p^1), \ldots, D_q(\p^n) \bigr) 
\Bigr)
=
D_q\bigl( \vc{w} \of (\p^1, \ldots, \p^n) \bigr)
\]
for all $\vc{w}$ and $\p^i$, by Example~\ref{eg:value-ch}.  Moreover,
$\sigma_q$ is homogeneous.  Hence by the uniqueness part of
Lemma~\ref{lemma:val-ext}, $\sigmaD = \sigma_q$.
\end{example}

We have now constructed from our diversity measure $D$ a value measure
$\sigmaD$.  From our standing assumption that $D$ has certain good
properties, it follows that $\sigmaD$ has good properties too:

\begin{lemma}
\lbl{lemma:sd-props} 
Suppose that $D \neq D_0$.  Then $\sigmaD$ is symmetric, absence-invariant,
increasing, homogeneous, continuous in positive probabilities and an
effective number, and satisfies the chain rule.
\end{lemma}

\begin{proof}
The symmetry, absence-invariance and effective number properties of $D$
imply the corresponding properties of $\sigmaD$.  The modular-monotonicity
of $D$ implies that $\rhoD$, hence $\sigmaD$, is increasing.  Homogeneity
is one of the defining properties of $\sigmaD$ (Lemma~\ref{lemma:val-ext}).
It remains to prove continuity in positive probabilities and the chain
rule.

To prove that $\sigmaD$ is continuous in positive probabilities, let
$\vc{v} \in (0, \infty)^n$; we wish to prove that
\[
\sigmaD(-, \vc{v}) \from \Delta_n^\circ \to (0, \infty)
\]
is continuous.  Choose $c \in (0, \infty)$ such that $c\vc{v} \in (\im
D)^n$.  Then $\sigmaD(-, \vc{v}) = \tfrac{1}{c}\rhoD(-, c\vc{v})$.  It
therefore suffices to prove that 
\[
\rhoD(-, \vc{x}) \from \Delta_n^\circ \to (0, \infty)
\]
is continuous for every $\vc{x} \in (\im D)^n$.  For each $i \in \{1, \ldots,
n\}$, choose $\p^i \in \Delta_{k_i}$ such that $x_i = D(\p^i)$.  By
absence-invariance, we may assume that each $\p^i$ has full support.
For all $\vc{w} \in \Delta_n$, we have
\[
\rhoD(\vc{w}, \vc{x}) = D\bigl( \vc{w} \of (\p^1, \ldots, \p^n) \bigr),
\]
and if $\vc{w}$ has full support then so does $\vc{w} \of (\p^1, \ldots,
\p^n)$.  Thus, the restriction of $\rhoD(-, \vc{x})$ to $\Delta_n^\circ$
is the composite of the continuous maps
\[
\xymatrix@C+1em@R-5ex@C-1em{
\Delta_n^\circ \ar[r]   &
\Delta_{k_1 + \cdots + k_n}^\circ \ar[r]^-D    &
(0, \infty)    \\
\vc{w}  
\ar@{|->}[r]    &
\vc{w} \of (\p^1, \ldots, \p^n).
}
\]
It is therefore continuous, as claimed.

To prove that $\sigmaD$ satisfies the chain rule, we first prove a chain
rule for $\rhoD$:
\begin{equation}
\lbl{eq:sd-props-1}
\rhoD\bigl( 
\vc{w} \of (\p^1, \ldots, \p^n), 
\vc{x}^1 \oplus\cdots\oplus \vc{x}^n
\bigr)  
=
\rhoD \Bigl( 
\vc{w}, \bigl(
\rhoD(\p^1, \vc{x}^1), \ldots, \rhoD(\p^n, \vc{x}^n) 
\bigr)
\Bigr)
\end{equation}
for all $\vc{w} \in \Delta_n$, $\p^i \in \Delta_{k_i}$, and $\vc{x}^i \in
(\im D)^{k_i}$.  To see this, begin by choosing for each $i \in \{1,
\ldots, n\}$ and $j \in \{1, \ldots, k_i\}$ a probability distribution
$\vc{r}^i_j$ such that $D(\vc{r}^i_j) = x^i_j$.  Then by definition of
$\rhoD$, the left-hand side of equation~\eqref{eq:sd-props-1} is equal to 
\[
D\Bigl( 
\bigl( \vc{w} \of (\p^1, \ldots, \p^n) \bigr)
\of
\bigl(\vc{r}^1_1, \ldots, \vc{r}^1_{k_1}, 
\ \ldots, \ 
\vc{r}^n_1, \ldots, \vc{r}^n_{k_n}\bigr)
\Bigr).
\]
By associativity of composition of distributions
(Remark~\ref{rmk:comp-dist-opd}), this is equal to
\[
D\Bigl(\vc{w} \of
\bigl( \p^1 \of \bigl(\vc{r}^1_1, \ldots, \vc{r}^1_{k_1}\bigr), 
\ \ldots, \ 
\p^n \of \bigl(\vc{r}^n_1, \ldots, \vc{r}^n_{k_n}\bigr)
\bigr)
\Bigr).
\]
By definition of $\rhoD$, this in turn is equal to
\[
\rhoD \biggl( \vc{w},
\Bigl(
D\bigl( \p^1 \of \bigl(\vc{r}^1_1, \ldots, \vc{r}^1_{k_1}\bigr) \bigr),
\ \ldots, \ 
D\bigl( \p^n \of \bigl(\vc{r}^n_1, \ldots, \vc{r}^n_{k_n}\bigr) \bigr)
\Bigr)
\biggr),
\]
which by definition of $\rhoD$ again is equal to the right-hand side
of~\eqref{eq:sd-props-1}.  This proves the claimed chain
rule~\eqref{eq:sd-props-1} for $\rhoD$.

We now want to prove the chain rule for $\sigma$:
\[
\sigmaD\bigl( 
\vc{w} \of (\p^1, \ldots, \p^n), 
\vc{v}^1 \oplus\cdots\oplus \vc{v}^n
\bigr)          
=
\sigmaD \Bigl( 
\vc{w}, \bigl(
\sigmaD(\p^1, \vc{v}^1), \ldots, \sigmaD(\p^n, \vc{v}^n) 
\bigr)
\Bigr)
\]
for all $\vc{w} \in \Delta_n$, $\p^i \in \Delta_{k_i}$, and $\vc{v}^i \in
(0, \infty)^{k_i}$.  We may choose $c \in (0, \infty)$ such that $cv^i_j
\in \im D$ for all $i, j$.  Then by definition of $\sigmaD$ and
the chain rule~\eqref{eq:sd-props-1} for $\rhoD$,
\begin{align*}
\sigmaD \bigl( \vc{w} \of (\p^1, \ldots, \p^n), 
\vc{v}^1 \oplus\cdots\oplus \vc{v}^n \bigr)     
&
=
\tfrac{1}{c} \rhoD \bigl( \vc{w} \of (\p^1, \ldots, \p^n),
c\vc{v}^1 \oplus\cdots\oplus c\vc{v}^n \bigr)   \\
&
=
\tfrac{1}{c} \rhoD \Bigl( \vc{w},
\bigl( \rhoD(\p^1, c\vc{v}^1), \ldots, \rhoD(\p^n, c\vc{v}^n) \bigr)
\Bigr)  \\
&
=
\tfrac{1}{c} \rhoD \Bigl( \vc{w},
\bigl( c\sigmaD(\p^1, \vc{v}^1), \ldots, c\sigmaD(\p^n, \vc{v}^n) \bigr)
\Bigr)  \\
&
=
\sigmaD\Bigl( \vc{w},
\bigl( \sigmaD(\p^1, \vc{v}^1), \ldots, \sigmaD(\p^n, \vc{v}^n) \bigr)
\Bigr),
\end{align*}
as required.
\end{proof}

We are now ready to prove that when a community is modelled as a
probability distribution, the Hill numbers are the only sensible measures
of diversity.

\begin{pfof}{Theorem~\ref{thm:total-hill}}
We have to show that $D = D_q$ for some $q \in [-\infty, \infty]$.  If $D =
D_0$, this is immediate.  Otherwise, by Lemma~\ref{lemma:sd-props},
$\sigmaD$ is a value measure satisfying the hypotheses of
Theorem~\ref{thm:val-char}.  By that theorem, $\sigmaD = \sigma_q$
for some $q \in [-\infty, \infty]$.  Let $\p \in \Delta_n$.  Then
\begin{align}
D(\p)   &
=
D\bigl(\p \of (\underbrace{\vc{u}_1, \ldots, \vc{u}_1}_n) \bigr)
\nonumber       \\
&
=
\rhoD\Bigl( \p, 
\bigl( D(\vc{u}_1), \ldots, D(\vc{u}_1) \bigr)
\Bigr)  
\lbl{eq:th-1} \\
&
=
\rhoD\bigl( \p, (1, \ldots, 1) \bigr)
\lbl{eq:th-2} \\
&
=
\sigma_q\bigl( \p, (1, \ldots, 1) \bigr)
\lbl{eq:th-3} \\
&
=
D_q(\p),
\lbl{eq:th-4}
\end{align}
where equation~\eqref{eq:th-1} is by definition of $\rhoD$,
equation~\eqref{eq:th-2} holds because $D$ is normalized,
equation~\eqref{eq:th-3} holds because $\sigmaD$ extends $\rhoD$ and
$\sigmaD = \sigma_q$, and equation~\eqref{eq:th-4} is from
Example~\ref{eg:value-hill}.  Hence $D = D_q$.
\end{pfof}

The theorem axiomatically characterizes the whole family $(\sigma_q)_{q \in
  [-\infty, \infty]}$ of Hill numbers.  But as argued in
Remark~\ref{rmks:hill-dec}\bref{rmk:hill-dec-neg}, $D_q$ probably does not
deserve to be called a measure of diversity when $q$ is negative.%
\index{Hill number!negative order@of negative order}%
\index{order!negative}
We may therefore wish to characterize the Hill numbers $D_q$ for which $q
\geq 0$, and the following result achieves this.

\begin{lemma}
\lbl{lemma:hill-nonneg-order}
Let $q \in [-\infty, \infty]$.  The following are equivalent:
\begin{enumerate}
\item 
\lbl{part:hno-umax}
$D_q(\p) \leq D_q(\vc{u}_n)$ for all $n \geq 1$ and $\p \in \Delta_n$;

\item
\lbl{part:hno-two}
$D_q(\p) \leq 2$ for all $\p \in \Delta_2$;

\item
\lbl{part:hno-form}
$q \in [0, \infty]$.
\end{enumerate}
\end{lemma}

\begin{proof}
\bref{part:hno-umax} implies~\bref{part:hno-two} trivially,
\bref{part:hno-two} implies~\bref{part:hno-form} by
Remark~\ref{rmks:hill-dec}\bref{rmk:hill-dec-neg}, and~\bref{part:hno-form}
implies~\bref{part:hno-umax} by
Lemma~\ref{lemma:div-max-min}\bref{part:div-max}.
\end{proof}

\begin{remark}
\lbl{rmk:total-hill-translate}
Our characterization theorem for the Hill numbers can easily be translated
into a characterization theorem for the R\'enyi or $q$-logarithmic entropies,
using the transformations of Section~\ref{sec:value-rel}.  However, the
hypotheses of Theorem~\ref{thm:total-hill} are particularly natural
in the context of diversity.  

When translated into terms of $q$-logarithmic entropy,
Theorem~\ref{thm:total-hill} is of the same general type as a result of
Forte%
\index{Forte, Bruno} 
and Ng~\cite{FoNg}%
\index{Ng, Che Tat} 
(also stated as Theorem~6.3.12 of Acz\'el and
Dar\'oczy~\cite{AcDa}).  Apart from some differences in hypotheses, Forte
and Ng's characterization excludes the case $q = 0$, which from the point
of view of diversity measurement is a serious drawback: the Hill number
$D_0$ is species richness,%
\index{species!richness} 
the most common diversity measure of all.
\end{remark}

%% file: mm.tex
\chapter{Mutual information and metacommunities}
\lbl{ch:mm}

From the viewpoint of information theory, there is a conspicuous omission
from this text so far.  Given a random variable $X$ taking values in a
finite set, we have a measure of the information associated with $X$:
its entropy $H(X)$.  But suppose that we are also given another
random variable $Y$, not necessarily independent of $X$, taking values in
another finite set.  If we know the value of $X$, how much information does
that give us about the value of $Y$?

For instance, $Y$ might be a function of $X$, in which case knowing the
value of $X$ gives complete information about $Y$.  Or, at the other
extreme, $X$ and $Y$ might be independent, in which case knowing the
value of $X$ tells us nothing about $Y$.  We would like to quantify the
dependence between the two variables.  The covariance and correlation
coefficients will not do, since they are usually only defined
for random variables taking values in $\R^n$; and while they can be defined
in greater generality, there is no definition for an arbitrary pair of
finite sets.

From the viewpoint of diversity measurement, there is also something
missing.  We know how to quantify the diversity of a single community.
But when we have several associated or adjacent communities~-- for instance,
the gut%
\index{gut microbiome}%
\index{microbial systems}
microbiomes of healthy and unhealthy adults, or the aquatic life in areas
of different salinity\index{salinity} near the mouth of a river~-- some
natural questions present themselves.  How much variation is there between
the communities?  Which contribute most to the overall diversity?  Which
are most or least typical in the context of the system as a whole?  The
diversity measures discussed so far give no answers to such questions.

We will see that these two problems, one information-theoretic and one
ecological, have the same solution. 

Our starting point is the classical information-theoretic concept of mutual
information (a measure of the dependence between two random variables) and
the closely related concepts of conditional and joint entropy.  These are
introduced in Section~\ref{sec:mm-info}.  Then we take exponentials of all
these quantities, which produces a suite of meaningful measures of an
ecological metacommunity (large community) divided into smaller
subcommunities.  The two random variables in play here correspond to a
choice of species and a choice of subcommunity.  Some of the measures
reflect features of individual subcommunities (Section~\ref{sec:mm-sub}),
while others encapsulate information about the entire metacommunity
(Section~\ref{sec:mm-meta}).  We establish the many good logical properties
of these measures in Section~\ref{sec:mm-props}.

All of the entropies and diversities in this chapter can be reduced to
relative entropy (Section~\ref{sec:all-ent-rel}).  In the diversity case,
they are also usefully expressed in terms of value, in the sense of
Chapter~\ref{ch:value}.  Reducing the various metacommunity and
subcommunity measures to one single concept provides new insights into
their ecological meaning.

The diversity measures treated in this chapter are a very special case of
those introduced in work of Reeve%
\index{Reeve, Richard|(} 
et al.~\cite{HPD}.  In the terminology of Chapter~\ref{ch:sim}, it is the
case $q = 1$ (no deformation) and $Z = I$ (no inter-species similarity).
The framework of Reeve et al.\ allows a general $q$ (variable emphasis on
rare or common species) and a general $Z$ (to model the varying
similarities between species).  Section~\ref{sec:beyond} is a sketch of the
development for a general $q$, the details of which lie outside the scope
of this book.

\section{Joint entropy, conditional entropy and mutual information}
\lbl{sec:mm-info}

Shannon entropy $H$ assigns a real number $H(\p)$ to each probability
distribution $\p$, but information theory also associates several
quantities with any \emph{pair} of probability distributions.  To organize
them, it is helpful to distinguish between two types of quantity:
those defined for a pair of distributions on the \emph{same} set, and those
defined for a pair of distributions on potentially \emph{different} sets.

We have already met two quantities of the first type: the relative entropy
$\relent{\p}{\vc{r}}$ and cross entropy $\crossent{\p}{\vc{r}}$ of two
distributions $\p$ and $\vc{r}$ on the same finite set
(Chapter~\ref{ch:rel}).  

We now introduce the standard information-theoretic quantities of the
second type.  The material in this section is all classical, and can be
found in texts such as Cover and Thomas (\cite{CoTh1}, Chapter~2) and
MacKay (\cite{MacKITI}, Chapter~8).  As usual, we only
consider probability distributions on \emph{finite} sets.  But it is
convenient to switch from the language of probability distributions to that
of random variables.%
\index{random!variable}

So, \femph{for the rest of this section}, we consider a random variable $X$
taking values in a finite set $\XX$, and another random variable $Y$ taking
values in a finite set $\YY$.  Assuming that $X$ and $Y$ have the same
sample space, we also have the random variable $(X, Y)$, which takes values
in $\XX \times \YY$.

Given $x \in \XX$ and $y \in \YY$, we write
$\Pr((X, Y) = (x, y))$ as $\Pr(X = x, Y = y)$.  We usually abbreviate
$\Pr(X = x)$ as $\Pr(x)$\ntn{Px}, etc.  Thus, by definition, $X$ and $Y$
are independent%
\index{independent!random variables} 
if and only if
\[
\Pr(x, y) = \Pr(x) \Pr(y)
\]
for all $x \in \XX$ and $y \in \YY$.  The 
conditional%
\index{conditional probability} 
probability of $x$ 
given $y$ is
\[
\Pr(x \given y)
=
\frac{\Pr(x, y)}{\Pr(y)},
\]
and is defined as long as $\Pr(y) > 0$.  

The \demph{Shannon%
\index{Shannon, Claude!entropy}%
\index{entropy!Shannon} 
entropy} of the random variable $X$ is the Shannon
entropy of its distribution:
\[
H(X) 
=
\sum_{x \csuch \Pr(x) > 0} \Pr(x) \log \frac{1}{\Pr(x)}.
\ntn{HRV}
\]
Here and below, the variable $x$ in summations is assumed to run over the
set $\XX$ unless indicated otherwise, and similarly for $y$ in $\YY$.

\subsection*{Joint entropy}

The general definition of the entropy of a random variable can
be applied to the random variable $(X, Y)$, giving
\[
H(X, Y)
=
\sum_{x, y \csuch \Pr(x, y) > 0}
\Pr(x, y) \log \frac{1}{\Pr(x, y)},
\ntn{jointent}
\]
the \demph{joint%
\index{joint entropy} 
entropy} of $X$ and $Y$.

\begin{examples}
\lbl{egs:joint}
\begin{enumerate}
\item
\lbl{eg:joint-ind}
Suppose that $X$ and $Y$ are independent.  If $X$ has distribution $\p$ and
$Y$ has distribution $\vc{r}$ then $(X, Y)$ has distribution $\p \otimes
\vc{r}$, so 
\[
H(X, Y) = H(X) + H(Y)
\]
by Corollary~\ref{cor:ent-log}.  

\item 
\lbl{eg:joint-one}
Suppose that $\YY$ is a one-element set.  Then the distribution of $Y$ is
uniquely determined, $H(Y) = 0$, and $H(X, Y) = H(X)$.

\item
\lbl{eg:joint-equal}
Suppose that $\XX = \YY$ and $X = Y$.  Then $H(X, Y) = H(X) = H(Y)$.

\item
\lbl{eg:joint-det}
Generalizing the last two examples, let us say that $Y$ is
\demph{determined%
\index{determined by} 
by} $X$ if for all $x \in \XX$ such that $\Pr(x) > 0$,
there is a unique $y \in \YY$ such that $\Pr(x, y) > 0$.  Writing this
element $y$ as $f(x)$, we then have $\Pr(x, f(x)) = \Pr(x)$, or
equivalently, $\Pr(f(x)\given x) = 1$.  The joint entropy is given by
\begin{align*}
H(X, Y) &
=
\sum_{x, y \csuch \Pr(x, y) > 0} 
\Pr(x, y) \log \frac{1}{\Pr(x, y)}      \\
&
=
\sum_{x \csuch \Pr(x) > 0} \Pr(x) \log \frac{1}{\Pr(x)} \\
&
=
H(X).
\end{align*}
\end{enumerate}
\end{examples}

\subsection*{Conditional entropy}

The definitions of the conditional entropies $\condent{X}{Y}$ and
$\condent{Y}{X}$ and the mutual information $I(X; Y)$ are suggested by the
schematic diagram of Figure~\ref{fig:venn-info}.
\begin{figure}
\centering
\lengths\setlength{\unitlength}{.8mm}%
\begin{picture}(120,45)
\cell{60}{25}{c}{\includegraphics[height=40\unitlength]{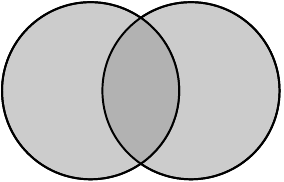}}
\cell{60}{0}{b}{$H(X, Y)$}
\cell{25}{35}{c}{$H(X)$}
\cell{95}{35}{c}{$H(Y)$}
\cell{39}{25}{c}{$\condent{X}{Y}$}
\cell{60}{25}{c}{$I(X; Y)$}
\cell{81}{25}{c}{$\condent{Y}{X}$}
\end{picture}
\caption{Venn diagram showing entropic quantities associated with a pair of
  random variables taking values in different sets: the Shannon entropies
  $H(X)$ and $H(Y)$, the joint entropy $H(X, Y)$, the conditional entropies
  $\condent{X}{Y}$ and $\condent{Y}{X}$, and the mutual information $I(X;
  Y)$.}
\lbl{fig:venn-info}
\end{figure}
The diagram depicts the joint entropy $H(X, Y)$ as the union of the two
discs and $\condent{X}{Y}$ as the complement of the second disc in the
union.  This suggests:

\begin{defn}
The \demph{conditional%
\index{conditional entropy} 
entropy} of $X$ given $Y$ is
\[
\condent{X}{Y} = H(X, Y) - H(Y).
\ntn{condent}
\]
\end{defn}

We now explore this definition.  For each $y \in Y$
such that $\Pr(y) > 0$, there is a random variable $X\given y$\ntn{givenRV}
taking values in $\XX$, with distribution
\[
\Pr\bigl( (X \given y) = x\bigr)
=
\Pr(x \given y)
\]
($x \in \XX$).  Like all random variables, it has an entropy, $H(X
\given y)$.  The name `conditional entropy' is explained by
part~\bref{part:cond-alts-cond} of the following result.

\begin{lemma}
\lbl{lemma:cond-alts}
\begin{enumerate}
\item
\lbl{part:cond-alts-exp}
$\displaystyle
\condent{X}{Y} = \sum_{x, y \csuch \Pr(x, y) > 0} \Pr(x, y) \log
\frac{1}{\Pr(x \given y)}$.
\item
\lbl{part:cond-alts-cond}
$\displaystyle
\condent{X}{Y} = \sum_{y \csuch \Pr(y) > 0} \Pr(y) H(X \given y)$.
\end{enumerate}
\end{lemma}

\begin{proof}
For~\bref{part:cond-alts-exp}, first note that $\Pr(y) = \sum_x \Pr(x, y)$
for each $y \in \YY$, and in particular, $\Pr(y) > 0$ if there exists an
$x$ such that $\Pr(x, y) > 0$.  Hence
%
\begin{equation}
H(Y)    
=
\sum_{x, y \csuch \Pr(x, y) > 0} \Pr(x, y) \log \frac{1}{\Pr(y)}.
\lbl{eq:ca-1}
\end{equation}
%
It follows that
\begin{align*}
\condent{X}{Y}  &
=
\sum_{x, y \csuch \Pr(x, y) > 0} \Pr(x, y) \log \frac{1}{\Pr(x, y)}
-
\sum_{x, y \csuch \Pr(x, y) > 0} \Pr(x, y) \log \frac{1}{\Pr(y)}      
\nonumber       \\
&
=
\sum_{x, y \csuch \Pr(x, y) > 0} 
\Pr(x, y) \log \frac{\Pr(y)}{\Pr(x, y)} \\
&
=
\sum_{x, y \csuch \Pr(x, y) > 0} 
\Pr(x, y) \log \frac{1}{\Pr(x \given y)},
\end{align*}
proving~\bref{part:cond-alts-exp}.  This in turn is equal to 
\[
\sum_{y \csuch \Pr(y) > 0} \Pr(y)
\sum_{x \csuch \Pr(x \given y) > 0} 
\Pr(x \given y) \log \frac{1}{\Pr(x \given y)}
=
\sum_{y \csuch \Pr(y) > 0} \Pr(y) H(X \given y),
\]
proving~\bref{part:cond-alts-cond}.
\end{proof}

The conditional entropy $\condent{X}{Y}$ is, therefore, the expected
entropy of the conditional random variable $X \given y$ when $y$ is chosen
at random.  It follows that $\condent{X}{Y} \geq 0$, or equivalently, that
$H(X, Y) \geq H(Y)$.

\begin{examples}
\lbl{egs:cond}
Consider again the four scenarios of Examples~\ref{egs:joint}.
\begin{enumerate}
\item
\lbl{eg:cond-ind}
Suppose that $X$ and $Y$ are independent.  Then by
Example~\ref{egs:joint}\bref{eg:joint-ind}, 
\[
\condent{X}{Y} = H(X),
\qquad
\condent{Y}{X} = H(Y).
\]
Knowing the value of $Y$ gives no information on the value of $X$, nor vice
versa.  This is the situation shown in
Figure~\ref{fig:venn-special}\hardref{(a)}.

\item 
\lbl{eg:cond-one}
Suppose that $\YY$ is a one-element set.  Then by
Example~\ref{egs:joint}\bref{eg:joint-one}, 
\begin{align*}
\condent{X}{Y}  &
=
H(X, Y) - H(Y) = H(X),  \\
\condent{Y}{X}  &
=
H(X, Y) - H(X) = 0.
\end{align*}

\item
\lbl{eg:cond-equal}
Suppose that $\XX = \YY$ and $X = Y$.  By
Example~\ref{egs:joint}\bref{eg:joint-equal}, $\condent{X}{Y} = 0$.  This
is intuitively plausible: once we know the value of $Y$, we know the value
of $X$ with certainty, so its probability distribution is concentrated on a
single element and therefore has entropy $0$.  Similarly, $\condent{Y}{X} =
0$.

\item
\lbl{eg:cond-det}
More generally, suppose that $Y$ is determined by $X$
(Figure~\ref{fig:venn-special}\hardref{(b)}).  Then by
Example~\ref{egs:joint}\bref{eg:joint-det},
\begin{align}
\condent{X}{Y}  & 
= 
H(X) - H(Y),
\lbl{eq:cond-det-diff}        \\
\condent{Y}{X}  &
=
0.
\nonumber
\end{align}
Since $\condent{X}{Y} \geq 0$, we have
\[
H(Y) \leq H(X)
\]
whenever $Y$ is determined by $X$.
\end{enumerate}
\end{examples}

\begin{figure}
\centering
\lengths
\begin{picture}(120,35)
\cell{30}{20}{c}{\includegraphics[width=50\unitlength]{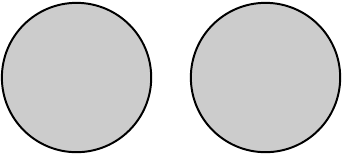}}
\cell{100}{20}{c}{\includegraphics[width=30\unitlength]{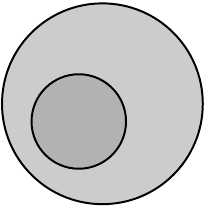}}
\cell{0}{31}{l}{$H(X)$}
\cell{55}{31}{c}{$H(Y)$}
\cell{30}{0}{b}{(a)}
\cell{105}{25}{c}{$H(Y)$}
\cell{115}{32}{c}{$H(X)$}
\cell{100}{0}{b}{(b)}
\end{picture}
\caption{Venn diagrams for the cases in which \hardref{(a)}~the random
  variables $X$ and $Y$ are independent; \hardref{(b)}~$Y$ is determined by
  $X$.} 
\lbl{fig:venn-special}
\end{figure}

\begin{remark} 
\lbl{rmk:chain-rules}
In most texts, Lemma~\ref{lemma:cond-alts}\bref{part:cond-alts-cond} is
taken as the \emph{definition} of conditional entropy, and the equation
$\condent{X}{Y} = H(X, Y) - H(Y)$ that we took as our definition is proved
as a theorem.  This theorem is called the 
\demph{chain%
\index{chain rule!naming of}%
\index{chain rule!conditional entropy@for conditional entropy}
rule}.  In other words, the chain rule states that
\begin{equation}
\lbl{eq:joint-cond}
H(X, Y) 
=
H(Y) + \sum_{y \csuch \Pr(y) > 0} \Pr(y) H(X \given y).
\end{equation}
It is essentially the same as what we have been calling the chain rule
throughout this text (beginning in Proposition~\ref{propn:ent-chain}).
This can be seen as follows.

Write $\XX = \{1, \ldots, k\}$ and $\YY = \{1, \ldots, n\}$.  Write $\vc{w}
= (w_1, \ldots, w_n) \in \Delta_n$ for the distribution of $Y$; thus, $w_i
= \Pr(Y = i)$ for each $i \in \YY$.  Also, for each $i \in \YY$ and $j \in
\XX$, define $p^i_j$ by
\[
w_i p^i_j = \Pr(X = j, Y = i),
\]
so that $\p^i = (p^i_1, \ldots, p^i_k) \in \Delta_k$ is the distribution of
the random variable $X \given i$.  (Here we have assumed that $\Pr(i) > 0$;
otherwise, choose $\p^i \in \Delta_k$ arbitrarily.)  Then $\vc{w} \of
(\p^1, \ldots, \p^n) \in \Delta_{nk}$ is the joint distribution of $X$ and
$Y$.  In this notation, equation~\eqref{eq:joint-cond} states that
\[
H\bigl( \vc{w} \of (\p^1, \ldots, \p^n) \bigr)
=
H(\vc{w}) + \sum_{i \csuch w_i > 0} w_i H(\p^i).
\]
This is exactly the chain rule in our usual sense.  
\end{remark}

\subsection*{Mutual information}

In Figure~\ref{fig:venn-info}, the intersection of the two discs is
labelled as $I(X; Y)$, and the inclusion-exclusion principle suggests the
formula
\[
H(X, Y) = H(X) + H(Y) - I(X; Y).
\]
We define $I(X; Y)$\ntn{mutinfo} to make this true:

\begin{defn}
The \demph{mutual%
\index{mutual information} 
information} of $X$ and $Y$ is
\[
I(X; Y) = H(X) + H(Y) - H(X, Y).
\]
\end{defn}

Evidently $I$ is symmetric:
\begin{equation}
\lbl{eq:mut-sym}
I(X; Y) = I(Y; X).
\end{equation}
Alternative expressions for $I$, in terms of conditional rather than joint
entropy, follow immediately from the definitions and are also suggested by
the Venn diagram:
\begin{equation}
\lbl{eq:mut-alt}
I(X; Y) = H(X) - \condent{X}{Y} = H(Y) - \condent{Y}{X}.
\end{equation}
Mutual information can be expressed in two further ways still:

\begin{lemma}
\lbl{lemma:mut-alts}
\begin{enumerate}
\item 
\lbl{part:mut-alts-exp}
$\displaystyle
I(X; Y) = \sum_{x, y \csuch \Pr(x, y) > 0} 
\Pr(x, y) \log \frac{\Pr(x, y)}{\Pr(x) \Pr(y)}$.

\item
\lbl{part:mut-alts-mean-rel}
$\displaystyle
I(X; Y) = \sum_{y \csuch \Pr(y) > 0} \Pr(y) 
\relEnt{(X \given y)}{X}$.
\end{enumerate}
\end{lemma}

The right-hand side of~\bref{part:mut-alts-mean-rel} refers to the 
random variables $X \given y$ and $X$ taking values in $\XX$, and the
relative entropy of the first with respect to the second.

\begin{proof}
For~\bref{part:mut-alts-exp}, we have
\begin{align*}
I(X; Y) &
=
H(X) - \condent{X}{Y}   \\
&
=
\sum_{x, y \csuch \Pr(x, y) > 0} \Pr(x, y) \log \frac{1}{\Pr(x)}
-
\sum_{x, y \csuch \Pr(x, y) > 0} \Pr(x, y) \log \frac{\Pr(y)}{\Pr(x, y)},
\end{align*}
by equation~\eqref{eq:ca-1} and
Lemma~\ref{lemma:cond-alts}\bref{part:cond-alts-exp}.  Collecting terms,
the result follows.

To prove~\bref{part:mut-alts-mean-rel}, we use~\bref{part:mut-alts-exp} and
the equation $\Pr(x, y) = \Pr(y) \Pr(x \given y)$:
\begin{align*}
I(X; Y) &
=
\sum_{x, y \csuch \Pr(y) > 0, \, \Pr(x \given y) > 0}
\Pr(y) \Pr(x \given y) \log\frac{\Pr(x \given y)}{\Pr(x)}       \\
&
=
\sum_{y \csuch \Pr(y) > 0} \Pr(y) 
\relEnt{(X \given y)}{X},
\end{align*}
as required.
\end{proof}

The formula in~\bref{part:mut-alts-mean-rel} can be interpreted as
follows.  For probability distributions $\p$ and $\vc{r}$ on the same
finite set, $\relent{\p}{\vc{r}}$ can be understood as the information
gained when learning that the distribution of a random variable is $\p$,
when one had previously believed that it was $\vc{r}$.  Thus, $\relent{(X
  \given y)}{X}$ is the information gained about $X$ by learning that $Y =
y$.  Consequently,
\[
\sum_{y \csuch \Pr(y) > 0} \Pr(y) \relent{(X \given y)}{X}
\]
is the expected information about $X$ gained by learning the value of $Y$.
This is the mutual information $I(X; Y)$.  Briefly put, it is the
information that $Y$ gives about $X$.

For instance, if $X$ and $Y$ are independent, then knowing the value of $Y$
gives us no clue as to the value of $X$, so one would expect that $I(X; Y)
= 0$.  And indeed, $X \given y$ has the same distribution as $X$ (for each
$y$), so $\relEnt{(X \given y)}{X} = 0$, giving $I(X; Y) = 0$.  We examine
the extremal cases more systematically in
Proposition~\ref{propn:pair-bounds}. 
  
Of course, Lemma~\ref{lemma:mut-alts}\bref{part:mut-alts-mean-rel} has a
counterpart with $X$ and $Y$ interchanged, and the symmetry property $I(X;
Y) = I(Y; X)$ of mutual information (equation~\eqref{eq:mut-sym}) implies
that
\[
\sum_{y \csuch \Pr(y) > 0} \Pr(y) \relEnt{(X \given y)}{X}
=
\sum_{x \csuch \Pr(x) > 0} \Pr(x) \relEnt{(Y \given x)}{Y}.
\]
That is: the information that $Y$ gives about $X$ is equal to the
information that $X$ gives about $Y$.  This explains the word `mutual'.

\begin{examples}
\lbl{egs:mut}
We consider again the four cases of Examples~\ref{egs:joint}
and~\ref{egs:cond}, using the results derived there.

\begin{enumerate}
\item 
\lbl{eg:mut-ind}
If $X$ and $Y$ are independent then $I(X; Y) = 0$: neither variable gives
any information about the other.

\item 
\lbl{eg:mut-one}
If $\YY$ is a one-element set then $I(X; Y) = 0$.  From one viewpoint,
knowing the value of $X$ gives no information about the value of $Y$, 
since the value of $Y$ is predetermined anyway.  From the other, knowing the
value of $Y$ gives no information about the value of $X$ (or indeed,
about anything).

\item 
\lbl{eg:mut-equal} 
If $\XX = \YY$ and $X = Y$ then $I(X; Y) = H(X) = H(Y)$.  This is the
maximal value that $I(X; Y)$ can take (by
Proposition~\ref{propn:pair-bounds}\bref{part:mut-bounds} below), which is
intuitively plausible: knowing $X$ gives complete information about $Y$.

\item 
\lbl{eg:mut-det}
Generally, if $Y$ is determined by $X$, then $I(X; Y) = H(Y)$.  As
in~\bref{eg:mut-equal}, this tells us that knowledge of $X$ gives certain
knowledge of $Y$ (even though knowledge of $Y$ does not, in this case, give
certain knowledge of $X$).
\end{enumerate}
\end{examples}

The Venn diagram of Figure~\ref{fig:venn-info} is not merely a metaphor or
an analogy.  It depicts a specific example:

\begin{example}
\lbl{eg:random-subsets} For this example, first note that joint entropy,
conditional entropy and mutual information can be defined using
logarithms to any base.  Just as we write $\Hi(X) = H(X)/\log 2$
(Remark~\ref{rmk:ent-base}), let us write the base $2$ version of joint
entropy as $\Hi(X, Y)\ntn{jcmi} = H(X, Y)/\log 2$, and similarly for
$\condenti{X}{Y}$ and $I^{\binsym}(X; Y)$.

Fix finite subsets $K$ and $L$ of some set.  Let $Z$ denote a subset%
\index{random!subset}%
\index{subset, random}
of $K \cup L$ chosen uniformly at random, and put
\[
X = Z \cap K,
\qquad
Y = Z \cap L
\]
(Figure~\ref{fig:random-subsets}).  
\begin{figure}
\centering
\lengths\setlength{\unitlength}{.75mm}%
\begin{picture}(60,39)(0,5)
\cell{30}{24.7}{c}{\includegraphics[height=38.5\unitlength]{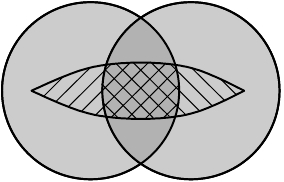}}
\cell{2}{40}{l}{$K$}
\cell{57.5}{40}{r}{$L$}
\cell{10}{30}{c}{$X$}
\cell{30}{16}{c}{$Z$}
\cell{50}{30}{c}{$Y$}
\end{picture}%
\caption{Random subsets (Example~\ref{eg:random-subsets}).
  The subset $Z$ is the whole lozenge, $X = Z \cap K$ is shaded
  as~\protect\raisebox{-1mm}{\protect\includegraphics[height=4mm]{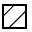}},
  and $Y = Z \cap L$ is shaded 
  as~\protect\raisebox{-1mm}{\protect\includegraphics[height=4mm]{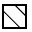}}.} 
\lbl{fig:random-subsets}
\end{figure}
Then $X$ and $Y$ are uniformly distributed random variables taking values
in the power sets $\pset(K)$ and $\pset(L)$ respectively, so
\begin{align*}
\Hi(X)  &= \log_2\bigl(2^{\mg{K}}\bigr) = \mg{K}, \\
\Hi(Y)  &= \log_2\bigl(2^{\mg{L}}\bigr) = \mg{L}.
\end{align*}
The random variable $(X, Y)$, which takes values in $\pset(K) \times
\pset(L)$, is uniformly distributed on the set of pairs
\[
\bigl\{ 
(A, B) \in \pset(K) \times \pset(L) 
\such 
A = C \cap K \text{ and } B = C \cap L
\text{ for some } C \sub K \cup L
\bigr\}.
\]
Such pairs are in one-to-one correspondence with subsets of $K \cup L$, so
the entropy of $(X, Y)$ is equal to the entropy of the uniform distribution
on $\pset(K \cup L)$.  Hence
\[
\Hi(X, Y)
=
\log_2 \bigl( 2^{\mg{K \cup L}} \bigr)
=
\mg{K \cup L}.
\]
Then by definition of conditional entropy and mutual information,
\begin{align*}
\condenti{X}{Y} &
=
\mg{K \cup L} - \mg{L} 
= 
\mg{K \without L},     \\
\condenti{Y}{X} &
=
\mg{K \cup L} - \mg{K} 
= 
\mg{L \without K},     \\
I^{\binsym}(X; Y)       &
=
\mg{K} + \mg{L} - \mg{K \cup L} 
=
\mg{K \cap L}.
\end{align*}
So, this example realizes the various entropies shown in the Venn diagram
of Figure~\ref{fig:venn-info} as actual cardinalities.%
\index{cardinality}
\end{example}

\subsection*{Extremal cases}

We finish this introduction to joint entropy, conditional entropy and
mutual information by finding their maximal and minimal values in terms of
ordinary entropy.  Here is the central fact.

\begin{lemma}
\lbl{lemma:mut-lb}
$I(X; Y) \geq 0$, with equality if and only if $X$ and $Y$ are independent.
\end{lemma}

\begin{proof}
Lemma~\ref{lemma:mut-alts}\bref{part:mut-alts-mean-rel} states that
\[
I(X; Y) 
= 
\sum_{y \csuch \Pr(y) > 0} \Pr(y) 
\relEnt{(X \given y)}{X}.
\]
Given $y \in \YY$ such that $\Pr(y) > 0$, Lemma~\ref{lemma:rel-ent-pos-def}
implies that $\relEnt{(X \given y)}{X} \geq 0$, with equality if and only
if $\Pr(x \given y) = \Pr(x)$ for all $x \in \XX$.  Thus, $I(X; Y) \geq 0$,
with equality if and only if $\Pr(x \given y) = \Pr(x)$ for all $x$, $y$
such that $\Pr(y) > 0$.  But this condition is equivalent to $X$ and $Y$
being independent.
\end{proof}

\begin{remark}
Given three random variables $X$, $Y$ and $Z$ with the same sample
space, one can define a threefold%
\index{mutual information!threefold}
mutual information $I(X; Y; Z)$ by the same inclusion-exclusion principle
that has guided us so far:
\[
I(X; Y; Z) 
=
\bigl( H(X) + H(Y) + H(Z) \bigr) 
-
\bigl( H(X, Y) + H(X, Z) + H(Y, Z) \bigr)
+
H(X, Y, Z).
\]
But in contrast to the binary case, $I(X; Y; Z)$ is sometimes
negative.  For example, this is the case when all three random variables
take values in $\{0, 1\}$ and $(X, Y, Z)$ is uniformly distributed on the
four triples
\[
(0, 0, 0), \ (0, 1, 1), \ (1, 0, 1), \ (1, 1, 0),
\]
with probability zero on the other four.  For discussion of this and other
multivariate information measures, see Timme et al.~\cite{TAFB}, especially
Section~4.2.
\end{remark}

The following proposition gathers together the various maximal and minimal
values and the conditions under which they are attained.  All the results
are as one would guess from the Venn diagrams of Figure~\ref{fig:venn-info}
and~\ref{fig:venn-special}. 

\begin{propn}
\lbl{propn:pair-bounds}
\begin{enumerate}
\item 
\lbl{part:joint-bounds}
Joint entropy is bounded as follows:
\begin{enumerate}
\item 
$\max\{H(X), H(Y)\} \leq H(X, Y) \leq H(X) + H(Y)$;

\item
$H(X, Y) = \max\{H(X), H(Y)\}$ if and only if $X$ is determined by $Y$ or
  $Y$ is determined by $X$;

\item
$H(X, Y) = H(X) + H(Y)$ if and only if $X$ and $Y$ are independent.
\end{enumerate}

\item
\lbl{part:cond-bounds}
Conditional entropy is bounded as follows:
\begin{enumerate}
\item 
$0 \leq \condent{X}{Y} \leq H(X)$;

\item
$\condent{X}{Y} = 0$ if and only if $X$ is determined by $Y$;

\item
$\condent{X}{Y} = H(X)$ if and only if $X$ and $Y$ are independent.
\end{enumerate}

\item
\lbl{part:mut-bounds}
Mutual information is bounded as follows:
\begin{enumerate}
\item 
$0 \leq I(X; Y) \leq \min \{H(X), H(Y)\}$;

\item
$I(X; Y) = 0$ if and only if $X$ and $Y$ are independent;

\item
$I(X; Y) = \min \{H(X), H(Y)\}$ if and only if $X$ is determined by $Y$ or
  $Y$ is determined by $X$.
\end{enumerate}
\end{enumerate}
\end{propn}

\begin{proof}
We begin with~\bref{part:cond-bounds}, using
Lemma~\ref{lemma:cond-alts}\bref{part:cond-alts-cond}: 
\[
\condent{X}{Y} 
=
\sum_{y \csuch \Pr(y) > 0} \Pr(y) H(X \given y).
\]
For each $y$ such that $\Pr(y) > 0$,
Lemma~\ref{lemma:ent-max-min}\bref{part:ent-min} implies that $H(X \given
y) \geq 0$, with equality if and only if there is some $x$ such that $\Pr(x
\given y) = 1$.  So, $\condent{X}{Y} \geq 0$, with equality if and only if
$X$ is determined by $Y$.  For the upper bound, Lemma~\ref{lemma:mut-lb}
gives 
\[
H(X) - \condent{X}{Y} = I(X; Y) \geq 0,
\]
with equality if and only if $X$ and $Y$ are independent.

For~\bref{part:joint-bounds}, we have
\[
H(X, Y) - H(X) = \condent{Y}{X} \geq 0,
\]
with equality if and only if $Y$ is determined by $X$
(by~\bref{part:cond-bounds}).  Hence $H(X, Y) \geq \max\{H(X), H(Y)\}$, and
if equality holds then $Y$ is determined by $X$ or vice versa.  Conversely,
suppose without loss of generality that $Y$ is determined by $X$.  We
have $H(X, Y) = H(X)$ by Example~\ref{egs:joint}\bref{eg:joint-det}
and $H(Y) \leq H(X)$ by Example~\ref{egs:cond}\bref{eg:cond-det}, so
$H(X, Y) = \max\{H(X), H(Y)\}$, as required.  For the upper bound on $H(X,
Y)$, we have
\[
H(X) + H(Y) - H(X, Y) = I(X; Y) \geq 0
\]
with equality if and only if $X$ and $Y$ are independent, by
Lemma~\ref{lemma:mut-lb}. 

For~\bref{part:mut-bounds}, the lower bound and its equality condition
were proved as Lemma~\ref{lemma:mut-lb}.  The upper bound follows from the
lower bound in~\bref{part:joint-bounds} by subtracting from $H(X) +
H(Y)$:
\begin{align*}
&
\max\{H(X), H(Y)\} \leq H(X, Y)         \\
\iff
&
H(X) + H(Y) - \max\{H(X), H(Y)\} 
\geq H(X) + H(Y) - H(X, Y)      \\
\iff
&
\min\{H(X), H(Y)\} \geq I(X; Y),
\end{align*}
with the same condition for equality as in~\bref{part:joint-bounds}.
\end{proof}

\begin{remark}
\lbl{rmk:ub-coupling} 
Given random variables $X$ and $Y$ taking values in finite sets $\XX$ and
$\YY$ respectively, there is a random variable $X \otimes Y$ taking values
in $\XX \times \YY$, the \demph{independent coupling}%
\index{coupling!independent}%
\index{independent!coupling}
of $X$ and $Y$, with
distribution
\[
\Pr\bigl( X \otimes Y = (x, y) \bigr)
=
\Pr(X = x) \Pr(Y = y)
\]
($x \in \XX$, $y \in \YY$).  That is, if $X$ has distribution $\p$ and $Y$
has distribution $\vc{r}$ then $X \otimes Y$ has distribution $\p \otimes
\vc{r}$.  Then
\[
H(X \otimes Y) = H(X) + H(Y)
\]
by Corollary~\ref{cor:ent-log}, so the upper bound in
Proposition~\ref{propn:pair-bounds}\bref{part:joint-bounds} is equivalent
to
\[
H(X, Y) \leq H(X \otimes Y).
\]
Another way to state this is as follows.  Take probability distributions
$\p$ on $\XX$ and $\vc{r}$ on $\YY$.  Then among all probability
distributions on $\XX \times \YY$ with marginals $\p$ and $\vc{r}$,
none has greater entropy than $\p \otimes \vc{r}$.

This is a special property of \emph{Shannon} entropy.  It does not hold for
any of the other R\'enyi entropies $H_q$ or $q$-logarithmic entropies $S_q$
except, trivially, when $q = 0$.  Counterexamples are given in
Appendix~\ref{sec:max-cpl}.  There is a substantial literature on the
entropy of couplings; see, for instance, Sason~\cite{Saso},
Kova\v{c}evi\'{c}, Stanojevi\'{c} and \v{S}enk~\cite{KSS}, and references
therein.
\end{remark}

\section{Diversity measures for subcommunities}
\lbl{sec:mm-sub}
\index{subcommunity!diversity measures}
\index{diversity measure!subcommunity}

In the next two sections, we introduce quantities measuring
features of a large community of organisms (a \dmph{metacommunity}) divided
into smaller communities (\demph{subcommunities}\index{subcommunity}), to
answer the ecological questions posed in the introduction to this
chapter.  As before, we use terminology inspired by ecology, even though
the mathematics applies far more generally to any types of object.

We have already discussed several times a special type of metacommunity,
namely, a group of islands%
\index{islands!subcommunities@as subcommunities} 
(Examples~\ref{eg:comp-islands}, \ref{eg:div1-chain-islands}, \ref{eg:oil},
etc.).  There, the subcommunities are the islands, the metacommunity is the
union of all of them, and a very strong assumption is made: that no species
are shared between islands. Although this is a useful hypothetical extreme
case, it is not realistic.  In the metacommunities that we are about to
consider, each species may be present in one, many, or all of the
subcommunities, in any proportions.

In ecology, there is established terminology for measures of metacommunity
diversity:
\begin{itemize}
\item 
the \dmph{alpha-diversity}\lbl{p:alpha-ave} is the \underline{a}verage
diversity of the subcommunities (in some sense of `average');

\item
the \dmph{beta-diversity}\lbl{p:beta-bet} is the variation \underline{b}etween
the subcommunities;%
\index{partitioning of diversity}%
\index{diversity!partitioning of}

\item
the \dmph{gamma-diversity} is the diversity of the whole metacommunity (the
\underline{g}lobal diversity), ignoring its division into subcommunities.
\end{itemize}
These terms were introduced by the ecologist Robert
Whittaker%
\index{Whittaker, Robert} 
in an influential paper of 1960 (\cite{WhitVSM}, p.~320).  As Tuomisto%
\index{Tuomisto, Hanna}
observed in a survey paper on beta-diversity, `Obviously, Whittaker~(1960)
did not have an exact definition of beta diversity in mind' (\cite{Tuom1},
p.~2).  However, a large number of specific proposals have been made for
defining these three quantities mathematically.  Some early work on the
subject may have been inspired by analysis%
\index{analysis of variance} 
of variance (ANOVA) in statistics, where one seeks to quantify
within-group and between-group variation.  But the broad concepts
of alpha-, beta- and gamma-diversity acquired their own independent
standing long ago.

This section and the next are largely based on a paper of Reeve
et al.~\cite{HPD} that sets
out a comprehensive and non-traditional suite of diversity measures for
metacommunities and their subcommunities.  The system of measures is highly
flexible, incorporating both the parameter $q$ (to allow different
emphasis on rare and common species) and the similarity matrix $Z$ (to
encode the different similarities between species).
Here, we confine ourselves to the very special case where $q = 1$ and
$Z = I$ (thus, ignoring inter-species similarity).  Even so, we will be
able to see some of the power and subtlety of the system.  

We begin by fixing our notation (Figure~\ref{fig:Ppw}).  The metacommunity
consists of a collection of individuals, each of which belongs to exactly
one of $S$ species (numbered as $1, \ldots, S$) and exactly one of $N$
subcommunities (numbered as $1, \ldots, N$).  We write $P_{ij}$\ntn{Pij}
for the proportion or relative abundance of individuals belonging to 
the $i$th species and the $j$th subcommunity.
\begin{figure}
\centering
\vspace*{8mm}
\begin{align*}
\lengths
\begin{picture}(0,0)
\put(19,14){\vector(-1,0){11.5}}
\put(19,14){\vector(1,0){11.5}}
\cell{19}{18}{c}{subcommunities}
\put(-6,1){\vector(0,-1){7}}
\put(-6,1){\vector(0,1){7}}
\cell{-10}{1}{c}{\rotatebox{90}{species}}
\end{picture}
P &
=
\begin{pmatrix}
P_{11}  &\cdots &P_{1N} \\
\vdots  &       &\vdots \\
P_{S1}  &\cdots &P_{SN}
\end{pmatrix}
\hspace*{5mm}
\p      
=
\begin{pmatrix}
p_1     \\
\vdots  \\
p_S     
\end{pmatrix}
\\[3mm]
\vc{w}  &
=
\,
\begin{pmatrix}
\,w_1    &\ \cdots\ &w_N\,
\end{pmatrix}
\end{align*}
\caption{Notation for the relative abundances in a metacommunity.}
\lbl{fig:Ppw}
\end{figure}
Thus, $\sum_{i, j} P_{ij} = 1$.  We adopt the convention that
the index $i$ ranges over the set $\{1, \ldots, S\}$ of species and the
index $j$ ranges over the set $\{1, \ldots, N\}$ of subcommunities.

For each species $i$, write
\[
p_i = \sum_j P_{ij},
\ntn{pi}
\]
which is the relative abundance of species $i$ in the whole metacommunity.
Then $\sum_i p_i = 1$.  For each subcommunity $j$, write
\[
w_j = \sum_i P_{ij},
\ntn{wj}
\]
which is the relative size of subcommunity $j$ in the metacommunity.  Then
$\sum_j w_j = 1$.

In purely mathematical terms, the matrix $P$\ntn{Pmx} defines a probability
distribution on the set $\{1, \ldots, S\} \times \{1, \ldots, N\}$, with
marginal distributions 
\[
\p = (p_1, \ldots, p_S)\ntn{pvec},
\qquad
\vc{w} = (w_1, \ldots, w_N)\ntn{wvec}.  
\]
To translate into the language of random variables, we will consider a
random variable $(X, Y)$ taking values in $\{1, \ldots, S\} \times \{1,
\ldots, N\}$, with distribution $P$.  Then $X$ is a random variable with
values in $\{1, \ldots, S\}$ and distribution $\p$, and $Y$ is a random
variable with values in $\{1, \ldots, N\}$ and distribution $\vc{w}$.
Thus, $X$ is a random species and $Y$ is a random subcommunity.

What are the ecological meanings of the joint entropy, conditional entropy
and mutual information of the random variables $X$ and $Y$?  And what are
the roles of relative and cross entropy?  We have seen that when measuring
diversity, it is more appropriate to use the \emph{exponential} of entropy
than entropy itself (Section~\ref{sec:ent-div}, especially
Example~\ref{eg:plague}).  So it is better to ask: what are the ecological
meanings of the exponentials of relative entropy, mutual information, and
so on?

We now proceed to answer these questions, following throughout the notation and
terminology of Reeve et al.~\cite{HPD}.  

First, consider the two entropies defined for a pair of
distributions on the \emph{same} set: relative and cross entropy.  The
$j$th subcommunity has species distribution
\[
P_{\Pdot j}/w_j
=
\bigl( P_{1j}/w_j, \ldots, P_{Sj}/w_j \bigr),
\ntn{Pdotj}
\]
which is the normalization of the $j$th column $P_{\Pdot j}$ of the matrix
$P$.  (Assume that the $j$th subcommunity is nonempty: $w_j > 0$.)
We write
\[
\ovln{\alpha}_j(P) 
=
D(P_{\Pdot j}/w_j)
=
\exp H(P_{\Pdot j}/w_j)
\ntn{alphabar}
\]
for the diversity of order $1$ of the $j$th subcommunity, and call it the
\demph{subcommunity alpha-diversity}.%
\index{subcommunity!alpha-diversity}%
\index{alpha-diversity!subcommunity}
Thus, $\ovln{\alpha}_j(P)$ depends on the $j$th subcommunity only, and is
unaffected by the rest of the metacommunity.  Here $D$ denotes the
Hill number $D_1$ of order $1$ (as in Section~\ref{sec:ent-div}); 
no other value of the parameter $q$ is under consideration.

As well as considering the $j$th subcommunity in isolation, we can compare
its species distribution $P_{\Pdot j}/w_j$ with the species distribution
$\p$ of the whole metacommunity, using the relative entropy
$\relent{P_{\Pdot j}/w_j}{\p}$.  Better, we can use the exponential of
relative entropy, which is a relative%
\index{relative diversity}
diversity in the sense of Section~\ref{sec:rel-div}.  Thus, we define the
\demph{subcommunity%
\index{beta-diversity!subcommunity}%
\index{subcommunity!beta-diversity}
beta-diversity} $\ovln{\beta}_j(P)$ by
\begin{equation}
\ovln{\beta}_j(P)       
=
\reldiv{P_{\Pdot j}/w_j}{\p}
=
\prod_i \Biggl( \frac{P_{ij}}{p_i w_j} \Biggr)^{P_{ij}/w_j}.
\lbl{eq:betaj-exp}      
\end{equation}
This is the diversity of the species distribution of the $j$th subcommunity
relative to that of the metacommunity.  As established in
Section~\ref{sec:rel-div}, it reflects the unusualness or atypicality of
the subcommunity in the context of the metacommunity as a whole.  For
example, if the subcommunity is exactly representative of the whole
metacommunity then $\ovln{\beta}_j(P)$ takes its minimum possible value, $1$. 

(Reeve et al.\ also defined quantities called $\alpha_j$ and $\beta_j$, not
discussed here.  The bars are used in that work to indicate normalization
by subcommunity size.)

Alternatively, we can compare a subcommunity with the metacommunity using
cross%
\index{cross entropy} 
entropy rather than relative entropy.  The exponential
$\crossdiv{P_{\Pdot j}/w_j}{\p}$ of the cross entropy is a cross%
\index{cross diversity} 
diversity (again in the sense of Section~\ref{sec:rel-div}), and is called
the \demph{subcommunity%
\index{gamma-diversity!subcommunity}%
\index{subcommunity!gamma-diversity}
gamma-diversity},
\begin{align}
\gamma_j(P)     &
=
\crossdiv{P_{\Pdot j}/w_j}{\p}
=
\prod_i \Biggl( \frac{1}{p_i} \Biggr)^{P_{ij}/w_j}.
\lbl{eq:gammaj-exp}
\end{align}
Thus, $\gamma_j(P)$ is the cross diversity of the species distribution of
the $j$th subcommunity with respect to that of the metacommunity.  It is
the average rarity of species in the subcommunity, measuring rarity by the
standards of the metacommunity (and using the geometric mean as our notion
of average).  For example, if the subcommunity is exactly representative of
the metacommunity then $\gamma_j(P)$ is just the diversity of the
metacommunity. 

Other examples of relative diversity and cross diversity were given in
Section~\ref{sec:rel-div}, illustrating the ecological meanings of high or
low values of $\ovln{\beta}_j(P)$ or $\gamma_j(P)$.

Equation~\eqref{eq:rco} (p.~\pageref{eq:rco}) implies that
\begin{equation}
\lbl{eq:abg}
\ovln{\alpha}_j(P) \cdot \ovln{\beta}_j(P) = \gamma_j(P).
\end{equation}
This identity can be understood as follows:
\begin{itemize}
\item
$\ovln{\alpha}_j(P)$ measures how unusual
the average individual is within the subcommunity;

\item
$\ovln{\beta}_j(P)$ measures how unusual the subcommunity is within the
  metacommunity;

\item
$\gamma_j(P)$ measures how unusual the average individual in the
  subcommunity is within the metacommunity.
\end{itemize}
Thus, equation~\eqref{eq:abg} partitions%
\index{partitioning of diversity}%
\index{diversity!partitioning of}
the global diversity measure $\gamma_j(P)$ into components measuring
diversity at different levels of resolution.

In the next section, we explain the connection between, on the one hand,
the \emph{subcommunity} alpha-, beta- and gamma-diversities just defined,
and, on the other, what ecologists usually call alpha-, beta- and
gamma-diversity, which are quantities associated with the
\emph{metacommunity}.  We will use the language of random variables.  In
that language, the subcommunity measures that we have just defined are
given by
\begin{align*}
\ovln{\alpha}_j(P)      &
=
\exp\bigl( H(X \given j) \bigr),        \\
\ovln{\beta}_j(P)  &
=
\exp\Bigl( \relEnt{(X \given j)}{X} \Bigr),     \\
\gamma_j(P)        &
=
\exp\Bigl( \crossEnt{(X \given j)}{X} \Bigr),
\end{align*}
since the distribution of the conditional random variable $X \given j$ is
the species distribution $P_{\Pdot j}/w_j$ in the $j$th subcommunity.

\section{Diversity measures for metacommunities}
\lbl{sec:mm-meta}
\index{diversity measure!metacommunity}
\index{metacommunity!diversity measures}

In the last section, the alpha-, beta- and gamma-diversities of the $j$th
subcommunity were defined by comparing two random variables taking values
in the set of species: $X \given j$, which is the species of an individual
chosen at random from the $j$th subcommunity, and $X$, which is the species
of an individual chosen at random from the whole metacommunity.

In this section, we derive measures of the metacommunity by comparing two
random variables taking values in different sets: the species $X$ and the
subcommunity $Y$ of an individual chosen at random from the metacommunity.
Specifically, we consider the exponentials of their joint entropy,
conditional entropies, and mutual information.

Figure~\ref{fig:mc-measures}\hardref{(a)} summarizes the situation, with
the previous Venn diagram for entropies (Figure~\ref{fig:venn-info})
reproduced in~\hardref{(b)} for reference.  The notation in~\hardref{(a)}
is again taken from Reeve et al.~\cite{HPD}, and each term is now explained in
turn.  Tables~\ref{table:mcm-desc} and~\ref{table:mcm-range} give 
summaries.

\begin{figure}
\centering
\lengths
\begin{picture}(54,35)(0,-5)
\cell{27}{17}{c}{\includegraphics[height=26\unitlength]{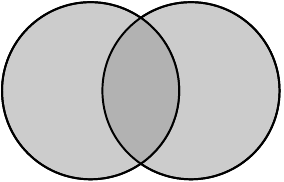}}
\cell{7}{25}{c}{$G$}
\cell{49.5}{25}{c}{$D(\vc{w})$}
\cell{14}{17}{c}{$\ovln{A}$}
\cell{27}{17}{c}{$\ovln{B}$}
\cell{40}{16.5}{c}{$R$}
\cell{27}{1}{b}{$A$}
\cell{27}{-5}{b}{(a)}
\end{picture}%
\hspace*{12mm}%
\begin{picture}(54,35)(0,-5)
\cell{27}{17}{c}{\includegraphics[height=26\unitlength]{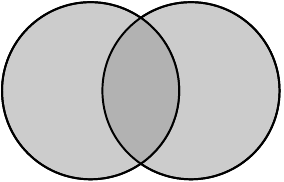}}
\cell{5}{25}{c}{$H(X)$}
\cell{49}{25}{c}{$H(Y)$}
\cell{14}{17}{c}{$\condent{X}{Y}$}
\cell{27}{17}{c}{$I(X; Y)$}
\cell{40}{17}{c}{$\condent{Y}{X}$}
\cell{27}{0.5}{b}{$H(X, Y)$}
\cell{27}{-5}{b}{(b)}
\end{picture}
\caption{The metacommunity gamma-diversities, shown in~\hardref{(a)}, are
  the exponentials of the entropies shown in~\hardref{(b)}.  For instance,
  $\ovln{A}(P) = \exp \condent{X}{Y}$.}  
\lbl{fig:mc-measures}
\end{figure}

\begin{table}
\centering
\begin{tabular}{lll@{\hspace{2.8mm}}l}
\hline
Quantity        &
Name            &
Formula &
Description     \\
\hline
$\exp H(X)$     &
$G(P)$          &
$\displaystyle\prod_i \Biggl( \frac{1}{p_i} \Biggr)^{p_i\vphantom{X^X_X}}$&
Effective no.\ of species in metacommunity,   \\[-1.9ex]
&
&
&
ignoring division into subcommunities   \\[1.7ex]
$\exp H(Y)$     &
$D(\vc{w})$     &
$\displaystyle\prod_j \Biggl( \frac{1}{w_j} \Biggr)^{w_j}$     &
Effective no.\ of subcommunities in meta-\\[-1.9ex]
&
&
&
community, ignoring division into species          \\[1.7ex]
$\exp H(X, Y)$  &
$A(P)$     &
$\displaystyle\prod_{i, j} \Biggl( \frac{1}{P_{ij}} \Biggr)^{P_{ij}}$  &
Effective no.\ of (species, subcommunity)       \\[-1.9ex]
&
&
&
pairs   \\[1.7ex]
$\exp \condent{X}{Y}$   &
$\ovln{A}(P)$      &
$\displaystyle\prod_{i, j} \Biggl( \frac{w_j}{P_{ij}} \Biggr)^{P_{ij}}$ &
Average effective no.\ of species per    \\[-1.9ex]
&
&
&
subcommunity        \\[1.7ex]
$\exp \condent{Y}{X}$   &
$R(P)$     &
$\displaystyle\prod_{i, j} \Biggl( \frac{p_i}{P_{ij}} \Biggr)^{P_{ij}}$ &
Redundancy of subcommunities \\[1.7ex]
$\exp I(X; Y)$  &
$\ovln{B}(P)$      &
$\displaystyle\prod_{i, j} \Biggl( \frac{P_{ij}}{p_i w_j} \Biggr)^{P_{ij}}$&
Effective no.\ of isolated subcommunities     \\
\hline
\end{tabular}
\index{effective number!species@of species}%
\index{effective number!subcommunities@of subcommunities}%
\caption{Formulas for and descriptions of the metacommunity diversity
  measures.  The first product is over the support of $\p$, the second is
  over the support of $\vc{w}$, and the others are over the support of
  $P$.}
\lbl{table:mcm-desc}
\end{table}

\begin{table}
\centering
\begin{tabular}{|l|l|l|l|}
\hline
Name            &
Range           &
Minimized when  &
Maximized when  \\
\hline
$G(P)$          &
$[1, S]\vphantom{X^{X^X}}$        &
Only one species in        &
Species in metacommunity    \\
&
&
metacommunity       &
are balanced        \\[1ex]
$D(\vc{w})$     &
$[1, N]$        &
Only one subcommunity in  &
Subcommunities are same \\
&
&
metacommunity        &
size    \\[1ex]
$A(P)$          &
$[1, SN]$       &
Only one species and one        &
Subcommunities are same     \\
&
&
subcommunity in&
size and all balanced        \\
&
&
metacommunity&
\\[1ex]
$\ovln{A}(P)$   &
$[1, S]$        &
Each subcommunity contains      &
Each subcommunity is  \\
&
&
only one species        &
balanced        \\[1ex]
$R(P)$          &
$[1, N]$        &
Subcommunities share no         &
Subcommunities have same    \\
&
&
species &
size and composition    \\[1ex]
$\ovln{B}(P)$   &
$[1, N]$        &
Subcommunities have same        &
Subcommunities have same   \\
&
&
composition     &
size and share no species    \\
\hline
\end{tabular}
\caption{Minimum and maximum values of the metacommunity diversity
  measures.  The bounds shown depend only on $S$ and $N$; tighter bounds
  are given in the text.
  \demph{Balanced}\index{balanced} means that all species have equal
  abundance.}  
\lbl{table:mcm-range}
\end{table}

\subsection*{Metacommunity gamma-diversity}

First consider the random variable $X$ for species.  The exponential of its
Shannon entropy $H(X)$ is 
\[
D(\p)
=
\prod_{i \in \supp(\p)} \Biggl( \frac{1}{p_i} \Biggr)^{p_i}.
\]
This is simply the diversity of order $1$ of the species distribution $\p$
of the whole metacommunity, ignoring its division into subcommunities.
(Throughout this section, all diversities are of order~$1$.)  We write
$G(P)\ntn{Gdiv} = D(\p)$ and call it the \demph{metacommunity
  gamma-diversity}.%
\index{gamma-diversity!metacommunity}%
\index{metacommunity!gamma-diversity}

The metacommunity gamma-diversity $G(P)$ is related to the subcommunity
gamma-diversities $\gamma_1(P), \ldots, \gamma_N(P)$ as follows:
\begin{align}
G(P)    &
=
\prod_{i, j \csuch i \in \supp(\p)}
\Biggl( \frac{1}{p_i} \Biggr)^{P_{ij}}  
\nonumber       \\
&
=
\prod_j \gamma_j(P)^{w_j}  
\nonumber       \\
&
=
M_0 \Bigl( \vc{w}, \bigl(\gamma_1(P), \ldots, \gamma_N(P)\bigr) \Bigr).
\lbl{eq:G-mean}
\end{align}
Here we have used the definition of $p_i$ as $\sum_j P_{ij}$ and the
formula~\eqref{eq:gammaj-exp} for $\gamma_j(P)$.  So, the metacommunity
gamma-diversity $G(P)$ is the geometric mean of the subcommunity
gamma-diversities $\gamma_j(P)$, weighted by the sizes of the
subcommunities. 

In this sense, the subcommunity gamma-diversity $\gamma_j(P)$ is the mean
contribution~\lbl{p:gamma-contrib} per individual of the $j$th subcommunity
to the metacommunity diversity.

The metacommunity gamma-diversity is constrained by the bounds
\[
1 \leq G(P) \leq S,
\]
by Lemma~\ref{lemma:div1-max-min}.  It attains the lower bound $G(P) = 1$
when the metacommunity consists of a single species, and the upper bound
$G(P) = S$ when all $S$ species have equal abundance in the metacommunity
(regardless of how they are distributed across the subcommunities).

Now consider the random variable $Y$ for subcommunities. The exponential of
$H(Y)$ is
\[
D(\vc{w}) 
=
\prod_{j \in \supp(\vc{w})} \Biggl( \frac{1}{w_j} \Biggr)^{w_j}.
\]
It measures how evenly the population is distributed across the
subcommunities (regardless of how it is distributed across species).  It is
bounded by
\[
1 \leq D(\vc{w}) \leq N
\]
(by Lemma~\ref{lemma:div1-max-min} again), with $D(\vc{w}) = 1$ when the
metacommunity contains only one nonempty subcommunity and $D(\vc{w}) =
N$ when the populations of the $N$ subcommunities are of equal size.

\subsection*{Metacommunity alpha-diversities}

The joint%
\index{joint entropy}
entropy $H(X, Y)$ has exponential
\[
D(P)
=
\prod_{(i, j) \in \supp(P)} 
\Biggl( \frac{1}{P_{ij}} \Biggr)^{P_{ij}}.
\]
Here we are treating the $S \times N$ matrix $P$ as a probability
distribution on the set $\{1, \ldots, S\} \times \{1, \ldots, N\}$.  So,
$D(P)$ is the effective number of (species, subcommunity) pairs, in the
sense of Section~\ref{sec:ent-div}.  It is the species
diversity that the metacommunity would have if individuals in different
subcommunities were decreed to be of different species (as in the island%
\index{islands!subcommunities@as subcommunities}
scenario).  We write $A(P)\ntn{Adiv} = D(P)$ and call it the \demph{raw
  metacommunity alpha-diversity}.%
\index{alpha-diversity!raw metacommunity}%
\index{metacommunity!alpha-diversity, raw}

Since $A(P)$ measures diversity as if no species were shared between
subcommunities, it overestimates the true diversity.  Indeed, taking
exponentials in the inequalities
\[
H(X) \leq H(X, Y) \leq H(X) + H(Y)
\]
of Proposition~\ref{propn:pair-bounds}\bref{part:joint-bounds} gives
\begin{equation}
\lbl{eq:A-bounds}
G(P) \leq A(P) \leq D(\vc{w}) G(P).
\end{equation}
The upper bound states that the factor of overestimation is at most
$D(\vc{w})$, the effective number of subcommunities. 

The minimum value $A(P) = G(P)$ occurs when $H(X, Y) = H(X)$, which by
Proposition~\ref{propn:pair-bounds}\bref{part:cond-bounds} means that $Y$
is determined by $X$: the subcommunity is determined by the species.  So,
$A(P) = G(P)$ when no species are shared between subcommunities.  

The maximum value $A(P) = D(\vc{w})G(P)$ occurs when $H(X, Y) = H(X) +
H(Y)$.  By Proposition~\ref{propn:pair-bounds}\bref{part:joint-bounds},
this is true just when $X$ and $Y$ are independent.  Equivalently, $A(P)$
attains its maximum when the metacommunity is \lbl{p:wm}\dmph{well-mixed},
meaning that each of the subcommunity species distributions $P_{\Pdot
  j}/w_j$ is equal to the metacommunity species distribution $\p$.

In summary: $A(P)$ does not overestimate $G(P)$ at all when the
subcommunities share no species, whereas the overestimation is most
pronounced when all of the subcommunities have identical composition.

Since $1 \leq G(P) \leq S$ and $1 \leq D(\vc{w}) \leq N$, the
inequalities~\eqref{eq:A-bounds} imply the cruder bounds
\[
1 \leq A(P) \leq SN.  
\]
This conforms with the interpretation of $A(P)$ as the effective number
of (species, subcommunity) pairs.  The minimum $A(P) = 1$ is attained when
there is only one species present and only one nonempty subcommunity.  The
maximum $A(P) = SN$ is attained when the metacommunity is well-mixed and the
subcommunities all have the same size.  

Next consider conditional%
\index{conditional entropy}
entropy.  By Lemma~\ref{lemma:cond-alts}\bref{part:cond-alts-exp}, the
conditional entropy $\condent{X}{Y}$ is given by
\[
\condent{X}{Y}
=
\sum_{i, j \csuch \Pr(i, j) > 0} 
\Pr(i, j) \log \frac{\Pr(j)}{\Pr(i, j)}.
\]
The \demph{normalized%
\index{alpha-diversity!normalized metacommunity}%
\index{metacommunity!alpha-diversity, normalized}
metacommunity alpha-diversity} $\ovln{A}(P)$ is its
exponential: 
\begin{equation}
\lbl{eq:Abar-defn}
\ovln{A}(P) 
=
\exp\condent{X}{Y}
=
\prod_{i, j \csuch P_{ij} > 0} 
\Biggl( \frac{w_j}{P_{ij}} \Biggr)^{P_{ij}}.
\end{equation}
To understand $\ovln{A}$, we use one of the other formulas for conditional
entropy:
\begin{equation}
\lbl{eq:Abar-cond}
\condent{X}{Y}
=
\sum_{j \csuch \Pr(j) > 0} \Pr(j) H(X \given j)
\end{equation}
(Lemma~\ref{lemma:cond-alts}\bref{part:cond-alts-cond}).  The random
variable $X \given j$ is the species of a random individual from the $j$th
subcommunity, so taking exponentials throughout
equation~\eqref{eq:Abar-cond} gives
\begin{align}
\ovln{A}(P)        &
=
\prod_{j \csuch w_j > 0} \ovln{\alpha}_j(P)^{w_j}  
\nonumber       \\
&
=
M_0 \Bigl( 
\vc{w}, \bigl(\ovln{\alpha}_1(P), \ldots, \ovln{\alpha}_N(P)\bigr) 
\Bigr).
\lbl{eq:A-mean}
\end{align}
Hence $\ovln{A}(P)$ is the geometric mean of the individual subcommunity
diversities $\ovln{\alpha}_j(P)$, weighted by their sizes.  It is therefore
an alpha-diversity in the traditional sense (Remark~\ref{rmk:abg-rel} and
p.~\pageref{p:alpha-ave}). 

To find the maximum and minimum values of $\ovln{A}$, we take exponentials
throughout the inequalities
\[
0 \leq \condent{X}{Y} \leq H(X)
\]
(Proposition~\ref{propn:pair-bounds}\bref{part:cond-bounds}).  This gives
\begin{equation}
\lbl{eq:Abar-bounds}
1 \leq \ovln{A}(P) \leq G(P),
\end{equation}
with $\ovln{A}(P) = 1$ when each subcommunity contains at most one species
and $\ovln{A}(P) = G(P)$ when the metacommunity is well-mixed.  Since $G(P)
\leq S$, we also have the cruder bounds
\[
1 \leq \ovln{A}(P) \leq S,
\]
with $\ovln{A}(P) = S$ when each subcommunity contains all $S$ species in
equal proportions.

The raw%
\index{alpha-diversity!raw metacommunity}%
\index{metacommunity!alpha-diversity, raw}
and normalized%
\index{alpha-diversity!normalized metacommunity}%
\index{metacommunity!alpha-diversity, normalized}
metacommunity alpha-diversities, $A(P)$ and $\ovln{A}(P)$, are linked by
the equation
\begin{equation}
\lbl{eq:AAw}
\ovln{A}(P) = A(P)/D(\vc{w}),
\end{equation}
which is the exponential of the definition
\[
\condent{X}{Y} = H(X, Y) - H(Y)
\]
of conditional entropy. 

\begin{examples}
\lbl{egs:mc-A}
Here we take the four running examples of Section~\ref{sec:mm-info} and
translate them into ecological terms.  The results below follow immediately
from those examples, and are summarized in the first few rows of
Table~\ref{table:mm-egs}.

\begin{table}
\centering
\begin{tabular}{llllll}
\hline 
Quantity        &Name   &
\bref{eg:mc-A-ind}      &
\bref{eg:mc-A-one}      &
\bref{eg:mc-A-equal}    &
\bref{eg:mc-A-det}      \\
        &       &
Well-mixed      &
Only one        &
Subcomms        &
Isolated        \\
                &       &
metacomm   &
subcomm    &
are species     &
subcomms  \\[0ex]
\hline
$\exp H(X)\vphantom{X^{X^X}}$     &
$G(P)$          &
$D(\p)$         &
$D(\p)$         &
$D(\p)$         &
$D(\p)$         \\[1ex]
$\exp H(Y)$     &
$D(\vc{w})$     &
$D(\vc{w})$     &
$1$             &
$D(\p)$         &
$D(\vc{w})$     \\[1ex]
$\exp H(X, Y)$  &
$A(P)$          &
$D(\p)D(\vc{w})$&
$D(\p)$         &
$D(\p)$         &
$D(\vc{w})\ovln{A}(P)$\\[1ex]
$\exp\condent{X}{Y}$&
$\ovln{A}(P)$   &
$D(\p)$         &
$D(\p)$         &
$1$             &
$\ovln{A}(P)$   \\[1ex]
$\exp\condent{Y}{X}$&
$R(P)$          &
$D(\vc{w})$     &
$1$             &
$1$             &
$1$             \\[1ex]
$\exp I(X; Y)$  &
$\ovln{B}(P)$   &
$1$             &
$1$             &
$D(\p)$         &
$D(\vc{w})$     \\
\hline
\end{tabular}
\caption{Summary of Examples~\ref{egs:mc-A} and~\ref{egs:mc-B}.}
\lbl{table:mm-egs}
\end{table}

\begin{enumerate}
\item
\lbl{eg:mc-A-ind}
Suppose that the metacommunity is well-mixed.
Then $P = \p \otimes \vc{w}$ and $A(P) = D(P) = D(\p)D(\vc{w})$.  Each
subcommunity has the same species composition as the metacommunity, so the
mean subcommunity diversity $\ovln{A}(P)$ is the same as the metacommunity
diversity $G(P)$.  

\item
\lbl{eg:mc-A-one}
Suppose that the metacommunity consists of a single subcommunity.  Then $N
= 1$, $\vc{w} = (1)$, and $P = \p$.  The effective number $A(P)$ of
(species, subcommunity) pairs is just the effective number $D(\p)$ of
species, and since there is only one subcommunity, the average subcommunity
diversity $\ovln{A}(P)$ is also $D(\p)$.

\item
\lbl{eg:mc-A-equal}
Suppose that the subcommunities are exactly the species.  Thus, $N = S$,
$\vc{w} = \p$, and $P$ is the diagonal matrix with entries $p_1, \ldots,
p_S$.  The effective number $A(P)$ of (species, subcommunity) pairs is again
just $D(\p)$, but since each subcommunity has a diversity of $1$, the
average subcommunity diversity $\ovln{A}(P)$ is now $1$.  

\item
\lbl{eg:mc-A-det}
Finally, suppose that all subcommunities are isolated (share no species).
Nothing special can be said about $\ovln{A}(P)$, the mean subcommunity
diversity, since it is unaffected by the degree of overlap of species
between subcommunities.  As always, $A(P) = D(\vc{w})\ovln{A}(P)$.
\end{enumerate}
\end{examples}

\subsection*{The redundancy of a metacommunity}

We have already considered one conditional%
\index{conditional entropy}
entropy, $\condent{X}{Y}$.  The other,
\[
\condent{Y}{X}
=
\sum_{i, j \csuch \Pr(i, j) > 0} 
\Pr(i, j) \log \frac{\Pr(i)}{\Pr(i, j)},
\]
has exponential
\begin{equation}
\lbl{eq:R-defn}
R(P) = 
\prod_{i, j \csuch P_{ij} > 0} 
\Biggl( \frac{p_i}{P_{ij}} \Biggr)^{P_{ij}}.
\end{equation}
This is the \dmph{redundancy} of the metacommunity.  The word is meant in
the following sense: if some subcommunities were to be destroyed,%
\index{subcommunity!loss}
how much of the diversity in the metacommunity would be preserved?  High
redundancy means that there is enough repetition of species across
subcommunities that loss of some subcommunities would probably not
cause great loss of diversity.

We now justify this interpretation.  For each species $i$, consider
the relative abundances $P_{i1}, \ldots, P_{iN}$ of that species within the
$N$ subcommunities, and normalize to obtain a probability distribution
\[
P_{i\,\Pdot}/p_i
=
(P_{i1}/p_i, \ldots, P_{iN}/p_i)
\]
on the set $\{1, \ldots, N\}$ of subcommunities.  (We assume in this
explanation that $p_i > 0$.)  Then $D(P_{i\,\Pdot}/p_i)$ measures the
extent to which the $i$th species is spread evenly across the
subcommunities; for instance, it takes its maximum value $N$ when there is
the same amount of species $i$ in every subcommunity.  This is the
`redundancy' of the $i$th species.  

To obtain a measure of the redundancy of the whole metacommunity, we take
the geometric mean
\begin{equation}
\lbl{eq:R-mean}
\prod_{i \in \supp(\p)} D(P_{i\,\Pdot}/p_i)^{p_i}
\end{equation}
of the redundancies of the species, weighted by their relative abundances.
But this is exactly $R(P)$, since, using
Lemma~\ref{lemma:cond-alts}\bref{part:cond-alts-cond}, 
\begin{align*}
\prod_{i \in \supp(\p)} D(P_{i\,\Pdot}/p_i)^{p_i}       &
=
\exp \Biggl( \sum_{i \in \supp(\p)} p_i H(P_{i\,\Pdot}/p_i) \Biggr) \\
&
=
\exp \Biggl( \sum_{i \csuch \Pr(i) > 0} \Pr(i) \condent{Y}{i} \Biggr)\\
&
=
\exp \condent{Y}{X}     \\
&
=
R(P).
\end{align*}
In conclusion, $R(P)$ is the average species redundancy~\eqref{eq:R-mean}:
the effective number of subcommunities across which a typical species is
spread.

A different way to understand redundancy is through the equation
\begin{equation}
\lbl{eq:GAR}
R = A/G,
\end{equation}
which is the exponential of the definition 
\[
\condent{Y}{X} = H(X, Y) - H(X)
\]
of conditional entropy.  The gamma-diversity $G(P)$ is the effective number
of species in the metacommunity, whereas $A(P)$ is the effective number of
species if we pretend that individuals in different subcommunities are
always of different species.  The amount by which $A(P)$ overestimates
$G(P)$ reflects the extent to which species are, in reality, shared across
subcommunities: thus, it measures redundancy.  

Bounds on the redundancy can be obtained by dividing
inequalities~\eqref{eq:A-bounds} by $G(P)$.  This gives
\[
1 \leq R(P) \leq D(\vc{w}),
\]
with the same extremal cases as for~\eqref{eq:A-bounds}: redundancy takes
its minimum value of $1$ when no species are shared between
subcommunities, and its maximum value of $D(\vc{w})$ when the species
distributions in the subcommunities are all the same.  It follows that
\[
1 \leq R(P) \leq N,
\]
with $R(P) = N$ when the subcommunities have not only the same composition,
but also the same size.

\subsection*{Metacommunity beta-diversity}

Finally, consider the mutual%
\index{mutual information} 
information $I(X; Y)$.  By
Lemma~\ref{lemma:mut-alts}\bref{part:mut-alts-exp}, its exponential
$\ovln{B}(P)$ is given by
\[
\ovln{B}(P)
=
\prod_{i, j \csuch P_{ij} > 0}
\Biggl( \frac{P_{ij}}{p_i w_j} \Biggr)^{P_{ij}}.
\ntn{Bbardiv}
\]
This is the \demph{metacommunity%
\index{beta-diversity!metacommunity}%
\index{metacommunity!beta-diversity}
beta-diversity}.  (In Reeve et al.~\cite{HPD}, it is called the
`normalized' beta-diversity, and there is also a `raw' beta-diversity
$B(P)$, not treated here.)  By the discussion of mutual information in
Section~\ref{sec:mm-info}, $\ovln{B}(P)$ can be understood as the
exponential of the amount of information that knowledge of an individual's
species gives us about its subcommunity~-- or equivalently, vice versa.

Loosely, then, $\ovln{B}(P)$ measures the alignment between subcommunity
structure and species structure.  It is a beta-diversity in the traditional
sense of Remark~\ref{rmk:abg-rel} and p.~\pageref{p:beta-bet}.

By Proposition~\ref{propn:pair-bounds}\bref{part:mut-bounds} on mutual
information,
\begin{equation}
\lbl{eq:Bbar-bounds}
1 \leq \ovln{B}(P) \leq \min \{G(P), D(\vc{w})\}.  
\end{equation}
The minimum $\ovln{B}(P) = 1$ is attained when $X$ and $Y$ are independent,
that is, the metacommunity is well-mixed.  In that case, knowing an
individual's species does not help us to guess its subcommunity,
nor vice versa.  
By Proposition~\ref{propn:pair-bounds}\bref{part:mut-bounds}, there are two
cases in which the maximum is attained.  One is where $X$ is determined by
$Y$, that is, there is at most one species in each subcommunity.  Then by
Example~\ref{egs:cond}\bref{eg:cond-det},
%
\[
\ovln{B}(P) = G(P) \leq D(\vc{w}).
\]
%
In this case, knowing the subcommunity to which an individual belongs
enables us to infer its species with certainty.  The other case in which
the maximum is attained is where $Y$ is determined by $X$, that is, the
subcommunities are isolated.  Then
\[
\ovln{B}(P) = D(\vc{w}) \leq G(P)
\]
by Example~\ref{egs:cond}\bref{eg:cond-det} again, and knowing an
individual's species enables us to infer its subcommunity with certainty.

We can also interpret $\ovln{B}$ as the effective%
\index{effective number!subcommunities@of subcommunities}%
\index{subcommunity!effective number of isolated}
number of isolated subcommunities.  Indeed, since $1 \leq D(\vc{w}) \leq
N$, the inequalities~\eqref{eq:Bbar-bounds} imply that
\begin{equation}
\lbl{eq:Bbar-crude}
1 \leq \ovln{B}(P) \leq N.
\end{equation}
The maximum $\ovln{B}(P) = N$ occurs when the $N$ subcommunities are
isolated and of equal size.  We will see in
Proposition~\ref{propn:mc-chain} and Corollary~\ref{cor:mc-rep} that
$\ovln{B}$ satisfies a chain rule and a replication principle, supporting
the effective number interpretation.

For yet another viewpoint on $\ovln{B}$, recall from
equations~\eqref{eq:mut-alt} that
\[
\condent{X}{Y} + I(X; Y) = H(X).
\]
Taking exponentials throughout gives
\begin{equation}
\lbl{eq:ABG}
\ovln{A} \, \ovln{B} = G,
\end{equation}
that is,
\[
\text{alpha-diversity} \times \text{beta-diversity}
=
\text{gamma-diversity}.
\]
This equation partitions%
\index{partitioning of diversity}%
\index{diversity!partitioning of}
the diversity of the metacommunity (gamma) into two components: the average
diversity within subcommunities (alpha) and the variation between
subcommunities (beta).  The general principle has a long history, going
back to the foundational work of Whittaker%
\index{Whittaker, Robert} 
(p.~321 of~\cite{WhitVSM} and p.~232 of~\cite{WhitEMS}).

From equations~\eqref{eq:ABG}, \eqref{eq:AAw} and~\eqref{eq:GAR}, it
follows that 
\[
\ovln{B}(P)
=
\frac{G(P)}{\ovln{A}(P)}
=
\frac{G(P)D(\vc{w})}{A(P)} 
=
\frac{D(\vc{w})}{R(P)}, 
\]
giving
\[
\ovln{B}(P) = \frac{D(\vc{w})}{R(P)}.
\]
So when the subcommunity sizes $\vc{w}$ are fixed, the effective number
$\ovln{B}(P)$ of isolated subcommunities is inversely proportional to the
redundancy $R(P)$.  This is reasonable: $R(P)$ measures overlap of species
between subcommunities, whereas $\ovln{B}(P)$ measures how disjoint the 
subcommunities are.

(Matters are more subtle outside the case $q = 1$, $Z = I$ to which we have
confined ourselves.  When $q \neq 1$ in the work of Reeve et al., the
dependency between $\ovln{B}$ and $R$ breaks down, in the strong sense that
the two quantities no longer determine one another; they convey
different information.)

\begin{examples}
\lbl{egs:mc-B}
We return to the four scenarios of Examples~\ref{egs:mc-A}, finding the
redundancy\index{redundancy} $R(P)$ and metacommunity beta-diversity
$\ovln{B}(P)$.  The results are summarized in Table~\ref{table:mm-egs}.
\begin{enumerate}
\item
\lbl{eg:mc-B-ind}
A well-mixed metacommunity is maximally redundant ($R(P) = D(\vc{w})$),
since all subcommunities are identical.  For the same reason, the effective
number $\ovln{B}(P)$ of isolated subcommunities is just $1$.

\item
\lbl{eg:mc-B-one}
If the metacommunity consists of a single subcommunity ($N = 1$) then the
redundancy $R(P)$ and effective number $\ovln{B}(P)$ of isolated
subcommunities both take their minimum values of $1$.  

\item
\lbl{eg:mc-B-equal}
Suppose that the subcommunities are exactly the species.  Then 
the metacommunity is minimally redundant ($R(P) = 1$), reflecting the fact
that each species is present in just one subcommunity: losing any of the
subcommunities means losing a species.  And since subcommunities are
species, the effective number $\ovln{B}(P)$ of isolated subcommunities is the
effective number $D(\p)$ of species.  

\item
\lbl{eg:mc-B-det}
More generally, suppose that all subcommunities are isolated.  The
redundancy is minimal ($R(P) = 1$), since no species is repeated across
subcommunities.  The effective number $\ovln{B}(P)$ of isolated
subcommunities is simply the diversity $D(\vc{w})$ of the subcommunity
distribution $\vc{w}$, which is reasonable since, in fact, the
subcommunities \emph{are} isolated.
\end{enumerate}
\end{examples}

Just as the metacommunity gamma-diversity $G(P)$ is the geometric mean of
the subcommunity gamma-diversities $\gamma_j(P)$, and just as the
metacommunity alpha-diversity $\ovln{A}(P)$ is the geometric mean of the
subcommunity alpha-diversities $\ovln{\alpha}_j(P)$, the metacommunity
beta-diversity $\ovln{B}(P)$ is the geometric mean of the subcommunity
beta-diversities $\ovln{\beta}_j(P)$.  Indeed, recall from
Lemma~\ref{lemma:mut-alts}\bref{part:mut-alts-mean-rel} that
\[
I(X; Y)
=
\sum_{j \csuch \Pr(j) > 0} \Pr(j) \relEnt{(X \given j)}{X}.
\]
Taking exponentials throughout gives
\begin{align}
\ovln{B}(P)        &
=
\prod_{j \csuch w_j > 0} \ovln{\beta}_j(P)^{w_j}   
\nonumber       \\
&
=
M_0 \Bigl( \vc{w}, \bigl(\ovln{\beta}_1(P), \ldots, \ovln{\beta}_N(P)\bigr) 
\Bigr),
\lbl{eq:B-mean}
\end{align}
as claimed.

We have seen that $\ovln{\beta}_j(P)$ measures how unusual the $j$th
subcommunity is in the context of the metacommunity.  Taking the geometric
mean over all subcommunities gives $\ovln{B}(P)$, which is therefore an
overall measure of the atypicality or isolation of the subcommunities
within the metacommunity.

Further connections between beta-diversity and information-theoretic
quantities are described in the first appendix of the paper~\cite{HPD} of
Reeve et al.

\section{Properties of the metacommunity measures}
\lbl{sec:mm-props}
\index{metacommunity!diversity measures}
\index{diversity measure!metacommunity}

In this chapter so far, we have introduced a system of measures of the
diversity and structure of a metacommunity, and explained their behaviour
in a variety of hypothetical examples.  But just as a measure of the
diversity of a \emph{single} community should not be accepted or used until
it can be shown to behave logically (Section~\ref{sec:ent-div}), the
metacommunity measures should also be required to have sensible logical and
algebraic properties.  Here we show that they do.

\subsection*{Independence}

We begin by showing that the alpha-diversity $\ovln{A}$ and beta-diversity
$\ovln{B}$ are independent~-- not in the sense of probability theory, but in
the sense of \emph{certain} knowledge.  An informal example illustrates the
idea.  Assume for simplicity that every person in the world is
either dark-haired%
\index{hair colour} 
or fair-haired, and either brown-eyed%
\index{eye colour}
or blue-eyed.  These two variables, hair colour and eye colour, are not
independent in the probabilistic sense: people with dark hair are more
likely to have dark eyes.  However, they are independent in a weaker sense:
knowing an individual's hair colour gives no \emph{certain} knowledge of
their eye colour, nor vice versa.  All four combinations occur.

The formal definition is as follows.

\begin{defn}
Let $J$, $K$ and $L$ be sets.  Functions
\[
\xymatrix@R-4ex{
        &K      \\
J \ar[ru]^\kappa \ar[rd]_\lambda   &       \\
        &L
}
\]
are \demph{independent}\index{independent!functions} if for all $k \in
\kappa J$ and $\ell \in \lambda J$, there exists $j \in J$ such that
$\kappa(j) = k$ and $\lambda(j) = \ell$.
\end{defn}

For $\kappa$ and $\lambda$ to be independent means that if I choose in
secret an element $j$ of $J$, and tell you the value of $\kappa(j) \in K$,
you gain no certain information about the value of $\lambda(j)$.  (For by
definition of independence, the image under $\lambda$ of the fibre
$\kappa^{-1}\{\kappa(j)\}$ is no smaller than the whole image $\lambda J$.)
Of course, the same is also true with the roles of $\kappa$ and $\lambda$
reversed.  In the informal example above, $J$ is the set of all people, $K
= \{ \text{dark hair}, \text{fair hair} \}$, and $L = \{ \text{brown eyes},
\text{blue eyes} \}$.

We have discussed the general goal of decomposing any measure of
metacommunity diversity (any `gamma-diversity') into within-group (alpha) and
between-group (beta) components.  The alpha-diversity and beta-diversity
should be independent, otherwise the word `decomposition' is not deserved:
certain values of alpha would exclude certain values of beta, and vice
versa.  This requirement has been recognized in ecology since at least the
1984 work of Wilson and Shmida~\cite{WiSh}.  As Jost put it:
\begin{quote}
\index{Jost, Lou}
Since [alpha- and beta-diversity] measure completely different aspects of
regional diversity, they must be free to vary independently; alpha should
not put mathematical constraints on the possible values of beta, and vice
versa.  If beta depended on alpha, it would be impossible to compare beta
diversities of regions whose alpha diversities differed.
\end{quote}
(\cite{JostPDI}, p.~2428.)

We will show that the decomposition
\[
\ovln{A} \, \ovln{B} = G
\]
(equation~\eqref{eq:ABG}) passes this test.  Since the number $N$ of
subcommunities is usually known in advance, but the number $S$ of species
may not be, we interpret independence as meaning that for each $N \geq 1$, the
functions
\[
\xymatrix@R-4ex{
        &\R     \\
\coprod\limits_{S \geq 1} \Delta_{SN} 
\ar[ru]^-{\ovln{A}} \ar[rd]_-{\ovln{B}}   &       \\
        &\R
}
\]
are independent.  Here $\Delta_{SN}$ is understood as the set of $S
\times N$ matrices $P$ of nonnegative reals summing to $1$, so that the
disjoint union $\coprod_{S \geq 1} \Delta_{SN}$ is the set of all such
matrices $P$ with $N$ columns and any number of rows.

Independence of these two functions means that given a metacommunity
divided into a known number $N$ of subcommunities, knowledge of the mean
diversity $\ovln{A}(P)$ of the subcommunities does not restrict the range
of possible values of $\ovln{B}(P)$, the effective number of isolated
subcommunities.  Thus, $\ovln{B}(P)$ can still take all the values that it
could have taken had we not known $\ovln{A}(P)$.  Equivalently,
independence means that knowing the value of $\ovln{B}(P)$ does not enable
us to deduce anything about $\ovln{A}(P)$.  We prove this now.

\begin{propn}[Independence of alpha- and beta-diversities]
\index{alpha-diversity!independent of beta-diversity}%
\index{beta-diversity!independent of alpha-diversity}%
\index{independence!alpha- and beta-diversities@of alpha- and beta-diversities}
For each $N \geq 1$, the functions
\[
\xymatrix@C+2em{
\coprod\limits_{S \geq 1} \Delta_{SN} 
\ar@<1ex>[r]^-{\ovln{A}}
\ar@<-1ex>[r]_-{\ovln{B}}       &
\R
}
\]
are independent.
\end{propn}

\begin{proof}
We have already shown that $\ovln{A}(P) \geq 1$ and $1 \leq \ovln{B}(P)
\leq N$ for all $P \in \coprod_{S \geq 1} \Delta_{SN}$
(inequalities~\eqref{eq:Abar-bounds} and~\eqref{eq:Bbar-crude}).  So it
suffices to show that given any $a \in [1, \infty)$ and $b \in [1, N]$,
  there exist some $S \geq 1$ and some $S \times N$ matrix $P \in
  \Delta_{SN}$ such that $\ovln{A}(P) = a$ and $\ovln{B}(P) = b$.

One way to do this is as follows.  Choose an integer $T \geq a$.  The
diversity measure $D \from \Delta_T \to \R$ is continuous
(Lemma~\ref{lemma:div1-cts}) with minimum $1$ and maximum $T$
(Lemma~\ref{lemma:div1-max-min}), so we can choose some $\vc{t} \in
\Delta_T$ such that $D(\vc{t}) = a$.  Similarly, we can choose some $\vc{w}
\in \Delta_N$ such that $D(\vc{w}) = b$.

Now consider a metacommunity made up of $N$ subcommunities of relative
sizes $w_1, \ldots, w_N$, with no shared species, where each subcommunity
contains $T$ species in proportions $t_1, \ldots, t_T$.  Thus, there are
$TN$ species in all, and 
\[
P
=
\begin{pmatrix}
w_1 t_1 &0      &       &       &       &0      \\
\vdots  &\vdots &       &\cdots &       &\vdots \\
w_1 t_T &0      &       &       &       &0      \\
0       &w_2 t_1&       &       &       &0      \\
\vdots  &\vdots &       &\cdots &       &\vdots \\
0       &w_2 t_T&       &       &       &0      \\
        &       &       &       &       &       \\
\vdots  &\vdots &       &       &       &\vdots \\
        &       &       &       &       &       \\
0       &0      &       &       &       &w_N t_1\\
\vdots  &\vdots &       &\cdots &       &\vdots \\
0       &0      &       &       &       &w_N t_T
\end{pmatrix}.
\]
The species distribution $P_{\Pdot j}/w_j$ in the $j$th subcommunity is 
\[
(0, \ldots, 0, t_1, \ldots, t_T, 0, \ldots, 0),
\]
so its diversity $\ovln{\alpha}_j(P)$ is $D(\vc{t}) = a$.  But
$\ovln{A}(P)$ is an average of $\ovln{\alpha}_1(P), \ldots,
\ovln{\alpha}_N(P)$ (equation~\eqref{eq:A-mean}), so $\ovln{A}(P) = a$.
Moreover, the subcommunities are isolated, so by
Example~\ref{egs:mc-B}\bref{eg:mc-B-det}, $\ovln{B}(P) = D(\vc{w}) =
b$.
\end{proof}

In the same sense, the average subcommunity diversity $\ovln{A}$ and
the redundancy $R$ are independent:

\begin{propn}[Independence of alpha-diversity and redundancy]
\index{alpha-diversity!independent of redundancy}%
\index{redundancy!independent of alpha-diversity}%
\index{independence!alpha-diversity and redundancy@of alpha-diversity and redundancy}
For each $N \geq 1$, the functions
\[
\xymatrix@C+2em{
\coprod\limits_{S \geq 1} \Delta_{SN} 
\ar@<1ex>[r]^-{\ovln{A}}
\ar@<-1ex>[r]_-{R}       &
\R
}
\]
are independent.
\end{propn}

\begin{proof}
The proof is similar to that of the last proposition.  We have already seen
that $\ovln{A}(P) \in [1, \infty)$ and $R(P) \in [1, N]$ for all $P \in
  \coprod_{S \geq 1} \Delta_{SN}$.  So it suffices to show that given $a
  \in [1, \infty)$ and $r \in [1, N]$, there exist an integer $S \geq 1$
  and an $S \times N$ matrix $P \in \Delta_{SN}$ such that $\ovln{A}(P) =
  a$ and $R(P) = r$.

To prove this, choose an integer $S \geq a$ and distributions $\p \in
\Delta_S, \vc{w} \in \Delta_N$ such that $D(\p) = a$ and $D(\vc{w}) = r$.
Consider a well-mixed metacommunity made up of $N$ subcommunities of
relative sizes $w_1, \ldots, w_N$, each with the same $S$ species in
proportions $p_1, \ldots, p_S$.  Thus, $P = \vc{p} \otimes \vc{w}$.  
By Examples~\ref{egs:mc-A}\bref{eg:mc-A-ind}
and~\ref{egs:mc-B}\bref{eg:mc-B-ind}, $\ovln{A}(P) = D(\p) = a$ and $R(P) =
D(\vc{w}) = r$.
\end{proof}

\subsection*{Identical subcommunities}
\index{identical subcommunities}
\index{subcommunity!identical}

When we were analysing the diversity of a single community, we argued that
any similarity-sensitive diversity measure should be unchanged if a species
is reclassified into two identical smaller species, and we proved that the
diversity measures $D_q^Z$ do indeed enjoy this property
(Lemma~\ref{lemma:dqz-id}, Example~\ref{eg:reclassification}, and the text
afterwards).  It follows by continuity that if a species is divided
into two nearly identical parts, the diversity increases only slightly.
This is sensible behaviour, given that diversity is intended to measure the
effective number of completely dissimilar species (p.~\pageref{p:en-cd}).

The same principle applies to $\ovln{B}$, the effective%
\index{effective number!subcommunities@of subcommunities}%
\index{subcommunity!effective number of isolated}
number of isolated subcommunities in a metacommunity.  Dividing a
subcommunity into two smaller subcommunities of identical composition
should not change $\ovln{B}$.  In other words, the effective number of
isolated subcommunities should not be changed by the presence or absence of
boundaries between subcommunities that are identical.  The average
subcommunity diversity $\ovln{A}$ should be similarly unaffected.

In summary, then, the decomposition of global diversity into
within-subcommunity and between-subcommunity components,
\begin{equation}
\lbl{eq:ABG2}
\ovln{A} \, \ovln{B} = G,
\end{equation}
should be unaffected by arbitrary decisions about where subcommunity
boundaries lie.  This is best explained by example.

\begin{example}
Suppose that we are interested in the tree%
\index{tree!diversity}
diversity of a country that is divided into administrative districts with
no ecological significance.  Suppose further that in a particular pair of
neighbouring districts, the distributions of tree species are identical.
In that case, the partitioning~\eqref{eq:ABG2} of the overall diversity
into within- and between-district components should be the same as if the
neighbouring districts had been merged into one.  The effective number of
isolated subcommunities should be invariant under the removal or addition
of ecologically irrelevant boundaries.
\end{example}

\begin{example}
Suppose that we are studying the various species of grass\index{grass} on a
hillside.%
\index{hill}
To investigate the varying abundances of different species at different
altitudes, we divide the hillside into height bands ($0$--$10$m,
$10$--$20$m, etc.)\ and regard them as subcommunities.

The beta-diversity $\ovln{B}$ measures the effective number of
isolated or disjoint subcommunities, so if it turns out that the bottom two
height bands have the same species distribution, then $\ovln{B}$ should be
the same as if they were considered as a single band ($0$--$20$m).
\end{example}

In short, we require that the decomposition~\eqref{eq:ABG2} of
metacommunity diversity into alpha and beta components is ecologically
meaningful, not an artefact of the particular subcommunity division chosen.
As far as possible, the decomposition should be independent of
resolution\index{resolution} (that is, how fine or coarse the subcommunity
division may be).  In general, the finer the division one uses, the more
variation between subcommunities one will observe.  But if a subcommunity
is ecologically uniform (has the same species distribution throughout) then
dividing it further should make no difference to $\ovln{B}$ or $\ovln{A}$.

We now give a formal statement and proof of the desired invariance property
of $\ovln{A}$, $\ovln{B}$ and $G$.  To minimize notational overhead, we
consider splitting a single subcommunity into two rather than splitting
every subcommunity into an arbitrary number of smaller parts; but the
general case follows by induction.

In the standard notation of this chapter, take an $S \times N$ matrix $P
\in \Delta_{SN}$ with species distribution $\p \in \Delta_S$ and
subcommunity size distribution $\vc{w} \in \Delta_N$, so that $p_i = \sum_j
P_{ij}$ and $w_j = \sum_i P_{ij}$.  Split the last subcommunity into two
parts of relative sizes $t$ and $1 - t$ (where $0 \leq t \leq 1$), and
suppose that the two parts have the same species distribution.  Then the
new relative abundance matrix is the $S \times (N + 1)$ matrix $P'$ given
by
\[
P'_{ij}
=
\begin{cases}
P_{ij}  &\text{if } 1 \leq j \leq N - 1,\\
tP_{iN} &\text{if } j = N,      \\
(1 - t)P_{iN}   &\text{if } j = N + 1.
\end{cases}
\]

\begin{propn}[Identical subcommunities]
\index{subcommunity!identical}%
\index{identical subcommunities}%
\lbl{propn:shattering}
In the situation described,
\[
\ovln{A}(P') = \ovln{A}(P),
\qquad
\ovln{B}(P') = \ovln{B}(P),
\qquad
G(P') = G(P).
\]
\end{propn}

(In Reeve et al.~\cite{HPD}, the splitting of a subcommunity into smaller
parts with the same species distribution is called `shattering',%
\index{shattering}
so the result is that $\ovln{A}$, $\ovln{B}$ and $G$ are invariant under
shattering.)

The idea behind the proof is that $\ovln{A}$ is the average diversity of
the subcommunity of an individual chosen at random, and this quantity is
unchanged if a well-mixed subcommunity is split into smaller parts.

\begin{proof}
Write $\p' \in \Delta_S$ and $\vc{w}' \in
\Delta_{N + 1}$ for the row- and column-sums of $P'$.  Then for each $i \in
\{1, \ldots, S\}$, 
\[
p'_i 
=
\sum_{j = 1}^{N - 1} P_{ij} + tP_{iN} + (1 - t)P_{iN} 
=
\sum_{j = 1}^N P_{ij} 
=
p_i,
\]
so $\p' = \p$, and for each $j \in \{1, \ldots, N + 1\}$, 
\[
w'_j 
=
\sum_{i = 1}^S P'_{ij}
=
\begin{cases}
w_j             &\text{if } j \leq N - 1,       \\
tw_N            &\text{if } j = N,              \\
(1 - t)w_N      &\text{if } j = N + 1.
\end{cases}
\]

First, $G(P') = D(\p') = D(\p) = G(P)$, so $G(P') = G(P)$.  (This is also
clear informally, since the definition of metacommunity gamma-diversity $G$
does not refer to the division into subcommunities.)

Next, to calculate $\ovln{A}(P')$, consider the subcommunity
alpha-diversities $\ovln{\alpha}_j(P')$.  For each $j \in \{1, \ldots, N + 1\}$
such that $w'_j > 0$, the species distribution of subcommunity $j$ is
\[
P'_{\Pdot j}/w'_j
=
\begin{cases}
P_{\Pdot j}/w_j &\text{if } 1 \leq j \leq N - 1,        \\
P_{\Pdot N}/w_N &\text{if } j \in \{N, N + 1\},
\end{cases}
\]
giving
\[
\ovln{\alpha}_j(P')
=
\begin{cases}
\ovln{\alpha}_j(P)      &\text{if } 1 \leq j \leq N - 1,        \\
\ovln{\alpha}_N(P)      &\text{if } j \in \{N, N + 1\}.
\end{cases}
\]
Equation~\eqref{eq:A-mean} then gives
\begin{align}
\ovln{A}(P')    &
=
M_0\Bigl( \vc{w}', 
\bigl( 
\ovln{\alpha}_1(P'), \ldots, \ovln{\alpha}_N(P'), 
\ovln{\alpha}_{N + 1}(P')
\bigr)
\Bigr)  
\nonumber       \\
&
=
M_0\Bigl( 
\bigl( w_1, \ldots, w_{N - 1}, tw_N, (1 - t)w_N \bigr), 
\bigl( 
\ovln{\alpha}_1(P), \ldots, \ovln{\alpha}_N(P), \ovln{\alpha}_N(P)
\bigr)
\Bigr)  
\nonumber       \\
&
=
M_0\Bigl(
( w_1, \ldots, w_{N - 1}, w_N ),
\bigl( \ovln{\alpha}_1(P), \ldots, \ovln{\alpha}_{N - 1}(P), 
\ovln{\alpha}_N(P) \bigr)
\Bigr)
\lbl{eq:shattering-1} \\
&
=
\ovln{A}(P),
\nonumber
\end{align}
where in equation~\eqref{eq:shattering-1} we used the repetition property
of the power means (Lemma~\ref{lemma:pwr-mns-elem}). Hence $\ovln{A}(P') =
\ovln{A}(P)$. 

Finally, by equation~\eqref{eq:ABG2},
\[
\ovln{B}(P')
=
\frac{G(P')}{\ovln{A}(P')}
=
\frac{G(P)}{\ovln{A}(P)}
=
\ovln{B}(P),
\]
completing the proof.
\end{proof}

Although $\ovln{A}$, $\ovln{B}$ and $G$ have the invariance property just
established, the redundancy $R$ and raw metacommunity alpha-diversity $A$
do not:

\begin{example}
Consider a metacommunity consisting of a single subcommunity, as in
Examples~\ref{egs:mc-A}\bref{eg:mc-A-one}
and~\ref{egs:mc-B}\bref{eg:mc-B-one}.  Suppose that it is ecologically
homogeneous, and split it arbitrarily into new subcommunities of relative
sizes $w_1, \ldots, w_N$.  Then the species distributions in the new
subcommunities are identical.  As we see from columns~\bref{eg:mc-A-ind}
and~\bref{eg:mc-A-one} of Table~\ref{table:mm-egs}, the global diversity
$G$, average subcommunity diversity $\ovln{A}$ and effective number
$\ovln{B}$ of isolated subcommunities are the same before and after the
division.

On the other hand, the effective number $A$ of (species, subcommunity)
pairs is greater by a factor of $D(\vc{w})$ in the newly divided
metacommunity.  This is because $A$ counts individuals of the same species
but different subcommunities as being in different groups, and therefore
depends directly on the subcommunity divisions, however arbitrary they
may be.  The redundancy $R$ is also greater by a factor of $D(\vc{w})$ in
the divided metacommunity, because it too measures properties of the
subcommunity division (namely, the effective number of subcommunities
that a typical species is spread across).  So it is reasonable
that $A$ and $R$ increase when a subcommunity is split into smaller units,
even when that subcommunity is well-mixed.
\end{example}

\subsection*{Chain rule, modularity and replication principle}

For measures of the diversity of a single community, we have seen that the
most important algebraic properties are the chain rule and the principles
of modularity and replication.  Here we show that versions of these
properties also hold for the metacommunity measures.

Consider a group of islands,%
\index{islands!metacommunities@as metacommunities}
each divided into several regions (Figure~\ref{fig:mc-islands}).
\begin{figure}
\centering
\lengths\setlength{\unitlength}{.8mm}
\begin{picture}(120,79)(0,-4)
\cell{60}{40}{c}{\includegraphics[height=60\unitlength]{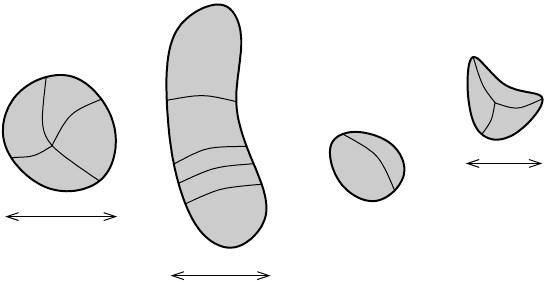}}
\cell{15}{58}{c}{metacommunity 1}
\cell{15}{21.5}{c}{$x_1$}
\cell{47}{73}{c}{metacommunity 2}
\cell{50}{9}{c}{$x_2$}
\cell{80}{46}{c}{$\cdots$}
\cell{80}{24}{c}{$\cdots$}
\cell{106}{62}{c}{metacommunity $m$}
\cell{110}{33}{c}{$x_m$}
\cell{60}{0}{c}{$\underbrace{\hspace*{118\unitlength}}_{\text{\large multicommunity}}$}
\end{picture}
\caption{Terminology for the chain rule, modularity principle and
  replication principle.  A multicommunity is divided into $m$
  metacommunities, with no species in common, of relative sizes $x_1,
  \ldots, x_m$.  Each metacommunity is further divided into subcommunities,
  which \emph{may} have species in common.  In the example shown, there are
  $m = 4$ metacommunities, divided into $N_1 = 4$, $N_2 = 5$, $N_3 = 2$ and
  $N_4 = 3$ subcommunities, respectively.  We may choose to ignore the
  metacommunity level and view the multicommunity as divided into $\sum_k
  N_k = 14$ subcommunities.}  \lbl{fig:mc-islands}
\end{figure}
Each island can be considered as a metacommunity, and has associated with
it all the metacommunity measures $A$, $\ovln{A}$, $R$, $\ovln{B}$, etc.,
discussed above.  On the other hand, the whole island group can be
considered as a metacommunity made up of regions, ignoring the intermediate
level of islands.  Can the redundancy of the whole island group be computed
from the redundancies and relative sizes of the individual islands?  If so,
how?  And the same questions can be asked for all the other metacommunity
measures.

To give a precise statement of the problem and its solution, we need some
notation and terminology.

We consider a \dmph{multicommunity} divided into $m$ metacommunities,
which have no species in common.  The $k$th metacommunity ($1 \leq k \leq
m$) is further divided into $N_k$ subcommunities and $S_k$ species; the
subcommunities of each metacommunity \emph{may} have species in common.
There are $N_1 + \cdots + N_m$ subcommunities and $S_1 + \cdots + S_m$
species in the multicommunity as a whole.  The relative sizes (that is,
relative population abundances) of the metacommunities are denoted by $x_1,
\ldots, x_m$, so that $\vc{x} = (x_1, \ldots, x_m) \in \Delta_m$.

Write $P^k$ for the relative abundance matrix of the $k$th metacommunity
divided into its subcommunities.  Thus, $P^k$ is an $S_k \times N_k$
matrix.  Write $\p^k \in \Delta_{S_k}$ for the relative abundance
distribution of species in the $k$th metacommunity, and $\vc{w}^k \in
\Delta_{N_k}$ for the relative sizes of its subcommunities.  Since the
metacommunities share no species, the relative abundance matrix
$\muc{P}$\ntn{Pstar} of the multicommunity (with respect to its division
into subcommunities, ignoring the metacommunity level) is the matrix block
sum
\begin{align}
\muc{P} &
=
x_1 P^1 \oplus \cdots \oplus x_m P^m
\lbl{eq:muc-dsum}
\\
&
=
\begin{pmatrix}
x_1 P^1 &0      &\cdots &0      \\
0       &x_2 P^2&\ddots &\vdots \\
\vdots  &\ddots &\ddots &0      \\
0       &\cdots &0      &x_m P^m
\end{pmatrix}.
\nonumber
\end{align}
The relative abundance distribution $\muc{\p}$ of species in the
multicommunity is given by
\[
\muc{\p} 
=
x_1 \p^1 \oplus \cdots \oplus x_m \p^m
=
\vc{x} \of (\p^1, \ldots, \p^m)
\in 
\Delta_{S_1 + \cdots + S_m},
\ntn{pstar}
\]
and the relative size distribution $\muc{\vc{w}}$ of the $N_1 + \cdots
+ N_k$ subcommunities in the multicommunity is given by
\[
\muc{\vc{w}}
=
x_1 \vc{w}^1 \oplus\cdots\oplus x_m \vc{w}^m
=
\vc{x} \of (\vc{w}^1, \ldots, \vc{w}^m).
\ntn{wstar}
\]
The main result is:

\begin{propn}[Chain rule]
\lbl{propn:mc-chain}%
\index{chain rule!metacommunity diversities@for metacommunity diversities}%
\index{metacommunity!chain rule}
With notation as above, 
\begin{align*}
G(\muc{P})      &
=
D(\vc{x}) \cdot \prod_k G(P^k)^{x_k},       \\
D(\muc{\vc{w}}) &
=
D(\vc{x}) \cdot \prod_k D(\vc{w}^k)^{x_k},  \\
A(\muc{P})      &
=
D(\vc{x}) \cdot \prod_k A(P^k)^{x_k},       \\
\ovln{A}(\muc{P})&
=
\prod_k \ovln{A}(P^k)^{x_k},    \\
R(\muc{P})      &
=
\prod_k R(P^k)^{x_k},   \\
\ovln{B}(\muc{P})       &
=
D(\vc{x}) \cdot \prod_k \ovln{B}(P^k)^{x_k},
\end{align*}
where the products are over all $k \in \{1, \ldots, m\}$ such that $x_k >
0$.  
\end{propn}

\begin{proof}
The statement on gamma-diversity is simply the chain rule for the
diversity of a single community (Corollary~\ref{cor:div1-chain}):
\begin{align*}
G(\muc{P})      &
=
D(\muc{\p})     \\
&
=
D\bigl(\vc{x} \of (\p^1, \ldots, \p^m)\bigr)    \\
&
=
D(\vc{x}) \cdot \prod_k D(\p^k)^{x_k}       \\
&
=
D(\vc{x}) \cdot \prod_k G(P^k)^{x_k}.
\end{align*}
The same argument gives the formula for $D(\muc{\vc{w}})$.  It also gives
the formula for $A(\muc{P})$, as follows.  When the matrices $\muc{P}$ and
$P^1, \ldots, P^m$ are regarded as finite probability distributions,
equation~\eqref{eq:muc-dsum} implies that $\muc{P}$ can be obtained from $\vc{x} \of
(P^1, \ldots, P^m)$ by permutation of its entries and insertion of
zeros.  By the symmetry and absence-invariance of $D$, it follows that
\[
D(\muc{P}) = D\bigl( \vc{x} \of (P^1, \ldots, P^m) \bigr).
\]
The chain rule for $D$ then gives 
\[
D(\muc{P}) = 
D(\vc{x}) \cdot \prod_k D(P^k)^{x_k},
\]
or equivalently,
\[
A(\muc{P}) = 
D(\vc{x}) \cdot \prod_k A(P^k)^{x_k}.
\]
This proves the first three equations.  Since the last three left-hand
sides can be calculated from the first three
(Figure~\ref{fig:mc-measures}), the rest of the proof is routine. Indeed,
by equation~\eqref{eq:AAw},  
\[
\ovln{A}(\muc{P})
=
\frac{A(\muc{P})}{D(\muc{\vc{w}})}
=
\prod_k \Biggl( \frac{A(P^k)}{D(\vc{w}^k)} \Biggr)^{x_k}
=
\prod_k \ovln{A}(P^k)^{x_k},
\]
and similarly, by equation~\eqref{eq:GAR},
\[
R(\muc{P})
=
\frac{A(\muc{P})}{G(\muc{P})}
=
\prod_k \Biggl( \frac{A(P^k)}{G(P^k)} \Biggr)^{x_k}
=
\prod_k R(P^k)^{x_k}.
\]
Finally, by equation~\eqref{eq:ABG}, 
\[
\ovln{B}(\muc{P})
=
\frac{G(\muc{P})}{\ovln{A}(\muc{P})}
=
D(\vc{x}) \cdot \prod_k \Biggl( \frac{G(P^k)}{\ovln{A}(P_k)} \Biggr)^{x_k}
=
D(\vc{x}) \cdot \prod_k \ovln{B}(P^k)^{x_k},
\]
completing the proof.
\end{proof}

In particular, each multicommunity measure (such as $A(\muc{P})$) is
determined by the corresponding metacommunity measures (such as $A(P^1),
\ldots, A(P^m)$) and the relative sizes of the metacommunities 
($x_1, \ldots, x_m$).  This is the \demph{modularity}%
\index{modularity!metacommunity diversities@of metacommunity diversities}%
\index{metacommunity!modularity principle}
property of the measures $G$, $A$, $\ovln{A}$, $R$ and $\ovln{B}$.

The formulas in Proposition~\ref{propn:mc-chain} can be restated more
compactly in terms of the value measure $\sigma_1$ (defined in
Section~\ref{sec:value-defn}) and the geometric mean $M_0$.  Write
\[
A(P^\blb) = \bigl( A(P^1), \ldots, A(P^m) \bigr) \in \R^m, 
\]
and similarly for the other measures.  Then Proposition~\ref{propn:mc-chain}
states:

\begin{cor}
\lbl{cor:mc-chain-val}
In the notation of Proposition~\ref{propn:mc-chain},
\begin{align*}
G(\muc{P})      &       
= \sigma_1\bigl( \vc{x}, G(P^\blb) \bigr),      &
D(\muc{\vc{w}}) &
= \sigma_1\bigl( \vc{x}, D(\vc{w}^\blb) \bigr), &
A(\muc{P})      &
= \sigma_1\bigl( \vc{x}, A(P^\blb) \bigr),      \\
\ovln{A}(\muc{P})       &
= M_0\bigl( \vc{x}, \ovln{A}(P^\blb) \bigr),    &
R(\muc{P})      &
= M_0\bigl( \vc{x}, R(P^\blb) \bigr),   \\
\ovln{B}(\muc{P})       &
= \sigma_1\bigl( \vc{x}, \ovln{B}(P^\blb) \bigr).
\end{align*}
\qed
\end{cor}

Here, the first row consists of exponentials of ordinary and joint
entropies, the second of exponentials of conditional entropies, and the
last of the exponential of mutual information.

An essential distinction is clear.  The formulas in the
first and third rows are value measures, meaning that the multicommunity
measure $G(\muc{P})$ \emph{aggregates} the measures $G(P^1), \ldots,
G(P^m)$ of the islands%
\index{islands!metacommunities@as metacommunities} 
of which the multicommunity is comprised, and similarly for
$D(\muc{\vc{w}})$, $A(\muc{P})$ and $\ovln{B}(\muc{P})$.  Those in the
second row are means: $\ovln{A}(\muc{P})$ \emph{averages} the island
measures $\ovln{A}(P^1), \ldots, \ovln{A}(P^m)$, and similarly for
$R(\muc{P})$.

The point is clarified by considering a specific case.  Suppose that the
$m$ islands are identical in almost every way: they have the same size, the
same number $S$ of species, the same division into subcommunities, and the
same species distribution within each subcommunity.  The only difference is
that each island uses a disjoint set of species.  Thus, $G(P^k)$,
$D(\vc{w}^k)$, $A(P^k)$, $\ovln{A}(P^k)$, $R(P^k)$ and $\ovln{B}(P^k)$ are
all independent of $k \in \{1, \ldots, m\}$, and $\vc{x} = \vc{u}_m$.  

\begin{cor}[Replication]
\lbl{cor:mc-rep}%
\index{replication principle!metacommunity diversities@for metacommunity diversities}%
\index{metacommunity!replication principle}%
In this situation,
\begin{align*}
G(\muc{P})      &= mG(P^1),     &
D(\muc{\vc{w}}) &= mD(\vc{w}^1),&
A(\muc{P})      &= mA(P^1),     \\
\ovln{A}(\muc{P})       &= \ovln{A}(P^1),       &
R(\muc{P})              &= R(P^1),              \\
\ovln{B}(\muc{P})       &= m\ovln{B}(P^1).
\end{align*}
\qed
\end{cor}

\begin{example}
Take a single island%
\index{islands!metacommunities@as metacommunities} 
divided into subcommunities, then make a copy of it, using a disjoint set
of species for the copy.  In the new, larger system consisting of both
islands, four of the metacommunity measures are twice what they were for a
single island: the effective number $G$ of species, the diversity of the
relative size distribution of the subcommunities, the effective number $A$
of (species, subcommunity) pairs, and the effective number $\ovln{B}$ of
isolated subcommunities.

But the other two remain the same.  The mean diversity of the
subcommunities, $\ovln{A}$, is unchanged, because the subcommunities on the
second island have the same abundance distributions as those on the first.
The redundancy, $R$, is also unchanged, because the two islands have no
species in common.  Put another way, the average spread of species across
subcommunities is the same in the two-island system as on either
island individually.
\end{example}

\section{All entropy is relative}
\lbl{sec:all-ent-rel}
\index{all entropy is relative}

The title of this section has two meanings.  First, the definition of the
entropy of a probability distribution on a finite set is implicitly
relative to the uniform distribution.  Hence on a general measurable space,
ordinary entropy does not even make sense; only relative entropy does.
This point was discussed in Section~\ref{sec:rel-misc}.

Here we explore a different meaning: that all of the entropies associated
with a pair of random variables~-- cross, joint, conditional, and mutual
information~-- can be reduced to relative entropy.  This reduction sheds new
light on the subcommunity and metacommunity diversity measures.

We examine each type of entropy in turn, beginning with ordinary Shannon
entropy.  Let $X$ be a random variable taking values in a finite set $\XX$,
and let $U_{\XX}$\ntn{uset} denote a random variable uniformly distributed
in $\XX$.  We have already seen that
\begin{equation}
\lbl{eq:rel-to-ord}
H(X) = \log\mg{\XX} - \relent{X}{U_{\XX}}
\end{equation}
(Example~\ref{eg:rel-ent-ufm}), which expresses ordinary entropy in terms
of relative entropy together with the cardinality $\mg{\XX}$ of the set $\XX$.

For cross%
\index{cross entropy} 
entropy, let $X_1$ and $X_2$ be random variables taking values in the same
finite set $\XX$.  By equations~\eqref{eq:rco} and~\eqref{eq:rel-to-ord},
\begin{align}
\crossent{X_1}{X_2}     &
=
\relent{X_1}{X_2} + H(X_1)      
\nonumber       \\
&
=
\log\mg{\XX} + \relent{X_1}{X_2} - \relent{X_1}{U_{\XX}},
\lbl{eq:rel-to-cross}
\end{align}
expressing cross entropy in terms of relative entropy and $\mg{\XX}$. 

Now let $X$ and $Y$ be random variables, not necessarily independent,
taking values in finite sets $\XX$ and $\YY$ respectively.  Thus, the
random variable $(X, Y)$ takes values in $\XX \times \YY$.  By
equation~\eqref{eq:rel-to-ord}, the joint%
\index{joint entropy} 
entropy of $X$ and $Y$ is
\begin{equation}
\lbl{eq:rel-to-joint}
H(X, Y) 
=
\log\mg{\XX} + \log\mg{\YY} - \relEnt{(X, Y)}{U_{\XX} \otimes U_{\YY}},
\end{equation}
where $\otimes$ denotes the independent coupling of random variables (as in
Remark~\ref{rmk:ub-coupling}).  Here we have used the observation that
$U_{\XX} \otimes U_{\YY}$ is uniformly distributed on $\XX \times \YY$.

By the explicit formula for mutual%
\index{mutual information}
information in Lemma~\ref{lemma:mut-alts}\bref{part:mut-alts-exp},
\begin{equation}
\lbl{eq:rel-to-mut}
I(X; Y) = \relEnt{(X, Y)}{X \otimes Y}.
\end{equation}
Thus, mutual information is not merely expressible in terms of relative
entropy; it is an \emph{instance} of relative entropy.  By
Lemma~\ref{lemma:rel-ent-pos-def} on relative entropy, $I(X; Y) \geq 0$
with equality if and only if $(X, Y)$ and $X \otimes Y$ are identically
distributed, that is, $X$ and $Y$ are independent.  This gives another
proof of the lower bound in
Proposition~\ref{propn:pair-bounds}\bref{part:mut-bounds}.

It remains to consider conditional%
\index{conditional entropy} 
entropy.

\begin{lemma}
\lbl{lemma:rel-pyth}
Take random variables $V$ and $W$ on the same sample space, with values in
finite sets $\VV$ and $\WW$ respectively.  Also take random variables $V'$
and $W'$, with values in $\VV$ and $\WW$ respectively.  Then
\[
\relEnt{(V, W)}{V' \otimes W'}
=
\relEnt{(V, W)}{V \otimes W'} + \relent{V}{V'}.
\]
\end{lemma}
Equations of this type are called \demph{Pythagorean%
\index{Pythagorean identity} 
identities} (as in Theorem~4.2 of Csisz\'ar and Shields~\cite{CsSh}),
because of the features that relative%
\index{relative entropy!metric@as metric} 
entropy shares with a squared distance (Section~\ref{sec:rel-misc}).

\begin{proof}
By definition, the right-hand side is
\begin{multline*}
\sum_{v, w \csuch \Pr(V = v, W = w) > 0}
\Pr(V = v, W = w) \log \frac{\Pr(V = v, W = w)}{\Pr(V = v)\Pr(W' = w)}
\\
+
\sum_{v \csuch \Pr(V = v) > 0} 
\Pr(V = v) \log \frac{\Pr(V = v)}{\Pr(V' = v)}.
\end{multline*}
But since $\Pr(V = v) = \sum_w \Pr(V = v, W = w)$, the second term is equal
to
\[
\sum_{v, w \csuch \Pr(V = v, W = w) > 0} 
\Pr(V = v, W = w) \log \frac{\Pr(V = v)}{\Pr(V' = v)}.
\]
Collecting terms and cancelling gives the result.
\end{proof}

We return to our setting of random variables $X$ and $Y$ taking values in
finite sets $\XX$ and $\YY$.  By equations~\eqref{eq:rel-to-ord}
and~\eqref{eq:rel-to-mut}, we can express the conditional entropy as
\begin{align*}
\condent{X}{Y}   &
=
H(X) - I(X; Y)  \\
&
=
\log\mg{\XX} - 
\bigl\{ \relent{X}{U_{\XX}} + \relEnt{(X, Y)}{X \otimes Y} \bigr\},
\end{align*}
which by Lemma~\ref{lemma:rel-pyth} gives a formula for conditional entropy
in terms of relative entropy:
\begin{equation}
\lbl{eq:rel-to-cond}
\condent{X}{Y} 
=
\log\mg{\XX} - \relEnt{(X, Y)}{U_{\XX} \otimes Y}.
\end{equation}
For example, if we fix $\XX$, $\YY$ and $Y$ but allow $X$ to vary, the
conditional entropy $\condent{X}{Y}$ is greatest when the relative entropy
$\relEnt{(X, Y)}{U_{\XX} \otimes Y}$ is least.  This happens when $(X, Y)$
has the same distribution as $U_{\XX} \otimes Y$, that is, when $X$ is
independent of $Y$ and uniformly distributed.

We have now reduced each of the various kinds of entropy to relative
entropy.  The purpose of this reduction is to illuminate the various
measures of subcommunity%
\index{subcommunity!diversity measures!relative diversity@and relative diversity} 
and metacommunity%
\index{metacommunity!diversity measures!relative diversity@and relative diversity} 
diversity.  In this setting, relative entropy is replaced by relative%
\index{relative diversity}%
\index{relative entropy!diversity@and diversity}
diversity (introduced in Section~\ref{sec:rel-div}), and the concept of
value (Chapter~\ref{ch:value}) also plays an important part.  The results
are summarized in Table~\ref{table:rel-reduction}.

\begin{table}
\centering%
\parbox{.98\textwidth}{%
\normalsize%
\begin{align*}
\ovln{\alpha}_j(P)      &= \sigma\bigl(\ovln{P}_{\Pdot j}, \One_S\bigr)&
\ovln{\beta}_j(P)       &= \relDiv{\ovln{P}_{\Pdot j}}{\p}      &
\gamma_j(P)             &
= \sigma\bigl(\ovln{P}_{\Pdot j}, \ovln{P}_{\Pdot j}/\p\bigr)   \\
\ovln{A}(P)             &= \sigma(P, \One_S \otimes \vc{w})     &
\ovln{B}(P)             &= \reldiv{P}{\p \otimes \vc{w}}        &
G(P)                    &= \sigma(\p, \One_S)           \\
A(P)                    &= \sigma(P, \One_S \otimes \One_N)     &
R(P)                    &= \sigma(P, \p \otimes \One_N)
\end{align*}}
\caption{The subcommunity and metacommunity measures expressed in terms of
  relative diversity $\reldiv{-}{-}$ and value $\sigma$.}
\lbl{table:rel-reduction}
\end{table}

As before in this chapter, let $P \in \Delta_{SN}$ be an $S \times N$
matrix representing the relative abundances of $S$ species in $N$
subcommunities, write $\p \in \Delta_S$ for the overall relative abundance
vector of the species, and write $\vc{w} \in \Delta_N$ for the relative
sizes of the subcommunities.  
We will often want to refer to the relative abundance
distribution $P_{\Pdot j}/w_j$ of species in the $j$th subcommunity, so let
us write
\[
\ovln{P}_{\Pdot j} 
= 
P_{\Pdot j}/w_j
=
( P_{1j}/w_j, \ldots, P_{Sj}/w_j )
\in 
\Delta_S
\ntn{Pbardotj}
\]
for each $j \in \{1, \ldots, N\}$ such that $w_j > 0$.

We begin with beta-diversity.  The subcommunity measure $\ovln{\beta}_j(P)$
is the relative%
\index{relative diversity}
diversity
\[
\ovln{\beta}_j(P) = \relDiv{\ovln{P}_{\Pdot j}}{\p}
\]
by definition (equation~\eqref{eq:betaj-exp}); its
interpretation was discussed in Sections~\ref{sec:rel-div}
and~\ref{sec:mm-sub}.  The metacommunity measure $\ovln{B}(P)$, which is
the effective number of isolated subcommunities, is given by
\begin{equation*}
\ovln{B}(P) = \reldiv{P}{\p \otimes \vc{w}}.
\end{equation*}
This is simply the exponential of equation~\eqref{eq:rel-to-mut} (taking
$(X, Y)$ to have distribution $P$).  Here $P$ is the distribution of
(species, subcommunity) pairs, $\p \otimes \vc{w}$ is the hypothetical
distribution of (species, subcommunity) pairs in which the overall
proportions of species and subcommunities are correct but the
subcommunities all have the same composition, and $\ovln{B}(P)$ is the
diversity of the first relative to the second.  It is the divergence of our
metacommunity from being well-mixed.  

The minimal value of $\ovln{B}(P)$, which is $1$, is taken when the
metacommunity \emph{is} well-mixed.  Fixing $\vc{w}$ and letting $P$ and
$\p$ vary, the maximum value of $\ovln{B}(P)$ is $D(\vc{w})$
(inequality~\eqref{eq:Bbar-bounds}).  This maximum is attained when the
metacommunity is as far as possible from being well-mixed, that is, when
the subcommunities share no species.

To interpret the other subcommunity and metacommunity measures in terms of
relative diversity, we use the value measure 
\[
\begin{array}{cccc}
\sigma_1\from   &\Delta_n \times [0, \infty)^n  &\to            &
[0, \infty)     \\
                &(\p, \vc{v})                   &\mapsto        &
\displaystyle\prod_{i = 1}^n \biggl( \frac{v_i}{p_i} \biggr)^{p_i},
\end{array}
\]
defined in Section~\ref{sec:value-defn}.  Since we are currently
abbreviating $D_1$ as $D$ and working exclusively with the parameter value
$q = 1$, we also abbreviate $\sigma_1$ as $\sigma$\ntn{sigmaimplicit}.
Note that when $\vc{v}$ is a probability distribution on $\{1, \ldots,
n\}$,
\begin{equation}
\lbl{eq:val-reldiv-1}
\index{value!relative diversity@and relative diversity}
\sigma(\p, \vc{v}) = \frac{1}{\reldiv{\p}{\vc{v}}}.
\end{equation}

We now consider the metacommunity gamma-diversity $G(P)$.  By
equation~\eqref{eq:rel-to-ord} for ordinary entropy in terms of relative
entropy,
\[
G(P)
=
D(\p)
=
\frac{S}{\reldiv{\p}{\vc{u}_S}}.
\]
But the reciprocal of $\reldiv{-}{-}$ is $\sigma$
(equation~\eqref{eq:val-reldiv-1}), which is homogeneous in its second
argument, so
\[
G(P) = \sigma(\p, \One_S)
\]
where 
\[
\One_S = (1, \ldots, 1) \in [0, \infty)^S.  
\ntn{vecofones}
\]
The same conclusion also follows from Example~\ref{eg:value-hill}, where we
showed that the diversity of a species distribution $\p$ is the value of
the community when each species is given value $1$.

The gamma-diversity of the $j$th subcommunity, $\gamma_j(P)$, is by
definition the cross diversity
\[
\gamma_j(P) = \crossDiv{\ovln{P}_{\Pdot j}}{\p}.
\]
Directly from the definition of $\sigma$, we also have
\begin{equation}
\lbl{eq:gam-alt}
\index{value!gamma-diversity@and gamma-diversity}
\gamma_j(P)
=
\sigma\bigl(\ovln{P}_{\Pdot j}, \ovln{P}_{\Pdot j}/\p\bigr).
\end{equation}
In this expression, the value $\ovln{P}_{ij}/p_i$ of species $i$ is high if
it is common in subcommunity $j$ but rare in the metacommunity as a whole.
Thus, $\gamma_j(P)$ is high if subcommunity $j$ is rich in species that are
globally rare.  This supports the earlier interpretation of $\gamma_j(P)$
as the contribution of subcommunity $j$ to metacommunity diversity
(p.~\pageref{p:gamma-contrib}).

Taking the exponential of the formula~\eqref{eq:rel-to-joint} for joint
entropy in terms of relative entropy gives
\[
\index{value!alpha-diversity@and alpha-diversity}
A(P)
=
\frac{SN}{\reldiv{P}{\vc{u}_S \otimes \vc{u}_N}}
=
\sigma(P, \One_S \otimes \One_N).
\]
This is the effective number of (species, subcommunity) pairs.  It takes no
account of the extent to which the same species appear in different
subcommunities, simply treating these $SN$ pairs as separate classes.  This
formula is another instance of Example~\ref{eg:value-hill}, which expressed
the diversity of a \emph{single} community in terms of value.  So too is
the value expression for the diversity $\ovln{\alpha}_j(P)$ of subcommunity
$j$ in isolation:
\[
\ovln{\alpha}_j(P)
=
D\bigl(\ovln{P}_{\Pdot j}\bigr)
=
\sigma\bigl(\ovln{P}_{\Pdot j}, \One_S\bigr).
\]

The average subcommunity diversity $\ovln{A}(P)$ and the redundancy $R(P)$
are both exponentials of conditional entropies, so by
equation~\eqref{eq:rel-to-cond},
\begin{align}
\index{value!alpha-diversity@and alpha-diversity}
\ovln{A}(P)     &
=
\frac{S}{\reldiv{P}{\vc{u}_S \otimes \vc{w}}}
=
\sigma(P, \One_S \otimes \vc{w}),       
\lbl{eq:Abar-val}       \\
\index{value!redundancy@and redundancy}
R(P)            &
=
\frac{N}{\reldiv{P}{\p \otimes \vc{u}_N}}
=
\sigma(P, \vc{p} \otimes \One_N).
\lbl{eq:R-val}
\end{align}
Hence by Lemma~\ref{lemma:sigma-sum-ineq}, the average subcommunity
diversity $\ovln{A}(P)$ is greatest when $P_{ij}$ is proportional to
$(\One_S \otimes \vc{w})_{ij} = w_j$, that is, when each subcommunity has a
uniform species distribution. Similarly, the redundancy $R(P)$ is greatest
when $P_{ij}$ is proportional to $(\p \otimes \One_N)_{ij} = p_i$, that is,
when each species is distributed uniformly across subcommunities.  These
observations confirm the upper bounds on $\ovln{A}(P)$ and $R(P)$ obtained
in Section~\ref{sec:mm-meta}.

\section{Beyond}
\lbl{sec:beyond}

The entropies and diversities discussed in this chapter so far are all
situated in the case $q = 1$ and $Z = I$ (hence, not incorporating any
notion of similarity or distance between species).  In this short section,
we sketch the definitions for a general $q$, omitting proofs and details.
A more detailed development can be found in Reeve et al.~\cite{HPD}, on
which this section is based.

In generalizing from $q = 1$ to an arbitrary $q \in [0, \infty]$, we
replace the Shannon entropy $H$ by the R\'enyi entropy $H_q$, and its
exponential $D$ by the Hill number $D_q$.  The R\'enyi analogue of relative
entropy has already been discussed (Section~\ref{sec:value-rel}), and
R\'enyi-type analogues of conditional entropy and mutual information have
appeared in other works such as Arimoto~\cite{Arim} and
Csisz\'ar~\cite{CsisGCR}.  

In terms of diversity, $q$ controls the comparative importance attached to
rare and common species, and to smaller and larger subcommunities.
(See the discussion at the end of Section~\ref{sec:ren-hill}.)  We
obtain the $q$-analogues%
\index{metacommunity!diversity measures!order q@of order $q$}%
\index{subcommunity!diversity measures!order q@of order $q$}
of each of $\ovln{\alpha}_j$, $\gamma_j$, $A$, $\ovln{A}$, $R$ and $G$ by
taking its expression in terms of value $\sigma$
(Table~\ref{table:rel-reduction}) and replacing $\sigma$ by $\sigma_q$.
The $q$-analogue of $\ovln{\beta}_j(P)$ is $1/\sigma_q\bigl(\ovln{P}_{\Pdot
  j}, \p\bigr)$, as Table~\ref{table:rel-reduction} and
equation~\eqref{eq:val-reldiv-1} would lead one to expect.  But the
situation for $\ovln{B}$ is more subtle, and for this we refer to Reeve et
al.~\cite{HPD}.

The previously-established relationships
\begin{align*}
\ovln{A}        &
=
M_0 \bigl( \vc{w}, 
(\ovln{\alpha}_1, \ldots, \ovln{\alpha}_N) 
\bigr),   \\
\ovln{B}        &
=
M_0 \bigl( \vc{w}, 
(\ovln{\beta}_1, \ldots, \ovln{\beta}_N) 
\bigr),     \\
G               &
=
M_0 \bigl( \vc{w}, 
(\gamma_1, \ldots, \gamma_N) 
\bigr)
\end{align*}
(equations~\eqref{eq:A-mean}, \eqref{eq:B-mean} and~\eqref{eq:G-mean})
continue to hold with $M_{1 - q}$ in place of $M_0$.  Moreover, all of the
bounds and extremal cases established in Section~\ref{sec:mm-meta} and
listed in Table~\ref{table:mcm-range} remain true without alteration for
general $q$, as proved in the second appendix of Reeve et al.~\cite{HPD}.

\begin{example}
In the $q = 1$ setting, we proved that
\[
1 \leq \frac{A(P)}{G(P)} \leq D(\vc{w})
\]
(equation~\eqref{eq:A-bounds}), or equivalently,
\[
0 \leq \frac{A(P)}{G(P)} - 1 \leq D(\vc{w}) - 1.
\]
These inequalities persist for arbitrary $q$, and in
particular for $q = 0$, where they reduce to the elementary statement that
\begin{equation}
\lbl{eq:jac-N}
0 \leq
\frac{\mg{\supp(P)}}{\mg{\supp(\p)}} - 1
\leq N - 1.
\end{equation}
Here, $\mg{\supp(\p)}$ is the number of species present in the
metacommunity, $\mg{\supp(P)}$ is the number of pairs $(i, j)$ such that
species $i$ is present in subcommunity $j$, and we are assuming that no
subcommunity is empty (so that $\mg{\supp(\vc{w})} = N$).  

For instance, suppose that our metacommunity is divided into just two
subcommunities ($N = 2$), so that~\eqref{eq:jac-N} reads
\begin{equation}
\lbl{eq:jac-2}
0 \leq
\frac{\mg{\supp(P)}}{\mg{\supp(\p)}} - 1
\leq 1.
\end{equation}
The middle term in~\eqref{eq:jac-2} is known as the \demph{Jaccard%
\index{Jaccard index} 
index}, after the early twentieth-century botanist Paul Jaccard~\cite{Jacc}.
(For a modern reference, see p.~\mbox{172--3} of Magurran~\cite{Magu}.)
Traditionally, one writes $a$ for the number of species present in both
subcommunities, $b$ for the number present in the first only, and $c$ for
the number present in the second only; then the middle term
in~\eqref{eq:jac-2} is
\[
\frac{(a + b) + (a + c)}{a + b + c} - 1
=
\frac{a}{a + b + c}.
\]
In other words, the Jaccard index is the proportion of species in the
metacommunity that are present in both subcommunities.  It is, therefore, a
simple measure of how much the two subcommunities overlap.  The
$q$-analogue $\frac{A(P)}{G(P)} - 1$ therefore functions as a
generalization of Jaccard's index to an arbitrary number of subcommunities
and an arbitrary degree $q$ of emphasis on rare or common species.  (I
thank Richard Reeve for this observation.)
\end{example}

Several good properties of the metacommunity measures were proved in
Section~\ref{sec:mm-props}: independence of $\ovln{A}$ and $\ovln{B}$,
independence of $\ovln{A}$ and $R$, the identical subcommunities property,
chain rules for the various
metacommunity measures, and the consequent modularity and replication
principles.  All of these results extend without change to an arbitrary $q
\in [0, \infty]$, as shown in the second appendix of Reeve et
al.~\cite{HPD}.

In contrast, the equations
\[
\index{partitioning of diversity}
\index{diversity!partitioning of}
\ovln{\alpha}_j \ovln{\beta}_j = \gamma_j,
\qquad
\ovln{A} \, \ovln{B} = G
\]
are a special feature of the case $q = 1$.  These relationships ultimately
derive from the identity
\[
M_0(\p, \vc{x}\vc{y})
=
M_0(\p, \vc{x}) M_0(\p, \vc{y})
\]
($\p \in \Delta_n$, $\vc{x}, \vc{y} \in [0, \infty)^n$), which becomes
  false when $M_0$ is replaced by $M_{1 - q}$.  For arbitrary $q$, there
  appears to be \emph{no} formula for $G$ in terms of $\ovln{A}$ and
  $\ovln{B}$.  That is, although $\ovln{A}$ and $\ovln{B}$ are canonical
  measures of average diversity within subcommunities and of variation
  between them, they do not together determine the diversity $G$ of the
  metacommunity.  As we have seen many times, and as Shannon%
\index{Shannon, Claude}
himself recognized, entropy of order $1$ has uniquely good properties.

The challenge of partitioning metacommunity diversity into 
within- and between-subcommunity components, for arbitrary $q$, was 
taken up by Jost~\cite{JostPDI,JostIAB},%
\index{Jost, Lou} 
who proposed formulas for alpha- and beta-diversities.  When $q = 1$, they
are equal to our $\ovln{A}$ and $\ovln{B}$, but for $q \neq 1$, they
disagree.  Jost's measures satisfy the relationship
\[
\text{alpha} \times \text{beta} = \text{gamma}
\]
for arbitrary $q$, but his beta-diversity does not have the `identical%
\index{identical subcommunities}%
\index{subcommunity!identical}
subcommunities' property of Proposition~\ref{propn:shattering}.  (The
second appendix of Reeve et al.~\cite{HPD} gives a counterexample.)  That
is, an artificial division of a subcommunity into two identically-composed
smaller subcommunities can cause a change in the alpha- and
beta-diversities that Jost proposed.

In summary, the generalization of the metacommunity and subcommunity
diversity measures from $q = 1$ to an arbitrary $q \in [0, \infty]$ is
mostly straightforward, as long as we abandon the idea that metacommunity
gamma-diversity must be determined by metacommunity alpha- and
beta-diversities.

However, to incorporate a species similarity matrix $Z$ into the measures
requires more care.  We do not discuss this generalization here; again, the
reader is referred to Reeve et al.~\cite{HPD}.%
\index{Reeve, Richard|)}

%% file: prob.tex
\chapter{Probabilistic methods}
\lbl{ch:prob}
\index{Aubrun, Guillaume}
\index{Nechita, Ion}

Much of this book is about characterization theorems for entropies,
diversities and means, and the conditions that characterize these
quantities are mostly functional%
\index{functional equation}
equations.  In this chapter, we will see how to solve certain
functional equations using results from probability theory, following the
pioneering~2011 work of Aubrun and Nechita~\cite{AuNe}.  The technique is
demonstrated first with their startlingly simple characterization of the
$\ell^p$ norms, and then with a similar theorem for the power means,
different from the characterizations in Chapter~\ref{ch:mns}.

Functional equations are completely deterministic entities, with no
stochastic element.  How, then, can the power of probability theory be
brought to bear?

A simple analogy demonstrates the general idea.  Suppose that we want to
multiply out the expression
\[
(x + y)^{1000} = (x + y)(x + y) \cdots (x + y)
\]
as a sum of terms $x^a y^b$.  Which terms $x^a y^b$%
\index{binomial expansion} 
appear, and how many of them are there?

The standard answer is, of course, that all the terms in the expansion
satisfy $a + b = 1000$ with $a, b \geq 0$, and that the number of such
terms is exactly $1000!/a!b!$.  But there is a different kind of answer:
that \emph{most} of the terms are of the form $x^a y^b$ where $a$ and $b$
are each \emph{about} $500$.  To see this, we can contemplate
the process of multiplying out the brackets, in which one has to go through
all $2^{1000}$ ways of making $1000$ choices between $x$ and $y$.  If we
flip a fair coin%
\index{coin!toss}
$1000$ times, we usually obtain about $500$ each of
heads and tails, and this is the reason why most values of $a$ and $b$ are
about $500$.

This alternative answer has several distinguishing features.  It is
approximate, and the approximation is obtained by probabilistic reasoning.
Depending on the degree of precision required for the purpose at hand, and
depending on the meanings of `most' and `about', this approximation may be
all that we need.  It is also simpler than the first, precise, answer.  All
of these features are also displayed by the probabilistic method described
in this chapter.

For us, the key theorem from probability theory is a variational formula
for the moment generating function (Section~\ref{sec:mgfs}).  Conceptually,
this formula can be understood as the convex conjugate of Cram\'er's large
deviation theorem (Section~\ref{sec:large}).  The probabilistic method is
applied to characterize the $\ell^p$ norms in Section~\ref{sec:mult-norms}
and the power means in Section~\ref{sec:mult-means}.

This chapter assumes some basic probability theory, but not much more than
the language of random variables.  The most technically sophisticated part,
Section~\ref{sec:large}, is for context only and is not logically necessary
for anything that follows.

\section{Moment generating functions}
\lbl{sec:mgfs}

In this short section, we give a variational formula for the moment
generating function of any real random variable.  It can be found in Cerf%
\index{Cerf, Rapha\"el}
and Petit~\cite{CePe},%
\index{Petit, Pierre}
who call it a `dual equality', a name explained in
the next section.  The proof given here is different from theirs.

Let $X$ be a real random variable.  The \demph{moment%
\index{moment generating function}
generating function}
of $X$ is the function 
\[
\begin{array}{cccc}
m_X\from        &\R             &\to            &[0, \infty]            \\
                &\lambda        &\mapsto        &\Ex\bigl(e^{\lambda X}\bigr),
\ntn{mgf}
\end{array}
\]
where $\Ex$ denotes expected value.

\begin{thm}
\lbl{thm:cp}
Let $X, X_1, X_2, \ldots$ be independent identically distributed real
random variables.  Write
\[
\ovln{X}_r
=
\tfrac{1}{r}(X_1 + \cdots + X_r)
\ntn{runningmean}
\]
($r \geq 1$).  Then
\begin{equation}
\lbl{eq:cp}
m_X(\lambda)
=
\sup_{x \in \R, \ r \geq 1}
e^{\lambda x} \, \Pr\bigl(\ovln{X}_r \geq x\bigr)^{1/r} 
\end{equation}
for all $\lambda \geq 0$, where the supremum is over all real $x$ and
positive integers $r$.
\end{thm}

We allow infinite values on either side of equation~\eqref{eq:cp}.

The proof, given below, uses the elementary result of probability theory
known as \demph{Markov's inequality} (Grimmett and Stirzaker~\cite{GrSt},
Lemma~7.2(7)):

\begin{lemma}[Markov] 
\lbl{lemma:markov}%
\index{Markov's inequality}%
Let $Z$ be a random variable taking nonnegative real values.  Then for all
$z \in \R$,
\[
\Ex(Z) \geq z \cdot \Pr(Z \geq z).
\]
\end{lemma}

This is intuitively clear: if one third of the people in a room are at
least 60 years old, then the mean age\index{age} is at least 20.

For the proof, we use some standard notation: given $S \sub \R$, let
$I_S \from \R \to \R$ denote the \demph{indicator%
\index{indicator function}
function} (or \demph{characteristic%
\index{characteristic function}
function}) of $S$, defined by
\[
I_S(x)
=
\begin{cases}
1       &\text{if } x \in S,    \\
0       &\text{otherwise}.
\end{cases}
\ntn{indicator}
\]

\begin{proof}
We have
\[
Z 
\geq 
z \cdot I_{[z, \infty)}(Z),
\]
by considering the cases $Z \geq z$ and $Z < z$ separately.  Hence
\[
\Ex(Z) 
\geq
\Ex\bigl( 
z \cdot I_{[z, \infty)}(Z)
\bigr)
=
z \cdot \Pr(Z \geq z).
\]
\end{proof}

\begin{pfof}{Theorem~\ref{thm:cp}}
Let $\lambda \geq 0$.  We prove equation~\eqref{eq:cp} by showing that each
side is greater than or equal to the other.

First we show that 
\[
m_X(\lambda) 
\geq
\sup_{x \in \R, \ r \geq 1}
e^{\lambda x} \, \Pr\bigl(\ovln{X}_r \geq x\bigr)^{1/r}.
\]
Let $x \in \R$ and $r \geq 1$; we must show that
\[
\Ex\bigl(e^{\lambda X}\bigl)^r
\geq
e^{r\lambda x} \, \Pr\bigl(\ovln{X}_r \geq x\bigr).
\]
And indeed, 
\begin{align}
\Ex\bigl(e^{\lambda X}\bigr)^r    &
=
\Ex\bigl(e^{\lambda(X_1 + \cdots + X_r)}\bigr)
\lbl{eq:cp-1} \\
&
=
\Ex\bigl(e^{r\lambda\ovln{X}_r}\bigr)
\nonumber       \\
&
\geq
e^{r\lambda x} \,
\Pr\bigl(
e^{r\lambda\ovln{X}_r} \geq e^{r\lambda x}
\bigr)
\lbl{eq:cp-3} \\
&
\geq
e^{r\lambda x} \, \Pr\bigl(\ovln{X}_r \geq x\bigr),
\lbl{eq:cp-4}
\end{align}
where~\eqref{eq:cp-1} holds because $X, X_1, X_2, \ldots$ are independent
and identically distributed, \eqref{eq:cp-3}~follows from Markov's
inequality, and \eqref{eq:cp-4}~holds because $e^{r\lambda y}$ is
increasing in $y \in \R$.

Now we prove the opposite inequality, 
\begin{equation}
\lbl{eq:cp-hard}
m_X(\lambda) 
\leq
\sup_{x \in \R, \ r \geq 1}
e^{\lambda x} \, \Pr\bigl(\ovln{X}_r \geq x\bigr)^{1/r}.
\end{equation}
The strategy is to show that $\Ex\bigl( e^{\lambda X} I_{[-a, a]}(X)
\bigr)$ is bounded above by the right-hand side of~\eqref{eq:cp-hard} for
each $a > 0$, then to deduce that the same is true of $\Ex(e^{\lambda X}) =
m_X(\lambda)$ itself.

Let $a > 0$ and $\delta > 0$ be real numbers.  We can choose an integer $d
\geq 1$ and real numbers $v_0, \ldots, v_d$ such that
\[
-a = v_0 < v_1 < \cdots < v_d = a
\]
and $v_k \leq v_{k - 1} + \delta$ for all $k \in \{1, \ldots, d\}$.  Then
for all integers $s \geq 1$, 
\begin{align}
\Ex\bigl(
e^{\lambda X} I_{[-a, a]}(X) \bigr)^s   &
=
\Ex\Bigl(
e^{\lambda X_1} I_{[-a, a]}(X_1) 
\cdots
e^{\lambda X_s} I_{[-a, a]}(X_s)
\Bigr)
\lbl{eq:cph-1}        \\
&
\leq
\Ex\Bigl(
e^{s\lambda\ovln{X}_s} I_{[-a, a]}\bigl(\ovln{X}_s\bigr)
\Bigr)
\lbl{eq:cph-2}        \\
&
\leq
\sum_{k = 1}^d 
\Pr\bigl(v_{k - 1} \leq \ovln{X}_s \leq v_k\bigr) e^{s\lambda v_k}
\lbl{eq:cph-3}        \\
&
\leq
\sum_{k = 1}^d \Pr\bigl(\ovln{X}_s \geq v_{k - 1}\bigr) 
e^{s\lambda v_{k - 1}} e^{s \lambda \delta}     
\lbl{eq:cph-4}        \\
&
\leq
e^{s\lambda\delta} d \, \sup_{x \in \R} \, 
e^{s\lambda x}\, \Pr\bigl(\ovln{X}_s \geq x\bigr),
\nonumber
\end{align}
where~\eqref{eq:cph-1} holds because $X, X_1, X_2, \ldots$ are independent
and identically distributed, \eqref{eq:cph-2}~holds because the mean of a
collection of numbers in the interval $[-a, a]$ also belongs to that
interval, \eqref{eq:cph-3}~follows from the definition of expected value,
and~\eqref{eq:cph-4} follows from the defining properties of $v_0, \ldots,
v_d$.  Hence
\begin{align*}
\Ex\bigl(
e^{\lambda X} I_{[-a, a]}(X) 
\bigr)  &
\leq
e^{\lambda\delta} d^{1/s} \,
\,\sup_{x \in \R}\,
e^{\lambda x} \, \Pr\bigl(\ovln{X}_s \geq x\bigr)^{1/s}      \\
&
\leq
e^{\lambda \delta} d^{1/s}\,
\,\sup_{x \in \R, \ r \geq 1} \,
e^{\lambda x} \, \Pr\bigl(\ovln{X}_r \geq x\bigr)^{1/r}.
\end{align*}
This holds for all real $\delta > 0$ and integers $s \geq 1$, so we can let
$\delta \to 0$ and $s \to \infty$, which gives
%
\[
\Ex\bigl(
e^{\lambda X} I_{[-a, a]}(X) 
\bigr)  
\leq
\sup_{x \in \R, \ r \geq 1} 
e^{\lambda x} \, \Pr\bigl(\ovln{X}_r \geq x\bigr)^{1/r}.
\]
%
Finally, letting $a \to \infty$ and using the monotone convergence theorem
gives the desired inequality~\eqref{eq:cp-hard}.
%
%
\end{pfof}

The following example is the only instance of Theorem~\ref{thm:cp} that we
will need.

\begin{example}
\lbl{eg:cp-fin}
Let $n \geq 1$ and $c_1, \ldots, c_n \in \R$.  Let $X, X_1, X_2, \ldots$ be
independent random variables with distribution $\tfrac{1}{n} \sum_{i = 1}^n
\delta_{c_i}$, where $\delta_c$ denotes the point mass at a real number
$c$. Thus, the random variables take values $c_1, \ldots, c_n$ with
probability $1/n$ each, adding probabilities if there are repeats among the
$c_i$.  For instance, if $n = 3$ and $(c_1, c_2, c_3) = (7, 7, 8)$, then
$X$ takes value $7$ with probability $2/3$ and $8$ with probability $1/3$.

The moment generating function of $X$ is given by
\[
m_X(\lambda)
=
\tfrac{1}{n}(e^{c_1 \lambda} + \cdots + e^{c_n \lambda})
\]
($\lambda \in \R$).  On the other hand, for $r \geq 1$,
\[
\Pr\bigl(\ovln{X}_r \geq x\bigr) 
=
\tfrac{1}{n^r} 
\bigl|\bigl\{
(i_1, \ldots, i_r) \such c_{i_1} + \cdots + c_{i_r} \geq rx
\bigr\}\bigr|.
\]
Hence by Theorem~\ref{thm:cp}, for all $\lambda \geq 0$, 
\begin{equation}
\lbl{eq:cp-fin}
e^{c_1\lambda} + \cdots + e^{c_n\lambda}
=
\sup_{x \in \R, \ r \geq 1}
e^{\lambda x} 
\bigl|\bigl\{
(i_1, \ldots, i_r) \such c_{i_1} + \cdots + c_{i_r} \geq rx
\bigr\}\bigr|^{1/r}.
\end{equation}
This is a completely deterministic statement about real numbers $c_1,
\ldots, c_n, \lambda$.
\end{example}

Theorem~\ref{thm:cp} gives the value of $m_X(\lambda)$ for $\lambda \geq 0$
only, but it is easy to deduce the value for negative $\lambda$:

\begin{cor}
\lbl{cor:cp-neg}
In the context of Theorem~\ref{thm:cp},
\[
m_X(\lambda)
=
\sup_{x \in \R, \ r \geq 1} 
e^{\lambda x} \, \Pr\bigl(\ovln{X}_r \leq x\bigr)^{1/r}
\]
for all $\lambda \leq 0$.
\end{cor}

\begin{proof}
Apply Theorem~\ref{thm:cp} to $-X$ and $-\lambda$, renaming $x$ as $-x$.
\end{proof}

\section{Large deviations and convex duality}
\lbl{sec:large}
\index{large deviations}

This section is not logically necessary for anything that follows, but
places the moment generating function formula of Theorem~\ref{thm:cp} into
a wider context.  Briefly put, that formula is the convex conjugate of
Cram\'er's large deviation theorem.  Here we explain what this means and
why it is true.

\subsection*{Cram\'er's theorem}
\index{Cramer, Harald@Cram\'er, Harald}

Let $X, X_1, X_2, \ldots$ be independent identically distributed real
random variables, with mean $\mu$, say.  Given $x \in \R$, what can be said
about $\Pr\bigl(\ovln{X}_r \geq x\bigr)$ for large integers $r$?

The law%
\index{law of large numbers} 
of large numbers implies that
\[
\Pr\bigl(\ovln{X}_r \geq x\bigr)
\to
\begin{cases}
1       &\text{if } x < \mu,    \\
0       &\text{if } x > \mu
\end{cases}
\]
as $r \to \infty$ (assuming that $\Ex(\mg{X})$ is finite).  However, it is
silent on the question of how \emph{fast} $\Pr\bigl(\ovln{X}_r \geq
x\bigr)$ converges as $r \to \infty$.

Consider, then, the central%
\index{central limit theorem} 
limit theorem.  Loosely, this states that when $r$ is large, the
distribution of $\ovln{X}_r$ is approximately normal.%
\index{normal distribution}  
This enables us to
estimate $\Pr\bigl(\ovln{X}_r \geq x\bigr)$ for large $r$; but again, it
does not help us with the \emph{rate} of convergence.

More exactly, assume without loss of generality that $\mu = 0$.  Then for
each $r \geq 1$, the random variable
\[
\sqrt{r}\,\ovln{X}_r
=
\frac{1}{\sqrt{r}}(X_1 + \cdots + X_r)
\]
has mean $0$ and the same variance ($\sigma^2$, say) as $X$.  The central
limit theorem states that as $r \to \infty$, the distribution of
$\sqrt{r}\,\ovln{X}_r$ converges to the normal distribution with mean $0$
and variance $\sigma^2$.  This gives a way of estimating the probability
\[
\Pr\Bigl(
\tfrac{1}{\sqrt{r}} (X_1 + \cdots + X_r) \geq x
\Bigr)
=
\Pr\Bigl( \ovln{X}_r \geq \tfrac{x}{\sqrt{r}} \Bigr)
\]
for any $x \in \R$ and large integer $r$.  But the original question was
about $\Pr\bigl(\ovln{X}_r \geq x\bigr)$, not $\Pr\bigl(\ovln{X}_r \geq
x/\sqrt{r}\bigr)$.  In other words, we are interested in larger deviations
from the mean than those addressed by the central limit theorem.

So, neither the law of the large numbers nor the central limit theorem
tells us the rate of convergence of $\Pr\bigl(\ovln{X}_r \geq x\bigr)$ as
$r \to \infty$.  But large deviation theory does.  Roughly speaking, the
basic fact is that for each $x \in \R$ there is a constant $k(x) \in [0,
  1]$ such that
\[
\Pr\bigl(\ovln{X}_r \geq x\bigr) 
\approx
k(x)^r
\]
when $r$ is large.  If $x > \mu$ then $k(x) < 1$, so the decay of
$\Pr\bigl(\ovln{X}_r \geq x\bigr)$ as $r \to \infty$ is exponential.  The
precise result is this.

\begin{thm}[Cram\'er]
\lbl{thm:cramer}%
\index{Cramer, Harald@Cram\'er, Harald}%
Let $X, X_1, X_2, \ldots$ be independent identically distributed real
random variables, and let $x \in \R$.  Then the limit
\[
\lim_{r \to \infty} \Pr\bigl(\ovln{X}_r \geq x\bigr)^{1/r}
\]
exists and is equal to
\[
\inf_{\lambda \geq 0} \frac{\Ex(e^{\lambda X})}{e^{\lambda x}}.
\]
\end{thm}

Part of this statement is an easy consequence of Markov's%
\index{Markov's inequality} 
inequality.  Indeed, we used Markov's inequality in
equations~\eqref{eq:cp-1}--\eqref{eq:cp-4} to show that
\[
\Pr\bigl(\ovln{X}_r \geq x\bigr)^{1/r} 
\leq 
\frac{\Ex(e^{\lambda X})}{e^{\lambda x}}
\]
for each $r \geq 1$ and $\lambda \geq 0$, so if the limit in Cram\'er's
theorem does exist then it is at most the stated infimum.  We do not prove
Cram\'er's theorem here, but a short proof can be found in Cerf%
\index{Cerf, Rapha\"el}
and Petit~\cite{CePe}%
\index{Petit, Pierre} 
(who deduce it from Theorem~\ref{thm:cp} using the convex duality that we
are about to discuss), or see standard probability texts such as Grimmett
and Stirzaker (\cite{GrSt}, Theorem~5.11(4)).

\begin{example}
When $X$ is distributed normally%
\index{normal distribution}
with mean $\mu$ and variance $\sigma^2$,
its moment generating function is
\[
\Ex\bigl(e^{\lambda X}\bigr)
=
\exp\bigl(\lambda\mu + \hlf \lambda^2 \sigma^2\bigr).
\]
Hence
\[
\frac{\Ex\bigl(e^{\lambda X}\bigr)}{e^{\lambda x}}
=
\exp\bigl(
\hlf \sigma^2 \cdot \lambda^2 - (x - \mu) \cdot \lambda
\bigr).
\]
Minimizing $\Ex(e^{\lambda X})/e^{\lambda x}$ over $\lambda \geq 0$
therefore reduces to the routine task of minimizing a quadratic.  This
done, Cram\'er's theorem gives
\[
\lim_{r \to \infty} \Pr\bigl(\ovln{X}_r \geq x\bigr)^{1/r}
=
\begin{cases}
\displaystyle
1       &\text{if } x \leq \mu, \\
\displaystyle
\exp \Biggl( - \frac{(x - \mu)^2}{2\sigma^2} \Biggr)      &
\text{if } x \geq \mu.
\end{cases}
\]
As one would expect, this is a decreasing function of $x$ but an increasing
function of both $\mu$ and $\sigma$.
\end{example}

As this example suggests, it is natural to split Cram\'er's theorem into
two cases, according to whether $x$ is greater than or less than $\Ex(X)$:

\begin{cor}
\lbl{cor:cramer-cases}
Let $X, X_1, X_2, \ldots$ be independent identically distributed real
random variables.  
\begin{enumerate}
\item 
\lbl{part:cramer-cases-R}
For all $x \geq \Ex(X)$,
\[
\lim_{r \to \infty} \Pr\bigl(\ovln{X}_r \geq x\bigr)^{1/r}
=
\inf_{\lambda \in \R} 
\frac{\Ex\bigl(e^{\lambda X}\bigr)}{e^{\lambda x}},
\]
and for all $x \leq \Ex(X)$,
\[
\lim_{r \to \infty} \Pr\bigl(\ovln{X}_r \leq x\bigr)^{1/r}
=
\inf_{\lambda \in \R} 
\frac{\Ex\bigl(e^{\lambda X}\bigr)}{e^{\lambda x}}.
\]
(Note that both infima are over \emph{all} $\lambda \in \R$, in contrast to 
Theorem~\ref{thm:cramer}.)

\item
\lbl{part:cramer-cases-1}
For all $x \leq \Ex(X)$,
\[
\lim_{r \to \infty} \Pr\bigl(\ovln{X}_r \geq x\bigr)^{1/r} 
=
1,
\]
and for all $x \geq \Ex(X)$,
\[
\lim_{r \to \infty} \Pr\bigl(\ovln{X}_r \leq x\bigr)^{1/r} 
=
1.
\]
\end{enumerate}
\end{cor}

\begin{proof}
For both parts, we use the inequality $e^x \geq 1 + x$, which implies that
\begin{equation}
\lbl{eq:cc-1}
\frac{\Ex\bigl(e^{\lambda X}\bigr)}{e^{\lambda x}}
=
\Ex\bigl(e^{\lambda(X - x)}\bigr)
\geq
\Ex\bigl(1 + \lambda(X - x)\bigr)
=
1 + \lambda\bigl(\Ex(X) - x\bigr)  
\end{equation}
for all $\lambda, x \in \R$.  We also use the fact that
\begin{equation}
\lbl{eq:cc-2}
\frac{\Ex\bigl(e^{0X}\bigr)}{e^{0x}} = 1
\end{equation}
for all $x \in \R$.

For~\bref{part:cramer-cases-R}, let $x \geq \Ex(X)$.  When $\lambda \leq
0$, \eqref{eq:cc-1} and~\eqref{eq:cc-2} give
\begin{equation}
\lbl{eq:cc-3}
\frac{\Ex\bigl(e^{\lambda X}\bigr)}{e^{\lambda x}}
\geq
1 
=
\frac{\Ex\bigl(e^{0 X}\bigr)}{e^{0 x}},
\end{equation}
so the infimum in Theorem~\ref{thm:cramer} is unchanged if we allow
$\lambda$ to range over all of $\R$.  This gives the first equation
of~\bref{part:cramer-cases-R}.  The second follows by applying the first to
$-X$ and $-x$, renaming $\lambda$ as $-\lambda$.

For~\bref{part:cramer-cases-1}, let $x \leq \Ex(X)$.  When $\lambda \geq
0$, \eqref{eq:cc-1} and~\eqref{eq:cc-2} again imply~\eqref{eq:cc-3}, so the
infimum in Theorem~\ref{thm:cramer} is $1$.  This gives the first equation
of~\bref{part:cramer-cases-1}, and again, the second follows by applying the
first to $-X$ and $-x$.
\end{proof}

\subsection*{Convex duality}
\index{convex!duality}

To relate the formula for moment generating functions in
Theorem~\ref{thm:cp} to Cram\'er's theorem, we use the principle of convex
duality.

\begin{defn}
Let $f \from \R \to [-\infty, \infty]$ be a function.  Its
\demph{convex%
\index{convex!conjugate}
conjugate} or \demph{Legendre--Fenchel%
\index{Legendre--Fenchel transform}
transform} is the function $f^*
\from \R \to [-\infty, \infty]$ defined by
\begin{equation}
\lbl{eq:defn-conj}
f^*(\lambda) = \sup_{x \in \R} \bigl( \lambda x - f(x) \bigr).
\end{equation}
\end{defn}

The theory of convex conjugates is developed thoroughly in texts such as
Borwein and Lewis~\cite{BoLe} and Rockafellar~\cite{Rock}.  Here we give a
brief summary tailored to our needs.

\begin{examples}
\lbl{egs:conj}
\begin{enumerate}
\item 
\lbl{eg:conj-diff}
Let $f \from \R \to \R$ be a differentiable function such that $f' \from \R
\to \R$ is an increasing bijection. 
Then for each
$\lambda \in \R$, the function 
\[
x \mapsto \lambda x - f(x)
\]
has a unique critical point $x_\lambda = {f'}^{-1}(\lambda)$, which is also
the unique global maximum.  Hence
\[
f^*(\lambda) = \lambda x_\lambda - f(x_\lambda).
\]
In graphical terms, for each real number $\lambda$, there is a unique
tangent line to the graph of $f$ with gradient (slope) $\lambda$, and its
equation is
\[
y = \lambda x - f^*(\lambda)
\]
(Figure~\ref{fig:conj-diff}).  Thus, $f^*(\lambda)$ is the negative of the
$y$-intercept of this tangent line.  The convex conjugate $f^*$ therefore
describes $f$ in terms of its envelope of tangent lines.

\begin{figure}
\centering
\lengths
\begin{picture}(120,50)
\cell{60}{25}{c}{\includegraphics[height=50\unitlength]{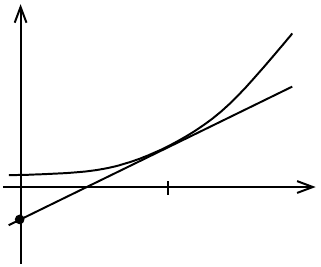}}
\cell{88}{11.7}{c}{$x$}
\cell{76}{40}{c}{$f(x)$}
\cell{82}{30}{l}{gradient $\lambda$}
\cell{44}{7}{c}{$(0, -f^*(\lambda))$}
\cell{62}{11.5}{c}{$x_\lambda$}
\end{picture}
\caption{The relationship between a differentiable function $f$ and its
  convex conjugate $f^*$ (Example~\ref{egs:conj}\bref{eg:conj-diff}).}
\lbl{fig:conj-diff}
\end{figure}

\item
Let $p, q \in (1, \infty)$ be conjugate exponents, that is, $1/p + 1/q =
1$.  Then the functions $x \mapsto \mg{x}^p/p$ and $x \mapsto
\mg{x}^q/q$ are convex conjugate to one another, as can be shown
using~\bref{eg:conj-diff}.

\item
More generally, let $f, g \from \R \to \R$ be differentiable functions such
that $f'$ and $g'$ are increasing and $f'(0) = 0 = g'(0)$.  It can be shown
that if $f', g' \from \R \to \R$ are mutually inverse then $f$ and $g$ are
mutually convex conjugate (Section~I.9 of Zygmund~\cite{Zygm}).  At this
level of generality, convex duality has also been called Young
complementarity or Young%
\index{Young duality} 
duality, as in~\cite{Zygm} or Section~14D of Arnold~\cite{Arno}.
\end{enumerate}
\end{examples}


\begin{lemma}
\lbl{lemma:conj-cvx}
For every function $f \from \R \to [-\infty, \infty]$, the convex conjugate
$f^* \from \R \to [-\infty, \infty]$ is convex.\index{convex!function}
\end{lemma}

Before we can prove this, we need to state what it means for a function
into $[-\infty, \infty]$ to be convex.  If the function takes only finite
values, or takes $\infty$ but not $-\infty$ as a value, or vice versa, then
the meaning is clear.  If it takes both $-\infty$ and $\infty$ as values
then the matter is more delicate; a careful treatment can be found in
Section~2.2.2 of Willerton~\cite{WillLFT} (ultimately derived from
Lawvere~\cite{LawvSCC}).  Fortunately, we can avoid the issue here.  If $f
\equiv \infty$ then $f^* \equiv -\infty$, in which case $f^*$ is convex by
any reasonable definition.  Otherwise, $f^*$ never takes the value
$-\infty$, so the problem does not arise.

\begin{proof}
Let $\lambda, \mu \in \R$ and $p \in [0, 1]$.  Then
\begin{align*}
f^*\bigl(p\lambda + (1 - p)\mu\bigr)    &
=
\sup_{x \in \R} \bigl(
p\lambda x + (1 - p)\mu x - f(x)
\bigr)  \\
&
=
\sup_{x \in \R} \bigl(
p [\lambda x - f(x)] + (1 - p)[\mu x - f(x)]
\bigr)  \\
&
\leq
\sup_{y, z \in \R} \bigl(
p [\lambda y - f(y)] + (1 - p)[\mu z - f(z)]
\bigr)  \\
&
=
p f^*(\lambda) + (1 - p)f^*(\mu),
\end{align*}
as required.
\end{proof}

The examples above suggest that often $f^{**} = f$.  By
Lemma~\ref{lemma:conj-cvx}, this cannot be true unless $f$ is convex.  For
finite-valued $f$, that is the \emph{only} restriction:

\begin{thm}[Legendre--Fenchel]
\lbl{thm:lf}%
\index{Legendre--Fenchel transform}%
\index{Fenchel, Werner}%
Let $f \from \R \to \R$ be a convex function.  Then $f^{**} = f$.
\end{thm}

\begin{proof}
This standard result can be found in textbooks on convex analysis; see
Theorem~4.2.1 of Borwein and Lewis~\cite{BoLe} or Section~14C of
Arnold~\cite{Arno}, for instance.  A proof is also included as
Appendix~\ref{sec:cvx-du}.
\end{proof}

\begin{remarks}
\begin{enumerate}
\item 
Theorem~\ref{thm:lf} is a very special case of the full Legendre--Fenchel
theorem.  For a start, we restricted to finite-valued functions, thus
avoiding the semicontinuity requirement on $f$ that is needed when
values of $\pm\infty$ are allowed.  But much more significantly, the
duality can be generalized beyond functions on $\R$ to functions on a
finite-dimensional real vector space $X$.

In that context, the convex conjugate of a function $f \from X \to
[-\infty, \infty]$ is a function $f^* \from X^* \to [-\infty, \infty]$ on
the dual vector space $X^*$.  The function $f^*$ is defined by the same
formula~\eqref{eq:defn-conj} as before, now understanding the term $\lambda
x$ to mean the functional $\lambda \in X^*$ evaluated at the vector $x \in
X$.  For the Legendre--Fenchel theorem at this level of generality, see
Theorem~4.2.1 of Borwein and Lewis~\cite{BoLe}, Theorem~12.2 of
Rockafellar~\cite{Rock}, or Fenchel~\cite{Fenc}.

\item
The Legendre--Fenchel theorem for vector spaces is itself an instance of
a more general duality still, recently discovered by
Willerton~\cite{WillLFT}.%
\index{Willerton, Simon}  
It is framed in terms of enriched%
\index{enriched category} 
categories,%
\index{category theory} 
as follows.

Let $\cat{V}$ be a complete symmetric monoidal closed category.  In the
rest of this remark, all categories, functors, adjunctions, etc., are taken
to be enriched in $\cat{V}$.  For any categories $\scat{A}$ and $\scat{B}$
and functor $M \from \scat{A}^\op \times \scat{B} \to \cat{V}$, there
is an induced adjunction
\[
[\scat{A}^\op, \cat{V}]
\oppairu
[\scat{B}, \cat{V}]^\op
\]
between functor categories, in which both functors are defined by mapping
into $M$.  For instance, given $X \in [\scat{A}^\op, \cat{V}]$, the
resulting functor $\scat{B} \to \cat{V}$ is 
\[
b \mapsto [\scat{A}^\op, \cat{V}]\bigl(X, M(-, b)\bigr)
\]
($b \in \scat{B}$).
On the other hand, any adjunction restricts canonically to an equivalence
between full subcategories, consisting of its fixed points.  Here, this
gives a dual equivalence
\begin{equation}
\lbl{eq:fmw-adjn}
\cat{C} \oppairu \cat{D}^\op
\end{equation}
between a full subcategory $\cat{C}$ of $[\scat{A}^\op, \cat{V}]$ and a
full subcategory $\cat{D}$ of $[\scat{B}, \cat{V}]$.  Pavlovic calls
either of the categories $\cat{C}$ and $\cat{D}^\op$ the \dmph{nucleus} of
$M$ (Definition~3.9 of~\cite{Pavl}).%
\index{Pavlovic, Dusko}  

Willerton showed that the Legendre--Fenchel theorem is a special case of
this very general categorical construction.  Let $\cat{V}$ be the ordered
set $([-\infty, \infty], \geq)$, regarded as a category in the standard
way, and with monoidal structure defined by addition.  Any real vector
space $X$ gives rise to a category enriched in $\cat{V}$: the objects are
the elements of $X$, and $\Hom(x, y) \in \cat{V}$ is $0$ if $x = y$
and $\infty$ otherwise.  The usual pairing between a vector space and its
dual gives a canonical functor $M \from (X^*)^\op \times X \to \cat{V}$.
Applying the general construction above then gives a dual
equivalence~\eqref{eq:fmw-adjn} between two enriched categories.  As
Willerton showed, this is precisely the convex duality established by the
classical Legendre--Fenchel theorem for $[-\infty, \infty]$-valued
functions on finite-dimensional vector spaces.
\end{enumerate}
\end{remarks}

\subsection*{The dual of Cram\'er's theorem}
\index{Cramer, Harald@Cram\'er, Harald!dual of theorem}

As before, let $X, X_1, X_2, \ldots$ be independent identically distributed
real random variables.  In
Corollary~\ref{cor:cramer-cases}\bref{part:cramer-cases-R}, Cram\'er's
theorem was restated as
\[
\inf_{\lambda \in \R} \frac{\Ex(e^{\lambda X})}{e^{\lambda x}}
=
\begin{cases}
\lim_{r \to \infty} \Pr\bigl(\ovln{X}_r \geq x\bigr)^{1/r}        &
\text{if } x \geq \Ex(X),       \\
\lim_{r \to \infty} \Pr\bigl(\ovln{X}_r \leq x\bigr)^{1/r}        &
\text{if } x \leq \Ex(X).
\end{cases}
\]
Taking logarithms and changing sign, an equivalent statement is that
\begin{equation}
\lbl{eq:cram-log}
(\log m_X)^*(x)
=
\begin{cases}
- \lim_{r \to \infty} \tfrac{1}{r} \log \Pr\bigl(\ovln{X}_r \geq x\bigr)  &
\text{if } x \geq \Ex(X),       \\
- \lim_{r \to \infty} \tfrac{1}{r} \log \Pr\bigl(\ovln{X}_r \leq x\bigr)  &
\text{if } x \leq \Ex(X).       
\end{cases}
\end{equation}
It is a general fact that $\log m_X$, called the
\demph{cumulant%
\index{cumulant generating function} 
generating function} of $X$, is a convex function
(Appendix~\ref{sec:mgfs-log-cvx}).  So by taking convex conjugates on each
side of~\eqref{eq:cram-log} and using the Legendre--Fenchel theorem, we
will obtain an expression for $\log m_X$ and, therefore, the moment
generating function $m_X$ itself.

Specifically, equation~\eqref{eq:cram-log} and the Legendre--Fenchel
theorem imply that for all $\lambda \in \R$,
\begin{multline*}
\log m_X(\lambda)       
= \\
\max \Biggl\{
\sup_{x \geq \Ex(X)} \Bigl(
\lambda x + \lim_{r \to \infty} \tfrac{1}{r} \log 
\Pr\bigl(\ovln{X}_r \geq x\bigr)
\Bigr), 
\sup_{x \leq \Ex(X)} \Bigl(
\lambda x + \lim_{r \to \infty} \tfrac{1}{r} \log 
\Pr\bigl(\ovln{X}_r \leq x\bigr)
\Bigr)
\Biggr\},
\end{multline*}
or equivalently,
\begin{equation}
\lbl{eq:cram-max}
m_X(\lambda)
=
\max \Biggl\{
\sup_{x \geq \Ex(X)} 
e^{\lambda x} \lim_{r \to \infty} 
\Pr\bigl(\ovln{X}_r \geq x\bigr)^{1/r},
\
\sup_{x \leq \Ex(X)} 
e^{\lambda x} \lim_{r \to \infty} 
\Pr\bigl(\ovln{X}_r \leq x\bigr)^{1/r}
\Biggr\}.
\end{equation}
Let $\lambda \geq 0$.  We analyse the second supremum in
equation~\eqref{eq:cram-max}.  The quantity $e^{\lambda x} \lim_{r \to
  \infty} \Pr\bigl(\ovln{X}_r \leq x\bigr)^{1/r}$ is increasing in $x$, so
the supremum is attained when $x = \Ex(X)$.  But by
Corollary~\ref{cor:cramer-cases}\bref{part:cramer-cases-1},
\[
\lim_{r \to \infty} 
\Pr\bigl(
\ovln{X}_r \leq \Ex(X)
\bigr)^{1/r}
=
1,
\]
so the second supremum is just $e^{\lambda\Ex(X)}$.  On the other hand,
Corollary~\ref{cor:cramer-cases}\bref{part:cramer-cases-1} also states that
for all $x \leq \Ex(X)$, 
\[
\lim_{r \to \infty} 
\Pr\bigl(
\ovln{X}_r \geq x
\bigr)^{1/r}
=
1,
\]
so the second supremum can be expressed as
\[
\sup_{x \leq \Ex(X)} e^{\lambda x} 
\lim_{r \to \infty} \Pr\bigl(\ovln{X}_r \geq x\bigr)^{1/r}.
\]
Hence by~\eqref{eq:cram-max},
\begin{equation}
\lbl{eq:cram-sup-lim}
m_X(\lambda)
=
\sup_{x \in \R} \ e^{\lambda x} \lim_{r \to \infty} 
\Pr\bigl(\ovln{X}_r \geq x\bigr)^{1/r}.
\end{equation}
We have derived equation~\eqref{eq:cram-sup-lim} as the convex dual of
Cram\'er's theorem.  It is very nearly the moment generating function
formula of Theorem~\ref{thm:cp}.  The only difference is that
where~\eqref{eq:cram-sup-lim} has a limit as $r \to \infty$,
Theorem~\ref{thm:cp} has a supremum over $r \geq 1$.  However, Cerf and
Petit showed that the two forms are equivalent:
\begin{equation}
\lbl{eq:cp-sup-lim}
\lim_{r \to \infty} \Pr\bigl(\ovln{X}_r \geq x\bigr)^{1/r}
=
\sup_{r \geq 1} \Pr\bigl(\ovln{X}_r \geq x\bigr)^{1/r}
\end{equation}
(\cite{CePe}, p.~928).  In this sense, Theorem~\ref{thm:cp} can also be
regarded as the dual of Cram\'er's theorem.

\begin{remark}
\lbl{rmk:cp-reverse}
In their work, Cerf%
\index{Cerf, Rapha\"el}
and Petit~\cite{CePe}%
\index{Petit, Pierre}
travelled the opposite path from
the one just described.  They started by proving Theorem~\ref{thm:cp},
took convex conjugates, and thus, with the aid of~\eqref{eq:cp-sup-lim}, 
deduced Cram\'er's theorem.
\end{remark}

\section{Multiplicative characterization of the $p$-norms}
\lbl{sec:mult-norms}
\index{Aubrun, Guillaume|(}
\index{Nechita, Ion|(}
\index{multiplicative!characterization of p-norms@characterization of $p$-norms}

Here we show how probabilistic methods can be used to solve functional
equations, following Aubrun and Nechita~\cite{AuNe}.  We give a version of
their theorem that among all coherent ways of putting a norm on each of the
vector spaces $\R^0, \R^1, \R^2, \ldots$, the only ones satisfying a
certain multiplicativity condition are the $p$-norms.

\begin{defn}
\lbl{defn:norm}
Let $n \geq 0$.  A \dmph{norm} $\|\cdot\|$ on $\R^n$ is a function $\R^n
\to [0, \infty)$, written as $\vc{x} \mapsto \|\vc{x}\|$, with the
  following properties:
\begin{enumerate}
\item 
$\|\vc{x}\| = 0 \implies \vc{x} = 0$;

\item
$\|c\vc{x}\| = \mg{c}\,\|\vc{x}\|$ for all $c \in \R$ and $\vc{x} \in \R^n$;

\item
$\|\vc{x} + \vc{y}\| \leq \|\vc{x}\| + \|\vc{y}\|$ for all $\vc{x}, \vc{y}
  \in \R^n$ (the \demph{triangle%
\index{triangle inequality} 
inequality}).
\end{enumerate}
\end{defn}

\begin{example}
\lbl{eg:norm-p} 
Let $n \geq 0$ and $p \in [1, \infty]$.  The
\demph{$p$-norm}\index{pnorm@$p$-norm} or 
\demph{$\ell^p$ norm} $\|\cdot\|_p$\ntn{pnorm} on $\R^n$ is defined by
\[
\|\vc{x}\|_p
=
\Biggl( \sum_{i = 1}^n \mg{x_i}^p \Biggr)^{1/p}
\]
for $p < \infty$, and for $p = \infty$ by
\[
\|\vc{x}\|_\infty
=
\max_{1 \leq i \leq n} \mg{x_i}
\]
($\vc{x} \in \R^n$).  Then $\|\vc{x}\|_\infty = \lim_{p \to \infty}
\|\vc{x}\|_p$, by Lemma~\ref{lemma:pwr-mns-cts-t} on power
means: writing $\mg{\vc{x}} = (\mg{x_1}, \ldots, \mg{x_n})$,
\[
\|\vc{x}\|_p
=
n^{1/p} M_p(\vc{u}_n, \mg{\vc{x}})
\to
M_\infty(\vc{u}_n, \mg{\vc{x}})
=
\|\vc{x}\|_\infty
\]
as $p \to \infty$.
\end{example}

\begin{example}
\lbl{eg:norm-cvx}
Let $\phi \from [0, \infty) \to [0, \infty)$ be an increasing convex
function such that $\phi^{-1}\{0\} = \{0\}$.  For $n \geq 0$, put
\[
K_n
=
\biggl\{ 
\vc{x} \in \R^n \such \sum_{i = 1}^n \phi(\mg{x_i}) \leq 1 
\biggr\},
\]
which is a convex subset of $\R^n$.  Then for $\vc{x} \in \R^n$, put
\[
\|\vc{x}\| = \inf \{ \lambda \geq 0 \such \vc{x} \in \lambda K_n \}.
\]
It can be shown that $\|\cdot\|$ is a norm on $\R^n$ (known as an
\demph{Orlicz%
\index{Orlicz norm}%
\index{norm!Orlicz}
norm}), whose unit ball $\{ \vc{x} \in \R^n \such \|\vc{x}\| \leq 1 \}$
is $K_n$.  For instance, taking $\phi(x) = x^p$ for some $p \in [1,
  \infty)$ gives the $p$-norm of Example~\ref{eg:norm-p}.
\end{example}

Fix $p \in [1, \infty]$.  The $p$-norms on the sequence of spaces $\R^0,
\R^1, \R^2, \ldots$ are compatible with one another in the following two
ways. 

First, the $p$-norm of a vector is unchanged by permuting its entries or
inserting zeros.  For instance,
\begin{equation}
\lbl{eq:p-norm-func-eg}
\|(x_1, x_2, x_3)\|_p
=
\|(x_2, 0, x_3, x_1)\|_p.
\end{equation}
Generally, writing $\lwr{n} = \{1, \ldots, n\}$, any injection $f \from
\lwr{n} \to \lwr{m}$ induces an injective linear map $f_* \from \R^n \to
\R^m$, defined by
\[
(f_* \vc{x})_j
=
\begin{cases}
x_i     &\text{if } j = f(i) \text{ for some } i \in \{1, \ldots, n\},  \\
0       &\text{otherwise} 
\end{cases}
\ntn{pfwdinj}
\]
($\vc{x} \in \R^n$, $j \in \{1, \ldots, m\}$).  Then the $p$-norm has the
property that 
\begin{equation}
\lbl{eq:p-norm-func}
\|f_* \vc{x}\|_p = \|\vc{x}\|_p
\end{equation}
for all injections $f \from \lwr{n} \to \lwr{m}$ and all $\vc{x} \in \R^n$.
For example, equation~\eqref{eq:p-norm-func-eg} is the instance of
equation~\eqref{eq:p-norm-func} where $f$ is the map $\{1, 2, 3\} \to \{1,
2, 3, 4\}$ defined by $f(1) = 4$, $f(2) = 1$, and $f(3) = 3$.

Second, the $p$-norm satisfies a multiplicativity law.  Let $\vc{x} \in
\R^n$ and $\vc{y} \in \R^m$, and recall from
equation~\eqref{eq:defn-real-tensor} (p.~\pageref{eq:defn-real-tensor})
the definition of $\vc{x} \otimes \vc{y} \in \R^{nm}$.  Then
%
\[
\|\vc{x} \otimes \vc{y}\|_p
=
\|\vc{x}\|_p \, \|\vc{y}\|_p.
\]
%
For instance,
\[
\|(Ax, Ay, Az, Bx, By, Bz)\|_p
=
\|(A, B)\|_p \|(x, y, z)\|_p
\]
for all $A, B, x, y, z \in \R$.

These two properties of the $p$-norms determine them completely, as we
shall see.

\begin{defn}
\lbl{defn:son}
\begin{enumerate}
\item 
\lbl{defn:son-son}
A \demph{system%
\index{system of norms} 
of norms} consists of a norm $\|\cdot\|$ on $\R^n$ for each
$n \geq 0$, such that for each $n, m \geq 0$ and injection $f \from \lwr{n}
\to \lwr{m}$,
\[
\|f_*\vc{x}\| = \|\vc{x}\|
\]
for all $\vc{x} \in \R^n$.

\item
A system of norms $\|\cdot\|$ is \demph{multiplicative}%
\index{multiplicative!system of norms}
if 
\[
\|\vc{x} \otimes \vc{y}\| = \|\vc{x}\| \, \|\vc{y}\|
\]
for all $n, m \geq 0$, $\vc{x} \in \R^n$, and $\vc{y} \in \R^m$.  
\end{enumerate}
\end{defn}

\begin{examples}
\begin{enumerate}
\item 
For each $p \in [1, \infty]$, the $p$-norm $\|\cdot\|_p$ is a
multiplicative system of norms.

\item
Fix a function $\phi$ as in Example~\ref{eg:norm-cvx}.  The norms
$\|\cdot\|$ defined there always form a system of norms, but it is not in
general multiplicative.
\end{enumerate}
\end{examples}

\begin{remark}
The notion of a system of norms can be recast in two equivalent ways.  
First, instead of only considering $\R^n$ for natural numbers $n$, we can
consider
\[
\R^I 
=
\{ \text{functions } I \to \R \}
=
\{ \text{families } (x_i)_{i \in I} \text{ of reals} \}
\]
for arbitrary finite sets $I$.  (This was the approach taken in
Leinster~\cite{MCPM}.)  We then require the equation $\|f_* \vc{x}\| =
\|\vc{x}\|$ to hold for every injection $f \from I \to J$ between finite
sets.  In particular, taking $f$ to be a bijection, the norm on $\R^J$
determines the norm on $\R^I$ for all sets $I$ of the same cardinality as
$J$.  So, the norm on $\R^n$ determines the norm on $\R^I$ for all
$n$-element sets $I$.  It follows that this apparently more general notion
of a system of norms is equivalent to the original one.

In the opposite direction, we can construe a system of norms as a norm on
the single space $c_{00}$ of infinite real sequences with only finitely
many nonzero entries, subject to a symmetry axiom.  (This was the approach
taken in Aubrun and Nechita~\cite{AuNe}.)  To state the multiplicativity
property, we have to choose a bijection between the set of nonnegative
integers and its cartesian square, but by symmetry, the definition of
multiplicativity is unaffected by that choice.
\end{remark}

We now come to the main theorem of this section.  In its present form, it
was first stated by Aubrun and Nechita~\cite{AuNe}.  The result also
follows from Theorem~3.9 of an earlier paper of Fern\'andez-Gonz\'alez,%
\index{Fernandez-Gonzalez@Fern\'andez-Gonz\'alez, Carlos}
Palazuelos%
\index{Palazuelos, Carlos}
and P\'erez-Garc\'{i}a~\cite{FGPPG}%
\index{Perez-Garcia@P\'erez-Garc\'{i}a, David}
(at least, putting aside some delicacies concerning $\|\cdot\|_\infty$).
The arguments in~\cite{FGPPG} are very different, coming as they do from
the theory of Banach spaces.  We will consider only Aubrun and Nechita's
method.

\begin{thm}
\lbl{thm:an}
\index{pnorm@$p$-norm!characterization of}
Every multiplicative system of norms is equal to $\|\cdot\|_p$ for some $p
\in [1, \infty]$.    
\end{thm}

The proof will rest on the moment generating function formula of
Theorem~\ref{thm:cp}.  Specifically, we will need the following consequence
of that theorem. 
Given $\vc{v} = (v_1, \ldots, v_n) \in \R^n$ and $t \in \R$, write
\[
N(\vc{v}, t) 
=
\bigl|\bigl\{ 
i \in \{1, \ldots, n\} 
\such
v_i \geq t
\bigr\}\bigr|.
\ntn{Ncount}
\]

\begin{propn}[Aubrun and Nechita]
\lbl{propn:p-norm-var}%
\index{Aubrun, Guillaume}%
\index{Nechita, Ion}%
Let $p \in [1, \infty)$, $n \geq 0$, and $\vc{x} \in (0, \infty)^n$.  Then
\[
\|\vc{x}\|_p
=
\sup_{u > 0, \ r \geq 1} 
u \cdot N\bigl(\vc{x}^{\otimes r}, u^r\bigr)^{1/rp},
\]
where the supremum is over real $u > 0$ and integers $r \geq 1$.
\end{propn}

This formula was central to Aubrun and Nechita's argument in~\cite{AuNe},
although not quite stated explicitly there.

\begin{proof}
In equation~\eqref{eq:cp-fin} (Example~\ref{eg:cp-fin}), put $c_i = \log x_i$
and $\lambda = p$.  Then
\begin{align*}
x_1^p + \cdots + x_n^p  &
=
\sup_{y \in \R, \ r \geq 1} 
e^{py} \, 
\bigl|\bigl\{
(i_1, \ldots, i_r) \such x_{i_1} \cdots x_{i_r} \geq e^{ry}
\bigr\}\bigr|^{1/r}     \\
&
=
\sup_{u > 0, \ r \geq 1} u^p 
N\bigl(\vc{x}^{\otimes r}, u^r\bigr)^{1/r},
\end{align*}
and the result follows by taking $p$th roots throughout.
\end{proof}

We now embark on the proof of Theorem~\ref{thm:an}, roughly following
Aubrun and Nechita~\cite{AuNe}, but with some simplifications described in
Remark~\ref{rmk:an-diff}.  In the words of Aubrun and Nechita, the proof
proceeds by%
\index{Aubrun, Guillaume}%
\index{Nechita, Ion}
`examining the statistical distribution of large coordinates of the $r$th
tensor power $\vc{x}^{\otimes r}$ ($r$ large)' (\cite{AuNe}, Section~1.1;
notation adapted).   

\femph{For the rest of this section}, let $\|\cdot\|$ be a multiplicative
system of norms.

\paragraph*{Step 1: elementary results}
We begin by deriving some elementary properties of the norms $\|\cdot\|$.
For $n \geq 0$, write $\One_n = (1, \ldots, 1) \in \R^n$.\ntn{Oneprob}

\begin{lemma}
\lbl{lemma:p-norm-elem}
Let $n \geq 0$ and $\vc{x}, \vc{y} \in \R^n$.
\begin{enumerate}
\item 
\lbl{part:p-norm-pm}
If $y_i = \pm x_i$ for each $i$ then $\|\vc{x}\| = \|\vc{y}\|$.

\item
\lbl{part:p-norm-mono}
If $\vc{0} \leq \vc{x} \leq \vc{y}$ then $\|\vc{x}\| \leq \|\vc{y}\|$.

\item
\lbl{part:p-norm-one}
$\|\One_m\| \leq \|\One_n\|$ whenever $0 \leq m \leq n$.
\end{enumerate}
\end{lemma}

\begin{proof}
For~\bref{part:p-norm-pm}, the vector $\vc{x} \otimes (1, -1)$ is a
permutation of $\vc{y} \otimes (1, -1)$, so by definition of system of
norms, 
\[
\|\vc{x} \otimes (1, -1)\| 
=
\|\vc{y} \otimes (1, -1)\|.
\]
But by multiplicativity, this equation is equivalent to 
\[
\|\vc{x}\| \, \|(1, -1)\|
=
\|\vc{y}\| \, \|(1, -1)\|.
\]
Hence $\|\vc{x}\| = \|\vc{y}\|$.  

For~\bref{part:p-norm-mono}, let $S$ be the set of vectors of the form
$(\epsln_1 y_1, \ldots, \epsln_n y_n) \in \R^n$ with $\epsln_i = \pm 1$.
Recall that the \demph{convex\index{convex!hull} hull} of
$S$ is the set of vectors expressible as $\sum_{\vc{s} \in S}
\lambda_{\vc{s}} \vc{s}$ for some nonnegative reals
$(\lambda_{\vc{s}})_{\vc{s} \in S}$ summing to $1$.  A straightforward
induction shows that the convex hull of $S$ is
\[
\prod_{i = 1}^n [-y_i, y_i] 
=
[-y_1, y_1] \times \cdots \times [-y_n, y_n]
\]
(Figure~\ref{fig:hull}).  
\begin{figure}
\centering
\lengths
\begin{picture}(120,50)
\cell{60}{25}{c}{\includegraphics[height=50\unitlength]{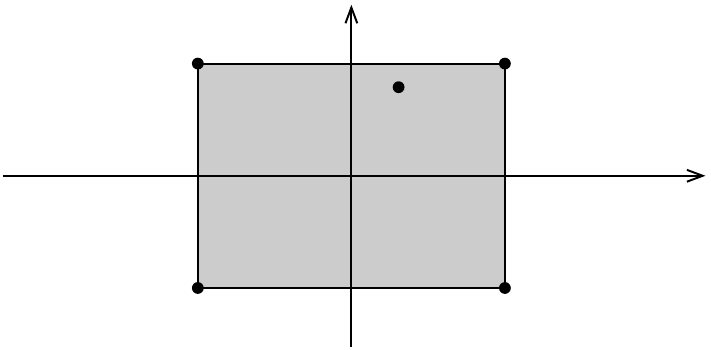}}
\cell{71}{34}{c}{$\vc{x} = (x_1, x_2)$}
\cell{93}{41}{c}{$\vc{y} = (y_1, y_2)$}
\cell{90.5}{9}{c}{$(y_1, -y_2)$}
\cell{27.5}{9}{c}{$(-y_1, -y_2)$}
\cell{28.5}{41}{c}{$(-y_1, y_2)$}
\end{picture}
\caption{The vector $\vc{x}$ in the convex hull of the set $S$, as in the
  proof of Lemma~\ref{lemma:p-norm-elem}\bref{part:p-norm-mono}, shown for
  $n = 2$.}
\lbl{fig:hull}
\end{figure}
But $\vc{x} \in \prod
[-y_i, y_i]$, and $\|\vc{s}\| = \|\vc{y}\|$ for each $\vc{s} \in S$ by
part~\bref{part:p-norm-pm}.  Hence, writing $\vc{x} = \sum
\lambda_{\vc{s}} \vc{s}$ and using the triangle inequality, 
\[
\|\vc{x}\|
\leq
\sum_{\vc{s} \in S} \lambda_{\vc{s}} \|\vc{s}\|
=
\sum_{\vc{s} \in S} \lambda_{\vc{s}} \|\vc{y}\|
=
\|\vc{y}\|.
\]

For~\bref{part:p-norm-one}, let $0 \leq m \leq n$.  We have
\[
\|\One_m\|
=
\|(\overbrace{\underbrace{1, \ldots, 1}_m,
0, \ldots, 0}^n) \|
\leq
\|\One_n\|,
\]
where the equality follows from the definition of system of norms and the
inequality follows from part~\bref{part:p-norm-mono}.
\end{proof}

\paragraph*{Step 2: finding $p$}
The idea now is that since $\|\One_n\|_p = n^{1/p}$ for all $p \in [1,
  \infty]$ and $n \geq 1$, we should be able to recover $p$ from
$\|\cdot\|$ by examining the sequence $\bigl(\|\One_n\|\bigr)_{n \geq 1}$.

Indeed, for all $m, n \geq 1$, multiplicativity gives
\[
\|\One_{mn}\| 
= 
\|\One_m \otimes \One_n\| 
=
\|\One_m\| \, \|\One_n\|.
\]
Moreover, Lemma~\ref{lemma:p-norm-elem}\bref{part:p-norm-one} implies that
the sequence $\bigl( \|\One_n\| \bigr)_{n \geq 1}$ is increasing.  Hence by
Theorem~\ref{thm:erdos-inc} applied to the sequence $\bigl( \log \|\One_n\|
\bigr)_{n \geq 1}$, there exists $c \geq 0$ such that $\|\One_n\| = n^c$
for all $n \geq 1$.  Now
\[
2^c
=
\|(1, 1)\|      
\leq
\|(1, 0)\| + \|(0, 1)\| 
=
2 \cdot \|(1)\| 
=
2 \cdot 1^c
=
2,
\]
so $c \in [0, 1]$.  Put $p = 1/c \in [1, \infty]$.  Then
\[
\|\One_n\| = n^{1/p} = \|\One_n\|_p
\]
for all $n \geq 1$.

We will show that $\|\vc{x}\| = \|\vc{x}\|_p$ for all $n \geq 0$ and
$\vc{x} \in \R^n$.  By definition of system of norms and
Lemma~\ref{lemma:p-norm-elem}\bref{part:p-norm-pm}, it is enough to prove
this when $\vc{x} \in (0, \infty)^n$.  The case $n = 0$ is trivial, so we
can also restrict to $n \geq 1$.

\paragraph*{Step 3: the case $p = \infty$}
This case needs separate handling, and is straightforward anyway.  We show
directly that if $p = \infty$ (that is, if $\|\One_n\| = 1$ for all $n \geq
1$) then $\|\cdot\| = \|\cdot\|_\infty$.

Let $\vc{x} \in (0, \infty)^n$, and choose $j$ such that $x_j =
\|\vc{x}\|_\infty$.  Then by
Lemma~\ref{lemma:p-norm-elem}\bref{part:p-norm-mono}, 
\[
\|\vc{x}\|
\leq
\|(x_j, \ldots, x_j)\|
=
x_j \|\One_n\|
=
x_j.
\]
But also 
\[
\|\vc{x}\|
\geq
\|(\underbrace{0, \ldots, 0}_{j - 1}, x_j, 
\underbrace{0, \ldots, 0}_{n - j})\|
=
\|(x_j)\|
=
x_j \|\One_1\|
=
x_j.
\]
Hence $\|\vc{x}\| = x_j = \|\vc{x}\|_\infty$, as required.

So, we may assume henceforth that $p \in [1, \infty)$.

\paragraph*{Step 4: exploiting the variational formula for $p$-norms}
We now use the formula for $p$-norms in Proposition~\ref{propn:p-norm-var}:
for $\vc{x} \in (0, \infty)^n$,
\[
\|\vc{x}\|_p 
=
\sup_{u > 0, \ r \geq 1} 
\Bigl(
u^r \, N(\vc{x}^{\otimes r}, u^r)^{1/p}
\Bigr)^{1/r}.
\]
(This is where the probability theory is used, as
Proposition~\ref{propn:p-norm-var} was derived from the variational
formula for moment generating functions.)  Since $m^{1/p} =
\|\One_m\|$ for all $m$, an equivalent statement is that
\begin{equation}
\lbl{eq:p-norm-norm}
\|\vc{x}\|_p
=
\sup_{u > 0, \ r \geq 1}
\bigl\|
\bigl( 
N(\vc{x}^{\otimes r}, u^r) \mc u^r
\bigr)
\bigr\|^{1/r}.
\end{equation}
Here we have used the notation $\mc$ introduced after
Definition~\ref{defn:decomp}.

The expression~\eqref{eq:p-norm-norm} for $\|\vc{x}\|_p$ has the feature
that it makes no mention of $p$.  We will use it to prove first that
$\|\vc{x}\| \geq \|\vc{x}\|_p$, then that $\|\vc{x}\| \leq \|\vc{x}\|_p$.

\paragraph{Step 5: the lower bound} 
Let $\vc{x} \in (0, \infty)^n$.  We show that $\|\vc{x}\| \geq
\|\vc{x}\|_p$.  By~\eqref{eq:p-norm-norm} and multiplicativity, it is
equivalent to show that
\[
\|\vc{x}^{\otimes r}\|
\geq
\bigl\|
\bigl( 
N(\vc{x}^{\otimes r}, u^r) \mc u^r
\bigr)
\bigr\|
\]
for all real $u > 0$ and integers $r \geq 1$.  But this is clear, since by
Lemma~\ref{lemma:p-norm-elem}\bref{part:p-norm-mono} and the definition of
system of norms, 
\[
\|\vc{x}^{\otimes r}\|
\geq 
\bigl\|\bigl(
\overbrace{
\underbrace{u^r, \ldots, u^r}_{N(\vc{x}^{\otimes r}, u^r)},
0, \ldots, 0}^{n^r}
\bigr)\bigr\|
=
\bigl\| \bigl(
N(\vc{x}^{\otimes r}, u^r) \mc u^r
\bigr) \bigr\|.
\]

\paragraph*{Step 6: the upper bound}
Let $\vc{x} \in (0, \infty)^n$.  We show that $\|\vc{x}\| \leq
\|\vc{x}\|_p$.  The argument is structurally very similar to the 
second part of the proof of Theorem~\ref{thm:cp}, and uses the tensor%
\index{tensor power trick} 
power trick (Tao~\cite{TaoSR}, Section~1.9).

Let $\theta \in (1, \infty)$.  We will prove that $\|\vc{x}\| \leq
\theta\|\vc{x}\|_p$.  Since $\min_i x_i > 0$, we can choose an integer $d
\geq 1$ and real numbers $u_0, \ldots, u_d$ such that
\[
\min_i x_i = u_0 < u_1 < \cdots < u_d = \max_i x_i
\]
and $u_k/u_{k - 1} < \theta$ for all $k \in \{1, \ldots, d\}$.  

Let $r \geq 1$.  We have the vector $\vc{x}^{\otimes r} \in \R^{n^r}$, and
we define a new vector $\vc{y}_r \in \R^{n^r}$ by rounding up each
coordinate of $\vc{x}^{\otimes r}$ to the next element of the set $\{u_1^r,
\ldots, u_d^r\}$.  (Formally, define a map $f_r \from [u_0^r, u_d^r] \to
      [u_0^r, u_d^r]$ by $f_r(w) = u_k^r$, where $k \in \{1, \ldots, d\}$
      is least such that $w \leq u_k^r$. Then $\vc{y}_r$ is obtained from
      $\vc{x}^{\otimes r}$ by applying $f_r$ in each coordinate.)

By construction, $\vc{x}^{\otimes r} \leq \vc{y}_r$, and the number $n_{k,
  r}$ of coordinates of $\vc{y}_r$ equal to $u_k^r$ is at most
$N(\vc{x}^{\otimes r}, u_{k - 1}^r)$.  Hence
\begin{align}
\|\vc{x}^{\otimes r}\|   &
\leq
\|\vc{y}_r\|    
\lbl{eq:ub-1} \\
&
=
\bigl\|
\bigl(
n_{1, r} \mc u_1^r, \ldots, n_{d, r} \mc u_d^r
\bigr)
\bigr\|
\lbl{eq:ub-2} \\
&
\leq
\sum_{k = 1}^d 
\bigl\|
\bigl(
(n_{1, r} + \cdots + n_{k - 1, r}) \mc 0,
n_{k, r} \mc u_k^r,
(n_{k + 1, r} + \cdots + n_{d, r}) \mc 0
\bigr)
\bigr\|
\lbl{eq:ub-3} \\
&
=
\sum_{k = 1}^d \| (n_{k, r} \mc u_k^r) \|
\lbl{eq:ub-4} \\
&
\leq
d \max_{1 \leq k \leq d} \| (n_{k, r} \mc u_k^r) \|
\lbl{eq:ub-5} \\
&
\leq
d \theta^r \max_{1 \leq k \leq d} \| (n_{k, r} \mc u_{k - 1}^r) \|
\lbl{eq:ub-6} \\
&
\leq
d \theta^r \max_{1 \leq k \leq d} 
\bigl\| \bigl(
N(\vc{x}^{\otimes r}, u_{k - 1}^r) \mc u_{k - 1}^r
\bigr) \bigr\|  
\lbl{eq:ub-7} \\
&
\leq
d \theta^r \|\vc{x}\|_p^r,
\lbl{eq:ub-8}
\end{align}
where \eqref{eq:ub-1} is by
Lemma~\ref{lemma:p-norm-elem}\bref{part:p-norm-mono}, \eqref{eq:ub-2} is by
symmetry and definition of $n_{k, r}$, \eqref{eq:ub-3} is by the triangle
inequality, \eqref{eq:ub-4} is by definition of system of norms,
\eqref{eq:ub-5} is elementary, \eqref{eq:ub-6} is by hypothesis on $u_0,
\ldots, u_d$ and Lemma~\ref{lemma:p-norm-elem}\bref{part:p-norm-mono},
\eqref{eq:ub-7} uses Lemma~\ref{lemma:p-norm-elem}\bref{part:p-norm-one},
and~\eqref{eq:ub-8} follows from~\eqref{eq:p-norm-norm}.  Hence by
multiplicativity, 
\[
\|\vc{x}\| 
= 
\|\vc{x}^{\otimes r}\|^{1/r} 
\leq 
d^{1/r} \theta \|\vc{x}\|_p.
\]
This holds for all integers $r \geq 1$ and real numbers $\theta > 1$.
Letting $r \to \infty$ and $\theta \to 1$ gives $\|\vc{x}\| \leq
\|\vc{x}\|_p$, completing the proof of Theorem~\ref{thm:an}.

\begin{remark}
\lbl{rmk:an-diff} 
The proof of Theorem~\ref{thm:an} originally given by Aubrun%
\index{Aubrun, Guillaume}
and Nechita%
\index{Nechita, Ion}
relied on both Cram\'er's theorem and the Legendre--Fenchel theorem.
Effectively, they used Cram\'er's theorem and convex duality to derive the
moment generating function formula of Theorem~\ref{thm:cp} in the specific
case required.

However, Cerf%
\index{Cerf, Rapha\"el}
and Petit~\cite{CePe}%
\index{Petit, Pierre} 
showed how the moment generating function formula can be proved without
these tools. (In fact, they used it as part of their proof of Cram\'er's
theorem.)  The proof of the moment generating function formula given in
Section~\ref{sec:mgfs} is similarly elementary.  Our proof of
Theorem~\ref{thm:an} works directly from the moment generating function
formula, and does not, therefore, need Cram\'er's theorem, the
Legendre--Fenchel theorem, or even the notion of convex conjugate.

Aubrun and Nechita went on to prove similar characterizations of the $L^p$
norms (Theorem~1.2 of~\cite{AuNe}) and the Schatten $p$-norms (their
Theorem~4.2).  The main focus of the article of Fern\'andez-Gonz\'alez,
Palazuelos and P\'erez-Garc\'{i}a~\cite{FGPPG} was also the $L^p$ norms
(their Theorem~3.1).  We do not discuss these results further.
\end{remark}
\index{Aubrun, Guillaume|)}
\index{Nechita, Ion|)}

\section{Multiplicative characterization of the power means}
\lbl{sec:mult-means}

From the multiplicative characterization of the $p$-norms, we derive a
multiplicative characterization of the power means of order at least~$1$.
It differs from the characterizations of power means in
Section~\ref{sec:w-mns} in that it does not assume modularity.  Instead, it
uses the multiplicativity condition of Definition~\ref{defn:w-mult}, as
well as a convexity axiom that provides the connection with norms.

\begin{defn}
\lbl{defn:mean-cvx}
A sequence of functions $\bigl( M \from \Delta_n \times [0, \infty)^n \to
[0, \infty) \bigr)_{n \geq 1}$ is \demph{convex}%
\index{convex!mean}%
\index{mean!convex} 
if
\[
M\bigl( \vc{p}, \hlf(\vc{x} + \vc{y}) \bigr)
\leq
\max\bigl\{ M(\vc{p}, \vc{x}), M(\vc{p}, \vc{y}) \bigr\}
\]
for all $n \geq 1$, $\vc{p} \in \Delta_n$, and $\vc{x}, \vc{y} \in [0,
\infty)^n$. 
\end{defn}

\begin{example}
\lbl{eg:mean-mc}
The power mean $M_t$ is multiplicative for all $t \in [-\infty, \infty]$
(Corollary~\ref{cor:pwr-mns-mult}).  If $t \in [1, \infty]$ then $M_t$ is
also convex.  To show this, it suffices to prove the inequality in
Definition~\ref{defn:mean-cvx} in the case where $\vc{p}$ has full support.
In that case, $M_t$ can be expressed in terms of $\|\cdot\|_t$ by the
formula
\[
M_t(\vc{p}, \vc{x})
=
\| \vc{p}^{1/t} \vc{x} \|_t
\]
($\vc{x} \in [0, \infty)^n$), where both the power and the product of
  vectors are defined coordinatewise.  Now, for $\vc{x}, \vc{y} \in [0,
    \infty)^n$,
\begin{align*}
M_t\bigl( \vc{p}, \hlf(\vc{x} + \vc{y}) \bigr)  &
=
\bigl\| \hlf \vc{p}^{1/t}\vc{x} + \hlf \vc{p}^{1/t}\vc{y} \bigr\|_t     \\
&
\leq
\hlf \bigl\| \vc{p}^{1/t} \vc{x} \bigr\|_t 
+
\hlf \bigl\| \vc{p}^{1/t} \vc{y} \bigr\|_t      \\
&
=
\hlf \bigl( M_t(\vc{p}, \vc{x}) + M_t(\vc{p}, \vc{y}) \bigr)    \\
&
\leq
\max \bigl\{ M_t(\vc{p}, \vc{x}), M_t(\vc{p}, \vc{y}) \bigr\},
\end{align*}
by the triangle inequality for $\|\cdot\|_t$.  Thus, $M_t$ is convex for $t
\in [1, \infty]$.   

On the other hand, $M_t$ is not convex for $t \in [-\infty, 1)$, since then
\[
M_t \Bigl( 
\bigl( \hlf, \hlf \bigr),
\hlf \bigl( (1, 0) + (0, 1) \bigr) 
\Bigr)
=
\hlf
\]
but
\begin{align*}
\max \Bigl\{
M_t \bigl( \bigl(\hlf, \hlf\bigr), (1, 0) \bigr),
M_t \bigl( \bigl(\hlf, \hlf\bigr), (0, 1) \bigr)
\Bigr\} &
=
M_t\bigl( \bigl(\hlf, \hlf\bigr), (1, 0) \bigr)   \\
&
=
\begin{cases}
\bigl( \hlf \bigr)^{1/t}        &\text{if } t \in (0, 1),       \\
0                               &\text{if } t \in [-\infty, 0],
\end{cases}
\end{align*}
which is strictly less than $1/2$.
\end{example}

The multiplicative characterization of the power means is as follows.  For
a review of the terminology used in~\bref{part:mcpm-condns}, see
Appendix~\ref{app:condns}.

\begin{thm}
\lbl{thm:mcpm}%
\index{power mean!characterization of!multiplicative}%
\index{power mean!characterization of!weighted on $[0, \infty)$}%
\index{multiplicative!characterization of power means}
Let $\bigl( M \from \Delta_n \times [0, \infty)^n \to [0, \infty) \bigr)_{n
\geq 1}$ be a sequence of functions.  The following are equivalent:
\begin{enumerate}
\item 
\lbl{part:mcpm-condns}
$M$ is natural, consistent, increasing, multiplicative, and convex;

\item
\lbl{part:mcpm-form}
$M = M_t$ for some $t \in [1, \infty]$.
\end{enumerate}
\end{thm}

The proof follows shortly.

\begin{remarks}
\lbl{rmks:mcpm-matisos}
\begin{enumerate}
\item 
\lbl{rmk:mcpm-m-condns} 
We have already made some elementary inferences from combinations of the
properties in part~\bref{part:mcpm-condns} of the theorem.  In the proof of
Lemma~\ref{lemma:pwr-mns-elem} (p.~\pageref{p:lpme-pf}), we showed that
naturality implies symmetry, absence-invariance and repetition.  Since $M$
is increasing, Lemma~\ref{lemma:transfer} then implies that $M$ also has
the transfer property.  Moreover, $M$ is homogeneous, since for all $\p \in
\Delta_n$, $\vc{x} \in [0, \infty)^n$, and $c \in [0, \infty)$,
\begin{align*}
M(\vc{p}, c\vc{x})      &
=
M\bigl(\vc{u}_1 \otimes \vc{p}, (c) \otimes \vc{x}\bigr)        \\
&
=
M\bigl( \vc{u}_1, (c) \bigr) M(\vc{p}, \vc{x})  \\
&
=
c M(\vc{p}, \vc{x}), 
\end{align*}
by definition of $\otimes$, multiplicativity, and consistency.

\item
\lbl{rmk:mcpm-m-cons}
Theorem~\ref{thm:mcpm} first appeared as Theorem~1.3 of
Leinster~\cite{MCPM}.  There, the result was stated with a superficially
weaker consistency axiom: that $M(\vc{u}_1, (x)) = x$ for all $x \in [0,
\infty)$.  But in the presence of the naturality property, this easily
implies full consistency: for naturality implies repetition (as just
noted), which in turn implies that
\[
M\bigl(\vc{p}, (x, \ldots, x)\bigr)
=
M\bigl((p_1 + \cdots + p_n), (x)\bigr)
=
M\bigl(\vc{u}_1, (x)\bigr)
=
x
\]
for all $\p \in \Delta_n$ and $x \in [0, \infty)$. 
\end{enumerate}
\end{remarks}

We now embark on the proof of Theorem~\ref{thm:mcpm}.
Certainly~\bref{part:mcpm-form} implies~\bref{part:mcpm-condns}, by
Lemmas~\ref{lemma:pwr-mns-nat}, \ref{lemma:pwr-mns-con}
and~\ref{lemma:pwr-mns-inc}, Corollary~\ref{cor:pwr-mns-mult}, and
Example~\ref{eg:mean-mc}.  For the converse, and \femph{for the rest of
  this section}, let $M$ be a sequence of functions satisfying the
conditions in Theorem~\ref{thm:mcpm}\bref{part:mcpm-condns}.  We will prove
that $M = M_t$ for some $t \in [1, \infty]$.

\paragraph*{Step 1: finding $t$} 
The observation behind this step is that 
\[
M_t \bigl((p, 1 - p), (1, 0)\bigr) = p^{1/t}
\]
for all $p \in (0, 1)$. 

Define a function $f \from (0, 1) \to [0, \infty)$ by
\[
f(p) = M\bigl( (p, 1 - p), (1, 0) \bigr).
\]
By multiplicativity and repetition (proved in
Remark~\ref{rmks:mcpm-matisos}\bref{rmk:mcpm-m-condns}),
\begin{align*}
f(p)f(r)        &
=
M\bigl( (p, 1 - p), (1, 0) \bigr)
\cdot
M\bigl( (r, 1 - r), (1, 0) \bigr)       \\
&
=
M\bigl( 
(p, 1 - p) \otimes (r, 1 - r), \, (1, 0) \otimes (1, 0)
\bigr)  \\
&
=
M\Bigl(
\bigl(pr, p(1 - r), (1 - p)r, (1 - p)(1 - r)\bigr), \, 
(1, 0, 0, 0)
\Bigr)  \\
&
=
M\bigl( (pr, 1 - pr), \, (1, 0) \bigr)  \\
&
=
f(pr)
\end{align*}
for all $p, r \in (0, 1)$.  By transfer
(Remark~\ref{rmks:mcpm-matisos}\bref{rmk:mcpm-m-condns}),  $f$ is
increasing.  If $f(r) = 0$ for some $r \in (0, 1)$ then for all $p \in (0,
1)$, 
\[
f(p)
=
f(p/r) f(r)
=
0 
=
p^\infty.
\]
If not, then $f$ defines an increasing multiplicative function $(0, 1) \to
(0, \infty)$, so by Corollary~\ref{cor:cauchy-mult-01}, there is some
constant $c \in [0, \infty)$ such that $f(p) = p^c$ for all $p \in (0, 1)$.
  So in either case, there is a constant $c \in [0, \infty]$ such that
  $f(p) = p^c$ for all $p \in (0, 1)$.  But
\[
f\bigl(\hlf\bigr)
=
M\bigl( \vc{u}_2, (1, 0) \bigr)
=
M\bigl( \vc{u}_2, (0, 1) \bigr)
\]
by symmetry, so 
\begin{align*}
\bigl(\hlf\bigr)^c        &
=
f\bigl(\hlf\bigr) \\
&
=
\max\Bigl\{
M\bigl( \vc{u}_2, (1, 0) \bigr),
M\bigl( \vc{u}_2, (0, 1) \bigr) 
\Bigr\} \\
&
\geq
M\bigl( \vc{u}_2, \bigl(\hlf, \hlf\bigr) \bigr)
=
\hlf
\end{align*}
by convexity and consistency.  It follows that $c \in [0, 1]$.  Put $t =
1/c \in [1, \infty]$: then
\begin{equation}
\lbl{eq:agree-10}
M\bigl( (p, 1 - p), (1, 0) \bigr)
=
p^{1/t}
=
M_t \bigl( (p, 1 - p), (1, 0) \bigr)
\end{equation}
for all $p \in (0, 1)$.

\paragraph*{Step 2: constructing a system of norms} 
Here we take our inspiration from the relationship
\[
\|\vc{x}\|_t
=
n^{1/t} 
M_t\bigl(
\vc{u}_n, (\mg{x_1}, \ldots, \mg{x_n})
\bigr)
\]
($\vc{x} \in \R^n$) between the $t$-norm and the power mean of order $t$.

For each $n \geq 1$, define a function $\|\cdot\| \from \R^n \to [0,
  \infty)$ by
\[
\|\vc{x}\|
=
n^{1/t} 
M\bigl(
\vc{u}_n, (\mg{x_1}, \ldots, \mg{x_n})
\bigr)
\]
($\vc{x} \in \R^n$).  To cover the case $n = 0$, let $\|\cdot\| \from \R^0
\to [0, \infty)$ be the function whose single value is $0$.  The next few
lemmas show that $\|\cdot\|$ is a multiplicative system of norms.

\begin{lemma}
\lbl{lemma:mcpm-pwr}
$n^{-1/t} = M\bigl(\vc{u}_n, (1, 0, \ldots, 0)\bigr)$ for all $n \geq 1$.  
\end{lemma}

\begin{proof}
By the defining property of $t$ (equation~\eqref{eq:agree-10}) and the
repetition property of $M$, both sides are equal to $M\bigl( (1/n, 1 -
1/n), (1, 0) \bigr)$.
\end{proof}

\begin{lemma}
\lbl{lemma:mcpm-yes-norm}
For each $n \geq 0$, the function $\|\cdot\| \from \R^n \to [0, \infty)$
  is a norm.
\end{lemma}

\begin{proof}
This is trivial when $n = 0$; suppose that $n \geq 1$.  We verify the three
conditions in the definition of norm (Definition~\ref{defn:norm}).

First, we have to prove that if $\vc{0} \neq \vc{x} \in \R^n$ then
$\|\vc{x}\| \neq 0$.  We may assume by symmetry that $x_1 \neq 0$, and then
\begin{align*}
\|\vc{x}\|      &
\geq
n^{1/t} M\bigl(\vc{u}_n, (\mg{x_1}, 0, \ldots, 0) \bigr)   \\
&
=
n^{1/t} \mg{x_1} M\bigl(\vc{u}_n, (1, 0, \ldots, 0)\bigr)  \\
&
=
\mg{x_1} > 0
\end{align*}
by definition of $\|\vc{x}\|$, the increasing and homogeneity properties of
$M$, and 
Lemma~\ref{lemma:mcpm-pwr}. 

The homogeneity of $M$ implies that $\|c\vc{x}\| = \mg{c}\,\|\vc{x}\|$ for
all $\vc{x} \in \R^n$ and $c \in \R$.

It remains to prove the triangle inequality, which we do in stages.  First
let $\vc{x}, \vc{y} \in \R^n$ with $\|\vc{x}\|, \|\vc{y}\| \leq 1$ and
$x_i, y_i \geq 0$ for all $i$.  Using the convexity of $M$,
\begin{align*}
\bigl\| \hlf \vc{x} + \hlf \vc{y} \bigr\|       &
=
n^{1/t} M\bigl( \vc{u}_n, \hlf(\vc{x} + \vc{y}) \bigr)  \\
&
\leq
n^{1/t} \max\bigl\{ M(\vc{u}_n, \vc{x}), M(\vc{u}_n, \vc{y}) \bigr\}    \\
&
=
\max \bigl\{ \|\vc{x}\|, \|\vc{y}\| \bigr\}    \\
&
\leq
1.
\end{align*}
It follows that
\begin{equation}
\lbl{eq:tri-cvx}
\|\lambda \vc{x} + (1 - \lambda) \vc{y}\| \leq 1
\end{equation}
for all dyadic rationals $\lambda = k/2^\ell \in [0, 1]$, by induction on
$\ell$.  We now show that~\eqref{eq:tri-cvx} holds for all $\lambda \in [0,
  1]$.  Indeed, given $\lambda \in [0, 1]$ and $\epsln > 0$, we can choose
a dyadic rational $\lambda' \in [0, 1]$ such that
\[
\lambda \leq (1 + \epsln) \lambda',
\qquad
1 - \lambda \leq (1 + \epsln)(1 - \lambda'),
\]
and then
\begin{align*}
\| \lambda \vc{x} + (1 - \lambda) \vc{y} \|   &
\leq
\| 
(1 + \epsln) \lambda' \vc{x} + (1 + \epsln)(1 - \lambda') \vc{y} 
\| \\
&
=
(1 + \epsln) \| \lambda' \vc{x} + (1 - \lambda') \vc{y} \|    \\
&
\leq
1 + \epsln,
\end{align*}
where in the first inequality, we used the assumptions that $M$ is
increasing and $x_i, y_i \geq 0$.  This holds for all $\epsln > 0$, proving
the claimed inequality~\eqref{eq:tri-cvx}.

Now take any $\vc{x}, \vc{y} \in \R^n$ with $x_i, y_i \geq 0$ for all $i$.
We will prove that 
\begin{equation}
\lbl{eq:tri-mn}
\|\vc{x} + \vc{y}\| \leq \|\vc{x}\| + \|\vc{y}\|.
\end{equation}
This is immediate if $\vc{x} = \vc{0}$ or $\vc{y} = \vc{0}$.  Supposing
otherwise, put
\[
\hat{\vc{x}} = \frac{\vc{x}}{\|\vc{x}\|},
\qquad
\hat{\vc{y}} = \frac{\vc{y}}{\|\vc{y}\|},
\qquad
\lambda = \frac{\|\vc{x}\|}{\|\vc{x}\| + \|\vc{y}\|}.
\]
Then $\|\hat{\vc{x}}\| = \|\hat{\vc{y}}\| = 1$, so by 
inequality~\eqref{eq:tri-cvx} applied to $\hat{\vc{x}}$, $\hat{\vc{y}}$ and
$\lambda$, 
\[
\|\vc{x} + \vc{y}\|     
=
( \|\vc{x}\| + \|\vc{y}\| ) \,
\| \lambda \hat{\vc{x}} + (1 - \lambda) \hat{\vc{y}} \|     
\leq
\|\vc{x}\| + \|\vc{y}\|.
\]

Finally, take any $\vc{x}, \vc{y} \in \R^n$.  To prove the triangle
inequality~\eqref{eq:tri-mn}, put $\vc{x}' = (\mg{x_1}, \ldots, \mg{x_n})$
and $\vc{y}' = (\mg{y_1}, \ldots, \mg{y_n})$.  Then $\|\vc{x}\| =
\|\vc{x}'\|$ and $\|\vc{y}\| = \|\vc{y}'\|$ by definition of $\|\cdot\|$,
and
\[
\|\vc{x} + \vc{y}\| \leq \|\vc{x}' + \vc{y}'\|
\]
since $M$ is increasing.  By the inequality proved in the previous
paragraph, 
\[
\|\vc{x}' + \vc{y}'\| \leq \|\vc{x}'\| + \|\vc{y}'\|,
\]
and the triangle inequality~\eqref{eq:tri-mn} follows.
\end{proof}

\begin{lemma}
\lbl{lemma:mcpm-yes-sys}
$\|\cdot\|$ is a multiplicative system of norms.  
\end{lemma}

\begin{proof}
We have just shown that $\|\cdot\|$ is a norm on $\R^n$ for each individual
$n$.  Symmetry of $M$ implies symmetry of $\|\cdot\|$, so to show that
$\|\cdot\|$ is a system of norms, it suffices to prove that
\begin{equation}
\lbl{eq:sys-pad}
\| (x_1, \ldots, x_n) \|
=
\| (x_1, \ldots, x_n, 0) \|
\end{equation}
for all $n \geq 1$ and $\vc{x} \in \R^n$.  By definition of $\|\cdot\|$ and
Lemma~\ref{lemma:mcpm-pwr}, equation~\eqref{eq:sys-pad} is equivalent to
\[
\frac{M\bigl( \vc{u}_n, (\mg{x_1}, \ldots, \mg{x_n}) \bigr)}%
{M\bigl(\vc{u}_n, (1, 0, \ldots, 0) \bigr)}
=
\frac{M\bigl( \vc{u}_{n + 1}, (\mg{x_1}, \ldots, \mg{x_n}, 0) \bigr)}%
{M\bigl(\vc{u}_{n + 1}, (1, 0, \ldots, 0, 0) \bigr)},
\]
or equivalently,
\begin{align*}
&
M\bigl( \vc{u}_{n + 1}, (1, 0, \ldots, 0, 0) \bigr)
\cdot
M\bigl( \vc{u}_n, (\mg{x_1}, \ldots, \mg{x_n}) \bigr) \\
=\ 
&
M\bigl( \vc{u}_n, (1, 0, \ldots, 0) \bigr)
\cdot
M\bigl( \vc{u}_{n + 1}, (\mg{x_1}, \ldots, \mg{x_n}, 0) \bigr).
\end{align*}
But by multiplicativity and symmetry, both sides are equal to
\[
M\bigl(
\vc{u}_{n(n + 1)},
(\mg{x_1}, \ldots, \mg{x_n}, \underbrace{0, \ldots, 0}_{n^2})
\bigr),
\]
proving~\eqref{eq:sys-pad}.

Finally, the system of norms $\|\cdot\|$ is multiplicative, by
multiplicativity of $M$.
\end{proof}

\paragraph*{Step 3: using the norm theorem}
It now follows from Theorem~\ref{thm:an} that $\|\cdot\| = \|\cdot\|_s$ for
some $s \in [1, \infty]$.  Thus, $\|\vc{x}\|_s = n^{1/t} M(\vc{u}_n,
\vc{x})$ for all $n \geq 1$ and $\vc{x} \in [0, \infty)^n$.  But also,
  $\|\vc{x}\|_s = n^{1/s} M_s(\vc{u}_n, \vc{x})$, so
\[
n^{1/t} M(\vc{u}_n, \vc{x})
=
n^{1/s} M_s(\vc{u}_n, \vc{x})
\]
for all $n \geq 1$ and $\vc{x} \in [0, \infty)$.  Putting $n = 2$ and
  $\vc{x} = (1, 1)$, and using the consistency of both $M$ and $M_s$, gives
  $s = t$.  Hence for all $n \geq 1$ and $\vc{x} \in [0, \infty)^n$,
\[
M(\vc{u}_n, \vc{x})
=
M_t(\vc{u}_n, \vc{x}).
\]

\paragraph*{Step 4: arbitrary weights}
We have now shown that $M(\vc{u}_n, -) = M_t(\vc{u}_n, -)$ for all $n \geq
1$.  To extend the equality to arbitrary weights, we use
Proposition~\ref{propn:u-to-w}, taking $M' = M_t$ there.  The hypotheses of
that proposition are satisfied, by
Remark~\ref{rmks:mcpm-matisos}\bref{rmk:mcpm-m-condns} and
Lemma~\ref{lemma:pwr-mns-cts-px}\bref{part:pwr-mns-cts-px-1}.  Hence $M =
M_t$.

This completes the proof of Theorem~\ref{thm:mcpm}, the multiplicative
characterization of the power means.

%% file: loss.tex
\chapter{Information loss}
\lbl{ch:loss}
\index{information loss}

\begin{quote}%
\index{Grothendieck, Alexander}%
\index{Katz, Nicholas}
Grothendieck came along and said, `No, the Riemann--Roch%
\index{Riemann, Bernhard!Roch theorem@--Roch theorem} 
theorem is \emph{not} a theorem about varieties, it's a theorem about
morphisms between varieties.'  
\hfill 
-- Nicholas Katz (quoted in \cite{JackCAN}, p.~1046).
\end{quote}

\noindent
This short chapter tells the following story.
A measure-preserving map between finite probability spaces can be regarded
as a deterministic%
\index{deterministic process} 
process.  As such, it loses information.  We can attempt to quantify how
much information is lost.  It turns out that as soon as we impose a few
reasonable requirements on this quantity, it is highly constrained: up to a
constant factor, it must be the difference between the entropies of the
domain and the codomain.  That is our main theorem.

This result is essentially another characterization of Shannon entropy, and
first appeared in a 2011 paper of Baez, Fritz and Leinster~\cite{CETIL}.
The broad idea is to shift the focus from \emph{objects}%
\index{objects vs.\ maps} 
(finite probability spaces) to \emph{maps}%
\index{maps vs.\ objects} 
between objects (measure-preserving maps).  Entropy is an invariant of
finite probability spaces; information loss is an invariant of
measure-preserving maps.  The shift of emphasis from objects to maps is
integral to category theory, and has borne fruit such as the
Grothendieck--Riemann--Roch theorem alluded to in the opening quotation, as
well as the considerably more humble characterization of information loss
described here.

In full categorical generality, a map $\Xx \toby{f} \Yy$ of any kind can be
viewed as an object $\Xx$ parametrized by another object $\Yy$.
An object $\Xx$ can be viewed as a map of a special kind, namely, the
unique map $\Xx \toby{!_\Xx} 1$ to the terminal object $1$ of the category
concerned.  In the case at hand, we associate with any probability
space $\Xx$ the unique measure-preserving map $\Xx \toby{!_\Xx} 1$ to the
one-point space $1$, and the information loss of the map $!_\Xx$ is equal
to the entropy of the space $\Xx$.  Thus, entropy is a special case of
information loss.

An advantage of working with information loss rather than entropy (that is,
maps rather than objects) is that the characterization theorems take on a
new simplicity.  For instance, the conditions in our main result
(Theorem~\ref{thm:cetil}) look just like the linearity or homomorphism
conditions that appear throughout mathematics.  In contrast, the
chain%
\index{chain rule!complicated nature of} 
rule for entropy, while justifiable in many other ways, has a more
complicated algebraic form.

We begin with a review of measure-preserving maps, then define information
loss (Section~\ref{sec:meas-pres}).  After recording a few simple
properties of information loss, we prove that they characterize it uniquely
(Section~\ref{sec:cil}).  An analogous and even simpler result is then
proved for $q$-logarithmic information loss ($q \neq 1$).  Both of these
theorems first appeared in the 2011 paper of Baez, Fritz and
Leinster~\cite{CETIL}.

\section{Measure-preserving maps}
\lbl{sec:meas-pres}

So far in this text, we have focused on probability distributions on finite
sets of the special form $\{1, \ldots, n\}$.  Here, it is convenient to use
arbitrary finite sets.  The difference is cosmetic, but does cause some
shifts in notation, as follows.

\begin{defn}
\begin{enumerate}
\item 
Let $\Xx$ be a finite set.  A \demph{probability%
\index{probability distribution} 
distribution} $\p$ on $\Xx$ is a family $(p_i)_{i \in \Xx}$ of nonnegative
real numbers such that $\sum_{i \in \Xx} p_i = 1$.  We write
$\Delta_\Xx$\ntn{DeltaX} for the set of probability distributions on $\Xx$.

\item
A \demph{finite probability%
\index{probability space} 
space} is a pair $(\Xx, \p)$ where $\Xx$ is a finite set and $\p \in
\Delta_\Xx$.
\end{enumerate}
\end{defn}

The set $\Delta_\Xx$ is topologized as a subspace of the product space
$\R^\Xx$. 

\begin{defn}
\lbl{defn:meas-pres}
Let $(\Yy, \vc{s})$ and $(\Xx, \p)$ be finite probability spaces.  A
\demph{measure-preserving%
\index{measure-preserving map} 
map} $(\Yy, \vc{s}) \to (\Xx, \p)$ is a function $f \from \Yy \to \Xx$ such
that
\begin{equation}
\lbl{eq:meas-pres}
p_i = \sum_{j \in f^{-1}(i)} s_j
\end{equation}
for all $i \in \Xx$.  
\end{defn}

An equivalent statement is that $f \from (\Yy, \vc{s}) \to (\Xx, \p)$ is
measure-preserving if and only if
\begin{equation}
\lbl{eq:meas-pres-full}
\sum_{i \in \VV} p_i = \sum_{j \in f^{-1}\VV} s_j
\end{equation}
for all $\VV \sub \Xx$.  Indeed, \eqref{eq:meas-pres}~is the case
of~\eqref{eq:meas-pres-full} where $\VV = \{i\}$,
and~\eqref{eq:meas-pres-full} follows from~\eqref{eq:meas-pres} by summing
over all $i \in \VV$.  

\begin{remarks}
\lbl{rmks:finprob}
\begin{enumerate}
\item 
\lbl{rmk:finprob-pf} 
For any finite probability space $(\Yy, \vc{s})$ and function $f$ from
$\Yy$ to another finite set $\Xx$, there is an induced probability
distribution $f \vc{s}$ on $\Xx$, the \dmph{pushforward} of $\vc{s}$ along
$f$.  It is defined by the obvious generalization of
Definition~\ref{defn:pfwd}:
\[
(f \vc{s})_i = \sum_{j \in f^{-1}(i)} s_j
\ntn{pfwdgen}
\]
($i \in \Xx$).  In these terms, a function $f \from (\Yy, \vc{s}) \to (\Xx, \p)$
is measure-preserving if and only if $f \vc{s} = \p$.

\item
Finite probability spaces and measure-preserving maps form a category
$\FinProb$\ntn{FinProb}.  We note in passing that by~\bref{rmk:finprob-pf},
the forgetful functor $\FinProb \to \FinSet$ is a discrete opfibration.  In
fact, $\FinProb$ is the category of elements of the functor $\FinSet \to
\Set$ defined on objects by $\Xx \mapsto \Delta_\Xx$ and on maps by
pushforward. (For the categorical terminology used here, see for instance 
Riehl~\cite{RiehCTC}, Definition~2.4.1 and Exercise~2.4.viii.)
\end{enumerate}
\end{remarks}

Although a measure-preserving map need not be literally surjective, it is
essentially so, in the sense that all elements not in the image have
probability zero.

\begin{example}
\lbl{eg:meas-pres-surj}
Let $\Yy = \{ \as{a}, \as{\`a}, \as{\^a}, \as{b}, \as{c}, \as{\c{c}},
\ldots \}$ be the set of symbols in the French%
\index{French language}
language, and let $\vc{s} \in \Delta_\Yy$ be their frequency distribution
(as in Example~\ref{eg:comp-french}).  Let $\Xx = \{ \as{a}, \as{b},
\as{c}, \ldots \}$ be the 26-element set of letters, and $\p \in
\Delta_\Xx$ their frequency distribution.  There is a function $f \from \Yy
\to \Xx$ that forgets accents; for instance, $f(\as{a}) = f(\as{\`a}) =
f(\as{\^a}) = \as{a}$.  Then $f \from (\Yy, \vc{s}) \to (\Xx, \p)$ is
measure-preserving and surjective.
\end{example}

\begin{example}
\lbl{eg:meas-pres-incl}
Let $\ell$ be the inclusion function $\{1\} \incl \{1, 2\}$.  Give $\{1\}$
its unique probability distribution $(1) = \vc{u}_1$, and give $\{1, 2\}$
the distribution $(1, 0)$.  Then $\ell$ is measure-preserving but not
surjective. 
\end{example}

Any measure-preserving map between finite probability spaces can be
factorized%
\index{measure-preserving map!factorization of} 
canonically into maps of the two types in these two examples: a
surjection followed by a subset inclusion, where the subset concerned has
total probability~$1$.  Specifically, $f \from (\Yy, \vc{s}) \to (\Xx, \p)$
factorizes as
\[
(\Yy, \vc{s}) \toby{f'} (f\Yy, \p') \toby{\ell} (\Xx, \p),
\]
where $\p'$ is the probability distribution on $f\Yy$ defined by
$p'_i = p_i$ for all $i \in f\Yy$, the surjection $f'$ is defined by $f'(j)
= f(j)$ for all $j \in \Yy$, and $\ell$ is inclusion.  

A measure-preserving surjection simply discards information (such as the
accents in Example~\ref{eg:meas-pres-surj}).  It is a coarse-graining, in
the sense of taking finely-grained information (such as letters with
accents) and converting it into more coarsely-grained information (such as
mere letters).
A measure-preserving inclusion is essentially trivial, simply
appending some events of probability zero.

For any measure-preserving bijection $f \from (\Yy, \vc{s}) \to (\Xx, \p)$
between finite probability spaces, the inverse $f^{-1}$ is also
measure-preserving.  We call such an $f$ an \demph{isomorphism},%
\index{isomorphism of probability spaces}%
\index{probability space!isomorphism of}
and write $(\Yy, \vc{s}) \iso (\Xx, \p)$.

An important feature of probability spaces is that we can take convex%
\index{convex!combination}%
\index{probability space!convex combination of}
combinations of them.  Given $\vc{w} \in \Delta_n$ and finite probability
spaces $(\Xx_1, \p^1), \ldots, (\Xx_n, \p^n)$, we obtain a new probability
space
%
\[
\Biggl(
\coprod_{i = 1}^n \Xx_i, 
\,
\coprod_{i = 1}^n w_i \p^i
\Biggr),
\ntn{bigccdists}
\]
%
where $\coprod \Xx_i$\ntn{disjtu} is the disjoint union of sets $\Xx_1 \cpd
\cdots \cpd \Xx_n$ and $\coprod w_i \p^i$ is the probability distribution
on $\coprod \Xx_i$ that gives probability $w_i p^i_j$ to an element $j \in
\Xx_i$.

Convex combination of probability spaces is just composition of probability
distributions, translated into different notation.  More exactly, if $\Xx_i
= \{1, \ldots, k_i\}$ then $\coprod \Xx_i$ is in canonical bijection with
$\{1, \ldots, k_1 + \cdots + k_n\}$, and under this bijection, $\coprod w_i
\p^i$ corresponds to the composite distribution $\vc{w} \of (\p^1, \ldots,
\p^n)$.

The construction of convex combinations is functorial,%
\lbl{p:conv-comb-func} 
that is, applies not only to probability spaces but also to maps between
them.  Indeed, take measure-preserving maps
\begin{eqnarray*}
(\Yy_1, \vc{s}^1) &\toby{f_1}& (\Xx_1, \p^1)   \\
\vdots& &\vdots \\
(\Yy_n, \vc{s}^n) &\toby{f_n}& (\Xx_n, \p^n)
\end{eqnarray*}
between finite probability spaces, and a probability distribution $\vc{w}
\in \Delta_n$.  There is a function
\[
\xymatrix@C+1em{
\displaystyle
\coprod_{i = 1}^n \Yy_i 
\ar[r]^{\coprod\limits_{i = 1}^n f_i} &
\displaystyle
\coprod_{i = 1}^n \Xx_i
}
\]
that maps $j \in \Yy_i$ to $f_i(j) \in \Xx_i$, and it is easily checked
that $\coprod f_i$ is a measure-preserving map
\begin{equation}
\lbl{eq:conv-comb-map}
\xymatrix@C+1em{
\displaystyle
\Biggl( \coprod_{i = 1}^n \Yy_i, \, \coprod_{i = 1}^n w_i \vc{s}^i \Biggr)
\ar[r]^{\coprod\limits_{i = 1}^n f_i} &
\displaystyle
\Biggl( \coprod_{i = 1}^n \Xx_i, \, \coprod_{i = 1}^n w_i \vc{p}^i \Biggr).
}
\end{equation}
It will be convenient to use the alternative notation
\[
\coprod_{i = 1}^n w_i f_i
\qquad
\text{or} 
\qquad
w_1 f_1 \cpd \cdots \cpd w_n f_n
\ntn{ccmaps}
\]
for the measure-preserving map $\coprod_{i = 1}^n f_i$
of~\bref{eq:conv-comb-map}. 

We defined Shannon%
\index{Shannon, Claude!entropy}%
\index{entropy!Shannon}
entropy only for probability distributions on sets of
the form $\{1, \ldots, n\}$, but, of course, the definition for general
finite probability spaces $(\Xx, \p)$ is
\[
H(\p) 
=
- \sum_{i \in \supp(\p)} p_i \log p_i,
\ntn{Hgen}
\]
where \ntn{suppgen}$\supp(\p) = \{ i \in \Xx \such p_i > 0\}$.  Shannon
entropy is
\demph{isomorphism-invariant},%
\index{isomorphism invariant@isomorphism-invariant}
meaning that $H(\p) = H(\vc{s})$ whenever $(\Xx, \vc{p})$ and $(\Yy,
\vc{s})$ are isomorphic finite probability spaces.

Translated into this notation, the chain rule for Shannon entropy
states that
\begin{equation}
\lbl{eq:g-ent-chain}
H\Biggl( \coprod_{i = 1}^n w_i \p^i \Biggr)
=
H(\vc{w}) + \sum_{i = 1}^n w_i H(\p^i)
\end{equation}
for all $\vc{w} \in \Delta_n$ and finite probability spaces $(\Xx_1, \p^1),
\ldots, (\Xx_n, \p^n)$. The continuity property of entropy is that for
each finite set $\Xx$, the function
\begin{equation}
\lbl{eq:g-cts}
\begin{array}{ccc}
\Delta_\Xx        &\to            &\R     \\
\p              &\mapsto        &H(\p)
\end{array}
\end{equation}
is continuous.

We now set out to quantify the information lost by a measure-preserving map
$f$, first exploring through examples how a reasonable definition of
information loss ought to behave.

\begin{example}
If $f$ is an isomorphism then $f$ should lose no information at all.
More generally, the same should be true if $f$ is injective. 
\end{example}

\begin{example}
\lbl{eg:info-loss-toss}
The unique measure-preserving map $(\{1, 2\}, \vc{u}_2) \to (\{1\},
\vc{u}_1)$ forgets the result of a fair coin%
\index{coin!toss} 
toss.  Intuitively, then, it loses one bit of information.
\end{example}

\begin{example}
\lbl{eg:info-loss-bang}
More generally, for any finite probability space $(\Xx, \p)$, consider the
unique measure-preserving map
\[
f \from (\Xx, \p)
\to 
\bigl(\{1\}, \vc{u}_1\bigr),
\]
which forgets the result of an observation drawn from the distribution $\p$.
Such an observation contains $\Hi(\p)$ bits of information (in the sense
of Section~\ref{sec:ent-coding}), so the information lost by $f$ should be
$\Hi(\p)$ bits.
\end{example}

\begin{example}
Suppose that I draw fairly from a pack of playing cards,%
\index{cards, playing} 
and tell you only the rank (number) of the card chosen.  The
information that I am withholding is the suit, which needs $\log_2 4 = 2$
bits to encode.  Thus, if $f \from \Yy \to \Xx$ is a four-to-one map from a
$52$-element set $\Yy$ to a $13$-element set $\Xx$, and if we equip $\Yy$
and $\Xx$ with their uniform distributions $\vc{u}_\Yy$ and $\vc{u}_\Xx$,
then the information loss of the measure-preserving map $f \from (\Yy,
\vc{u}_\Yy) \to (\Xx, \vc{u}_\Xx)$ should be $2$~bits.
\end{example}

\begin{example}
\lbl{eg:info-loss-french}
Take the measure-preserving map
\[
f \from
\bigl(
\{ \as{a}, \as{\`a}, \as{\^a}, \as{b}, \ldots \}, \vc{s}
\bigr)
\to
\bigl(
\{ \as{a}, \as{b}, \ldots \},
\p
\bigr)
\]
of Example~\ref{eg:meas-pres-surj}, representing the process of forgetting
the accent on a letter in the French%
\index{French language} 
language.  There are two quantities that we could reasonably call the
`amount of information lost' by the process $f$.

First, we could condition on the underlying letter.  To do this, we go
through the $26$ letters, we take for each letter the amount of information
lost by forgetting the accent on that letter, and we form the weighted
mean.  Write
\[
\vc{r}^1 \in \Delta_3, \ 
\vc{r}^2 \in \Delta_1, \ 
\ldots, \ 
\vc{r}^{26} \in \Delta_1
\]
for the accent distributions on each letter, so that $\vc{s} = \coprod_{i =
  1}^{26} p_i \vc{r}^i$.  As in Example~\ref{eg:info-loss-bang}, the
amount of information lost by forgetting the accent on an $\as{a}$ (for
instance) should be $\Hi(\vc{r}^1)$ bits.  So, the expected amount of
information lost by forgetting the accent on a random letter should be
\begin{equation}
\lbl{eq:french-loss}
\sum_{i = 1}^{26} p_i \Hi(\vc{r}^i).
\end{equation}
This is one possible definition of the amount of information lost by $f$. 

Alternatively, we could define the information loss to be
the amount of information we had at the start of the process minus the
amount of information that remains at the end.  This is
\begin{equation}
\lbl{eq:french-diff}
\Hi(\vc{s}) - \Hi(\vc{p}).
\end{equation}
But since $\vc{s} = \coprod p_i \vc{r}^i$, the chain
rule~\eqref{eq:g-ent-chain} tells us that the two
quantities~\eqref{eq:french-loss} and~\eqref{eq:french-diff} are equal.
So, our two ways of quantifying information loss are equivalent.
\end{example}

Motivated by these examples, we make the following definition.

\begin{defn}
\lbl{defn:loss}
Let 
\[
f \from (\Yy, \vc{s}) \to (\Xx, \p)
\]
be a measure-preserving map of finite probability spaces.  The
\demph{information%
\index{information loss}
loss} of $f$ is
\[
L(f) = H(\vc{s}) - H(\vc{p}).
\ntn{loss}
\]
\end{defn}

As with other entropic quantities that we have encountered, the definition
of information loss depends on a choice of logarithmic base, and
changing that base scales the quantity by a constant factor.

A deterministic%
\index{deterministic process}
process cannot create new information, and correspondingly,
information loss is always nonnegative:%
\index{information loss!nonnegative@is nonnegative}

\begin{lemma}
\lbl{lemma:loss}
Let $f \from (\Yy, \vc{s}) \to (\Xx, \p)$ be a measure-preserving map of finite
probability spaces.  Then:
\begin{enumerate}
\item
\lbl{part:loss-formula}
$\displaystyle 
L(f)
=
\sum_{j \in \supp(\vc{s})} s_j \log \frac{p_{f(j)}}{s_j}$;

\item
\lbl{part:loss-ineq}
$L(f) \geq 0$.
\end{enumerate}
\end{lemma}

\begin{proof}
By definition of measure-preserving map
(Definition~\ref{defn:meas-pres}), $p_{f(j)} \geq s_j$ for all $j \in \Yy$.
It follows that 
\begin{equation}
\lbl{eq:loss-supp}
j \in \supp(\vc{s}) \implies f(j) \in \supp(\p).
\end{equation}
It also follows that $\log(p_{f(j)}/s_j) \geq 0$ for all $j \in
\supp(\vc{s})$, so part~\bref{part:loss-ineq} will follow once
we have proved~\bref{part:loss-formula}.

To prove~\bref{part:loss-formula}, first note that by definition of
measure-preserving map,
\begin{align*}
H(\p)   &
=
\sum_{i \in \supp(\p)} p_i \log \frac{1}{p_i}   \\
&
=
\sum_{i \in \supp(\p), \ j \in \Yy\csuch f(j) = i}
s_j \log \frac{1}{p_i}  \\
&
=
\sum_{j \csuch f(j) \in \supp(\p)} s_j \log \frac{1}{p_{f(j)}}.
\end{align*}
By~\eqref{eq:loss-supp}, this sum is unchanged if we take $j$ to range over
$\supp(\vc{s})$ instead.  Hence
\begin{align*}
L(f)    &
=
H(\vc{s}) - H(\p)       \\
&
=
\sum_{j \in \supp(\vc{s})} s_j \log \frac{1}{s_j}
- \sum_{j \in \supp(\vc{s})} s_j \log \frac{1}{p_{f(j)}}        \\
&
=
\sum_{j \in \supp(\vc{s})} s_j \log \frac{p_{f(j)}}{s_j},
\end{align*}
as claimed.
\end{proof}

\begin{remark}
This result is also an instance of
Lemma~\ref{lemma:cond-alts}\bref{part:cond-alts-exp} on conditional%
\index{conditional entropy!information loss@and information loss}%
\index{information loss!conditional entropy@and conditional entropy}
entropy, as follows.  Let $V$ be a random variable taking values in $\Yy$,
with distribution $\vc{s}$.  Put $U = f(V)$, which is a random variable
taking values in $\Xx$, with distribution $f\vc{s} = \vc{p}$.  Then $U$ is
determined by $V$, so by Example~\ref{egs:cond}\bref{eg:cond-det},
\[
0 
\leq 
\condent{V}{U} 
= 
H(V) - H(U) 
=
H(\vc{s}) - H(\p)
=
L(f).
\]
On the other hand, by Lemma~\ref{lemma:cond-alts}\bref{part:cond-alts-exp},
\[
\condent{V}{U}
=
\sum_{j, i \csuch \Pr(j, i) > 0} \Pr(j, i) \log \frac{\Pr(i)}{\Pr(j, i)}
=
\sum_{j \csuch s_j > 0} s_j \log \frac{p_{f(j)}}{s_j}.
\]
Comparing the two expressions for $\condent{V}{U}$ gives another proof of
Lemma~\ref{lemma:loss}.

This argument shows that information loss is a special case of conditional
entropy.  But conditional entropy is also a special case of information
loss.  Indeed, let $U$ and $V$ be random variables with the same sample
space, taking values in finite sets $\Xx$ and $\Yy$ respectively.  Equip
$\Xx \times \Yy$ with the distribution of $(U, V)$ and $\Xx$ with the
distribution of $U$.  Then the projection map
\[
\begin{array}{cccc}
\pr_1 \from     &\Xx \times \Yy &\to            &\Xx    \\
                &(i, j)         &\mapsto        &i
\end{array}
\]
is measure-preserving.  By definition, its information loss is 
\[
L(\pr_1) = H(U, V) - H(U) = \condent{V}{U}.
\]
Hence $\condent{V}{U} = L(\pr_1)$,
expressing conditional entropy in terms of information loss.
\end{remark}

\section{Characterization of information loss}
\lbl{sec:cil}
\index{information loss!characterization of}

In this section, we prove that information loss is uniquely characterized
(up to a constant factor) by four basic properties.  

First, a reversible process%
\index{deterministic process!reversible} 
loses no information: $L(f) = 0$ for all isomorphisms $f$.  This follows
from the definition of $L$ and the isomorphism-invariance of $H$.

Second, the amount of information lost by two processes%
\index{deterministic process}
in series is the sum of the amounts of information lost by each
individually.  Formally,
\begin{equation}
\lbl{eq:series-sum}
L(g \of f) = L(g) + L(f)
\end{equation}
whenever
%
\[
(\Yy, \vc{s}) \toby{f} (\Xx, \p) \toby{g} (W, \vc{t})
\]
%
are measure-preserving maps of finite probability spaces.  This is
immediate from the definition of information loss.

Third, given $n$ measure-preserving maps
\begin{eqnarray*}
(\Yy_1, \vc{s}^1) &\toby{f_1}& (\Xx_1, \p^1)   \\
\vdots& &\vdots \\
(\Yy_n, \vc{s}^n) &\toby{f_n}& (\Xx_n, \p^n)
\end{eqnarray*}
and a distribution $\vc{w} \in \Delta_n$, the amount of information lost
by the convex combination $\coprod w_i f_i$ is given by
\begin{equation}
\lbl{eq:parallel-sum}
L\Biggl( \coprod_{i = 1}^n w_i f_i \Biggr)
=
\sum_{i = 1}^n w_i L(f_i).
\end{equation}
This follows from the chain rule~\eqref{eq:g-ent-chain}:
\begin{align*}
L\Bigl( \coprod w_i f_i \Bigr)  &
=
H\Bigl( \coprod w_i \vc{s}^i \Bigr) 
- H\Bigl( \coprod w_i \vc{p}^i \Bigr)   \\
&
=
\biggl\{ H(\vc{w}) + \sum w_i H(\vc{s}^i) \biggr\}
-
\biggl\{ H(\vc{w}) + \sum w_i H(\vc{p}^i) \biggr\}        \\
&
=
\sum w_i L(f_i).
\end{align*}
In particular, given measure-preserving maps
\begin{eqnarray*}
(\Yy, \vc{s}) &\toby{f}& (\Xx, \p),   \\
(\Yy', \vc{s}') &\toby{f'}& (\Xx', \p')
\end{eqnarray*}
and a constant $\lambda \in [0, 1]$,
\[
L\bigl( \lambda f \cpd (1 - \lambda) f' \bigr)
=
\lambda L(f) + (1 - \lambda) L(f').
\]
Intuitively, this means that if we flip a probability-$\lambda$ coin%
\index{coin!toss} 
and, depending on the outcome, do either the process $f$ or the process
$f'$, then the expected information loss is $\lambda$ times the information
loss of $f$ plus $1 - \lambda$ times the information loss of $f'$.  So,
while the previous property of $L$ (equation~\eqref{eq:series-sum})
concerned the information lost by two processes in \emph{series}, this
property (equation~\eqref{eq:parallel-sum}) concerns the information lost
by two or more processes in \emph{parallel}.

Fourth and finally, information loss is continuous,%
\index{continuous!measure of information loss}%
\index{information loss!continuity of}
in the following sense.  Let $f \from \Yy \to \Xx$ be a map of finite sets.
For each probability distribution $\vc{s}$ on $\Yy$, we have the
pushforward distribution $f \vc{s}$ on $\Xx$, and $f$ defines a
measure-preserving map
\[
f \from (\Yy, \vc{s}) \to (\Xx, f \vc{s})
\]
(Remark~\ref{rmks:finprob}\bref{rmk:finprob-pf}).  The statement is
that the map 
\[
\begin{array}{ccc}
\Delta_\Yy      &\to            &\R     \\
\vc{s}          &\mapsto        &
L \Bigl( (\Yy, \vc{s}) \toby{f} (\Xx, f\vc{s}) \Bigr)
\end{array}
\]
is continuous.  This follows from the fact that all the maps in the
(noncommutative) triangle
\[
\xymatrix{
\Delta_\Yy \ar[rr]^{\vc{s} \mapsto f\vc{s}} \ar[rd]_H        &
&
\Delta_\Xx \ar[ld]^H                      \\
&
\R
} 
\]
are continuous.

An equivalent way to state continuity is as follows.  Let us say that an
infinite sequence
\[
\Bigl(
(\Yy_m, \vc{s}^m) \toby{f_m} (\Xx_m, \vc{p}^m)
\Bigr)_{m \geq 1}
\]
of measure-preserving maps of finite probability spaces
\demph{converges}\index{convergence} to a map
\[
(\Yy, \vc{s}) \toby{f} (\Xx, \p)
\]
if 
\[
\Bigl( \Yy_m \toby{f_m} \Xx_m \Bigr) = \Bigl( \Yy \toby{f} \Xx \Bigr)
\]
for all sufficiently large $m$, and $\vc{s}^m \to \vc{s}$ and $\p^m \to \p$
as $m \to \infty$.  Then continuity of $L$ is equivalent to the statement
that for any such convergent sequence,
\[
L
\Bigl(
(\Yy_m, \vc{s}^m) \toby{f_m} (\Xx_m, \vc{p}^m)
\Bigr)
\to
L
\Bigl(
(\Yy, \vc{s}) \toby{f} (\Xx, \vc{p})
\Bigr)
\text{ as } m \to \infty.
\]
The equivalence between these two formulations of continuity follows from
the elementary fact that a map of metrizable spaces is continuous if and
only if it preserves convergence of sequences.

We now state the main theorem, which first appeared as Theorem~2 of
Baez, Fritz and Leinster~\cite{CETIL}.

\begin{thm}[Baez, Fritz and Leinster]
\lbl{thm:cetil}%
\index{Baez, John}%
\index{Fritz, Tobias}%
\index{information loss!characterization of}%
Let $K$ be a function assigning a real number $K(f)$ to each
measure-preserving map $f$ of finite probability spaces.  The following
are equivalent:
\begin{enumerate}
\item 
\lbl{part:cetil-condns}
$K$ has these four properties:
\begin{enumerate}
\item 
$K(f) = 0$ for all isomorphisms $f$;

\item
$K(g \of f) = K(g) + K(f)$ for all composable pairs $(f, g)$ of
  measure-preserving maps;

\item
$K\bigl(\lambda f \cpd (1 - \lambda) f'\bigr) = \lambda K(f) + (1 -
  \lambda) K(f')$ for all measure-preserving maps $f$ and $f'$ and all
  $\lambda \in [0, 1]$;

\item
$K$ is continuous;
\end{enumerate}

\item
\lbl{part:cetil-form}
$K = cL$ for some $c \in \R$.
\end{enumerate}
\end{thm}

The proof, given below, will use a version of Faddeev's theorem:

\begin{thm}[Faddeev, version~2]
\lbl{thm:g-faddeev}%
\index{Faddeev, Dmitry!entropy theorem}%
Let $I$ be a function assigning a real number $I(\p)$ to each finite
probability space $(\Xx, \p)$.  The following are equivalent:
\begin{enumerate}
\item
\lbl{part:g-faddeev-condns}
$I$ is isomorphism-invariant, satisfies the chain
rule~\eqref{eq:g-ent-chain}, and is continuous in the sense
of~\eqref{eq:g-cts} (with $I$ in place of $H$); 

\item
\lbl{part:g-faddeev-form}
$I = cH$ for some $c \in \R$.
\end{enumerate}
\end{thm}

\begin{proof}
%
We have already observed that $H$ satisfies the conditions
in~\bref{part:g-faddeev-condns}, and it follows
that~\bref{part:g-faddeev-form} implies~\bref{part:g-faddeev-condns}.

Conversely, take a function $I$ satisfying~\bref{part:g-faddeev-condns}.
Restricting $I$ to finite sets of the form $\{1, \ldots, n\}$ defines, for
each $n \geq 1$, a continuous function $I \from \Delta_n \to \R$ satisfying
the chain rule.  Hence by Faddeev's Theorem~\ref{thm:faddeev}, there is
some constant $c \in \R$ such that $I(\p) = cH(\p)$ for all $n \geq 1$ and
$\p \in \Delta_n$.  Next, take any finite probability space $(\Yy,
\vc{s})$.  We have
\[
(\Yy, \vc{s}) 
\iso
\bigl( \{1, \ldots, n\}, \p \bigr)
\]
for some $n \geq 1$ and $\p \in \Delta_n$, and then by
isomorphism-invariance of both $I$ and $H$,
\[
I(\vc{s})
=
I(\vc{p})
=
cH(\vc{p})
=
cH(\vc{s}),
\]
as required.
\end{proof}

\begin{remark}
\lbl{rmk:g-faddeev-sym}
The version of Faddeev's theorem just stated is slightly weaker than the
earlier version, Theorem~\ref{thm:faddeev}.  To see this, take $\p \in
\Delta_n$ and a permutation $\sigma$ of $\{1, \ldots, n\}$.  Then $\sigma$
defines a measure-preserving bijection 
\[
\sigma \from
\bigl( \{1, \ldots, n\}, \p\sigma \bigr) 
\to
\bigl( \{1, \ldots, n\}, \p \bigr).
\]
In Theorem~\ref{thm:g-faddeev}, therefore, the isomorphism-invariance axiom
on $I$ includes as a special case that $I(\p\sigma) = I(\p)$ for all $\p
\in \Delta_n$ and permutations $\sigma$.  This is the symmetry%
\index{symmetry in Faddeev-type theorems}%
%
axiom that is traditionally included in statements of Faddeev's theorem,
but is not in fact necessary, as observed in
Remark~\ref{rmks:faddeev}\bref{rmk:faddeev-sym}.  So,
Theorem~\ref{thm:g-faddeev} is a restatement of that traditional, weaker
form of Faddeev's theorem.  The analogous restatement of the stronger
Theorem~\ref{thm:faddeev} would involve \emph{ordered}%
\index{order!probability space@on probability space}%
\index{probability space!ordered}
probability spaces.
\end{remark}

We can now prove the characterization theorem for information loss.

\begin{pfof}{Theorem~\ref{thm:cetil}}
We have already shown that information loss $L$ satisfies the four
conditions of~\bref{part:cetil-condns}, and it follows
that~\bref{part:cetil-form} implies~\bref{part:cetil-condns}.

For the converse, suppose that $K$ satisfies~\bref{part:cetil-condns}.
Given a finite probability space $(\Xx, \p)$, write $!_{\p}$ for the unique
measure-preserving map
\[
!_{\p} \from (\Xx, \p) \to (\{1\}, \vc{u}_1),
\]
and define $I(\p) = K(!_{\p})$.  For any measure-preserving map $f \from
(\Yy, \vc{s}) \to (\Xx, \p)$, the triangle
\[
\xymatrix{
(\Yy, \vc{s}) \ar[rr]^f \ar[rd]_{!_{\vc{s}}}      &
&
(\Xx, \p) \ar[ld]^{!_{\p}}        \\
&
(\{1\}, \vc{u}_1)
}
\]
commutes, so by the composition condition on $K$,
\[
K(!_{\vc{s}}) = K(!_{\p}) + K(f).
\]
Equivalently,
\begin{equation}
\lbl{eq:K-I-diff}
K(f) = I(\vc{s}) - I(\vc{p}).
\end{equation}
So in order to prove the theorem, it suffices to show that $I = cH$ for
some constant $c$; and for this, it is enough to prove that $I$ satisfies
the hypotheses of Theorem~\ref{thm:g-faddeev}.

First, $I$ is isomorphism-invariant, since if $f\from (\Yy, \vc{s}) \to (\Xx,
\p)$ is an isomorphism then $K(f) = 0$, so $I(\vc{s}) = I(\p)$
by~\eqref{eq:K-I-diff}.  

Second, $I$ satisfies the chain rule~\eqref{eq:g-ent-chain}; that is,
\begin{equation}
\lbl{eq:cetil-ch}
I\Biggl( \coprod_{i = 1}^n w_i \p^i \Biggr)
=
I(\vc{w}) + \sum_{i = 1}^n w_i I(\p^i)
\end{equation}
for all $\vc{w} \in \Delta_n$ and finite probability spaces $(\Xx_1, \p^1),
\ldots, (\Xx_n, \p^n)$.  To see this, write
\[
f \from
\coprod_{i = 1}^n \Xx_i \to \{1, \ldots, n\}
\]
for the function defined by $f(j) = i$ whenever $j \in \Xx_i$.  Then $f$
defines a measure-preserving map
\[
f \from
\Bigl( \coprod \Xx_i, \coprod w_i \p^i \Bigr)
\to
\bigl( \{1, \ldots, n\}, \vc{w} \bigr).
\]
We now evaluate $K(f)$ in two ways.  On the one hand, by
equation~\eqref{eq:K-I-diff},
\[
K(f) =
I\Bigl( \coprod w_i \p^i \Bigr) - I(\vc{w}).
\]
On the other, 
\[
f = \coprod w_i \, !_{\p^i},
\]
so by hypothesis on $K$ and induction,
\[
K(f)
=
\sum w_i K(!_{\p^i})
=
\sum w_i I(\p^i).
\]
Comparing the two expressions for $K(f)$ gives the chain
rule~\eqref{eq:cetil-ch} for $I$.

Third and finally, for each finite set $\Xx$, the function $I \from
\Delta_\Xx \to \R$ is continuous, by continuity of $K$.

Theorem~\ref{thm:g-faddeev} can therefore be applied, giving $I = cH$ for some
$c \in \R$. It follows from equation~\eqref{eq:K-I-diff} that $K = cL$.
\end{pfof}

As observed in~\cite{CETIL} (p.~1947), the charm of Theorem~\ref{thm:cetil}
is that the axioms on the information loss function $K$ are entirely
linear.  They give no hint of any special role for the function
\[
p \mapsto -p \log p.
\]
And yet, this function emerges in the conclusion.

Another striking feature of Theorem~\ref{thm:cetil} is that the natural
conditions imposed on $K$ force $K(f)$ to depend only on the domain and
codomain of $f$.  This is a consequence of condition~\hardref{(b)} alone
(on the information lost by a composite process), as can be seen from the
argument leading up to equation~\eqref{eq:K-I-diff}.  It is an instance of
a general categorical fact: for any functor $K$ from a category $\cat{P}$
with a terminal object to a groupoid, $K(f) = K(f')$ whenever $f$
and $f'$ are maps in $\cat{P}$ with the same domain and the same codomain.

Theorem~\ref{thm:cetil} has several variants.  We can drop the condition
that $K(f) = 0$ for isomorphisms $f$ if we instead require that $K(f) \geq
0$ for all $f$.  (This was the version stated in Baez, Fritz and
Leinster~\cite{CETIL}.)  There is another version of
Theorem~\ref{thm:cetil} for finite sets equipped with arbitrary finite
measures instead of probability measures (Corollary~4 of~\cite{CETIL}).
And there is a further variant for the $q$-logarithmic entropies $S_q$,
which we give now.

For a measure-preserving map
\[
f \from (\Yy, \vc{s}) \to (\Xx, \p)
\]
between finite probability spaces, define the \demph{$q$-logarithmic%
\index{information loss!q-logarithmic@$q$-logarithmic}%
\index{q-logarithmic entropy@$q$-logarithmic entropy!information loss@and information loss}
information loss} of $f$ as
\[
L_q(f) = S_q(\vc{s}) - S_q(\p).
\ntn{lossq}
\]
The following characterization of $L_q$ 
is identical to Theorem~\ref{thm:cetil} except for a change in the rule for
the information lost by two processes in parallel (condition~\hardref{(c)}
below) and the absence of a continuity condition.  With some minor
differences, it first appeared as Theorem~7 of Baez, Fritz and
Leinster~\cite{CETIL}.

\begin{thm}[Baez, Fritz and Leinster]
\lbl{thm:cetil-q}%
\index{information loss!characterization of}%
\index{Baez, John}%
\index{Fritz, Tobias}%
Let $1 \neq q \in \R$.  Let $K$ be a function assigning a real number
$K(f)$ to each measure-preserving map $f$ of finite probability spaces.
The following are equivalent:
\begin{enumerate}
\item 
\lbl{part:cetil-q-condns}
$K$ has these three properties:
\begin{enumerate}
\item 
$K(f) = 0$ for all isomorphisms $f$;

\item
$K(g \of f) = K(g) + K(f)$ for all composable pairs $(f, g)$ of
  measure-preserving maps;

\item
$K\bigl(\lambda f \cpd (1 - \lambda) f'\bigr) = \lambda^q K(f) + (1 -
  \lambda)^q K(f')$ for all measure-preserving maps $f$ and $f'$ and all
  $\lambda \in (0, 1)$;
\end{enumerate}

\item
\lbl{part:cetil-q-form}
$K = cL_q$ for some $c \in \R$.
\end{enumerate}
\end{thm}

No continuity or other regularity condition is needed, in contrast to
Theorem~\ref{thm:cetil}. 

\begin{proof}
As for the proof of Theorem~\ref{thm:cetil}, but using the characterization
theorem for $S_q$ (Theorem~\ref{thm:q-ent-char}) instead of Faddeev's
characterization of $H$ (Theorem~\ref{thm:faddeev}).
\end{proof}

%% file: p.tex
\chapter{Entropy modulo a prime}
\lbl{ch:p}
\index{entropy!modulo a prime}

\begin{quote}
\emph{Conclusion:} If we have a random variable $\xi$ which takes
finitely many values with all probabilities in $\Q$ then we can define
not only the transcendental number $H(\xi)$ but also its `residues%
\index{residue class}
modulo $p$' for almost all primes $p$\,!  \quad%
\index{Kontsevich, Maxim}
\hfill -- Maxim Kontsevich~\cite{KontOHL}.
\end{quote}

\noindent
In this chapter, we define the entropy of any probability distribution
whose `probabilities' are not real numbers, but integers modulo a prime $p$.
Its entropy, too, is an integer mod~$p$.  We justify the definition by
proving a characterization theorem very similar to Faddeev's theorem on
real entropy (Theorem~\ref{thm:faddeev}), and by a characterization theorem
for information loss mod~$p$ that is also closely analogous to the real
case.

In earlier chapters, we reached our axiomatic characterization of real
information loss in three steps:
\begin{list}{}{}
\item[\phantom{(III)}]\makebox[0em][r]{(I)} 
characterize the sequence $(\log n)_{n \geq 1}$
(Theorem~\ref{thm:erdos-liminf});  

\item[\phantom{(III)}]\makebox[0em][r]{(II)} 
using~(I), characterize entropy (Theorem~\ref{thm:faddeev});

\item[\phantom{(III)}]\makebox[0em][r]{(III)} 
using~(II), characterize information loss (Theorem~\ref{thm:cetil}).  
\end{list}
Here, we follow three analogous steps to characterize entropy and
information loss modulo~$p$ (Sections~\ref{sec:p-defn}
and~\ref{sec:p-char}).  The analytic subtleties disappear, but instead we
encounter a number-theoretic obstacle.  

With the definition of entropy mod~$p$ in place, we implement the idea
proposed by Kontsevich in the quotation above.  That is, we define a sense
in which certain real numbers can be said to have residues mod~$p$
(Section~\ref{sec:p-res}).  The residue map establishes a direct
relationship between entropy over $\R$ and entropy over $\Zp$,
supplementing the analogy between the Faddeev-type theorems over $\R$ and
$\Zp$. 

We finish by developing an alternative but equivalent approach to entropy
modulo a prime (Section~\ref{sec:p-poly}).  It takes place in the ring of
polynomials over the field of $p$ elements.  It is related more closely
than the rest of this chapter to the subject of
polylogarithms\index{polylogarithm}, which formed the context of
Kontsevich's note~\cite{KontOHL} and of subsequent related work such as
that of Elbaz-Vincent and Gangl~\cite{EVGOPI,EVGFPM}.

The results of this chapter first appeared in~\cite{EMP}.  While~\cite{EMP}
seems to have been the first place where the theory of entropy mod~$p$ was
developed in detail, many of the ideas had been sketched or at least hinted
at in Kontsevich's note~\cite{KontOHL}, which itself was preceded by
related work of Cathelineau~\cite{CathSHS,CathRDP}.%
\index{Cathelineau, Jean-Louis}
The introduction to Elbaz-Vincent and Gangl~\cite{EVGOPI} relates some of
the history, including the connection with polylogarithms; see also
Remark~\ref{rmk:kont-comp} below.

\section{Fermat quotients and the definition of entropy}
\lbl{sec:p-defn}

For the whole of this chapter, fix a prime $p$.  To avoid confusion between
the prime $p$ and a probability distribution $\p$, we now denote a typical
probability distribution by $\ppi = (\pi_1, \ldots, \pi_n)$.

Our first task is to formulate the correct definition of the entropy of a
probability%
\index{probability distribution!modulo a prime} 
distribution $\ppi$ in which $\pi_1, \ldots, \pi_n$ are not real numbers,
but elements of the field $\Zp$ of integers modulo~$p$.

A problem arises immediately.  Real probabilities are ordinarily required to
be nonnegative, and the logarithms in the definition of entropy over $\R$
would be undefined if any probability were negative.  So in the familiar
real setting, the notion of positivity seems to be needed in order to state
a definition of entropy.  But in $\Zp$, there is no sense of positive or
negative.  How, then, are we to imitate the definition of entropy in
$\Zp$?

This problem is solved by a simple observation.  Although Shannon entropy%
\index{entropy!negative probabilities@with negative probabilities}
is usually only defined for sequences $\ppi = (\pi_1, \ldots, \pi_n)$ of
\emph{nonnegative} reals summing to $1$, it can just as easily be defined
for sequences $\ppi$ of \emph{arbitrary} reals summing to $1$.  One simply
puts
\begin{equation}
\lbl{eq:defn-ent-hyp}
H(\ppi) = - \sum_{i \in \supp(\ppi)} \pi_i \log \,\mg{\pi_i},
\end{equation}
where \ntn{suppall}$\supp(\ppi) 
\index{support}
= \{ i \such \pi_i \neq 0\}$.  (See Kontsevich~\cite{KontOHL},%
\index{Kontsevich, Maxim} 
for instance.)  This extended entropy is still continuous and symmetric,
and still satisfies the chain rule.  So, real entropy can in fact be
defined without reference to the notion of positivity.  (And generally
speaking, negative%
\index{negative!probability}
probabilities are not as outlandish as they might seem; see
Feynman~\cite{Feyn} and Blass and Gurevich~\cite{BlGuNP1,BlGuNP2}.)

Thus, writing
\[
\Pi_n = \{ \ppi \in (\Zp)^n \such \pi_1 + \cdots + \pi_n = 1 \},
\ntn{Pin}
\]
it is reasonable to attempt to define the entropy of any element of
$\Pi_n$.  We will refer to elements $\ppi = (\pi_1, \ldots, \pi_n)$ of
$\Pi_n$ as \demph{probability%
\index{probability distribution!modulo a prime}
distributions mod $p$}, or simply \demph{distributions}.%
\index{distribution}
Geometrically, the set $\Pi_n$ of
distributions on $n$ elements is a hyperplane in the $n$-dimensional vector
space $(\Zp)^n$ over the field $\Zp$.

The function $x \mapsto \log\mg{x}$ is a homomorphism from the
multiplicative group $\R^\times$ of nonzero reals to the additive
group $\R$.  But when we look for an analogue over $\Zp$, we run into
an obstacle:

\begin{lemma}
\lbl{lemma:no-log}
\index{logarithm modulo a prime}
There is no nontrivial homomorphism from the multiplicative group
$(\Zp)^\times$ of nonzero integers modulo~$p$ to the additive group $\Zp$.
\end{lemma}

\begin{proof}
Let $\phi\from (\Zp)^\times \to \Zp$ be a homomorphism.  The image of
$\phi$ is a subgroup of $\Zp$, which by Lagrange's theorem has order
$1$ or $p$.  Since $(\Zp)^\times$ has order $p - 1$, the image of
$\phi$ has order at most $p - 1$.  It therefore has order $1$; that is,
$\phi = 0$.
\end{proof}

In this sense, there is no logarithm for the integers modulo~$p$.
Nevertheless, there is an acceptable substitute.  For integers $n$ not
divisible by $p$, Fermat's little theorem implies that $p$ divides $n^{p -
  1} - 1$.  The \demph{Fermat%
\index{Fermat quotient} 
quotient} of $n$ modulo~$p$ is defined as   
\[
\fq{p}(n) = \frac{n^{p - 1} - 1}{p} \in \Zp.
\ntn{fq}
\]
The resemblance between the formulas for the Fermat quotient and the
$q$-logarithm (equation~\eqref{eq:q-log-general}) hints that the
Fermat quotient might function as some kind of logarithm, and
part~\bref{part:fq-elem-log} of the following lemma confirms that this is
so.

\begin{lemma}
\lbl{lemma:fq-elem}
The map $\fq{p} \from \{ n \in \Z \such p \ndvd n \} \to \Zp$ has the
following properties:
\begin{enumerate}
\item 
\lbl{part:fq-elem-log} 
$\fq{p}(mn) = \fq{p}(m) + \fq{p}(n)$ for all $m, n \in
\Z$ not divisible by~$p$, and $\fq{p}(1) = 0$;

\item
\lbl{part:fq-elem-shift}
$\fq{p}(n + rp) = \fq{p}(n) - r/n$ for all $n, r \in \Z$ such that $n$ is not
divisible by $p$;

\item
\lbl{part:fq-elem-per}
$\fq{p}(n + p^2) = \fq{p}(n)$ for all $n \in \Z$ not divisible by $p$.
%
\end{enumerate}
\end{lemma}

\begin{proof}
For~\bref{part:fq-elem-log}, certainly $\fq{p}(1) = 0$.  We now have to
show that
\[
m^{p - 1}n^{p - 1} - 1 
\equiv
\bigl( m^{p - 1} - 1 \bigr) + \bigl( n^{p - 1} - 1 \bigr)
\pmod{p^2},
\]
or equivalently,
\[
\bigl( m^{p - 1} - 1 \bigr)\bigl( n^{p - 1} - 1 \bigr)
\equiv
0
\pmod{p^2}.
\]
Since both $m^{p - 1} - 1$ and $n^{p - 1} - 1$ are integer multiples of
$p$, this is true.

For~\bref{part:fq-elem-shift}, we have
\begin{align*}
(n + rp)^{p - 1}        &
=
n^{p - 1} + (p - 1)n^{p - 2}rp
+ \sum_{i = 2}^{p - 1} \binom{p - 1}{i} n^{p - i - 1} r^i p^i       \\
&
\equiv
n^{p - 1} + p(p - 1)r n^{p - 2}
\pmod{p^2}.
\end{align*}
Subtracting $1$ from each side and dividing by $p$ gives 
\[
\fq{p}(n + rp)
\equiv
\fq{p}(n) + (p - 1)rn^{p - 2}
\pmod{p},
\]
and~\bref{part:fq-elem-shift} then follows from the fact that $n^{p - 1}
\equiv 1 \pmod{p}$.  Taking $r = p$ in~\bref{part:fq-elem-shift}
gives~\bref{part:fq-elem-per}. 
\end{proof}

It follows that $\fq{p}$ defines a group homomorphism
\[
\fq{p} \from (\Zps)^\times \to \Zp,
\]
where $(\Zps)^\times$\ntn{Zpsm} is the multiplicative group of integers
modulo~$p^2$.  (The elements of $(\Zps)^\times$ are the congruence
classes modulo~$p^2$ of the integers not divisible by $p$.)  Moreover, the
homomorphism $\fq{p}$ is surjective, since the lemma implies that
\[
\fq{p}(1 - rp) = \fq{p}(1) + r \equiv r \pmod{p}
\]
for all integers $r$.

Lemma~\ref{lemma:no-log} states that there is no logarithm mod~$p$, in the
sense that there is no nontrivial group homomorphism $(\Zp)^\times \to
\Zp$.  But the Fermat quotient is the next best thing, being a homomorphism
$(\Zps)^\times \to \Zp$.  It is essentially the only such homomorphism:

\begin{propn}
\lbl{propn:fq-mult}
Every group homomorphism $(\Zps)^\times \to \Zp$ is a scalar multiple of
the Fermat quotient.
\end{propn}

\begin{proof}
It is a standard fact that the group $(\Zps)^\times$ is cyclic
(Theorem~10.6 of Apostol~\cite{AposIAN}, for instance).  Choose a generator
$e$.  Since $\fq{p}$ is surjective, it is not identically zero, so
$\fq{p}(e) \neq 0$.

Let $\phi\from (\Zps)^\times \to \Zp$ be a group homomorphism.  Put $c =
\phi(e)/\fq{p}(e) \in \Zp$.  Then for all $n \in \Z$,
\[
\phi(e^n) = n\phi(e) = nc\fq{p}(e) = c\fq{p}(e^n).
\]
Since $e$ is a generator, it follows that $\phi = c\fq{p}$.
\end{proof}

In Section~\ref{sec:log-seqs}, we proved characterization theorems for the
sequence $(\log(n))_{n \geq 1}$.  The next result plays a similar role for
$(\fq{p}(n))_{n \geq 1,\:p \ndvd n}$.  

\begin{thm}
\lbl{thm:fq-char}
\index{Fermat quotient!characterization of}
Let $f \from \{ n \in \nat \such p \ndvd n \} \to \Zp$ be a function.  The
following are equivalent:
\begin{enumerate}
\item 
\lbl{part:fq-char-condns}
$f(mn) = f(m) + f(n)$ and $f(n + p^2) = f(n)$ for all $m, n \in \nat$ not
divisible by $p$;

\item
\lbl{part:fq-char-form}
$f = c\fq{p}$ for some $c \in \Zp$.
\end{enumerate}
\end{thm}

\begin{proof}
We have already shown that $\fq{p}$ satisfies the conditions
in~\bref{part:fq-char-condns}, so~\bref{part:fq-char-form}
implies~\bref{part:fq-char-condns}.  For the converse, suppose that $f$
satisfies the conditions in~\bref{part:fq-char-condns}.  Then $f$ induces a
group homomorphism $(\Zps)^\times \to \Zp$, which by
Proposition~\ref{propn:fq-mult} is a scalar multiple of $\fq{p}$.  The
result follows.
\end{proof}

In terms of the three-step plan in the introduction to this chapter, we
have now completed step~(I): defining and characterizing the appropriate
notion of logarithm.  We now begin step~(II): defining and characterizing
the appropriate notion of entropy.

To state the definition, we will need an elementary lemma.

\begin{lemma}
\lbl{lemma:pps}
Let $a, b \in \Z$.  If $a \equiv b \pmod{p}$ then $a^p \equiv b^p
\pmod{p^2}$. 
\end{lemma}

\begin{proof}
If $b = a + rp$ with $r \in \Z$, then
\begin{align*}
b^p     &
= 
(a + rp)^p      \\
&
=
a^p + p a^{p - 1} rp + \sum_{i = 2}^p \binom{p}{i} a^{p - i} r^i p^i    \\
&
\equiv
a^p
\pmod{p^2}.
\end{align*}
%
\end{proof}


\begin{defn}
\lbl{defn:ent-p}
Let $n \geq 1$ and $\ppi \in \Pi_n$.  The
\demph{entropy}%
\index{entropy!modulo a prime} 
of $\ppi$ is
\[
H(\ppi) 
= 
\frac{1}{p} \Biggl( 1 - \sum_{i = 1}^n a_i^p \Biggr)
\in \Zp,
\ntn{Hp}
\]
where $a_i \in \Z$ represents $\pi_i \in \Zp$ for each $i \in \{1,
\ldots, n\}$. 
\end{defn}

Lemma~\ref{lemma:pps} guarantees that the definition is independent of the
choice of $a_1, \ldots, a_n$.  

We now explain and justify the definition of entropy mod~$p$.  In
particular, we will prove a theorem characterizing the sequence of
functions $\bigl( H \from \Pi_n \to \Zp \bigr)$ uniquely up to a scalar
multiple.  This result is plainly analogous to Faddeev's theorem for real
entropy, and as such, is the strongest justification for the definition.
But the analogy with the real case can also be seen in terms of
derivations, as follows.

The entropy of a real probability distribution $\ppi$ is equal
to $\sum_i \partial(\pi_i)$, where
\begin{equation}
\lbl{eq:real-deriv}
\partial(x) = 
\begin{cases}
-x\log x        &\text{if } x > 0,   \\
0               &\text{if } x = 0
\end{cases}
\end{equation}
(as in Section~\ref{sec:ent-defn}).  What is the analogue of $\partial$ over
$\Zp$?  Given the analogy between the logarithm and the Fermat quotient, it
is natural to consider $-n\fq{p}(n)$ as a candidate.  For integers $n$ not
divisible by $p$, 
\[
-n \fq{p}(n) 
=
\frac{n - n^p}{p}.
\]
The right-hand side is a well-defined integer even if $n$ is divisible by
$p$.  We therefore define a map $\partial \from \Z \to \Zp$ by
\begin{equation}
\lbl{eq:p-deriv}
\partial(n) = \frac{n - n^p}{p} \in \Zp.
\end{equation}
Thus,
\[
\partial(n)
=
\begin{cases}
-n\fq{p}(n)        &\text{if } p \ndvd n,  \\
n/p             &\text{if } p \dvd n.
\end{cases}
\]
If $n \equiv m \pmod{p^2}$ then $\partial(n) = \partial(m)$, so $\partial$
can also be regarded as a map $\Zps \to \Zp$.  Like its real counterpart,
$\partial$ satisfies a version of the Leibniz rule (and for this reason, is
essentially what is called a
\demph{$p$-derivation}%
\index{derivation!p-@$p$-}%
\index{pderivation@$p$-derivation}%
~\cite{BuiuDCA,BuiuAAD}):

\begin{lemma}
\lbl{lemma:p-deriv-elem}
$\partial(mn) = m\partial(n) + \partial(m)n$ for all $m, n \in \Z$.
\end{lemma}

\begin{proof}
The proof is similar to that of
Lemma~\ref{lemma:fq-elem}\bref{part:fq-elem-log}.  The statement to be
proved is equivalent to
\[
mn - m^p n^p \equiv m(n - n^p) + (m - m^p)n \pmod{p^2}.
\]
Rearranging, this in turn is equivalent to
\[
0 \equiv (m - m^p)(n - n^p) \pmod{p^2},
\]
which is true since $m \equiv m^p \pmod{p}$ and $n \equiv n^p \pmod{p}$.
\end{proof}

Using this lemma, we derive an equivalent expression for entropy mod~$p$:

\begin{lemma}
\lbl{lemma:ent-partial}
For all $n \geq 1$ and $\ppi \in \Pi_n$,
\[
H(\ppi)
=
\sum_{i = 1}^n \partial(a_i) 
- \partial\Biggl( \sum_{i = 1}^n a_i \Biggr),
\]
where $a_i \in \Z$ represents $\pi_i \in \Zp$ for each $i \in \{1, \ldots,
n\}$. 
\end{lemma}

\begin{proof}
An equivalent statement is that
\[
1 - \sum a_i^p
\equiv
\sum (a_i - a_i^p) - \biggl\{ \sum a_i - \Bigl( \sum a_i \Bigr)^p \biggr\}
\pmod{p^2}.
\]
Cancelling, this reduces to
\[
1 \equiv \Bigl( \sum a_i \Bigr)^p \pmod{p^2}.
\]
But $\sum \pi_i = 1$ in $\Zp$ by definition of $\Pi_n$, so $\sum a_i \equiv
1 \pmod{p}$, so $\bigl( \sum a_i \bigr)^p \equiv 1 \pmod{p^2}$ by
Lemma~\ref{lemma:pps}. 
\end{proof}

Thus, $H(\ppi)$ measures the extent to which the nonlinear derivation
$\partial$ fails to preserve the sum $\sum a_i$.

The analogy with entropy over $\R$ is now evident.  For a real probability
distribution $\ppi$, and defining $\partial \from [0, \infty) \to \R$ as
  in equation~\eqref{eq:real-deriv}, we also have
\[
H(\ppi) = \sum \partial(\pi_i) - \partial\Bigl(\sum \pi_i\Bigr).
\]
In the real case, since $\sum \pi_i = 1$, the second term on the right-hand
side vanishes.  But over $\Zp$,
it is not true in general that $\partial(\sum a_i) = 0$, so it is not true
either that $H(\ppi) = \sum \partial(a_i)$.  (Indeed, $\sum \partial(a_i)$,
unlike $H(\ppi)$, depends on the choice of representatives $a_i$.)
So in the formula
\[
H(\ppi) = \sum \partial(a_i) - \partial\Bigl(\sum a_i\Bigr)
\]
for entropy mod~$p$, the second summand is indispensable.

\begin{example}
\lbl{eg:ent-p-ufm}
Let $n \geq 1$ with $p \ndvd n$.  Since $n$ is invertible mod~$p$, there is
a \demph{uniform%
\index{uniform distribution!modulo a prime} 
distribution}
\[
\vc{u}_n = (\underbrace{1/n, \ldots, 1/n}_n) \in \Pi_n.
\ntn{unp}
\]
Choose $a \in \Z$ representing $1/n \in \Zp$.  By
Lemma~\ref{lemma:ent-partial} and then the derivation property of
$\partial$, 
\[
H(\vc{u}_n)
=
n\partial(a) - \partial(na)
=
-a\partial(n).
\]
But $\partial(n) = -n\fq{p}(n)$, so $H(\vc{u}_n) = \fq{p}(n)$.  This result over
$\Zp$ is analogous to the formula $H(\vc{u}_n) = \log n$ for the real entropy
of a uniform distribution.
\end{example}

\begin{example}
\lbl{eg:ent-p-2}
Let $p = 2$.  Any distribution $\ppi \in \Pi_n$ has an odd
number of elements in its support, since $\sum \pi_i = 1$.  Directly from
the definition of entropy, $H(\ppi) \in \Z/2\Z$ is given by 
\[
H(\ppi) 
= 
\hlf\bigl(\mg{\supp(\ppi)} - 1\bigr) 
=
\begin{cases}
0       &\text{if } \mg{\supp(\ppi)} \equiv 1 \!\!\!\!\pmod{4}, \\
1       &\text{if } \mg{\supp(\ppi)} \equiv 3 \!\!\!\!\pmod{4}.
\end{cases}
\]
\end{example}

In preparation for the next example, we record a useful standard lemma:

\begin{lemma}
\lbl{lemma:p-binom}
$\binom{p - 1}{s} \equiv (-1)^s \pmod{p}$ for all $s \in \{0, \ldots, p -
  1\}$. 
\end{lemma}

\begin{proof}
In $\Zp$, we have equalities
\begin{align*}
\binom{p - 1}{s}        &
=
\frac{(p - 1)(p - 2) \cdots (p - s)}{s!}    \\
&
=
\frac{(-1)(-2) \cdots (-s)}{s!}   \\
&
=
(-1)^s.
\end{align*}
%
\end{proof}

\begin{example}
\lbl{eg:p-bin} Here we find the entropy of a distribution $(\pi, 1 - \pi)$
on two elements.  Choose an integer $a$ representing $\pi \in \Zp$.  From the
definition of entropy, assuming that $p \neq 2$,
\[
H(\pi, 1 - \pi)
=
\frac{1}{p} \bigl( 1 - a^p - (1 - a)^p \bigr)
=
\sum_{r = 1}^{p - 1} (-1)^{r + 1} \frac{1}{p} \binom{p}{r} a^r.
\]
But $\tfrac{1}{p} \binom{p}{r} = \tfrac{1}{r} \binom{p - 1}{r - 1}$, so by
Lemma~\ref{lemma:p-binom}, the coefficient of $a^r$ in the sum is simply
$\tfrac{1}{r}$.  We can now replace $a$ by $\pi$, giving
\[
H(\pi, 1 - \pi) = \sum_{r = 1}^{p - 1} \frac{\pi^r}{r}.
\]
The function on the right-hand side was the starting point of Kontsevich's
note~\cite{KontOHL}, and we return to it in Section~\ref{sec:p-poly}.  

In the case $p = 2$, we have $H(\pi, 1 - \pi) = 0$ for both values of $\pi
\in \Ztwo$.
\end{example}

\begin{example}
Appending zero probabilities to a distribution does not change its entropy:
\[
H(\pi_1, \ldots, \pi_n, 0, \ldots, 0) = H(\pi_1, \ldots, \pi_n).
\]
This is immediate from the definition.  But a subtlety of distributions
mod~$p$, absent in the standard real setting, is that nonzero
probabilities can sum to zero.  So, one might ask whether
\[
H(\pi_1, \ldots, \pi_n, \tau_1, \ldots, \tau_m) = H(\pi_1, \ldots, \pi_n)
\]
whenever $\tau_1, \ldots, \tau_m \in \Zp$ with $\sum \tau_j = 0$.  The
answer is trivially yes for $m = 0$ and $m = 1$, and it is also yes for $m
= 2$ as long as $p \neq 2$.  (For if we choose an integer $a$ to represent
$\tau_1$ then $-a$ represents $\tau_2$, and $a^p + (-a)^p = 0$.)  But the
answer is no for $m \geq 3$.  For instance, when $p = 3$, we have
\[
H(1, 1, 1, 1) = H(\vc{u}_4) = \fq{3}(4) = \tfrac{1}{3}(4^2 - 1) = -1
\]
by Example~\ref{eg:ent-p-ufm}, which is not equal to $H(1) = 0$, even
though $1 + 1 + 1 = 0$.
\end{example}

Distributions over $\Zp$ can be composed, using the same formula as in the
real case (Definition~\ref{defn:comp-dist}).  As in the real case, entropy
mod~$p$ satisfies the chain rule:

\begin{propn}[Chain rule]
\lbl{propn:chn-p}%
\index{chain rule!modulo a prime}%
We have
\[
H\bigl( \ggamma \of (\ppi^1, \ldots, \ppi^n) \bigr)
=
H(\ggamma) + \sum_{i = 1}^n \gamma_i H(\ppi^i)
\]
for all $n, k_1, \ldots, k_n \geq 1$, all $\ggamma = (\gamma_1, \ldots,
\gamma_n) \in \Pi_n$, and all $\ppi^i \in \Pi_{k_i}$.
\end{propn}

\begin{proof}
Write $\ppi^i = \bigl(\pi^i_1, \ldots, \pi^i_{k_i}\bigr)$.  Choose $b_i \in
\Z$ representing $\gamma_i \in \Zp$ and $a^i_j \in \Z$ representing
$\pi^i_j \in \Zp$, for each $i$ and $j$.  Write $A^i = a^i_1 + \cdots +
a^i_{k_i}$.

We evaluate in turn the three terms $H\bigl( \ggamma \of (\ppi^1, \ldots,
\ppi^n) \bigr)$, $H(\ggamma)$, and $\sum \gamma_i H(\ppi^i)$.
First, by Lemma~\ref{lemma:ent-partial} and the derivation property of
$\partial$ (Lemma~\ref{lemma:p-deriv-elem}),
\begin{align*}
H\bigl( \ggamma \of (\ppi^1, \ldots, \ppi^n) \bigr)     &
=
\sum_{i = 1}^n \sum_{j = 1}^{k_i} \partial\bigl(b_i a^i_j\bigr)
-
\partial \Biggl( \sum_{i = 1}^n \sum_{j = 1}^{k_i} b_i a^i_j \Biggr)    \\
&
=
\sum_{i = 1}^n \sum_{j = 1}^{k_i}
\bigl( \partial(b_i) a^i_j + b_i \partial\bigl(a^i_j\bigr) \bigr)
-
\partial \Biggl( \sum_{i = 1}^n b_i A^i \Biggr) \\
&
=
\sum_{i = 1}^n \partial(b_i) A^i 
+ \sum_{i = 1}^n b_i \sum_{j = 1}^{k_i} \partial\bigl(a^i_j\bigr)
-
\partial \Biggl( \sum_{i = 1}^n b_i A^i \Biggr). 
\end{align*}
Second, $A^i \equiv 1 \pmod{p}$ since $\ppi^i \in \Pi_{k_i}$, so
$b_i A^i \in \Z$ represents $\gamma_i \in \Zp$.  Hence
\begin{align*}
H(\ggamma)      &
=
\sum_{i = 1}^n \partial\bigl(b_i A^i\bigr) 
- \partial\Biggl( \sum_{i = 1}^n b_i A^i \Biggr)        \\
&
=
\sum_{i = 1}^n \partial(b_i) A^i
+ \sum_{i = 1}^n b_i \partial(A^i)
- \partial\Biggl( \sum_{i = 1}^n b_i A^i \Biggr).        
\end{align*}
Third, 
\[
\sum_{i = 1}^n \gamma_i H(\ppi^i)       
=
\sum_{i = 1}^n b_i \sum_{j = 1}^{k_i} \partial\bigl(a^i_j\bigr)
- \sum_{i = 1}^n b_i \partial(A^i).
\]
The result follows.
\end{proof}

There is a tensor%
\index{tensor product} 
product for distributions mod~$p$, defined as in the real case
(p.~\pageref{p:tensor}), and entropy mod~$p$ has the familiar logarithmic%
\index{entropy!modulo a prime!logarithmic property}%
\index{logarithmic sequence} 
property:

\begin{cor}
\lbl{cor:ent-p-loglike}
$H(\ggamma \otimes \ppi) = H(\ggamma) + H(\ppi)$ for all $\ggamma \in
  \Pi_n$ and $\ppi \in \Pi_m$.
\qed
\end{cor}

\section{Characterizations of entropy and information loss}
\lbl{sec:p-char}

We now state our characterization theorem for entropy mod~$p$, whose close
resemblance to the characterization theorem for real entropy
(Theorem~\ref{thm:faddeev}) is the main justification for the definition.

\begin{thm}
\lbl{thm:fad-p}%
\index{entropy!modulo a prime!characterization of}%
\index{Faddeev, Dmitry!mod p analogue of entropy theorem@mod $p$ analogue of entropy theorem}%
Let $\bigl( I \from \Pi_n \to \Zp \bigr)_{n \geq 1}$ be a sequence of
functions.  The following are equivalent:
\begin{enumerate}
\item 
\lbl{part:fad-p-condns}
$I$ satisfies the chain rule (that is, satisfies the conclusion of
Proposition~\ref{propn:chn-p} with $I$ in place of $H$);

\item
\lbl{part:fad-p-form}
$I = cH$ for some $c \in \Zp$.
\end{enumerate}
\end{thm}

As in our sharper version of Faddeev's theorem over $\R$
(Theorem~\ref{thm:faddeev}), no symmetry condition is needed.

Since $H$ satisfies the chain rule, so does any constant multiple of $H$.
Hence~\bref{part:fad-p-form} implies~\bref{part:fad-p-condns}.
We now begin the proof of the converse.  \femph{For the rest of the proof},
let $\bigl( I \from \Pi_n \to \Zp \bigr)_{n \geq 1}$ be a sequence of
functions satisfying the chain rule.

\begin{lemma}
\lbl{lemma:ent-p-elem}
\begin{enumerate}
\item
\lbl{part:epe-bin}
$I(\vc{u}_{mn}) = I(\vc{u}_m) + I(\vc{u}_n)$ for all $m, n \in \nat$ not
  divisible by $p$;

\item 
\lbl{part:epe-null}
$I(\vc{u}_1) = 0$.
\end{enumerate}
\end{lemma}

\begin{proof}
Both parts are proved exactly as in the real case
(Lemma~\ref{lemma:fad-log}).  
\end{proof}

\begin{lemma}
\lbl{lemma:p-10}
$I(1, 0) = I(0, 1) = 0$.
\end{lemma}

\begin{proof}
The proof that $I(1, 0) = 0$ is identical to the proof in the real case
(Lemma~\ref{lemma:fad-10}), and $I(0, 1) = 0$ is proved similarly.
\end{proof}

\begin{lemma}
\lbl{lemma:ent-p-abs}
For all $\ppi \in \Pi_n$ and $i \in \{0, \ldots, n\}$, 
\[
I(\pi_1, \ldots, \pi_n)
=
I(\pi_1, \ldots, \pi_i, 0, \pi_{i + 1}, \ldots, \pi_n).
\]
\end{lemma}

\begin{proof}
First suppose that $i \neq 0$.  Then
\[
(\pi_1, \ldots, \pi_i, 0, \pi_{i + 1}, \ldots, \pi_n)
=
\ppi \of 
\bigl(\underbrace{\vc{u}_1, \ldots, \vc{u}_1}_{i - 1}, (1, 0), 
\underbrace{\vc{u}_1, \ldots, \vc{u}_1}_{n - i}\bigr).
\]
Applying $I$ to both sides, then using the chain rule and that $I(\vc{u}_1)
= I(1, 0) = 0$, gives the result.  The case $i = 0$ is proved similarly,
now using $I(0, 1) = 0$.
\end{proof}

As in the real case, we will prove the characterization theorem by
analysing $I(\vc{u}_n)$ as $n$ varies.  And as in the real case, the chain
rule will allow us to deduce the value of $I(\ppi)$ for more general
distributions $\ppi$:

\begin{lemma}
\lbl{lemma:p-ufm-to-gen}
Let $\ppi \in \Pi_n$ with $\pi_i \neq 0$ for all $i$.  For each $i$, let
$k_i \geq 1$ be an integer representing $\pi_i \in \Zp$, and write $k =
\sum_{i = 1}^n k_i$.  Then 
\[
I(\ppi) = I(\vc{u}_k) - \sum_{i = 1}^n k_i I(\vc{u}_{k_i}).
\]
\end{lemma}

\begin{proof}
First note that none of $k_1, \ldots, k_n$ is a multiple of $p$, and since
$k$ represents $\sum \pi_i = 1 \in \Zp$, neither is $k$.  Hence
$\vc{u}_{k_i}$ and $\vc{u}_k$ are well-defined.  By definition of
composition,
\[
\ppi \of (\vc{u}_{k_1}, \ldots, \vc{u}_{k_n}) 
= 
(\underbrace{1, \ldots, 1}_k)
=
\vc{u}_k.
\]
Applying $I$ and using the chain rule gives the result.
\end{proof}

We now come to the most delicate part of the argument.  Since $H(\vc{u}_n)
= \fq{p}(n)$, and since $\fq{p}(n)$ is $p^2$-periodic in $n$, if $I$ is to
be a constant multiple of $H$ then $I(\vc{u}_n)$ must also be
$p^2$-periodic in $n$.  We show this directly.

\begin{lemma}
\lbl{lemma:p-per}
$I(\vc{u}_{n + p^2}) = I(\vc{u}_n)$ for all natural numbers $n$ not
  divisible by $p$.
\end{lemma}

\begin{proof}
First we prove the existence of a constant $c \in \Zp$ such that for all $n
\in \nat$ not divisible by $p$,
\begin{equation}
\lbl{eq:per-one}
I(\vc{u}_{n + p}) = I(\vc{u}_n) - c/n.
\end{equation}
(Compare Lemma~\ref{lemma:fq-elem}\bref{part:fq-elem-shift}.)  An
equivalent statement is that $n(I(\vc{u}_{n + p}) - I(\vc{u}_n))$ is
independent of $n$.  For any $n_1$ and $n_2$, we can
choose some $m \geq \max\{n_1, n_2\}$ with $m \equiv 1 \pmod{p}$, so it is
enough to show that whenever $0 \leq n \leq m$ with $n \not\equiv 0
\pmod{p}$ and $m \equiv 1 \pmod{p}$,
\begin{equation}
\lbl{eq:per-nm}
n \bigl( I(\vc{u}_{n + p}) - I(\vc{u}_{n}) \bigr)
=
I(\vc{u}_{m + p}) - I(\vc{u}_{m}).
\end{equation}
To prove this, consider the distribution
\[
\ppi = (n, \underbrace{1, \ldots, 1}_{m - n}).
\]
By Lemma~\ref{lemma:p-ufm-to-gen} and the fact that $I(\vc{u}_1) = 0$,
\[
I(\ppi) = I(\vc{u}_m) - nI(\vc{u}_n).
\]
But also
\[
\ppi = (n + p, \underbrace{1, \ldots, 1}_{m - n}),
\]
so by the same argument,
\begin{align*}
I(\ppi) &
=
I(\vc{u}_{m + p}) - (n + p)I(\vc{u}_{n + p})    \\
&
= 
I(\vc{u}_{m + p}) - nI(\vc{u}_{n + p}).
\end{align*}
Comparing the two expressions for $I(\ppi)$ gives
equation~\eqref{eq:per-nm}, thus proving the initial claim.

By induction on equation~\eqref{eq:per-one},
\[
I(\vc{u}_{n + rp}) = I(\vc{u}_n) - cr/n
\]
for all $n, r \in \nat$ with $n$ not divisible by $p$.  The result 
follows by setting $r = p$.
\end{proof}

We can now prove the characterization theorem for entropy modulo~$p$.

\begin{pfof}{Theorem~\ref{thm:fad-p}}
Define $f \from \{ n \in \nat \such p \ndvd n\} \to \Zp$ by $f(n) =
I(\vc{u}_n)$.  By Lemma~\ref{lemma:ent-p-elem}, $f(mn) = f(m) + f(n)$ for
all $m, n$ not divisible by $p$.  By Lemma~\ref{lemma:p-per}, $f(n + p^2) =
f(n)$ for all $n$ not divisible by $p$.  Hence by
Theorem~\ref{thm:fq-char}, $f = c\fq{p}$ for some $c \in \Zp$.  It follows
from Example~\ref{eg:ent-p-ufm} that $I(\vc{u}_n) = cH(\vc{u}_n)$ for all
$n$ not divisible by $p$.

Since both $I$ and $cH$ satisfy the chain rule,
Lemma~\ref{lemma:p-ufm-to-gen} applies to both; and since $I$ and $cH$ are
equal on uniform distributions, they are also equal on all distributions
$\ppi$ such that $\pi_i \neq 0$ for all $i$.  Finally, applying
Lemma~\ref{lemma:ent-p-abs} to both $I$ and $cH$, we deduce by 
induction that $I(\ppi) = cH(\ppi)$ for all $\ppi \in \Pi_n$.
\end{pfof}

In the real case, the characterization theorem for entropy leads to a
characterization of information loss involving only linear conditions
(Theorem~\ref{thm:cetil}).  The same holds for entropy mod~$p$, and the
argument can be copied over from the real case nearly verbatim.

Thus, given a finite set $\XX$, we write $\Pi_\XX$ for the set of families
$\ppi = (\pi_i)_{i \in \XX}$ of elements of $\Zp$ such that $\sum_{i \in \XX}
\pi_i = 1$.  A \demph{finite probability space mod~$p$}%
\index{probability space!modulo a prime}
is a finite set $\XX$ together with an element $\ppi \in \Pi_\XX$.  A
\demph{measure-preserving%
\index{measure-preserving map!modulo a prime} 
map} $f \from (\YY, \ssigma) \to (\XX, \ppi)$
between such spaces is a function $f \from \YY \to \XX$ such that
\[
\pi_i = \sum_{j \in f^{-1}(i)} s_j
\]
for all $i \in \XX$.  

As in the real case, we can take convex combinations of both
probability spaces and maps between them.  Given two finite probability
spaces mod~$p$, say $(\XX, \ppi)$ and $(\XX', \ppi')$, and given also a
scalar $\lambda \in \Zp$, we obtain another such space, $\bigl(\XX \cpd
\XX', \lambda\ppi \cpd (1 - \lambda)\ppi'\bigr)$\ntn{smallccdistsp}.  Given
two measure-preserving maps
\begin{eqnarray*}
f \from         (\YY, \ssigma)   &\to&    (\XX, \ppi), \\
f' \from        (\YY', \ssigma') &\to&    (\XX', \ppi')
\end{eqnarray*}
and an element $\lambda \in \Zp$, we obtain a new measure-preserving map
\[
\lambda f \cpd (1 - \lambda) f'
\from
\bigl(\YY \cpd \YY', \lambda\ssigma \cpd (1 - \lambda)\ssigma'\bigr)
\to
\bigl(\XX \cpd \XX', \lambda\ppi \cpd (1 - \lambda)\ppi'\bigr),
\ntn{smallccmapsp}
\]
exactly as in Section~\ref{sec:meas-pres}.

The \demph{entropy}%
\index{entropy!modulo a prime} 
of $\ppi \in \Pi_\XX$ is, naturally, 
\[
H(\ppi) = \frac{1}{p} \Biggl( 1 - \sum_{i \in \XX} a_i^p \Biggr),
\ntn{entgenp}
\]
where $a_i \in \Z$ represents $\pi_i \in \Zp$ for each $i \in \XX$.  The
\demph{information loss}%
\index{information loss!modulo a prime} 
of a measure-preserving map $f \from (\YY, \ssigma)
\to (\XX, \ppi)$ between finite probability spaces mod~$p$ is
\[
L(f) = H(\ssigma) - H(\ppi) \in \Zp.
\ntn{Lp}
\]

\begin{thm}
\lbl{thm:cetil-p}
\index{information loss!modulo a prime!characterization of}
Let $K$ be a function assigning an element $K(f) \in \Zp$ to each
measure-preserving map $f$ of finite probability spaces mod~$p$.  The
following are equivalent:
\begin{enumerate}
\item 
\lbl{part:cetil-p-condns}
$K$ has these three properties:
\begin{enumerate}
\item 
$K(f) = 0$ for all isomorphisms $f$;

\item
$K(g \of f) = K(g) + K(f)$ for all composable pairs $(f, g)$ of
  measure-preserving maps;

\item
$K\bigl(\lambda f \cpd (1 - \lambda) f'\bigr) = \lambda K(f) + (1 -
  \lambda) K(f')$ for all measure-preserving maps $f$ and $f'$ and all
  $\lambda \in \Zp$;
\end{enumerate}

\item
\lbl{part:cetil-p-form}
$K = cL$ for some $c \in \Zp$.
\end{enumerate}
\end{thm}

\begin{proof}
The proof is identical to that of the real case, Theorem~\ref{thm:cetil},
but with $\Zp$ in place of $\R$, Theorem~\ref{thm:fad-p} in place of
Faddeev's theorem, and all mention of continuity removed.
\end{proof}

\section{The residues of real entropy}
\lbl{sec:p-res}
\index{entropy!residue modulo a prime}

Having found a satisfactory definition of the entropy of a probability
distribution mod~$p$, we are now in a position to develop 
Kontsevich's%
\index{Kontsevich, Maxim} 
suggestion about the residues mod~$p$ of real entropy, quoted at the start of
this chapter.  (That quotation was the sum total of what he wrote on the
subject.)

Let $\ppi \in \Delta_n$ be a probability distribution with
rational probabilities, say $\ppi = (a_1/b_1, \ldots, a_n/b_n)$ with $a_i,
b_i \in \Z$.  There are only finitely many primes that divide one or more
of the denominators $b_i$.  If $p$ is not in that exceptional set then
$\ppi$ defines an element of $\Pi_n$, and therefore has a mod~$p$ entropy
$H(\ppi) \in \Zp$.  Kontsevich invites us to think of this as the residue%
\index{residue class} 
class modulo~$p$ of the real entropy $H(\ppi) \in \R$.

Kontsevich phrased his suggestion playfully, but there is more to it than
meets the eye.  To explain it, let us write $H_\R$\ntn{HR} for
entropy over the reals, $H_p$\ntn{Hpp} for entropy mod~$p$, and
$\Rat{n}{p}$\ntn{Deltanp} for the set of real probability distributions
$\ppi \in \Delta_n$ such that each $\pi_i$ can be expressed as a rational
number with denominator not divisible by $p$.  The proposal is that given
$\ppi \in \Rat{n}{p}$, we regard $H_p(\ppi) \in \Zp$ as the residue mod~$p$
of $H_\R(\ppi) \in \R$.

Now, different distributions can have the same entropy over $\R$.  For
instance, 
\[
H_\R\bigl(\tfrac{1}{2}, 
\tfrac{1}{8}, \tfrac{1}{8}, \tfrac{1}{8}, \tfrac{1}{8}\bigr)
=
H_\R\bigl(\tfrac{1}{4}, \tfrac{1}{4}, \tfrac{1}{4}, \tfrac{1}{4}\bigr).
\]
There is, therefore, a question of consistency: Kontsevich's proposal only
makes sense if
\[
H_\R(\ppi) = H_\R(\ggamma) \implies H_p(\ppi) = H_p(\ggamma)
\]
for all $\ppi \in \Rat{n}{p}$ and $\ggamma \in \Rat{m}{p}$.  We now show
that this is true.

\begin{lemma}
\lbl{lemma:p-prod-par}
Let $n, m \geq 1$ and let $a_1, \ldots, a_n, b_1, \ldots, b_m \geq 0$ be
integers.  Then
\[
\prod_{i = 1}^n a_i^{a_i} = \prod_{j = 1}^m b_j^{b_j}
\implies
\sum_{i = 1}^n \partial(a_i) = \sum_{j = 1}^m \partial(b_j),
\]
where the first equality is in $\Z$, the second is in $\Zp$, and we use the
convention that $0^0 = 1$.
\end{lemma}

Here $\partial$ is the map $\Z \to \Zp$ defined in
equation~\eqref{eq:p-deriv}.  The analogue of this lemma for the
real-valued map $\partial$ of equation~\eqref{eq:real-deriv} is trivial:
simply discard the factors in the products for which $a_i$ or $b_j$ is $0$,
then take logarithms.  But over $\Zp$, it is not so simple.  The subtlety
arises from the possibility that some $a_i$ or $b_j$ is not zero but is
divisible by $p$.  In that case, $\prod a_i^{a_i} = \prod b_j^{b_j}$ is
divisible by $p$, so its Fermat quotient~-- the analogue of the
logarithm~-- is undefined.  A more detailed analysis is therefore
required. 

\begin{proof}
Since $0^0 = 1$ and $\partial(0) = 0$, it is enough to prove the result in
the case where each of the integers $a_i$ and $b_j$ is strictly positive.
We may then write $a_i = p^{\alpha_i} A_i$ with $\alpha_i \geq 0$ and $p
\ndvd A_i$, and similarly $b_j = p^{\beta_j} B_j$.  We adopt the convention
that unless mentioned otherwise, the index $i$ ranges over $1, \ldots, n$
and the index $j$ over $1, \ldots, m$.

Assume that $\prod a_i^{a_i} = \prod b_j^{b_j}$.  We have
\[
\prod a_i^{a_i}
=
p^{\sum \alpha_i a_i} \prod A_i^{a_i}
\]
with $p \ndvd \prod A_i^{a_i}$, and similarly for $\prod b_j^{b_j}$.
It follows that
\begin{align}
\prod A_i^{a_i}       &
=
\prod B_j^{b_j},       
\lbl{eq:ppp-prod}       \\
\sum \alpha_i a_i &
=
\sum \beta_j b_j
\lbl{eq:ppp-sum}        
\end{align}
in $\Z$. We consider each of these equations in turn.

First, since $p \ndvd \prod A_i^{a_i}$, the Fermat quotient
$\fq{p}\bigl(\prod A_i^{a_i}\bigr)$ is well-defined, and the logarithmic
property of $\fq{p}$ (Lemma~\ref{lemma:fq-elem}\bref{part:fq-elem-log})
gives 
\[
-\fq{p}\Bigl( \prod A_i^{a_i} \Bigr)
=
\sum - a_i \fq{p}(A_i).
\]
Consider the right-hand side as an element of $\Zp$.  When $p \dvd a_i$,
the $i$-summand vanishes.  When $p \ndvd a_i$, the $i$-summand is $-a_i
\fq{p}(a_i) = \partial(a_i)$.  Hence
\[
-\fq{p}\Bigl( \prod A_i^{a_i} \Bigr)
=
\sum_{i \csuch \alpha_i = 0} \partial(a_i)
\]
in $\Zp$.  A similar result holds for $\prod B_j^{b_j}$, so
equation~\eqref{eq:ppp-prod} gives
\begin{equation}
\lbl{eq:ppp-zero}
\sum_{i \csuch \alpha_i = 0} \partial(a_i)
=
\sum_{j \csuch \beta_j = 0} \partial(b_j).
\end{equation}

Second,
\[
\sum_{i = 1}^n \alpha_i a_i 
= 
\sum_{i \csuch \alpha_i \geq 1} \alpha_i a_i,
\]
so $p \dvd \sum \alpha_i a_i$.  Now
\[
\frac{1}{p} \sum \alpha_i a_i 
=
\sum_{i \csuch \alpha_i \geq 1} \alpha_i p^{\alpha_i - 1} A_i
\equiv
\sum_{i \csuch \alpha_i = 1} A_i \pmod{p},
\]
and if $\alpha_i = 1$ then $A_i = a_i/p = \partial(a_i)$.  
A similar result holds for $\sum \beta_j b_j$, so
equation~\eqref{eq:ppp-sum} gives
\begin{equation}
\lbl{eq:ppp-one}
\sum_{i \csuch \alpha_i = 1} \partial(a_i)
=
\sum_{j \csuch \beta_j = 1} \partial(b_j)
\end{equation}
in $\Zp$.

Finally, for each $i$ such that $\alpha_i \geq 2$, we have $p^2 \dvd a_i$
and so $\partial(a_i) = 0$ in $\Zp$.  The same holds for $b_j$, so 
\begin{equation}
\lbl{eq:ppp-geqtwo}
\sum_{i \csuch \alpha_i \geq 2} \partial(a_i)
=
\sum_{j \csuch \beta_j \geq 2} \partial(b_j), 
\end{equation}
both sides being $0$.  Adding equations~\eqref{eq:ppp-zero},
\eqref{eq:ppp-one} and~\eqref{eq:ppp-geqtwo} gives the result.
\end{proof}

We deduce that the real entropy of a rational distribution determines its
entropy mod $p$:

\begin{thm}
\lbl{thm:R-p}
\index{entropy!residue modulo a prime}
\index{residue class}
Let $n, m \geq 1$, $\ppi \in \Rat{n}{p}$, and $\ggamma \in \Rat{m}{p}$.
Then
\[
H_\R(\ppi) = H_\R(\ggamma) \implies H_p(\ppi) = H_p(\ggamma).
\]
\end{thm}

\begin{proof}
We can write
\[
\ppi = (r_1/t, \ldots, r_n/t),  
\qquad
\ggamma = (s_1/t, \ldots, s_m/t),
\]
where $r_i$, $s_j$ and $t$ are nonnegative integers such that $p \ndvd t$
and
\[
r_1 + \cdots + r_n = t = s_1 + \cdots + s_m.
\]
By multiplying all of these integers by a constant if necessary, we may
assume that $t \equiv 1 \pmod{p}$.  

By definition,
\[
e^{-H_\R(\ppi)} = \prod_i (r_i/t)^{r_i/t},
\]
with the convention that $0^0 = 1$.  Multiplying both sides by $t$ and then
raising to the power of $t$ gives
\[
t^t e^{-t H_\R(\ppi)} = \prod_i r_i^{r_i}.
\]
By the analogous equation for $\ggamma$ and the assumption that $H_\R(\ppi)
= H_\R(\ggamma)$, it follows that
\[
\prod_i r_i^{r_i} = \prod_j s_j^{s_j}.
\]
Lemma~\ref{lemma:p-prod-par} now gives
\[
\sum_i \partial(r_i) = \sum_j \partial (s_j)
\]
in $\Zp$.  Since $\sum r_i = t = \sum s_j$, it follows that
\[
\sum_i \partial(r_i) - \partial \Biggl( \sum_i r_i \Biggr)
=
\sum_j \partial(s_j) - \partial \Biggl( \sum_j s_j \Biggr).
\]
But $t \equiv 1 \pmod{p}$, so $r_i$ represents the element $r_i/t = \pi_i$
of $\Zp$, so by Lemma~\ref{lemma:ent-partial}, the left-hand side of this
equation is $H_p(\ppi)$.  Similarly, the right-hand side is
$H_p(\ggamma)$. Hence $H_p(\ppi) = H_p(\ggamma)$.
\end{proof}

It follows that Kontsevich's residue classes of real entropies are
well-defined.  That is, writing
\[
\Rentp
=
\bigcup_{n = 1}^\infty
\Bigl\{ H_\R(\ppi) \such \ppi \in \Rat{n}{p} \Bigr\}
\sub \R,
\]
there is a unique map of sets
\[
[\,\cdot\,] \from \Rentp \to \Zp
\]
such that $[H_\R(\ppi)] = H_p(\ppi)$ for all $\ppi \in
\Rat{n}{p}$ and $n \geq 1$.  We now show that this map is additive, as the
word `residue' leads one to expect.

\begin{propn}
\index{residue class!additivity of}
The set $\Rentp$ is closed under addition, and the residue map
\[
\begin{array}{cccc}
{}[\,\cdot\,] \from     &     
\Rentp  &
\to     &
\Zp     \\
&
H_\R(\ppi)      &
\mapsto &
H_p(\ppi)
\end{array}
\]
preserves addition.
\end{propn}

\begin{proof}
Let $\ppi \in \Rat{n}{p}$ and $\ggamma \in \Rat{m}{p}$.  We must
show that $H_\R(\ppi) + H_\R(\ggamma) \in \Rentp$ and 
\[
[H_\R(\ppi) + H_\R(\ggamma)] 
=
[H_\R(\ppi)] + [H_\R(\ggamma)].
\]
Evidently $\ppi \otimes \ggamma \in \Rat{nm}{p}$, so by the logarithmic
property of $H_\R$,
\[
H_\R(\ppi) + H_\R(\ggamma)
=
H_\R(\ppi \otimes \ggamma)
\in \Rentp. 
\]
Now also using the logarithmic property of $H_p$
(Corollary~\ref{cor:ent-p-loglike}),
\begin{align*}
[H_\R(\ppi) + H_\R(\ggamma)]    
&
= 
[H_\R(\ppi \otimes \ggamma)]    \\
&
=
H_p(\ppi \otimes \ggamma)       \\
&
=
H_p(\ppi) + H_p(\ggamma)        \\
&
=
[H_\R(\ppi)] + [H_\R(\ggamma)],
\end{align*}
as required.
\end{proof}

\section{Polynomial approach}
\lbl{sec:p-poly}

There is an alternative approach to entropy modulo a prime.  It repairs a
defect of the approach above: that in order to define the entropy of a
distribution $\ppi$ over $\Zp$, we had to step outside $\Zp$ to make
arbitrary choices of integers representing the probabilities $\pi_i$, then
show that the definition was independent of those choices.  We now show how
to define $H(\ppi)$ directly as a function of $\pi_1, \ldots, \pi_n$.

Inevitably, that function is a polynomial, by the classical fact that every
function $\fld^n \to \fld$ on a finite field $\fld$ is induced by some
polynomial in $n$ variables.  Indeed, there is a \emph{unique} such
polynomial whose degree in each variable is strictly less than the order of
the field:

\begin{lemma}
\lbl{lemma:fn-fin-fld}%
\index{finite field, functions on}%
\index{polynomial!finite field@over finite field}%
Let $\fld$ be a finite field with $q$ elements, let $n \geq 0$, and let $F
\from \fld^n \to \fld$ be a function.  Then there is a unique polynomial
$f$ of the form
\begin{equation*}
f(x_1, \ldots, x_n)
=
\sum_{0 \leq r_1, \ldots, r_n < q} c_{r_1, \ldots, r_n} 
x_1^{r_1} \cdots x_n^{r_n}
\end{equation*}
($c_{r_1, \ldots, r_n} \in \fld$) such that
\[
f(\pi_1, \ldots, \pi_n) = F(\pi_1, \ldots, \pi_n)
\]
for all $\pi_1, \ldots, \pi_n \in \fld$.
\end{lemma}

\begin{proof}
See Appendix~\ref{sec:fff}.
\end{proof}

In particular, taking $\fld = \Zp$, entropy modulo~$p$ can be expressed as
a polynomial of degree less than $p$ in each variable.  We now 
identify such a polynomial.

For each $n \geq 1$, define $h(x_1, \ldots, x_n) \in (\Zp)[x_1, \ldots,
  x_n]$ by 
\[
\ntn{hpoly}
h(x_1, \ldots, x_n)
=
-\sum_{\substack{0 \leq r_1, \ldots, r_n < p\\ r_1 + \cdots + r_n = p}}
\frac{x_1^{r_1} \cdots x_n^{r_n}}{r_1! \cdots r_n!}.
\]

\begin{propn}
\lbl{propn:ent-p-eqv}%
\index{entropy!modulo a prime!polynomial form}%
For all $n \geq 1$ and $(\pi_1, \ldots, \pi_n) \in \Pi_n$,
\[
H(\pi_1, \ldots, \pi_n) 
=
h(\pi_1, \ldots, \pi_n).
\]
\end{propn}

\begin{proof}
Let $\pi_1, \ldots, \pi_n \in \Zp$.  We will show that whenever $a_1,
\ldots, a_n$ are integers representing $\pi_1, \ldots, \pi_n$, then
\begin{equation}
\lbl{eq:Hh}
\frac{1}{p} \Biggl( 
\Biggl( \sum_{i = 1}^n a_i \Biggr)^p - \sum_{i = 1}^n a_i^p
\Biggr)
\end{equation}
is an integer representing $h(\pi_1, \ldots, \pi_n)$.  The result will
follow, since if $\ppi \in \Pi_n$ then $\sum \pi_i = 1$, so $(\sum a_i)^p
\equiv 1 \pmod{p^2}$ by Lemma~\ref{lemma:pps}.

We have to prove that
\[
\Biggl( \sum_{i = 1}^n a_i \Biggr)^p - \sum_{i = 1}^n a_i^p 
\equiv
-p
\sum_{\substack{0 \leq r_1, \ldots, r_n < p\\ r_1 + \cdots + r_n = p}}
\frac{a_1^{r_1} \cdots a_n^{r_n}}{r_1! \cdots r_n!}
\pmod{p^2}.
\]
Since $(p - 1)!$ is invertible in $\Zps$, an equivalent statement is that
\begin{equation}
\lbl{eq:ent-p-eqv-1}
(p - 1)! \Biggl( \sum_{i = 1}^n a_i^p  
- \Biggl( \sum_{i = 1}^n a_i \Biggr)^p \Biggr)
\equiv
\sum_{\substack{0 \leq r_1, \ldots, r_n < p\\ r_1 + \cdots + r_n = p}}
\frac{p!}{r_1! \cdots r_n!}
a_1^{r_1} \cdots a_n^{r_n}
\pmod{p^2}.
\end{equation}
The right-hand side is $\bigl(\sum a_i\bigr)^p - \sum a_i^p$, so
equation~\eqref{eq:ent-p-eqv-1} reduces to
\[
\bigl( (p - 1)! + 1 \bigr)
\Biggl( \sum_{i = 1}^n a_i^p - \Biggl( \sum_{i = 1}^n a_i \Biggr)^p \Biggr)
\equiv
0
\pmod{p^2}.
\]
And since $(p - 1)! \equiv - 1 \pmod{p}$ and $\sum a_i^p \equiv \sum a_i
\equiv (\sum a_i)^p \pmod{p}$, this is true.
\end{proof}


We have shown that $h(\pi_1, \ldots, \pi_n) = H(\pi_1, \ldots, \pi_n)$
whenever $\sum \pi_i = 1$.  Lemma~\ref{lemma:fn-fin-fld} does not imply
that $h$ is the unique such polynomial of degree less than $p$ in each
variable, since this equation is only stated (and $H$ is only defined) in
the case where the arguments sum to $1$.  However, $h$ has further good
properties.  It is homogeneous of degree $p$, which implies that the
polynomial function $\bar{H} \from (\Zp)^n \to \Zp$ induced by $h$ is
homogeneous of degree $1$.  In fact,
\[
\bar{H}(\ppi) 
= 
\sum_{i = 1}^n \partial(a_i) 
- \partial\Biggl( \sum_{i = 1} a_i \Biggr).
\]
(This follows from the fact that the integer~\eqref{eq:Hh} represents
$h(\pi_1, \ldots, \pi_n)$.)  So in the light of
Lemma~\ref{lemma:ent-partial} and the explanation that follows it, $h$ is
the natural choice of polynomial representing entropy mod~$p$.

We now establish several polynomial identities satisfied by $h$, which are
stronger than the functional equations previously proved for $H$.  The
first is closely related to the chain rule, as we shall see.

\begin{thm}
\lbl{thm:p-grouping}
Let $n, k_1, \ldots, k_n \geq 0$.  Then $h$ satisfies the following
identity of polynomials in commuting variables $x_{ij}$ over $\Zp$:
\begin{multline*}
h(x_{11}, \ldots, x_{1k_1}, 
\ \ldots, \ 
x_{n1}, \ldots, x_{nk_n})
\\
=
h\bigl( x_{11} + \cdots + x_{1k_1}, \ \ldots, \
x_{n1} + \cdots + x_{nk_n} \bigr)
+
\sum_{i = 1}^n h(x_{i1}, \ldots, x_{ik_i}).
\end{multline*}
\end{thm}

\begin{proof}
The left-hand side of this equation is equal to
\begin{equation}
\lbl{eq:ch-poly-1}
-\sum_{\substack{0 \leq s_1, \ldots, s_n \leq p\\ s_1 + \cdots + s_n = p}}
\sum
\frac{x_{11}^{r_{11}} \cdots x_{1k_1}^{r_{1k_1}} \ \cdots \ 
x_{n1}^{r_{n1}} \cdots x_{nk_n}^{r_{nk_n}}}%
{r_{11}! \cdots r_{1k_1}! \ \cdots \ r_{n1}! \cdots r_{nk_n}!},
\end{equation}
where the inner sum is over all $r_{11}, \ldots, r_{nk_n}$ such
that $0 \leq r_{ij} < p$ and 
\[
r_{11} + \cdots + r_{1k_1} = s_1,
\ \ldots,\ 
r_{n1} + \cdots + r_{nk_n} = s_n.
\]
Split the outer sum into two parts, the
first consisting of the summands in which none of $s_1, \ldots, s_n$ is
equal to $p$, and the second consisting of the summands in which one $s_i$
is equal to $p$ and the others are zero.  Then the
polynomial~\eqref{eq:ch-poly-1} is equal to $A + B$, where
\begin{align*}
A       &
=
-\sum_{\substack{0 \leq s_1, \ldots, s_n < p\\ s_1 + \cdots + s_n = p}}
\prod_{i = 1}^n
\sum_{\substack{r_{i1}, \ldots, r_{ik_i} \geq 0\\ 
r_{i1} + \cdots + r_{ik_i} = s_i}}
\frac{x_{i1}^{r_{i1}} \cdots x_{ik_i}^{r_{ik_i}}}%
{r_{i1}! \cdots r_{ik_i}!},     \\
B       &
=
-\sum_{i = 1}^n 
\sum_{\substack{0 \leq r_{i1}, \ldots, r_{ik_i} < p\\ 
r_{i1} + \cdots + r_{ik_i} = p}}
\frac{x_{i1}^{r_{i1}} \cdots x_{ik_i}^{r_{ik_i}}}%
{r_{i1}! \cdots r_{ik_i}!}.
\end{align*}
We have
\begin{align*}
A       &
=
-\sum_{\substack{0 \leq s_1, \ldots, s_n < p\\ s_1 + \cdots + s_n = p}}
\frac{1}{s_1! \cdots s_n!}
\prod_{i = 1}^n
\sum_{\substack{r_{i1}, \ldots, r_{ik_i} \geq 0\\ 
r_{i1} + \cdots + r_{ik_i} = s_i}}
\frac{s_i!}{r_{i1}! \cdots r_{ik_i}!}      
x_{i1}^{r_{i1}} \cdots x_{ik_i}^{r_{ik_i}}  \\
&
=
-\sum_{\substack{0 \leq s_1, \ldots, s_n < p\\ s_1 + \cdots + s_n = p}}
\frac{1}{s_1! \cdots s_n!}
\prod_{i = 1}^n
(x_{i1} + \cdots + x_{ik_i})^{s_i}      \\
&
=
h(x_{11} + \cdots + x_{1k_1}, \ \ldots,\ 
x_{n1} + \cdots + x_{nk_n})
\end{align*}
and
\[
B = 
\sum_{i = 1}^n h(x_{i1}, \ldots, x_{ik_i}).
\]
The result follows.
\end{proof}

We easily deduce the polynomial form of the chain rule.

\begin{cor}[Chain rule]
\lbl{cor:p-chn-poly}%
\index{chain rule!modulo a prime}%
Let $n, k_1, \ldots, k_n \geq 0$.  Then $h$ satisfies the following
identity of polynomials in commuting variables $y_i$, $x_{ij}$ over $\Zp$:
\begin{multline*}
h(y_1 x_{11}, \ldots, y_1 x_{1k_1}, 
\ \ldots, \ 
y_n x_{n1}, \ldots, y_n x_{nk_n})
\\
=
h\bigl( y_1(x_{11} + \cdots + x_{1k_1}), \ \ldots, \
y_n(x_{n1} + \cdots + x_{nk_n}) \bigr)
+
\sum_{i = 1}^n y_i^p h(x_{i1}, \ldots, x_{ik_i}).
\end{multline*}
\end{cor}

\begin{proof}
This follows from Theorem~\ref{thm:p-grouping} by substituting $y_i x_{ij}$
for $x_{ij}$ then using the degree $p$ homogeneity of $h$.
\end{proof}

This polynomial identity provides another proof of the chain rule for
entropy mod~$p$: given $\ggamma \in \Pi_n$ and $\ppi^i \in \Pi_{k_i}$ as in
Proposition~\ref{propn:chn-p}, substitute $y_i = \gamma_i$ and $x_{ij} =
\pi^i_j$, then use the facts that $\sum_j \pi^i_j = 1$ and $\gamma_i^p =
\gamma_i$ for each $i$.  (Here $i$ is a superscript and $p$ is a power.)

The entropy polynomial $h(x)$ in a single variable is $0$, by definition.
But the entropy polynomial in two variables has important properties:

\begin{cor}
\lbl{cor:cocycle}
The two-variable entropy polynomial $h$ satisfies the cocycle%
\index{cocycle identity}
condition
\[
h(x, y) - h(x, y + z) + h(x + y, z) - h(y, z) = 0
\]
as a polynomial identity.
\end{cor}

Similar results appear in Cathelineau~\cite{CathSHS} (p.~58--59),%
\index{Cathelineau, Jean-Louis}
Kontsevich~\cite{KontOHL},%
\index{Kontsevich, Maxim} 
and Elbaz--Vincent%
\index{Elbaz-Vincent, Philippe} 
and Gangl~\cite{EVGFPM}%
\index{Gangl, Herbert}
(Section~2.3), and can be understood through the information
cohomology of Baudot, Bennequin and Vigneaux~\cite{BaBe,VignTSS}.

\begin{proof}
Theorem~\ref{thm:p-grouping} with $n = 2$ and $(k_1, k_2) = (2, 1)$ gives
\[
h(x, y, z) = h(x + y, z) + h(x, y)
\]
(since $h(z)$ is the zero polynomial), and similarly,
\[
h(x, y, z) = h(x, y + z) + h(y, z).
\]
The result follows.
\end{proof}

We are especially interested in the case where the arguments of the entropy
function sum to $1$, and under that restriction, $h(x, y)$ reduces to a
simple form:

\begin{lemma}
\lbl{lemma:p-to-two}
If $p \neq 2$, there is an identity of polynomials
\[
h(x, 1 - x) = \sum_{r = 1}^{p - 1} \frac{x^r}{r},
\]
and if $p = 2$, there is an identity of polynomials 
\[
h(x, 1 - x) = x + x^2.
\]
\end{lemma}

\begin{proof}
The case $p = 2$ is trivial.  Suppose, then, that $p > 2$.  
The result can be proved by direct calculation, but we shorten the proof 
using Example~\ref{eg:p-bin}, which implies that
\[
h(\pi, 1 - \pi) = \sum_{r = 1}^{p - 1} \frac{\pi^r}{r}
\]
for all $\pi \in \Zp$.  We now want to prove that this is a
\emph{polynomial} identity, not just an equality of functions.  By
Lemma~\ref{lemma:fn-fin-fld}, it suffices to show that the polynomial
\[
h(x, 1 - x) 
= 
-\sum_{r = 1}^{p - 1} \frac{x^r(1 - x)^{p - r}}{r!(p - r)!}
\]
has degree strictly less than $p$.  Since it plainly has degree at most
$p$, we need only show that the coefficient of $x^p$ vanishes.  That
coefficient is 
\[
-\sum_{r = 1}^{p - 1} \frac{(-1)^{p - r}}{r! (p - r)!}.
\]
For $1 \leq r \leq p - 1$,
\[
- \frac{(-1)^{p - r}}{r! (p - r)!}
=
(-1)^{p - r} \frac{(p - 1)!}{r! (p - r)!}
=
(-1)^{p - r} \frac{1}{r} \binom{p - 1}{r - 1}
=
(-1)^{p - 1} \frac{1}{r}
\]
in $\Zp$, using first the fact that $(p - 1)! = -1$ and then
Lemma~\ref{lemma:p-binom}.  Hence the coefficient of $x^p$ in $h(x, 1 - x)$
is
\[
(-1)^{p - 1} \sum_{r \in (\Zp)^\times} \frac{1}{r}.
\]
But $r \mapsto 1/r$ defines a permutation of $(\Zp)^\times$, so the sum
here is equal to $\sum_{r = 1}^{p - 1} r$, which is $0$ since $p$ is odd.
\end{proof}

Following Elbaz-Vincent%
\index{Elbaz-Vincent, Philippe} 
and Gangl~\cite{EVGOPI},%
\index{Gangl, Herbert}
we write
\begin{equation}
\ntn{sterling}
\pounds_1(x) 
= 
h(x, 1 - x)
=
\begin{cases}
\sum_{r = 1}^{p - 1} x^r/r      &\text{if } p \neq 2,   \\
x + x^2                         &\text{if } p = 2.
\end{cases}
\end{equation}
(Elbaz-Vincent and Gangl omitted the case $p = 2$.)  The
function $\pounds_1$ is the mod~$p$ analogue of the real function
\begin{equation}
\lbl{eq:real-two}
x 
\mapsto 
H_\R(x, 1 - x)
=
- x \log x - (1 - x) \log(1 - x).
\end{equation}
This may be a surprise, given the lack of formal resemblance between the
expressions~\eqref{notn:sterling} and~\eqref{eq:real-two}.  Indeed, the
polynomial $\sum_{r = 1}^{p - 1} x^r/r$ is the truncation of the power
series of $-\log(1 - x)$, not~\eqref{eq:real-two}.  Nevertheless, the
Faddeev theorem and its mod~$p$ counterpart (Theorems~\ref{thm:faddeev}
and~\ref{thm:fad-p}) establish a tight analogy between the entropy
functions over $\R$ and $\Zp$.

It is immediate from the definition of $h$ that it is a symmetric
polynomial, so there is a polynomial identity
\begin{equation}
\lbl{eq:pounds-sym}
\pounds_1(x) = \pounds_1(1 - x).
\end{equation}
The polynomial $\pounds_1$ also satisfies a more complicated identity,
whose significance will be explained shortly.  Following
Kontsevich~\cite{KontOHL}, Elbaz-Vincent and Gangl proved: 
  
\begin{propn}[Elbaz-Vincent and Gangl]
\lbl{propn:p-feith}%
\index{Elbaz-Vincent, Philippe}%
\index{Gangl, Herbert}%
There is a polynomial identity
\[
\pounds_1(x) + (1 - x)^p \pounds_1\biggl( \frac{y}{1 - x} \biggr)
=
\pounds_1(y) + (1 - y)^p \pounds_1\biggl( \frac{x}{1 - y} \biggr).
\]
\end{propn}

Both sides of this equation are indeed polynomials, since $\pounds_1$ has
degree at most $p$.  Elbaz-Vincent and Gangl proved it using differential
equations (Proposition~5.9(2) of~\cite{EVGOPI}), but it also follows easily
from the cocycle identity:

\begin{proof}
We work in the field of rational expressions over $\Zp$ in commuting
variables $x$ and $y$.  Since $h$ is homogeneous of degree $p$,
\[
h(x, y) = (x + y)^p \pounds_1\biggl( \frac{x}{x + y} \biggr).
\]
The identity to be proved is, therefore, equivalent to
\[
h(x, 1 - x) + h(y, 1 - x - y)
=
h(y, 1 - y) + h(x, 1 - x - y).
\]
Since $h$ is symmetric, this in turn is equivalent to
\[
h(x, 1 - x - y) - h(x, 1 - x) + h(1 - y, y) - h(1 - x - y, y) = 0,
\]
which is an instance of the cocycle identity
of Corollary~\ref{cor:cocycle}. 
\end{proof}

Proposition~\ref{propn:p-feith} can be understood as follows.  Any
probability distribution mod~$p$ can be expressed as an iterated composite
of distributions on two elements.  Hence, the entropy of any distribution
can be computed in terms of entropies $H(\pi, 1 - \pi)$ of distributions on
two elements, using the chain rule.  In this sense, the sequence of
functions
\[
\bigl( H \from \Pi_n \to \Zp \bigr)_{n \geq 1}
\]
reduces to the single function $H \from \Pi_2 \to \Zp$, which is
effectively a function in one variable:
\[
\begin{array}{cccc}
F\from  &\Zp    &\to            &\Zp    \\
        &\pi    &\mapsto        &H(\pi, 1 - \pi).
\end{array}
\]
The same is true over $\R$: the sequence of functions $(H \from \Delta_n
\to \R)_{n \geq 1}$ reduces to the single function $F \from [0, 1] \to \R$
defined by $F(\pi) = H(\pi, 1 - \pi)$.

On the other hand, given an arbitrary function $F \from \Zp \to \Zp$, one
cannot generally extend it to a sequence of functions $(\Pi_n \to \Zp)_{n
  \geq 1}$ satisfying the chain rule (nor, similarly, in the real case).
Indeed, by expressing a distribution $(\pi, 1 - \pi - \tau, \tau)$ as a
composite in two different ways, we obtain an equation that $F$ must
satisfy if such an extension is to exist.  Assuming the symmetry property
$F(\pi) = F(1 - \pi)$, that equation is
\begin{equation}
\lbl{eq:feith}
F(\pi) + (1 - \pi) F\biggl(\frac{\tau}{1 - \pi}\biggr)
=
F(\tau) + (1 - \tau) F\biggl(\frac{\pi}{1 - \tau}\biggr)
\end{equation}
($\pi, \tau \neq 1$).  When the function $F$ is $\pi \mapsto H(\pi, 1 -
\pi)$, equation~\eqref{eq:feith} also follows from
Proposition~\ref{propn:p-feith}.

Equation~\eqref{eq:feith} is sometimes called the \demph{fundamental%
\index{fundamental equation of information theory} 
equation of information theory}.  (Over $\R$, this functional equation has
been studied since at least the 1958 work of Tverberg~\cite{Tver}. The name
seems to have come later, and appears in Acz\'el and Dar\'oczy's 1975
book~\cite{AcDa}.)  Assuming that $F$ is symmetric, it is the only obstacle
to the extension problem, in the sense that if $F$ satisfies the
fundamental equation then the extension can be performed.

In the real case, the function~\eqref{eq:real-two} is a solution of the
fundamental equation.  In fact, up to a scalar multiple, it is the
\emph{only} measurable solution $F$ of the fundamental equation such that
$F(0) = F(1)$ (Corollary~3.4.22 of Acz\'el and Dar\'oczy~\cite{AcDa}).  It
can be deduced that up to a constant factor, Shannon entropy for finite
real probability distributions is characterized uniquely by measurability,
symmetry and the chain rule.  This is the 1964 theorem of Lee%
\index{Lee, Pan-Mook} 
mentioned in Remark~\ref{rmks:faddeev}\bref{rmk:faddeev-lee},
proofs of which can be found in Lee~\cite{Lee} and Acz\'el and
Dar\'oczy~\cite{AcDa} (Corollary~3.4.23).

In the mod~$p$ case, we know that the function $F = \pounds_1$ is symmetric
and satisfies the fundamental equation.  Since any such function $F$ can be
extended to a sequence of functions $\Pi_n \to \Zp$ satisfying the chain
rule, it follows from Theorem~\ref{thm:fad-p} that up to a constant factor,
$\pounds_1$ is unique with these properties.

Kontsevich~\cite{KontOHL}%
\index{Kontsevich, Maxim} 
proposed calling $\pounds_1$ the \demph{$1\hlf$-logarithm},%
%
\index{one-logarithm@$1\hlf$-logarithm} 
because the ordinary logarithm satisfies a three-term functional equation
($\log(xy) = \log x + \log y$), the dilogarithm satisfies a five-term
functional equation (as in Section~2 of Zagier~\cite{Zagi} or
Proposition~3.5 of Elbaz-Vincent and Gangl~\cite{EVGOPI}), and the
$1\hlf$-logarithm satisfies the four-term functional
equation~\eqref{eq:feith}.

\begin{remark}
\lbl{rmk:kont-comp}
In his seminal note~\cite{KontOHL}, Kontsevich%
\index{Kontsevich, Maxim} 
unified the real and mod~$p$ cases with a homological argument, using a
cocycle%
\index{cocycle identity}
identity equivalent to the one in Corollary~\ref{cor:cocycle}.  In doing so,
he established that $\sum_{0 < r < p} \pi^r/r$ is the correct formula for
the entropy of a distribution $(\pi, 1 - \pi)$ mod $p$ on two
elements (assuming, as he did, that $p \neq 2$).  Although he gave no
definition of the entropy of a distribution mod $p$ on an
arbitrary number of elements, his arguments showed that a unique reasonable
such definition must exist.

In this chapter, we have developed the framework hinted at by Kontsevich,
and also provided the definition and characterization of information
loss mod~$p$.  Two further features of this theory are apparent. The first
is the streamlined inclusion of the case $p = 2$.  The second is the
dropping of all symmetry%
\index{Faddeev, Dmitry!mod p analogue of entropy theorem@mod $p$ analogue of entropy theorem!role of symmetry in}%
\index{symmetry in Faddeev-type theorems}
requirements.  In axiomatic approaches to entropy based on the fundamental
equation of information theory~\eqref{eq:feith}, such as those of
Lee~\cite{Lee} and Kontsevich, the symmetry axiom $F(\pi) = F(1 - \pi)$ is
essential.  Indeed, $F(\pi) = \pi$ is also a solution of~\eqref{eq:feith},
and the polynomial identity of Proposition~\ref{propn:p-feith} is also
satisfied by $x^p$ in place of $\pounds_1(x)$.  The symmetry axiom is used
to rule out these and other undesired solutions.  This is why Lee's
characterization of real entropy $H$ needed the assumption that $H$ is a
symmetric function.  In contrast, symmetry is needed nowhere in the
approach taken here.
\end{remark}

%% file: cat.tex
\chapter{The categorical origins of entropy}
\lbl{ch:cat}
\index{category theory}

In this chapter, we describe a general category-theoretic construction
which, when given as input the real line and the notion of finite
probability distribution, automatically produces as output the notion of
Shannon entropy (Figure~\ref{fig:machines}).

\begin{figure}
\centering
\lengths
\begin{picture}(54,60)
\cell{27}{30}{c}{\includegraphics[height=40\unitlength]{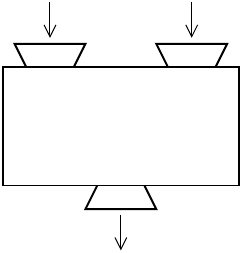}}
\cell{27}{35}{c}{internal algebras}
\cell{27}{30}{c}{in a categorical algebra}
\cell{27}{25}{c}{for an operad}
\cell{16}{52}{c}{$(\Delta_n)$}
\cell{38}{52}{c}{$\R$}
\cell{27}{7}{c}{Shannon entropy}
\cell{27}{-1}{b}{(a)}
\end{picture}%
\hspace*{12mm}%
\begin{picture}(54,60)
\cell{27}{30}{c}{\includegraphics[height=40\unitlength]{machinem}}
\cell{27}{35}{c}{internal algebras}
\cell{27}{30}{c}{in a categorical algebra}
\cell{27}{25}{c}{for an operad}
\cell{16}{52}{c}{$(1)$}
\cell{38}{52}{c}{$\cat{V}$}
\cell{27}{7}{c}{monoid in $\cat{V}$\vphantom{y}}
\cell{27}{-1}{b}{(b)}
\end{picture}%
\caption{Schematic illustration of the main result of this chapter,
  Theorem~\ref{thm:cat-ent}.  There is a general categorical machine
  which \hardref{(a)}~when given as input the simplices $(\Delta_n)$ and
  the real line, produces as output the notion of Shannon entropy, and
  \hardref{(b)}~when given as input the one-point set $1$ and a monoidal
  category $\cat{V}$, produces as output the notion of monoid in $\cat{V}$.}
\lbl{fig:machines}
\index{internal algebra!schematic illustration}
\end{figure}

The moral of this result is that even in the pure-mathematical heartlands
of algebra and topology, entropy is inescapable.%
\index{entropy!inescapable@is inescapable}
This may come as a surprise: for although entropy is a major concept in
many branches of science, an algebraist, topologist or category
theorist can easily go a lifetime without encountering entropy of any
kind.

Yet the categorical construction described here is entirely general and
natural.  It is not tailor-made for this particular purpose.  Other
familiar inputs produce familiar outputs.  And the inputs that we give to
the construction here, the real line $\R$ and the standard topological
simplices $\Delta_n$, are fundamental objects of pure mathematics.  So, we
are all but forced to accept Shannon entropy as a natural concept in
pure mathematics too~-- quite independently of any motivation in terms of
information, diversity, thermodynamics, and so on.

The categorical construction involves operads and their algebras.  The
first two sections set out some standard definitions, beginning
with operads and algebras themselves in Section~\ref{sec:opds}.  For an
operad $P$, there are notions of categorical $P$-algebra $\scat{A}$ (a
category acted on by $P$) and of internal algebra in $\scat{A}$.  With
these definitions in place (Section~\ref{sec:cat-int-alg}), we can fulfil
the promise of the first paragraph above (Theorem~\ref{thm:cat-ent}).
Specifically, we show that for the operad $\Delta$ of simplices and the
categorical $\Delta$-algebra $\R$, the internal algebras in $\R$ are
precisely the scalar multiples of Shannon entropy.

In the final section, we describe the free categorical $\Delta$-algebra
containing an internal algebra.  The result proved is analogous to the
classical theorem that the free monoidal category containing a monoid is
the category of finite totally ordered sets.  To reach this result involves
a further climb up the mountain of categorical abstraction.  But at the end
of the path is a characterization of information loss that is entirely
concrete. It is almost exactly the characterization theorem of
Chapter~\ref{ch:loss}.

This chapter assumes some knowledge of category theory, including the
concepts of product in a category, monoid in a monoidal category, and
internal category in a category with finite limits.

\section{Operads and their algebras}
\lbl{sec:opds}

An operad is a system of abstract operations, somewhat like an algebraic
theory in the sense of universal algebra, but more restricted in nature.
Operads first emerged in algebraic topology (Boardman and Vogt~\cite{BoVo};
May~\cite{MayGIL}), while independently, the more general notion of
multicategory was being developed in categorical logic
(Lambek~\cite{LambDSC2}).  Nowadays, operads (like many other categorical
structures) have found application in a very wide range of subjects, from
algebra to theoretical physics.  Some samples of such applications can be
found in Kontsevich~\cite{KontOMD}, Loday and Vallette~\cite{LoVa}, and
Markl, Shnider and Stasheff~\cite{MSS}.

Many introductions to operads are available, such as \cite{LoVa},
\cite{MSS}, and Chapter~2 of~\cite{HOHC}.  Here we give only the definitions
and results that are needed in order to reach our goal.

An operad consists of a sequence $(P_n)_{n \geq 0}$ of sets equipped with
certain algebraic structure obeying certain laws.  It is
useful to view the elements $\theta$ of $P_n$ as abstract operations with
$n$ inputs and one output, as in Figure~\ref{fig:opd-opn}.
\begin{figure}
\centering
\lengths
\begin{picture}(120,16)
\cell{60}{8}{c}{\includegraphics[height=16\unitlength]{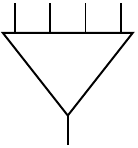}}
\cell{60}{9}{c}{\normalsize$\theta$}
\end{picture}
\caption{An element $\theta \in P_4$ of an operad $P$.}
\lbl{fig:opd-opn}
\end{figure}
A typical example will be given by $P_n = \cat{A}(A^{\otimes n}, A)$, for
any object $A$ of a monoidal category $\cat{A}$.  The algebraic structure
on the sequence of sets $(P_n)_{n \geq 0}$, and the equational laws that
this structure obeys, are exactly those suggested by this example.

\begin{figure}
\centering
\lengths
\begin{picture}(120,41.4)
\cell{35}{20.7}{c}{\includegraphics[height=41.4\unitlength]{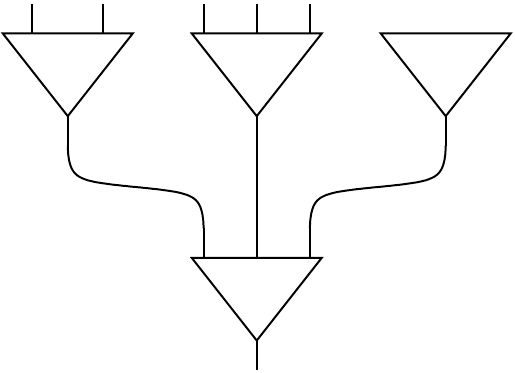}}
\cell{80}{20.7}{c}{\Large$\mapsto$}
\cell{105}{20.7}{c}{\includegraphics[height=16\unitlength]{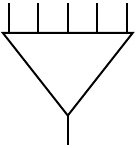}}
\cell{35}{9}{c}{\normalsize$\theta$}
\cell{15}{34}{c}{\normalsize$\phi^1$}
\cell{36}{34}{c}{\normalsize$\phi^2$}
\cell{56.5}{34}{c}{\normalsize$\phi^3$}
\cell{105}{10}{c}{\normalsize$\theta\of(\phi^1, \phi^2, \phi^3)$}
\end{picture}
\caption{Composition in an operad: $\theta \in P_3$ composes with $\phi^1
  \in P_2$, $\phi^2 \in P_3$ and $\phi^3 \in P_0$ to give $\theta \of
  (\phi^1, \phi^2, \phi^3) \in P_5$.}  
\lbl{fig:opd-defn}
\end{figure}

\begin{defn}
\lbl{defn:opd}
An \dmph{operad} $P$ consists of:
\begin{itemize}
\item 
a sequence $(P_n)_{n \geq 0}$ of sets;

\item
for each $n, k_1, \ldots, k_n \geq 0$, a function
\begin{equation}
\lbl{eq:opd-comp}
P_n \times P_{k_1} \times \cdots \times P_{k_n} 
\to
P_{k_1 + \cdots + k_n}
\end{equation}
(Figure~\ref{fig:opd-defn}), called \demph{composition}%
\index{composition!operad@in operad}
and written as
\[
(\theta, \phi^1, \ldots, \phi^n) \mapsto 
\theta \of (\phi^1, \ldots, \phi^n);
\ntn{compopd}
\]

\item
an element $1_P\ntn{idopd} \in P_1$, called the \demph{identity},%
\index{identity in operad}
\end{itemize}
satisfying the following axioms:
\begin{itemize}
\item 
\demph{associativity}:%
\index{associativity!operad@in operad}%
\index{composition!associativity of}
for each $n, k_i, \ell_{i j} \geq 0$ and $\theta
\in P_n$, $\phi^i \in P_{k_i}$, $\psi^{i j} \in P_{\ell_{i j}}$, 
\begin{multline*}
\Bigl( \theta \of \bigl(\phi^1, \ldots, \phi^n\bigr) \Bigr) \of
\bigl(\psi^{1 1}, \ldots, \psi^{1 k_1}, 
\ \ldots, \ 
\psi^{n 1}, \ldots, \psi^{n k_n}\bigr)\\
=
\theta \of \Bigl(
\phi^1 \of \bigl(\psi^{1 1}, \ldots, \psi^{1 k_1}\bigr), 
\ \ldots, \ 
\phi^n \of \bigl(\psi^{n 1}, \ldots, \psi^{n k_n}\bigr) 
\Bigr);
\end{multline*}

\item
\demph{identity}:%
\index{identity in operad}
for each $n \geq 0$ and $\theta \in P_n$,
\[
\theta \of (\underbrace{1_P, \ldots, 1_P}_n) 
= 
\theta
=
1_P \of (\theta).
\]
\end{itemize}
\end{defn}

Every tree%
\index{tree!operad@in operad} 
of operations such as that shown in Figure~\ref{fig:opd-tree}
has an unambiguous composite, obtained by repeatedly using the composition
and identity of the operad.  The associativity and identity axioms
guarantee that the order in which this is done makes no difference to the
outcome.
\begin{figure}
\centering
\lengths\setlength{\unitlength}{.8mm}%
\begin{picture}(120,60)
\cell{60}{30}{c}{\includegraphics[height=60\unitlength]{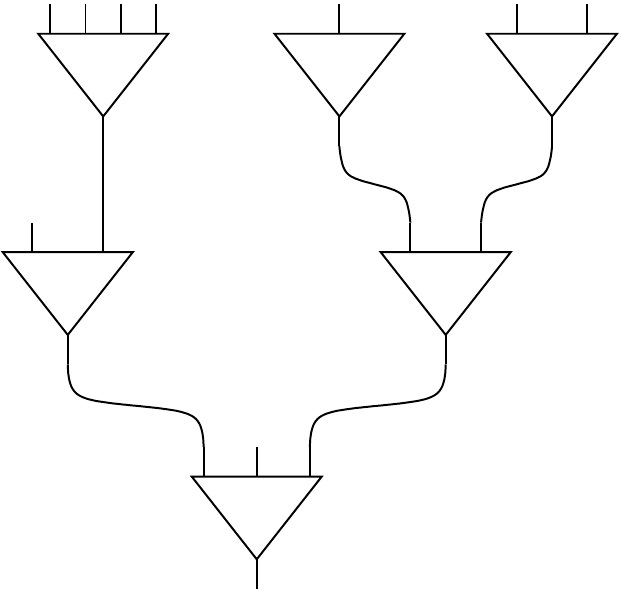}}
\cell{55}{8}{c}{\normalsize$\theta$}
\cell{35.5}{31}{c}{\normalsize$\phi$}
\cell{74}{31}{c}{\normalsize$\psi$}
\cell{39}{53}{c}{\normalsize$\chi$}
\cell{63}{53}{c}{\normalsize$\xi$}
\cell{85}{53}{c}{\normalsize$\omega$}
\end{picture}
\caption{Every tree of operations in an operad $P$ has a well-defined
  composite (in this case, an element of $P_9$).}
\lbl{fig:opd-tree}
\end{figure}

\begin{examples}
\lbl{egs:opd}
\begin{enumerate}
\item 
There is an operad $\One$\ntn{termopd} in which $\One_n$ is the one-element
set for each $n \geq 0$.  The composition and identities are uniquely
determined.  With the obvious notion of map of operads, $\One$ is the
terminal%
\index{operad!terminal}%
\index{terminal operad}
operad.

\item
\lbl{eg:opd-M}
Fix a monoid%
\index{monoid!operad from}%
\index{operad!monoid@from monoid}
$M$.  There is an operad $P(M)$\ntn{monopd} given by
\[
P(M)_n
=
\begin{cases}
M               &\text{if } n = 1,      \\
\emptyset       &\text{otherwise}
\end{cases}
\]
($n \geq 0$).  There is no choice in how to define the composition of
$P(M)$ except when $n = k_1 = 1$ (in the notation of~\eqref{eq:opd-comp}),
and in that case it is defined to be the multiplication of $M$.  Similarly,
the identity of the operad $P(M)$ is the identity of the monoid $M$.

\item
\lbl{eg:opd-Delta} 
There is an operad%
\index{simplex!operad}%
\index{probability distribution!operad}%
\index{operad!simplex}
\ntn{simpopd}$\Delta = (\Delta_n)_{n \geq 0}$, where as
usual $\Delta_n$ is the set of probability distributions on $\{1, \ldots,
n\}$.  The composition of the operad is composition of distributions,
and the identity is the unique distribution $\vc{u}_1$ on $\{1\}$.  We
already noted in Remark~\ref{rmk:comp-dist-opd} that the associativity and
identity axioms are satisfied.

\item
There is a larger operad $\measopd$ consisting of not just the
\emph{probability} measures on finite sets, but all finite measures%
\index{measure operad}%
\index{operad!measure}
on finite sets.  Thus, $\measopd_n = [0, \infty)^n$.  The composition is
  given by the same formula as for $\Delta$
  (Definition~\ref{defn:comp-dist}), and the identity is $(1) \in
  \measopd_1$.

\item
\lbl{eg:opd-End}
Let $\cat{A}$ be a monoidal category and $A \in \cat{A}$. Then there is an
operad $\End(A)$%
\index{endomorphism operad}%
\index{operad!endomorphism} 
with 
\[
\End(A)_n = \cat{A}(A^{\otimes n}, A),
\ntn{Endopd}
\]
and with composition and identities defined using the composition,
identities and monoidal structure of $\cat{A}$.  For a general operad $P$,
we have suggested that elements of $P_n$ be thought of as operations, but
when $P = \End(A)$, this is true in a concrete sense: $\End(A)_n$ is the set
of maps $A^{\otimes n} \to A$.

\item
Fix a field $k$, and let $P_n = k[x_1, \ldots, x_n]$ be the set of
polynomials%
\index{polynomial!operad}%
\index{operad!polynomial} 
over $k$ in $n$ variables.  Then $P = (P_n)_{n \geq 0}$ has the
structure of an operad, with composition given by substitution and
reindexing of variable names.  For instance, if
\[
\theta = x_1^2 + x_2^3 \in P_2,
\quad
\phi = 2x_1 x_3 - x_2 \in P_3,
\quad
\psi = x_1 + x_2 x_3 x_4 \in P_4,
\]
then
\[
\theta \of (\phi, \psi)
=
(2x_1 x_3 - x_2)^2 + (x_4 + x_5 x_6 x_7)^3 \in P_7.
\]
This example of an operad is just one of a large family.  In this case,
$P_n$ is the free $k$-algebra on $n$ generators.  There are similar
examples where $P_n$ is the free group, free Lie algebra, free distributive
lattice, etc., on $n$ generators.  In all cases, composition is by
substitution and reindexing.

\item
\lbl{eg:opd-little-discs}
Fix $d \geq 1$.  The \demph{little%
\index{operad!little discs}%
\index{little discs operad} 
$d$-discs operad} $P$ is defined as follows.  Let $P_n$ be the set of
configurations of $n$ $d$-dimensional discs inside the unit disc, numbered
in order and with disjoint interiors (Figure~\ref{fig:little-discs}).
\begin{figure}
\centering
\lengths
\begin{picture}(54,58)
\cell{27}{30}{c}{\includegraphics[width=54\unitlength]{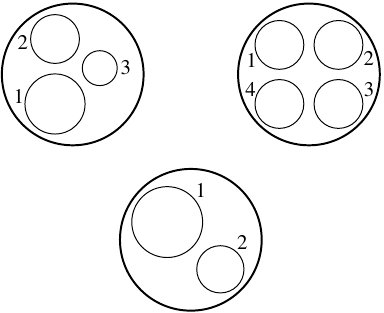}}
\cell{18}{9}{c}{\normalsize$\theta$}
\cell{3}{30}{c}{\normalsize$\phi$}
\cell{51}{30}{c}{\normalsize$\psi$}
\cell{27}{0}{b}{(a)}
\end{picture}%
\hspace*{12mm}%
\begin{picture}(54,58)
\cell{27}{34}{c}{\includegraphics[width=48\unitlength]{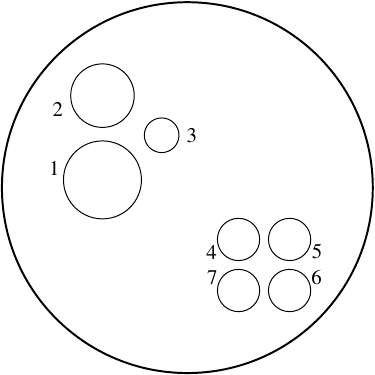}}
\cell{44}{8}{c}{\normalsize$\theta\of(\phi, \psi)$}
\cell{27}{0}{b}{(b)}
\end{picture}%
\caption{(a)~Operations $\theta \in P_2$, $\phi \in P_3$ and $\psi \in
P_4$ in the little $2$-discs operad $P$
(Example~\ref{egs:opd}\bref{eg:opd-little-discs}); (b)~the composite
operation $\theta \of (\phi, \psi) \in P_7$.}  
\lbl{fig:little-discs}
\end{figure}
Composition is by substitution (using affine transformations) and
reindexing, as suggested by the figure.  

The little discs operad and its close relative, the little cubes operad,
were some of the very first operads to be defined (Boardman%
\index{Boardman, Michael} 
and Vogt~\cite{BoVo}%
\index{Vogt, Rainer} 
and Section~4 of May~\cite{MayGIL}).%
\index{May, J. Peter}
\end{enumerate}
\end{examples}

In more precise terminology, operads as defined in
Definition~\ref{defn:opd} are called nonsymmetric operads of sets.  Just as
the definition of monoidal category has symmetric and nonsymmetric
variants, so too does the definition of operad.  We will concentrate on the
nonsymmetric variant.

However, we will not only need operads \emph{of sets}.  Let $\cat{E}$ be
any category with finite products (or indeed, any symmetric monoidal
category, a level of generality that we will not need).  An \demph{operad%
\index{operad!internal} 
in $\cat{E}$} is a sequence $(P_n)_{n \geq 0}$ of objects of $\cat{E}$
together with maps~\eqref{eq:opd-comp} in $\cat{E}$ (encoding the
composition) and a map $1 \to P_1$ in $\cat{E}$ (encoding the identity),
all subject to commutative diagrams expressing the associativity and
identity equations of Definition~\ref{defn:opd}.

Details of this more general definition can be found in May~\cite{MayDOA},
for instance, but we will need only two cases.  The first is $\cat{E} =
\Set$\ntn{Set}, the category of sets.  In that case, an operad in $\cat{E}$
is just an operad as in Definition~\ref{defn:opd}.  The second is $\cat{E}
= \Tp$\ntn{Tp}, the category of topological spaces.  An operad in $\Tp$ is
just an operad $P$ of sets in which each set $P_n$ is equipped with a
topology and the composition maps~\eqref{eq:opd-comp} are continuous. We
will refer to operads in $\Tp$ as \demph{topological%
\index{topological operad} 
operads}.

\begin{examples}
\begin{enumerate}
\item 
The terminal%
\index{operad!terminal}%
\index{terminal operad}
operad $\One$ is a topological operad in a unique way.

\item
For a topological monoid%
\index{monoid!operad from}%
\index{operad!monoid@from monoid}
$M$, the operad $P(M)$ of Example~\ref{egs:opd}\bref{eg:opd-M} is a
topological operad in an evident way.

\item
Putting the standard topology on the simplices%
\index{simplex!operad}%
\index{probability distribution!operad}%
\index{operad!simplex}
$\Delta_n$ gives $\Delta$
the structure of a topological operad.

\item
The little%
\index{operad!little discs}%
\index{little discs operad} 
discs operad is also naturally a topological operad.
\end{enumerate}
\end{examples}

An operad $P$ is a system of abstract operations.  An algebra for $P$ is an
interpretation of the elements of $P$ as \emph{actual} operations:

\begin{defn}
\lbl{defn:alg}
Let $P$ be an operad of sets.  A \demph{$P$-algebra}%
\index{algebra for operad} 
is a set $A$ together with a map
\[
\begin{array}{cccc}
\alpha_n \from  &P_n \times A^n &\to    &A      \\
&
\bigl(\theta, (a^1, \ldots, a^n)\bigr)  &\mapsto        &
\ovln{\theta}(a^1, \ldots, a^n)
\end{array}
\ntn{algbar}
\]
for each $n \geq 0$, satisfying two axioms:
\begin{multline}
\lbl{eq:alg-comp}
\ovln{\theta \of (\phi^1, \ldots, \phi^n)}
\bigl(
a^{1 1}, \ldots, a^{1 k_1}, \ \ldots, \ a^{n 1}, \ldots, a^{n k_n}
\bigr)  \\
=
\ovln{\theta} \Bigl(
\ovln{\phi^1} \bigl( a^{1 1}, \ldots, a^{1 k_1} \bigr),
\ \ldots, \ 
\ovln{\phi^n} \bigl( a^{n 1}, \ldots, a^{n k_n} \bigr)
\Bigr)
\end{multline}
for all $\theta \in P_n$, $\phi^i \in P_{k_i}$ and $a^{i j} \in A$; and
\begin{equation}
\lbl{eq:alg-id}
\ovln{1_P}(a) = a
\end{equation}
for all $a \in A$.
\end{defn}

The definition of algebra extends easily from operads of sets to operads in
any category $\cat{E}$ with finite products: then $A$ is an object of
$\cat{E}$ and $\alpha_n$ is a map in $\cat{E}$, while the
equations~\eqref{eq:alg-comp} and~\eqref{eq:alg-id} are expressed as
commutative diagrams in $\cat{E}$ (May~\cite{MayDOA}).  In the only other
case that concerns us here, $\cat{E} = \Tp$, an algebra%
\index{algebra for operad!topological}%
\index{topological operad!algebra for}
for a topological operad $P$ is a topological space $A$ together with a
sequence of continuous maps
\[
\Bigl( P_n \times A^n \toby{\alpha_n} A \Bigr)_{n \geq 0}
\]
satisfying equations~\eqref{eq:alg-comp} and~\eqref{eq:alg-id}.

\begin{examples}
\lbl{egs:opd-alg}
\begin{enumerate}
\item
\lbl{eg:opd-alg-one} 
Consider the terminal%
\index{operad!terminal}%
\index{terminal operad}
operad $\One$ of sets.  A $\One$-algebra is a set $A$ together with a map
$\alpha_n \from A^n \to A$ for each $n \geq 0$, satisfying
equations~\eqref{eq:alg-comp} and~\eqref{eq:alg-id}.  One easily deduces
that a $\One$-algebra is exactly a monoid, with $\alpha_n$ as its $n$-fold
multiplication.  If $\One$ is regarded as a topological operad then a
$\One$-algebra is exactly a topological monoid.

\item
Fix a monoid%
\index{monoid!operad from}%
\index{operad!monoid@from monoid}
$M$.  An algebra for the operad $P(M)$ is simply a set with a
left $M$-action.  If $M$ is a topological monoid then a $P(M)$-algebra is a
topological space with a continuous left $M$-action.

\item
\lbl{eg:opd-alg-simp}
Now consider the topological operad $\Delta$ of simplices.%
\index{simplex!operad}%
\index{probability distribution!operad}%
\index{operad!simplex}
Any convex subset $A$ of $\R^d$, for any $d \geq 0$, is a $\Delta$-algebra
in a natural way: given $\p \in \Delta_n$ and $\vc{a}^1, \ldots, \vc{a}^n
\in A$, put
\[
\ovln{\p}(\vc{a}^1, \ldots, \vc{a}^n)
=
p_1 \vc{a}^1 + \cdots + p_n \vc{a}^n
\in 
A.
\]
Equations~\eqref{eq:alg-comp} and~\eqref{eq:alg-id} express elementary
facts about convex combinations.  

We refer to this as the \demph{standard}%
\index{standard Delta-algebra structure@standard $\Delta$-algebra structure} 
$\Delta$-algebra structure on $A$.

\item
\lbl{eg:opd-alg-def}
The previous example admits a family of deformations,%
\index{deformed Delta-algebra structure@deformed $\Delta$-algebra structure} 
at least when $A$ is a linear subspace of $\R^d$.  For each $q \in \R$,
there is a $\Delta$-algebra structure on $A$ given by
\[
\ovln{\p}(\vc{a}^1, \ldots, \vc{a}^n)
=
\sum_{i \in \supp{\p}} p_i^q \vc{a}^i.
\]
(Here the superscript $q$ is a power but the superscript $i$ is an
index.)  The previous example is the case $q = 1$.

\item
The chain%
\index{chain rule!means@for means} 
rule for a weighted mean $M$ on an interval $I$ nearly states
that the maps
\[
\bigl(M \from \Delta_n \times I^n \to I\bigr)_{n \geq 0}
\]
give $I$ the structure of an algebra for the operad $\Delta$.  More
exactly, the chain rule is the composition axiom~\eqref{eq:alg-comp} for
a $\Delta$-algebra.  For $I$ to be a $\Delta$-algebra, it must also satisfy
the identity axiom~\eqref{eq:alg-id}, which is a special case of
consistency: $M(\vc{u}_1, (x)) = x$ for each $x \in I$.

\item
For any operad $P$ of sets, a $P$-algebra amounts to a set $A$ together
with a map $P \to \End(A)$ of operads.  Here, $\End(A)$%
\index{endomorphism operad}%
\index{operad!endomorphism} 
is the operad defined in Example~\ref{egs:opd}\bref{eg:opd-End}, with
$\cat{A} = \Set$.  
This makes precise the earlier assertion that an algebra for $P$ is an
interpretation of the elements of $P$ as actual operations.

\item
Any $d$-fold loop%
\index{loop space}
space is an algebra for the little%
\index{operad!little discs}%
\index{little discs operad} 
$d$-discs operad in a natural way.  This was one of the first examples of
an algebra for an operad, the details of which can be found in Section~5 of
May~\cite{MayGIL}.%
\index{May, J. Peter}
\end{enumerate}
\end{examples}

Let $P$ be an operad in a finite product category $\cat{E}$, and let $A =
(A, \alpha)$ and $B = (B, \beta)$ be $P$-algebras.  A \demph{map of
  $P$-algebras}%
\index{algebra for operad!map of} 
from $A$ to $B$ is a map $f \from A \to B$ in $\cat{E}$
such that the square
\[
\xymatrix{
P_n \times A^n \ar[r]^{1 \times f^n} \ar[d]_{\alpha_n}  &
P_n \times B^n \ar[d]^{\beta_n} \\
A \ar[r]_{f}    &
B
}
\]
commutes for each $n \geq 0$.  This defines a category $\Alg(P)$\ntn{Alg}
of $P$-algebras.

\begin{remarks}
\lbl{rmks:opds-thys}
The language of operations and algebras invites comparison with other
categorical formulations of the concept of algebraic theory.  Systematic
comparisons can be found in Kelly~\cite{KellOOJ}, Section~2.8 of
Gould~\cite{GoulCCO}, and Chapter~3 of Avery~\cite{AverSS}.  Here, we just
make the following observations.
\begin{enumerate}
\item 
\lbl{rmk:opds-thys-mnds}
Let $\cat{E}$ be a finite product category satisfying the further mild
condition that it has countable coproducts over which the product
distributes.  ($\Set$ and $\Tp$ are examples.)  Then any operad $P$ in
$\cat{E}$ induces a monad%
\index{operad!monad from}%
\index{monad!operad@from operad}
$T_P$ on $\cat{E}$, with functor part given by
\[
T_P(A) = \coprod_{n \geq 0} P_n \times A^n
\]
($A \in \cat{E}$).  The category of algebras for the
operad $P$ is exactly the category of algebras for the monad $T_P$.
Non-isomorphic operads $P$ sometimes induce the same monad
$T_P$~\cite{AOAT}, although many aspects of an operad can still be
understood through its induced monad.

\item
\lbl{rmk:opds-thys-thys}
Remark~\bref{rmk:opds-thys-mnds} provides a semantic connection between
operads and a different conception of algebraic theory, monads.  On the
syntactic side, the definition of operad can easily be adapted to give a
definition of finitary algebraic theory, equivalent to any of the usual
definitions (as given in Manes~\cite{ManeAT}, for instance).  Indeed,
a finitary algebraic theory can be defined as an operad $P$ together with,
for each map of sets
\[
f \from \{1, \ldots, m\} \to \{1, \ldots, n\},
\]
a map $f_* \from P_m \to P_n$, subject to equations expressing
compatibility between the operad structure and these maps $f_*$
(Tronin~\cite{TronACO,TronOVA}).  The idea is that $f_*$ transforms an
$m$-ary operation into an $n$-ary operation by reindexing the variables
according to $f$.  For instance, in the theory $P$ of groups, $P_n$ is the
underlying set of the free group on $n$ generators (which can be regarded
as the set of $n$-ary operations defined on any group), and if $f$ is the
unique map $\{1, 2\} \to \{1\}$ then $f_*\from P_2 \to P_1$ sends the
operation of multiplication to the operation of squaring.

If we take the definition of finitary algebraic theory sketched in the
previous paragraph but restrict $f$ to be a bijection, we obtain the
definition of
\demph{symmetric%
\index{operad!symmetric}%
\index{symmetric!operad} 
operad}.  (In much of the literature, `operad' is taken to mean `symmetric
operad' by default.)  If we further restrict $f$ to be an identity, we
recover the definition of nonsymmetric operad.

\item
As the previous remark suggests, most algebraic theories cannot be
described by an operad.  For instance, there is no operad $P$ of sets such
that $\Alg(P)$ is equivalent to the category of groups.%
\index{operad!groups@for groups, nonexistent}%
\index{groups, theory of}  
For a proof of a strong version of this statement, see Lin~\cite{LinAGA}.
\end{enumerate}
\end{remarks}

\section{Categorical algebras and internal algebras}
\lbl{sec:cat-int-alg}

Let $P$ be an operad of sets.  An algebra for $P$ is a set acted on by $P$,
but more generally, we can consider \emph{categories} acted on by $P$.
Such a structure is called a categorical $P$-algebra.  

More generally still, let $\cat{E}$ be a category with finite limits.  Then
there is the notion of internal category in $\cat{E}$ (as in Chapter~2 of
Johnstone~\cite{JohnTT}), and when $P$ is an operad in
$\cat{E}$, we can consider actions of $P$ on such an internal category.


\begin{defn}
\lbl{defn:cat-alg}
Let $\cat{E}$ be a category with finite limits and let $P$ be an operad in
$\cat{E}$.  A \demph{categorical%
\index{categorical algebra for operad} 
$P$-algebra} is an internal category in $\Alg(P)$.
\end{defn}

It is straightforward to verify that $\Alg(P)$ has finite limits, computed
as in $\cat{E}$, so this definition does make sense.  But it is also
helpful to have at hand a more explicit form, as follows.

A categorical $P$-algebra $\scat{A}$ can be described as a
pair of ordinary $P$-algebras, $\scat{A}_0$\ntn{A0} and
$\scat{A}_1$\ntn{A1}, together with domain and codomain maps
\[
\scat{A}_1 \parpairu \scat{A}_0
\]
and composition and identity maps
\[
\scat{A}_1 \times_{\scat{A}_0} \scat{A}_1 \to \scat{A}_1,
\qquad
\scat{A}_0 \to \scat{A}_1,
\]
all of which are required to be maps of $P$-algebras, as well as obeying
the usual axioms for an internal category.  Here $\scat{A}_0$ is to be
thought of as the object of objects of $\scat{A}$, and $\scat{A}_1$ as the
object of maps in $\scat{A}$.

Equivalently, a categorical $P$-algebra is an internal category in
$\cat{E}$ on which $P$ acts functorially.  To see this, first note that for
any object $X$ of $\cat{E}$ and internal category $\scat{A}$ in $\cat{E}$,
we can define another internal category
\[
X \times \scat{A}
\ntn{XA}
\]
in $\cat{E}$.  This is the product $D(X) \times \scat{A}$, where $D(X)$ is
the discrete internal category on $X$.  Thus, its object of objects
and object of maps are given by
\[
(X \times \scat{A})_0 = X \times \scat{A}_0,
\qquad
(X \times \scat{A})_1 = X \times \scat{A}_1.
\]
In this notation, a categorical $P$-algebra consists of an internal
category $\scat{A}$ in $\cat{E}$ together with internal functors
\begin{equation}
\lbl{eq:int-ftr-act}
\alpha_n \from P_n \times \scat{A}^n \to \scat{A}
\end{equation}
($n \geq 0$), satisfying analogues of the usual algebra axioms.  


As usual, we are principally concerned with the cases $\cat{E} = \Set$ and
$\cat{E} = \Tp$, where categorical algebras for an operad can be
understood as follows.
\begin{examples}
\lbl{egs:cat-alg-gen}
\begin{enumerate}
\item 
\lbl{eg:cat-alg-gen-set}
Let $P$ be an operad in $\cat{E} = \Set$.  A categorical $P$-algebra
consists of a small category $\scat{A}$ together with a functor
\[
\ovln{\theta} \from \scat{A}^n \to \scat{A}
\]
for each $n \geq 0$ and $\theta \in P_n$.  These functors are required to
satisfy equations~\eqref{eq:alg-comp} and~\eqref{eq:alg-id} for objects
$a^{ij}$ and $a$ of $\scat{A}$, and analogous equations for maps in
$\scat{A}$.

\item
\lbl{eg:cat-alg-gen-top}
Let $P$ be an operad%
\index{categorical algebra for operad!topological}%
\index{topological operad!categorical algebra for}
in $\cat{E} = \Tp$.  The categorical $P$-algebras can be described as
in~\bref{eg:cat-alg-gen-set}, but with the following additions.  $\scat{A}$
is now a topological category, so that $\scat{A}_0$ and $\scat{A}_1$ carry
topologies with the property that the domain, codomain, composition and
identity operations are continuous.  Moreover, the structure maps
\[
P_n \times \scat{A}_0^n \to \scat{A}_0,
\quad
P_n \times \scat{A}_1^n \to \scat{A}_1
\]
of the $P$-algebras $\scat{A}_0$ and $\scat{A}_1$ are required to be
continuous. 

\item
\lbl{eg:cat-alg-gen-one} 
Here we consider the special case of categorical algebras with only one%
\index{categorical algebra for operad!one-object} 
object, over both $\Set$ and $\Tp$.

First, let $P$ be an operad of sets.  Let $A$ be a monoid, viewed as a
one-object category $\scat{A}$.  To give $\scat{A}$ the structure of a
categorical $P$-algebra is to give the set $A$ the structure of an ordinary
$P$-algebra in such a way that for each $n \geq 0$ and $\theta \in P_n$,
the structure map
\[
\ovln{\theta} \from A^n \to A
\]
is a monoid homomorphism.  In short, a one-object categorical $P$-algebra
is a monoid on which $P$ acts by homomorphisms.

Similarly, when $P$ is a topological operad, a one-object categorical
$P$-algebra is a topological monoid on which $P$ acts by continuous
homomorphisms. 
\end{enumerate}
\end{examples}

Some specific examples now follow.

\begin{examples}
\lbl{egs:cat-alg-spec}
\begin{enumerate}
\item 
Consider the terminal%
\index{operad!terminal}%
\index{terminal operad}
operad of sets, $\One$.  By the description preceding
equation~\eqref{eq:int-ftr-act}, a categorical $\One$-algebra is a category
on which $\One$ acts functorially, that is, a category $\scat{A}$ together
with a functor $\scat{A}^n \to \scat{A}$, subject to certain axioms.  These
axioms give $\scat{A}$ the structure of a monoid in $\Cat$.  Thus, a
categorical $\One$-algebra is exactly a strict monoidal category.

Alternatively, working directly from Definition~\ref{defn:cat-alg}, a
categorical $\One$-algebra is an internal category in $\Mon$, the category
of monoids.  Again, this is just a strict monoidal category.

\item
\lbl{eg:cat-alg-spec-act}
Let $M$ be a monoid,%
\index{monoid!operad from}%
\index{operad!monoid@from monoid}
and form the operad $P(M)$ of sets (Example~\ref{egs:opd}\bref{eg:opd-M}).
A categorical $P(M)$-algebra is a category equipped with a left $M$-action.

\item
\lbl{eg:cat-alg-spec-simp}
Consider the topological operad $\Delta$%
\index{simplex!operad}%
\index{probability distribution!operad}%
\index{operad!simplex}
of simplices.  Let $A$ be a linear subspace (or more generally, a convex
additive submonoid) of $\R^d$.  Then $A$ is a topological monoid under
addition.  We have already considered the standard%
\index{standard Delta-algebra structure@standard $\Delta$-algebra structure} 
$\Delta$-algebra structure on the topological space $A$, given for
$\p \in \Delta_n$ by
\[
\begin{array}{cccc}
\ovln{\p}\from  &
A^n                             &\to            &
A      \\
&
(\vc{a}^1, \ldots, \vc{a}^n)    &\mapsto        &
\sum_{i = 1}^n p_i \vc{a}^i
\end{array}
\]
(Example~\ref{egs:opd-alg}\bref{eg:opd-alg-simp}).  Each of these maps
$\ovln{\p}$ is a monoid homomorphism. Hence
by Example~\ref{egs:cat-alg-gen}\bref{eg:cat-alg-gen-one}, $A$ is a
one-object categorical $\Delta$-algebra.

\item
\lbl{eg:cat-alg-spec-def}
The same is true for the $q$-deformed%
\index{deformed Delta-algebra structure@deformed $\Delta$-algebra structure} 
algebra structure
\[
(\vc{a}^1, \ldots, \vc{a}^n)
\mapsto
\sum_{i \in \supp(\p)} p_i^q \vc{a}^i
\]
of Example~\ref{egs:opd-alg}\bref{eg:opd-alg-def}, for any $q \in \R$.  
\end{enumerate}
\end{examples}

For ordinary algebras for an operad, there is only one sensible notion of
map between algebras, but for \emph{categorical} algebras, there are
several.  Indeed, let $\cat{E}$ be a category with finite limits, let $P$
be an operad in $\cat{E}$, and let $\scat{B}$ and $\scat{A}$ be categorical
$P$-algebras.  Then $\scat{B}$ and $\scat{A}$ are, by definition, internal
categories in $\Alg(P)$, and a \demph{strict%
\index{strict map}%
\index{categorical algebra for operad!strict map of}
map} from $\scat{B}$ to $\scat{A}$ is an internal functor $\scat{B} \to
\scat{A}$ in $\Alg(P)$.  Equivalently, it is an internal functor
\[
G \from \scat{B} \to \scat{A}
\]
in $\cat{E}$ such that for all $n \geq 0$, the square
\[
\xymatrix{
P_n \times \scat{B}^n \ar[r]^{1 \times G^n} \ar[d]_{\beta_n}  &
P_n \times \scat{A}^n \ar[d]^{\alpha_n} \\
\scat{B} \ar[r]_{G}    &
\scat{A}
}
\]
commutes, where $\beta_n$ and $\alpha_n$ are the structure maps of
$\scat{B}$ and $\scat{A}$ (as in equation~\eqref{eq:int-ftr-act}).

However, this is a square of (internal) categories and functors, so we can
also consider variants in which the square is only required to commute up
to a specified natural isomorphism, or a natural transformation in one
direction or the other~-- subject, as usual, to coherence axioms.  The
particular variant that will concern us is the following.

\begin{defn}
\lbl{defn:lax-map-cat-alg}
Let $\cat{E}$ be a category with finite limits and let $P$ be an operad in
$\cat{E}$.  Let $\scat{B}$ and $\scat{A}$ be categorical $P$-algebras, with
structure maps $(\beta_n)$ and $(\alpha_n)$ respectively.  A
\demph{lax%
\index{lax map}%
\index{categorical algebra for operad!lax map of}
map} $\scat{B} \to \scat{A}$ of categorical $P$-algebras consists of a
functor $G \from \scat{B} \to \scat{A}$ (internal to $\cat{E}$) together
with a natural transformation
\[
\xymatrix{
P_n \times \scat{B}^n 
\ar[r]^{1 \times G^n} \ar[d]_{\beta_n} \ar@{}[dr]|-{\swnt\gamma_n}      &
P_n \times \scat{A}^n \ar[d]^{\alpha_n} \\
\scat{B} \ar[r]_G       &
\scat{A}
}
\]
(again internal to $\cat{E}$) for each $n \geq 0$, satisfying the following
two axioms:
\begin{enumerate}
\item 
For each $n, k_1, \ldots, k_n \geq 0$, writing $k = \sum k_i$, 
the composite natural transformation
\[
\xymatrix@C+6em{
\diagstack{P_n \times P_{k_1} \times \scat{B}^{k_1} \times \cdots 
\hspace*{8mm}}%
{\hspace*{20mm}{}\times P_{k_n} \times \scat{B}^{k_n}}
\ar@{}[dr]|-{\swnt 1 \times \gamma_{k_1} \times \cdots \times \gamma_{k_n}}
\ar[d]_{1 \times \beta_{k_1} \times \cdots \times \beta_{k_n}} 
\ar[r]^{1 \times 1 \times G^{k_1} \times \cdots \times 1 \times G^{k_n}}
&
\diagstack{P_n \times P_{k_1} \times \scat{A}^{k_1} \times \cdots
\hspace*{8mm}}%
{\hspace*{20mm}{}\times P_{k_n} \times \scat{A}^{k_n}}
\ar[d]^{1 \times \alpha_{k_1} \times \cdots \times \alpha_{k_n}}   \\
P_n \times \scat{B}^n
\ar@{}[dr]|-{\swnt \gamma_n}
\ar[d]_{\beta_n}
\ar[r]|{\:1 \times G^n\:}       &
P_n \times \scat{A}^n   
\ar[d]^{\alpha_n}        \\
\scat{B} \ar[r]_G       &
\scat{A}
}
\]
is equal to 
\[
\xymatrix@C+6em{
\diagstack{P_n \times P_{k_1} \times \scat{B}^{k_1} \times \cdots
\hspace*{8mm}}%
{\hspace*{20mm}{}\times P_{k_n} \times \scat{B}^{k_n}}
\ar[d]_{\iso} 
\ar[r]^{1 \times 1 \times G^{k_1} \times \cdots \times 1 \times G^{k_n}}
&
\diagstack{P_n \times P_{k_1} \times \scat{A}^{k_1} \times \cdots
\hspace*{8mm}}%
{\hspace*{20mm}{} \times P_{k_n} \times \scat{A}^{k_n}}
\ar[d]^{\iso}   \\
P_n \times P_{k_1} \times \cdots \times P_{k_n} \times \scat{B}^k
\ar[d]_{\of \times 1}
\ar@{}[r]|-{\neeq} &
P_n \times P_{k_1} \times \cdots \times P_{k_n} \times \scat{A}^k       
\ar[d]^{\of \times 1}   \\
P_k \times \scat{B}^k
\ar@{}[dr]|-{\swnt \gamma_k}
\ar[d]_{\beta_k}
\ar[r]|{\:1 \times G^k\:}   &
P_k \times \scat{A}^k   
\ar[d]^{\alpha_k}        \\
\scat{B} \ar[r]_G       &
\scat{A}.
}
\]

\item
The composite natural transformation
\[
\xymatrix@C+3em{
\scat{B} \ar[d]_{\iso} \ar[r]^G &
\scat{A} \ar[d]^{\iso}  \\
1 \times \scat{B} \ar[d]_{1_P \times 1} \ar@{}[r]|-{\neeq}      &
1 \times \scat{A} \ar[d]^{1_P \times 1} \\
P_1 \times \scat{B} 
\ar@{}[dr]|-{\swnt \gamma_1} 
\ar[d]_{\beta_1} 
\ar[r]|{\:1 \times G\:} &
P_1 \times \scat{A} \ar[d]^{\alpha_1}   \\
\scat{B} \ar[r]_G       &
\scat{A}
}
\]
is equal to the identity.  (Here, $1_P \from 1 \to P_1$ denotes the map
encoding the identity of the operad $P$.)
\end{enumerate}
\end{defn}

A strict map of $P$-algebras can equivalently be viewed as a lax map $(G,
\gamma)$ in which each of the maps $\gamma_n$ is an identity.

\begin{remark}
Definition~\ref{defn:lax-map-cat-alg} can also be derived from the theory
of 2-monads,%
\index{monad!two-@2-}%
\index{two-monad@2-monad}%
\index{operad!two-monad from@2-monad from} 
as follows.  We observed in
Remark~\ref{rmks:opds-thys}\bref{rmk:opds-thys-mnds} that any operad $P$ in
$\cat{E}$ induces a monad $T_P$ on $\cat{E}$ (under mild hypotheses on the
category $\cat{E}$).  In the same way, it induces a 2-monad on
$\Cat(\cat{E})$, the 2-category of internal categories in $\cat{E}$.  An
algebra for that 2-monad is exactly a categorical $P$-algebra, and a lax
map of algebras for the 2-monad (in the sense of Blackwell, Kelly and
Power~\cite{BKP}) is exactly a lax map of categorical $P$-algebras.
\end{remark}

As a general categorical principle, it is often worth considering the maps
into an object from the terminal object.  (In categories of spaces, this
gives the notion of point.)  For any operad $P$ in any category $\cat{E}$
with finite limits, there is a terminal categorical $P$-algebra $\One$.  We
consider the lax maps from $\One$ to other categorical $P$-algebras.

\begin{defn}
\lbl{defn:int-alg}
Let $\cat{E}$ be a category with finite limits, let $P$ be an operad in
$\cat{E}$, and let $\scat{A}$ be a categorical $P$-algebra.  An
\demph{internal%
\index{internal algebra}%
\index{categorical algebra for operad!internal algebra in}
algebra} in $\scat{A}$ is a lax map $\One \to \scat{A}$ of categorical
$P$-algebras.
\end{defn}

This definition is due to Batanin%
\index{Batanin, Michael}  
(\cite{BataEHA}, Definition~7.2).  We will see that it generalizes the
notion of internal monoid in a monoidal category.  But first, we give an
explicit description of internal algebras in the cases $\cat{E} = \Set$ and
$\cat{E} = \Tp$.

\begin{examples}
\lbl{egs:int-alg-gen}
\begin{enumerate}
\item 
\lbl{eg:int-alg-gen-Set}
Let $\cat{E} = \Set$.  Take an operad $P$ of sets and a categorical
$P$-algebra $\scat{A}$.  An internal algebra in $\scat{A}$ consists of, first
of all, a functor $G \from \One \to \scat{A}$.  This simply picks out an
object $a$ of $\scat{A}$.  Next, the natural transformations $\gamma_n$ in
Definition~\ref{defn:lax-map-cat-alg} amount to a family of
maps 
\[
\gamma_\theta \from 
\ovln{\theta}(\underbrace{a, \ldots, a}_n)
\to 
a,
\]
one for each $n \geq 0$ and $\theta \in P_n$.  The first coherence axiom in
Definition~\ref{defn:lax-map-cat-alg} states that the diagram
\[
\xymatrix@C+4em{
\ovln{\theta}\Bigl( 
\ovln{\phi^1}(a, \ldots, a), \ldots, \ovln{\phi^n}(a, \ldots, a)
\Bigr)
\ar[r]^-{\ovln{\theta}(\gamma_{\phi^1}, \ldots, \gamma_{\phi^n})}
\ar@{=}[d]      &
\ovln{\theta}(a, \ldots, a)
\ar[d]^{\gamma_\theta}  \\
\ovln{\theta \of (\phi^1, \ldots, \phi^n)}(a, \ldots, a)
\ar[r]_-{\gamma_{\theta \of (\phi^1, \ldots, \phi^n)}}  &
a
}
\]
commutes for all $\theta \in P_n$ and $\phi^i \in P_{k_i}$, and the second
states that
\[
\gamma_{1_P} \from \ovln{1_P}(a) \to a
\]
is equal to the identity on $a$.

\item
An identical description of internal algebras applies when $\cat{E} = \Tp$,
with the additional condition that for each $n \geq 0$, the function
\[
\begin{array}{ccc}
P_n     &\to            &\scat{A}_1     \\
\theta  &\mapsto        &\gamma_\theta
\end{array}
\]
is continuous.  (Here $\scat{A}_1$ denotes the space of maps in
the topological category $\scat{A}$, as in
Example~\ref{egs:cat-alg-gen}\bref{eg:cat-alg-gen-top}.) 

\item
\lbl{eg:int-alg-gen-mon}
Now let $\cat{E} = \Set$, and let $\scat{A}$ be a one-object%
\index{categorical algebra for operad!one-object}
categorical $P$-algebra.  As we saw in
Example~\ref{egs:cat-alg-gen}\bref{eg:cat-alg-gen-one}, $\scat{A}$ amounts
to a monoid $A$ on which $P$ acts by homomorphisms.  An internal
$P$-algebra in $\scat{A}$ consists of an element $\gamma_\theta \in A$ for
each $n \geq 0$ and $\theta \in P_n$, satisfying the coherence axioms
in~\bref{eg:int-alg-gen-Set}. 

Writing $\gamma_\theta$ as $\gamma(\theta)$, we have a sequence of
functions
\[
\bigl( \gamma \from P_n \to A \bigr)_{n \geq 0}.
\]
The coherence axioms in~\bref{eg:int-alg-gen-Set} state that 
\begin{equation}
\lbl{eq:iagm-1}
\gamma(\theta) \cdot 
\ovln{\theta}\bigl(\gamma(\phi^1), \ldots, \gamma(\phi^n)\bigr)
=
\gamma\bigl(\theta \of (\phi^1, \ldots, \phi^n)\bigr)
\end{equation}
for all $n, k_1, \ldots, k_n \geq 0$, $\theta \in P_n$ and $\phi^i \in
P_{k_i}$, and that
\begin{equation}
\lbl{eq:iagm-2}
\gamma(1_P) = 1.
\end{equation}
In summary, when $\scat{A}$ is a one-object categorical $P$-algebra
corresponding to a monoid $A$, an internal $P$-algebra in $\scat{A}$
amounts to a sequence of maps $\gamma \from P_n \to A$ satisfying
equations~\eqref{eq:iagm-1} and~\eqref{eq:iagm-2}.

\item
\lbl{eg:int-alg-gen-top-mon}
For a topological operad $P$, internal algebras in a one-object categorical
$P$-algebra admit exactly the same explicit description as in the previous
example, with the added requirement that the maps $\gamma \from P_n \to A$
are continuous.
\end{enumerate}
\end{examples}

We now give some specific examples.

\begin{examples}
\lbl{egs:int-alg}
\begin{enumerate}
\item
Let $\One$ be the terminal%
\index{operad!terminal}%
\index{terminal operad}
operad of sets.  As we have seen, a categorical
$\One$-algebra is just a monoidal category.  By the explicit
description in Example~\ref{egs:int-alg-gen}\bref{eg:int-alg-gen-Set}, an
internal algebra in a categorical $\One$-algebra $\scat{A}$ consists
of an object $a \in \scat{A}$ together with a map 
\[
\gamma_n \from a^{\otimes n} \to a
\]
for each $n \geq 0$, satisfying the equations given there.  It follows
easily that an internal algebra in $\scat{A}$ is exactly a monoid%
\index{monoid!monoidal category@in monoidal category}
in the monoidal category $\scat{A}$.

As an alternative proof, note that for strict monoidal
categories $\scat{B}$ and $\scat{A}$, a lax map $\scat{B} \to \scat{A}$ of
categorical $\One$-algebras is precisely a lax monoidal functor.  This is
immediate from the definitions.  Hence an internal algebra in a strict
monoidal category $\scat{A}$ is a lax monoidal functor $\One \to \scat{A}$,
and it is well-known that such functors correspond naturally to monoids in
$\scat{A}$ (paragraph~(5.4.1) of B\'enabou~\cite{BenIB}).

An algebra for $\One$ is exactly a monoid
(Example~\ref{egs:opd-alg}\bref{eg:opd-alg-one}), so it is logical
terminology that an internal algebra is exactly an internal monoid.

\item
Fix a monoid%
\index{monoid!operad from}%
\index{operad!monoid@from monoid}
$M$.  We saw in Example~\ref{egs:cat-alg-spec}\bref{eg:cat-alg-spec-act}
that a categorical $P(M)$-algebra is a category $\scat{A}$ with a left
$M$-action; let us write the action as
\[
\begin{array}{ccc}
M \times \scat{A}       &\to            &\scat{A}       \\
(m, a)                  &\mapsto        &m \cdot a.
\end{array}
\]
By the explicit description in
Example~\ref{egs:int-alg-gen}\bref{eg:int-alg-gen-Set}, an internal
$P(M)$-algebra in $\scat{A}$ consists of an object $a \in \scat{A}$
together with a map
\[
\gamma_m \from m \cdot a \to a
\]
for each $m \in M$, satisfying natural coherence axioms.  
\end{enumerate}
\end{examples}

Missing from this list of examples is the case of internal algebras in a
categorical algebra for $\Delta$, the operad of simplices.  This is the
subject of the next section, and will transport us directly to the concept
of entropy.

\section{Entropy as an internal algebra}
\lbl{sec:cat-ent}
\index{entropy!internal algebra@as internal algebra}
\index{internal algebra!entropy as}

In this chapter so far, we have reviewed some established general
concepts in the theory of operads.  We now apply them to the topological
operad $\Delta$ of simplices.

We saw in Example~\ref{egs:cat-alg-spec}\bref{eg:cat-alg-spec-simp} that
the real line $\R$, as a topological monoid under addition, is a
categorical $\Delta$-algebra in a standard way:
\begin{equation}
\lbl{eq:R-cat-alg}
\bigl(\p, (x_1, \ldots, x_n)\bigr) \mapsto \sum_{i = 1}^n p_i x_i
\end{equation}
($\p \in \Delta_n$, $x_1, \ldots, x_n \in \R$).  What are the internal
algebras in the categorical $\Delta$-algebra $\R$?

By Example~\ref{egs:int-alg-gen}\bref{eg:int-alg-gen-top-mon}, an internal
$\Delta$-algebra in $\R$ amounts to a sequence of functions $\bigl( \gamma
\from \Delta_n \to \R \bigr)_{n \geq 0}$ satisfying certain axioms.  It is
in this sense that the following theorem holds.

\begin{thm}
\lbl{thm:cat-ent}%
\index{R@$\R$, internal algebras in}
Let $\Delta$ be the topological operad of simplices, and equip $\R$ with
its standard%
\index{standard Delta-algebra structure@standard $\Delta$-algebra structure} 
categorical $\Delta$-algebra structure~\eqref{eq:R-cat-alg}.
Then the internal algebras in $\R$ are precisely the real scalar multiples
of Shannon entropy.
\end{thm}

In other words, a sequence of functions $\bigl( \gamma \from \Delta_n \to
\R \bigr)_{n \geq 0}$ defines an internal algebra in $\R$ if and only if
$\gamma = cH$ for some $c \in \R$.

\begin{proof}
By Example~\ref{egs:int-alg-gen}\bref{eg:int-alg-gen-top-mon}, an internal
algebra in $\R$ is a sequence of functions $\bigl( \gamma \from \Delta_n
\to \R \bigr)_{n \geq 0}$ with the following properties:
\begin{enumerate}
\item 
\lbl{part:cat-ent-chain}
for all $n, k_1, \ldots, k_n \geq 0$ and $\vc{w} \in \Delta_n$, $\p^1
\in \Delta_{k_1}, \ldots, \p^n \in \Delta_{k_n}$,
\[
\gamma(\vc{w}) + \sum_{i = 1}^n w_i \gamma(\p^i)
=
\gamma\bigl(\vc{w} \of (\p^1, \ldots, \p^n)\bigr);
\]

\item
\lbl{part:cat-ent-unit}
$\gamma(\vc{u}_1) = 0$;

\item
$\gamma \from \Delta_n \to \R$ is continuous for each $n \geq 0$.
\end{enumerate}
Condition~\bref{part:cat-ent-unit} is redundant, since it follows
from~\bref{part:cat-ent-chain} by taking $n = k_1 = 1$ and $\vc{w} = \p^1 =
\vc{u}_1$.  Hence by Faddeev's Theorem~\ref{thm:faddeev}, $\gamma$ defines
an internal algebra if and only if $\gamma = cH$ for some $c \in \R$.
\end{proof}

This theorem can be deformed.%
\index{deformed Delta-algebra structure@deformed $\Delta$-algebra structure} 
In Example~\ref{egs:cat-alg-spec}\bref{eg:cat-alg-spec-def}, we defined 
a one-parameter family of categorical $\Delta$-algebra structures
on $\R$, where for a real parameter $q$, the action of $\Delta$ on
$\R$ is 
\begin{equation}
\lbl{eq:R-q-cat-alg}
\bigl(\p, (x_1, \ldots, x_n)\bigr) 
\mapsto 
\sum_{i \in \supp(\p)} p_i^q x_i
\end{equation}
($\p \in \Delta_n$, $x_1, \ldots, x_n \in \R$).  

\begin{thm}
\lbl{thm:q-cat-ent}%
\index{R@$\R$, internal algebras in}
Let $1 \neq q \in \R$.  Let $\Delta$ be the operad of simplices, considered
as an operad of sets, and equip $\R$ with its $q$-deformed categorical
$\Delta$-algebra structure~\eqref{eq:R-q-cat-alg}.  Then the internal
algebras in $\R$ are precisely the real scalar multiples of $q$-logarithmic
entropy.%
\index{internal algebra!q-logarithmic entropy as@$q$-logarithmic entropy as}
\index{q-logarithmic entropy@$q$-logarithmic entropy!internal algebra@as internal algebra}
\end{thm}

\begin{proof}
By Example~\ref{egs:int-alg-gen}\bref{eg:int-alg-gen-mon}, an internal
algebra in $\R$ is a sequence of functions $\bigl( \gamma \from \Delta_n
\to \R \bigr)_{n \geq 0}$ with the following properties:
\begin{enumerate}
\item 
\lbl{part:q-cat-ent-chain}
for all $n, k_1, \ldots, k_n \geq 0$ and all $\vc{w} \in \Delta_n$, $\p^1
\in \Delta_{k_1}, \ldots, \p^n \in \Delta_{k_n}$,
\[
\gamma(\vc{w}) + \sum_{i \in \supp(\vc{w})} w_i^q \gamma(\p^i)
=
\gamma\bigl(\vc{w} \of (\p^1, \ldots, \p^n)\bigr);
\]

\item
\lbl{part:q-cat-ent-unit}
$\gamma(\vc{u}_1) = 0$.
\end{enumerate}
Condition~\bref{part:q-cat-ent-unit} is redundant, for the same reason as
in the proof of Theorem~\ref{thm:cat-ent}.
Condition~\bref{part:q-cat-ent-chain} is satisfied if $\gamma = cS_q$, by
the chain rule~\eqref{eq:q-ent-chain} for $q$-logarithmic entropies.
Conversely, condition~\bref{part:q-cat-ent-chain} implies that
\[
\gamma(\vc{w} \otimes \p) 
= 
\gamma(\vc{w}) + 
\Biggl(\sum_{i \in \supp(\vc{w})} w_i^q\Biggr) \gamma(\vc{p})
\]
for all $\vc{w} \in \Delta_n$ and $\p \in \Delta_k$, by taking $\p^1 =
\cdots = \p^n = \p$.  Hence by Theorem~\ref{thm:q-ent-char}, if $\gamma$
defines an internal algebra then $\gamma = cS_q$ for some $c \in
\R$.
\end{proof}

Continuity was not needed in this theorem, and in fact the structure maps
$\Delta_n \times \R^n \to \R$ of the $q$-deformed $\Delta$-algebra $\R$ are
discontinuous when $q \leq 0$.  But they are evidently continuous when $q >
0$, so we have:

\begin{cor}
Let $q \in (0, \infty)$.  Let $\Delta$ be the topological operad of
simplices, and equip $\R$ with its $q$-deformed categorical
$\Delta$-algebra structure~\eqref{eq:R-q-cat-alg}.  Then the internal
algebras in $\R$ are precisely the real scalar multiples of $q$-logarithmic
entropy. 
\end{cor}

\begin{proof}
The case $q = 1$ is Theorem~\ref{thm:cat-ent}, and all other cases
follow from Theorem~\ref{thm:q-cat-ent}.
\end{proof}

\section{The universal internal algebra}
\lbl{sec:univ-int}
\index{internal algebra!universal}
\index{universal internal algebra}
\index{universal property}

In algebra, an important role is played by free algebraic structures
(groups, modules, etc.).  But since one forms the free algebraic structure
on a \emph{set}, and a set is merely a cardinality (for these purposes at
least), the possibilities are in a sense limited.  Greater riches are to be
found one categorical level up, where one can speak of the free categorical
structure containing some specified internal algebraic structure.  This
leads to categorical characterizations of some important
mathematical objects.
\begin{examples}
\begin{enumerate}
\item 
The free monoidal category containing a monoid%
\index{monoid!monoidal category@in monoidal category} 
is equivalent to the
category of finite totally ordered%
\index{order!finite total}
sets (Mac~Lane~\cite{MacLCWM}, Proposition~VII.5.1).  We will return to
this example shortly.  Informally, the statement is that if we build a
monoidal category by starting from nothing, putting in an internal monoid,
then adjoining no more other objects and maps than are forced by the
definitions, and making no unnecessary identifications, then the result is
the category of finite totally ordered sets.

\item
The free monoidal category containing an object $A$ and an isomorphism $A
\otimes A \to A$ is equivalent to the disjoint union of the terminal
category and Thompson's%
\index{Thompson's group $F$} 
group $F$, viewed as a one-object category (Fiore%
\index{Fiore, Marcelo}
and Leinster~\cite{ACTGF}).  

(Thompson's group is an infinite group with remarkable properties; it
has been rediscovered multiple times in diverse contexts.  Cannon, Floyd and
Parry~\cite{CFP} provide a survey.  A major open question, which
has attracted an exceptional number of opposing claims and retractions, is
whether $F$ is amenable.  Cannon and Floyd~\cite{CaFl} report that even
among experts, opinion is evenly split.)

\item
The free symmetric monoidal category containing a commutative Frobenius
algebra is the category of compact oriented 1-manifolds and 2-dimensional
cobordisms between them (Theorem~3.6.19 of Kock~\cite{KockFA2}, for
instance).  This result lies at the foundations of topological%
\index{topological quantum field theory} 
quantum field theory.

\item
The free finite product category containing a group is the
Lawvere%
\index{Lawvere, F. William!theory} 
theory of groups.  The same statement holds for any other algebraic
structure in place of groups (Lawvere~\cite{LawvFSA}).  This is essentially
a tautology, but expresses a fundamental insight of categorical universal
algebra: an algebraic theory can be understood as a finite product
category, and a model of a theory as a finite-product-preserving functor.
\end{enumerate}
\end{examples}

In this section, we construct the free categorical $P$-algebra containing
an internal algebra, where $P$ is any given operad.  We proceed as follows.
First, we construct a certain categorical $P$-algebra $FP$.  Then, we make
precise what it means for a categorical $P$-algebra to be `free containing
an internal algebra'.  Next, we prove that $FP$ has that property.  This
last result, applied in the case $P = \Delta$, leads to a characterization
of information loss.

We begin by constructing the categorical $P$-algebra $FP$\ntn{freecat}, for
an operad $P$ of sets.

The objects of $FP$ are the pairs $(n, \theta)$ with $n \geq 0$ and $\theta
\in P_n$.  Where confusion will not arise, we write $(n, \theta)$ as just
$\theta$.  For objects $\psi = (k, \psi)$ and $\theta = (n, \theta)$, a map
$\psi \to \theta$ in $FP$ consists of integers $k_1, \ldots, k_n \geq 0$
and operations $\phi^1 \in P_{k_1}, \ldots, \phi^n \in P_{k_n}$ such that
\[
k = k_1 + \cdots + k_n, 
\qquad
\psi = \theta \of (\phi^1, \ldots, \phi^n).
\]
We write this map as
\begin{equation}
\lbl{eq:typical-FP-map}
\Fmap{\phi^1, \ldots, \phi^n}{\theta} \from \psi \to \theta.
\end{equation}
Thus, the set of objects of the category $FP$ and the set of maps in $FP$
are, respectively,
\begin{equation}
\lbl{eq:FP-oa}
\coprod_{n \geq 0} P_n,
\qquad
\coprod_{n, k_1, \ldots, k_n \geq 0} 
P_n \times P_{k_1} \times \cdots \times P_{k_n}.
\end{equation}
Composition and identities in the category $FP$ are defined using the
composition and identity of the operad $P$.

To give the category $FP$ the structure of a categorical $P$-algebra, we
must construct from each operation $\pi \in P_m$ a functor
\[
\ovln{\pi}\from (FP)^m \to FP.
\]
On objects, $\ovln{\pi}$ is defined by
\[
\ovln{\pi} (\theta^1, \ldots, \theta^m)
=
\pi \of (\theta^1, \ldots, \theta^m).
\]
To define the action of $\ovln{\pi}$ on maps, take an $m$-tuple of maps 
\begin{align*}
\Fmap{\phi^{1 1}, \ldots, \phi^{1 n_1}}{\theta^1} \from       &
\psi^1 \to \theta^1    \\
\vdots  \\
\Fmap{\phi^{m 1}, \ldots, \phi^{m n_m}}{\theta^m} \from       &
\psi^m \to \theta^m
\end{align*}
in $FP$.  Then
\begin{multline}
\lbl{eq:FP-act-maps} 
\ovln{\pi}\Bigl(
\Fmap{\phi^{1 1}, \ldots, \phi^{1 n_1}}{\theta^1},
\ \ldots, \ 
\Fmap{\phi^{m 1}, \ldots, \phi^{m n_m}}{\theta^m}
\Bigr)  \\
=
\Fmap{\phi^{1 1}, \ldots, \phi^{1 n_1},
\ \ldots, \
\phi^{m 1}, \ldots, \phi^{m n_m}}%
{\pi\of(\theta^1, \ldots, \theta^m)},
\end{multline}
which is a map $\ovln{\pi}(\psi^1, \ldots, \psi^m) \to
\ovln{\pi}(\theta^1, \ldots, \theta^m)$ in $FP$. 

Verifying that $FP$
satisfies the axioms for a categorical $P$-algebra is routine.


\begin{lemma}
\lbl{lemma:FP-term}
Let $P$ be an operad of sets.  
\begin{enumerate}
\item 
\lbl{part:FP-term-term}
The object $1_P$ of $FP$ is terminal.

\item
\lbl{part:FP-term-decomp}
Write $\uqq{\phi}\ntn{toterm} \from \phi \to 1_P$ for the unique map from
an object $\phi$ of $FP$ to $1_P$.  Then for any map
\[
\Fmap{\phi^1, \ldots, \phi^n}{\theta} \from \psi \to \theta
\]
in $FP$, we have
\[
\Fmap{\phi^1, \ldots, \phi^n}{\theta}
=
\ovln{\theta}(\uqq{\phi^1}, \ldots, \uqq{\phi_n}).
\]
\end{enumerate}
\end{lemma}

The notation in~\bref{part:FP-term-term} refers to the identity element
$1_P \in P_1$ of the operad $P$, which corresponds to the object
$1_P\ntn{idopdobj} = (1, 1_P)$ of the category $FP$.  It is this object
that is terminal.

\begin{proof}
For~\bref{part:FP-term-term}, given any object $\phi$ of $FP$, it is
immediate from the definition of $FP$ that there is a unique map $\phi \to
1_P$, namely, 
\[
\uqq{\phi} = \Fmap{\phi}{1_P} \from \phi \to 1_P.
\]

For~\bref{part:FP-term-decomp}, take a map
\[
\Fmap{\phi^1, \ldots, \phi^n}{\theta} \from \psi \to \theta
\]
in $FP$.  Since $\uqq{\phi^i}$ is a map $\phi^i \to 1_P$, the map
$
\ovln{\theta}(\uqq{\phi^1}, \ldots, \uqq{\phi_n})
$
has domain 
\[
\ovln{\theta}(\phi^1, \ldots, \phi^n)
=
\theta \of (\phi^1, \ldots, \phi^n)
=
\psi
\]
and codomain
\[
\ovln{\theta}(1_P, \ldots, 1_P)
=
\theta \of (1_P, \ldots, 1_P)
=
\theta,
\]
matching the domain and codomain of $\Fmap{\phi^1, \ldots,
  \phi^n}{\theta}$.  Now by definition of $\uqq{\phi^i}$ and by
definition~\eqref{eq:FP-act-maps} of the $P$-action on maps in $FP$,
\begin{align*}
\ovln{\theta}(\uqq{\phi^1}, \ldots, \uqq{\phi_n})   &
=
\ovln{\theta}\bigl(\Fmap{\phi^1}{1_P}, \ldots, \Fmap{\phi^n}{1_P}\bigr) \\
&
=
\Fmap{\phi^1, \ldots, \phi^n}{\theta\of(1_P, \ldots, 1_P)}      \\
&
=
\Fmap{\phi^1, \ldots, \phi^n}{\theta},
\end{align*}
as required.
\end{proof}

The categorical $P$-algebra $FP$ contains a canonical internal algebra.  To
specify it, we use the description of internal algebras in
Example~\ref{egs:int-alg-gen}\bref{eg:int-alg-gen-Set}.  Its underlying
object is the terminal object $1_P$.  To give $1_P$ the structure of an
internal algebra, we have to specify, for each $n \geq 0$ and $\theta \in
P_n$, a map
\[
\ovln{\theta}
\bigl(\,
\underbrace{1_P, \ldots, 1_P}_n
\,\bigr)
\to 
1_P.
\]
The domain here is $\theta$, and the codomain is terminal, so the only
possible choice is the unique map $\uqq{\theta} \from \theta \to 1_P$.
This gives $1_P$ the structure of an internal algebra in the categorical
$P$-algebra $FP$.  We refer to this internal algebra as $(1_P, !)$.

When $P$ is a topological operad, the set of objects of $FP$ and the set of
maps in $FP$ (both given in~\eqref{eq:FP-oa}) each carry a natural
topology.  For instance, the set of maps in $FP$ is a coproduct of product
spaces.  In this way, $FP$ is an internal category in $\Tp$.
Indeed, $FP$ is a categorical $P$-algebra in the topological sense (by the
description in Example~\ref{egs:cat-alg-gen}\bref{eg:cat-alg-gen-top}) and
$(1_P, !)$ is an internal algebra in $FP$ in the topological sense (by the
description in Example~\ref{egs:int-alg-gen}\bref{eg:int-alg-gen-top-mon}.)

\begin{remark}
As for all of the operadic definitions and constructions in this chapter,
the construction of $FP$ can be generalized to an operad $P$ in an
arbitrary category $\cat{E}$ with suitable properties (in this case, finite
products and countable coproducts over which the products distribute).  The
general definition is exactly as suggested by the case $\cat{E} = \Tp$.
\end{remark}

\begin{examples}
\lbl{egs:F}
\begin{enumerate}
\item 
\lbl{eg:F-mon}
Consider the terminal%
\index{operad!terminal}%
\index{terminal operad}
operad $\One$ of sets.  The objects of the category
$\scat{D} = F\One$ are the natural numbers $0, 1, \ldots$\:\  A map $k \to n$
in $\scat{D}$ is an ordered $n$-tuple of natural numbers summing to $k$, or
equivalently, an order-preserving map $\{1, \ldots, k\} \to \{1, \ldots,
n\}$.  Thus, $\scat{D}$ is equivalent to the category of finite totally
ordered%
\index{order!finite total}
sets.  It is almost the same as the category usually denoted by $\Delta$ in
algebraic topology, the only difference being that it also contains the
object $0$ (corresponding to the empty ordered set).

By construction, $\scat{D}$ is a categorical $\One$-algebra, that is, a
strict monoidal category.  The monoidal structure is defined on objects by
addition and on maps by disjoint union.  Moreover, $\scat{D}$ contains a
canonical internal algebra, that is, internal monoid.  It is the object $1
\in \scat{D}$ with its unique monoid structure: the multiplication is the
unique map $1 + 1 = 2 \to 1$ in $\scat{D}$, and the identity is the unique
map $0 \to 1$.

\item
Fix a monoid%
\index{monoid!operad from}%
\index{operad!monoid@from monoid}
$M$ and consider the operad $P(M)$.  Since $P(M)_n$ is empty for all $n
\neq 1$, the objects of the category $FP(M)$ are just the elements $\theta
\in M$.  A map $\psi \to \theta$ in $FP(M)$ is an element $\phi \in M$ such
that $\psi = \theta\phi$.  In other words, regarding the monoid $M$ as a
category with a single object $\star$, the category $FP(M)$ is the slice 
$M/\star$.  For instance, when the monoid $M$ is cancellative, $FP(M)$ is
the poset of elements of $M$ ordered by divisibility.

\item
\lbl{eg:F-prob}
Now take the topological operad $\Delta$.%
\index{simplex!operad}%
\index{probability distribution!operad}%
\index{operad!simplex}
The objects of the category $F\Delta$ are the pairs $(n, \p)$ with $n \geq
0$ and $\p \in \Delta_n$.  A map $(k, \vc{s}) \to (n, \p)$ consists of
natural numbers $k_1, \ldots, k_n$ summing to $k$ together with probability
distributions $\vc{r}^i \in \Delta_{k_i}$ satisfying
\begin{equation}
\lbl{eq:F-prob-comp}
\vc{s} = \p \of (\vc{r}^1, \ldots, \vc{r}^n).
\end{equation}
This category has a more familiar description. As
in~\bref{eg:F-mon} above, the $n$-tuple $(k_1, \ldots, k_n)$ amounts to an
order-preserving map
\[
f \from \{1, \ldots, k\} \to \{1, \ldots, n\}.  
\]
Then $\vc{p}$ is equal to the pushforward $f \vc{s}$ of the probability
measure $\vc{s}$ along $f$. (See Definition~\ref{defn:pfwd}.)  Thus, $f$ is
a measure-preserving map
\[
\bigl( \{1, \ldots, k\}, \vc{s} \bigr)
 \to 
\bigl( \{1, \ldots, n\}, \vc{p} \bigr).
\]
In Lemma~\ref{lemma:decomp}, we showed that given $\vc{s}$, $\p$ and $k_1,
\ldots, k_n$ (or equivalently $\vc{s}$, $\p$ and $f$), it is always
possible to find distributions $\vc{r}^i$ satisfying
equation~\eqref{eq:F-prob-comp}.  Furthermore, we showed that for $i \in
\supp(\p)$, the distribution $\vc{r}^i$ is uniquely determined, and for $i
\not\in \supp(\p)$, we can choose $\vc{r}^i$ freely in $\Delta_{k_i}$.

These observations together imply that up to equivalence,
$F\Delta$ is the category whose objects are finite totally ordered%
\index{order!probability space@on probability space}
probability spaces $(\XX, \p)$, in which a map
$(\YY, \vc{s}) \to (\XX, \vc{p})$ is an order-preserving,
measure-preserving map $f$ together with a probability distribution on
$f^{-1}(i)$ for each $i \in \XX$ such that $p_i = 0$.

By construction, $F\Delta$ has the structure of a categorical
$\Delta$-algebra.  On objects, the $\Delta$-action takes convex combinations
of finite probability spaces, as in Section~\ref{sec:meas-pres}.
The one-element probability space $(1, \vc{u}_1)$ has a unique internal
algebra structure in $F\Delta$.
\end{enumerate}
\end{examples}

\begin{remark}
\lbl{rmk:zero-bayes} 
The category $F\Delta$ just described is nearly the category
$\fcat{FinOrdProb}$ of finite totally ordered%
\index{order!probability space@on probability space}%
\index{probability space!ordered}
probability spaces.  There is a forgetful functor $F\Delta \to
\fcat{FinOrdProb}$, but it is not an equivalence, because of the
complication associated with zero probabilities.

From the point of view of Bayesian%
\index{Bayesian inference} 
inference, it is broadly unsurprising that such a complication arises.  In
that subject, special caution is reserved for probabilities of exactly
zero.  The Bayesian statistician Dennis Lindley%
\index{Lindley, Dennis}
wrote:
\begin{quote}
leave a little probability for the moon%
\index{moon} 
being made of green%
\index{green cheese} 
cheese; it can be as small as 1 in a million, but have it there since
otherwise an army of astronauts returning with samples of the said cheese
will leave you unmoved.  [\ldots] So never believe in anything absolutely,
leave some room for doubt.
\end{quote}
(\cite{LindMD}, p.~104.)  He named this principle \demph{Cromwell's%
\index{Cromwell, Oliver}
rule},
after the English Lord Protector Oliver Cromwell, who wrote to the Church
of Scotland in 1650:
\begin{quote}
I beseech you, in the bowels of Christ, think it possible you may be
mistaken. 
\end{quote}
Further discussion can be found in Section~6.8 of Lindley~\cite{LindUU}.
\end{remark}

We now make precise, and prove, the statement that $FP$ is the `free%
\index{categorical algebra for operad!free on internal algebra}%
\index{internal algebra!universal}
categorical $P$-algebra containing an internal algebra'.

Let $P$ be an operad of sets or topological spaces, and let $E \from
\scat{B} \to \scat{A}$ be a strict map of categorical $P$-algebras.  An
internal algebra in $\scat{B}$ is a lax map $\One \to \scat{B}$, and can be
composed with $E$ to obtain a lax map $\One \to \scat{A}$.  In this way,
$E$ maps internal algebras in $\scat{B}$ to internal algebras in
$\scat{A}$.

It will be convenient to use the explicit description of internal algebras
derived in Example~\ref{egs:int-alg-gen}\bref{eg:int-alg-gen-Set}.  There,
we showed that an internal algebra $(b, \delta)$ in $\scat{B}$ consists of
an object $b$ and a family of maps $\delta_\theta \from \ovln{\theta}(b,
\ldots, b) \to b$ subject to certain equations.  In these terms, the
induced internal algebra $E(b, \delta)$ in $\scat{A}$ consists of the
object $E(b)$ and the maps $E(\delta_\theta)$.

We now state and prove the universal%
\index{universal property} 
property of the categorical $P$-algebra $FP$ equipped with its internal
algebra $(1_P, \uq)$.

\begin{thm}
\lbl{thm:uia}
Let $P$ be an operad of either sets or topological spaces, let $\scat{A}$
be a categorical $P$-algebra, and let $(a, \gamma)$ be an internal algebra
in $\scat{A}$.  Then there is a unique strict map $E \from FP \to \scat{A}$
of categorical $P$-algebras such that $E(1_P, \uq) = (a, \gamma)$.
\end{thm}

This is a universal property of $FP$ together with its internal algebra,
and therefore determines them uniquely up to isomorphism.

\begin{proof}
To prove uniqueness, let $E$ be a map with the properties stated.  Let
$\theta = (n, \theta)$ be an object of $FP$; thus, $n \geq 0$ and $\theta
\in P_n$.  By definition of the categorical $P$-algebra structure on $FP$,
\[
\theta
= 
\ovln{\theta}(1_P, \ldots, 1_P).
\]
Applying $E$ to both sides gives
\[
E(\theta)
=
E\bigl(\ovln{\theta}(1_P, \ldots, 1_P)\bigr)
=
\ovln{\theta}\bigl(E(1_P), \ldots, E(1_P)\bigr)
=
\ovln{\theta}(a, \ldots, a),
\]
where the second equality holds because $E$ is a strict map of categorical
$P$-algebras, and the last is by hypothesis.  Hence
\begin{equation}
\lbl{eq:uia-obj}
E(\theta) = \ovln{\theta}(a, \ldots, a),
\end{equation}
which determines $E$ uniquely on the objects of $FP$.

To show the same for maps, take a map 
\[
\Fmap{\phi^1, \ldots, \phi^n}{\theta} \from \psi \to \theta
\]
in $FP$.  By Lemma~\ref{lemma:FP-term}\bref{part:FP-term-decomp}, 
\[
\Fmap{\phi^1, \ldots, \phi^n}{\theta}
=
\ovln{\theta}(\uqq{\phi^1}, \ldots, \uqq{\phi^n}).
\]
Applying $E$ to both sides gives
\begin{align*}
E\bigl( \Fmap{\phi^1, \ldots, \phi^n}{\theta} \bigr)    &
=
E\bigl( \ovln{\theta}(\uqq{\phi^1}, \ldots, \uqq{\phi^n}) \bigr)    \\
&
=
\ovln{\theta}\bigl(
E(\uqq{\phi^1}), \ldots, E(\uqq{\phi^n})
\bigr)  \\
&
=
\ovln{\theta}(\gamma_{\phi^1}, \ldots, \gamma_{\phi^n}),
\end{align*}
for the same reasons as in the argument for objects.  Hence
\begin{equation}
\lbl{eq:uia-map}
E\bigl( \Fmap{\phi^1, \ldots, \phi^n}{\theta} \bigr)    
=
\ovln{\theta}(\gamma_{\phi^1}, \ldots, \gamma_{\phi^n}),
\end{equation}
which determines $E$ uniquely on the maps in $FP$.  We have therefore
proved uniqueness.

To prove existence, we define $E$ on objects by equation~\eqref{eq:uia-obj}
and on maps by equation~\eqref{eq:uia-map}.  Verifying that $E$ satisfies
the stated conditions (including continuity in the topological case) is a
series of routine checks. 
\end{proof}

\begin{cor}
\lbl{cor:uia-bjn}
Let $P$ be an operad of sets or topological spaces.  Let $\scat{A}$ be a
categorical $P$-algebra.  Then there is a canonical bijection between
internal algebras in $\scat{A}$ and strict maps $FP \to \scat{A}$ of
categorical%
\index{categorical algebra for operad!strict map of} 
$P$-algebras.
\qed
\end{cor}

Thus, an internal algebra in $\scat{A}$ can be described as either a lax
map $\One \to \scat{A}$ or a strict map $FP \to \scat{A}$.

\begin{example}
In the case $P = \One$, Theorem~\ref{thm:uia} states that for any strict
monoidal category $\scat{A}$ and monoid $a$ in $\scat{A}$, there is exactly
one strict monoidal functor $E: \scat{D} \to \scat{A}$ that maps the
trivial monoid $1$ in $\scat{D}$ to the given monoid $a$ in $\scat{A}$.

Hence, Corollary~\ref{cor:uia-bjn} implies that given just a monoidal
category $\scat{A}$, the monoids%
\index{monoid!monoidal category@in monoidal category} 
in $\scat{A}$ correspond naturally to the strict monoidal functors
$\scat{D} \to \scat{A}$.  We have therefore recovered the classical fact
that a monoid in $\scat{A}$ can be described as either a lax monoidal
functor $\One \to \scat{A}$ or a strict monoidal functor $\scat{D} \to
\scat{A}$ (paragraph~(5.4.1) of B\'enabou~\cite{BenIB} and
Proposition~VII.5.1 of Mac~Lane~\cite{MacLCWM}, for instance).
\end{example}

Now consider Theorem~\ref{thm:uia} in the case where $P$ is the topological
operad $\Delta$ and $\scat{A}$ is the topological monoid $\R$.  By
Corollary~\ref{cor:uia-bjn}, the strict maps $F\Delta \to \scat{A}$ of
categorical $\Delta$-algebras are in natural bijection with the internal
$\Delta$-algebras in $\scat{A}$.  By Theorem~\ref{thm:cat-ent}, these in
turn correspond to real scalar multiples of Shannon entropy.  Together,
these results imply that the strict maps $F\Delta \to \scat{A}$ are
naturally parametrized by $\R$.

We now make this parametrization explicit.  Since $\scat{A}$ has only one
object, a strict map $F\Delta \to \scat{A}$ of categorical
$\Delta$-algebras amounts to a function
\[
E \from \{ \text{maps in } F\Delta \} \to \R
\]
satisfying certain conditions.  Our final theorem classifies such functions.

\begin{thm}
\lbl{thm:fd-semi}
Let $E$ be a function $\{ \text{maps in } F\Delta \} \to \R$.  The
following are equivalent:
\begin{enumerate}
\item
\lbl{part:fd-semi-func}
$E$ defines a strict map $F\Delta \to \R$ of categorical $\Delta$-algebras
in $\Tp$ (with respect to the standard%
\index{standard Delta-algebra structure@standard $\Delta$-algebra structure} 
categorical $\Delta$-algebra structure on $\R$);

\item
\lbl{part:fd-semi-form}
there is some $c \in \R$ such that for all maps $f \from \vc{s} \to \p$
in $F\Delta$,
\[
E(f)
= 
c\bigl(H(\vc{s}) - H(\vc{p})\bigr).
\]
\end{enumerate}
\end{thm}

\begin{proof}
First assume~\bref{part:fd-semi-func}.  Applying $E$ to the internal
algebra $(\vc{u}_1, \uq)$ in $F\Delta$ gives an internal algebra
$E(\vc{u}_1, \uq)$ in $\R$ (whose underlying object is necessarily the
unique object of the category $\R$).  So by Theorem~\ref{thm:cat-ent},
there is some constant $c \in \R$ such that $E(\uqq{\p}) =
cH(\vc{p})$ for all $n \geq 1$ and $\p \in \Delta_n$.

Now take any map
\begin{equation}
\lbl{eq:fd-map} 
\Fmap{\vc{r}^1, \ldots, \vc{r}^n}{\p} \from \vc{s} \to \p
\end{equation}
in $F\Delta$.  Since $\vc{u}_1$ is terminal in $F\Delta$, there is a
commutative triangle
\begin{equation}
\lbl{eq:H-triangle}
\xymatrix{
\vc{s} \ar[rr]^{\Fmap{\vc{r}^1, \ldots, \vc{r}^n}{\p}} 
\ar[rd]_{\uqq{\vc{s}}}   &       &\vc{p} \ar[ld]^{\uqq{\vc{p}}}   \\
                                &\vc{u}_1
}
\end{equation}
in $F\Delta$.  Applying the functor $E$ to this triangle gives
\begin{equation}
\lbl{eq:E-comp}
E(\uqq{\vc{s}}) 
=
E(\uqq{\vc{p}}) 
+
E\bigl( \Fmap{\vc{r}^1, \ldots, \vc{r}^n}{\p} \bigr),
\end{equation}
which by the result of the last paragraph gives
\[
cH(\vc{s}) = cH(\vc{p}) + 
E\bigl(\Fmap{\vc{r}^1, \ldots, \vc{r}^n}{\p}\bigr),
\]
proving~\bref{part:fd-semi-form}.

To show that~\bref{part:fd-semi-form} implies~\bref{part:fd-semi-func},
let $c \in \R$.  By Theorem~\ref{thm:cat-ent}, $cH$ defines an internal
algebra structure on the unique object of the category $\R$.  Now take a
map~\eqref{eq:fd-map} in $F\Delta$.  By definition of $E$,
\[
E\bigl(\Fmap{\vc{r}^1, \ldots, \vc{r}^n}{\p}\bigr)
= 
c\bigl(H(\vc{s}) - H(\vc{p})\bigr).
\]
But $\vc{s} = \vc{p} \of (\vc{r}^1, \ldots, \vc{r}^n)$ by definition of the
maps in $F\Delta$, so by the chain rule,
\[
E\bigl(\Fmap{\vc{r}^1, \ldots, \vc{r}^n}{\p}\bigr)
= 
c \sum_{i = 1}^n p_i H(\vc{r}^i)
=
\ovln{\vc{p}}\bigl(cH(\vc{r}^1), \ldots, cH(\vc{r}^n)\bigr).
\]
It follows from the proof of Theorem~\ref{thm:uia} that $E$ is a strict map
$F\Delta \to \R$ of categorical $\Delta$-algebras.
\end{proof}

A result similar to Theorem~\ref{thm:fd-semi} can also be proved for
the $q$-logarithmic entropies, using the $q$-deformed categorical
$\Delta$-algebra structure on $\R$ and Theorem~\ref{thm:q-cat-ent}.

Theorem~\ref{thm:fd-semi} bears a striking resemblance to the
characterization of information loss in Theorem~\ref{thm:cetil}.  It states
that the strict maps $F\Delta \to \R$ are the scalar multiples of the
information loss function.  But where one theorem uses the category
$F\Delta$, the other uses the category $\FinProb$ of finite probability
spaces. The explicit description of $F\Delta$ in
Example~\ref{egs:F}\bref{eg:F-prob} shows that there are three differences
between $F\Delta$ and $\FinProb$. First, the maps in $F\Delta$ are required
to be order-preserving,%
\index{order!probability space@on probability space}%
\index{probability space!ordered}
whereas in $\FinProb$ there is no notion of ordering at all.  Second, the
category $F\Delta$ is skeletal (isomorphic objects are equal), but
$\FinProb$ is not.  Third, the maps in the category $F\Delta$ are not
merely measure-preserving maps; they also come equipped with a probability
distribution on the fibre over each zero-probability element of the codomain.

There is an analogue of Theorem~\ref{thm:fd-semi} that comes close to
Theorem~\ref{thm:cetil}; we sketch it now. It uses \emph{symmetric}%
\index{operad!symmetric}%
\index{symmetric!operad} 
operads. As indicated in
Remark~\ref{rmks:opds-thys}\bref{rmk:opds-thys-thys}, a symmetric operad is
an operad $P$ together with an action of the symmetric group $S_n$ on $P_n$
for each $n \geq 0$, satisfying suitable axioms.  For example, if $A$ is
an object of a symmetric monoidal category then the operad $\End(A)$
of Example~\ref{egs:opd}\bref{eg:opd-End} has the structure of a symmetric
operad.  The operad $\Delta$ is also symmetric in a natural way.

At the cost of some further complications, the notions of categorical
$P$-algebra and internal algebra, and the construction of the free
categorical $P$-algebra on an internal algebra, can be extended to
symmetric operads $P$.  The free categorical $\Delta$-algebra $\Fsym
\Delta$ on an internal algebra is much like $F\Delta$, but the maps are no
longer required to be order-preserving.  In other words, the first of the
three differences between $F\Delta$ and $\FinProb$ vanishes for
$\Fsym\Delta$.  The second, skeletality, is categorically unimportant.  So,
the only substantial difference between $\Fsym\Delta$ and $\FinProb$ is the
third: a map in $\Fsym\Delta$ between finite probability spaces is a
measure-preserving map \emph{together with} a probability distribution on
each fibre over an element of probability zero.

The symmetric analogue of Theorem~\ref{thm:fd-semi} states that the strict
maps $\Fsym\Delta \to \R$ of symmetric categorical $\Delta$-algebras are
precisely the scalar multiples of information loss.  Translated into
explicit terms, this theorem is nearly the same as the
characterization of information loss in Theorem~\ref{thm:cetil}.  The only
difference is in the handling of zero probabilities.
But the result can easily be adapted in an ad hoc way to discard the
extra data associated with elements of probability zero, and it then becomes
exactly Theorem~\ref{thm:cetil}.  Historically, this
categorical argument was, in fact, how the wholly elementary
and concrete Theorem~\ref{thm:cetil} was first obtained.

%% file: proofs.tex
\chapter{Proofs of background facts}
\lbl{app:proofs}

This appendix consists of proofs deferred from the main text.

\section{Forms of the chain rule for entropy}
\lbl{sec:chain}
\index{chain rule!forms of}

In Remark~\ref{rmk:ent-chain-simp}, it was asserted that although the
chain rule
\[
H\bigl( \vc{w} \of (\p^1, \ldots, \p^n)\bigr)
=
H(\vc{w}) + \sum_{i = 1}^n w_i H(\p^i)
\]
for Shannon entropy appears to be more general (that is, stronger) than the
versions used by some previous authors, straightforward inductive arguments
show that it is equivalent to those special cases.
Remark~\ref{rmk:q-ent-char-hist} made a similar assertion for the
$q$-logarithmic entropies $S_q$, where the equation becomes
\[
S_q\bigl( \vc{w} \of (\p^1, \ldots, \p^n)\bigr)
=
S_q(\vc{w}) + \sum_{i \in \supp(\vc{w})} w_i^q S_q(\p^i).
\]
Here we prove those claims.

In Lemma~\ref{lemma:ch-forms} below, part~\bref{part:cf-full} is the
general form of the chain rule, parts~\bref{part:cf-l}
and~\bref{part:cf-twotop} are the special cases used by other authors, and
part~\bref{part:cf-int} is an intermediate case that is helpful for the
proof.  Each of the four parts corresponds to a certain type of composition
of probability distributions, depicted as a tree in
Figure~\ref{fig:comp-trees}.  

Rather than working with sums over the support of $\vc{w}$, in this lemma
we adopt the convention that $0^q = 0$ for all $q \in \R$.

\begin{lemma}
\lbl{lemma:ch-forms}
Let $q \in \R$.  Let $( I \from \Delta_n \to \R)_{n \geq 1}$ be a sequence
of symmetric functions.  The following are equivalent:
\begin{enumerate}
\item 
\lbl{part:cf-full}
for all $n, k_1, \ldots, k_n \geq 1$, $\vc{w} \in \Delta_n$, and $\p^i \in
\Delta_{k_i}$, 
\[
I\bigl( \vc{w} \of (\p^1, \ldots, \p^n)\bigr) 
= 
I(\vc{w}) +
\sum_{i = 1}^n w_i^q I(\p^i);
\]

\item
\lbl{part:cf-l}
for all $n \geq 1$, $\vc{w} \in \Delta_n$, and $p \in [0, 1]$,
\[
I\bigl(w_1 p, w_1(1 - p), w_2, \ldots, w_n)
=
I(\vc{w}) + w_1^q I(p, 1 - p);
\]

\item
\lbl{part:cf-int}
for all $n, k \geq 1$, $\vc{w} \in \Delta_n$, and $\vc{p} \in \Delta_k$,
\[
I(w_1 p_1, \ldots, w_1 p_k, w_2, \ldots, w_n)
=
I(\vc{w}) + w_1^q I(\vc{p});
\]

\item
\lbl{part:cf-twotop}
for all $k, \ell \geq 1$, $\p \in \Delta_k$, $\vc{r} \in \Delta_\ell$, and
$w \in [0, 1]$, 
\begin{multline*}
I(wp_1, \ldots, wp_k, (1 - w)r_1, \ldots, (1 - w)r_\ell)        \\
=
I(w, 1 - w) + w^q I(\p) + (1 - w)^q I(\vc{r}).
\end{multline*}
\end{enumerate}
\end{lemma}
Much of the following argument goes back to Feinstein
(\cite{Fein}, p.~5--6).%
\index{Feinstein, Amiel}

\begin{figure}
\lengths
\begin{picture}(60,44)(30,-1)
\cell{60}{23}{c}{\includegraphics[height=40\unitlength]{composite2}}
\cell{60}{25}{c}{$\cdots\cdots$}
\cell{60}{10}{c}{$\cdots$}
\cell{41}{5}{c}{$\cdots$}
\cell{79}{5}{c}{$\cdots$}
\cell{42}{28}{c}{$w_1$}
\cell{78}{28}{c}{$w_n$}
\cell{33}{9}{c}{$p^1_1$}
\cell{51}{9}{c}{$p^1_{k_1}$}
\cell{71}{9}{c}{$p^n_1$}
\cell{88}{9}{c}{$p^n_{k_n}$}
\cell{30}{23}{l}{\bref{part:cf-full}}
\end{picture}%
\begin{picture}(60,44)(22,-1)
\cell{57.5}{23}{c}{\includegraphics[height=40\unitlength]{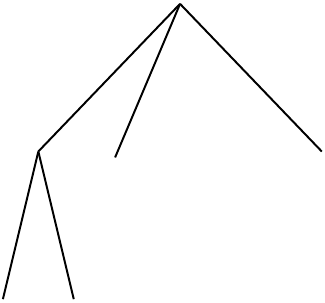}}
\cell{65}{25}{c}{$\cdots\cdots$}
\cell{42}{28}{c}{$w_1$}
\cell{51}{28}{c}{$w_2$}
\cell{78}{28}{c}{$w_n$}
\cell{36}{9}{r}{$p$}
\cell{46}{9}{l}{$1 - p$}
\cell{30}{23}{l}{\bref{part:cf-l}}
\end{picture}
\begin{picture}(60,40)(30,3)
\cell{56.2}{23}{c}{\includegraphics[height=40\unitlength]{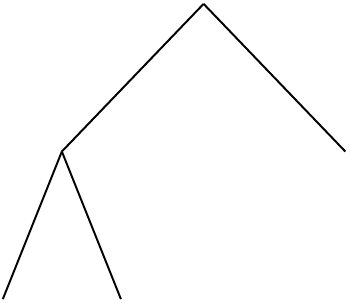}}
\cell{60}{25}{c}{$\cdots\cdots$}
\cell{41}{5}{c}{$\cdots$}
\cell{42}{28}{c}{$w_1$}
\cell{78}{28}{c}{$w_n$}
\cell{33}{9}{c}{$p_1$}
\cell{51}{9}{c}{$p_k$}
\cell{30}{23}{l}{\bref{part:cf-int}}
\end{picture}%
\begin{picture}(60,40)(22,3)
\cell{60}{23}{c}{\includegraphics[height=40\unitlength]{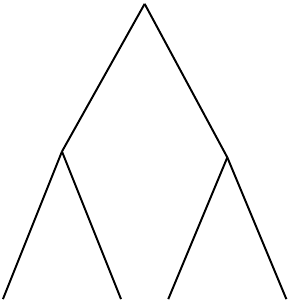}}
\cell{48}{5}{c}{$\cdots$}
\cell{72}{5}{c}{$\cdots$}
\cell{50}{28}{r}{$w$}
\cell{70}{28}{l}{$1 - w$}
\cell{41}{9}{c}{$p_1$}
\cell{57.5}{9}{c}{$p_k$}
\cell{63.5}{9}{c}{$r_1$}
\cell{79.5}{9}{c}{$r_\ell$}
\cell{30}{23}{l}{\bref{part:cf-twotop}}
\end{picture}%
\caption{Shapes of composites used in the four parts of
  Lemma~\ref{lemma:ch-forms}.}
\lbl{fig:comp-trees}
\end{figure}

\begin{proof}
%

Trivially, \bref{part:cf-full} implies~\bref{part:cf-l}.

Assuming~\bref{part:cf-l}, we prove~\bref{part:cf-int} by induction on
$k$.  The case $k = 1$ reduces to the statement that $I(\vc{u}_1) = 0$,
which follows by taking $n = 1$ in~\bref{part:cf-l}.  Now let $k \geq 2$,
and assume the result for $k - 1$.

Let $n \geq 1$, $\vc{w} \in \Delta_n$, and $\vc{p} \in \Delta_k$.  By
symmetry, we can assume that $p_k < 1$.  Using the inductive hypothesis, we
have
\begin{align*}
&
I(w_1 p_1, \ldots, w_1 p_k, w_2, \ldots, w_n)   \\
&
=
I\biggl( 
w_1(1 - p_k) \cdot \frac{p_1}{1 - p_k}, \ldots,
w_1(1 - p_k) \cdot \frac{p_{k - 1}}{1 - p_k}, 
w_1 p_k, w_2, \ldots, w_n
\biggr) \\
&
=
I\bigl( w_1(1 - p_k), w_1 p_k, w_2, \ldots, w_n\bigr)
+
\bigl( w_1 (1 - p_k) \bigr)^q
I\biggl( \frac{p_1}{1 - p_k}, \ldots, \frac{p_{k - 1}}{1 - p_k} \biggr),
\end{align*}
which by~\bref{part:cf-l} is equal to
\[
I(\vc{w}) 
+ 
w_1^q \biggl\{
I(1 - p_k, p_k) 
+
(1 - p_k)^q
I\biggl( \frac{p_1}{1 - p_k}, \ldots, \frac{p_{k - 1}}{1 - p_k} \biggr)
\biggr\}.
\]
But by the inductive hypothesis again, the term $\{\cdots\}$ is equal to
\[
I\biggl(
(1 - p_k) \cdot \frac{p_1}{1 - p_k}, \ldots, 
(1 - p_k) \cdot \frac{p_{k - 1}}{1 - p_k},
p_k
\biggr)
=
I(\p),
\]
completing the induction.

Now assuming~\bref{part:cf-int}, we prove~\bref{part:cf-twotop}.  Let $k,
\ell \geq 1$, $\p \in \Delta_k$, $\vc{r} \in \Delta_\ell$, and $w \in [0,
  1]$.  Using~\bref{part:cf-int}, we have
\begin{multline*}
I(wp_1, \ldots, wp_k, (1 - w)r_1, \ldots, (1 - w)r_\ell)        \\
=
I\bigl(w, (1 - w)r_1, \ldots, (1 - w)r_\ell\bigr)
+
w^q I(\p).
\end{multline*}
By symmetry and~\bref{part:cf-int} again, this in turn is equal to 
\[
I(w, 1 - w) + (1 - w)^q I(\vc{r}) + w^q I(\vc{p}),
\]
proving~\bref{part:cf-twotop}.

Finally, assume~\bref{part:cf-twotop}. We prove~\bref{part:cf-full} by
induction on $n$.  The case $n = 1$ just states that $I(\vc{u}_1) = 0$,
which follows from~\bref{part:cf-twotop} by taking $k = \ell = 1$.  Now let
$n \geq 2$, and assume the result for $n - 1$.

Let $k_1, \ldots, k_n \geq 1$, $\vc{w} \in \Delta_n$, and $\p^i \in
\Delta_{k_i}$.  By symmetry, we can assume that $w_1 > 0$.  Write
\[
\vc{p}^{12} 
=
\biggl( 
\frac{w_1}{w_1 + w_2} p^1_1, \ldots, \frac{w_1}{w_1 + w_2} p^1_{k_1},
\frac{w_2}{w_1 + w_2} p^2_1, \ldots, \frac{w_2}{w_1 + w_2} p^2_{k_2}
\biggr)
\in 
\Delta_{k_1 + k_2}.
\]
Then
\[
\vc{w} \of (\p^1, \ldots, \p^n)
=
(w_1 + w_2, w_3, \ldots, w_n) \of (\p^{12}, \p^3, \ldots, \p^n),
\]
so by inductive hypothesis,
\begin{multline}
\lbl{eq:41-mid}
I\bigl( \vc{w} \of (\p^1, \ldots, \p^n) \bigr) \\
=
I(w_1 + w_2, w_3, \ldots, w_n) + (w_1 + w_2)^q I(\p^{12}) 
+ \sum_{i = 3}^n w_i^q I(\p^i).
\end{multline}
On the other hand, by~\bref{part:cf-twotop},
\[
I(\p^{12})
=
I\biggl( \frac{w_1}{w_1 + w_2}, \frac{w_2}{w_1 + w_2} \biggr)
+
\biggl( \frac{w_1}{w_1 + w_2} \biggr)^q I(\p^1)
+
\biggl( \frac{w_2}{w_1 + w_2} \biggr)^q I(\p^2).
\]
Substituting this into~\eqref{eq:41-mid}, we deduce that $I\bigl( \vc{w}
\of (\p^1, \ldots, \p^n) \bigr)$ is equal to
\begin{equation}
\lbl{eq:cf-last}
I(w_1 + w_2, w_3, \ldots, w_n) 
+ 
(w_1 + w_2)^q I\biggl( \frac{w_1}{w_1 + w_2}, \frac{w_2}{w_1 + w_2} \biggr)
+
\sum_{i = 1}^n w_i^q I(\p^i).
\end{equation}
But applying the inductive hypothesis to the composite
\[
\vc{w}
=
(w_1 + w_2, w_3, \ldots, w_n)
\of
\biggl( \biggl(  \frac{w_1}{w_1 + w_2}, \frac{w_2}{w_1 + w_2} \biggr),
\vc{u}_1, \ldots, \vc{u}_1 \biggr)
\]
gives
\[
I(\vc{w})
=
I(w_1 + w_2, w_3, \ldots, w_n)
+ 
(w_1 + w_2)^q I\biggl( \frac{w_1}{w_1 + w_2}, \frac{w_2}{w_1 + w_2} \biggr)
\]
(recalling that $I(\vc{u}_1) = 0$).  Hence the
expression~\eqref{eq:cf-last} reduces to
\[
I(\vc{w}) + \sum_{i = 1}^n w_i^q I(\p^i),
\]
proving~\bref{part:cf-full}.
\end{proof}

\section{The expected number of species in a random sample}
\lbl{sec:hsg}
\index{expected number of species in sample}
\index{Hurlbert, Stuart!Smith--Grassle@--Smith--Grassle index}

Here we prove the result stated in Example~\ref{eg:hsg}, which expresses the
diversity index of Hurlbert, Smith and Grassle in terms of the Hill numbers
$D_q(\p)$.

Recall that we are modelling an ecological community with $n$ species via
its relative abundance distribution, and that $\hsg_m(\p)$ denotes the
expected number of different species represented in a random sample with
replacement of $m$ individuals.  The claim is that
\[
\hsg_m(\p)
=
\sum_{q = 1}^m (-1)^{q - 1} \binom{m}{q} D_q(\p)^{1 - q}.
\]

Define random variables $X_1, \ldots, X_n$ by
\[
X_i     
=
\begin{cases}
1       &\text{if species } i \text{ is present in the sample}, \\
0       &\text{otherwise}.
\end{cases}
\]
Then $\sum_{i = 1}^n X_i$ is the number of different species in the sample,
so
\begin{align*}
\hsg_m(\p)      &
=
\Ex\Biggl( \sum_{i = 1}^n X_i \Biggr)  
=
\sum_{i = 1}^n \Ex(X_i) \\
&
=
\sum_{i = 1}^n \Pr(\text{species } i \text{ is present in the sample})  \\
&
=
\sum_{i = 1}^n \bigl( 1 - (1 - p_i)^m \bigr),
\end{align*}
as Hurlbert%
\index{Hurlbert, Stuart} 
observed (equation~(14) of~\cite{Hurl}).  It follows that
\begin{align*}
\hsg_m(\p)      &
=
n - \sum_{i = 1}^n \sum_{q = 0}^m \binom{m}{q} (-p_i)^q \\
&
=
n - \sum_{q = 0}^m (-1)^q \binom{m}{q} \sum_{i = 1}^n p_i^q     \\
&
=
n - \Biggl\{ 
\binom{m}{0} n - \binom{m}{1} 1 + 
\sum_{q = 2}^m (-1)^q \binom{m}{q} D_q(\p)^{1 - q}
\Biggr\}        \\
&
=
m - \sum_{q = 2}^m (-1)^q \binom{m}{q} D_q(\p)^{1 - q}  \\
&
=
\sum_{q = 1}^m (-1)^{q - 1} \binom{m}{q} D_q(\p)^{1 - q},
\end{align*}
as claimed.

\section{The diversity profile determines the distribution}
\lbl{sec:prof}
\index{diversity profile!determines distribution}

Here we prove the result claimed in Remark~\ref{rmk:prof-perm}: two
probability distributions on the same finite set have the same diversity
profile if and only if one is a permutation of the other.  Formally:
\pagebreak

\begin{lemma}
\lbl{lemma:dp-perm}
Let $n \geq 1$ and $\p, \vc{r} \in \Delta_n$.  The following are
equivalent:
\begin{enumerate}
\item 
\lbl{part:dpp-full-prof}
$D_q(\p) = D_q(\vc{r})$ for all $q \in [-\infty, \infty]$;

\item
\lbl{part:dpp-part-prof}
there exists a subset $Q \sub [-\infty, \infty)$, unbounded above, such
that $D_q(\p) = D_q(\vc{r})$ for all $q \in Q$;

\item
\lbl{part:dpp-perm}
$\p = \vc{r}\sigma$ for some permutation $\sigma$ of $\{1, \ldots, n\}$.
\end{enumerate}
\end{lemma}

This result first appeared as Proposition~A22 of the appendix to Leinster
and Cobbold~\cite{MDISS}.

\begin{proof}
\bref{part:dpp-perm} implies~\bref{part:dpp-full-prof} by the symmetry of
the Hill numbers (Lemma~\ref{lemma:hill-elem}), and
\bref{part:dpp-full-prof} implies~\bref{part:dpp-part-prof} trivially.  Now
assuming~\bref{part:dpp-part-prof}, we prove~\bref{part:dpp-perm}
by induction on $n$.  It is trivial for $n = 1$.  Let $n
\geq 2$, assume the result for $n - 1$, and take $\p, \vc{r} \in \Delta_n$
such that $D_q(\p) = D_q(\vc{r})$ for all elements $q$ of some set $Q \sub
[-\infty, \infty)$ that is unbounded above.  We may assume that $-\infty
  \not\in Q$ and $1 \not\in Q$ (for if not, remove them).

We know that $D_q(\p)$ is continuous in $q \in [-\infty, \infty]$, by
Lemma~\ref{lemma:pwr-mns-cts-t} or
Lemma~\ref{lemma:dqz-cts}\bref{part:dqz-cts-q}.  Since $Q$ is unbounded
above,
\[
\lim_{q \in Q, \, q \to \infty} D_q(\p) 
= 
D_\infty(\p)
=
1\Big/\!\max_{1 \leq i \leq n} p_i.
\]
The same is true for $D_q(\vc{r})$.  Hence by assumption,
$\max_i p_i = \max_i r_i$.  Choose $k$ and $\ell$ such that $p_k = \max_i
p_i$ and $r_\ell = \max_i r_i$.  Then $p_k = r_\ell$.

If $p_k = r_\ell = 1$ then $\p$ and $\vc{r}$ are both of the form $(0,
\ldots, 0, 1, 0, \ldots, 0)$, so one is a permutation of the other.
Assuming otherwise, define $\p', \vc{r}' \in \Delta_{n - 1}$ by
\[
\p' 
= 
\biggl( 
\frac{p_1}{1 - p_k}, \ldots, \frac{p_{k - 1}}{1 - p_k},
\frac{p_{k + 1}}{1 - p_k}, \ldots, \frac{p_n}{1 - p_k} \biggr)
\]
and similarly for $\vc{r}'$.  Then for all $q \in Q$,
\begin{align*}
D_q(\p')        &
=
(1 - p_k)^{q/(q - 1)} 
\Biggl( \sum_{i \neq k} p_i^q \Biggr)^{1/(1 - q)}  \\
&
=
(1 - p_k)^{q/(q - 1)} 
\bigl( D_q(\p)^{1 - q} - p_k^q \bigr)^{1/(1 - q)}.
\end{align*}
Similarly,
\[
D_q(\vc{r}') 
=
(1 - r_\ell)^{q/(q - 1)} 
\bigl( D_q(\vc{r})^{1 - q} - r_\ell^q \bigr)^{1/(1 - q)}.
\]
But $p_k = r_\ell$ and $D_q(\p) = D_q(\vc{r})$, so
$D_q(\p') = D_q(\vc{r}')$.  This holds for all $q \in Q$, so by inductive
hypothesis, $\p'$ is a permutation of $\vc{r}'$.  It follows that $\p$ is a
permutation of $\vc{r}$, completing the induction.
\end{proof}

\section{Affine functions}
\lbl{sec:aff}
\index{affine function}

Here we prove Lemma~\ref{lemma:aff}, which is restated here for convenience.

\begin{lemmaaff}
Let $\alpha \from I \to J$ be a function between real intervals.  The
following are equivalent:
\begin{enumerate}
\item
$\alpha$ is affine;

\item
$\alpha\bigl( \sum \lambda_i x_i \bigr) = \sum \lambda_i \alpha(x_i)$ for
all $n \geq 1$, $x_1, \ldots, x_n \in I$ and $\lambda_1, \ldots,
\lambda_n \in \R$ such that $\sum \lambda_i = 1$ and $\sum \lambda_i x_i
\in I$;

\item 
there exist constants $a, b \in \R$ such that $\alpha(x) = ax + b$ for all
$x \in I$; 

\item
$\alpha$ is continuous and $\alpha\bigl(\tfrac{1}{2}(x_1 + x_2)\bigr) =
\tfrac{1}{2}\bigl(\alpha(x_1) + \alpha(x_2)\bigr)$ for all $x_1, x_2 \in I$.
\end{enumerate}
\end{lemmaaff}

\begin{proof}
First we assume~\bref{part:aff-cvx} and prove~\bref{part:aff-aff}.  By
induction,
\begin{equation}
\lbl{eq:aff-cvx-gen}
\alpha \Biggl( \sum_{i = 1}^n p_i x_i \Biggr)
=
\sum_{i = 1}^n p_i \alpha(x_i)
\end{equation}
for all $n \geq 1$, $\vc{p} \in \Delta_n$, and $\vc{x} \in I^n$.  Now let
$n \geq 1$, $x_1, \ldots, x_n \in I$ and $\lambda_1, \ldots, \lambda_n \in
\R$ with $\sum \lambda_i = 1$ and $\sum \lambda_i x_i \in I$.  Assume
without loss of generality that
\[
\lambda_1, \ldots, \lambda_k \geq 0,
\quad
\lambda_{k + 1}, \ldots, \lambda_n < 0
\]
for some $k \in \{1, \ldots, n\}$.  Write
\[
\mu 
=
\sum_{i = 1}^k \lambda_i
=
1 - \sum_{i = k + 1}^n \lambda_i
\geq 
1,
\qquad
w = \sum_{i = 1}^n \lambda_i x_i \in I.
\]
Then
\[
\sum_{i = 1}^k \frac{\lambda_i}{\mu} x_i
=
\frac{1}{\mu} w 
+ 
\sum_{i = k + 1}^n \frac{-\lambda_i}{\mu} x_i.
\]
The coefficients $\lambda_1/\mu, \ldots, \lambda_k/\mu$ on the left-hand
side are nonnegative and sum to $1$, and the same is true of the
coefficients $1/\mu, -\lambda_{k + 1}/\mu, \ldots, -\lambda_n/\mu$ on the
right-hand side.  Hence we can apply $\alpha$ throughout and use
equation~\eqref{eq:aff-cvx-gen} on both sides, giving
\[
\sum_{i = 1}^k \frac{\lambda_i}{\mu} \alpha(x_i)
=
\frac{1}{\mu} \alpha(w) 
+ 
\sum_{i = k + 1}^n \frac{-\lambda_i}{\mu} \alpha(x_i).
\]
Rearranging gives
\[
\alpha(w) 
=
\sum_{i = 1}^n \lambda_i \alpha(x_i),
\]
proving~\bref{part:aff-aff}.

Next we assume~\bref{part:aff-aff} and prove~\bref{part:aff-explicit}.  If
$I$ is trivial, the result is trivial.  Otherwise, we can choose distinct
$x_1, x_2 \in I$.  Put
\[
a = \frac{\alpha(x_2) - \alpha(x_1)}{x_2 - x_1},
\qquad
b = \frac{\alpha(x_1) x_2 - \alpha(x_2) x_1}{x_2 - x_1},
\]
and define $\alpha' \from \R \to \R$ by $\alpha'(x) = ax + b$.  We show
that $\alpha(x) = \alpha'(x)$ for all $x \in I$.  First, this is true when
$x \in \{x_1, x_2\}$, by direct calculation.  Second, every element of $I$
can be written as $\lambda_1 x_1 + \lambda_2 x_2$ for some $\lambda_1,
\lambda_2 \in \R$ with $\lambda_1 + \lambda_2 = 1$.  Since both $\alpha$
and $\alpha'$ satisfy~\bref{part:aff-aff}, the result follows.

Trivially, \bref{part:aff-explicit} implies~\bref{part:aff-cts}.

Finally, assuming~\bref{part:aff-cts}, we prove~\bref{part:aff-cvx}.  By
continuity, it is enough to prove that
\[
\alpha\bigl( px_1 + (1 - p)x_2 \bigr)
=
p\alpha(x_1) + (1 - p)\alpha(x_2)
\]
whenever $x_1, x_2 \in I$ and $p \in [0, 1]$ is a dyadic rational, that is,
$p = m/2^n$ for some integers $n \geq 0$ and $0 \leq m \leq 2^n$.  We do
this by induction on $n$.  It is trivial for $n = 0$.  Now let $n \geq 1$
and assume the result for $n - 1$.  Let $x_1, x_2 \in I$, let $0 \leq m
\leq 2^n$, and assume without loss of generality that $m \leq 2^{n - 1}$
(otherwise we can reverse the roles of $x_1$ and $x_2$).  Then
\begin{align}
\alpha \Biggl(
\frac{m}{2^n} x_1 + \biggl( 1 - \frac{m}{2^n} \biggr) x_2
\Biggr) &
=
\alpha \Biggl(
\frac{m}{2^{n - 1}} \cdot \frac{1}{2}(x_1 + x_2)
+
\biggl( 1 - \frac{m}{2^{n - 1}} \biggr) x_2
\Biggr) 
\lbl{eq:aff-big-1}    \\
&
=
\frac{m}{2^{n - 1}} \alpha\biggl( \frac{1}{2} (x_1 + x_2) \biggr)
+
\biggl( 1 - \frac{m}{2^{n - 1}} \biggr) \alpha(x_2)
\lbl{eq:aff-big-2}    \\
&
=
\frac{m}{2^{n - 1}} \cdot \frac{1}{2} 
\bigl( \alpha(x_1) + \alpha(x_2) \bigr) 
+
\biggl( 1 - \frac{m}{2^{n - 1}} \biggr) \alpha(x_2)       
\lbl{eq:aff-big-3}    \\
&
=
\frac{m}{2^n} \alpha(x_1) + \biggl( 1 - \frac{m}{2^n} \biggr) \alpha(x_2),
\lbl{eq:aff-big-4}
\end{align}
where~\eqref{eq:aff-big-1} and~\eqref{eq:aff-big-4} are elementary,
\eqref{eq:aff-big-2} is by inductive hypothesis, and~\eqref{eq:aff-big-3}
is by~\eqref{part:aff-cts}.  This completes the induction
and, therefore, the proof.
\end{proof}

\section{Diversity of integer orders}
\lbl{sec:q-int}
\index{Hill number!integer order@of integer order}
\index{diversity!integer order@of integer order}
\index{order!integer}

Here we prove the statement made in Example~\ref{eg:dqz-integers}
on computation of the diversity $D_q^Z(\p)$ for integers $q \geq 2$: that
in the notation defined there,
\[
D_q^Z(\p) = \mu_q^{1/(1 - q)}.
\]
Indeed, adopting the convention that all sums run over $1, \ldots, n$, 
\begin{align*}
D_q^Z(\p)^{1 - q}       &
=
\sum_i p_i 
\Biggl( \sum_j Z_{ij} p_j \Biggr)^{q - 1}     \\
&=
\sum_{i, j_1, \ldots, j_{q - 1}}
p_i Z_{i j_1} p_{j_1} Z_{i j_2} p_{j_2} 
\cdots 
Z_{i j_{q - 1}} p_{j_{q - 1}}  \\
&= 
\sum_{i_1, i_2, \ldots, i_q}
p_{i_1} p_{i_2} \cdots p_{i_q} 
Z_{i_1 i_2} Z_{i_1  i_3} \cdots Z_{i_1  i_q}   \\
&
=
\mu_q,
\end{align*}
as required.

\section{The maximum entropy of a coupling}
\lbl{sec:max-cpl}
\index{coupling!maximum entropy of}
\index{maximum entropy!coupling@of coupling}

Let $\vc{p}$ and $\vc{r}$ be probability distributions on finite sets $\XX$
and $\YY$, respectively.  We showed in Remark~\ref{rmk:ub-coupling} that
among all distributions on $\XX \times \YY$ with marginals $\vc{p}$ and
$\vc{r}$, none has greater entropy than $\p \otimes \vc{r}$.  In other
words,
\begin{equation}
\lbl{eq:cpl-H}
H(P) \leq H(\vc{p} \otimes \vc{r})
\end{equation}
for all probability distributions $P$ on $\XX \times \YY$ whose marginal
distributions are $\p$ and $\vc{r}$.  It was also claimed there that unless
$q = 0$ or $q = 1$, the inequality~\eqref{eq:cpl-H} fails when $H$ is
replaced by the R\'enyi entropy $H_q$ or the $q$-logarithmic entropy $S_q$.
Here we prove this claim.

Since $H_q$ and $S_q$ are increasing, invertible transformations of one
another, it suffices to prove it for $H_q$.  And since R\'enyi entropy is
logarithmic (equation~\eqref{eq:ren-log}), the inequality in question can
be restated as
\begin{equation}
\lbl{eq:cpl-q-sum}
H_q(P) \leq H_q(\vc{p}) + H_q(\vc{r}).
\end{equation}
This is true for $q = 0$: 
\[
\supp(P) \sub \supp(\p) \times \supp(\vc{r}),
\]
so
\[
\mg{\supp(P)} \leq \mg{\supp(\p)} \cdot \mg{\supp(\vc{r})},
\]
giving
\[
H_0(P)
=
\log\mg{\supp(P)}
\leq 
\log\mg{\supp(\p)} + \log\mg{\supp(\vc{r})}
=
H_0(\vc{p}) + H_0(\vc{r}).
\]
Our task now is to show that except in the cases $q = 0$ and $q =
1$, the inequality~\eqref{eq:cpl-q-sum} is false.
Thus, we prove that for each $q \in (0, 1) \cup (1, \infty]$, there
exist finite sets $\XX$ and $\YY$ and a probability distribution $P$ on
$\XX \times \YY$ such that
\[
H_q(P) > H_q(\vc{p}) + H_q(\vc{r}),
\]
where $\vc{p}$ and $\vc{r}$ are the marginal distributions of $P$.  

We will treat separately the cases $q \in (0, 1)$, \,$q \in (1, \infty)$,
and $q = \infty$.  In all cases, we will take $\XX = \YY = \{1, \ldots,
N\}$ for some $N$.  A probability distribution $P$ on $\XX \times \YY$ is
then an $N \times N$ matrix of nonnegative real numbers whose entries sum 
to $1$, and its marginals $\p$ and $\vc{r}$ are given by the row-sums and
column-sums: 
\[
p_i = \sum_{j = 1}^N P_{ij},
\qquad
r_j = \sum_{i = 1}^N P_{ij}
\]
($i, j \in \{1, \ldots, N\}$).  

First let $q \in (0, 1)$.  For each $N \geq 2$, define an $N \times N$
matrix $P$ by
\[
P
=
\begin{pmatrix}
1 - (N - 1)^{(q - 1)/q} &
0                       &\cdots &0      \\
0                       &
(N - 1)^{(-q-1)/q}      &\cdots &(N - 1)^{(-q-1)/q}      \\
\vdots                  &
\vdots                  &       &\vdots \\
0                       &
(N - 1)^{(-q-1)/q}      &\cdots &(N - 1)^{(-q-1)/q}      
\end{pmatrix}.
\]
The entries of $P$ sum to $1$, and $1 - (N - 1)^{(q - 1)/q} \geq 0$ since
$q \in (0, 1)$, so $P \in \Delta_{N^2}$.  We have
\begin{align*}
H_q(P)  &
=
\frac{1}{1 - q}
\log
\Bigl(
\bigl( 1 - (N - 1)^{(q - 1)/q} \bigr)^q
+
(N - 1)^2 (N - 1)^{-q - 1}
\Bigr)  \\
&
\geq
\frac{1}{1 - q}
\log
\bigl(
(N - 1)^{-q + 1}
\bigr)  \\
&
=
\log(N - 1).
\end{align*}
The marginals of $P$ are
\[
\vc{p} = \vc{r} = 
\Bigl( 
1 - (N - 1)^{(q - 1)/q}, 
\underbrace{(N - 1)^{-1/q}, \ldots, (N - 1)^{-1/q}}_{N - 1}
\Bigr),
\]
so
\begin{align*}
H_q(\p) = H_q(\vc{r}) &
=
\frac{1}{1 - q} 
\log
\Bigl( 
\bigl( 1 - (N - 1)^{(q - 1)/q} \bigr)^q
+
(N - 1) \cdot (N - 1)^{-1}
\Bigr)  \\
&
<
\frac{1}{1 - q} \log 2.
\end{align*}
Hence
\begin{align*}
H_q(P) - \bigl( H_q(\vc{p}) + H_q(\vc{r}) \bigr)        &
>
\log(N - 1) - \frac{2}{1 - q} \log 2    
\to 
\infty
\end{align*}
as $N \to \infty$.  In particular, $H_q(P) > H_q(\vc{p}) + H_q(\vc{r})$
when $N$ is sufficiently large.

Now let $q \in (1, \infty)$.  For each $N \geq 2$, define an $N \times N$
matrix $P$ by
\[
P
=
\begin{pmatrix}
0               &1/2(N - 1)     &\cdots &1/2(N - 1)     \\
1/2(N - 1)      &0              &\cdots &0              \\
\vdots          &\vdots         &       &\vdots         \\
1/2(N - 1)      &0              &\cdots &0              
\end{pmatrix}.
\]
The entries of $P$ are nonnegative and sum to $1$, and 
\[
H_q(P)
=
H_q(\vc{u}_{2(N - 1)})
=
\log\bigl(2(N - 1)\bigr)
\to 
\infty
\]
as $N \to \infty$.  The marginals of $P$ are
\[
\vc{p} = \vc{r} =
\bigl( 1/2, 
\underbrace{1/2(N - 1), \ldots, 1/2(N - 1)}_{N - 1}
\bigr),
\]
and
\begin{align*}
H_q(\vc{p}) = H_q(\vc{r}) 
&
= 
\frac{1}{1 - q}
\log
\Bigl(
(1/2)^q + (N - 1) \cdot \bigl(1/2(N - 1)\bigr)^q 
\Bigr)  \\
&
=
\frac{1}{1 - q} \log\bigl( (1/2)^q \bigr)
+
\frac{1}{1 - q} \log\bigl( 1 + (N - 1)^{1 - q} \bigr)   \\
&
\to 
\frac{1}{1 - q} \log\bigl( (1/2)^q \bigr)
\end{align*}
as $N \to \infty$, since $q > 1$.  Hence
\[
H_q(P) - \bigl( H_q(\vc{p}) + H_q(\vc{r}) \bigr)
\to 
\infty
\]
as $N \to \infty$, which again implies that $H_q(P) > H_q(\vc{p}) +
H_q(\vc{r})$ when $N$ is sufficiently large.

Finally, let $q = \infty$.  The same matrix $P$ as in the previous case has
\begin{align*}
H_\infty(P)     &
=
\log\bigl(2(N - 1)\bigr),       \\
H_\infty(\p) = H_\infty(\vc{r}) &
=
\log 2.
\end{align*}
Hence
\[
H_\infty(P) - \bigl( H_\infty(\vc{p}) + H_\infty(\vc{r}) \bigr) 
=
\log\bigl(2(N - 1)\bigr) - 2\log 2      
\to 
\infty
\]
as $N \to \infty$.  Once again, this implies that $H_\infty(P) >
H_\infty(\vc{p}) + H_\infty(\vc{r})$ for sufficiently large $N$.

\section{Convex duality}
\lbl{sec:cvx-du}
\index{convex!conjugate}
\index{convex!duality}

Here we prove Theorem~\ref{thm:lf}, which is restated here for convenience.

\begin{thmlf}[Legendre--Fenchel]%
\index{Legendre--Fenchel transform}%
\index{Fenchel, Werner}
Let $f \from \R \to \R$ be a convex function.  Then $f^{**} = f$.
\end{thmlf}

\begin{proof}
Let $x \in \R$.  By definition of convex conjugate,
\begin{align}
f^{**}(x)       &
=
\sup_{\lambda \in \R}\bigl(\lambda x - f^*(\lambda)\bigr) 
\nonumber       \\
&
=
\sup_{\lambda \in \R} \inf_{y \in \R} 
\bigl(\lambda(x - y) + f(y)\bigr).
\lbl{eq:fm-1}
\end{align}
In particular,
\[
f^{**}(x)
\leq
\sup_{\lambda \in \R} \bigl(
\lambda(x - x) + f(x)
\bigr)
=
f(x),
\]
so it remains to prove that $f^{**}(x) \geq f(x)$.  In fact, we will show
that there exists $\lambda \in \R$ such that
\begin{equation}
\lbl{eq:fm-2}
\lambda(x - y) + f(y) \geq f(x) 
\ \text{ for all } y \in \R.
\end{equation}
By~\eqref{eq:fm-1}, this will suffice.  Now, a real number $\lambda$
satisfies~\eqref{eq:fm-2} if and only if
\[
\sup_{y \in (-\infty, x)} \frac{f(x) - f(y)}{x - y}
\leq
\lambda
\leq
\inf_{z \in (x, \infty)} \frac{f(z) - f(x)}{z - x},
\]
so such a $\lambda$ exists if and only if
\begin{equation}
\lbl{eq:fm-3}
\frac{f(x) - f(y)}{x - y}
\leq
\frac{f(z) - f(x)}{z - x}
\end{equation}
for all $y < x < z$.  We now prove this.  Take $y$ and $z$ such that $y < x
< z$.  Then $x = py + (1 - p)z$ for some $p \in (0, 1)$, and the
inequality~\eqref{eq:fm-3} to be proved states that
\[
\frac{f(x) - f(y)}{(1 - p)(z - y)}
\leq
\frac{f(z) - f(x)}{p(z - y)},
\]
or equivalently,
\[
f(x) \leq pf(y) + (1 - p)f(z).
\]
This is true by convexity of $f$.
\end{proof}

\section{Cumulant generating functions are convex}
\lbl{sec:mgfs-log-cvx}
\index{cumulant generating function}

In Section~\ref{sec:large}, we used the fact that the cumulant generating
function of any real random variable is convex.  Here we prove this.  

If we are willing to assume that the cumulant generating function is twice
differentiable, then the result can be deduced from the Cauchy--Schwarz
inequality, as in Section~5.11 of Grimmett and Stirzaker~\cite{GrSt}.  But
there is no need to make this assumption.  Instead, we use a more general
standard inequality:

\begin{thm}[H\"older's inequality]%
\index{Holder's inequality@H\"older's inequality}
Let $\Omega$ be a measure space, let $p, q \in (1, \infty)$ with $1/p + 1/q
= 1$, and let $f, g \from \Omega \to [0, \infty)$ be measurable functions.  Then
\[
\int_\Omega fg
\leq
\Biggl( \int_\Omega f^p \Biggr)^{1/p}
\Biggl( \int_\Omega g^q \Biggr)^{1/q}.
\] 
\end{thm}

Here we allow the possibility that one or more of the integrals is
$\infty$.

\begin{proof}
This is Theorem~6.2 of Folland~\cite{FollRA}, for instance.
\end{proof}

\begin{cor}
Let $X$ be a real random variable.  Then the function
\[
\begin{array}{ccc}
\R      &\to            &[0, \infty]    \\
\lambda &\mapsto        &\log\Ex(e^{\lambda X})
\end{array}
\]
is convex.
\end{cor}

\begin{proof}
We have to prove that for all $\lambda, \mu \in \R$ and $t \in [0, 1]$,
\[
\log\Ex\Bigl(e^{(t\lambda + (1 - t)\mu)X}\Bigr)
\leq
t\log\Ex\bigl(e^{\lambda X}\bigr) + (1 - t)\log\Ex\bigl(e^{\,\mu X}\bigr),
\]
or equivalently,
\[
\Ex\Bigl(e^{t\lambda X} e^{(1 - t)\mu X}\Bigr)
\leq
\Ex\bigl(e^{\lambda X}\bigr)^t \, \Ex\bigl(e^{\,\mu X}\bigr)^{1 - t}.
\]
This is trivial if $t = 0$ or $t = 1$.  Supposing otherwise, write $p =
1/t$, $q = 1/(1 - t)$, $U = e^{t \lambda X}$, and $V = e^{(1 - t)\mu X}$.
Thus, $p, q \in (1, \infty)$ with $1/p + 1/q = 1$, and $U$ and $V$ are
nonnegative real random variables on the same sample space.  The
inequality to be proved is that
\[
\Ex(UV) \leq \Ex(U^p)^{1/p} \, \Ex(V^q)^{1/q},
\]
which is just H\"older's inequality in probabilistic notation.
\end{proof}

\section{Functions on a finite field}
\lbl{sec:fff}
\index{finite field, functions on}
\index{polynomial!finite field@over finite field}

Here we prove Lemma~\ref{lemma:fn-fin-fld}, which is restated here for
convenience.

\begin{lemmafff}
Let $\fld$ be a finite field with $q$ elements, let $n \geq 0$, and let $F
\from \fld^n \to \fld$ be a function.  Then there is a unique polynomial
$f$ of the form
\begin{equation}
\lbl{eq:poly-small-deg}
f(x_1, \ldots, x_n)
=
\sum_{0 \leq r_1, \ldots, r_n < q} c_{r_1, \ldots, r_n} 
x_1^{r_1} \cdots x_n^{r_n}
\end{equation}
($c_{r_1, \ldots, r_n} \in \fld$) such that
\[
f(\pi_1, \ldots, \pi_n) = F(\pi_1, \ldots, \pi_n)
\]
for all $\pi_1, \ldots, \pi_n \in \fld$.
\end{lemmafff}

This result is standard.  For instance, Section~10.3 of Roman~\cite{Roma}
gives a proof in the case $n = 1$.

\begin{proof}
Write $\fld^{< q}[x_1, \ldots, x_n]$ for the set of polynomials of degree
less than $q$ in each variable, that is, of the
form~\eqref{eq:poly-small-deg}.  Write $R(f) \from \fld^n \to \fld$ for the
function induced by a polynomial $f$ in $n$ variables.  Then $R$ defines a
map
\[
R \from \fld^{< q}[x_1, \ldots, x_n] \to 
\{\text{functions } \fld^n \to \fld\}.
\]
We have to prove that $R$ is bijective.
Both domain and codomain have $q^{q^n}$ elements, so it suffices to prove
that $R$ is surjective.

First define a polynomial $\delta$ by 
\[
\delta(x_1, \ldots, x_n) = (1 - x_1^{q - 1}) \cdots (1 - x_n^{q - 1}).
\]
Then $\delta$ has degree $q - 1$ in each variable, and for $a_1, \ldots,
a_n \in K$,
\[
R(\delta)(a_1, \ldots, a_n) 
=
\begin{cases}
1       &\text{if } a_1 = \cdots = a_n = 0,     \\
0       &\text{otherwise}.
\end{cases}
\]
Now, given a function $F \from K^n \to K$, define a polynomial $f$ by
\[
f(x_1, \ldots, x_n)
=
\sum_{a_1, \ldots, a_n \in K} 
F(a_1, \ldots, a_n) \delta(x_1 - a_1, \ldots, x_n - a_n).
\]
Then $f$ has degree at most $q - 1$ in each variable and $R(f) = F$, as
required.
\end{proof}

There are other proofs.  For instance, one can prove that $R$ is injective
rather than surjective, showing by induction on $n$ that its kernel is
trivial.  I thank Todd Trimble%
\index{Trimble, Todd} 
for pointing out to me the Lagrange interpolation argument above.

%% file: condns.tex
\chapter{Summary of conditions}
\lbl{app:condns}

Here we list the main conditions on means, diversity measures and value
measures used in the text.  For each condition, we give an abbreviated form
of the definition and a reference to the point(s) in the text where
it is defined in full.

\subsection*{Weighted means}
\index{mean!conditions on}

The following conditions apply to a sequence 
$
\bigl(M \from \Delta_n \times I^n \to I\bigr)_{n \geq 1}
$
of functions,
where $I$ is a real interval.  For the homogeneity and multiplicativity
conditions, $I$ is assumed to be closed under multiplication.

\begin{center}
\begin{tabular}{l@{\hspace{2.9mm}}l@{\hspace{2.9mm}}l}
\hline
Name    &Abbreviated definition      &Reference           \\
\hline  
Absence-        &
$M((\ldots, p_{i - 1}, 0, p_{i + 1}), 
(\ldots, x_{i - 1}, x_i, x_{i + 1}, \ldots))$   &
\\
invariant       &
$
=
M((\ldots, p_{i - 1}, p_{i + 1}, \ldots), 
(\ldots, x_{i - 1}, x_{i + 1}, \ldots))$        
&
Def~\ref{defn:pwr-mn-elem}       \\[\tsk]
Chain rule      &
$M(\vc{w} \of (\p^1, \ldots, \p^n), 
\vc{x}^1 \oplus\cdots\oplus \vc{x}^n)$     
&
\\
&
$= M(\vc{w},
( M(\p^1, \vc{x}^1), \ldots, M(\p^n, \vc{x}^n) )
)$
&
Def~\ref{defn:mns-chn}   \\[\tsk]
Consistent      &
$M(\p, (x, \ldots, x)) = x$     &
Def~\ref{defn:w-cons}    \\[\tsk]
Convex          &
$M(\vc{p}, \hlf(\vc{x} + \vc{y}))
\leq
\max\{ M(\vc{p}, \vc{x}), M(\vc{p}, \vc{y})\}$
&
Def~\ref{defn:mean-cvx}  \\[\tsk]
Homogeneous     &
$M(\p, c\vc{x}) = cM(\p, \vc{x})$       &
Def~\ref{defn:w-mn-hgs}  \\[\tsk]
Increasing      &
$\vc{x} \leq \vc{y} \implies M(\p, \vc{x}) \leq M(\p, \vc{y})$    &
Def~\ref{defn:w-isi}     \\[\tsk]
Modular         &
$M\bigl(\vc{w} \of (\p^1, \ldots, \p^n), 
\vc{x}^1 \oplus\cdots\oplus \vc{x}^n\bigr)$ 
depends  
&
\\
&
only on $\vc{w}$ and $M(\p^1, \vc{x}^1), \ldots, M(\p^n, \vc{x}^n)$
&
Def~\ref{defn:mns-mod}   \\[\tsk]
Multiplicative          &
$M(\vc{p} \otimes \vc{p}', \vc{x} \otimes \vc{x}')
=
M(\vc{p}, \vc{x}) M(\vc{p}', \vc{x}')$
&
Def~\ref{defn:w-mult}    \\[\tsk]
Natural         &
$M(f\p, \vc{x}) = M(\p, \vc{x}f)$       &
Def~\ref{defn:mn-nat}    \\[\tsk]
Quasiarithmetic &
$M(\vc{p}, \vc{x}) = \phi^{-1}\bigl(\sum p_i \phi(x_i)\bigr)$ 
for some $\phi$    
&
Def~\ref{defn:qam}      \\[\tsk]
\hline
\end{tabular}

\begin{tabular}{lll}
\hline
Repetition      &
$M((\ldots, p_i, p_{i + 1}, \ldots), 
(\ldots, x_i, x_i, \ldots))$
&
\\
&
$= M((\ldots, p_i + p_{i + 1}, \ldots), 
(\ldots, x_i, \ldots))$
&
Def~\ref{defn:pwr-mn-elem}       \\[\tsk]
Strictly increasing     &
$\vc{x} \leq \vc{y}$ and $x_i < y_i$ for some $i \in \supp(\p)$   &
\\
&
$\implies M(\p, \vc{x}) < M(\p, \vc{y})$     &
Def~\ref{defn:w-isi}     \\[\tsk]
Symmetric       & 
$M(\p, \vc{x}) = M(\p\sigma, \vc{x}\sigma)\vphantom{X^{X^{X^X}}}$   &
Def~\ref{defn:pwr-mn-elem}       \\[\tsk]
\hline
\end{tabular}
\end{center}

\subsection*{Unweighted means}
\index{mean!conditions on}

The following conditions apply to a sequence 
$
\bigl(M \from I^n \to I\bigr)_{n \geq 1}
$
of functions,
where $I$ is a real interval.  For the homogeneity and multiplicativity
conditions, $I$ is assumed to be closed under multiplication.

\begin{center}

\begin{tabular}{lll}
\hline
Name    &Abbreviated definition      &Reference           \\
\hline  
Consistent      &
$M(x, \ldots, x) = x$   &
Def~\ref{defn:u-cons}    \\[\tsk]
Decomposable    &
$M\bigl(x^1_1, \ldots, x^1_{k_1}, \ldots, x^n_1, \ldots, x^n_{k_n}\bigr)$
&
\\
&
$=
M(a_1, \ldots, a_1, \ldots, a_n, \ldots, a_n)$
&
\\
&
where $a_i = M\bigl(x^i_1, \ldots, x^i_{k_i}\bigr)$       
&
Def~\ref{defn:decomp}    \\[\tsk]
Homogeneous     &
$M(c\vc{x}) = cM(\vc{x})$       &
Def~\ref{defn:u-hgs}     \\[\tsk]
Increasing      &
$\vc{x} \leq \vc{y} \implies M(\vc{x}) \leq M(\vc{y})$  &
Def~\ref{defn:u-isi}     \\[\tsk]
Modular         &
$M\bigl(x^1_1, \ldots, x^1_{k_1}, \ldots, x^n_1, \ldots, x^n_{k_n}\bigr)$
&
\\
&
depends only on $k_1, \ldots, k_n$ and 
&
\\
&
$M\bigl(x^1_1, \ldots, x^1_{k_1}\bigr), \ldots, 
M\bigl(x^n_1, \ldots, x^n_{k_n}\bigr)$      
&
Def~\ref{defn:u-mod}     \\[\tsk]
Multiplicative  &
$M(\vc{x} \otimes \vc{y}) = M(\vc{x}) M(\vc{y})$        &
Def~\ref{defn:u-mult}    \\[\tsk]
Quasiarithmetic &
$\vphantom{X^{X^{X^X}}}$%
$M(\vc{x}) = \phi^{-1}\bigl(\sum \tfrac{1}{n} \phi(x_i)\bigr)$ 
for some $\phi$    
&
p.~\pageref{p:is-qam}        \\[\tsk]
Strictly increasing     &
$\vc{x} \leq \vc{y} \neq \vc{x} \implies M(\vc{x}) < M(\vc{y})$        &
Def~\ref{defn:u-isi}     \\[\tsk]
Symmetric       &
$M(\vc{x}) = M(\vc{x}\sigma)$   &
Def~\ref{defn:u-sym}    \\[\tsk]
\hline
\end{tabular}
\end{center}

\subsection*{Diversity measures}
\index{diversity measure!conditions on}

The following conditions apply to a sequence
$
\bigl(D \from \Delta_n \to (0, \infty) \bigr)_{n \geq 1}
$
of functions,
that is, a diversity measure for communities modelled as finite probability
distributions (without incorporating species similarity).

\begin{center}
\begin{tabular}{l@{\hspace{2.3mm}}l@{\hspace{2.3mm}}l}
\hline
Name    &Abbreviated definition      &Reference           \\
\hline  
Absence-invariant    &
$D(\ldots, p_{i - 1}, 0, p_{i + 1}, \ldots)$
&
\\
&
$
=
D(\ldots, p_{i - 1}, p_{i + 1}, \ldots)$
&
Def~\ref{defn:hill-abs}  \\[\tsk]
Continuous      &
$D \from \Delta_n \to (0, \infty)$ is continuous
&
Def~\ref{defn:hill-cts}  \\[\tsk]
Continuous in &       
\\
positive probabilities   &
$D \from \Delta_n^\circ \to (0, \infty)$ is continuous
&
Def~\ref{defn:hill-cts}  \\[\tsk]
Effective number        &
$\vphantom{X^{X^{X^X}}}$%
$D(\vc{u}_n) = n$
&
Def~\ref{defn:div-eff}          \\[\tsk]
Modular         &
$D(\vc{w} \of (\p^1, \ldots, \p^n))$ 
depends  &
p.~\pageref{p:D-mod}, \\
&
only on $\vc{w}$ and $D(\p^1), \ldots, D(\p^n)$
&
Def~\ref{defn:hill-mod}     \\[\tsk]
Modular-monotone        &
$D(\p^i) \leq D(\twid{\p}^i)$ for all $i$ $\implies$      &
\\
&
$D( \vc{w} \!\of\! (\p^1, \ldots, \p^n) ) \leq
D( \vc{w} \!\of\! (\twid{\p}^1, \ldots, \twid{\p}^n) )$
&
Def~\ref{defn:mod-mono}  \\[\tsk]
Multiplicative          &
$D(\p \otimes \vc{r}) = D(\p) D(\vc{r})$
&
Def~\ref{defn:hill-mult} \\[\tsk]
Normalized      &
$D(\vc{u}_1) = 1$
&
p.~\pageref{p:hill-norm} \\[\tsk]
Replication principle   &
$D(\vc{u}_n \otimes \p) = nD(\p)$
&
p.~\pageref{p:D-rep},   \\
&
&
Def~\ref{defn:hill-rep} \\[\tsk]
Symmetric       &
$D(\p) = D(\p\sigma)$   &
p.~\pageref{p:q-ent-sym} \\[\tsk]
\hline
\end{tabular}
\end{center}

\subsection*{Value measures}
\index{value measure!conditions on}

The following conditions apply to a sequence
$
\bigl(\sigma \from 
\Delta_n \times (0, \infty)^n \to (0, \infty)\bigr)_{n \geq 1}
$
of functions.  Such a sequence is of the same type as a weighted mean on
$(0, \infty)$, and the same terminology applies.  We also use two further
conditions.\\

\begin{center}
\begin{tabular}{lll}
\hline
Name    &Abbreviated definition      &Reference           \\
\hline  
Continuous in &
\\
positive probabilities    &
$\vphantom{X^{X^{X^X}}}$%
$\sigma(-, \vc{v}) \from \Delta_n^\circ \to (0, \infty)$ 
is continuous   &
Def~\ref{defn:val-cipp} \\[\tsk]
Effective number        &
$\sigma(\vc{u}_n, (1, \ldots, 1)) = n$
&
Def~\ref{defn:val-eff}  \\[\tsk]
\hline
\end{tabular}
\end{center}

%% file: biblio.tex
\bibliography{mathrefs}

%% file: notn.tex
\chapter*{Index of notation}
\smallish

\subsection*{Standard notation}

\setlength{\parindent}{0em}%
\ \\
\begin{tabular}{ll}
$\nat$& the set $\{0, 1, 2, \ldots\}$ of natural numbers\\
$\Z$& the set $\{ \ldots, -1, 0, 1, \ldots \}$ of integers\\
$\Q, \R, \C$& the sets of rational, real and complex numbers\\
$[a, b]$& the interval $\{ x \in \R \such a \leq x \leq b \}$\\
$[a, b)$& the interval $\{ x \in \R \such a \leq x < b \}$\\
$(a, b]$& the interval $\{ x \in \R \such a < x \leq b \}$\\
$(a, b)$& the interval $\{ x \in \R \such a < x < b \}$\\
$\lfloor x \rfloor$& the greatest integer less than or equal to $x$\\
$\lceil x \rceil$& the least integer greater than or equal to $x$\\
$f|_A$& the restriction of a function $f \from X \to Y$ to a subset $A \sub X$\\
$M^\transp$& the transpose of a matrix $M$\\
$\Pr(A)$& the probability of an event $A$\\
$\Ex(X)$& the expected value of a real random variable $X$\\
$\log$& natural (base $e$) logarithm\\
$d\nu/d\mu$& Radon--Nikodym derivative\\
$I$& identity matrix (see also $I$ below)\\
$\mg{S}$& the cardinality of a finite set $S$ (see also $\mg{\,\cdot\,}$ below)
\end{tabular}

\subsection*{Notation defined in the text}

\setlength{\parindent}{0em}%
\begin{multicols}{3}
\nuse{\nref{Adiv}}{$A$}
\nuse{\pageref{eq:Abar-defn}}{$\ovln{A}$}
\nuse{\pageref{p:An}}{$A_n$}
\nuse{\nref{Alg}}{$\Alg$}
\nuse{\nref{Bbardiv}}{$\ovln{B}$}
\nuse{\nref{D}}{$D$}
\nuse{\nref{Dq}}{$D_q$}
\nuse{\nref{reldiv}}{$\reldiv{-}{-}$}
\nuse{\nref{crossdiv}}{$\crossdiv{-}{-}$}
\nuse{\nref{Dmax}}{$D_\textup{max}$}
\nuse{\nref{dimM}}{$\dim_{\textup{M}}$}
\nuse{\nref{Endopd}}{$\End$}
\nuse{\nref{expq}}{$\exp_q$}
\nuse{\nref{freecat}}{$F$}
\nuse{\nref{FinProb}}{$\FinProb$}
\nuse{\nref{Gdiv}}{$G$}
\nuse{\nref{hpoly}}{$h$}
\nuse{\nref{H}, \nref{HRV}, \nref{Hgen}, \nref{Hp}, \nref{entgenp}}{$H$}
\nuse{\nref{Hpp}}{$H_p$}
\nuse{\pageref{eq:defn-renyi}}{$H_q$}
\nuse{\nref{HR}}{$H_\R$}
\nuse{\nref{Hi}}{$\Hi$}
\nuse{\nref{jointent}}{$H(-, -)$}
\nuse{\nref{jcmi}}{$\Hi(-, -)$}
\nuse{\nref{condent}}{$\condent{-}{-}$}
\nuse{\nref{jcmi}}{$\condenti{-}{-}$}
\nuse{\pageref{eq:defn-rel-ent}}{$\relent{-}{-}$}
\nuse{\nref{hrelent}}{$\hrelent{q}{-}{-}$}
\nuse{\pageref{eq:crossent}}{$\crossent{-}{-}$}
\nuse{\nref{relenti}}{$\relenti{-}{-}$}
\nuse{\nref{crossenti}}{$\crossenti{-}{-}$}
\nuse{\nref{HSG}}{$\hsg$}
\nuse{\pageref{eq:fisher-info}, \nref{indicator}}{$I$}
\nuse{\nref{mutinfo}}{$I(-; -)$}
\nuse{\nref{jcmi}}{$I^{\binsym}(-; -)$}
\nuse{\nref{im}}{$\im$}
\nuse{\nref{loss}, \nref{Lp}}{$L$}
\nuse{\nref{lossq}}{$L_q$}
\nuse{\nref{liminf}}{$\liminf$}
\nuse{\nref{lnq}}{$\ln_q$}
\nuse{\nref{mgf}}{$m_X$}
\nuse{\nref{Mtx}}{$M_t(-)$}
\nuse{\pageref{eq:defn-pwr-mn}}{$M_t(-, -)$}
\nuse{\nref{qam}, \pageref{eq:w-u-abbr}}{$M_\phi$}
\nuse{\nref{Ncount}}{$N(-, -)$}
\nuse{\nref{pvec}}{$\p$}
\nuse{\nref{pstar}}{$\muc{\p}$}
\nuse{\nref{Pmx}}{$P$}
\nuse{\nref{Pstar}}{$\muc{P}$}
\nuse{\nref{pi}}{$p_i$}
\nuse{\nref{Pij}}{$P_{ij}$}
\nuse{\nref{Pdotj}}{$P_{\Pdot j}$}
\nuse{\nref{Pbardotj}}{$\ovln{P}_{\Pdot j}$}
\nuse{\nref{monopd}}{$P(M)$}
\nuse{\nref{Px}}{$\Pr(x)$}
\nuse{\nref{fq}}{$\fq{p}$}
\nuse{\pageref{eq:R-defn}}{$R$}
\nuse{\nref{Sq}}{$S_q$}
\nuse{\nref{Sqi}}{$\Si_q$}
\nuse{\nref{srelent}}{$\srelent{q}{-}{-}$}
\nuse{\nref{Set}}{$\Set$}
\nuse{\nref{supp}, \nref{suppgen}, \nref{suppall}}{$\supp$}
\nuse{\nref{Tp}}{$\Tp$}
\nuse{\nref{un}, \nref{unp}}{$\vc{u}_n$}
\nuse{\nref{uset}}{$U_{\XX}$}
\nuse{\nref{Vi}}{$V_i$}
\nuse{\nref{ivellone}}{$V'_i$}
\nuse{\nref{wvec}}{$\vc{w}$}
\nuse{\nref{wstar}}{$\muc{\vc{w}}$}
\nuse{\nref{wj}}{$w_j$}
\nuse{\nref{runningmean}}{$\ovln{X}_r$}
\nuse{\nref{ZB}}{$Z_B$}
\nuse{\nref{Zpsm}}{$(\Zps)^\times$}

\nuse{\nref{indnum}}{$\alpha$}
\nuse{\nref{alphabar}}{$\ovln{\alpha}_j$}
\nuse{\pageref{eq:betaj-exp}}{$\ovln{\beta}_j$}
\nuse{\pageref{eq:gammaj-exp}}{$\gamma_j$}
\nuse{\nref{partial}, \pageref{eq:p-deriv}}{$\partial$}
\nuse{\nref{simpopd}}{$\Delta$}
\nuse{\nref{Deltan}}{$\Delta_n$}
\nuse{\nref{Deltano}}{$\Delta_n^\circ$}
\nuse{\nref{Deltanp}}{$\Delta^{(p)}_n$}
\nuse{\nref{DeltaX}}{$\Delta_{\Xx}$}
\nuse{\nref{lambdaLeb}}{$\lambda$}
\nuse{\nref{Pin}}{$\Pi_n$}
\nuse{\nref{sigmaimplicit}}{$\sigma$}
\nuse{\nref{sigmaq}}{$\sigma_q$}
\nuse{\nref{ECsp}}{$\chi$}
\nuse{\nref{clinum}}{$\omega$}

\nuse{\nref{comp}, \nref{compopd}}{$\circ$}
\nuse{\nref{pfwd}}{$\pi\vc{r}$}
\nuse{\nref{tensorprob}, \pageref{eq:defn-real-tensor}}{$\otimes$}
\nuse{\nref{tensorpwr}}{$(-)^{\otimes d}$}
\nuse{\pageref{eq:p-perm}}{$\p\sigma$}
\nuse{\nref{emp}, \nref{pext}}{$\hat{\ }$}
\nuse{\nref{vecops}}{$\vc{x}\vc{y}$, $\vc{x}/\vc{y}$, etc.{}}
\nuse{\nref{xf}}{$\vc{x}f$}
\nuse{\nref{leq}}{$\leq$}
\nuse{\nref{oplusvec}, \nref{oplusmx}}{$\oplus$}
\nuse{\nref{wq}}{$(-)^{(q)}$}
\nuse{\nref{hash}}{${}^{\#}$}
\nuse{\pageref{eq:mc}}{$\mc$}
\nuse{\nref{dumn}, \nref{algbar}}{$\ovln{\blank}$}
\nuse{\nref{adjc}}{$\adjc$}
\nuse{\nref{Ztheta}}{$Z\theta$}
\nuse{\nref{magmx}, \nref{magcat}, \nref{gauge}, \nref{magcpt}}{$\mg{\,\cdot\,}$}
\nuse{\nref{tA}}{$tA$}
\nuse{\nref{descd}}{$\descd$}
\nuse{\nref{givenRV}}{$X \given y$}
\nuse{\nref{vecofones}, \nref{Oneprob}}{$\One_n$}
\nuse{\pageref{eq:defn-conj}}{$f^*$}
\nuse{\nref{pnorm}}{$\|\cdot\|_p$}
\nuse{\nref{pfwdinj}}{$f_*$}
\nuse{\nref{idopd}, \nref{idopdobj}}{$1_P$}
\nuse{\nref{termopd}}{$\One$}
\nuse{\nref{A0}}{$\blank_0$}
\nuse{\nref{A1}}{$\blank_1$}
\nuse{\nref{XA}}{$X \times \scat{A}$}
\nuse{\pageref{eq:typical-FP-map}}{$\Fmap{\cdots}{\theta}$}
\nuse{\nref{toterm}}{$!$}
\nuse{\nref{pfwdgen}}{$f\vc{s}$}
\nuse{\nref{bigccdists}}{$\coprod$}
\nuse{\nref{ccmaps}, \nref{smallccmapsp}}{$\cpd$}
\nuse{\nref{sterling}}{$\pounds_1$}

\end{multicols}

%% file: see.tex
\index{Tsallis, Constantino!entropy|see{$q$-logarithmic entropy}}
\index{mean!power|see{power mean}}
\index{diversity!similarity-sensitive|see{similarity-sensitive diversity}}
\index{lpnorm@$\ell^p$ norm|see{$p$-norm}}
\index{quermassintegral|see{intrinsic volume}}
\index{Minkowski, Hermann!functional|see{intrinsic volume}}
\index{generalized mean|see{power mean}}
\index{Kullback--Leibler distance|see{relative entropy}}
\index{divergence|see{relative entropy}}
\index{information gain|see{relative entropy}}
\index{cross entropy|seealso{relative entropy}}
\index{diversity measure|seealso{Hill number}}
\index{unweighted mean|see{mean, unweighted}}
\index{weighted mean|see{mean, weighted}}
\index{q@$q$|see{order \textit{and} viewpoint parameter}}
\index{homology|seealso{magnitude homology}}
\index{norm|seealso{$p$-norm}}
\index{permutation-invariance|seealso{symmetric \textit{and} symmetry}}
\index{symmetric|seealso{permutation-invariance}}
\index{taxicab metric|see{$\ell^1$ geometry}}
\index{independence|seealso{independent}}
\index{relative entropy!Renyi@R\'enyi|see{R\'enyi relative entropy}}
\index{relative entropy!q-logarithmic@$q$-logarithmic|see{$q$-logarithmic relative entropy}}
\index{tree|seealso{forest}}
\index{operad!topological|see{topological operad}}
\index{algebra for operad!categorical|see{categorical algebra for operad}}
\index{Leibniz rule|see{derivation}}
\index{diversity|seealso{Hill number}}
\index{grouping rule|seealso{chain rule}}
\index{recursivity|seealso{chain rule}}
\index{value!measure|see{value measure}}